\documentclass[a4paper, 11pt, oneside]{book} % fleqn % twoside % openright
\usepackage{mystyle2}

\begin{document}

\thispagestyle{empty}
\includepdf[pages=1,pagecommand={},offset=1.6cm -1.2cm,width=1.35\textwidth]{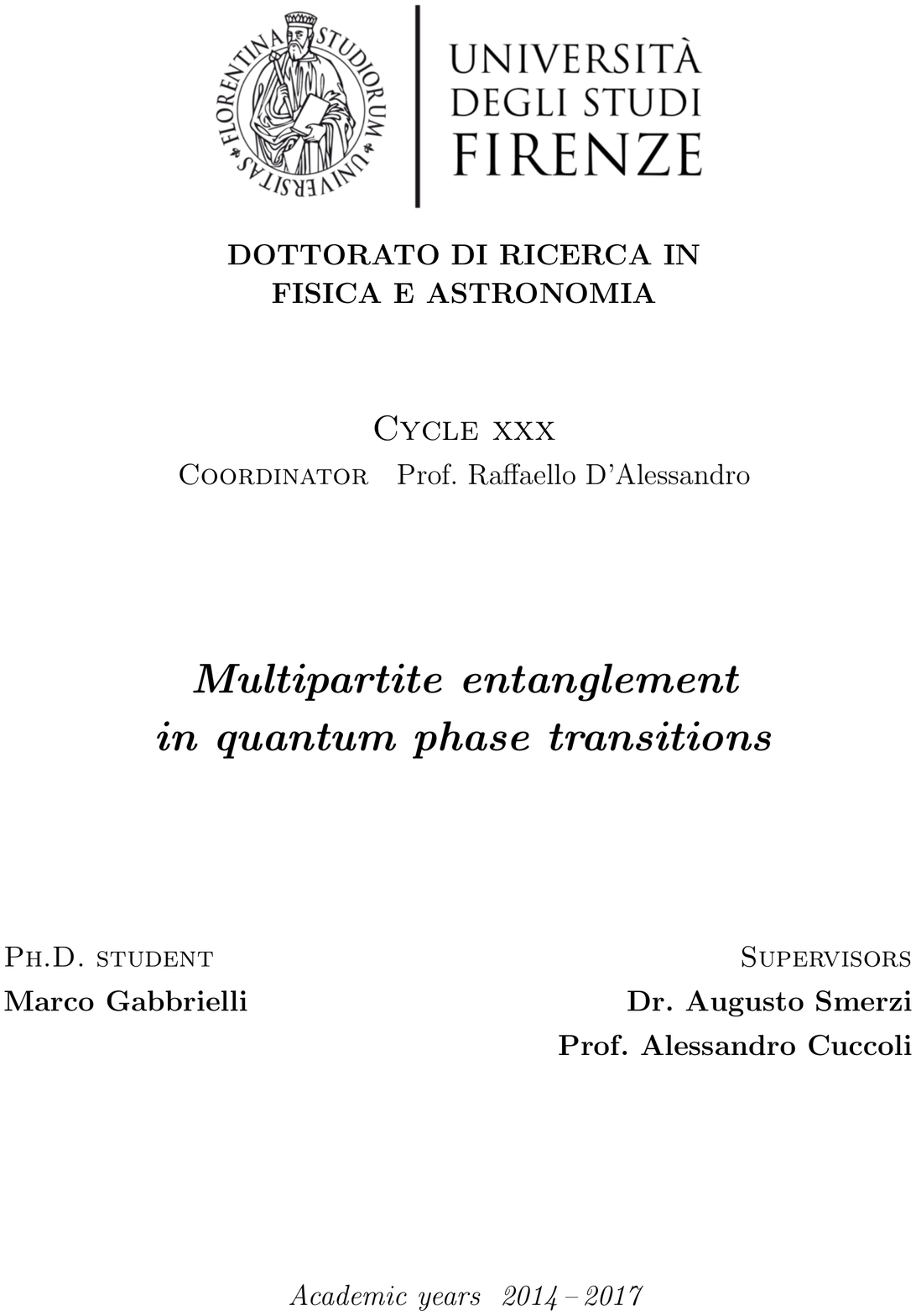}
\newpage

\frontmatter
{\hypersetup{linkcolor=black!45!blue} \tableofcontents}
      
\newpage
\thispagestyle{empty}

\begin{table}[!h] 
\begin{center}
\begin{tabular}{lcl}
\hline
\raisebox{0pt}[15pt]{\textbf{\small{Symbol}}} &  & \textbf{\small{Meaning}} \\[3pt]
\hline\hline
\raisebox{0pt}[15pt]{} $\hat a_k, \ \hat a_k^\dag$  & & annihilation and creation operators of a particle in mode $k$ \\
\raisebox{0pt}[15pt]{} $\alpha$  & & range of the two-body interaction \\
\raisebox{0pt}[15pt]{} $\beta$ & & order-parameter critical exponent \\
\raisebox{0pt}[15pt]{} $C^{\,(i,j)}$ & & two-body correlation function between sites $i$ and $j$ \\
\raisebox{0pt}[15pt]{} $\chi$ & & fidelity susceptibility \\
\raisebox{0pt}[15pt]{} $\Delta$ & & lowest nonvanishing energy separation in the many-body spectrum \\
\raisebox{0pt}[15pt]{} $\eta$ & & Fisher critical exponent \\
\raisebox{0pt}[15pt]{} $\hat \eta_k, \ \hat \eta_k^\dag$  & & annihilation and creation operators of a quasiparticle in mode $k$ \\
\raisebox{0pt}[15pt]{} $\pazocal{F}$ & & quantum fidelity \\
\raisebox{0pt}[15pt]{} $F_Q$ & & quantum Fisher information \\
\raisebox{0pt}[15pt]{} $f_Q$ & & quantum Fisher density \\
\raisebox{0pt}[15pt]{} $\mathbb{F}_Q$ & & quantum Fisher matrix \\
\raisebox{0pt}[15pt]{} $\gamma$ & & susceptibility critical exponent \emph{or} anisotropy of the LMG model \\
\raisebox{0pt}[15pt]{} $\pazocal{H}$  & & Hilbert space of a quantum system \\
% \raisebox{0pt}[15pt]{} $\pazocal{H}_\mathsf{A}$  & & Hilbert space associated to the quantum system \textsf{A} \\
\raisebox{0pt}[15pt]{} $\hat{H}$  & & Hamiltonian operator of a quantum system \\
\raisebox{0pt}[15pt]{} $J$ & & microscopic energy scale \\
\raisebox{0pt}[15pt]{} $\hat{\mathbf{J}}$ & & collective pseudospin vector operator \\
\raisebox{0pt}[15pt]{} $\lambda$ & & coupling constant \\
\raisebox{0pt}[15pt]{} $\mu$ & & degeneracy of the {1}$^{\rm st}$ energy eigenstate \emph{or} chemical potential \\
\raisebox{0pt}[15pt]{} $n$ & & quantum number labelling the energy levels \\
\raisebox{0pt}[15pt]{} $\bf{n}$ & & unit vector \\
\raisebox{0pt}[15pt]{} $N$  & & number of parties of a quantum system \\
\raisebox{0pt}[15pt]{} $\nu$ & & degeneracy of the {2}$^{\rm nd}$ energy eigenstate \emph{or} correlation critical exponent \\
\raisebox{0pt}[15pt]{} $\ket{\psi}$ & & % vector in the Hilbert space $\pazocal{H}_\mathsf{A}$ 
% pure state describing the quantum system \textsf{A} \\
ket describing the pure state of a quantum system \\
\raisebox{0pt}[15pt]{} $\hat{\rho}$ & & density matrix describing the mixed state of a quantum system \\
\raisebox{0pt}[15pt]{} $\pazocal{S}$ & & dynamic structure factor \\
\raisebox{0pt}[15pt]{} $T$ & & temperature \\
\raisebox{0pt}[15pt]{} $W$ & & winding number \\
\raisebox{0pt}[15pt]{} $\hat{W}$  & & entanglement witness \\
\raisebox{0pt}[15pt]{} $\xi$ & & correlation length \\
\raisebox{0pt}[15pt]{} $\xi^2_{\rm R}$ & & Wineland spin-squeezing parameter \\
\raisebox{0pt}[15pt]{} $z$ & & dynamic critical exponent \emph{or} population imbalance \\
% \raisebox{0pt}[15pt]{} $\theta$ & & generic angle ... interaction parameter \\
% \raisebox{0pt}[15pt]{} $\varphi$ & & azimuthal angle  ... order parameter \\
\raisebox{0pt}[15pt]{} $\pazocal{Z}$ & & canonical partition function \\[3pt]
\hline
\end{tabular}
\caption*{\textit{List of some symbols frequently used in the text}}
\end{center}
\end{table}

\clearpage
\thispagestyle{empty}

\begin{table}[!h] 
\begin{center}
\begin{tabular}{lcl}
\hline
\raisebox{0pt}[15pt]{\textbf{\small{Abbreviation}}} &  & \textbf{\small{Definition}} \\[3pt]
\hline\hline
\raisebox{0pt}[15pt]{} AFM & & antiferromagnetic \\
\raisebox{0pt}[15pt]{} BEC  & & Bose-Einstein condensate \emph{or} Bose-Einstein condensation \\
\raisebox{0pt}[15pt]{} BJJ & & bosonic Josephson junction \\
\raisebox{0pt}[15pt]{} BS & & beam splitter \emph{or} beam splitting \\
\raisebox{0pt}[15pt]{} CPT & & classical phase transition \\
\raisebox{0pt}[15pt]{} DMRG & & density-matrix renormalization group \\
\raisebox{0pt}[15pt]{} FM & & ferromagnetic \\
\raisebox{0pt}[15pt]{} GHZ & & Greenberger-Horne-Zeilinger \\
\raisebox{0pt}[15pt]{} LMG & & Lipkin-Meshkov-Glick \\
\raisebox{0pt}[15pt]{} PS & & phase shift \\
\raisebox{0pt}[15pt]{} PM & & paramagnetic \\
\raisebox{0pt}[15pt]{} QC & & quantum criticality \emph{or} quantum critical \\
\raisebox{0pt}[15pt]{} QFI  & & quantum Fisher information \\
\raisebox{0pt}[15pt]{} QPT & & quantum phase transition \\
\raisebox{0pt}[15pt]{} SN & & shot noise \\
\raisebox{0pt}[15pt]{} WSS & & Wineland spin squeezing \\[3pt]
\hline
\end{tabular}
\caption*{\textit{Table of the acronyms and abbreviations used throughout the text}}
\end{center}
\end{table}

\thispagestyle{empty}

\chapter{Abstract}
%\section*{\huge Abstract}

The characterization of quantum phases and quantum phase transitions through entanglement measures and witnesses 
is an intriguing problem at the verge of quantum information and many-body physics.
The study of critical systems from the perspective of information science advances our understanding of criticality 
beyond standard approaches developed in statistical mechanics, 
and sheds new light on the preparation and processing of useful resources for quantum technologies.
%%%
Current studies have mainly focused on bipartite or pairwise entanglement in the ground state of critical Hamiltonians: 
these studies have emphasized a growth of entanglement in the vicinity of quantum critical points. 
However, bipartite and pairwise correlations are hardly accessible in systems with a large number of particles, 
that are the preferred platforms for quantum sensors and the natural targets of quantum simulators.
Moreover, they cannot fully capture the richness of multiparticle correlations
and the complex structure of a many-body quantum state.

Much less attention has been devoted to witnessing multipartite entanglement in critical systems 
and it has been mainly limited to spin models.
Yet, multipartite entanglement among hundreds of particles has been detected experimentally in atomic ensembles so far,
and a variety of witnesses are available in the literature.
Among these witnesses, the quantum Fisher information has proved to be especially powerful:
it extends the class of entangled states detectable by popular methods such as the spin squeezing, 
it can be extracted from experimental data and 
it has an appealing physical meaning in terms of distinguishability of quantum states under external parametric transformations.
Here, we investigate the behaviour of multipartite entanglement as detected and quantified by the quantum Fisher information, 
both at zero and finite temperature, in a collection of benchmark models displaying quantum phase transitions. 
All the selected models describe exemplary systems in the field of condensed-matter or nuclear physics 
and can be realized and tested in the laboratory as well.

Most of the results presented here are based on the scientific papers \cite{Gabbrielli2018b}, \cite{Gabbrielli2018a} and \cite{Pezze2017}
and constitute the main work carried out during my three-year--long PhD experience.
Another line of research, undertaken during the first year in continuity with my Master's degree thesis, 
explored the feasibility of a new scheme for quantum-enhanced nonlocal atom interferometry.
This work yielded the publication \cite{GabbrielliPRL2015} but finds no space in this venue.
% nonostante un lavoro recente abbia connesso QPT and SMD.
% AND COLLABORATIONS under preparation or evaluation ...

% All of this research is in full progress.

\mainmatter
\chapter[Introduction]{Introduction} \label{ch:Intro}
% Introduction on \\ multipartite entanglement
% Introduction on \\ quantum phase transitions
% Introduction on \\ multipartite entanglement \\ and quantum phase transitions

{\sl The investigation of multipartite entanglement close to quantum phase transitions provides a novel perspective 
for understanding the role of quantum correlations in interacting critical matter, 
either confined in its ground state or in the presence on thermal excitations.
% in interacting critical matter in phases and on their border.
We open this chapter with a bird's eye view on basic concepts about entanglement and phase transitions.
Then, we mix these two ingredients and describe our key results by presenting the behaviour 
of the quantum Fisher information in different phases of general models.} % and at criticality.}

\section[Overview on entanglement]{Overview on entanglement}  \label{sec:MultipartiteEntanglement}
% Compendium/Synopsis on entanglement

The concept of quantum entanglement was introduced by Schr\"odinger in the mid Thirties of the last century
and it was immediately indicated as the fundamental responsible 
for the divergence from the classical lines of reasoning~\cite{Schroedinger1935}.
Entanglement is a purely quantum feature of Nature, having no classical counterpart. % as well as the quintessence of the quantum theory 
Even though its phenomenology originally appeared arcane and its interpretation problematic~\cite{EinsteinPR1935}, 
entanglement is nowadays conceptually and operationally well understood in the framework of quantum mechanics~\cite{HorodeckiRMP2009}:
formally, it directly stems from the superposition principle of quantum states.
In its simplest manifestation, it arises whenever the interaction between two physical systems 
generates some correlations that are conserved even after the interaction is turned off and the two systems are spatially separated,
and they are responsible for the instantaneous state collapse of one system as a consequence of a local measurement on the other one.
%Entanglement has been validated in many ways and in different contexts. observed by numerous experiments.

In the last thirty years, a conceptual revolution has generated a new standpoint in the study of entanglement: 
not only it is regarded as an amazing peculiar aspect of reality at the quantum level, 
but it also has been recognized as a potential resource for improving the performance of certain classical tasks
or even for performing tasks that cannot be carried out classically~\cite{nielsen2000}.
Entanglement-based technologies range from reliable secure data transmission (quantum cryptography)
to the futuristic proposal of computing machines that are faster than any other classical machine (quantum computer),
passing through the design of measurement schemes with precision at the limit imposed by 
the fundamental laws of the microscopical world (quantum metrology and sensing). 

Owing to its crucial role in physics as well as its intellectual implications,
and in view of the emerging applications in the field of quantum technologies, 
it is strikingly important to mathematically characterize, classify and quantify entanglement, 
and experimentally produce entangled states in physical platforms available in the laboratory.
In this section, we intend to provide a basic introduction on entanglement,
starting from the simplest case of bipartite entanglement (paragraph \ref{subsec:BipartiteEntanglement}) 
and then moving to the more general multipartite scenario 
(paragraphs~\ref{subsec:MultipartiteEntanglement} and \ref{subsec:QuantumFisherInformation}).
In particular, we will linger on its application in the field of quantum interferometry (paragraph~\ref{subsec:QuantumInterferometry}). 
%% For this brief review, we will mainly follow the exposition done Refs.~\cite{pezze2014,PezzeRMP}
%% The questions whether a state is entangled and to which extent 

\subsection{Bipartite entanglement} \label{subsec:BipartiteEntanglement}
Let us consider a composite system of two independent quantum physical systems, labelled by \textsf{A} and \textsf{B},
described by states in the Hilbert spaces $\pazocal{H}_\mathsf{A}$ and $\pazocal{H}_\mathsf{B}$ respectively.
In the Dirac--von\;Neumann formulation of quantum mechanics~\cite{Jauch1968}, it is simply postulated that
the state space of the composite system is the tensor product of the states associated to the two components: $\pazocal{H}_\mathsf{AB}=\pazocal{H}_\mathsf{A}\otimes\pazocal{H}_\mathsf{A}$.
If subsystems \textsf{A} and \textsf{B} are in pure states $\ket{\psi^{(\mathsf{A})}}$ and $\ket{\psi^{(\mathsf{B})}}$, 
the composite system is described by the pure \emph{product} state
$\ket{\psi^{(\mathsf{AB})}}=\ket{\psi^{(\mathsf{A})}}\otimes\ket{\psi^{(\mathsf{B})}}$.
Such a state is an example of a separable (namely, factorizable) state: 
any local operation acting on each subsystem does not affect the other one. 
A more general definition is possible when the two subsystems communicate with each other:
this communication creates classical correlations such that \textsf{A} and \textsf{B}
occupy the states $\ket{\psi^{(\mathsf{A})}_k}$ and $\ket{\psi^{(\mathsf{B})}_k}$ with a certain probability $p_k$.
Thus, a mixed state of the composite system is called \emph{separable} (or classically correlated) if it can be written
as a convex combination of product states~\cite{WernerPRA1989}
\be \label{BipartiteDefinition}
\hat{\rho}^{(\mathsf{AB})}_{\rm sep}=\sum_k p_k\,\ket{\psi^{(\mathsf{A})}_k} \bra{\psi^{(\mathsf{A})}_k} \otimes \ket{\psi^{(\mathsf{B})}_k} \bra{\psi^{(\mathsf{B})}_k} \, .
\ee
Any state of the composite system that cannot be written in the above form is called \emph{entangled}:
the subsystems share quantum correlations which can be stronger than classical correlations,
since they cannot be created nor destroyed by local operations and classical communication. 
These quantum correlations go by the name of ``entanglement''.

\paragraph{Criteria and measures} 
In the bipartite scenario, several necessary and sufficient conditions for \emph{qualifying} entangled state are available.
Moreover, a vast number of ``entanglement measures'' can be defined in order to \emph{quantify} entanglement: 
loosely speaking, they are continuous real functions of the state that quantify ``how much entanglement'' is contained in the state;
more formally, they are zero for separable states and positive for entangled states,
are invariant under unitary local transformations performed on the single subsystems and
cannot increase under local operations and classical communication performed on the composite system.

For entangled \emph{pure} states $\ket{\psi^{(\mathsf{AB})}}\neq\ket{\psi^{(\mathsf{A})}}\otimes\ket{\psi^{(\mathsf{B})}}$, 
an operative way to check if the correlations between the two parties have quantum (or only classical) nature
is observing (or not) the violation of Bell inequalities, certifying the existence of some form of nonlocality \cite{ClauserPRL1969}.
The natural expectation that entanglement is related to the degree of state mixedness 
of the individual subsystems has led to the definition of the ``entanglement entropy'' 
as the von Neumann entropy of the reduced density matrix~\cite{BennettPRA1996}.
%$S_\mathsf{AB}=$

For \emph{mixed} state, instead, the issue of qualifying bipartite entanglement is much more complicated:
without the complete knowledge of the preparation of the state, we are not able to unambiguously decide whether correlations 
between the parties are quantum (arising from a quantum interaction) or classical (generated by the preparation).
A mixed state would be labelled as entangled only if there is no possible way to engineer it using local classical preparation, 
but a similar test is not practicable. Fortunately, some operational criteria can be found~\cite{Adesso2007}.
Many measures have been developed as well, according to the specific system and task under consideration~\cite{PlenioQIC2007}:
tangle, concurrence, entanglement of formation, logarithmic negativity, just to cite few of them.

\subsection{Multipartite entanglement} \label{subsec:MultipartiteEntanglement}
The definition of separability and entanglement can be easily generalized to the case of systems 
composed by $N\geq2$ distinguishable parties labelled by the index $i=1,\,2,\dots N$.
A quantum state in the global Hilbert space $\pazocal{H}_1\otimes\pazocal{H}_2\otimes\dots\pazocal{H}_N$
is said to be \emph{fully separable} when it can be expressed as a statistical mixture of product states
\be \label{MultipartiteDefinition}
\hat{\rho}_{\rm sep} = \sum_k p_k\,\ket{\psi_{{\rm prod},\,k}} \bra{\psi_{{\rm prod},\,k}} = 
\sum_k p_k \bigotimes_{i=1}^N \ket{\psi_k^{(i)}} \bra{\psi_k^{(i)}} \, ,
\ee
where $\ket{\psi_{{\rm prod},\,k}}=\ket{\psi_k^{(1)}}\otimes\ket{\psi_k^{(2)}}\otimes\dots\ket{\psi_k^{(N)}}$ 
is the full product state (occurring with probability $p_k$ in the preparation) 
and $\ket{\psi_k^{(i)}}$ is the pure state of the $i$th party.
States that cannot be cast in the form (\ref{MultipartiteDefinition}) are \emph{entangled}~\cite{GuhnePR2009}.
%\be \label{MultipartiteDefinition}
%\hat{\rho}_{\rm sep} = \sum_k p_k\,\hat{\rho}^{(1)}_k\otimes\hat{\rho}^{(2)}_k\otimes\dots\hat{\rho}^{(N)}_k  \, ,
%\ee
Differently from the bipartite case $N=2$, where any state is either separable or entangled, 
the classification of multipartite entanglement is much richer and it is associated with 
the largest subset of parties whose state cannot be factorized, 
also known in literature as the ``entanglement depth''~\cite{sorensenPRL2001}, as described below.
% Only a fraction of the N parties are in a entangled state 

\paragraph{Classification}
A possibility for classifying multipartite entangled state relies on the factorizability of the many-party state into smaller parts.
In particular, the concept of $\kappa$-separability permits to discriminate among the multipartite correlations present in the system.
We want to outline here the essential definitions.
A pure state of $N>2$ parties is called $\kappa$-\emph{producible}~\cite{GuhneNJP2005,GuhnePR2009} 
if there exist a decomposition of the $N$ parties into $m\leq N$ clusters such that
$\ket{\psi_{\kappa\textrm{-}{\rm prod}}}=\ket{\psi_{N_1}}\otimes\ket{\psi_{N_2}}\otimes\dots\ket{\psi_{N_m}}$,
where $\ket{\psi_{N_l}}$ is the state of $N_l\leq\kappa$ parties and $\sum_{l=1}^m N_l=N$.
Note that a $\kappa$-producible state is the factorization of the $m$ states for the $m$ clusters;
each cluster may still contain some entanglement. 
The extension of this definition to states containing classical correlations is straightforward:
a mixed state is called $\kappa$-\emph{separable} whenever it is a statistical mixture of $\kappa$-producible states
\be \label{kappaSeparable}
\hat{\rho}_{\kappa\textrm{-}{\rm sep}} = \sum_k p_k\,\ket{\psi_{\kappa\textrm{-}{\rm prod},\,k}}
\bra{\psi_{\kappa\textrm{-}{\rm prod},\,k}} \, .
\ee
A state is $\kappa$-\emph{partite entangled} if it is $\kappa$-separable but not $(\kappa\!-\!1)$-separable.
In other words, it has the form of a classical mixture of partially-factorized pure states
which contain at least one $\kappa$-party state that does not factorize.
For example, a pure $\kappa$-partite entangled state can be written as 
$\ket{\psi_{\kappa\textrm{-}{\rm ent}}}=\bigotimes_{l=1}^m\ket{\psi_{N_l}}$
with at least one state $\ket{\psi_{N_l}}$ of $N_l=\kappa$ parties that cannot be further factorized.
Thus, $\kappa$ is the number of entangled parties in the largest nonseparable cluster: 
we say that the $\kappa$-partite entangled state (or the overall system described by it)
has an entanglement depth larger than $\kappa-1$.
For instance, the 1-producible state is the fully separable state in Eq.~(\ref{MultipartiteDefinition});
for $\kappa=N$, instead, we have the opposite limit of genuine $N$-partite entangled (or \emph{maximally entangled}) state;
in the special case $\kappa=2$, we retrieve bipartite entanglement.

\paragraph{Witnessing entanglement}
A definitive characterization and quantification of multipartite entanglement, both for pure and mixed states, 
is currently much less satisfactory than in the bipartite scenario and cannot be considered completely reached yet~\cite{Osterloh2007}.
In fact, for $N>3$ a separability criterion, able to specify if a given general $N$-party state is separable or not, is still missing.
For pure states, a canonical way to circumvent this lack is quantifying entanglement by means of a geometric measure~\cite{WeiPRA2003}, 
that is zero for fully separable state and monotonically increases up to its maximum value for $N$-partite entangled states.
Essentially, it provides the distance of the tested state from the closest separable state in the Hilbert space,
according to the metric induced by the inner product. Yet, the evaluation of the geometric measure of entanglement is a demanding task.

Alternative approaches have been chased for testifying the presence of entanglement in a state (and possibly quantify it). 
One of these is based on the violation of certain bounds posed on the expectation value of suitable operators, 
called ``entanglement witnesses''~\cite{SperlingPRL2013}. 
A witness is a bounded observable $\hat{W}$ such that $\tr\big(\hat{\rho}\,\hat{W}\big)\leq b$ for all separable states $\hat{\rho}$
(with $b$ a number that depends on the choice of $\hat{W}$). 
Thus, the violation $\tr\big(\hat{\rho}\,\hat{W}\big)>b$ certifies that the state under consideration is entangled.
We notice that the condition $\tr\big(\hat{\rho}\,\hat{W}\big)\leq b$ does not guarantee that $\hat{\rho}$ is separable.
This method relies on the fact that, for each multipartite entangled state, 
there always exist at least one entanglement witness that recognizes it~\cite{HorodeckiPRA1996}
and provides a powerful tool from an experimental point of view: with some preliminary knowledge of the state, 
it is possible to systematically track down an experimentally accessible witness~\cite{BarbieriPRL2003,BourennanePRL2004}.
The major drawback is that entanglement witnesses are device-dependent: experimental imperfections may lead to false positives.

A different approach to detect entanglement (and quantify its depth) based on the violation of certain bounds 
but not requiring the construction of a witness operator has been proposed~\cite{PezzePRL2009,HyllusPRA2012,TothPRA2012,PezzePNAS2016}.
It exploits the quantum Fisher information.

\subsection{Quantum Fisher information} \label{subsec:QuantumFisherInformation}
Suppose we have prepared a system in the quantum state $\hat{\rho}_0$ and we perturb it by means of 
a generic transformation $\hat{\mathcal{T}}(\phi)$ depending on the real parameter $\phi$, such that 
$\hat{\mathcal{T}}\,$:~$\hat{\rho}_0\mapsto\hat{\rho}_\phi$.
After choosing an observable $\measure_\varepsilon$, whose eigenvalues are denoted as $\varepsilon$'s,
we can perform a measurement on the transformed state $\hat{\rho}_\phi$.
The conditional probability of finding the outcome\footnote{\ In the simplest case, $\varepsilon$ is the eigenvalue of $\measure_\varepsilon$.
In a wider sense, the generalized quantum measurement is described by a set of positive observables $\{\measure(\varepsilon)\}$ 
parametrized by $\varepsilon$, the so-called ``positive operator valued measure''~\cite{nielsen2000}.\\[-9pt]} 
$\varepsilon$ for a given value of $\phi$ 
is given by the \emph{likelihood} $P(\varepsilon|\phi)=\tr\big(\hat{\rho}_\phi\measure_\varepsilon\big)$.
The amount of information on the parameter $\phi$ that we can extract from the measure is quantified by the 
``Fisher information''~\cite{Fisher1925}
\be \label{classicalFisherInformation}
F(\phi) = \sum_\varepsilon \frac{1}{P(\varepsilon|\phi)} \left( \frac{\partial P(\varepsilon|\phi)}{\partial \phi} \right)^2 \, ,
\ee
that depends on the initial state $\hat{\rho}_0$, the value of the parameter $\phi$, the way it is encoded into the transformed state by
the transformation $\hat{\mathcal{T}}$ and the chosen observable $\measure_\varepsilon$.
In Eq.~(\ref{classicalFisherInformation}), the sum extends over all possible measurement results $\varepsilon$.

Physically, the Fisher information is intimately related to the degree of statistical distinguishability 
between the two states $\hat{\rho}_0$ and $\hat{\rho}_\phi$ -- provided the measurement of $\measure$ -- 
when the perturbation $\phi$ is small. For a simple heuristic proof, we recall the canonical Hellinger distance 
$\ud_{H}^2(P_0,P_\phi)=1-\sum_\varepsilon\sqrt{P(\varepsilon|0)P(\varepsilon|\phi)}$ 
between two probability distributions $P(\varepsilon|0)$ and $P(\varepsilon|\phi)$ of observing the result $\varepsilon$ 
after a measurement of $\measure_\varepsilon$ on $\hat{\rho}_0$ and $\hat{\rho}_\phi$ respectively~\cite{Hellinger1909}: 
it is straightforward to show that, at least for unitary transformations $\hat{\mathcal{T}}\hat{\mathcal{T}}^{-1}=\identity$ 
and small $\phi\ll1$, we have
\be \label{HellingerDistance}
\ud_H^2(P_0,P_\phi) = \frac{F(\phi)}{8}\phi^2 + \pazocal{O}(\phi^3) \, .
\ee
Equation~(\ref{HellingerDistance}) reveals that the Fisher information $F(\phi)=8\,\big(\partial\ud_H/\partial\phi\big)^2$ 
can be understood as the square of the \emph{statistical speed} of variation of the distribution probability 
when tuning the parameter $\phi$~\cite{PezzeRMP}. This ``kinematic interpretation'' permits to extract the Fisher information 
from experimental data~\cite{StrobelSCIENCE2014,PezzePNAS2016}.

\paragraph{Definition and main properties} 
An upper bound to the Fisher information can be derived by maximizing Eq.~(\ref{classicalFisherInformation}) 
over all possible measurements~\cite{BraunsteinPRL1994}: 
\be \label{quantumFisherInformation}
F(\phi) \,\leq\, \max_{\measure} F(\phi) \,\equiv\, F_Q[\hat{\rho}_\phi] \, .
\ee
The result of the maximization is called ``quantum Fisher information'' (henceforth QFI). 
It depends on the initial state $\hat{\rho}_0$ and the transformation $\hat{\mathcal{T}}$ through the transformed state $\hat{\rho}_\phi$. 
In the simple case of unitary transformations, 
it can also be related to a state distance in the Hilbert space, namely the Bures distance~\cite{Bures1969}
and operationally interpreted as the maximal degree of distinguishability of the two states $\hat{\rho}_0$ and $\hat{\rho}_\phi$.

Moreover, it can be shown~\cite{Helstrom1967} that the QFI can be cast in the form 
\be \label{quantumFisherInformationSLD}
F_Q[\hat{\rho}_\phi] = \tr \big(\hat{\rho}_\phi \hat{L}_\phi^2 \big) \, , % \qquad {\rm with} \quad \frac{\partial\hat{\rho}_\phi}{\partial\phi} = \frac{\hat{\rho}_\phi\,\hat{L}_\phi+\hat{L}_\phi\,\hat{\rho}_\phi}{2} \, ,
\ee
where the Hermitian operator $\hat{L}_\phi$ is the solution of the equation 
$\partial_\phi\hat{\rho}_\phi = \frac{1}{2}\big\{\hat{\rho}_\phi,\hat{L}_\phi\big\}$ 
and is usually known as the ``symmetric logarithmic derivative''. 
Note that the theoretical form of the optimal measurement that permits to saturate the equality in Eq.~(\ref{quantumFisherInformation}) 
is known~\cite{BraunsteinPRL1994} and involves the projectors onto the eigenstates of $\hat{L}_\phi$.
Finally, the QFI is found to be \emph{convex} in the states~\cite{pezze2014}:
\be F_Q\Big[\sum_k p_k\,\hat{\rho}_{\phi,\,k}\Big]\leq\sum_k p_k F_Q[\hat{\rho}_{\phi,\,k}]
\ee
as long as $p_k\geq0$ and $\sum_k p_k = 1$.

In the special case of a \emph{unitary transformation} $\hat{\mathcal{T}}(\phi)=\neper^{-\ii\phi\hat{O}}$ 
generated by the observable $\hat{O}$, an explicit expression for the QFI can be obtained~\cite{BraunsteinPRL1994} 
in terms of the eigenvalues $p_k$ and eigenvectors $\ket{k}$ of the initial state $\hat{\rho}_0=\sum_k p_k\ket{k}\bra{k}$:
\be \label{QFIunitaryMixed}
F_Q\big[\hat{\rho}_0,\hat{O}\big] = 2 \, \sum_{k\,k'} \frac{(p_k-p_{k'})^2}{p_k+p_{k'}}\,\Big|\bra{k} \hat{O} \ket{k'} \Big|^2 \, ,
\ee
where the sum is limited to indices $k$ and $k'$ such that $p_k+p_{k'}\neq0$. 
Notably, the QFI does not depend on the amplitude $\phi$ of the transformation: 
it only depends on the initial state and the generator of the transformation. 
If the initial state is pure $\hat{\rho}_0=\ket{\psi_0}\bra{\psi_0}$, Eq.~(\ref{QFIunitaryMixed}) further simplifies into 
the variance of the operator $\hat{O}$ on the pure state:
\be \label{QFIunitaryPure}
F_Q\big[\ket{\psi_0},\hat{O}\big] = 4\,\big(\Delta\hat{O}\big)^2 = 4 \left[ \bra{\psi_0} \hat{O}^2 \ket{\psi_0} - \bra{\psi_0} \hat{O} \ket{\psi_0}^2 \right] \, .
\ee
In the following we will always consider unitary transformations.

In principle, the QFI is experimentally accessible. A Kubo-like formula relates the QFI to a measurable 
response function~\cite{HaukeNATPHYS2016}: in particular, the QFI is given by a weighted integral 
-- across the full spectrum of frequencies -- of the dissipative part of the dynamic susceptibility 
of the initial state $\hat{\rho}$ with respect to the generator $\hat{O}$, see Eq.~(\ref{IsingDynamicFactor}).
Alternatively, the statistical distance method~(\ref{HellingerDistance}) for the Fisher information 
and the lower bound in definition~(\ref{quantumFisherInformation}) provide a way to deduce the qualitative scaling behaviour of the QFI 
when increasing the number of parties $N$: if $F(\phi)$ diverges for $N\to\infty$, then $F_Q[\hat{\rho}_\phi]$ diverges as well. 
Other measurable lower bounds to the QFI have been discussed in Refs.~\cite{FrerotPRB2016,ApellanizPRA2017}.

\subsection{Upper bounds for separable states} \label{QFIbounds}
We want to establish the relation between the QFI and the criteria for recognizing multipartite entanglement.
How can the QFI come to our help in the difficult task of witnessing entangled states?
We consider a system composed by $N$ \emph{distinguishable} parties and assume, 
for simplicity but with no stern loss of generality, that all the parties share the \emph{same} \emph{finite}-dimensional Hilbert space 
($\pazocal{H}_i=\pazocal{H}$ $\forall\,i=1,\dots N$ and $\dim\pazocal{H}<\infty$).
Thus, we are requiring that the constituents of the system -- among which we are interested to detect entanglement, if present --
have equal physical nature and their degrees of freedom have finite spectra.
In general, the convexity of the QFI permits to state that
$F_Q\big[\hat{\rho}_{\kappa\textrm{-}{\rm sep}},\hat{O}\big] \leq b_{\kappa,\,\hat{O}}$ for any $\kappa$-separable state. 
The bound $b_{\kappa,\,\hat{O}}$ is the maximal variance $(\Delta\hat{O})^2$ 
over all the pure $\kappa$-producible states $\ket{\psi_{\kappa\textrm{-}{\rm prod}}}$,
and its calculation for a given operator $\hat{O}$ is still an open theoretical challenge.
Notwithstanding, the closed expression of $b_{\kappa,\,\hat{O}}$ can be found for some special (and practically relevant) 
classes of probe operators $\hat{O}$.

\paragraph{Local probe operators}
Suppose to act on the system via a unitary transformation 
$\hat{\mathcal{T}}(\phi)=\neper^{-\ii\phi\hat{O}}=\bigotimes_{i=1}^N\neper^{-\ii\phi\hat{o}^{(i)}}$, 
generated by a \emph{collective local} observable $\hat{O}=\sum_{i=1}^N\hat{o}^{(i)}$ 
(meaning linear in the local single-party operators $\hat{o}^{(i)}$), 
and suppose that the operator $\hat{o}^{(i)}$ acting on the $i$th party 
is the same observable for all the parties ($\hat{o}^{(i)}=\hat{o}$ $\forall\,i=1,\dots N$). 
Thereby, each party is perturbed individually, and the perturbation $\hat{o}$ and its amplitude $\phi$ are the same on each party.
The requirement of locality for $\hat{O}$ is important for the unitary operation $\hat{\mathcal{T}}$
not to introduce fictional correlations previously not present in the system.

If the state of the system is the fully separable state in Eq.~(\ref{MultipartiteDefinition}), 
the convexity of the QFI allows to set the following upper bound~\cite{PezzePRL2009, HyllusPRA2012, TothPRA2012}:
\be \label{boundSeparable}
F_Q\big[\hat{\rho}_{\rm sep},\hat{O}\big] \leq \sum_k p_k F_Q\big[\ket{\psi_{{\rm prod},\,k}},\hat{O}\big]
= \sum_{k,\,i} p_k F_Q\big[\ket{\psi_k^{(i)}},\hat{O}\big] \leq N(o_{\rm max}-o_{\rm min})^2 \, ,
\ee
where $o_{\rm min}$ and $o_{\rm max}$ are the minimum and maximum eigenvalues of the operator $\hat{o}$; 
the middle equality stems from the additivity of the QFI evaluated for a local operator~$\hat{O}$;
the last inequality is a consequence of Eq.~(\ref{QFIunitaryPure}) 
and $4(\Delta\hat{o})^2\leq(o_{\rm max}-o_{\rm min})^2$~\cite{pezze2014}.
The violation of Eq.~(\ref{boundSeparable}) is a \emph{sufficient} condition for entanglement among the parties:
\be \label{conditionEntanglement}
F_Q\big[\hat{\rho},\hat{O}\big]>N(o_{\rm max}-o_{\rm min})^2 \quad \Rightarrow \quad \hat{\rho} \textrm{\, is entangled} \, .
\ee
It is crucial to note that the QFI does not recognize the entire class of entangled states~\cite{PezzePRL2009}, 
but provides a criterion for ruling out the separability of the state.
The formulation~(\ref{quantumFisherInformationSLD}) of the QFI as the ensemble average of a Hermitian operator $\hat{L}_\phi^2$, 
along with the existence of an upper bound~(\ref{boundSeparable}) for separable states, 
expressly suggests the similarity to the witness approach; 
nevertheless, evaluating the QFI on unitary transformations~(\ref{QFIunitaryMixed}) 
does not imply the knowledge of the witness~$\hat{L}_\phi^2$.

The condition in Eq.~(\ref{conditionEntanglement}) certifies the presence on quantum correlations among the $N$ constituents,
but it tells nothing about the depth of entanglement: namely, the number of parties in the largest not factorizable cluster.
The solution is found by seeking the upper bound~\cite{HyllusPRA2012,TothPRA2012,pezze2014} 
for the more general $\kappa$-separable state in Eq.~(\ref{kappaSeparable}):
\be \label{boundKpartite}
F_Q\big[\hat{\rho}_{\kappa\textrm{-}{\rm sep}},\hat{O}\big] \leq 
\bigg\{\!\left\lfloor\frac{N}{\kappa}\right\rfloor\kappa^2+\bigg(N-\left\lfloor\frac{N}{\kappa}\right\rfloor\kappa\bigg)^2\bigg\}
\,(o_{\rm max}-o_{\rm min})^2 \, .
\ee
The bound on the right-side of Eq.~(\ref{boundKpartite}) monotonically increases for increasing $\kappa$.
Again, if the bound (\ref{boundKpartite}) is surpassed, then the state must host at least $(\kappa+1)$-partite entanglement.
For $\kappa=1$, we recover the weaker bound (\ref{boundSeparable}) for fully separable states 
and its violation witnesses bipartite entanglement.
Of course, different local operators $\hat{O}$ can lead to different bounds (\ref{boundKpartite}): 
whereas there is no systematic method to choose the \emph{optimal} operator $\hat{O}$ 
(i.e. the one that allows the detection of the largest cluster of entangled parties), one can rely on some model-dependent choice.

The definitions of separability and entanglement depth given in paragraph~\ref{subsec:MultipartiteEntanglement}
are completely general, regardless of the physical meaning of the partition of the global system: 
entanglement can arise among particles carrying a spin, momentum modes, lattice sites...
The case of systems composed of $N$ constituents whose degree of freedom are restricted to only two modes
\textsf{a} and \textsf{b} ($\dim\pazocal{H}=2$) is particularly relevant: 
examples are linearly polarized photons, two-level atoms, fermionic occupation modes... 
The identification of these two modes with the states ``spin up'' and ``spin down''
allows for an intuitive description of the system in terms of an ensemble of $N$ spins $\sfrac{1}{2}$~\cite{nielsen2000}.
In this case, the optimal probe operator to witness multipartite entanglement close to a quantum phase transition 
involving finite magnetization is a suitable component of the spin operator $\hat{o}=\frac{1}{2}\hat{\sigma}_\varrho$, 
with $\varrho=x,\,y,\,z$ (see section~\ref{sec:QuantumPhaseTransitions}). 
Hence $o_{\rm max}-o_{\rm min}=1$, and the upper bound~(\ref{boundSeparable}) justifies the criterion
$F_Q\big[\hat{\rho},\hat{O}\big]>N$ employed in chapters~\ref{ch:Ising} and \ref{ch:LMG} for claiming entanglement among the spins.
Moreover, from the inequality~(\ref{boundKpartite}) we find $F_Q\big[\hat{\rho}_{\kappa\textrm{-}{\rm sep}},\hat{O}\big]<\kappa N$, 
that justifies the name ``multipartiteness'' given to the quantity $F_Q/N$ throughout the thesis, 
since the condition $F_Q\big[\hat{\rho},\hat{O}\big]/N>\kappa$ witnesses $(\kappa+1)$-partite entanglement.

\paragraph{Quasi-local probe operators}
We find that the general bound (\ref{boundKpartite}) holds true not only for local operators, 
but even for the special class of \emph{quasi-local} operators defined as
\be
\hat{O} = \sum_{j=1}^N \cos\Big[\hat{\Theta}^{\{j\}}_{\small\mbox{\boldmath$\alpha$}} - \varphi\identity^{\{j\}}\Big] \, \hat{o}^{(j)}
\quad \textrm{with} \ \ \hat{\Theta}^{\{j\}}_{\mbox{\small\boldmath$\alpha$}} = \sum_{i=1}^{j-1} \alpha_i\,\hat{p}^{(i)} 
\ \ \textrm{and} \ \ \identity^{\{j\}} = \identity^{(1)}\otimes\dots\identity^{(j-1)}
\ee
where $\hat{o}^{(i)}$ and $\hat{p}^{(i)}$ are local observables acting on the subspace $\pazocal{H}_i$, 
the operators $\hat{p}$'s satisfy 
$\hat{p}^{(i)}\ket{\psi_{\kappa\textrm{-}{\rm prod},\,k}}=p_k^{(i)}\ket{\psi_{\kappa\textrm{-}{\rm prod},\,k}}$, 
lastly $\mbox{\boldmath$\alpha$}=(\alpha_1,\dots \alpha_{j-1})\in\mathbb{R}^{\otimes(j-1)}$ and $\varphi\in\mathbb{R}$.
Obviously, the shape of the quasi-local operator $\hat{O}$ relies on a reasoned guess for the $\hat{p}$'s 
based on some \emph{a priori} knowledge about the system.
For the Kitaev chain of chapter~\ref{ch:Kitaev}, $\hat{o}^{(i)}=\hat{a}_i$ is the fermionic annihilation operator 
and $\hat{p}^{(i)}=\hat{a}_i^\dag\hat{a}_i$ counts the number of fermions is each lattice site; 
thus, $o_{\rm max}-o_{\rm min}=1$ and the condition $F_Q\big[\hat{\rho},\hat{O}\big]>N$ guarantees entanglement among the sites.

\subsection{Application to the quantum theory of phase estimation} \label{subsec:QuantumInterferometry}
Measure is the key for our understanding of Nature: thanks to increasingly accurate measurements we have been able to investigate more and more recondite and fundamental phenomena. 
Historically, interferometry has revealed to be one among the most powerful and precise techniques in metrology. 
In the second half of the XIX century, the pioneering experiments by Michelson and Morley \cite{michelson} ruled out the hypothesis of a stationary aether, paving the way for new research that finally resulted in the theory of special relativity.
Since then, interferometric performance has continuously grown.
Nowadays interferometry provides the most precise measurements of fundamental constants~\cite{Hansch}, including the Newtonian constant of gravity \cite{Fixler,Rosi}, the local acceleration of gravity~\cite{Peters}, the fine-structure constant~\cite{Bouchendira} and the electric polarizability of atoms~\cite{Ekstrom1995}.
It also promises to give relevant contributions to general relativity: 
by means of interferometric techniques the first detection of gravitational waves has been recently achieved~\cite{abbott2016}, experiments have been proposed to test the universality of free fall~\cite{Hartwig} and other contributions to cosmology lie in establishing constraints on dark energy~\cite{Hamilton}.

Since the early Eighties, it started to become clear that the creation of suitable quantum correlations among the particles 
entering (or circulating in) the interferometer can tremendously boost its sensitivity~\cite{Caves1981,Yurke1986,Kitagawa1993,Wineland1994}.
Only recently, proof-of-principle experiments confirmed this quantum 
enhancement~\cite{NagataSCIENCE2007,KacprowiczNATPHOT2010,GrossNATURE2010,rugway2011,berrada2013,StrobelSCIENCE2014,MuesselPRL2014}.
We want to illustrate here how an interferometer works, briefly explain why a statistical inference of the phase is needed,
outline the connection between entanglement and interferometric enhancement
and show that the sole entanglement \emph{useful} for overcoming classical interferometry 
is the one recognized by the quantum Fisher information~\cite{PezzePRL2009}.

\paragraph{Review on interferometers} 
Broadly speaking, an interferometer is an apparatus that exploits the interference between waves to perform the estimation of an unknown phase shift. In a generic interferometric experiment, a source produces two ``wave beams'' which initially are combined by a first ``beam splitter'', then accumulate -- through different ``arms'' (or \emph{modes}) -- a phase difference, and finally are recombined by a second ``beam splitter''. A detecting device is put downline of the second beam splitter and a suitable observable is measured (usually particle current). In \emph{two-mode} interferometers only two arms are involved. 
A general cataloguing of two-mode interferometers is made according to the physical nature of the two beams.

\begin{figure}[!t]
\begin{center}
\includegraphics[width=1\textwidth]{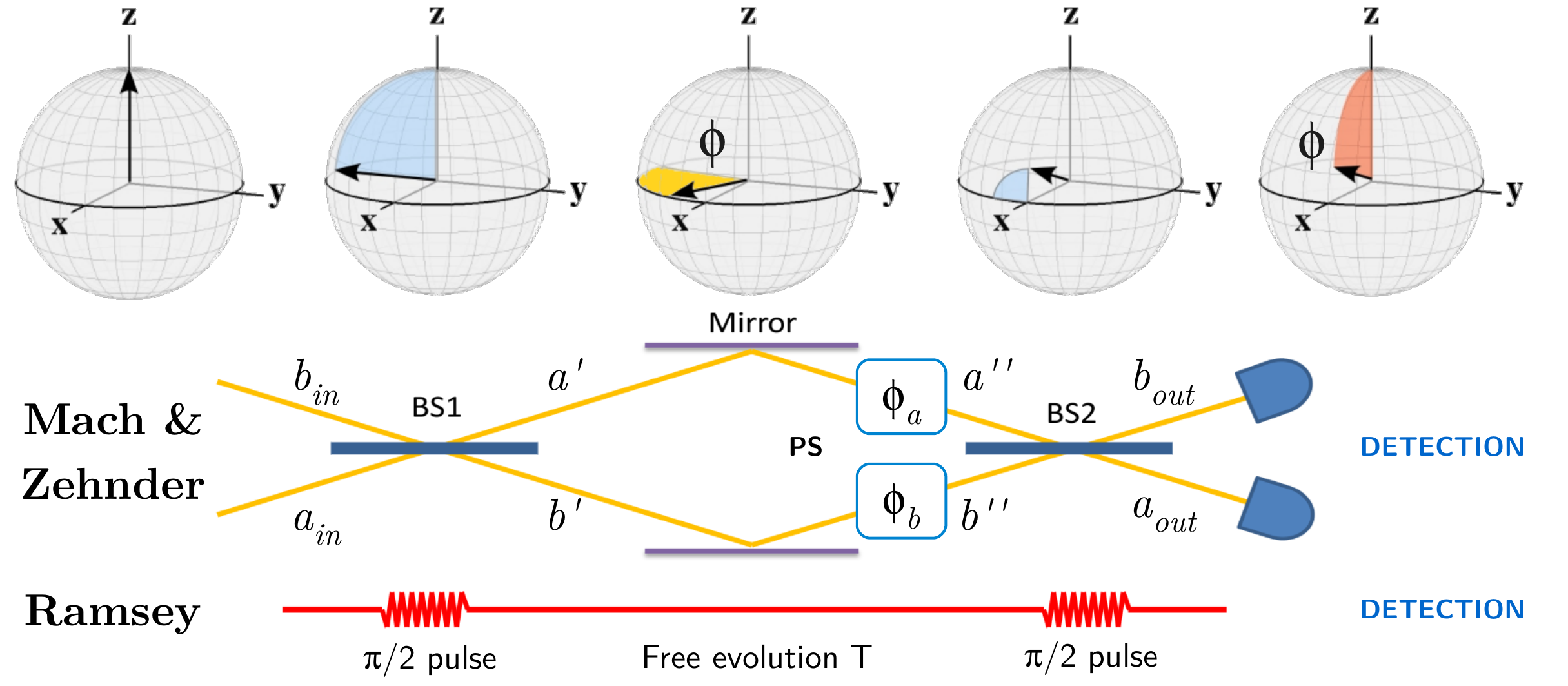}
\end{center}
\caption{Schematic diagram of Mach-Zehnder and Ramsey interferometers, 
compared to the sequence of rotations of the mean-spin direction on the Bloch sphere. 
The former device uses photons circulating in two spatially separated arms \textsf{a} and \textsf{b}, 
whereas the latter exploits internal states of atoms, but the working principle is the same. 
If the probe is initially prepared in a state having $N$ particles in mode \textsf{a} and no particles in mode \textsf{b}, 
the mean-spin direction is $\langle\hat{\mathbf{J}}\rangle_{\rm in}=\frac{N}{2}\,\mathbf{z}$. 
The action of the passive lossless components can be visualized 
as a sequence of rotations of the vector representing the state on the Bloch sphere: 
the vector is rotated by an angle $\pi/2$ around the \emph{x} axis by the first ``beam splitter'' 
(a balanced semitransparent mirror for light or a $\pi/2$-pulse for atoms); 
then it rotates by an angle $\phi=\phi_\textsf{a}-\phi_\textsf{b}$ around the \emph{z} axis during the ``phase accumulation'' 
(due to a phase shifter or a time of free evolution); 
finally it is rotated by an angle $-\pi/2$ around the \emph{x} axis.
The overall transformation is a rotation around the \emph{y} axis by an angle $\phi$. 
Measurements of the imbalance $\hat{J}_z$ of particle number between the two modes on the output of the interferometer 
easily provide information about the shift $\phi$. Picture adapted from Ref.~\cite{ma2011}.} 
\label{figMZR}
\end{figure}

In the \emph{optical} context, the beams are light waves. The first interferometers were purely optical~\cite{Thompson}. As a notable example of this class, we mention the traditional Mach-Zehnder interferometer~\cite{zehnder, mach}, which has been subject of extensive theoretical study and still is a standard configuration in current experiments: it is composed by passive lossless components such as beam splitters, mirrors and phase shifters (see Fig.~\ref{figMZR}). In a ``double-path interferometer'' (for instance, the Mach-Zehnder or Michelson ones) the two arms are spatially separated paths and the responsible for the phase difference is usually a difference of optical length. On the other hand, in a ``common-path interferometer'' such as the Sagnac ring~\cite{sagnac1913} the two arms share the same spatial path, but are travelled in opposite directions by the two beams, and the device is sensitive to the angular velocity of the apparatus. Other schemes are possible, in which the actual arms are the two modes of the internal degree of freedom (polarization) instead of the external paths. The measurement performed on the output can be either the rate of photons at the two ports (as in the Mach-Zehnder interferometer) or the spatial distribution of intensity on a screen (as in a double-slit experiment).

Similarly, an \emph{atom} interferometer takes advantage of the wave nature of matter. A large variety of atom interferometers have been realized \cite{cronin2009}. Interferometers exhibiting spatially separated paths as arms are made up by elements acting on atoms in a similar way usual splitters and mirrors act on light: the beam splitters are implemented by material diffraction gratings, that separate and recombine the paths of the atomic matter waves \cite{Godun}. Internal-structure modes such as hyperfine states can supply the arms for the interferometer: in this case, the splitting is provided by the interaction with an electromagnetic radiation, which can coherently change the population of the internal states, as in the Ramsey interferometer: see again Fig.~\ref{figMZR}. In atom interferometers, the external causes for the phase shift might be -- depending on the degree of freedom chosen for realizing the arms -- a mechanical force or a magnetic field.

%%%%%%%%%%%%
%% FIGURA %%
%%%%%%%%%%%%

The necessity for an interferometer using atoms instead of light mainly comes from the general fact that the interferometric probe has to interact with the external perturbation producing the phase shift. Photons do not couple with ordinary inertial forces since they are massless. On the contrary, atoms have mass and so are sensitive to accelerations and can be employed for realizing gyroscopes or gravimeters. The current generation of atom interferometers nowadays reach unprecedented precisions in the measurement of local gravity~\cite{Peters} and inertial forces~\cite{Geiger}.
Moreover, atom interferometers revealed to be powerful tools for measuring magnetic field \cite{MuesselPRL2014} 
and -- if the two modes supporting the interferometric transformation are two internal levels of the atoms -- the measurement of atomic transition frequencies with Ramsey interferometry can be exploited for spectroscopic purposes and for determining a frequency standard for atomic clocks~\cite{diddams2004}. 
In the last few years, the use of ultracold degenerate atoms as a probe in atom-interferometry experiments 
has become an established practice~\cite{GrossNATURE2010, rugway2011, berrada2013, StrobelSCIENCE2014}.

\paragraph{Phase estimation} 
A phase shift \emph{cannot} be measured: in quantum mechanics, 
there exist no Hermitian operator corresponding to the phase~\cite{LynchPR1995}. 
Rather, it must be inferred statistically. How is this estimation achieved?
From a quantum-mechanical point of view, we call ``interferometer'' any unitary\footnote{\ We are considering here an \emph{ideal} lossless interferometer, for which no source of dissipation is present.\\[-9pt]} transformation 
$\neper^{-\ii\phi\hat{O}}\,$: $\hat{\rho}_0 \mapsto \hat{\rho}_\phi$, generated by the collective operator $\hat{O}$, that encodes the true value of a real parameter\footnote{\ The present treatise is valid for \emph{single-parameter} estimation, dealing with the estimation of a scalar parameter. The generalization to multidimensional fields is also possible~\cite{baumgratz2016, ciampini2016}.} $\phi$ onto an input state $\hat{\rho}_0$, which is transformed into the output state $\hat{\rho}_\phi$. 
% This relative phase between the two ``beams'' is induced by some external perturbation $\Phi$ interacting with the waves.
The estimation of the phase shift $\phi$~\cite{helstrom1976, pezze2014} is done by choosing an observable $\measure$; 
performing $m$ independent measurements of $\measure$ on the output state of the interferometer $\hat{\rho}_\phi$,
one per each of the $m$ identical copies of the input state  $\hat{\rho}_0$;
constructing the conditional probability $P(\varepsilon|\phi)$ of finding the outcome $\varepsilon$ for a given $\phi$;
and finally decrypting the statistics $P(\varepsilon|\phi)$ by means of an \emph{unbiased estimator} $\Phi$, 
whose statistical average equals the true value of the unknown parameter: 
$\langle\Phi\rangle = \sum_\varepsilon P(\varepsilon|\phi)\,\Phi(\varepsilon) = \phi$. 
The uncertainty associated with such an inference is the standard deviation 
$\Delta\phi = \big[ \sum_\varepsilon  P(\varepsilon|\phi)\,(\Phi(\varepsilon) - \langle\Phi\rangle)^2 \big]^{1/2}$
and its inverse is usually named the ``phase sensitivity of the interferometer''. 

Since the phase shift is generated by the interaction of the probe with an external perturbation 
(a difference of optical path length for light; mechanical or magnetic fields for atoms), knowing the phase shift as best we can means to extract precise information about the external perturbation that generated the shift itself. 
% So, the aim of interferometry is measuring these effects in order to estimate the phase shift with the \emph{smallest possible uncertainty}. 
Given the interferometric transformation $\hat{O}$, the central goal of interferometry is choosing the input state $\hat{\rho}_0$, the observable to be measured $\measure$ and the specific estimator $\Phi$ in order to infer the phase shift $\phi$ with the \emph{smallest} possible uncertainty $\Delta\phi$ for the given finite available resources in input (for example, the number of particles $N$ the interferometer processes). 
% In this sense, interferometry is the central tool for phase estimation.

Which is the best precision achievable? 
The interferometric core of the estimation procedure described above has a strong resemblance 
to the operative definition of the QFI given in paragraph~\ref{subsec:QuantumFisherInformation}.
In light of this resemblance, we expect a notable contribution of the QFI in determining the interferometric sensitivity.
Indeed, the quantum Cram\'er-Rao theorem~\cite{rao1945,cramer1946,Helstrom1967} sets a lower bound on the variance of an arbitrary
unbiased estimator:
\begin{equation} \label{QCRB}
\Delta\phi \geq \frac{1}{\sqrt{m\,F_Q\big[\hat{\rho}_0,\hat{O}\big]}} \, .
\end{equation}
Heuristically, the relation (\ref{QCRB}) states that parameter estimation is naturally related to the problem of distinguishing states
along a path in the parameter space: the phase sensitivity $\Delta\phi$ can be understood as the smallest phase shift 
for which the output state $\hat{\rho}_\phi$ of the interferometer can be distinguished from the input state $\hat{\rho}_0$.
Here, the multipartite entanglement of the input state $\hat{\rho}_0$ comes into play: the larger the QFI, 
the higher both the witnessed entanglement depth and the sensitivity of the interferometer. 
For an explanation of the standard bounds to the phase sensitivity, we limit to the class of ``SU(2) interferometers''.

\paragraph{Local two-mode interferometers} % Mapping particle <---> spin
Whenever we study systems involving $N$ distinguishable noninteracting particles that access two arms, 
we can restrict to consider only the two active degrees of freedom associated to the arms 
(spatial path, polarization of photons, magnetization of atoms...) and represent each particle as an effective spin $\sfrac{1}{2}$
(being up or down according to the arm it occupies).
Under the assumptions that each particle experiences the same perturbation and the number of particles is conserved, 
any unitary collective transformation takes the form of a rotation 
$\neper^{-\ii\theta\hat{J}_{\textbf{n}}}$ of amplitude $\theta$ around the axis $\textbf{n}$,
generated by the pseudospin operator $\hat{J}_{\textbf{n}}=\hat{\mathbf{J}}\cdot\textbf{n}$~\cite{Yurke1986},
where the components of $\hat{\mathbf{J}}=\{\hat{J}_x,\hat{J}_y,\hat{J}_z\}$ are expressed in terms of 
Pauli matrices associated to the particles
\be \label{AMO}
\hat{J}_x = \frac{1}{2} \sum_{i=1}^N \hat{\sigma}_x^{(i)} \ , \quad \hat{J}_y = \frac{1}{2} \sum_{i=1}^N \hat{\sigma}_y^{(i)} \ , \quad \hat{J}_z = \frac{1}{2} \sum_{i=1}^N \hat{\sigma}_z^{(i)}
\ee
and satisfy the SU(2) algebra of the angular momentum: 
$\big[\hat{J}_\varrho,\hat{J}_\varsigma\big]=\ii\,\varepsilon_{\varrho\varsigma\tau}\hat{J}_\tau$.
For example, in the conventional spin jargon, a balanced 50-50 beam splitter is modelled by 
$\,\exp\big(\!-\ii\frac{\pi}{2}\hat{J}_x\big)\,$ and 
the phase-shift accumulation by $\exp\big(\!-\ii\phi\hat{J}_z\big)$. 
The action of the whole interferometer $\,\exp\big(\!-\ii\phi\hat{J}_{\textbf{n}}\big)\,$ % $\neper^{-\ii\phi\hat{J}_{\textbf{n}}}$ 
on the input state $\hat{\rho}_0$ 
is visualized by the composition of rotations on the \emph{mean spin direction}
$\langle\hat{\mathbf{J}}\rangle=\tr\big(\hat{\rho}_0\,\hat{\mathbf{J}}\big)$
on the generalized Bloch sphere: see Fig.~\ref{figMZR}.
Ordinary interferometers, like the Mach-Zehnder and Ramsey ones, are ideally modelled as local two-mode interferometers.
According to Eq.~(\ref{QCRB}), their sensitivity is lower bounded by 
$\Delta\phi\geq\big(m\,F_Q[\hat{\rho}_0,\hat{J}_{\textbf{n}}]\big)^{-1/2}$.

Given an input state $\hat{\rho}_0$, it is possible to optimize the rotation direction $\mathbf{n}$ of the unitary transformation 
$\neper^{-\ii\phi\hat{J}_{\textbf{n}}}$ in order to further maximize the QFI and obtain
$F_Q[\hat{\rho}_0] = \max_{\mathbf{n}}\,F_Q\big[\hat{\rho}_0,\hat{J}_{\mathbf{n}}\big]$. 
For a statistical mixture $\hat\rho_0 = \sum_k p_k |k\rangle\langle k|$, 
the optimization procedure for SU(2) interferometry reads~\cite{hyllus2010}
\be \label{SU2optimization}
F_Q[\hat{\rho}_0] = \max\,\textrm{eigval}\,\mathbb{F}_Q[\hat{\rho}_0] \quad {\rm with} \ \
\mathbb{F}_Q = 2\left[ \sum_{k\,k'} \frac{(p_k-p_{k'})^2}{p_k+p_{k'}}\,\langle k|\hat{J}_i|k'\rangle \langle k'|\hat{J}_j|k\rangle \right]_{i,\,j\,=\,x,\,y,\,z} \, .
\ee
The ``Fisher matrix'' $\mathbb{F}_Q$ reduces to the standard covariance matrix for pure states 
$\hat{\rho}_0=\ket{\psi_0}\bra{\psi_0}$:
%$\mathbb{F}_Q[\hat{\rho}_0]=4\big(\textrm{Re}\langle\hat{J}_i\hat{J}_j\rangle-\langle\hat{J}_i\rangle\langle\hat{J}_j\rangle$
$\mathbb{F}_Q\big[\ket{\psi_0}\big]=4\big(\textrm{Re}\bra{\psi_0}\hat{J}_i\hat{J}_j\ket{\psi_0}-
\bra{\psi_0}\hat{J}_i\ket{\psi_0}\bra{\psi_0}\hat{J}_j\ket{\psi_0}\big)$.
Now the optimal QFI depends just on the initial state $\hat{\rho}_0$ of the probe. 
%% Can we optimize the interferometric sensitivity with respect to the input state $\hat{\rho}_0$? 

Equations~(\ref{boundSeparable}) and (\ref{QCRB}) dictate that the maximum phase sensitivity attainable using separable input states
is $\Delta\phi_{\rm SN}=1\big/\sqrt{mN}$, indicated as ``shot-noise limit'' since it has been recognized to be
an imprint of corpuscolar nature of the particles in the probe~\cite{GiovannettiSCIENCE2004,GiovannettiPRL2006}.
Here, the number of particles $N$ as well as the number of independent repetitions $m$ play the role of a statistical gain. 
Classical interferometers are limited by this bound. 
Optimal fully separable states saturating the shot noise are the coherent spin states 
$|\vartheta,\varphi\rangle=\bigotimes_{j=1}^N\big[\cos\frac{\vartheta}{2}\,\spinup_j +
\neper^{\ii\varphi}\sin\frac{\vartheta}{2}\,\spindown_j\big]$, 
a product state of all the $N$ spins pointing along the same direction 
$(\sin\vartheta\cos\varphi,\,\sin\vartheta\sin\varphi,\,\cos\vartheta)$.

Conversely, Eq.~(\ref{conditionEntanglement}) prescribes that 
$F_Q[\hat{\rho}_0]>N$ is a necessary and sufficient condition for \emph{useful} entanglement in quantum interferometry: 
an entangled input state $\hat{\rho}_0$ feeding a two-mode local interferometer allows to overcome the shot-noise limit.
The violation of the sharper inequality Eq.~(\ref{boundKpartite}), witnessing $(\kappa\!+\!1)$-partite entanglement, 
defines the ultimate ``Heisenberg limit'' of phase sensitivity when $\kappa=N$: 
$\Delta\phi_{\rm H}=1\big/\sqrt{mN^2}$~\cite{GiovannettiSCIENCE2004,GiovannettiPRL2006}.
The scaling of the sensitivity with the number of particles is faster than in the classical shot-noise bound 
(in the next chapters we will sometimes refer to the quadratic scaling of the QFI with particle number $F_Q\sim N^2$ 
as ``Heisenberg-like scaling'').
Maximally entangled states saturate the Heisenberg limit~\cite{PezzeRMP}; 
no local interferometer can beat this fundamental scaling imposed by Nature.

\paragraph{Spin squeezing} 
Spin-squeezed states are a class of collective spin states having reduced spin variance (along a certain direction) 
with respect to the one exhibited by spin coherent states (at the expense of anti-squeezed variance along an orthogonal direction). 
Spin squeezing is one of the most successful witness for large-scale quantum correlations beating the shot noise. 
Amid the plethora of different definitions (each one tailored for a specific scope~\cite{ma2011}),
we just mention the Wineland spin-squeezing (WSS) parameter~\cite{Wineland1994}, 
defined in terms of first and second moments of collective pseudospin operators as
\be \label{WSS} 
\WSS = \frac{N(\Delta\hat{J}_{\mathbf{n}_\perp})^2}{\langle\hat{J}_{\mathbf{n}_\parallel}\rangle^2} \, ,
\ee
where $\mathbf{n}_\parallel$ is a versor proportional to the mean spin direction $\langle\hat{\mathbf{J}}\rangle$
and $\mathbf{n}_\perp$ refers to an arbitrary perpendicular versor.
A state is said to be \emph{spin squeezed} along the direction $\mathbf{n}_\perp$ if $\WSS<1$.
This inequality is also a criterion for multipartite entanglement~\cite{SorensenNATURE2001}:
in fact, the inequality $N/\WSS \leq F_Q$ holds~\cite{PezzePRL2009}, thus a spin-squeezed state is entangled, 
but there exist usefully entangled states that are not squeezed.

\section[Overview on phase transitions]{Overview on phase transitions} \label{sec:QuantumPhaseTransitions}
Phase transitions are marvellous collective phenomena that have been under the physicists' microscope 
for more than one century~\cite{Andrews1869,vanderWaals1873} and their appeal has never lost sheen over time.
They still are subject of extensive theoretical and experimental investigations,
since they appear to be omnipresent in Nature and arise within a huge variety of systems in disparate fields, 
ranging from condensed matter to ultracold atoms~\cite{Huang1987,sondhi1997,vojta2003,Dutta2015}, 
from particle physics to cosmology~\cite{Kolb1994}.

The purpose of the present section is defining phase transitions in a very broad sense. 
First, we will review the main concepts and set the terminology 
focusing on the subset of thermal classical phase transitions (paragraph~\ref{subsec:ClassicalPhaseTransitions}).
Then we will introduce quantum phase transitions in a pedagogical way 
(paragraphs~\ref{subsec:QuantumPhaseTransitions}, \ref{subsec:QuantumCriticality} and \ref{subsec:FidelitySusceptibility}).

\paragraph{General definition: an attempt}
From a physical perspective, a phase transition is any sudden change in the collective behaviour of a large system
that occurs whenever a control parameter is tuned across a transition point.
Let's try to make this intuitive definition more precise. 

A stable physical system at equilibrium is usually found in certain states, 
characterized by well-defined properties depending on the conditions externally imposed on the system: 
for instance, the temperature of the bath a liquid is in contact with, the doping of a semiconductor, 
the magnetic field determining the scattering properties of atoms in a dilute gas at low temperature, and so on.
According to the essential parameters defining the state, it is possible to draw a phase diagram, 
collecting all the state the system has access to.
The phase diagram is typically partitioned into disconnected regions: 
within each region, the physical state of the system can be changed smoothly by varying some parameters 
and no singularity in the ensemble average of any observables is encountered. 
These regular regions are called ``phases'': distinct phases exist or coexist at equilibrium.
On the contrary, the boundaries between phases are associated with the singular behaviour of some observables 
and are referred to as ``transition points''. 

Thus, more rigorously, at the transition point a \emph{nonanaliticity} of some macroscopic function of the parameters arises. 
Whereas the variation of the parameters does not affect qualitatively the properties of the system inside one phase, 
a minute variation across the transition point is accompanied by a clear discontinuous response.
% accompanied by a high susceptibility of the system to any change
This abrupt discontinuity is called ``phase transition'', and often it manifests with a rich variety of 
signatures regarding the susceptibility of the system, the length on which the constituents are correlated and so on.
The parameter driving the transition is named ``control parameter'': it should be tunable with continuity 
and can be a thermodynamic variable (such as temperature, pressure or chemical potential in classical transitions)
or an external field that couples to the constituents of the system 
(such as a magnetic field ruling the alignment of magnetic dipoles in quantum transitions)
%% or determining the scattering properties of atoms in a dilute gas at low temperature).
or time in the more exotic and still not completely explored dynamical transitions. 
%% when the transition happens during the dynamics of a system

At this level, no hypothesis has been made on the system and its constituents: 
phase transitions may concern molecules in a fluid, 
atoms carrying a magnetic moment in a magnet, electrons in a superconductor, 
or entire portions of the universe at its early age.
The crucial aspect of phase transitions is that their occurrence is not associated with any change 
in the nature of the microscopic constituents of the system and their interactions: 
the fundamental description of the system -- included in its effective Hamiltonian --
remains the same in each phase as well as at the transition point.
Furthermore, even if the involved systems are apparently enormously different under many physical respects, 
a large variety of transitions share common features, so that a classification is possible in terms of few essential ingredients, 
that are the nature of singularities encountered, the range of microscopic interactions
and the spatial dimensionality of the system.
%%regardless to the minute details of the systems 
%%(that anyway make the systems appear qualitatively different at our eyes, as water or a magnet).

\subsection{Thermal phase transitions} \label{subsec:ClassicalPhaseTransitions}
%%\think{Classical phase transitions regards macroscopic systems governed by the laws of thermodynamics, 
%%for which a quantum description is not necessary.} CONTROL PARAMETER?
%%CPT are a major topic of classical statistical mechanics at equilibrium.
A large and important family is the one of thermal classical phase transitions (CPTs),
that are accessed by tuning the temperature $T$ across a critical value $T_{\rm c}$.
They are well understood in terms of a competition between disordering thermal fluctuations and ordering internal interactions.
Here, the macroscopic function exhibiting a singularity at the transition point is the free energy.

When the transition contemplates a breaking of one symmetry of the underlying Hamiltonian, 
the powerful machinery of the standard Ginzburg-Landau theory can be employed~\cite{Ginzburg1950} and 
it is possible to build up the \emph{order parameter} $\varphi(T)$ compatible with the symmetry: 
namely, the statistical average $\varphi(T)=\langle\Phi\rangle$ of an observable 
that is finite in the ordered (less symmetric) phase $T<T_{\rm c}$ 
and null in the disordered (more symmetric) phase $T>T_{\rm c}$. 
Usually, we consider a collective local observable $\Phi=\sum_i\phi_i$ constructed from the $N$ variables $\phi_i$ 
describing each microscopical constituent ($i=1,\dots N$).
Two-body correlation functions of the order parameter $C^{(i,j)}=\langle\phi_i\phi_j\rangle$ can also be very useful 
for characterizing the different phase: apart from the details, 
for long distances they rapidly decay $C^{(i,j)}\sim\exp\big(\!-|i-j|/\xi\big)$
in the disordered phase -- defining a characteristic scale $\xi$ on which two constituents are correlated --
while they become approximately constant in the ordered phase.

It is customary to classify the transitions according to the lowest derivative of the free energy
that displays a singularity at the transition point.
In \emph{first-order} thermal CPTs, entropy shows a discontinuity as a function of $T$, the phases coexist at the transition point and
the order parameter undergoes a jump between two finite values, whereas the correlation length remains finite. 
These transitions are related to some form of release of the latent heat and commonly hysteresis loops are observed 
due to the existence of metastable states.
The ordinary changes of matter phase from solid to liquid and from liquid to gas are examples of first-order CPTs,
as well as the more sophisticated transition from the nematic phase to the isotropic phase in a liquid crystal.

Conversely, a continuously varying order parameter with a discontinuous first derivative announces
\emph{second-order} thermal CPTs, also known as ``critical phenomena''.
%where the order parameter varies with continuity from 
Here, for a system with an infinite number of constituents, at the critical point $T=T_{\rm c}$ the correlation length diverges 
as $\xi\sim|T-T_{\rm c}|^{-\nu}$, with $\nu$ a critical exponent, 
and the correlator falls off as a power law $C^{(i,j)}\sim|i-j|^{-(D+\eta-2)}$ at long distance,
where $D$ is the number of spatial dimension on which the system extends and $\eta$ is called the ``Fisher exponent''.
The divergence of the correlation length $\xi$ is the prominent responsible 
for the singular behaviour of all the response functions and other thermodynamic quantities:
critical exponents are introduced to describe these singularities.
Universality is observed within a region close to the transition point, 
in the sense that very different critical systems sharing the same set of exponents 
--~such as a volume of water at the end-point temperature and a three-dimensional uniaxial ferromagnet at the Curie temperature~--
end up belonging to the same universality class. Only few universality classes exist.
Universal properties at the critical point are understood in the framework of the renormalization group~\cite{Goldenfeld1992},
based on the self-similarity of the system under scale dilatation.
The transition between the paramagnetic and ferromagnetic phases, the superconducting transition in certain metals, 
the superfluid transition in liquid Helium and the Bose-Einstein condensation are classified as second-order CPTs.

\subsection{Quantum phase transitions} \label{subsec:QuantumPhaseTransitions}
Quantum phase transitions (henceforth QPTs) are cooperative phenomena that characterize the low 
temperature physics of a plethora of different systems~\cite{sondhi1997, vojta2003, Sachdev2011}. 
The widespread interest they attract lies in their appearance in a large variety of different fields 
and their intriguing features with potential applications in quantum information and materials science.
At the same time, also the exciting challenges QPTs pose for their complete theoretical portrayal
have drawn a significant amount of research in the last two decades and several issues are still open,
especially in characterizing their fingerprints at high experimentally-accessible temperature 
or in non-equilibrium processes~\cite{CuccoliPRB2007,BhattacharyyaSR2015}.
Understanding QPTs at zero and finite temperature constitutes a significant endeavor in the current agenda of condensed-matter physics.

% DEF NON GENERALE E CORR [Cuccoli]
% *** cooperatively interacting correlated many-body quantum systems ***

\paragraph{Definitions}
Whereas thermal phase transitions are driven by thermal fluctuations, QPTs take place at zero temperature 
and concern a qualitative change of the \emph{structure of the ground state} (either in its topology or symmetry).
Since at $T=0$ the disordering entropic contribution is absent, any possible order 
(either of topological nature or arising from a spontaneous symmetry breaking) can be destroyed by the zero-point fluctuations only. 
Thus, QPTs are driven by genuinely quantum fluctuations due to the Heisenberg uncertainty principle. 
Operationally, these transitions are accessed by tuning a nonthermal control parameter $\lambda$ 
that rules the competition between noncommuting terms in the many-body Hamiltonian describing the system.

In full generality, let the many-body system be governed by the family of Hamiltonians
\be \label{QPTbasicH}
\hat{H}(\lambda) = J\big(\HI + \lambda\,\HII\big)
\ee
parametrized by the real dimensionless coupling constant $\lambda$.
In paradigmatic systems, $\HI$ models the inner microscopic interactions among the constituent, with energy strength $J$, 
while $\HII$ describes their coupling with an external field; we will consider the nontrivial case $[\HI,\HII]\neq0$,
so that eigenvalues and eigenstates of Hamiltonian~(\ref{QPTbasicH}) do depend on $\lambda$. 
At zero temperature, the system lies in the ground state $\ket{\psi_{0}(\lambda)}$ of  $\hat{H}(\lambda)$ 
dictated by the value of the parameter $\lambda$. 
Each of the two terms $\HI$ and $\HII$, individually, induces a specific ground state and determines a specific phase in the phase diagram.
If $\bra{\psi_{0}}\HI\ket{\psi_{0}}\approx\bra{\psi_{0}}\HII\ket{\psi_{0}}$ 
(for example, they scale in the same fashion when increasing the number of microscopic constituents),
the limit condition $\lambda\ll1$ (or $\lambda\gg1$) makes the component $\HI$ (or $\HII$) to dominate over the other.
Therefore, there must exist (at least) one value $\lambda_{\rm c}\approx1$ of the parameter at which 
a boundary between different phases is encountered.

In the parameter space, points $\lambda_{\rm c}$ of nonanalyticity in the ground-state energy are called \emph{quantum critical points}.
Usually, they are associated with singular behaviours of the ground-state wave function, 
%%% INVERTIRE ORDINE ???
density of state in the lowest part of the energy spectrum and macroscopic observables.
We desire to stress here that, in a finite system, all such signatures manifest as smooth crossovers: 
no singularity is found for a finite number of constituents $N$,
similarly to what happens for classical phase transitions~\cite{LeePR1952,Huang1987}. 
A necessary requirement for labelling any emergent structural change of the ground state as a true QPT 
is the appearance of a singularity in the \emph{thermodynamic limit} $N\to\infty$~\cite{Sachdev2011}. 

Whenever it is possible to identify a (spontaneous or forced) symmetry-breaking mechanism,
%%\footnote{We are Setting aside topological QPTs for the moment, let's consider a QPT that involves symmetry breaking.} 
an \emph{order parameter} $\varphi(\lambda)=\bra{\psi_{0}}\,\hat{\Phi}\,\ket{\psi_{0}}$ can be constructed 
as the expectation value on the ground state of a collective local observable $\hat{\Phi}=\sum_i\hat{\phi}_i$, 
where the operator $\hat{\phi}_i$ acts on the $i$th constituent ($i=1,\dots N$).
The order parameter recognizes the ordered symmetry-broken phase ($\varphi\neq0$ at, say, $\lambda<\lambda_{\rm c}$) 
from the disordered symmetric phase ($\varphi=0$ at $\lambda>\lambda_{\rm c}$)
and allows for a simple classification of the QPTs.
Terminology is approximately borrowed from CPTs.
First-order (or \emph{discontinuous}) QPTs envisage an abrupt jump of the order parameter at $\lambda_{\rm c}$
and are associated with a finite discontinuity of the density of states.
In second-order (or \emph{continuous}) QPTs, instead, the order parameter vanishes continuously 
for $\lambda\to\lambda_{\rm c}^-$ and the density of states shows a divergent behaviour. 

The nonanalyticity in the ground-state energy can result both from a level crossing (which occurs even at finite size $N$, though) 
or from an avoided level crossing (that becomes sharper and sharper as the size $N$ increases).
Thus, strictly speaking, continuous QPTs only arise in the thermodynamic limit for some \emph{vanishing energy scale} $\Delta>0$:
the characteristic energy $\Delta$ shall be either 
the energy of the lowest excitation above the ground state when the spectrum is gapped 
or the lowest nonzero energy fluctuation when the spectrum is gapped 
(we will better specify the role of $\Delta$ in paragraph~\ref{subsec:QuantumCriticality}).
% for a \emph{vanishing energy gap} between the ground state and the first excited state in the many-body spectrum: 
In this case, the properties of the ground state of the system change fundamentally when crossing $\lambda_{\rm c}$:
the nature of the correlations $C^{(i,j)}=\bra{\psi_{0}}\hat{\phi}_i\hat{\phi}_j\ket{\psi_{0}}$ is a neat example, 
as explained below.

Examples of QPTs are the ferromagnetic or antiferromagnetic transitions from a disordered phase to an ordered phase
in spin systems subject to a transverse field~\cite{Dutta2015}
(like those discussed in chapters~\ref{ch:Ising} and \ref{ch:LMG}); 
transitions between phases with different topology~\cite{HasanRMP2010} 
% (a different class where no symmetry is broken and no local order parameter is found, BETTER EXPLAINED IN...) 
(like the one discussed in the toy model of topological superconductor discussed in chapter~\ref{ch:Kitaev});
the Mott transition from the insulator phase to the superfluid phase in an optical lattice~\cite{GreinerNAT2002};
and many others.\footnote{\ In this concise introduction, we have considered the simplest (but still relevant) level of sophistication.
Actually, the situation can be more complicated than this: 
for instance, it can be necessary to tune more than one control parameter $\lambda_1$, $\lambda_2$\dots for reaching the transition point (multidimensional parameter space~\cite{Dutta2015}); 
the order parameter can be a vector (multidimensional system~\cite{Chaikin1995}); 
there can be QPTs where the ground-state energy density is not necessarily singular~\cite{Dutta2015}; 
furthermore, it has been recently pointed out that ``excited-states QPTs'' can also be defined 
when a gap between higher-energy levels closes in the thermodynamic limit 
or undergoes an avoided crossing for finite particle number~\cite{cejnar2006,caprio2008}.
We have set all these complications aside.\\[-9pt]}

\paragraph{Critical exponents}
For QPTs of the second order, similarly to continuous CPT, %universal scaling behaviours are observed and
a set of critical exponent can be defined~\cite{sondhi1997,vojta2003,Sachdev2011}.
Following the notation used in paragraph~\ref{subsec:ClassicalPhaseTransitions},
in general we can state that the equal-time two-body correlator of the order parameter in the disordered phase 
quickly decays at long distances beyond a characteristic length scale $\xi$, defining a \emph{correlation length}.
As the critical point is approached, the correlation length diverges as $\xi\propto|\lambda-\lambda_{\rm c}|^{-\nu}$,
where $\nu>0$ is called the ``correlation exponent''.
This divergence implies that, sufficiently close to the transition, $\xi$ is the sole important length scale,
while, at the critical point, the system is scale invariant and the correlator acquires a slower power-law decay 
$C^{(i,j)}\sim|i-j|^{-(d+z+\eta-2)}$ at long distance, with $d$ the dimensionality of the system 
and $\eta$ and $z$ other two critical exponents. 
At $\lambda=\lambda_{\rm c}$, the ``dynamic exponent'' $z>0$ relates the diverging length scale $\xi$ 
to the vanishing energy scale $\Delta$: % between the ground state and the first excited state: 
$\Delta\propto J\,\xi^{-z}$, being $J$ the characteristic energy scale in Hamiltonian~(\ref{QPTbasicH}). 
As a consequence, as $\lambda\to\lambda_{\rm c}$, the energy scale $\Delta$ vanishes as 
$\Delta\propto J\,|\lambda-\lambda_{\rm c}|^{z\nu}$.
Critical exponents\footnote{\ Several other critical exponents exist, for example the exponent $\beta$ of 
the order parameter $\varphi\propto(\lambda_{\rm c}-\lambda)^\beta$ on the ordered side of the phase diagram. 
Nevertheless, they are not all independent: as in the CPT~\cite{Goldenfeld1992}, 
scaling relations connecting these exponents have been found.\\[-9pt]} 
$\eta$, $\nu$ and $z$ are universal: they only depend on the dimensionality of the system, the nature of its inner interactions 
and the symmetry of the order parameter. Very disparate systems sharing these few features exhibit the same exponents, 
regardless of the microscopic details of the respective Hamiltonians.

\subsection{Quantum criticality} \label{subsec:QuantumCriticality}
A significant contribution to designing new materials and unveiling their properties, 
as well as driving forward the long-standing quest for high-temperature superconductivity, comes from the study of matter 
at finite but sufficiently low temperature, where quantum mechanics determines the behaviour of the system.
We intend to briefly sketch out the role that quantum critical points can have in this area of research.

As defined above, QPTs concern the ground-state properties of many-body systems:
the smoking-gun evidences of their existence are expected to emerge from the lowest-energy states.\footnote{\ Recently, the possible existence of signatures of a QPT on higher excited levels -- featuring a much smaller density of states with respect to 
the low-lying part of the spectrum -- has been analyzed. 
Footprints of singular behaviours not limited to the ground state have been suggested in spin systems~\cite{RibeiroPRL2007} 
and might be probed by exciting the system towards strongly out-of-equilibrium states~\cite{BhattacharyyaSR2015}.\\[-9pt]}
Provided that actual experiments are always performed at finite temperature, 
it may seem that QPTs are just a theoretical notion, destined to be never observed,
and one may question about their relevance.
Surprisingly, the zero-temperature transition has a large impact on the physics of the system even at finite temperature~\cite{continentino2001,ColemanNATURE2005,Sachdev2011}.
This remarkable fact introduces the prospect to experimentally detect QPTs.

%%%%%%%%%%%%%%%%%%%%%%%%%%%%%%%%%%%%%%%%%%%%%%%%%%%%%%%%%
%%%%%%%%%%%%%%%%%%%%%%%%%%%%%%%%%%%%%%%%%%%%%%%%%%%%%%%%%
\begin{figure}[t!]
\centering
\includegraphics[width=0.49\textwidth]{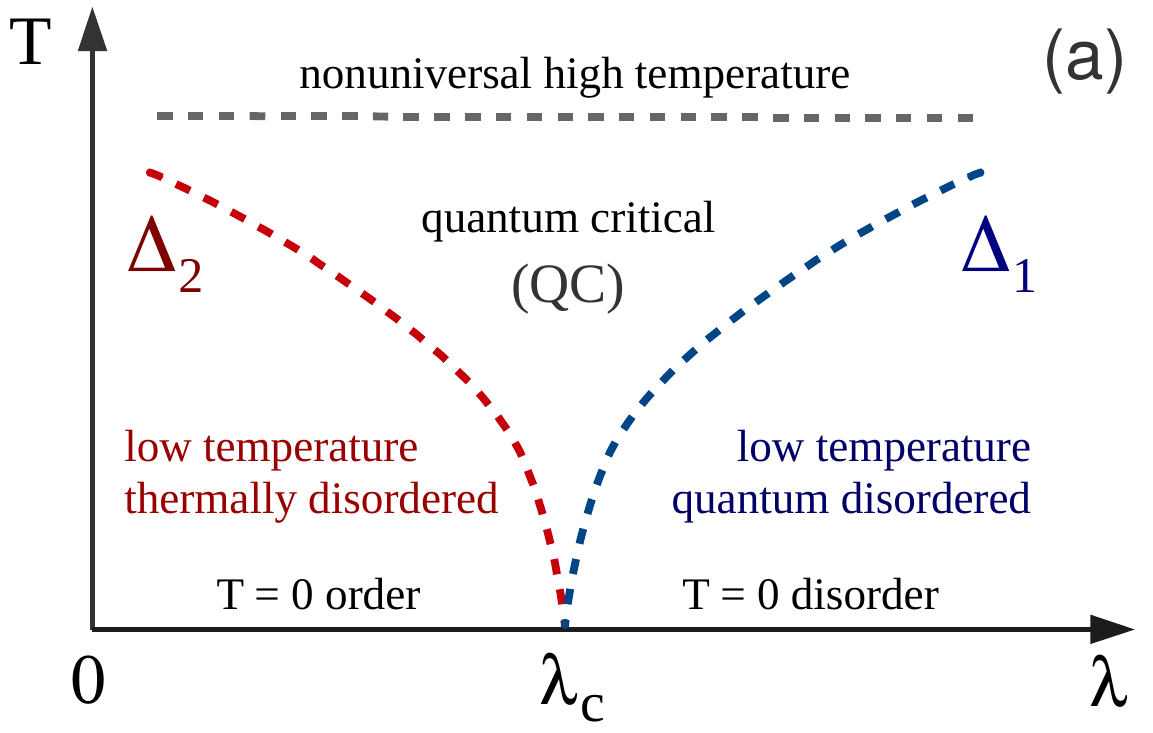} \hfill
\includegraphics[width=0.49\textwidth]{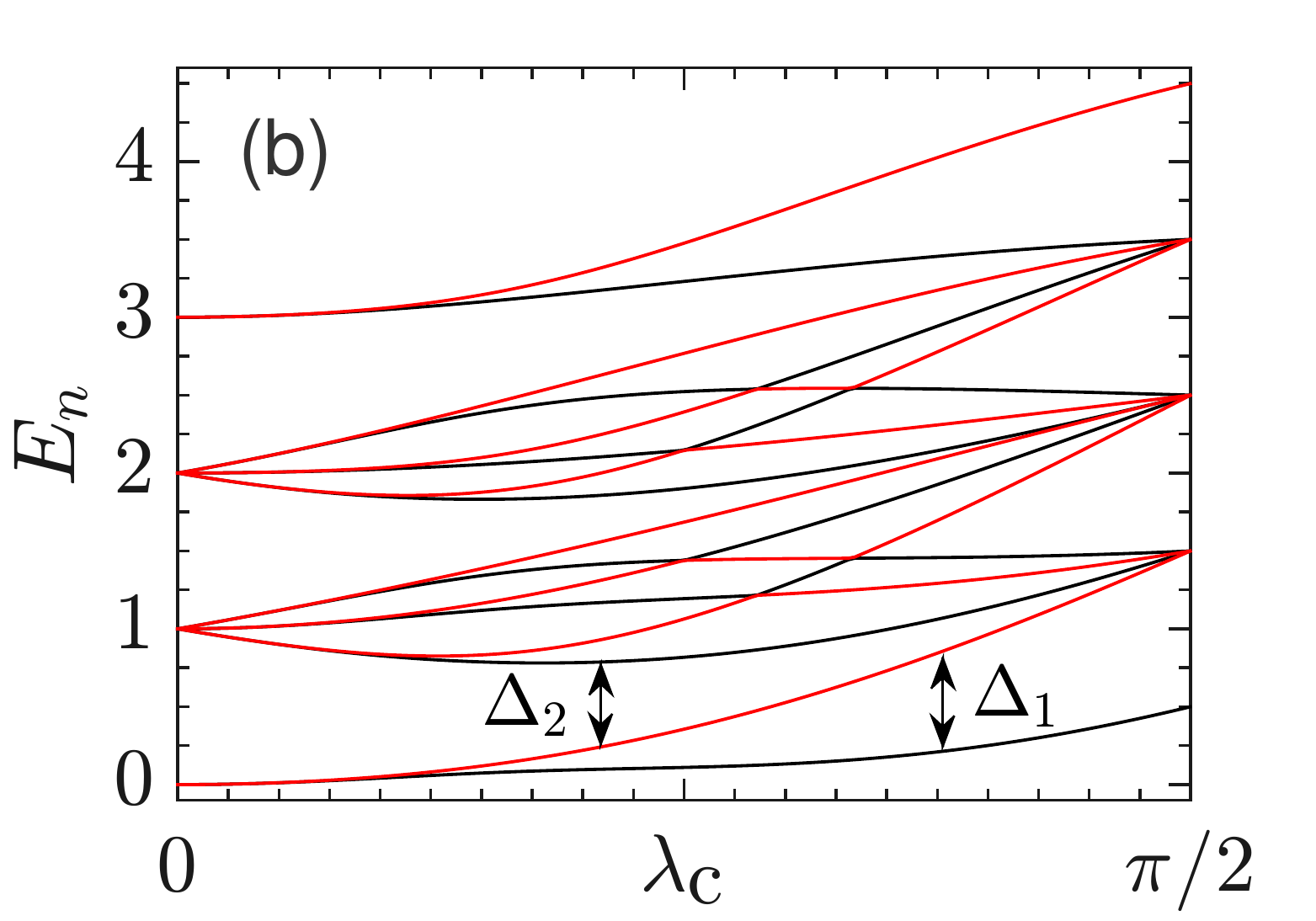}
\caption{\textbf{(a)} Schematic phase diagram control parameter versus temperature 
for a generic system having a quantum critical point at $\lambda=\lambdac$ 
and a long-range ordered phase ($\lambda<\lambda_{\rm c}$) only at $T=0$.
The crossover temperatures (coloured dashed lines) divide the diagram into the three regions discussed in the text;
they are of the order of the first nonzero energy separation in the thermodynamic limit, 
namely $\DeltaE=E_1-E_0$ or $\DeltaEE=E_2-E_1$ depending on whether the behaviour is dominated by thermal or quantum fluctuations:
$T\approx\DeltaE\propto J\,(\lambda-\lambdac)^{z\nu}$ for $\lambda>\lambdac$, while
$T\approx\DeltaEE\propto J\,(\lambdac-\lambda)^{z\nu}$ for $\lambda<\lambdac$.
The crossover towards the nonuniversal high-temperature limit where the microscopical details of the system emerge (gray dashed line)
is approximately located at $T\approx J$.
Paradigmatic examples described by this generic scenario are the one-dimensional nearest-neighbour Ising model in a transverse field 
and the two-dimensional short-range quantum spherical model; 
for both the models $z\nu=1$ and the crossover lines become straight lines radiating from the critical point $\lambdac$. 
For the fully-symmetric Lipkin-Meshkov-Glick model, these considerations still hold, with $z\nu=1/2$.
\textbf{(b)} Complete energy spectrum of the ferromagnetic transverse Ising chan
$\hat{H}=J\big(\!-\sum_{i,j}\hat{s}_z^{(i)}\hat{s}_z^{(i+1)}+\tan\lambda\,\sum_i\hat{s}_x^{(i)}\big)$, 
with control parameter $\lambda\in[0,\frac{\pi}{2}]$ and $N=4$. 
The $2^N=16$ eigenvalues $E_n$ (black lines for even $n$, red lines for odd $n$) are solutions of $\hat{H}\ket{\psi_n}=E_n\ket{\psi_n}$
and are measured in units of $J/\cos\lambda$. 
The ground state $\ket{\psi_0}$ and first excited state $\ket{\psi_1}$ are \emph{almost} degenerate at $\lambda\lesssim\lambdac$,
and the energy distance between the two levels becomes smaller and smaller in the thermodynamic limit $N\to\infty$. 
For finite $N$, the degeneracy is exact \emph{only} at the edge of the domain $\lambda=0$.}
\label{fig:QuantumCriticality}
\end{figure}
%%%%%%%%%%%%%%%%%%%%%%%%%%%%%%%%%%%%%%%%%%%%%%%%%%%%%%%%%
%%%%%%%%%%%%%%%%%%%%%%%%%%%%%%%%%%%%%%%%%%%%%%%%%%%%%%%%%

Seminal works, focused on the two-dimensional quantum Heisenberg model~\cite{ChakravartyPRB1989}
and the one-dimensional quantum Ising model~\cite{SachdevNPB1996,SachdevPRL1997}, 
suggested the possibility to delineate a \emph{universal finite-temperature phase diagram} $\lambda$--$T$ 
for systems exhibiting long-range order exclusively at $T=0$, as sketched in Fig.~\ref{fig:QuantumCriticality}\Panel{a}.
The diagram is divided into three different regions, separated by the crossover temperature 
$T\approx\Delta\propto J\,|\lambda-\lambdac|^{z\nu}$ on the two sides of the critical point $\lambdac$.

For low temperature $T\ll\Delta$, thermal excitations above the ground state (regardless of their nature) 
can be effectively described by a dynamical classical theory,
since they behave as particles whose average spacing is much larger than the thermal de~Broglie wavelength~\cite{SachdevPRL1997}.
After perturbing the system away from the thermal equilibrium state, 
the relaxation time $\tau$ it takes to recover local equilibrium is notably long, 
scaling as $\tau\sim T^{-1}\,\neper^{\,\Delta/T} \gg T^{-1}$,
and can be deduced from classical equations of motion for interacting particles~\cite{Sachdev2011}. 

On the two sides of the critical point $\lambdac$, we can distinguish two low-temperature regimes. 
First, the ``thermally disordered region'' lies above the zero-temperature ordered phase (${\lambda<\lambdac}$):
here the long-range order at $T=0$ is destroyed by thermal fluctuations at any arbitrarily small temperature;
the order-parameter correlation function $C^{(i,j)}$ decays exponentially on a large spatial scale 
$\xi\sim\big(\frac{\Delta}{T}\big)^{1/2}\,\neper^{\,\Delta/T}$, diverging as $T\to0$.
Second, the ``quantum disordered region'' lies above the zero-temperature disordered phase ($\lambda>\lambdac$):
here the physics is dominated by quantum fluctuations; 
the correlation function still decay exponentially, but the correlation length saturates the finite constant value 
$\xi\sim\Delta^{-1/z}$.
Most importantly, in the latter regime the correlation function factorizes into two separate contributions, 
arising from quantum and thermal fluctuations~\cite{SachdevPRL1997}. 
The physical interpretation of the excitations in the two low-temperature regimes depends on the specific model under consideration: 
for example, in the quantum Ising model excitations are domain walls (or kinks) on the magnetically ordered side ($\lambda<\lambdac$)
and flipped spins on the disordered side ($\lambda>\lambdac$).

For temperatures $\Delta \ll T \ll J$, the thermal de~Broglie wavelength is of the order of 
the average spacing between thermal excitations and the factorization of the correlation functions is not possible. 
Here, the physical properties of the system are dominated by a competition between quantum and thermal fluctuations:
both types of fluctuations are important and no effective classical model provides an accurate description 
of the thermalization after a local quench.
The equilibration time $\tau\sim T^{-1}$ is the shortest possible for the system and
does not depend on the microscopical energy scale $J$.
The correlation function decays exponentially and the correlation length $\xi\sim T^{-1/z}$~\cite{CuccoliPRB2007}
diverges as a power-law at $\lambdac$ when $T\to0$.

%% QC region has been recently observed via measuring relaxation time $\tau$~\cite{KinrossPRX2014}.

This regime where criticality strongly influences the properties of the system 
is often called ``quantum criticality'' (QC)~\cite{Sachdev2011}. 
Its investigation is crucial for interpreting a wide range of experiments~\cite{KeimerPRB1992, KobayashiPRB1999, SchroederNAT2000, GrigeraSCIENCE2001, LakeNATMAT2005, LorenzPRL2008, DaouNATPHYS2009, CooperSCIENCE2009, ColdeaSCIENCE2010, KinrossPRX2014}.
In the $\lambda$--$T$ plane, the QC region has a V-shape\footnote{\ A picturesque analogy with gravitation and general relativity 
has been proposed~\cite{ColemanNATURE2005}: one can image that the fabric of the phase diagram is distorted 
by the presence of the quantum critical point, and crossing $\Tcross$ is analogous to passing the event horizon of a black hole.\\[-9pt]}
centred at the critical point $\lambdac$: see Fig.~\ref{fig:QuantumCriticality}\Panel{a}.
The physics inside the QC region mirrors the physics of the quantum critical points: 
in this sense, finite-temperature properties of quantum critical matter do not depend on the microscopical details of the system 
and they are expected to be universal~\cite{ColemanNATURE2005}. 
A definition of QC region based on the notion of quantum coherence has recently been proposed~\cite{FrerotArxiv2018}.

The QC region fans out as $T$ increases up to temperatures $T \approx J$, 
at which the universal behaviour is lost and the system enters a generic high-temperature region where
physical properties (such as static correlations or thermalization dynamics) 
are nonuniversal and determined by the microscopic details~\cite{KoppNATPHYS2005,LonzarichNAT2005}.
Anyway, realistic values of the threshold $J$ allow to observe the QC region at surprisingly high temperatures.

\paragraph{Important remark on the energy scale}
We need to clarify the meaning we give to the energy scale $\Delta>0$, 
because it is used in the finite-temperature diagram of panel~\ref{fig:QuantumCriticality}\Panel{a}
and will be adopted throughout the thesis.
%%%
For this purpose, we refer to panel~\ref{fig:QuantumCriticality}\Panel{b}, where we plot the whole energy spectrum of the transverse Ising model discussed in chapter~\ref{ch:Ising}, for $N=4$ spins coupled by nearest-neighbour interaction.
The number of constituents is intentionally small to let the structure of the low-lying levels stand out clearly.
We regard this simple spectrum as a paradigm for the typical spectrum of interacting many-body systems at finite number of constituents $N$,
since it is qualitatively very similar to the energy level distribution in many other models 
(among which those discussed in the present thesis), such as
the transverse Ising chain with long-range interaction (chapter~\ref{ch:Ising}), the isotropic XY model,
the Lipkin-Meshkov-Glick model (chapter~\ref{ch:LMG}), the random transverse Ising model even in higher dimensions
and the realization of the cyclic Kitaev model on a chain (chapter~\ref{ch:Kitaev}) or a honeycomb lattice~\cite{Dutta2015}.
The spectrum for a much larger system and a different model is reported in Fig.~\ref{fig:spectrumLMG}.

%%At finite $N$, 
Tuning $\lambda$ permits to distinguish two different behaviours at low-energy levels. 
For $\lambda\gtrsim\lambdac$, the ground state $\ket{\psi_0}$ and the first excited state $\ket{\psi_1}$ are separated by an energy gap
$\DeltaE=E_1-E_0$. In the limit $N\to\infty$, the first gap $\DeltaE$ remains finite: 
the resulting disordered phase is termed \emph{gapped} and the proper scale of energy fluctuations is $\Delta=\DeltaE\textrm{\small$(N\to\infty)$}$.
For $\lambda\lesssim\lambdac$, the first two energy levels cluster into an almost degenerate doublet with energy $E_0 \approx E_1$;
% a similar bunching involve
and the same do the two higher levels corresponding to the second and third excited states with energy $E_2 \approx E_3$. 
The two doublets repel each other and are separated by an energy difference $\DeltaEE=E_2-E_1$.
In the limit $N\to\infty$, the two pairs of states truly become doubly degenerate
but their distance keeps to be finite: the resulting ordered phase is termed \emph{gapless} 
and the second energy gap is the proper finite energy scale $\Delta=\DeltaEE\textrm{\small$(N\to\infty)$}$ 
above the ground state.

To sum up, the scale of energy fluctuations $\Delta$ to be compared to thermal fluctuations at finite $T$
is identified as the lowest nonvanishing energy separation in the many-body spectrum: 
${\Delta=\Delta_n} \equiv E_{n}-E_{n-1}$ with $n=\min\{1,2,3\dots\}$
such that $\Delta_n\not\to0$ for $N\to\infty$.
In the thermodynamic limit --~where the universal properties of critical systems can be appropriately captured 
by a continuum quantum field theory and $\Delta$ is known as the``mass gap''~-- 
the ordinary convention in literature~\cite{Sachdev2011} uses $\Delta$ to indicate the energy separation 
between the ground and first excited states both when they are nondegenerate ($\lambda\gtrsim\lambdac$)
and when they are degenerate ($\lambda\lesssim\lambdac$). 
Nonetheless, since analyzing the finite-size scalings of spectral signatures of QPTs will be one of our major interest, 
we prefer to introduce the above distinction in the interpretation of $\Delta$ because we believe it is extremely useful at finite $N$, 
where the degeneracy of the states is lifted.

%%The qualitative change 
%%in the structure (symmetry, topology) of the ground state of the system,
%%as recognized by (order parameter, topological invariant).

\subsection{Fidelity susceptibility}  \label{subsec:FidelitySusceptibility}
A general and effective approach to QPTs is based on the concept of ``fidelity'', that 
it is rooted in quantum information theory and has a differential geometrical interpretation.
The fidelity between two pure states $|\psi\rangle$ and $|\phi\rangle$ in the same Hilbert space 
is defined as the amplitude of their overlap: $\pazocal{F}(|\psi\rangle,|\phi\rangle) = \big|\langle\psi|\phi\rangle\big|$.
Roughly speaking, it measures the closeness (that is, the similarity) between the two states;
since it is non-negative, symmetric and convex, the fidelity induces a metric in the Hilbert space, the so called ``Study-Fubini distance''. 
Provided that QPTs involve an abrupt change in the qualitative structure of the ground state, intuitively
the ground states of the many-body Hamiltonian~(\ref{QPTbasicH}) associated to infinitesimally close parameters
have an enhanced degree of orthogonality at the boundary between two phases: in fact,
$\pazocal{F}(\lambda,\lambda+\delta\lambda)=\big|\langle\psi_0(\lambda)|\psi_0(\lambda+\delta\lambda)\rangle\big|$
shows up a prominent drop at the critical points of a wide assortment of models~\cite{ZanardiPRE2006,GuIJMP2010}. 

In order to quantify the response of the fidelity to small changes of the driving parameter,
we define the ``fidelity susceptibility''~\cite{GuIJMP2010} as the first nontrivial leading term in the expansion of $\pazocal{F}$ 
in powers of $\delta\lambda$:
% up to the second order in $\delta\lambda$:
\be \label{QPTFidelitySusceptibility}
\chi_\lambda \,=\, -2 \lim_{\delta\lambda\to0}\frac{\log\pazocal{F}(\lambda,\lambda+\delta\lambda)}{(\delta\lambda)^2} \,\equiv\, -\frac{\ud^2\pazocal{F}}{\ud\lambda^2} 
\,=\, \sum_{n\neq0} \frac{\big|\bra{\psi_n(\lambda)}\HII\ket{\psi_0(\lambda)}\big|^2}{\big(E_n(\lambda)-E_0(\lambda)\big)^2} \, .
\ee
The last formulation in Eq.~(\ref{QPTFidelitySusceptibility}) establishes a meaningful relation with the structure of the spectrum:
since \emph{second-order} QPTs occur when the energy gap above the ground state vanishes ($E_1\to E_0$ in the thermodynamic limit),
a divergence of the fidelity susceptibility is expected at the critical points~\cite{VenutiPRL2007,ZanardiPRL2007,ZanardiPRB2007,GuIJMP2010}.
Away from criticality, if $\HII$ is a local operator (as it is usually the case for spin models, 
where it describes the coupling to an external field), the fidelity susceptibility grows extensively with the system size $N$
in gapped phases. In other words, if $\chi_\lambda$ is a superextensive quantity, the ground state of the system is gapless.

This approach to QPTs has been also applied to \emph{topological} QPTs~\cite{HammaPRB2008,AbastoPRA2008},
a class of zero-temperature transitions that take place between two different kinds of topological orders
and involve a change of some topological invariant.
Moreover, the definition of fidelity susceptibility has been extended to mixed states 
by replacing the standard fidelity with the Uhlmann fidelity between density matrices~\cite{Uhlmann1976,Jozsa1994},
and it has proven to recognize thermal critical points in CPTs~\cite{YouPRE2007,ZanardiPRL2007,ZanardiPRB2007}.
Thus, the fidelity susceptibility is able to detect disparate kinds of phase transitions 
with the huge advantage to need no \emph{a priori} knowledge of the local order parameter~\cite{ZanardiPRL2007,ZanardiPRB2007}:
this approach can be particularly convenient whenever the conceptual Ginzburg-Landau framework fails, 
due to the difficulty in identifying the suitable order parameter (such as in system where the broken symmetry is unknown) 
or to the actual absence of any local order parameter (such as in systems with topological order).

On the contrary, there is no general well-founded argument for expecting a singularity at \emph{first-order} QPTs, 
while it is still controversial whether the fidelity susceptibility 
can be helpful for systems undergoing \emph{higher-order} QPTs~\cite{YouPRE2007}.
In addition, fidelity susceptibility does not provide a direct quantification of the entanglement among the constituents of the system.

\section{Multipartite entanglement in critical systems} \label{sec:MEinQPT}
The study of entanglement close to quantum phase transitions sheds new light on these ubiquitous many-body phenomena 
and enhances our understanding of the puzzling behaviour of strongly-correlated interacting quantum systems 
beyond standard approaches of statistical mechanics. 
Merging the two concepts of entanglement and criticality is stimulating and still leave several open questions to be addressed or clarified.
A bridge between multipartiteness and critical points has recently been established: 
Ref.~\cite{HaukeNATPHYS2016} proposes an experimentally viable way for unveiling universal features of multipartiteness 
close to critical points, while Ref.~\cite{ZanardiPRA2008} describes the metrological potential of critical points 
as a resource for quantum-enhanced estimation of Hamiltonian coupling constants as well as temperature.

In what follows, we will briefly discuss, on a general ground, the central ideas of this thesis.
%% with no claim to be exhaustive.
In the subsequent chapters of the thesis we will test these ideas on a collection of benchmark models.
The original part of this section is contained in Ref.~\cite{Gabbrielli2018a}.

\subsection{Ground state} 
Albeit the ground state $\ket{\psi_0(\lambda)}$ lying close to the critical point is very susceptible to adiabatic deformation 
of the Hamiltonian $\hat{H}(\lambda)$ due to a variation of the control parameter $\lambda$, as in Eq.~(\ref{QPTFidelitySusceptibility}), 
in principle nothing guarantees a large susceptibility under unitary transformations, namely high values of the QFI, 
as in Eq.~(\ref{QFIunitaryMixed}).
However, the presence of large bipartite or pairwise entanglement in the ground state 
of a many-body system close to a quantum critical point
has been proved in different models~\cite{OsterlohNATURE2001,OsbornePRA2002,VidalPRL2003,WuPRL2004,AmicoRMP2008}.
\emph{A fortiori}, we clearly expect a growth of multipartite entanglement at QPTs once a preferred observable $\hat{O}$ is selected.

\paragraph{Optimal multipartiteness}
Unfortunately, there is no known method to identify the optimal operator $\hat{O}$, that is the one that allows the detection 
of the maximal multipartiteness $F_Q[\ket{\psi_0},\hat{O}]/N$. 
This is probably the major drawback in using the QFI as an entanglement witness.
A wise model-dependent choice can be suggested by the nature of the QPT itself.
%%%%%%%%
\\[-6pt]

\noindent \tripieno{MySky} \ For systems showing a \emph{symmetry-breaking pattern}, 
the transition is characterized by the divergence of fluctuations of a local order parameter:
we thus expect a large QFI at criticality when the optimal operator $\hat{O}$ 
is given by the order-parameter operator of the transition~\cite{HaukeNATPHYS2016}.
This is a collective local spin operator of the form discussed in paragraph~\ref{QFIbounds}
for the spin models of chapters~\ref{ch:Ising} and \ref{ch:LMG}, both with short-range and long-range interactions. 
One of our main findings is a distinguishable scaling for the optimal QFI in gapped disordered phases ($\DeltaE>0$, $\varphi=0$) 
and gapless ordered phases ($\DeltaE\to0$, $\varphi\neq0$):
%%%%%%%
\\[9pt]
%%%%%%%
\noindent \trivuoto{MyBlue} \ disordered phase \ $\Leftrightarrow$ \hspace{2pt} extensive Fisher information 
$F_Q\big[\ket{\psi_0},\hat{O}\big] \sim N$ \\
\phantom{\noindent \trivuoto{MyBlue} \ disordered phase \ $\Leftrightarrow$} \hspace{2pt} intensive multipartiteness
%$F_Q\big[\ket{\psi_0},\hat{O}\big]/N \sim \pazocal{O}(1)$
%%%%%%%
\\[3pt]
%%%%%%%
\noindent \trivuoto{MyBlue} \ ordered phase \hspace{13.4pt} $\Rightarrow$ \hspace{2pt} superextensive Fisher information  
$F_Q\big[\ket{\psi_0},\hat{O}\big] \sim N^b$ ($1<b\leq2$) \\
\phantom{\noindent \trivuoto{MyBlue} \ ordered phase \hspace{13.8pt} $\Rightarrow$} \hspace{1.6pt} subextensive or extensive multipartiteness 
%%%%%%%%
\\[-6pt]

%%%%%%%%%%%%
%% FIGURA %%
%%%%%%%%%%%%
\begin{figure}[!t]
\begin{center}
\includegraphics[width=0.9\textwidth]{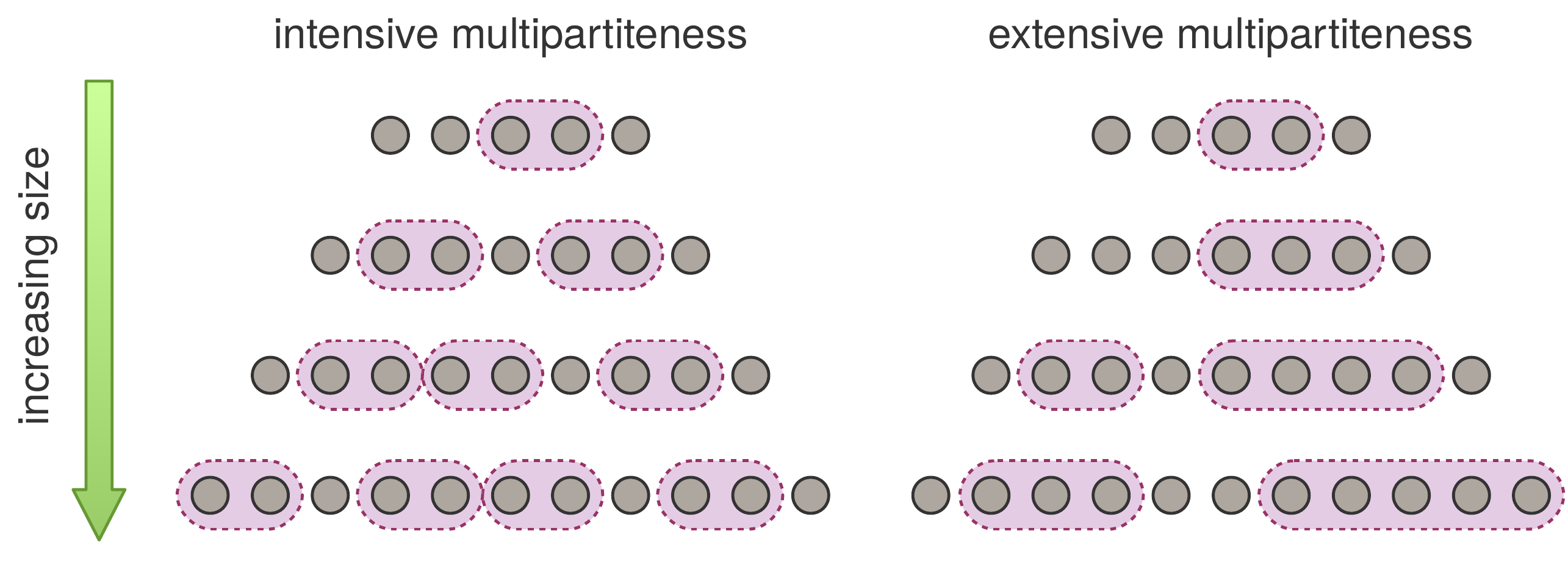}
\end{center}
\caption{Different scenarios for multipartiteness $F_Q[\hat{\rho}]/N$, 
that is the extension of the largest cluster of nonseparable parties contained in the state $\hat{\rho}$, 
when the total number $N$ of parties in the system increases. Gray circles can be particles, spins, momentum modes, lattice sites, 
or any other sort of physical parties among which entanglement can be established due to interactions, swapping, tunnelling and so on.
The purple shaded area denotes quantum correlations. For intensive multipartiteness $F_Q[\hat{\rho}]/N\sim\pazocal{O}(1)$, 
the entanglement depth is constant even in an infinite system. For extensive multipartiteness $F_Q[\hat{\rho}]/N\sim N$, 
the number of parties in the largest nonseparable cluster linearly grows with the system size.} \label{fig:GenericScaling}
\end{figure}

\noindent \tripieno{MySky} \ On the contrary, \emph{topological} QPTs are not detected by a local order parameter: 
in order to witness multipartite entanglement in topological models it is thus necessary to calculate the QFI 
with respect to nonlocal operators. For example, in the one-dimensional Kitaev chain of chapter~\ref{ch:Kitaev}, 
an optimal choice of operator is suggested by the correspondence to the Ising model via the Jordan-Wigner transformation.  
We phenomenologically observe 
%%%%%%%
\\[9pt]
%%%%%%%
\noindent \trivuoto{MyBlue} \ trivial phase \hspace{23.5pt} $\Leftrightarrow$ \hspace{2pt} extensive Fisher information 
$F_Q\big[\ket{\psi_0},\hat{O}\big] \sim N$ \\
\phantom{\noindent \trivuoto{MyBlue} \ trivial phase \hspace{24pt} $\Leftrightarrow$} \hspace{1.5pt} intensive multipartiteness
%$F_Q[\ket{\psi_0},\hat{O}]/N \sim \pazocal{O}(1)$
%%%%%%%
\\[3pt]
%%%%%%%
\noindent \trivuoto{MyBlue} \ topological phase \hspace{0.6pt} $\Rightarrow$ \hspace{2pt} superextensive Fisher information  
$F_Q\big[\ket{\psi_0},\hat{O}\big] \sim N^b$ ($1<b\leq2$) \\
\phantom{\noindent \trivuoto{MyBlue} \ topological phase \hspace{0.6pt} $\Rightarrow$} \hspace{2pt} subextensive or extensive multipartiteness 
%%%%%%%%
\\[-6pt]

\noindent Thus, at zero temperature, different quantum phases in ordinary models can be easily distinguished 
in terms of a different scaling of the entanglement content with the system size. 
The physical meaning of these different scalings is clarified pictorially in Fig.~\ref{fig:GenericScaling}.
%%%%%%%%
\\[-6pt]

%%%%%%%%
%and means that even the slightest perturbation of the ground state under the unitary transformation $\hat{U}=\neper^{-\ii\phi\hat{O}}$, 
%with arbitrary small $\phi$, can result in a high distinguishability of 
%\think{$\ket{\tilde{\psi}_0(\lambda)}=\hat{U}\ket{\psi_0(\lambda)}$} with respect to $\ket{\psi_0(\lambda)}$.
%In principle, crossing a critical point separating two phases should result in a singularity of $F_Q\big[\ket{\psi_0},\hat{O}\big]$.
%%%%%%%%

\noindent \tripieno{MySky} \ We believe the QFI is a sensitive probe for detecting quantum critical points 
when it is calculated via the optimal operator as discussed above.
In fact, \emph{at criticality} it must match the different scaling behaviours of the two phases
and typically this is accompanied by a change of concavity and a vertical tangent:
%Since the scaling is different in different phases, there must be a joint over the boundary be-
%tween two phases: this is only possible if multipartite entanglement scales differently at transition
%%%%%%%
\\[9pt]
%%%%%%%
\noindent \trivuoto{MyBlue} \ \ critical point \ \ $\Rightarrow$ \ $\partial_\lambda F_Q\big[\ket{\psi_0(\lambda)},\hat{O}\big]\to\infty$ 
in the thermodynamic limit 
%%%%%%%%
\\[-6pt]

\noindent This basic justification is corroborated by an explicit calculation using nondegenerate first-order perturbation theory:
\be \label{generalFirstDerivative}
\frac{\ud}{\ud\lambda}F_Q\big[\ket{\psi_0},\hat{O}\big] \,=\, 
8\,\mathrm{Re}\sum_{n\neq0} \,\frac{\bra{\psi_n}\HII\ket{\psi_0}\Big[ \bra{\psi_0}\hat{O}^2\ket{\psi_n} - 
2 \bra{\psi_0}\hat{O}\ket{\psi_0} \bra{\psi_0}\hat{O}\ket{\psi_n} \Big]
}{E_0-E_n} \, .
\ee
A vanishing denominator in Eq.~(\ref{generalFirstDerivative}) may lead to a singular behaviour: it occurs indeed at criticality, 
when $E_1\to E_0$ for $N\to\infty$.\footnote{\ Note that Eq.~(\ref{generalFirstDerivative}) 
cannot say anything about the gapless ordered phases (where the first gap is closed $\DeltaE=0$ in the thermodynamic limit)
since the nondegenerate perturbative theory cannot be applied.\\[-9pt]} 
%%%%%%%%
\\[-6pt]

\noindent For a $d$-dimensional model, the peculiar scaling behaviour of the QFI with respect to $N$ at criticality $\lambda = \lambdac$ follows from a standard scaling hypothesis~\cite{HaukeNATPHYS2016}: 
\be F_Q\big[\ket{\psi_0},\hat{O}\big] \sim N^{1 + \Delta_Q/d}, 
\ee
where $\Delta_Q$ is the scaling exponent of the two-body correlation function 
$\langle\hat{\phi}_i\hat{\phi}_j\rangle \sim |i-j|^{\Delta_Q-d}$ 
and can be related to critical exponents: $\Delta_Q = 2-\eta-z=\gamma/\nu-z$~\cite{Gabbrielli2018a}.

\subsection{Finite temperature}
\begin{figure}[t!]
\centering
\includegraphics[width=0.49\textwidth]{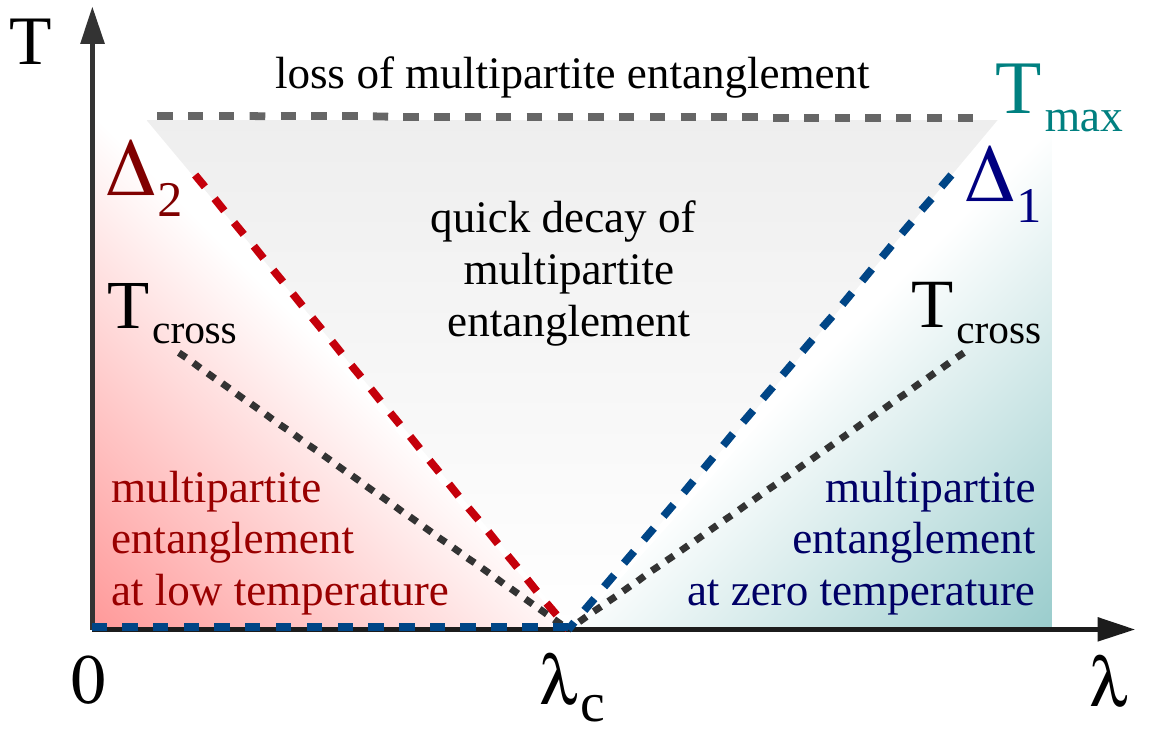} \hfill
\includegraphics[width=0.49\textwidth]{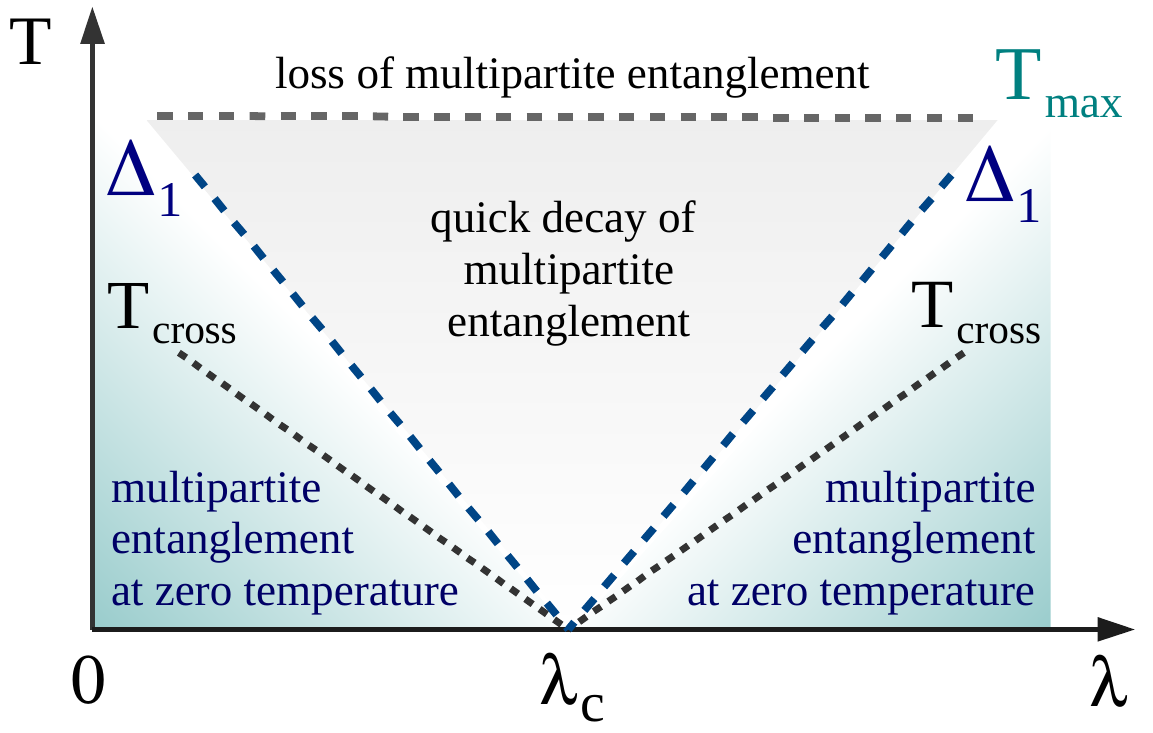}
\caption{Universal thermal phase diagram of multipartite entanglement as detected by the quantum Fisher information. 
The V-shaped structure with vertex at the critical point is generated by the crossover temperature 
$\Tcross\approx\Delta\propto J\,|\lambda-\lambdac|^{z\nu}$.
%We observe a decay of ME with temperature.
Wherever the ground state is nondegenerate, we have $\Delta=\DeltaE$ and the zero-temperature entanglement 
quantified by $F_Q[\ket{\psi_0}]$ is preserved on a quantum plateau $T\ll\DeltaE$ (blue area).
If the ground state is degenerate, we have $\Delta=\DeltaEE$ and the entanglement 
quantified by $F_Q[\hat{\rho}_T]$ at $T\to0$, when present, is preserved on a thermal plateau $T\ll\DeltaEE$ (red area).
Above a threshold $T\gtrsim\Tmax$ of the order of the inner interaction energy scale, no entanglement is witnessed.
\textbf{(Left)} In systems with spontaneous symmetry breaking (e.g. the Ising chain and the Lipkin-Meshkov-Glick model)
the thermal and quantum plateaux are found on the two sides of the critical point.
\textbf{(Right)} In absence of ground-state degeneracy (e.g. in the Kitaev cyclic chain), 
the thermal plateau is not present and the quantum plateau is found on both sides of the critical point.}
\label{fig:thermalDecay}
\end{figure}

An open question in quantum information concerns the interplay between zero-point quantum fluctuations and thermal excitations:
to what extent do quantum correlations survive at finite temperature? 
An answer to this question might spark a substantial breakthrough also in the field of quantum technologies.
We face the problem by calculating the quantum Fisher information 
to witness multipartite entanglement at finite temperature in the vicinity of a quantum critical point.
Critical points induce a characteristic behaviour in the landscape of multipartite entanglement,
that faithfully follows the finite-temperature texture of the phase-diagram. 
As a main result, we are able to draw the universal low-temperature phase diagram shown in Fig.~\ref{fig:thermalDecay}.

%%%%%%%%%%%%%%%%%%%%%%%%%%%%%%%%%%%%%%%%%%%%%%%%%%%%%%%%%%
%%%%%%%%%%%%%%%%%%%%%%%%%%%%%%%%%%%%%%%%%%%%%%%%%%%%%%%%%
%%%%%%%%%%%%%%%%%%%%%%%%%%%%%%%%%%%%%%%%%%%%%%%%%%%%%%%%%
%%%%%%%%%%%%%%%%%%%%%%%%%%%%%%%%%%%%%%%%%%%%%%%%%%%%%%%%%

\paragraph{Universality of the thermal phase diagram}
Let us consider a generic system modelled by the many-body Hamiltonian $\hat{H}(\lambda) = J\,\big(\HI + \lambda\,\HII\big)$, 
with $[\HI,\HII]\neq0$ like in Eq.~(\ref{QPTbasicH}), having eigenenergies $E_n$ and corresponding eigenstates $\ket{\psi_n}$
labelled by the integer quantum number $n\geq0$ and hosting a critical point at $\lambdac$. 
The thermal state describing the system at canonical equilibrium with a bath 
at temperature $T$ simply reads $\hat{\rho}_T = \pazocal{Z}^{-1}\,\neper^{-\hat{H}(\lambda)/T}$, 
% $\hat{\rho}_T = \pazocal{Z}^{-1}\,\neper^{-\hat{H}/T} = \pazocal{Z}^{-1}\sum_n \neper^{-E_n/T}\ket{\psi_n}\bra{\psi_n}$, 
where we are measuring temperature in units of energy ($k_{\rm B}=1$) 
and $\pazocal{Z}=\tr\big[\neper^{-\hat{H}(\lambda)/T}\big]$ is the canonical partition function.
According to the expression in Eq.~(\ref{QFIunitaryMixed}), we can write the QFI of the thermal state as
\be \label{thermalQFImixed}
F_Q\big[\hat{\rho}_T,\hat{O}\big] = \frac{2}{\pazocal{Z}} \sum_{n,\,m} \frac{\big(\neper^{-E_n/T} - \neper^{-E_m/T}\big)^2}{\neper^{-E_n/T} + \neper^{-E_m/T}} \,\Big|\langle \psi_n |\hat{O}| \psi_m \rangle\Big|^2 \, .
\ee

\noindent \tripieno{MySky} \ In the thermodynamic limit, at low enough temperature
-- such that we can take into account only the two lowest energy levels -- the QFI follows a simple universal law:\footnote{\ A closer inspection revealed that the right-hand side of Eq.~(\ref{QFIfactorization}) is actually a thermal lower bound for the QFI. The interested Reader can find details in Ref.~\cite{Gabbrielli2018a}.}
\be \label{QFIfactorization}
F_Q\big[\hat{\rho}_T,\hat{O}\big] = 
F_Q\big[\hat{\rho}_0,\hat{O}\big]\tanh^2\bigg(\frac{\Delta}{2T}\bigg)\,\mu\,\frac{1+\neper^{-\Delta/T}}{\mu+\nu\,\neper^{-\Delta/T}} \, ,
\ee
where $\mu$ and $\nu$ indicate the degeneracy of the ground state and first excited state respectively, 
$\Delta$~is the first nonzero energy separation in the many-body spectrum as discussed in paragraph~\ref{subsec:QuantumCriticality}
and $\hat{\rho}_0$ is the state of the system in the limit $T\to0$, i.e. the incoherent mixture of the $\mu$-fold degenerate ground state.
A spectacular feature of Eq.~(\ref{QFIfactorization}) is the \emph{factorization} of the QFI
into a quantum contribution and a thermal contribution. 
The former, $F_Q[\hat{\rho}_0,\hat{O}]$, is the amount of entanglement witnessed at low temperature. 
The latter, instead, provides the decay of witnessed entanglement for increasing temperature: it is due to a sort of thermal decoherence 
and depends only on the structure of the low-energy spectrum, namely the degeneracy of the two lowest-energy eigenstates 
and their separation.
This factorization holds regardless of the microscopical details of the system.
Remarkably, concurrence and mutual information seem not to exhibit a similar factorization~\cite{AmicoEPL2007,WilmsJSM2012}.
Notwithstanding, we underline that Eq.~(\ref{QFIfactorization}) only holds for temperature $T\lesssim\Delta$, 
sufficiently low to neglect higher excited states.
%%%%%%%
\\[6pt]
%%%%%%%
\noindent \trivuoto{MyBlue} \ If the ground state $\ket{\psi_0}$ is nondegenerate ($\mu=1$), 
like at $\lambda>\lambdac$ in Fig.~\ref{fig:QuantumCriticality},
the system is in the pure state $\hat{\rho}_0=\ket{\psi_0}\bra{\psi_0}$ in the limit $T\to0$, therefore
\be \label{QFI0nondegerate}
F_Q\big[\hat{\rho}_0,\hat{O}\big] \equiv F_Q\big[\ket{\psi_0},\hat{O}\big] = 
4\,\Big(\bra{\psi_0}\hat{O}^2\ket{\psi_0}-\bra{\psi_0}\hat{O}\ket{\psi_0}^2\Big) \, .
\ee 
Equation~(\ref{QFIfactorization}) shows that the QFI remains constant 
$F_Q[\hat{\rho}_T]\approx F_Q[\ket{\psi_0}]$ up to temperatures $T\propto\Delta$,
with a proportionality constant of the order of $\log(3+\nu)$. 
This fact permits to define a ``quantum plateau'' $\Tcross\approx\Delta$ where zero-temperature witnessed multipartite entanglement 
(if present) is insensitive to thermal fluctuations, protected by the finite energy gap $\Delta$.
%%%%%%%
\\[6pt]
%%%%%%%
\noindent \trivuoto{MyBlue} \ If the ground state $\ket{\psi_0}$ is degenerate ($\mu>1$),
like at $\lambda<\lambdac$ in left-hand panel of Fig.~\ref{fig:QuantumCriticality}, 
in the limit $T\to0$ the system is found in the equally weighted mixture 
$\hat{\rho}_0 = \frac{1}{\mu}\sum_{n=0}^{\mu-1}\ket{\psi_n}\bra{\psi_n}$, hence
\be \label{QFI0degerate}
F_Q\big[\hat{\rho}_0,\hat{O}\big] = \frac{4}{\mu}\,\sum_{n=0}^{\mu-1}\bigg[\bra{\psi_n}\hat{O}^2\ket{\psi_n}-\sum_{m=0}^{\mu-1}\Big|\bra{\psi_n}\hat{O}\ket{\psi_m}\Big|^2\bigg] \, .
\ee 
This value is, in general, much smaller that the QFI for the ground state $F_Q[\ket{\psi_0}]$.
This corresponds to a sharp decay\footnote{\ The multipartiteness $F_Q[\ket{\psi_0}]/N$ 
witnessed in the ground state $\ket{\psi_0}$ at $T=0$ decays to the value $F_Q[\hat{\rho}_0]/N$ 
exponentially fast in the system size $N$ at very small temperature. Here we are limiting to the thermodynamic limit:
see chapters~\ref{ch:Ising} and \ref{ch:LMG} for an extensive discussion of the finite-size behaviour.\\[-9pt]} 
of witnessed multipartite entanglement that becomes a discontinuity at $T=0$ in the thermodynamic limit. 
According to Eq.~(\ref{QFIfactorization}), the QFI of the incoherent mixture $\hat{\rho}_0$ remains constant for $T\lesssim\Tcross$, 
where $\Tcross\approx\Delta$ up to a numerical prefactor of order $\log(3/\mu+\nu\mu^2)$.
Thus, we can define a ``thermal plateau'', where thermal fluctuations strongly affect the multipartite entanglement of the ground state 
but not the one of the incoherent mixture.
%%%%%%%
\\[6pt]
%%%%%%%
\noindent \trivuoto{MyBlue} \ In the noticeable case the degeneracy of the two lowest states is the same $\mu=\nu$, 
as in all the models we will discuss in the following (where, in particular, $\mu=\nu=1$ or $\mu=\nu=2$), 
the peculiar decay function $\tanh^2(\Delta/2T)$ allows for a convenient identification of the border 
of the (thermal and quantum) plateaux as the curve of the inflection points of the QFI with respect to the temperature:
$\partial^2 F_Q/\partial T^2=0$ at $T=\Tcross$, signalling the maximum variation of $F_Q(T)$ for small variation of $T$. 
As expected, $\Tcross$ follows the energy gap $\Delta$ (i.e. $\Delta/\Tcross\approx2.70\approx{\rm e}$), 
thus allowing to identify the entanglement crossover $\Tcross$ as the finite-temperature crossover defining the QC region,
%where the tail of the decay function leads, 
and making it detectable in principle.
The universal low-temperature law~(\ref{QFIfactorization})
shows that the QFI only depends on the ratio $\DeltaE/T$ (if $\DeltaE>0$) or $\DeltaEE/T$ (if $\DeltaE=0$).
The numerical results presented in chapters~\ref{ch:Ising}, \ref{ch:LMG} and \ref{ch:Kitaev} show that 
$\DeltaE$ or $\DeltaEE$ are the proper energy scales also beyond the two-mode approximation. 
%%%%%%%%
\\[-6pt]

\noindent \tripieno{MySky} \ At higher temperature $T\gtrsim\Delta$ and close to the critical point, where $\Delta=0$, 
the calculation of Eq.~(\ref{thermalQFImixed}) involves many energy eigenstates 
and it is not possible to factorize a quantum and a thermal contribution: 
their interplay characterizes the QC region in the diagram of Fig.~\ref{fig:QuantumCriticality}.
In this regime, Eq.~(\ref{QFIfactorization}) does not hold 
and the behaviour of the QFI with temperature is a power-law decay ruled by critical exponents~\cite{Gabbrielli2018a,HaukeNATPHYS2016}:
\be \label{QFIScalingBehaviour}
F_Q\big[\hat{\rho}_T,\hat{O}\big] \sim T^{1+(\eta-2)/z} \, ,
% \,g\big(N^{-1/d}\,T^{-1/z},\,\tilde{\lambda}\,T^{-1/(z\nu)}\big) \, ,
\ee
where 
% $\tilde{\lambda}=|\lambda-\lambdac|/\lambdac$ is the normalized distance from the critical point, $d$ is the dimensionality of the system
$z$ and $\eta$ are the critical exponents introduced in paragraph~\ref{subsec:QuantumPhaseTransitions}. 
% and $g$ is a suitable scaling function.
The range of validity of Eq.~(\ref{QFIScalingBehaviour}) defines the width of quantum criticality in Fig.~\ref{fig:QuantumCriticality}.
%%%%%%%
\\[-6pt]

\noindent All these general considerations result into the phase diagrams of Fig.~\ref{fig:thermalDecay}. 
%Our main intention is to characterize the phase diagram \think{both at T zero and finite T} in terms of ME. 
%At the moment, no clear straightforward direct relation to the structure of QC unveiled in paragraph...  and charaterization in terms of $\xi$.
%We just claim that ME displays crossovers . Similar crossover can be found also looking at the concurrence [] or observables not directly related to entnaglment []. 
%THE ME has the advantage to show a separation of contributions, having direct relation to usefulness of entanglement for applications
%in quantum ...
%and we claim this is an experimentally-feasible method to detect quantum criticality.

%%%%%%%%%%%%%%%%%%%%%%%%%%%%%%%%%%%%%%%%%%%%%%%%%%%%%%%%%
%%%%%%%%%%%%%%%%%%%%%%%%%%%%%%%%%%%%%%%%%%%%%%%%%%%%%%%%%
\begin{figure}[t!]
\centering
\includegraphics[width=0.49\textwidth]{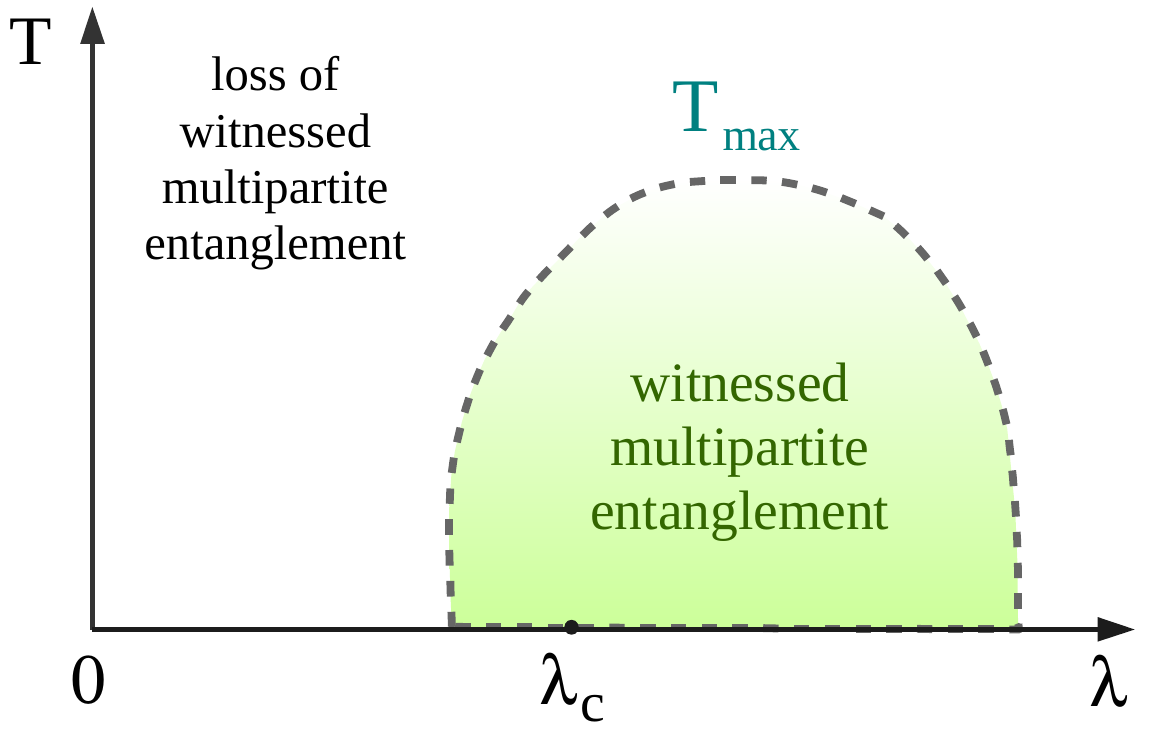} \hfill
\includegraphics[width=0.49\textwidth]{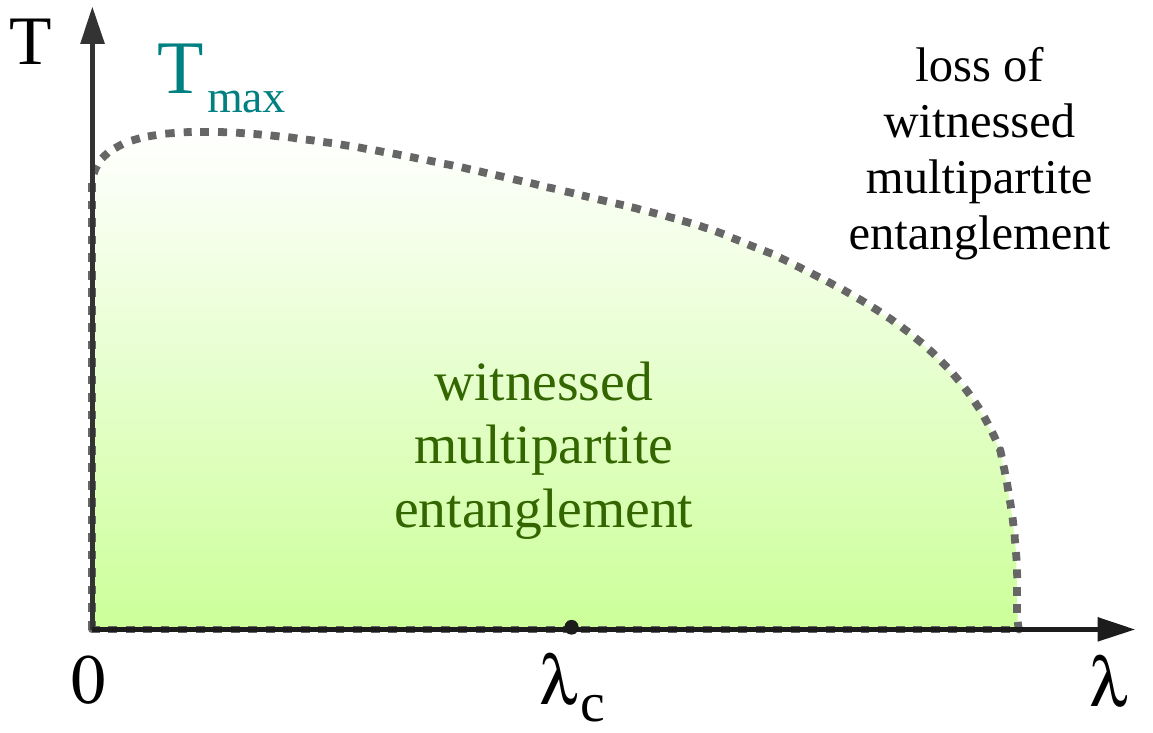}
\caption{Sketch of thermal multipartiteness $F_Q[\hat{\rho}_T]/N$ on the plane $\lambda$--$T$. 
Multipartite entanglement is witnessed in a limited region around the critical point (green area).
\textbf{(Left)} Systems with spontaneous symmetry breaking are gapless ($\DeltaE=0$) in the ordered phase $\lambda<\lambdac$:
the region of entanglement has a maximum extension over temperature $\Tmax$ at $\lambda\gtrsim\lambdac$ 
and then shrinks at $\lambda\lesssim\lambdac$.
\textbf{(Right)} System with gapped phases ($\DeltaE>0$) on the two sides of $\lambdac$, 
typically exhibit entanglement on a broader range of $\lambda$.
The two pictures should be compared to the two pictures of Fig.~\ref{fig:thermalDecay}, respectively.
Not necessarily the scales of the horizontal axis are the same:
usually, Fig.~\ref{fig:thermalDecay} focus on a narrower range of $\lambda$ around $\lambdac$.}
\label{fig:thermalMultipartiteness}
\end{figure}
%%%%%%%%%%%%%%%%%%%%%%%%%%%%%%%%%%%%%%%%%%%%%%%%%%%%%%%%%
%%%%%%%%%%%%%%%%%%%%%%%%%%%%%%%%%%%%%%%%%%%%%%%%%%%%%%%%%

\paragraph{Survival of thermal multipartiness}
Intriguingly, the persistence of the universal features of quantum criticality seems to be connected to 
the survival of multipartite entanglement witnessed by the QFI at finite temperature (i.e. $F_Q[\hat{\rho}_T]/N>1$).
In fact, on the one side, the QC region of Fig.~\ref{fig:QuantumCriticality} -- 
where a scaling analysis such as the one giving rise to Eq.~(\ref{QFIScalingBehaviour}) can be applied --
is expected to dissolve at high temperature~\cite{Sachdev2011}. 
Theoretical considerations~\cite{KoppNATPHYS2005} on prototypical short-range spin models
predicts this crossover to occur at about $T\approx J$, where $J$ is the scale of the interaction energy: see Eq.~(\ref{QPTbasicH}).
A recent experiment based on the measure of relaxation rate~\cite{KinrossPRX2014} managed to confirm this prediction. 
On the other side, multipartite entanglement is also expected to be lost 
above a maximal threshold temperature~\cite{NakataPRA2009,SadiekJPB2013,HaukeNATPHYS2016}, 
as illustrated in Fig.~\ref{fig:thermalMultipartiteness}.
A look into the different models studied in chapters~\ref{ch:Ising}, \ref{ch:LMG} and \ref{ch:Kitaev} reveals
that the Fisher information or the spin-squeezing parameter typically witness no entanglement for $T>\Tmax\approx J$. 

These observations suggest the exciting hypothesis that the QC region could be connected to
-- perhaps induced by, or at least probed with -- multipartite entanglement.
At this level, this is only a conjecture: a deep understanding is presently far to be well established 
and it is beyond the scope of this work.
It will be our primary aim in the next chapters to study multipartite entanglement both at zero and finite temperature
in order to 	strengthen this heuristic view.

\chapter[Ising model]{A benchmark model: \\ the Ising chain in a transverse field} \label{ch:Ising}
% A test-bed / prototypical / paradigmatic / simple model / example: the quantum Ising chain in a transverse field

{\sl % We witness multipartite entanglement in the transverse Ising chain, 
We investigate the behaviour of multipartite entanglement as quantified by the quantum Fisher information in the transverse Ising chain, 
a paradigmatic one-dimensional model exhibiting continuous symmetry-breaking quantum phase transitions. 
After providing a complete phase diagram for any strength and range of pairwise interaction, 
we prove that multipartiteness can recognize the ordered phases in terms of an extensive behaviour for increasing chain length.
Moreover, at critical points the quantum Fisher information is found to be singular. 
The effect of thermal fluctuations on the critical features of the nearest-neighbour chain is also studied:
multipartite entanglement in the ground state can survive at finite temperature in the vicinity of quantum critical points.}

\section*{Introduction}
The quantum Ising model in a transverse field represents the simplest example of interacting many-body system 
to exhibit zero-temperature phase transitions driven by zero-point quantum fluctuations.
Whenever the major properties of a quantum physical system arise from the competition between cooperative pairwise interactions 
among the microscopic constituents and a tunable noncommuting local external perturbation, 
a reasonable investigation of the system finds a good ally in the quantum Ising model or one of its many variants.
The easiest and most studied embodiment of such a model considers spins $\sfrac{1}{2}$.

The model can be formulated in any spatial dimension. In particular, it has been intensively studied -- and experimentally realized -- 
on a linear chain ($d=1$) and a square lattice ($d=2$)~\cite{Suzuki2013}. 
Interestingly, there exists a one-to-one mapping from the nondisordered $d$-dimensional quantum Ising model 
to the nondisordered $(d+1)$-dimensional classical Ising model, the dynamical exponent being $z=1$~\cite{Suzuki2013}.
This correspondence is responsible for the formal equivalence between the ground-state phase transition in dimension $d$
and the thermally-driven phase transition in dimension $(d+1)$: the two transitions share the same critical exponents~\cite{Suzuki1976}.

The \emph{one-dimensional} version for nearest-neighbour interactions first appeared more than fifty years ago, 
in the search for the exact solution of the two-dimensional ferromagnetic classical Ising model~\cite{Lieb1961}. 
Immediately after its introduction, the model was employed to theoretically describe the order-disorder transition 
in a real ferromagnetic crystal~\cite{deGennes1963}.
Since then, it never stopped providing insights and finding application in a large number of phenomena, including 
frustration~\cite{ChakravartyPRB1989, Chakrabarti1996}, nonequilibrium dynamics following a quench~\cite{ZurekPRL2005}, 
quantum phase transitions~\cite{Dutta2015} and quantum annealing~\cite{DasRMP2008}.
Furthermore, a plethora of condensed-matter systems can be described by this model~\cite{Suzuki2013}.

In this chapter we study the entanglement content in the Ising chain for arbitrary interaction range.
A basic introduction to the transverse Ising chain is provided in section~\ref{sec:IsingModel}.
In section~\ref{sec:IsingNN} we focus on the case of nearest-neighbour interaction, 
whose properties are well understood and widely available in literature.
After reviewing the analytical tools for exactly treating the ground-state properties, 
we discuss the connection between the quantum critical points and
multipartite entanglement witnessed by the quantum Fisher information computed via the order parameter, 
along the lines of Ref.~\cite{LiuJPA2013}. 
We also investigate the interplay of quantum and thermal fluctuations 
and study the multipartite entanglement at finite temperature in the vicinity of the critical points. 
Our findings are consistent with the general exposition carried out in chapter~\ref{ch:Intro} 
and agree with similar results presented in the seminal work of Ref.~\cite{HaukeNATPHYS2016}.
Additionally, we devote section~\ref{sec:IsingArbitrary} to the extension of the model to long-range interactions, 
modelled by means of suitable decaying power laws.
By extending the use of the fidelity susceptibility, initiated in Ref.~\cite{DamskiPRE2013} for the short-range model,
we draw a complete phase diagram and show an intriguing correspondence to the finite-size scaling of multipartite entanglement. 
The results for antiferromagnetic coupling agree with the ones previously determined 
in Refs.~\cite{KoffelPRL2012, VodolaNJP2016} by using bipartite entanglement.
We are not aware of previous studies on multipartite entanglement for arbitrary-range interactions.
The original parts of the present chapter are collected in the papers~\cite{Gabbrielli2018b} and \cite{Gabbrielli2018a}.

\section{The model} \label{sec:IsingModel}
The quantum transverse Ising model~\cite{Dutta2015} describes a one-dimensional equispaced\footnote{\ The length of the chain is $L=\textsl{a}N$, with \textsl{a} the lattice spacing. We set $\textsl{a}=1$ so that there is no difference between the system size and the number of spin upon it.\\[-9pt]} chain of $N$ spins $\sfrac{1}{2}$
interacting via a pairwise exchange coupling and subject to a transverse magnetic field. 
Considering an arbitrary range for the spin-spin interaction and adopting an angular parametrization, 
the Hamiltonian of the model reads
\be \label{IsingHam}
\hat H_{\alpha} = J\sin\theta\,\displaystyle \sum_{i=1}^N \sum_{j>i}^N \frac{\,\hat{\sigma}_z^{(i)} \hat{\sigma}_z^{(j)}}{\,|i-j|^\alpha} 
\, + \, J\cos\theta\,\displaystyle\sum_{i=1}^N \hat{\sigma}_x^{(i)} \, .
\ee
Here, $\hat{\sigma}^{(i)}_{\varrho}=2\hat{s}^{(i)}_{\varrho}$ (with $\varrho\in\{x,y,z\}$) are the Pauli matrices associated 
to the $i$th spin, % \think{operator $\hat{\mathbf{s}}^{(i)}$}, 
satisfying the standard angular-momentum commutation relations 
$\big[\hat{\sigma}_\varrho^{(j)},\hat{\sigma}_\varsigma^{(j)}\big]=2\,\ii\,\varepsilon_{\varrho\varsigma\tau}\hat{\sigma}_\tau^{(j)}$ 
at the same site $i$ (we are assuming natural units $\hbar=1$), and commuting at two different sites $i \neq j$.
The Hamiltonian is invariant under the global $Z_2$ transformation generated by $\prod_i\hat{\sigma}_x^{(i)}$, 
that flips all the spins along the $z$ direction: 
$\hat{\sigma}_x\mapsto\hat{\sigma}_x$, $\hat{\sigma}_y\mapsto-\hat{\sigma}_y$ and $\hat{\sigma}_z\mapsto-\hat{\sigma}_z$.
% Under time reversal, the direction of the field changes since the direction of the current in a loop that produces the magnetic field will get reversed. Spins S are also taken to flip under time reversal.
 
\begin{figure}[t!]
\centering
\includegraphics[width=1\textwidth]{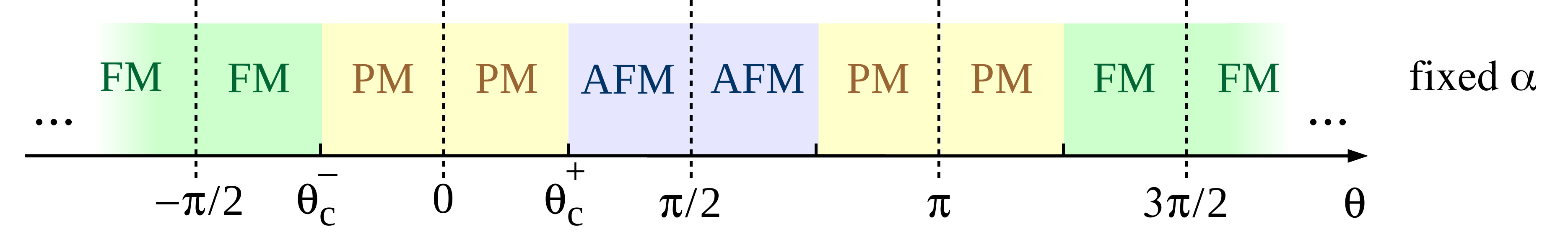}
\caption{Sketched phase diagram of the transverse Ising model for arbitrary fixed interaction range $\alpha$.
The symmetry around $\theta=\frac{\pi}{2}$ and the $2\pi$-periodicity are evident.
Symbols are explained in the text.}
\label{fig:IsingQualitativePhaseDiagram}
\end{figure}

The exchange energy has a strength $J\sin\theta$, where the constant $J>0$ sets the microscopic energy scale 
and $\theta\in\big[-\frac{\pi}{2},+\frac{\pi}{2}\big]$ is the parameter used to tune the system across the QPT.\footnote{\ The Hamiltonian is invariant under the transformation $\theta\mapsto\theta\pm\pi$ and $J\mapsto-J$, thus the phase diagram is symmetric around $\theta=\frac{\pi}{2}$, as shown in Fig.~\ref{fig:IsingQualitativePhaseDiagram}.
We limit our discussion to the interval $\theta\in\big[-\frac{\pi}{2},+\frac{\pi}{2}\big]$.\\[-9pt]}
The exchange energy decays with distance according to a power law determined by the exponent $\alpha\geq0$:
the interaction is said to be \emph{short ranged} when $\alpha>1$, 
while it is named \emph{long ranged} when $\alpha\leq1$, 
according to different distinctive properties that will be discussed later on.
In particular, $\alpha\to\infty$ corresponds to the nearest-neighbour interaction 
(each spin feels the presence of the spins on nearest-neighbour sites only), 
while $\alpha\to0$ corresponds to the infinite-range interaction 
(each spin of the system is coupled with all the other spins by the same coupling amplitude).
If $0<\theta\leq\frac{\pi}{2}$, the interaction is \emph{antiferromagnetic} (AFM): 
for $\alpha\to\infty$, neighboring spins prefer to be oriented in antiparallel directions. 
If $-\frac{\pi}{2}\leq\theta<0$, the interaction is \emph{ferromagnetic} (FM): 
spins tend to align in the same orientation.
The $z$ direction is chosen as the preferential axis along which the spins orient in absence of external field.

The potential energy\footnote{\ We will often refer to the potential energy as the amplitude of the transverse external magnetic field. 
To be precise, the two quantities are just proportional through the magnetic moment $\mu$ of the particle carrying the spin~$\sfrac{1}{2}$,
but it is customary to simplify the language by setting $\mu=1$.\\[-9pt]} 
$J\cos\theta$ is due to a field directed on the perpendicular plane: 
specifically along the negative $x$ direction, with no loss of generality.
The exchange and potential energies are not independent: their addition in quadrature must give the constant $J$.
This constrain results from the need of only one parameter for describing the entire class of Ising Hamiltonians:
usually in literature one of the two coefficients is discarded, being taken as unit for measuring energy.
In Hamiltonian~(\ref{IsingHam}), the ratio between the coefficient is the dimensionless coupling $\,\tan\theta\in(-\infty,+\infty)$. 
The limit case $\theta=0$ describes the system with dominant magnetic field;
conversely, at $\theta=\pm\frac{\pi}{2}$ the field is null and only spin-spin interaction is left.

The chain considered in the following is an \emph{open} one, with free end: we do not impose any boundary conditions.
In numerical simulations and exact calculations (as in section~\ref{sec:IsingNN}), we consider even $N$ for simplicity.

\begin{figure}[t!]
\begin{minipage}{0.51\textwidth}
\centering
\includegraphics[width=0.5\textwidth]{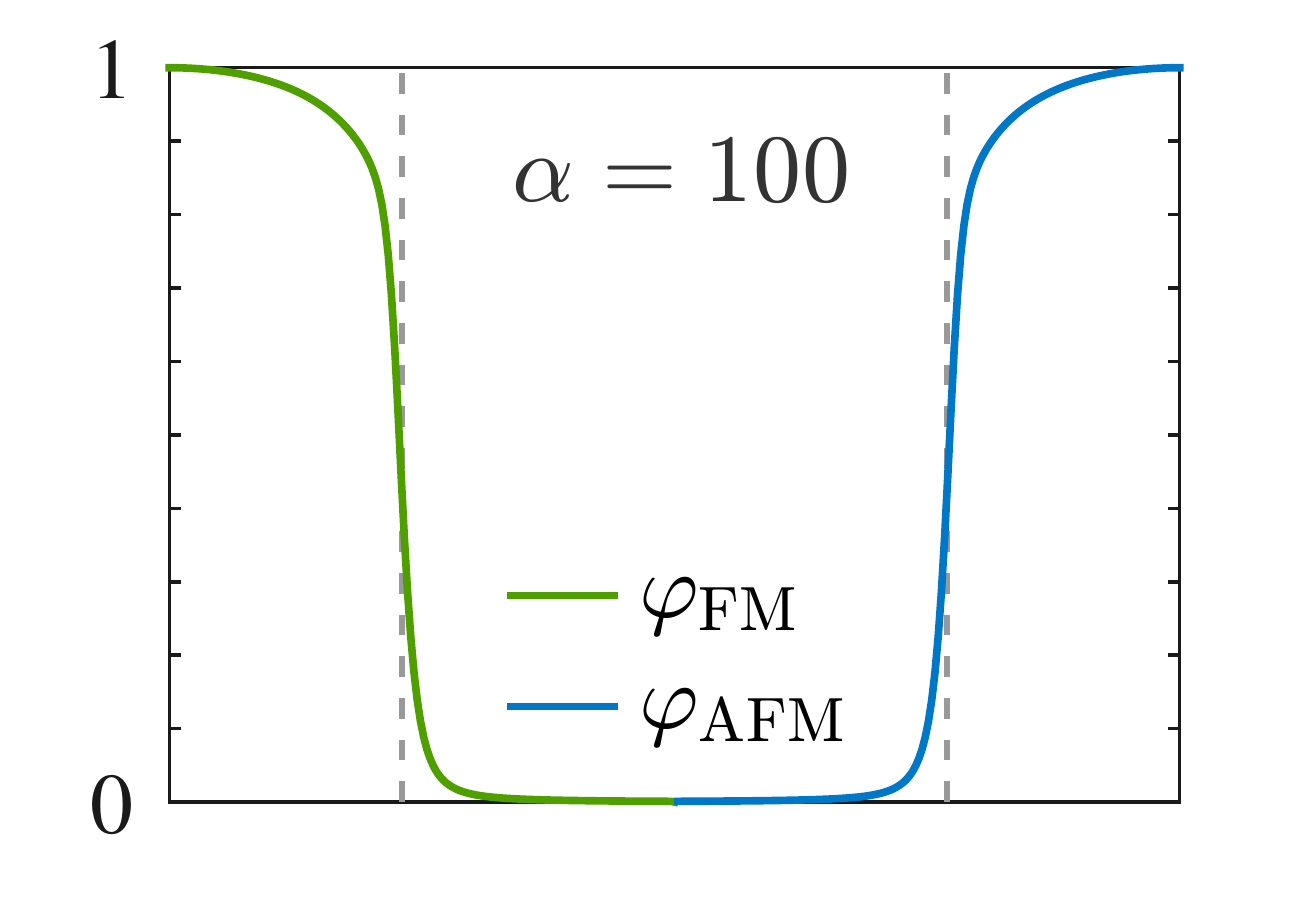} \hspace{-12pt}
\includegraphics[width=0.5\textwidth]{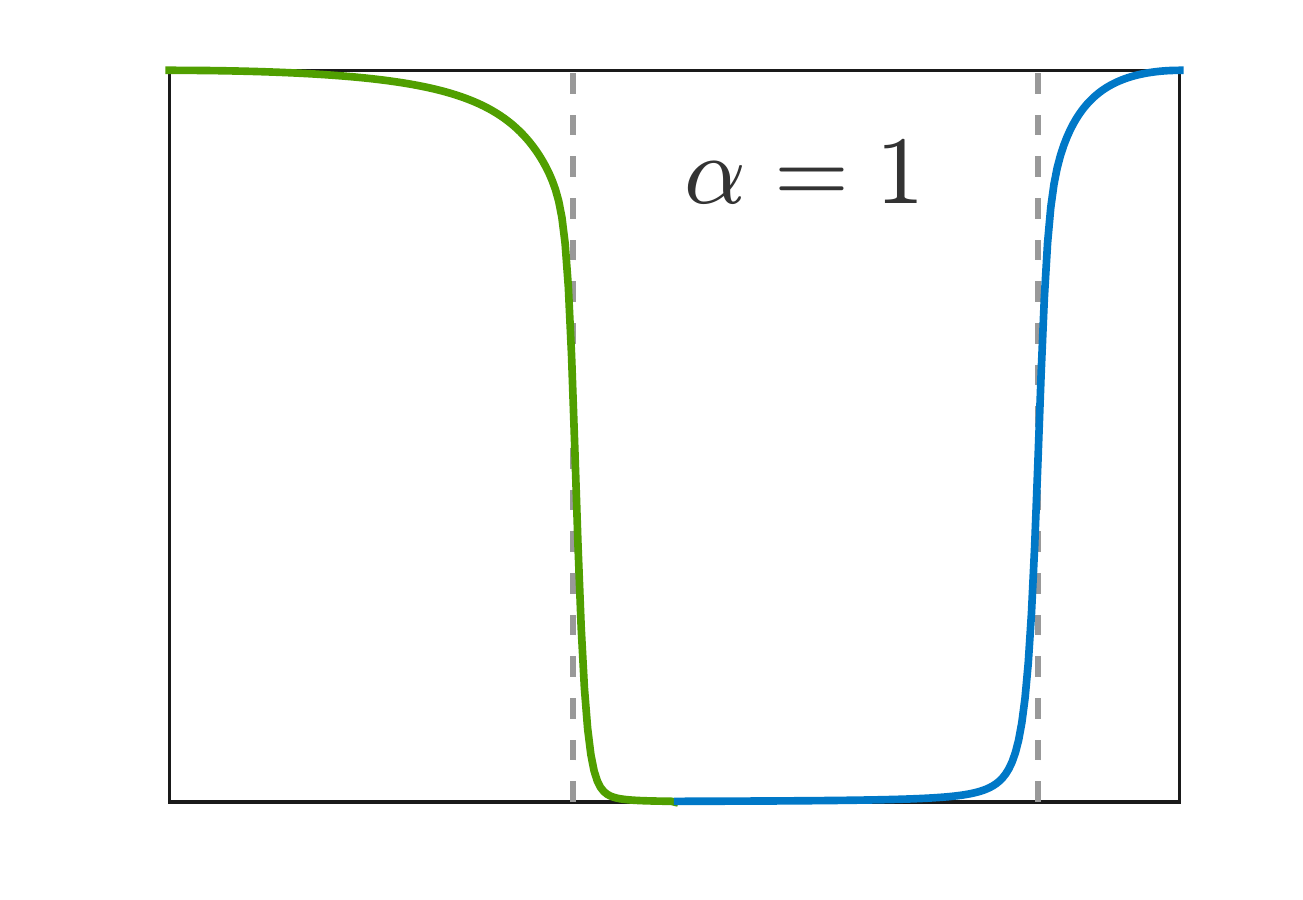} \\[-6pt]
\includegraphics[width=0.5\textwidth]{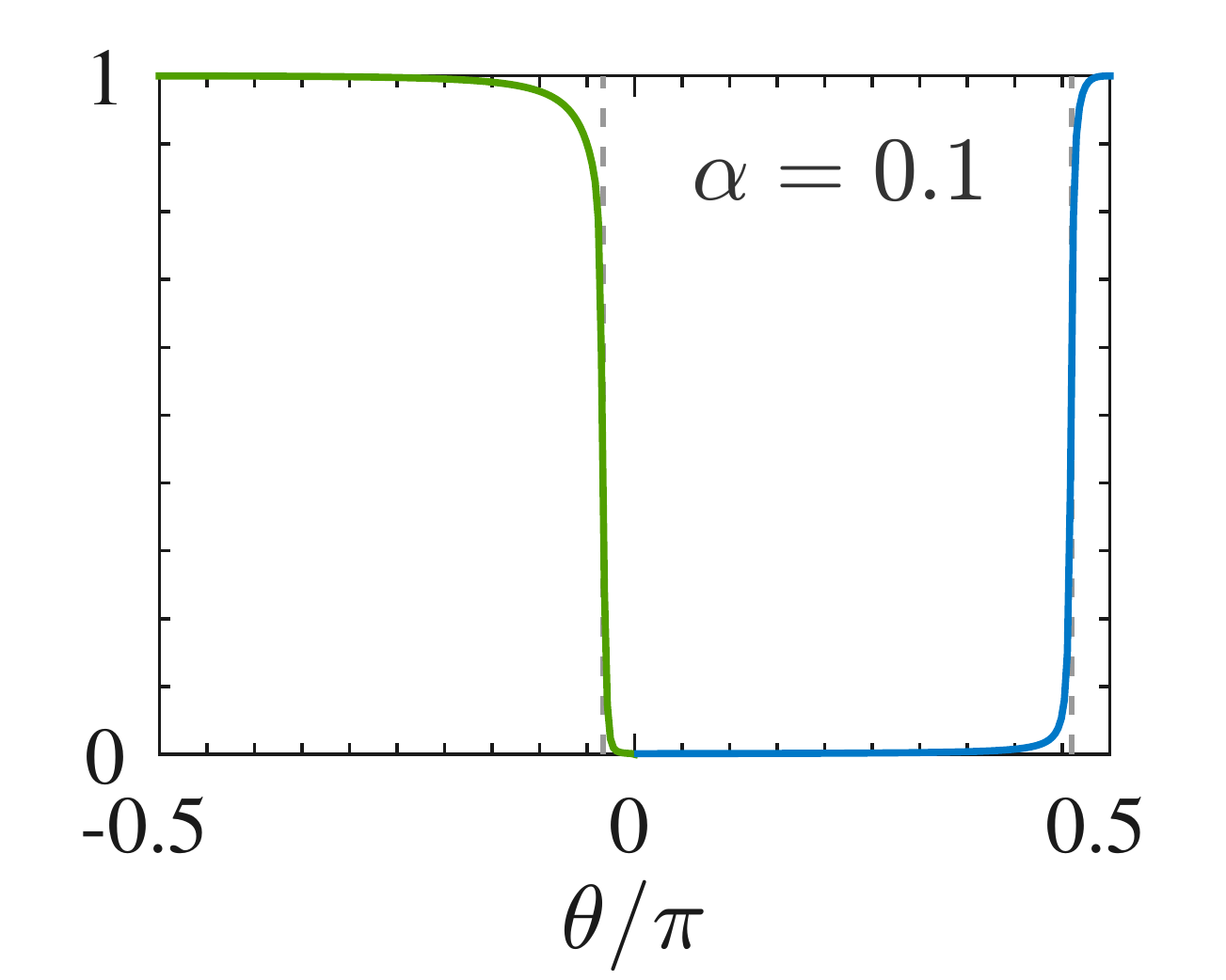} \hspace{-12pt}
\includegraphics[width=0.5\textwidth]{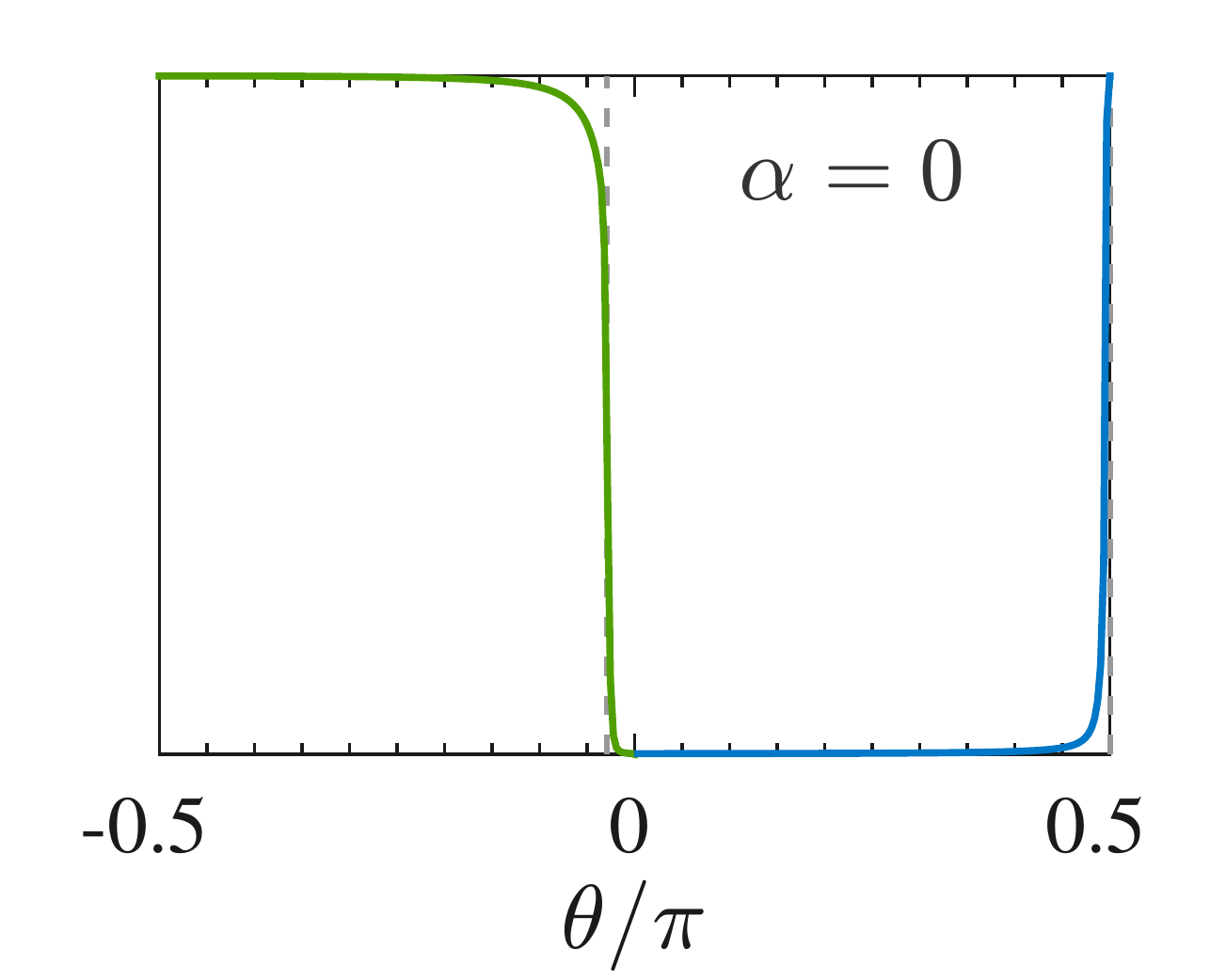}
\end{minipage}\hfill
\begin{minipage}{0.49\textwidth}
\centering
\includegraphics[width=1\textwidth]{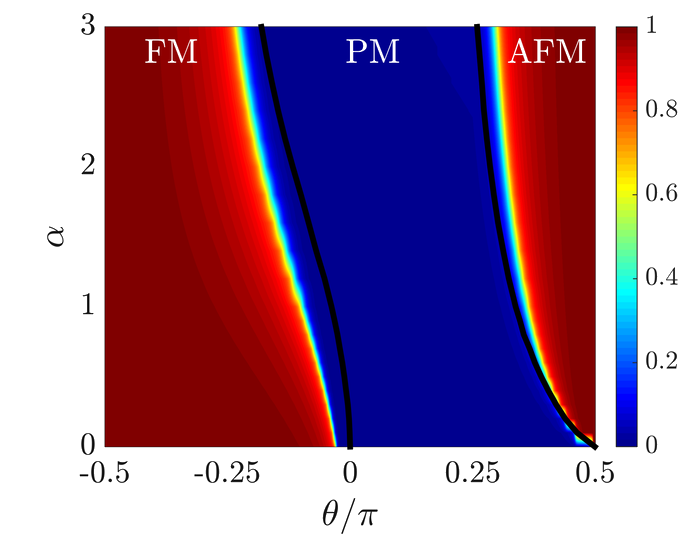}
\end{minipage}
\caption{\textbf{(Left)} Order parameters $\varphi_{\rm FM}=2\langle\hat{J}_z\rangle/N$ (for the FM order at $\theta<0$) 
and $\varphi_{\rm AFM}=2\langle\hat{J}_z^{\rm (st)}\rangle/N$ (for the AFM order at $\theta>0$),
calculated for a chain of $N=20$ spins, as functions of the control parameter $\theta$ for selected values of the decay range $\alpha$.
The dashed vertical lines indicates the point where concavity changes.
\textbf{(Right)} Density plot of the order parameters $\varphi_{\rm FM}$ $(\theta<0)$ and $\varphi_{\rm AFM}$ $(\theta>0)$ 
for $N=20$ on the $\theta$--$\alpha$ plane.
For fixed $\alpha$, the maximum of the first derivative (i.e., the inflection point) of the order parameters 
signals the proximity of a critical point; extrapolating its position for larger $N$ permits to approximately locate 
the critical point in the thermodynamic limit: the results for all values of $\alpha$ are superimposed as solid black lines.
Here, all data are obtained by forcing the symmetry breaking through a longitudinal field $\varepsilon=-10^{-3}\,J$. }
\label{fig:IsingOrderParameter}
\end{figure}

\paragraph{Qualitative phase diagram}
The noncommuting terms in Eq.~(\ref{IsingHam}) introduce quantum fluctuations in the ground state of the model, 
that lead to the appearance of critical points separating a ordered phase from a disordered phase 
according to a spontaneous symmetry breaking.
After fixing the interaction range $\alpha$, the Ising chain is known to host two transitions driven by the parameter $\theta$: 
a ferromagnetic QPT between a paramagnet and a ferromagnet at the critical point $\thetaFM<0$
and an antiferromagnetic QPT between a paramagnet and an antiferromagnet at the critical point $\thetaAFM>0$.
The exact aspect of the zero-temperature phase diagram for this model results from the competition 
between the spin-spin interaction and the ordering magnetic field, in addition to the sign of the interaction. 
The spin interaction favours the alignment (or antialignment) of the spins in the $z$ direction, 
while the magnetic field polarizes the spins along the $x$ direction. 
Anyway, we can outline the following general features, schematically illustrated in 
Figs.~\ref{fig:IsingQualitativePhaseDiagram} and \ref{fig:IsingOrderParameter}.

%%%%%%%%%%%%%%%%%%%%%%%%%%%%%%%%%%%%%%%%%%%%%%%%%%%%%%%%%
%%%%%%%%%%%%%%%%%%%%%%%%%%%%%%%%%%%%%%%%%%%%%%%%%%%%%%%%%
%%%%%%%%%%%%%%%%%%%%%%%%%%%%%%%%%%%%%%%%%%%%%%%%%%%%%%%%%
%%%%%%%%%%%%%%%%%%%%%%%%%%%%%%%%%%%%%%%%%%%%%%%%%%%%%%%%%

For small enough strength of interaction ($\thetaFM<\theta<\thetaAFM$), 
the system is in a \emph{quantum paramagnetic} (PM) phase. 
The adjective ``quantum'' is important to distinguish this paramagnet from the thermal paramagnet:
the former is obtained at zero temperature since the field along the $x$ direction forces each spin 
into a coherent superposition of the basis states ``up'' and ``down'', 
while the latter is obtained from a ferromagnet (or antiferromagnet) 
by raising the temperature above the critical Curie (or N\'eel) temperature so that the spins fluctuate in time 
because of the incoherent perturbation from the classical environment.
The paramagnetic phase possesses the $Z_2$ symmetry of the Hamiltonian; therefore it is named a \emph{disordered} phase, 
according to the adage ``the more symmetric, the less ordered''. 
In this phase, the system is gapped: all the excitations -- in the form of spin flips -- have nonzero energy cost, 
and the first excited state in the many-body energy spectrum is separated from the ground state by a finite gap 
$\DeltaE=E_2-E_1>0$.

For large strength of interaction ($-\frac{\pi}{2}\leq\theta<\thetaFM\,$ or $\,\thetaAFM<\theta\leq\frac{\pi}{2}$) 
the system is in an \emph{ordered} phase: there exist an operator -- local in the spins -- 
that recognizes the order in terms of finite expectation value on the ground state.
The order parameter is the expectation value of 
the longitudinal magnetization $\hat{J}_z=\frac{1}{2}\sum_{i=1}^N\hat{\sigma}_z^{(i)}$ for the FM order 
and the staggered longitudinal magnetization $\hat{J}_z^{\,\rm(st)}=\frac{1}{2}\sum_{i=1}^N(-1)^i\hat{\sigma}_z^{(i)}$ for the AFM order. These observables have zero expectation value in the disordered phase, as highlighted in Fig.~\ref{fig:IsingOrderParameter}.
In both the \emph{ferromagnetic} ($-\frac{\pi}{2}\leq\theta<\thetaFM$) 
and \emph{antiferromagnetic} phases ($\thetaAFM<\theta\leq\frac{\pi}{2}$), the system is gapless:
the ground state is doubly degenerate ($\DeltaE=0$) in the thermodynamic limit $N\to\infty$. 
This degeneracy is removed by the spontaneous breaking of the spin-flip $Z_2$ symmetry captured by the order parameters.

%%%%%%%%%%%%%%%%%%%%%%%%%%%%%%%%%%%%%%%%%%%%%%%%%%%%%%%%
%%%%%%%%%%%%%%%%%%%%%%%%%%%%%%%%%%%%%%%%%%%%%%%%%%%%%%%%
%%%%%%%%%%%%%%%%%%%%%%%%%%%%%%%%%%%%%%%%%%%%%%%%%%%%%%%%
%%%%%%%%%%%%%%%%%%%%%%%%%%%%%%%%%%%%%%%%%%%%%%%%%%%%%%%%

Actually, since the global Hamiltonian (\ref{IsingHam}) is symmetric under time reversal $Z_2$, 
the order parameters are rigorously null in both the phases, 
$\langle\hat{J}_z\rangle=\langle\hat{J}_z^{\,\rm(st)}\rangle=0 \ \forall\,\theta$, 
unless a symmetry-breaking term is introduced by hand in the model. 
Such a term has the form of the coupling with a weak longitudinal field $\varepsilon\sum_{i=1}^N\hat{\sigma}_z^{(i)}$
(or $\varepsilon\sum_{i=1}^N${\small$(-1)^i$}$\hat{\sigma}_z^{(i)}$) for the FM (or AFM) order. 
In the thermodynamic limit, whichever infinitesimal perturbation removes the double degeneracy of the ground state 
and forces the system into a well-defined ferromagnetic (or antiferromagnetic) state with finite (staggered) magnetization: 
$\langle\hat{J}_z\rangle \lessgtr 0$ for $-\frac{\pi}{2}\leq\theta<\thetaFM$
(or $\langle\hat{J}_z^{\rm (st)}\rangle \lessgtr 0$ for $\thetaAFM<\theta\leq\frac{\pi}{2}$) 
according to the sign of $\varepsilon\to0^\pm$. 
Physically, at finite size $N$, the condition $|\varepsilon| \sim \DeltaE^{-1} \ll J$ is sufficient  
for an experiment to observe finite magnetization in the ordered phase.
Under this assumption, the QPTs of the Ising model occurring at $\thetaFM$ and $\thetaAFM$ are second-order symmetry-breaking QPTs.
In the discussion about entanglement, we will not add any symmetry-breaking term: 
the two states generating the two-fold ground-state space are not splitted in energy: 
as a consequence, we can witness a large amount of entanglement in the ground state.
We will briefly show how the situation changes adding a longitudinal field with the example in Fig.~\ref{fig:IsingQFIT0}

In the thermodynamic limit, the order parameters are expected to exhibit a diverging derivative with respect to the control parameter
as the critical point is approached.
By inspecting the sudden variation of the order parameters as functions of $\theta$ at finite size $N$ 
and extrapolating the results to $N\to\infty$, we are able to approximately locate 
the position of the critical points $\thetaAFM$ and $\thetaFM$.
The complete phase diagram $\theta$--$\alpha$ for the Ising chain in a transverse field is reported in Fig.~\ref{fig:IsingOrderParameter}.
The thick curves are the lines of FM and AFM critical points, separating the ordered phases from the disordered one; 
we will qualitatively justify their aspect in section~\ref{sec:IsingArbitrary}.
Here we just notice that they seem to share a common behaviour only for large values of $\alpha$: 
in the long-range region $\alpha\lesssim1$ the geometries of the phase boundaries are considerably different.

\paragraph{Experimental realization}
Despite their apparent simplicity and the enormous number of applications in quantum computations~\cite{Suzuki2013}, 
the practical realization of models of the type in Eq.~(\ref{IsingHam}) has revealed to be an arduous task.
Historically, the standard platforms for this purpose have been quasi--one-dimensional magnetic anisotropic crystals: 
the magnetic moments of ions arranged on a linear pattern of the lattice experience a dominant interaction along an easy axis 
induced by the strong anisotropy of the crystal. 
In the last years, definitive experiments modelling the important case of nearest neighbours $\alpha=\infty$
has been performed on CoNb$_2$O$_6$~\cite{ColdeaSCIENCE2010}.
Another recent realization of the nearest-neighbour model was achieved using a degenerate Bose gas of spinless rubidium atoms 
confined in a tilted optical lattice, exploiting the mapping between the Mott insulator and the Ising chain~\cite{simon2011}.

Nowadays, the impressive control attained in addressing each single ion in a linear string of laser-cooled trapped ions
permits to engineer chains of effective spins whose internal states can be individually prepared and measured 
by changing intensity and polarization of the probe laser.
Recent experiments with cold trapped ions have demonstrated that 
laser-induced manipulation of the vibrational modes of the ion chain
can generate long-range spin interactions in Ising-like Hamiltonians with a tunable exponent $\alpha$ in the interval 
$0\lesssim\alpha\lesssim3$~\cite{SchneiderRPP2012,IslamSCIENCE2013,JurcevicNAT2014}.

\section{Chain with nearest-neighbour interaction} \label{sec:IsingNN}
\begin{figure}[t!]
\centering
\includegraphics[width=1\textwidth]{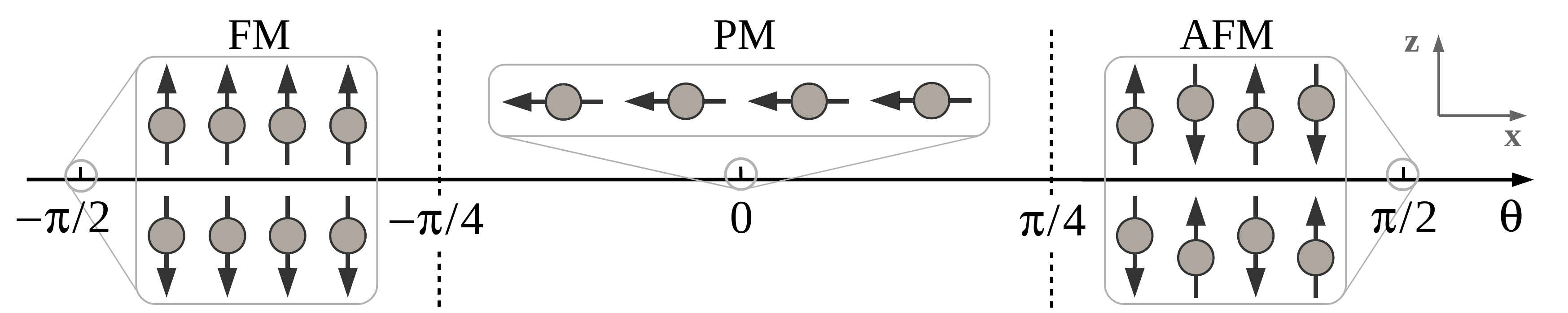}
\caption{Sketch of the phase diagram for the transverse Ising model with nearest-neighbour interaction $\alpha=\infty$. 
Here the critical points are $\thetaFM=-\frac{\pi}{4}$ and $\thetaAFM=+\frac{\pi}{4}$. 
The drawings represent the schematic spin configurations in the ground state of each phase: 
from left to right, they are strictly valid at $\theta=-\frac{\pi}{2}$ (doubly-degenerate GHZ state), 
at $\theta=0$ ($-x$-polarized state) and at $\theta=+\frac{\pi}{2}$ (doubly-degenerate N\'eel state).
The actual ground state for $\theta\simeq0$ in the disordered phase can be understood as a perturbation of the polarized product state: single spin flips arise and behave as quasiparticle excitations, according to the cartoon $\big|\!\cdots\!\leftarrow\leftarrow\!\textcolor{LightGray}{|}\mbox{\boldmath$\rightarrow$}\textcolor{LightGray}{|}\!\leftarrow\leftarrow\dots\big\rangle$.
The actual ground state for $\theta\simeq\pm\frac{\pi}{2}$ in the ordered phases is again explained perturbatively 
by the creation of pairs of domain walls enclosing contiguous flipped spins, such as
$\big|\!\cdots\!\uparrow\,\uparrow\,\uparrow\textcolor{LightGray}{|}\,\mbox{\boldmath$\downarrow$}\,\mbox{\boldmath$\downarrow$}\,\textcolor{LightGray}{|}\uparrow\uparrow\,\uparrow\dots\big\rangle$. 
The gray axes indicates the orthogonal exchange and field directions.}
\label{fig:IsingNNPhaseDiagram}
\end{figure}

The condition $\alpha\to\infty$ selects only $i-j=1$ in the double sum:
\be \label{IsingHamNN}
\hat H_{\infty} = J\sin\theta\,\displaystyle \sum_{i=1}^N \hat{\sigma}_z^{(i)} \hat{\sigma}_z^{(i+1)} 
\, + \, J\cos\theta\,\displaystyle\sum_{i=1}^N \hat{\sigma}_x^{(i)}
\ee
In the thermodynamic limit $N\to\infty$, this simple model exhibits a \emph{ferromagnetic} QPT at $\thetac^{(-)}=-\frac{\pi}{4}$, 
between a paramagnetic phase with $\DeltaE>0$ (for $\thetac^{(-)}<\theta\leq0$) 
and a ferromagnetic ordered phase with $\DeltaE=0$ (for $-\frac{\pi}{2}\leq\theta<\thetac^{(-)}$), 
and an \emph{antiferromagnetic} QPT at $\thetac^{(+)}=+\frac{\pi}{4}$, 
between a paramagnetic phase with $\DeltaE>0$ (for $0<\theta\leq\thetac^{(+)}$) 
and an antiferromagnetic ordered phase with $\DeltaE=0$ (for $\thetac^{(+)}<\theta\leq\frac{\pi}{2}$).
Since both the QPTs belongs to the universality class of the two-dimensional classical Ising model~\cite{Suzuki1976}, 
the exact location of the critical points can be found thanks to the Kramers-Wannier duality simmetry
$\cos\theta\leftrightarrow\sin\theta$~\cite{Kramers1941}, 
that maps the low-field regime to the high-field regime -- essentially, this transformation is the quantum generalization 
of the duality relation between the low-temperature and high-temperature phases of the two-dimensional classical Ising model.
Exploiting the self-duality at the critical points~\cite{FradkinPRD1978}, one finds $|\cos\theta|=|\sin\theta|$. 
This result can be justified intuitively: 
the transition happens when the competition between the two terms in Hamiltonian~(\ref{IsingHamNN})
is not balanced, namely when the weights of the noncommuting terms are equal, provided that 
$\langle\hat{\sigma}_z^{(i)}\hat{\sigma}_z^{(i+1)}\rangle\approx\langle\hat{\sigma}_x^{(i)}\rangle$.
Figure~\ref{fig:IsingNNPhaseDiagram} depicts the phase diagram for the nearest-neighbour Ising chain in a transverse field.

%%%%%%%%%%%%%%%%%%%%%%%%%%%%%%%%%%%%%%%%%%%%%%%%%%%%%%%%%
%%%%%%%%%%%%%%%%%%%%%%%%%%%%%%%%%%%%%%%%%%%%%%%%%%%%%%%%%
%%%%%%%%%%%%%%%%%%%%%%%%%%%%%%%%%%%%%%%%%%%%%%%%%%%%%%%%%
%%%%%%%%%%%%%%%%%%%%%%%%%%%%%%%%%%%%%%%%%%%%%%%%%%%%%%%%%

Magnetic order~\cite{Pfeuty1970} is quantified in terms of the full two-spin correlator in the longitudinal direction 
$C_{zz}^{(i,j)} = \bra{\psi_{0}}\hat{\sigma}_z^{(i)}\hat{\sigma}_z^{(j)}\ket{\psi_{0}}$ for the FM order,
or $C_{zz}^{(i,j)} = (-1)^{i-j}\bra{\psi_{0}}\hat{\sigma}_z^{(i)}\hat{\sigma}_z^{(j)}\ket{\psi_{0}}$ for the AFM order, 
where $\ket{\psi_{0}}$ indicates the many-body ground state of the system.
The longitudinal correlator is expected to be short ranged in the disordered phase $C_{zz}^{(i,j)}\sim\neper^{-|i-j|/\xi}\,$,
indicating the absence of any long-range order: the exponential fall of correlations occurs on a spatial scale
of the order of the correlation length $\xi\propto\Delta^{-1}$ (dynamical exponent being $z=1$), 
that becomes larger and larger for $\theta\to\thetac$.
At the critical point, the correlation length diverges as $\xi\propto\big(1-|\tan\theta|\big)^{-1}$ (critical exponent $\nu=1$)
and the correlation function decays as a power law $C_{zz}^{(i,j)}\sim|i-j|^{-\eta}$ for long distances $|i-j|\gg1$,
with $d=1$ the spatial dimensionality of the system and $\eta=\frac{1}{4}$ the Fisher exponent. 
Instead, the ordered phase exhibits long-range correlations signalled by distance-independent finite  value
$C_{zz}^{(i,j)}=\bra{\psi_{0}}\hat{\sigma}_z^{(i)}\ket{\psi_{0}}^2$, that permits to identify the order parameter 
as the spontaneous magnetization of the ground state 
$\bra{\psi_{0}}\hat{\sigma}_z^{(i)}\ket{\psi_{0}}\propto\big(1-\cot^2\theta\big)^{\beta}$ 
(with critical exponent $\beta=\frac{1}{8}$)~\cite{Pfeuty1970}.
In particular, at $\theta=-\frac{\pi}{2}$ we have $\langle\hat{J}_z\rangle=\frac{N}{2}$; at $\theta=+\frac{\pi}{2}$, we have $\langle\hat{J}_z^{\rm (st)}\rangle=\frac{N}{2}$.
Summarizing, the longitudinal correlator depends on long distance $|i-j|\gg1$ as
\be \label{SummaryCzz}
C_{zz}^{(i,j)} \sim \left\{
\begin{array}{ll} 
\neper^{-|i-j|/\xi} & {\rm for} \ \thetaFM<\theta<\thetaAFM \\ 
|i-j|^{-1/4} & {\rm for} \ \theta=\thetaFM   \, \vee \, \theta=\thetaAFM \\ 
{\rm const} & {\rm for} \ \theta<\thetaFM \, \vee \, \theta>\thetaAFM
\end{array}
\right. \, .
\ee

All the above considerations can be checked by direct calculations, 
because the model is exactly diagonalizable by means of a standard approach~\cite{Lieb1961}, that we review in the following.

\paragraph{Fermionization}
The whole energy spectrum and the eigenfunctions of Eq.~(\ref{IsingHamNN}) can be obtained by employing a Jordan-Wigner transformation, 
that provides a mapping from spin-$\sfrac{1}{2}$ degrees of freedom to spinless fermions:
$\hat{a}_i^\dag = \hat{s}_+^{(i)} \exp\big( -\ii\pi \sum_{j=1}^{i-1} \hat{s}_+^{(j)} \hat{s}_-^{(j)} \big)$ and
$\hat{a}_i = \exp\big( \ii\pi \sum_{j=1}^{i-1} \hat{s}_+^{(j)} \hat{s}_-^{(j)} \big) \hat{s}_-^{(i)}$, 
where $2\,\hat{s}_\pm^{(i)}\equiv\hat{\sigma}_\pm^{(i)}=\hat{\sigma}_x^{(i)}\pm\ii\,\hat{\sigma}_y^{(i)}$ 
and the operators $\hat{a}_i$ satisfy the canonical fermionic algebra 
$\{\hat{a}_i,\hat{a}_j^\dagger\}=\delta_{ij}$ and $\{\hat{a}_i,\hat{a}_j\}=\{\hat{a}_i^\dag,\hat{a}_j^\dag\}=0$.
In the fermionic representation, spins becomes highly nonlocal in the index $i$.
In order to handle Eq.~(\ref{IsingHamNN}), we rotate the spins of an angle $-\frac{\pi}{2}$ around the $y$ axis, using the canonical transformation 
$\hat{\sigma}_x\mapsto\hat{\sigma}_z$, $\hat{\sigma}_y\mapsto\hat{\sigma}_y$, $\hat{\sigma}_z\mapsto-\hat{\sigma}_x$. 
Recalling that $2\hat{\sigma}_z^{(i)}=\hat{\sigma}_+^{(i)}\hat{\sigma}_-^{(i)}-2 \identity^{(i)}$,
the resulting Hamiltonian reads
\beq \label{IsingHamNNJW}
\hat H_{\infty} & = & J\sin\theta\,\displaystyle \sum_{i=1}^N \left( \hat{a}_{i}^\dag\hat{a}_{i+1} + \hat{a}_{i+1}\hat{a}_{i} + {\rm H.c.}  \right) \, + \, J\cos\theta\,\displaystyle\sum_{i=1}^N \left( 2\hat{a}_{i}^\dag\hat{a}_{i} - \identity_i \right) = \nonumber \\
& = &\displaystyle\sum_{i,\,j=1}^\sites \bigg[\hat{a}^\dag_i A_{ij} \hat{a}_j \ + \, 
\frac{1}{2} \big( \hat{a}^\dag_i B_{ij} \hat{a}^\dag_j + {\rm H.c.} \big) \Big] \, ,
\eeq
with $ \, A_{ij} = 2J\cos\theta\,\delta_{ij} + J\sin\theta \big(\delta_{i,\,j+1}+\delta_{i,\,j-1}\big) \, $ 
as well as $ \, B_{ij} = J\sin\theta \big(\delta_{i,\,j+1}-\delta_{i,\,j-1}\big) \, $ for the open chain.
Notice that the number conservation is violated: the effective fermions can be created and annihilated in pairs.
The new Hamiltonian is quadratic in the fermionic operators: thus it can be easily diagonalized.
Physically, we aim to generate a new set of fermionic operators $\hat{\eta}_k$ 
as a linear combination of the $\hat{a}_k$'s and $\hat{a}_{k}^\dag$'s 
in order to remove terms that do not conserve the number of fermions:
we look for a linear canonical transformation of the form $\hat{\eta}_k=\frac{1}{2}\sum_{i=1}^N \big[(\Phi_{ki}+\Psi_{ki}) \hat{a}_i + (\Phi_{ki}-\Psi_{ki}) \hat{a}^\dag_i\big]$
with coefficients $\Phi_{ki},\,\Psi_{ki}\in\mathbb{R}$ such that 
\be \label{IsingHamNNB}
\hat H_{\infty} = \displaystyle \sum_{k} \Lambda_k \, \hat{\eta}_k^\dag\hat{\eta}_k \,+\, {\rm const} .
\ee
This turns out to be possible whether the columns of the matrices $\mathbf{\Phi}$ and $\mathbf{\Psi}$ are, respectively,
the left and right eigenvectors obtained from the singular value decomposition 
$\mathbb{A}+\mathbb{B}=\mathbf{\Phi}^T \mathbf{\Lambda} \mathbf{\Psi}$, 
where $\mathbf{\Lambda}$ is the diagonal matrix containing the eigenvalues $\Lambda_k$ 
and $\mathbb{A},\,\mathbb{B}$ are defined in Eq.~(\ref{IsingHamNNJW}).
Solutions for $\Lambda_k$ and $\mathbf{\Phi},\,\mathbf{\Psi}$ in terms of wave numbers $k$ can be found for the open chain~\cite{Lieb1961,Suzuki2013}, but it is much more complicated than for the closed chain, due to the absence of translational invariance.

% Magnetic order is measured by the correlators 
The longitudinal and transverse correlators 
$C_{\varrho\varrho}^{(i,j)} = \bra{\psi_{0}}\hat{\sigma}_\varrho^{(i)}\hat{\sigma}_\varrho^{(j)}\ket{\psi_{0}}$,
are now attainable by using the Wick's theorem, that allows to evaluate the vacuum expectation value of a product of 
anticommuting operators in terms of vacuum expectation values of the product of pairs of operators. 
Following the steps described above, one can find (let it be $j>i$, being $C_{\varrho\varrho}^{(j,i)}=C_{\varrho\varrho}^{(i,j)}$)
\be \label{IsingCorrelators}
C_{zz}^{(i,j)} = 
\det \begin{pmatrix}
G_{i,\,i+1} & \dots & G_{i,\,j} \\
\vdots & \ddots & \\
G_{j-1,\,i+1} & & G_{j-1,\,j} \\
\end{pmatrix} \, , \quad
C_{yy}^{(i,j)} = 
\det \begin{pmatrix}
G_{i+1,\,i} & \dots & G_{i+1,\,j-1} \\
\vdots & \ddots & \\
G_{j,\,i} & & G_{j,\,j-1} \\
\end{pmatrix}
\ee
are just subdeterminants of the Green matrix $\mathbb{G}=-\mathbf{\Psi}^T\mathbf{\Phi}$, 
while $C_{xx}^{(i,j)} = G_{i,\,i}G_{j,\,j} - G_{j,\,i}G_{i,\,j}$.

\begin{figure}[t!]
\centering
\includegraphics[height=0.38\textwidth]{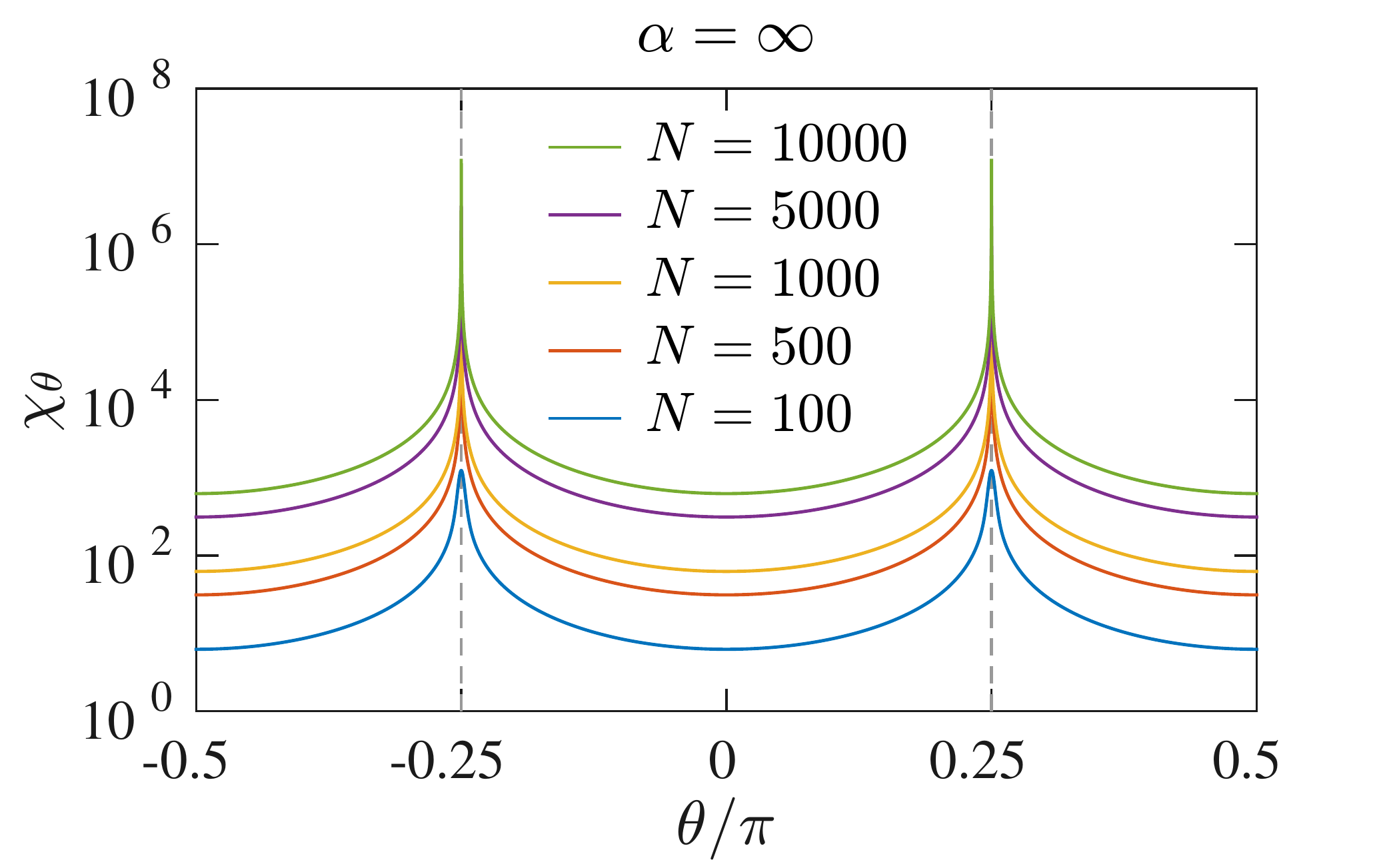}
\includegraphics[height=0.38\textwidth]{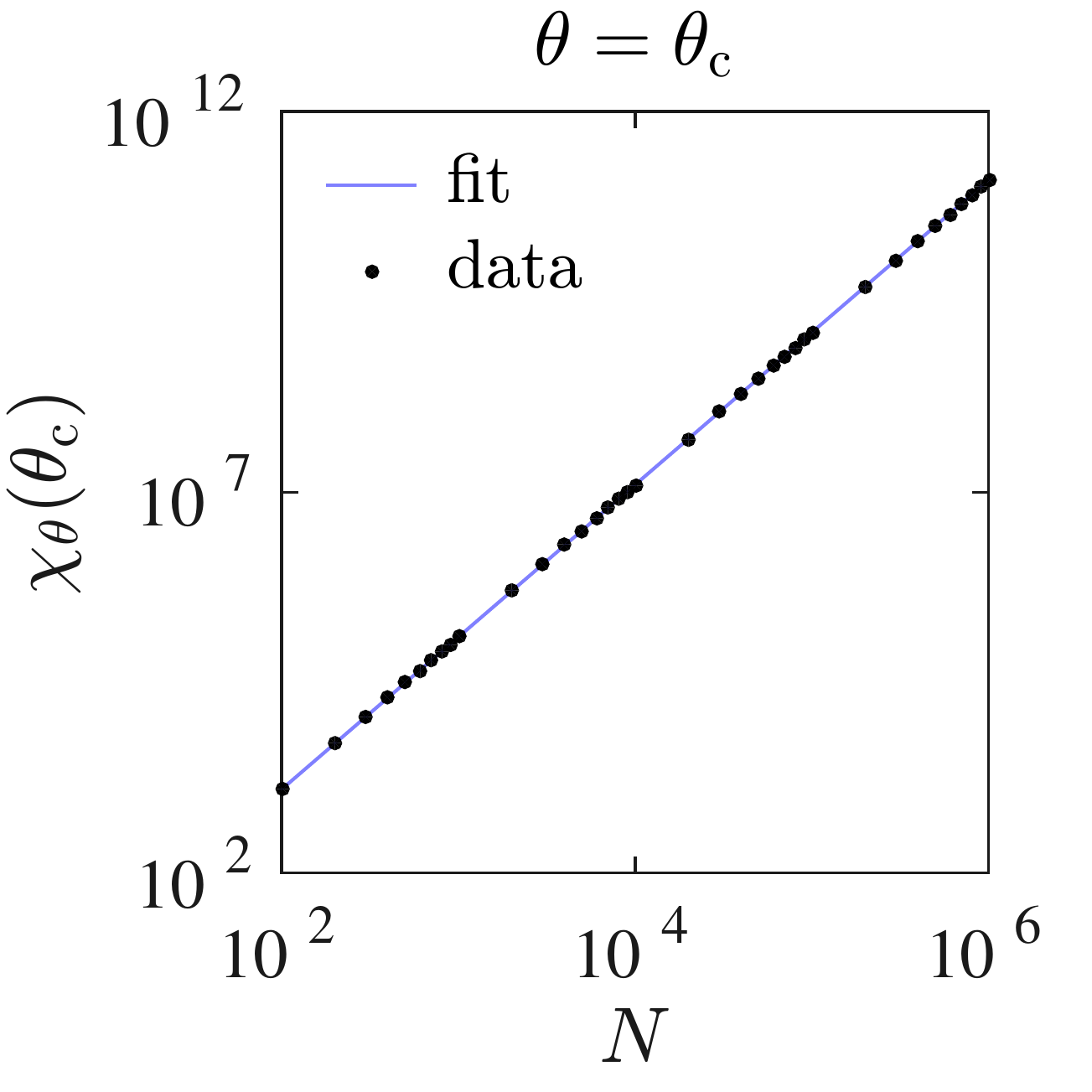}
\caption{\textbf{(Left)} Fidelity susceptibility Eq.~(\ref{IsingChiTheta}) for the nearest-neighbour Ising model as a function of $\theta$.
Different solid lines refer to different values of $N$ in the limit $N\to\infty$. 
The dashed gray line signals the critical points $\thetaFM=-\frac{\pi}{4}$ and $\thetaAFM=+\frac{\pi}{4}$.
\textbf{(Right)} Scaling of the maximum of $\chi_{\theta}$ for increasing $N$.
The solid line is the fit curve $\chi_\theta(\thetac)=N^2/8$.}
\label{fig:FidelitySusceptibilityNNTheta}
\end{figure}

\paragraph{Fidelity susceptibility}
Since the phase transition is a collective phenomenon that involves all the spins in the limit of infinite chain, 
and since any change at the edges does not modify the bulk properties of the system,
we expect the critical properties of the Ising chain to be the same regardless from the boundary conditions.
We can thus exploit the exact knowledge of the ground-state wave function for the closed chain
\be \label{IsingGS}
\ket{\psi_{0}} = \big(1+\sin2\theta\cos{k}\big)^{-N/2} \prod_{k} \Big[ (\sin\theta\cos{k}+\cos\theta) - \ii \sin\theta\sin{k} \, \hat{a}_{k}^\dag \hat{a}_{-k}^\dag \Big]\ket{0}\,,
\ee
where $\ket{0}$ is the vacuum of $\hat{a}_k$, to find the fidelity susceptibility of the closed chain, that has the advantage to be exact:
\be \label{IsingChiTheta}
\chi_\theta = \frac{1}{4} \sum_{k}
\frac{\sin^2{k}}{\big(1+\sin2\theta\cos{k}\big)^2} \, .
\ee
We plot the behaviour of $\chi_\theta$ in Fig.~\ref{fig:FidelitySusceptibilityNNTheta}. 
At finite size $N$, the fidelity susceptibility signals criticality in terms of a sharp peak, 
centred around $\theta_{\rm max}$ and scaling as $\max_\theta\chi_\theta \sim N^2$, that becomes,
in the limit $N\to\infty$, a divergence at $\theta_{\rm max}\to\thetac$.
The superextensive peak stands out from an extensive background.
Our analysis recovers well-established results in literature, already reported in Refs.~\cite{ZanardiPRE2006,DamskiPRE2013,DamskiJPA2014}.

%%%%%%%%%%%%%%%%%%%%%%%%%%%%%%%%%%%%%%%%%%%%%%%%%%%%%%%%%
%%%%%%%%%%%%%%%%%%%%%%%%%%%%%%%%%%%%%%%%%%%%%%%%%%%%%%%%%
%%%%%%%%%%%%%%%%%%%%%%%%%%%%%%%%%%%%%%%%%%%%%%%%%%%%%%%%%
%%%%%%%%%%%%%%%%%%%%%%%%%%%%%%%%%%%%%%%%%%%%%%%%%%%%%%%%%

\paragraph{Bipartite entanglement}
Recently, entanglement in the ground state of the transverse nearest-neighbour Ising model 
has been extensively considered~\cite{Dutta2015}.
At zero temperature, the analysis mainly focused on the behaviour of measures of bipartite entanglement,
both in the neighbourhood and at the quantum critical point.

A first remarkable example is shown by the concurrence~\cite{OsterlohNATURE2001}, measuring the pairwise quantum correlations 
between two spins in the chain and obtained from the ground-state wave function after tracing out all the spins 
except for the two selected spins. The derivative of the concurrence with respect to the control parameter $\theta$ displays 
clear signatures of singularity due to criticality: in a finite chain, it diverges with the size as $\log N$ at the critical points; 
in an infinite chain, it diverges logarithmically as $\log|1-\tan\theta|$ when the control parameter approaches the critical points. 

A second example is provided by the bipartite entanglement between a block of contiguous spins of arbitrary length 
and the rest of the chain, as quantified by the von Neumann entropy~\cite{VidalPRL2003}.
Also the entropy diverges logarithmically with the size $N$ for a finite chain at the critical points.

\subsection{Multipartite entanglement in the ground state} \label{subsec:IsingMEinGS}
The investigation of multipartite entanglement in the nearest-neighbour Ising chain at zero temperature 
has a meagre literature~\cite{GuhneNJP2005}, 
mainly owing to the lack of a monotone that can be of service as a many-party entanglement measure. Here we briefly cite the major attempts.
Geometric entanglement has been adopted as a measure of the distance between the actual ground state and the closest separable state: 
it behaves singularly over the phase transition~\cite{WeiPRA2003}. 
The genuine multipartite negativity, defined as optimization over the set of fully decomposable witnesses, is sensitive to 
the phase transition as concerns its first derivative with respect to the control parameter, that undergoes a logarithmic divergence 
and obeys a finite-size scaling behaviour around the critical points~\cite{HofmannPRB2014}.
Very similar features pertain to the global entanglement, even in dimension higher than one~\cite{MontakhabPRA2010}.
Multipartite entanglement has also been probed through the QFI and spin squeezing parameters~\cite{LiuJPA2013}.

We are interested in witnessing multipartite entanglement using the criterion based on the quantum Fisher information
for local operators acting on spins discussed in chapter~\ref{ch:Intro}.
As local observables we choose the (staggered) magnetization along the three spatial directions. 
Moreover, we optimize the QFI over the three directions to obtain the maximal value
$F_Q[\ket{\psi_{0}}]=\max_\varrho F_Q\big[\ket{\psi_{0}},\hat{J}_{\varrho}\big]$ for the FM coupling ($\theta<0$) 
and $F_Q[\ket{\psi_{0}}]=\max_\varrho F_Q\big[\ket{\psi_{0}},\hat{J}^{\rm (st)}_{\varrho}\big]$ for the AFM coupling ($\theta>0$),
with $\varrho=x,y,z$.
Truly, we are implementing the optimization procedure typical of SU(2) interferometry described in paragraph~\ref{subsec:QuantumInterferometry}.

\paragraph{Method}
Because of the symmetry of the ground state $\ket{\psi_{0}}$ inherited from the invariance of Hamiltonian~(\ref{IsingHamNN}) 
under global spin flip, we have
$\bra{\psi_{0}}\hat{\sigma}^{(i)}_y\ket{\psi_{0}} = \bra{\psi_{0}}\hat{\sigma}^{(i)}_z\ket{\psi_{0}} 
\allowbreak = \allowbreak
\bra{\psi_{0}}\hat{\sigma}^{(i)}_x\hat{\sigma}^{(j)}_y\ket{\psi_{0}} 
\allowbreak = \allowbreak
\bra{\psi_{0}}\hat{\sigma}^{(i)}_x\hat{\sigma}^{(j)}_z\ket{\psi_{0}} 
\allowbreak = \allowbreak 0$. 
Further, since the ground state for the one-dimensional chain can always be chosen to be real,
we have $\,{\rm Re}\bra{\psi_{0}}\hat{\sigma}^{(i)}_y\hat{\sigma}^{(j)}_z\ket{\psi_{0}} = 0$.
Hence, we just need to calculate the variances of the observables $\hat{J}_\varrho$ and $\hat{J}_\varrho^{\rm (st)}$:
%$F_Q[\ket{\psi_{0}}]=4\max\{(\Delta\hat{J}_x)^2,\,\langle\hat{J}_y^2\rangle,\,\langle\hat{J}_z^2\rangle\}$  for $\theta<0$ and
%$F_Q[\ket{\psi_{0}}]=4\max\{(\Delta\hat{J}^{\rm (st)}_x)^2,\,\langle\hat{J}_y^{\rm (st)\,2}\rangle,\,\langle\hat{J}_z^{\rm (st)\,2}\rangle\}$ 
%for $\theta>0$.
\be
F_Q[\ket{\psi_{0}}] = \left\{
\begin{array}{lr} 
4\max\big\{(\Delta\hat{J}_x)^2,\,\langle\hat{J}_y^2\rangle,\,\langle\hat{J}_z^2\rangle\big\} & {\rm for} \ \theta<0 \\
4\max\big\{(\Delta\hat{J}^{\rm (st)}_x)^2,\,\langle\hat{J}_y^{\rm (st)\,2}\rangle,\,\langle\hat{J}_z^{\rm (st)\,2}\rangle\big\} & {\rm for} \ \theta>0
\end{array}
\right. \, .
\ee
Numerical calculation shows clearly that the $z$ direction is the optimal one in the entire range of the parameter $\theta$, so that 
$F_Q[\ket{\psi_{0}}]=4\langle\hat{J}_z^{2}\rangle$ for $\theta<0$ and 
$F_Q[\ket{\psi_{0}}]=4\langle\hat{J}_z^{\rm (st)\,2}\rangle$ for $\theta>0$.
Defining the ``Fisher density'' as the maximal quantum Fisher information per spin $f_Q=F_Q/N$, optimization finally provides  
\be \label{IsingfQCorr}
f_Q[\ket{\psi_{0}}] = \frac{1}{N} \sum_{i,\,j=1}^N (\pm1)^{i-j} \bra{\psi_{0}} \hat{\sigma}^{(i)}_z\hat{\sigma}^{(j)}_z \ket{\psi_{0}} \equiv \frac{1}{N} \sum_{i,\,j=1}^N C_{zz}^{(i,j)} \quad {\rm for} \ \theta\lessgtr0 \, .
\ee
This finding has a clear interpretation: the optimal operator for witnessing entanglement is
the order parameter used to capture and describe the spontaneous symmetry breaking in the chain.
It sounds quite intuitive because the longitudinal magnetization is somehow sensitive to small changes 
of the control parameter $\theta$ across the transition point, and the spatial behaviour of its correlation function $C_{zz}^{(i,j)}$ 
has a sudden qualitative change reported in Eq.~(\ref{SummaryCzz}), from exponential confinement in the disordered phase 
to long-range delocalization on the entire chain in the ordered phase.
In chapter~\ref{ch:LMG} we will come into another example of system displaying a second-order QPT 
whose order parameter plays the role of optimal observable for evaluating the QFI.

In order to have a physical flavour of Eq.~(\ref{IsingfQCorr}), we can use the qualitative behaviours of the correlation function
in Eq.~(\ref{SummaryCzz}) to derive
\be \label{fQqualitative}
f_Q[\ket{\psi_{0}}] \sim \left\{
\begin{array}{ll} 
2\,(\neper^{1/\xi}-1)^{-1} & {\rm for} \ \thetaFM<\theta<\thetaAFM \\ 
N^{3/4} & {\rm for} \ \theta=\thetaFM   \, \vee \, \theta=\thetaAFM \\ 
N\,\big(1-\cot^2\theta\big)^{1/4} & {\rm for} \ \theta<\thetaFM \, \vee \, \theta>\thetaAFM
\end{array}
\right. \, .
\ee
These formulae openly show that the Fisher density has very different scalings with chain size $N$ 
in the disordered phase, critical point and ordered phase. 
Nevertheless, they require some comments. Above all, they have been deduced for a closed ring, a geometry that allows to evaluate easily
the sum (\ref{IsingfQCorr}); we expect the scalings not to be modified by different boundary conditions for $N\gg1$. 
The ring must be large enough also for a second reason: at $\theta=\thetac$ this condition permits to approximate
$f_Q=\sum_{r=1}^{N-1}r^{-1/4}$ with the generalized harmonic number $H_{N,\,1/4}$, 
whose asymptotic behaviour is known: $H_{N,\,1/4} \sim N^{3/4}$.

We judge that Eq.~(\ref{fQqualitative}) gives a brilliant insight to understand entanglement properties of the system, 
but it is by no means an exact result. 
In the following, rather than undertaking an analytical calculation of correlators, 
that would result in large and unilluminating equations,
we prefer to numerically calculate the correlation functions in Eq.~(\ref{IsingCorrelators}) for different values of the parameter $\theta$.
This can be done up to $N\approx10^3$: beyond this order of magnitude, computational cost becomes prohibitive.

%%%%%%%%%%%%%%%%%%%%%%%%%%%%%%%%%%%%%%%%%%%%%%%%%%%%%%%%%
%%%%%%%%%%%%%%%%%%%%%%%%%%%%%%%%%%%%%%%%%%%%%%%%%%%%%%%%%
\begin{figure}[t!]
\centering
\includegraphics[height=0.32\textwidth]{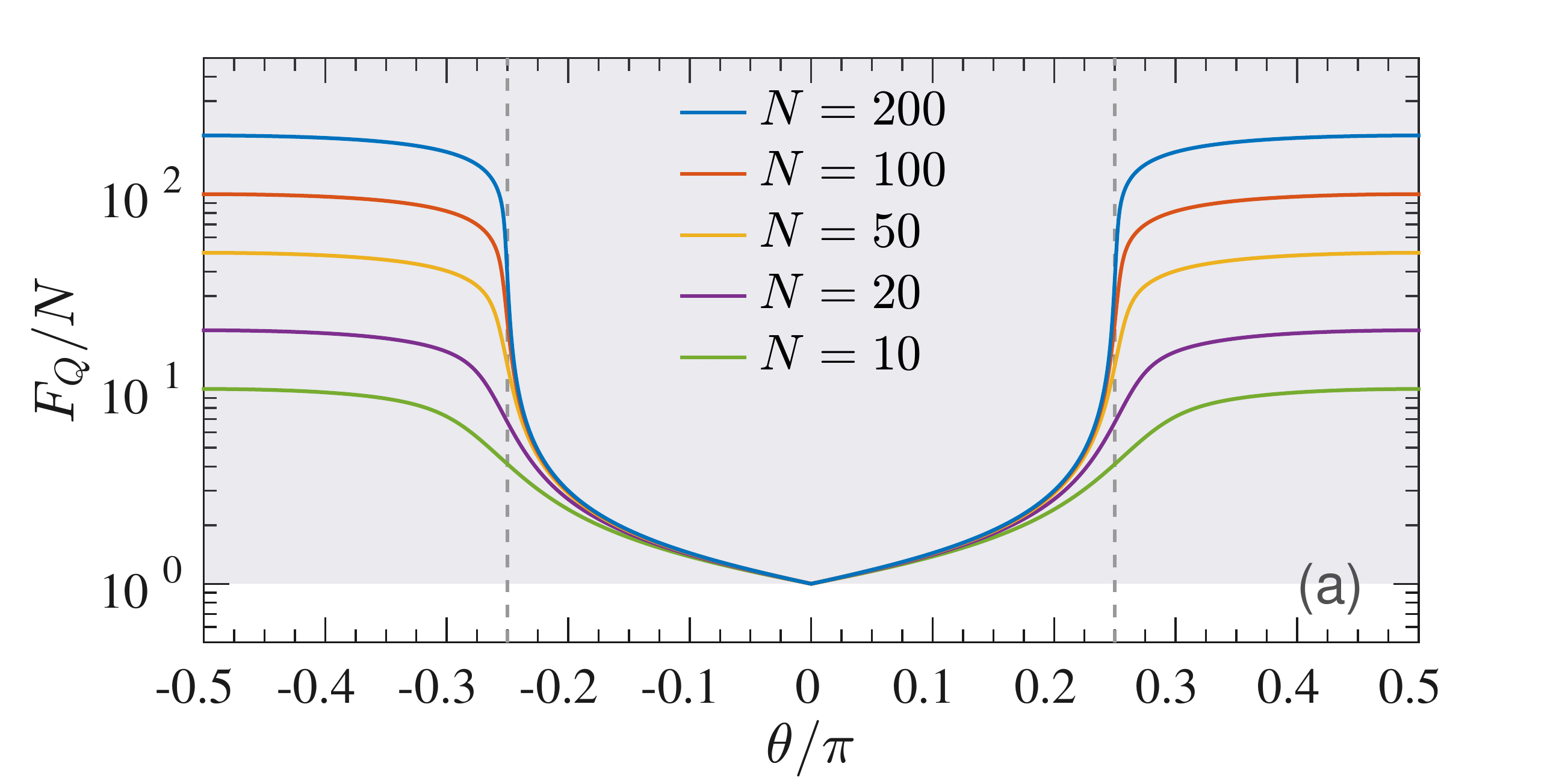} \hspace{-20pt}
\includegraphics[height=0.32\textwidth]{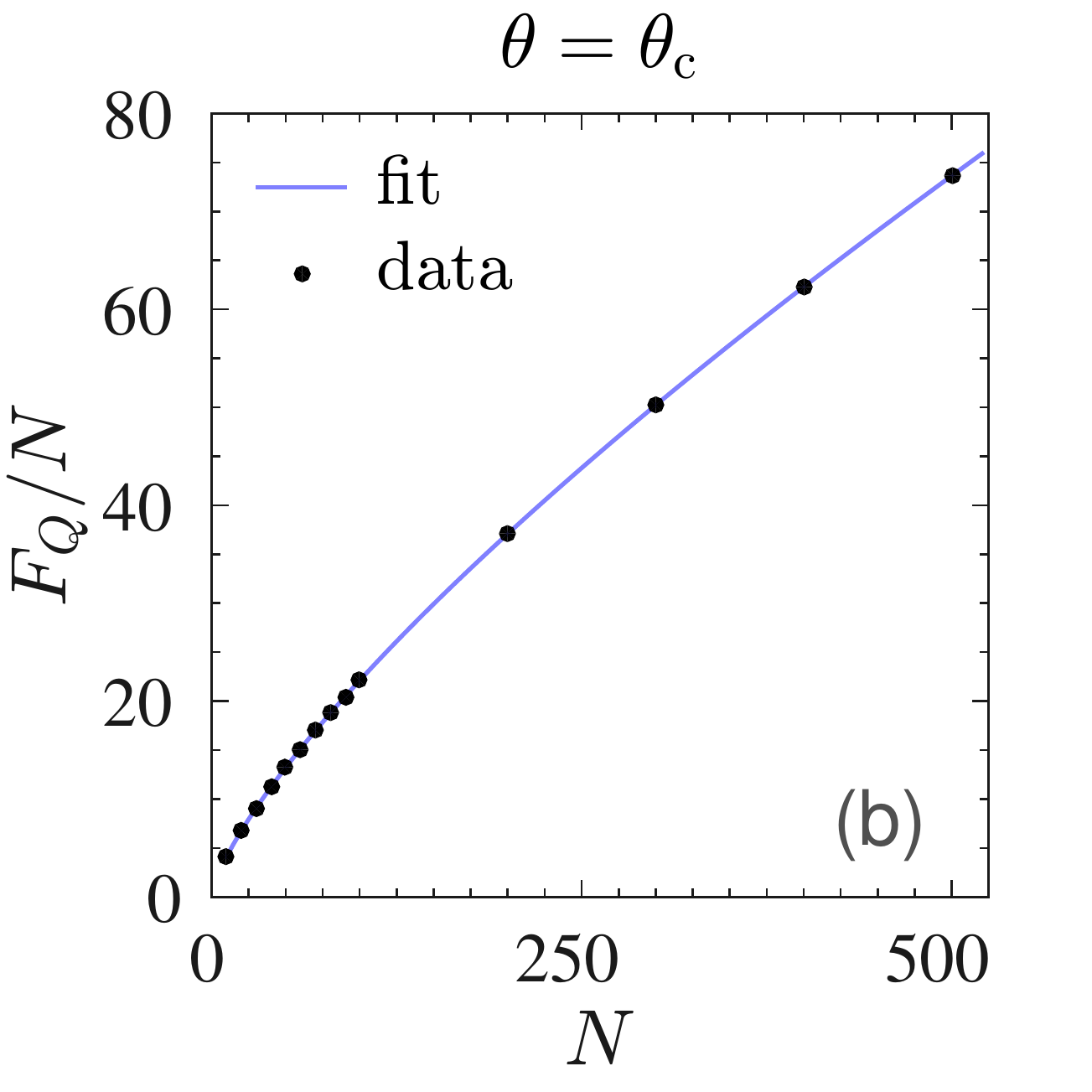} \\
\includegraphics[height=0.32\textwidth]{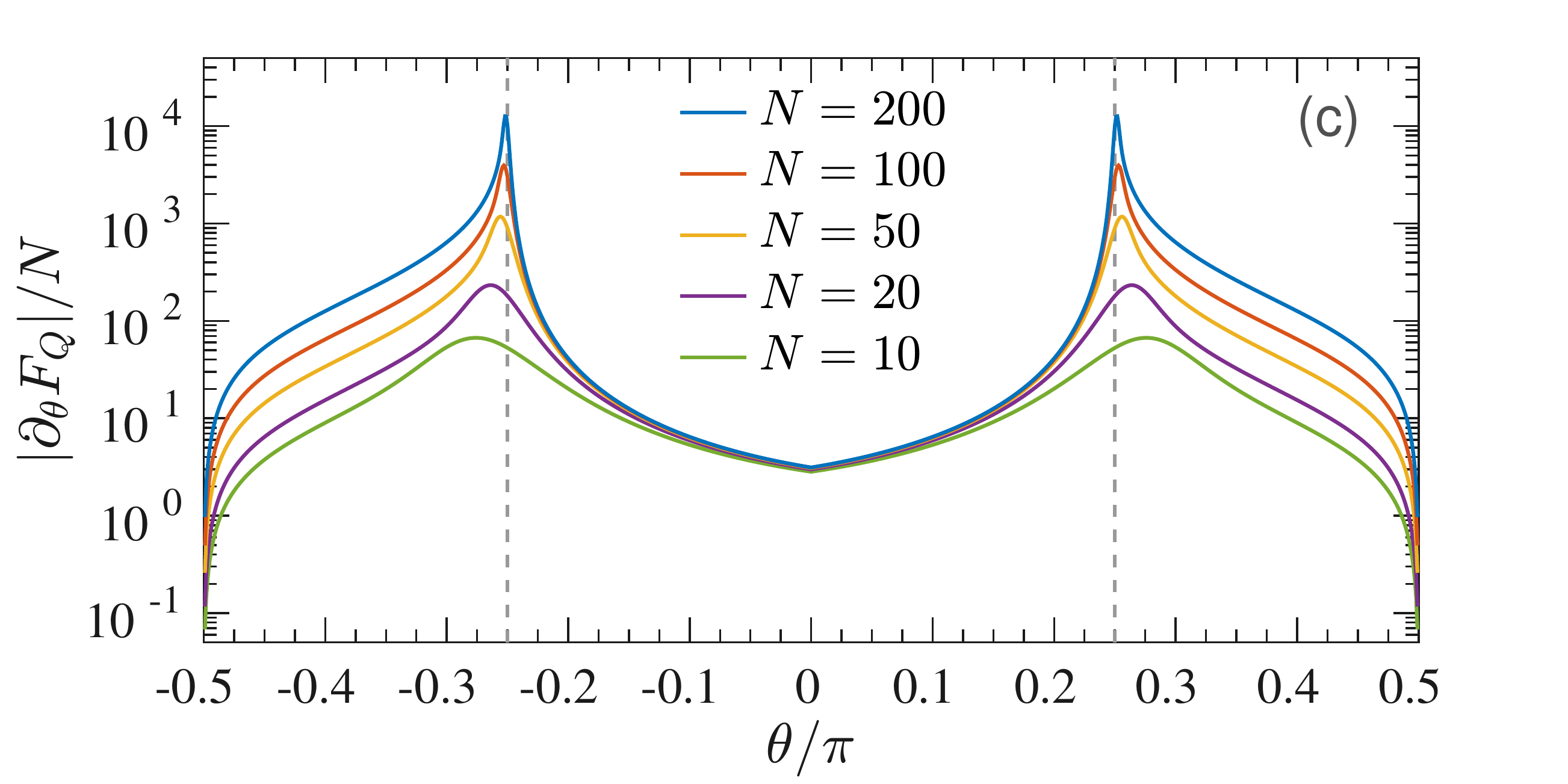} \hspace{-20pt}
\includegraphics[height=0.32\textwidth]{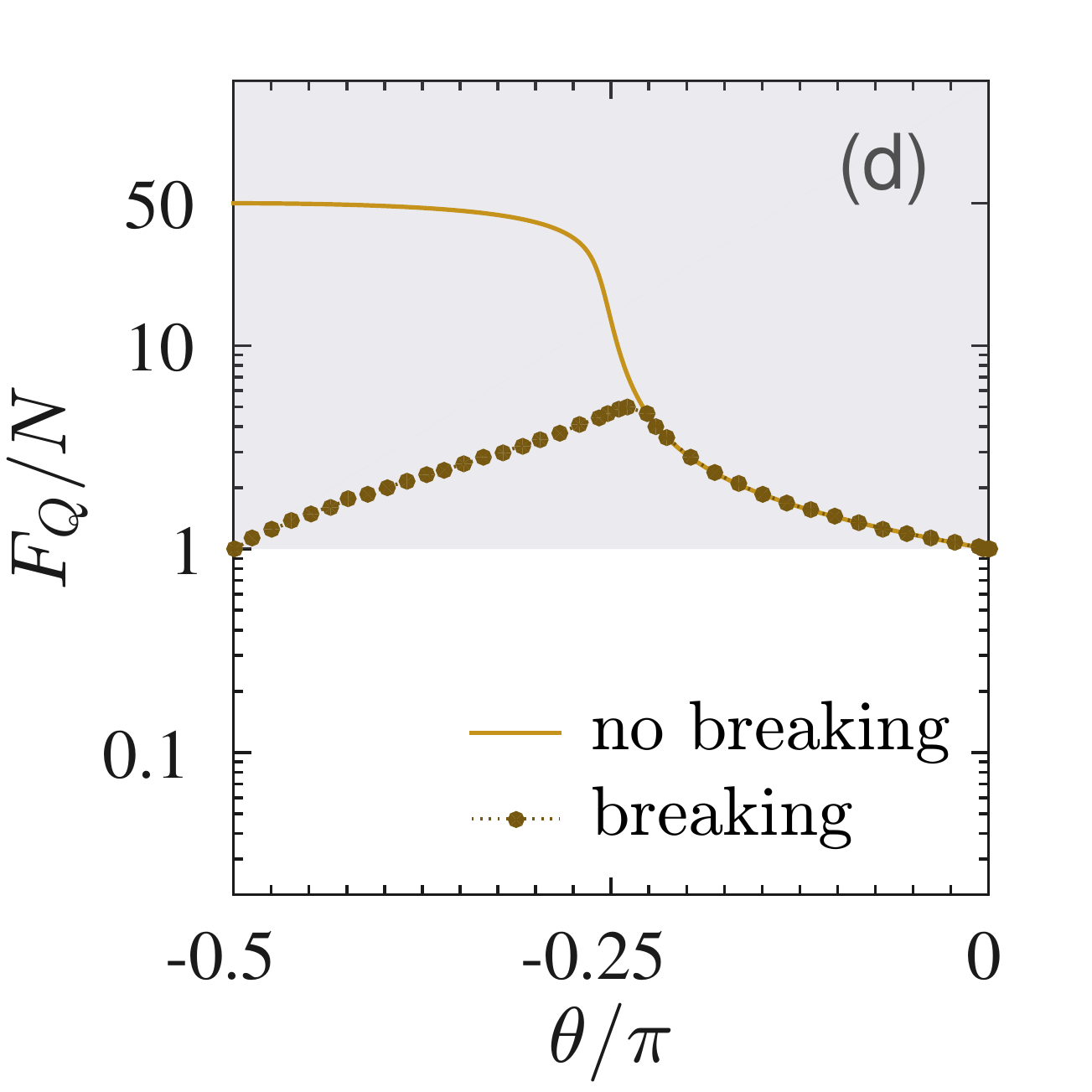} 
\caption{Behaviour of the quantum Fisher information evaluated using local magnetizations and optimized over the three spatial direction.
\textbf{(a)} Fisher density $f_Q[\ket{\psi_0},\hat{J}_z]$ for $\theta<0$ and $f_Q[\ket{\psi_0},\hat{J}^{\rm (st)}_z]$ for $\theta>0$: 
different colours refer to different chain lengths $N$; 
shaded area signals the region where $f_Q[\ket{\psi_0}]>1$ and entanglement is witnessed;
dashed gray lines indicate the critical points. 
\textbf{(b)} Finite-size scaling of the Fisher density at criticality $\theta=\thetaFM$ or $\theta=\thetaAFM$: 
dots are numerical data; the solid line is the fit curve $f_Q=aN^b$, with $a\simeq0.71$ and $b\simeq0.75$.
\textbf{(c)} First derivative of the Fisher density of panel (a) with respect to $\theta$. 
\textbf{(d)} Fisher density $f_Q[\ket{\psi_0},\hat{J}_z]$ for $\theta<0$ and $N=50$: the solid line is the same as in panel (a),
for the ground state of Hamiltonian~(\ref{IsingHamNN}) without the addition of symmetry-breaking terms; 
dots are obtained from the ground state of Hamiltonian~(\ref{IsingHamNN}) with an additional longitudinal field
$\varepsilon\hat{J}_z$ and $\varepsilon(\theta)=\DeltaE(\theta)$.}
\label{fig:IsingQFIT0}
\end{figure}
%%%%%%%%%%%%%%%%%%%%%%%%%%%%%%%%%%%%%%%%%%%%%%%%%%%%%%%%%
%%%%%%%%%%%%%%%%%%%%%%%%%%%%%%%%%%%%%%%%%%%%%%%%%%%%%%%%%

\paragraph{Results} 
The main results are contained in Fig.~\ref{fig:IsingQFIT0}, 
where we plot the behaviour of the maximal QFI as a function of the control parameter $\theta$ and its scaling for increasing size $N$.
The first aspect that stands out from panel~\ref{fig:IsingQFIT0}\Panel{a} is the symmetry 
of the QFI under change of the sign of interactions $\theta\mapsto-\theta$: 
 in fact, for nearest-neighbour interaction only, 
the ferromagnetic phase is mapped into the antiferromagnetic one through the canonical transformation
$\hat{\sigma}^{(i)}_x\mapsto\hat{\sigma}^{(i)}_x$ and 
$\hat{\sigma}^{(i)}_{y,\,z}\mapsto\textrm{{\small$(-1)^i$}}\,\hat{\sigma}^{(i)}_{y,\,z}$. 
% from which we know that the critical properties are the same for both. 
This allows us to refer indiscriminately to only one disordered phase and one generic ordered phase, 
comprising both the FM and AFM coupling.
Moreover, the multipartite entanglement witnessed by means of the local order parameters exhibits a variety of remarkable features.
%%%%%%%%
\\[-6pt]

\noindent \tripieno{MySky} \ Multipartite entanglement $f_Q[\ket{\psi_{0}}]>1$ is witnessed for any $\theta\neq0$ 
and it augments monotonically for increasing $|\theta|$. 
%%%%%%%
\\[9pt]
%%%%%%%
\trivuoto{MyBlue} \ From a geometrical point of view, this means that the spin-spin interaction -- regardless of its sign -- 
generates quantum correlations among the spins, 
making the ground state of the chain more distinguishable under a unitary transformation generated by the suitable optimal operator 
($\hat{J}_z$ for $\theta<0$ and $\hat{J}_z^{\rm (st)}$ for $\theta>0$) than any other classical state of the chain. 
The degree of distinguishability increases for increasing interaction strength.
%%%%%%%
\\[9pt]
%%%%%%%
\trivuoto{MyBlue} \ From a practical point of view, whichever finite interaction strength 
permits to obtain a good state for quantum interferometry. 
Suppose we want to implement an ideal interferometric scheme whose global transformation is generated by the order parameter: 
it is sufficient to let each spin rotate by an unknown angle (with the same amplitude $\phi$ for all the spins at $\theta<0$,
with staggered amplitude {\small$(-1)^i$}$\phi$ at $\theta>0$) around the $z$ direction.   
Then, it would be possible to estimate the amplitude $\phi$ (and trace back to the magnitude of the physical perturbation that induced it)
with sensitivity beyond the shot-noise limit.
%%%%%%%
\\[9pt]
%%%%%%%
\trivuoto{MyBlue} \ For $\theta=0$, instead, the ground state of the Ising chain is separable, 
given by the coherent spin state 
$\ket{\psi_{0}} \equiv|\vartheta=\frac{\pi}{2},\varphi=-\pi\rangle = 2^{-N/2}\,(\spinup_z - \spindown_z)^{\otimes N} $ 
polarized along the $-x$ direction by the magnetic field,
being $\spinup_z$ and $\spindown_z$ eigenstates of $\hat{\sigma}_z$. 
Consequently, the QFI is upper bounded by the classical limit of the shot noise $F_Q \leq N$ 
(see paragraph~\ref{subsec:QuantumInterferometry}), 
and the optimized Fisher density can only be $f_Q[\ket{\psi_{0}}]=1$.
%%%%%%%%
\\[-6pt]

\noindent \tripieno{MySky} \ The scaling behaviour manifested by multipartite entanglement when the number of spins grows
is different in different regions of the phase diagram and can be used to characterize it univocally. 
%%%%%%%
\\[9pt]
%%%%%%%
\trivuoto{MyBlue} \ In the \emph{disordered} phase the QFI is extensive: $f_Q[\ket{\psi_{0}}]$ does not increase with $N$, 
as predicted by Eq.~(\ref{fQqualitative}). 
This is justified by the intensive nature of the depth of entanglement in the ground state: 
it has been shown in Ref.~\cite{LiuJPA2013} that quantum correlations recognized by the QFI 
are essentially attributable to spin squeezing: for $\thetaFM\lesssim\theta\lesssim\thetaAFM$
the ground state is spin squeezed and $f_Q\approx1/\WSS$. 
As a consequence, correlations can only involve a finite number of spins
and no increase in multipartiteness results from an enlargement of the system.
%%%%%%%
\\[9pt]
%%%%%%%
\trivuoto{MyBlue} \ In the \emph{ordered} phases the QFI is superextensive, implying a divergent multipartiteness in the thermodynamic limit, 
and in particular $f_Q[\ket{\psi_{0}}]$ linearly increases with $N$ with a prefactor depending on $\theta$, 
as naively pointed out by Eq.~(\ref{fQqualitative}).
Here the entanglement depth grows for increasing system size:
although the model involves only pairwise interactions between neighbouring spins, arbitrarily distant spins become correlated. 
% leading to an entanglement depth growing
The metrological relevance is significant: 
the ground state of the transverse Ising chain is useful for performing phase estimation 
at the Heisenberg scaling: $\Delta\phi \sim N^{-1}$, with a prefactor larger than 1.
Nevertheless, criteria based on spin squeezing fail in detecting entanglement~\cite{LiuJPA2013}, 
signalling the non-Gaussianity of the ground state.
In fact, in proximity\footnote{\ Exactly on the borders $\theta=-\frac{\pi}{2}$ and $\theta=+\frac{\pi}{2}$, 
the ground state is given by the completely arbitrary superposition 
$\ket{\psi_0}=\alpha\spinup_z^{\otimes N} + \beta\spindown_z^{\otimes N}$ and 
$\ket{\psi_0}=\alpha(\spinup_z\spindown_z)^{\otimes N/2} + \beta(\spindown_z\spinup_z)^{\otimes N/2}$
respectively (with $\alpha,\,\beta\in\mathbb{R}$ and $|\alpha|^2+|\beta|^2=1$), 
because the degeneracy-lifting effect of the external field is rigorously null.\\[-9pt]}
of the borders of the ordered phases $\theta\to-\frac{\pi}{2}$ and $\theta\to+\frac{\pi}{2}$
the ground states of the Ising chain are, respectively, the Greenberger-Horne-Zeilinger (GHZ) state~\cite{GHZ1989}
$\ket{\psi_{0}} = \big(\spinup_z^{\otimes N} + \spindown_z^{\otimes N}\big)/\sqrt{2}$ and 
the N\'eel state $\ket{\psi_{0}} = \big[(\spinup_z\spindown_z)^{\otimes N/2} + (\spindown_z\spinup_z)^{\otimes N/2}\big]/\sqrt{2}$.
They both are highly-entangled highly--non-Gaussian spin states and allow for the saturation of the Heisenberg limit $f_Q=N$.
Unfortunately, these states are very fragile against temperature and it is currently impracticable 
to experimentally realize them in a scalable fashion.
%%%%%%%%
\\[-6pt]

\begin{figure}[t!]
\centering
\includegraphics[width=1\textwidth]{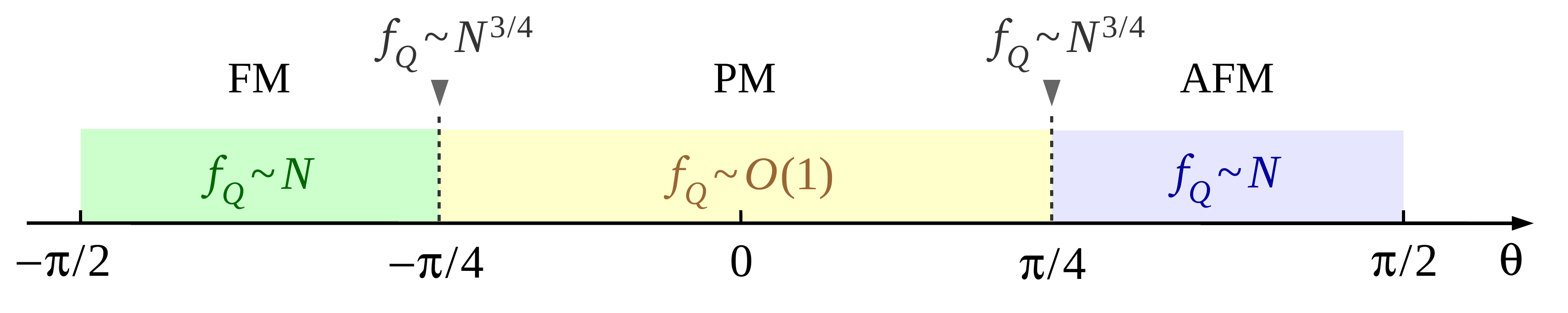}
\caption{Scaling of the Fisher density for increasing chain size in the different quantum phases and at the critical points 
of the transverse Ising model with nearest-neighbour interactions.}
\label{fig:IsingNNPhaseDiagramEntanglement}
\end{figure}

\noindent The major conclusion here is that \emph{multipartite entanglement detected by the quantum Fisher information 
manages to discriminate among ordered and disordered phases}.
This result is summarized in Fig.~\ref{fig:IsingNNPhaseDiagramEntanglement}.
%%%%%%%%
\\[-6pt]

\noindent \tripieno{MySky} \ Since the scaling is different in different phases, there must be a joint over the boundary between two phases:
this is only possible if multipartite entanglement scales differently at transition points $\theta=\thetac$. 
%%%%%%%
\\[9pt]
%%%%%%%
\trivuoto{MyBlue} \ In a finite chain, at criticality the QFI is still superextensive 
(multipartiteness diverges in the thermodynamic limit), but behaves as $f_Q \sim N^{3/4}$ with respect to the size.
This crucial result -- already envisaged by considerations on the cyclic chain in Eq.~(\ref{fQqualitative}) -- 
has be found from numerical calculations and is depicted in panel~\ref{fig:IsingQFIT0}\Panel{b}.
An argument based on scaling analysis and provided by Ref.~\cite{HaukeNATPHYS2016} rigorously demonstrated this algebraic growth.
%%%%%%%
\\[9pt]
%%%%%%%
\trivuoto{MyBlue} \ In an infinite chain, at criticality the first derivative of the QFI with respect to the parameter $\theta$ diverges: 
the critical points are inflection points with vertical tangent for the QFI, separating the convex disordered region from the concave ordered region. See panels~\ref{fig:IsingQFIT0}\Panel{a,\,c} for an exemplification. 
Numerical data are compatible with the finite-size divergence $\max_\theta |\partial_\theta f_Q| \sim N^{7/4}$ 
and the points of maximum derivative approaches the critical points according to the power law $|\theta_{\rm max}-\thetac| \propto N^{-1}$.
Moreover, we specify that the jump discontinuity displayed by the first derivative in $\theta=0$ is not associated to any transition at all: 
it is caused by the change of sign of the interaction combined to the linear dependence of the QFI on $\theta$ for small values of $\theta$, as pointed out by the perturbative analysis performed in paragraph~\ref{subsec:IsingDetection}.
We will encounter a similar finite discontinuity in chapter~\ref{ch:LMG}.
% the jump discontinuity displayed by the first derivative in $\theta=0$ has a physical meaning: 
% it is is caused by the sudden change of optimal operator ($\hat{J}_z$ for $\theta<0$ and $\hat{J}_z^{\rm (st)}$ for $\theta>0$)
% without change of operator, the discontinuity is removed and a smooth QFI is recovered.
%%%%%%%%
\\[-6pt]

\noindent Hence, \emph{the quantum Fisher information recognizes the critical points in term of a unique scaling 
and a sharp nonanalyticity in the thermodynamic limit}.
%%%%%%%%
\\[-6pt]

%%%%%%%%%%%%%%%%%%%%%%%%%%%%%%%%%%%%%%%%%%%%%%%%%%%%%%%%%
%%%%%%%%%%%%%%%%%%%%%%%%%%%%%%%%%%%%%%%%%%%%%%%%%%%%%%%%%
%%%%%%%%%%%%%%%%%%%%%%%%%%%%%%%%%%%%%%%%%%%%%%%%%%%%%%%%%
%%%%%%%%%%%%%%%%%%%%%%%%%%%%%%%%%%%%%%%%%%%%%%%%%%%%%%%%%

\noindent \tripieno{MySky} \ Finally, we want to clarify an important aspect related to our methodology. 
The high amount of particle entanglement detected in the ordered phases is ascribable to 
the double degeneracy of the ground state of the Ising Hamiltonian~(\ref{IsingHamNN}). 
The introduction of an additional term $\varepsilon\sum_{i=1}^N${\small$(\pm1)^i$}$\hat{\sigma}_z^{(i)}$ in the sector $\theta\lessgtr0$
breaks the $Z_2$ invariance and selects a nondegenerate ground state, as usual in experiments.
Whereas the ground state is not perturbed (thus the entanglement content does not change) in the PM phase if $\varepsilon\to0$,
a major effect pertains to the FM and AFM phases. 
In particular, in accordance with the sign of $\varepsilon\to0^\pm$,
the ground state would be the separable state $\,\spindown_z^{\otimes N}\,$ or $\,\spinup_z^{\otimes N}\,$ at $\,\theta=-\frac{\pi}{2}\,$,
as well as $\,(\spinup_z\spindown_z)^{\otimes N/2}\,$ or $\,(\spindown_z\spinup_z)^{\otimes N/2}\,$ at $\,\theta=+\frac{\pi}{2}\,$.
Hence, for zero field and FM or AFM coupling, we expect the QFI to saturate the classical bound $f_Q=1$,
precisely as for zero interaction $\theta=0$. In this case, at the critical point both the inflection point and the superextensive scaling 
are lost and criticality is signalled only by a maximum in the QFI, as illustrated in panel~\ref{fig:IsingQFIT0}\Panel{d}. 
This maximum of the QFI as a function of $\theta$ is preserved also at finite temperature in proximity of the critical point, 
which is what we are investigating in the following paragraph.

\subsection{Multipartite entanglement at finite temperature} \label{subsec:IsingthermalME} 
The Ising model with short range interaction has been the cradle 
of the concept of quantum criticality at finite temperature~\cite{ChakravartyPRB1989, SachdevNPB1996, Sachdev2011} 
and its first experimental testbed~\cite{ColdeaSCIENCE2010, KinrossPRX2014}, as discussed in paragraph~\ref{subsec:QuantumCriticality}.
Quantum criticality in the Ising model has been characterized by studying correlation lengths~\cite{ChakravartyPRB1989}, 
asymptotic correlation functions~\cite{SachdevPRL1997}, nature of the spin excitations~\cite{ColdeaSCIENCE2010} 
and relaxation dynamics~\cite{KinrossPRX2014}.
Moreover, signatures of the Ising transition at finite temperature has also been studied looking at 
the specific heat and the susceptibility~\cite{CuccoliPRB2007}.
These characterizations are not directly related to entanglement.

Recent works tried to draw the quantum phase diagram according to bipartite entanglement properties. 
References~\cite{OsbornePRA2002,AmicoEPL2007} explored the effect of temperature on concurrence~\cite{OsbornePRA2002,AmicoEPL2007}: 
they revealed that the largest amount of pairwise entanglement is present at low temperature in the vicinity of the critical points,
where it can survive for temperatures much larger than the energy gap.
Entanglement of formation at finite temperature in the Ising chain has also been discussed~\cite{SadiekJPB2013}: 
the threshold temperature at which the two-spin entanglement vanishes has been found to strongly depend on the energy gap, 
larger energy gap leading to stronger thermal robustness.
Notably, pairwise entanglement seems not to exhibit the simple factorization exhibited by the QFI in Eq.~\ref{QFIfactorization}.

However, the connection between multipartite entanglement and quantum phase digram is less scrutinized in literature.
Above all, investigations on the geometric entanglement reported the existence of a lower limit for the threshold temperature
below which the thermal state is guaranteed to be entangled~\cite{NakataPRA2009, SadiekJPB2013}.
Our work follows in the footsteps of Ref.~\cite{HaukeNATPHYS2016},
where the use of the QFI as a witness for finite-temperature many-party entanglement in the Ising chain was initiated. 
We corroborate and extend the results included therein.

\paragraph{Method}
If the spin chain exchanges energy with a bath at constant temperature $T$,
the equilibrium thermal density matrix of the system is given by the canonical ensamble $\hat{\rho}_T=\neper^{-\hat{H}_\infty}/\pazocal{Z}$, 
where $\pazocal{Z}$ is the canonical partition function and we choose natural units such that $k_{\rm B}=1$.
Due to the symmetry $\theta\leftrightarrow-\theta$, we can restrict our inspection to the ferromagnetic coupling, 
hence we limit $-\frac{\pi}{2}\leq\theta\leq0$.
We evaluate the QFI by means of the recipe given in Ref.~\cite{HaukeNATPHYS2016} 
and presented in paragraph~\ref{subsec:QuantumFisherInformation}, 
relating the QFI to the dynamic structure factor:
\be \label{IsingDynamicFactor}
F_Q\big[\hat{\rho}_T,\hat{J}_\varrho\big] = \frac{2}{\pi} \int_{-\infty}^{+\infty} \ud{\omega} \tanh^2\left(\frac{\omega}{2T}\right)\pazocal{S}_{\varrho\varrho}(\omega),
\ \ {\rm where} \ \
\pazocal{S}_{\varrho\varrho}(\omega)=\int_{-\infty}^{+\infty}\ud{t}\,\neper^{\ii\omega{t}}\,
{\rm Re}\,\big\langle\hat{J}_\varrho(t)\hat{J}_\varrho\big\rangle
\ee 
is the Fourier transform of the real part of the two-time correlation functions,
the brackets denote thermodynamic average $\langle\,\cdot\,\rangle=\tr(\hat{\rho}\,\,\cdot\,)/\pazocal{Z}$ 
and the time evolution of the operator in the Heisenberg picture is given by 
$\hat{J}_\varrho(t)=\neper^{\ii\hat{H}_{\infty}t}\,\hat{J}_\varrho\,\neper^{-\ii\hat{H}_{\infty}t}$. 

Investigation of multipartite entanglement at finite temperature $T>0$ relies upon the knowledge of the 
time-dependent two-spin correlator $\langle\hat{\sigma}_\varrho^{(i)}(t)\hat{\sigma}_\varrho^{(j)}\rangle$. 
This is exactly known in the free-fermion representation~\cite{Lieb1961} 
and can be expressed in terms of the Pfaffian of a $2(i+j-1)\times2(i+j-1)$ antisymmetric matrix 
via the Wick-Bloch-de~Dominicis theorem~\cite{DerzhkoPRB1997}.
Exploiting the analytical results provided by Ref.~\cite{DerzhkoPRB1997} 
together with the optimized algorithm reported in Ref.~\cite{Wimmer2012} for evaluating the Pfaffian,
the QFI can be efficiently computed up to $N\approx100$. 
Any finite-temperature analysis for $N\gtrsim100$ is hard to access due to severe computational costs.

\begin{figure}[t!]
\centering
\includegraphics[height=0.283\textwidth]{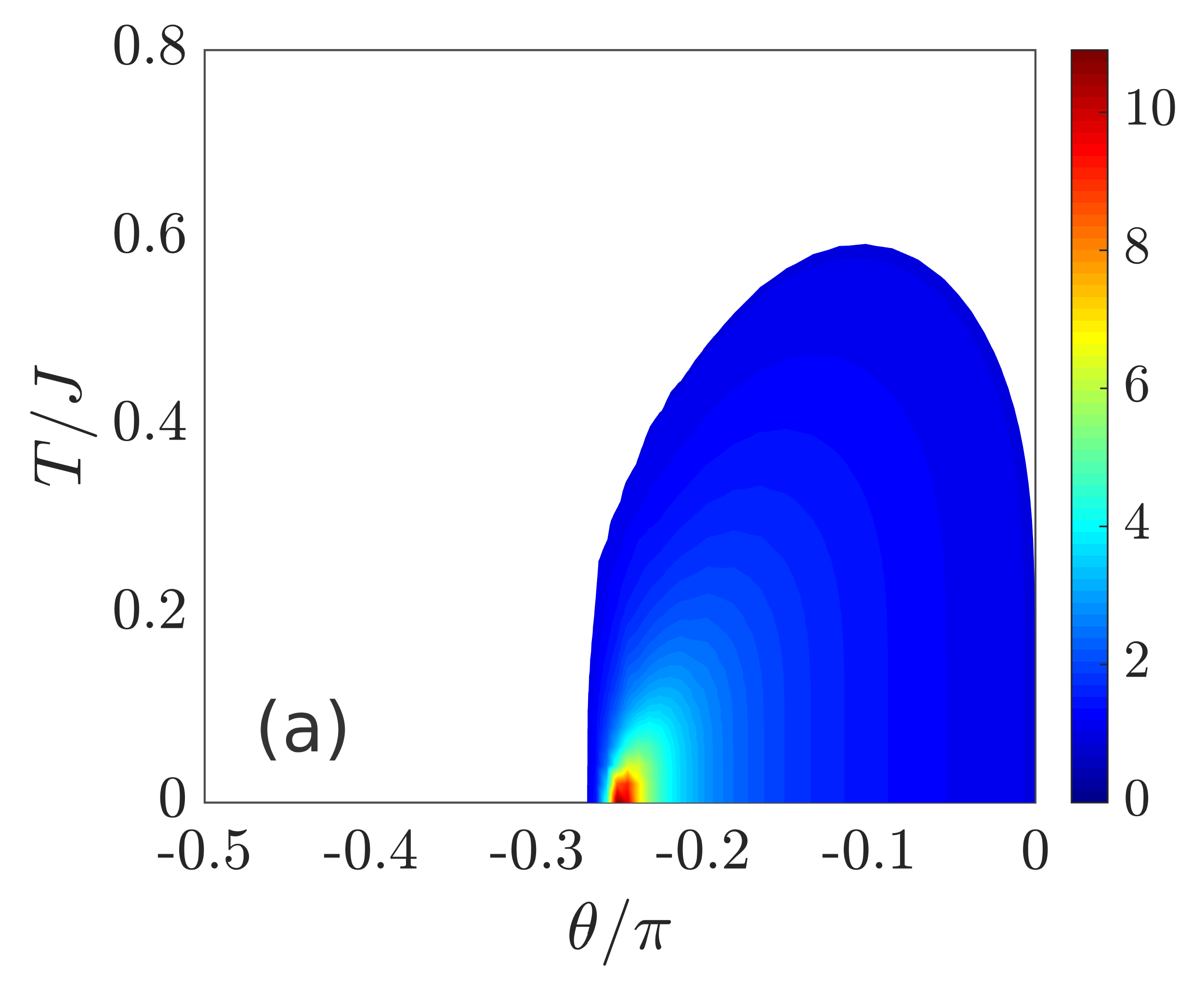} \hspace{-9pt}
\includegraphics[height=0.283\textwidth]{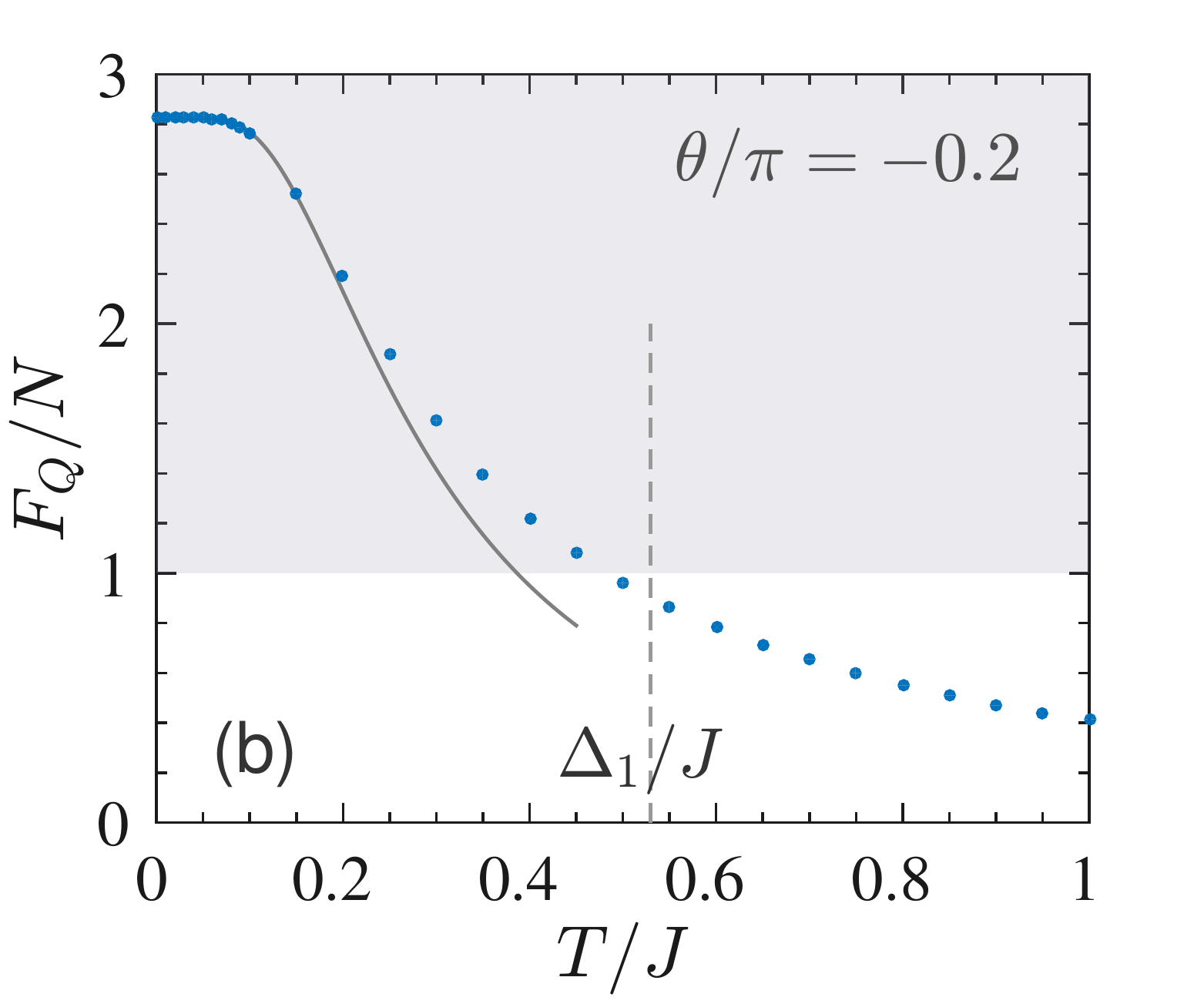} \hspace{-15pt}
\includegraphics[height=0.283\textwidth]{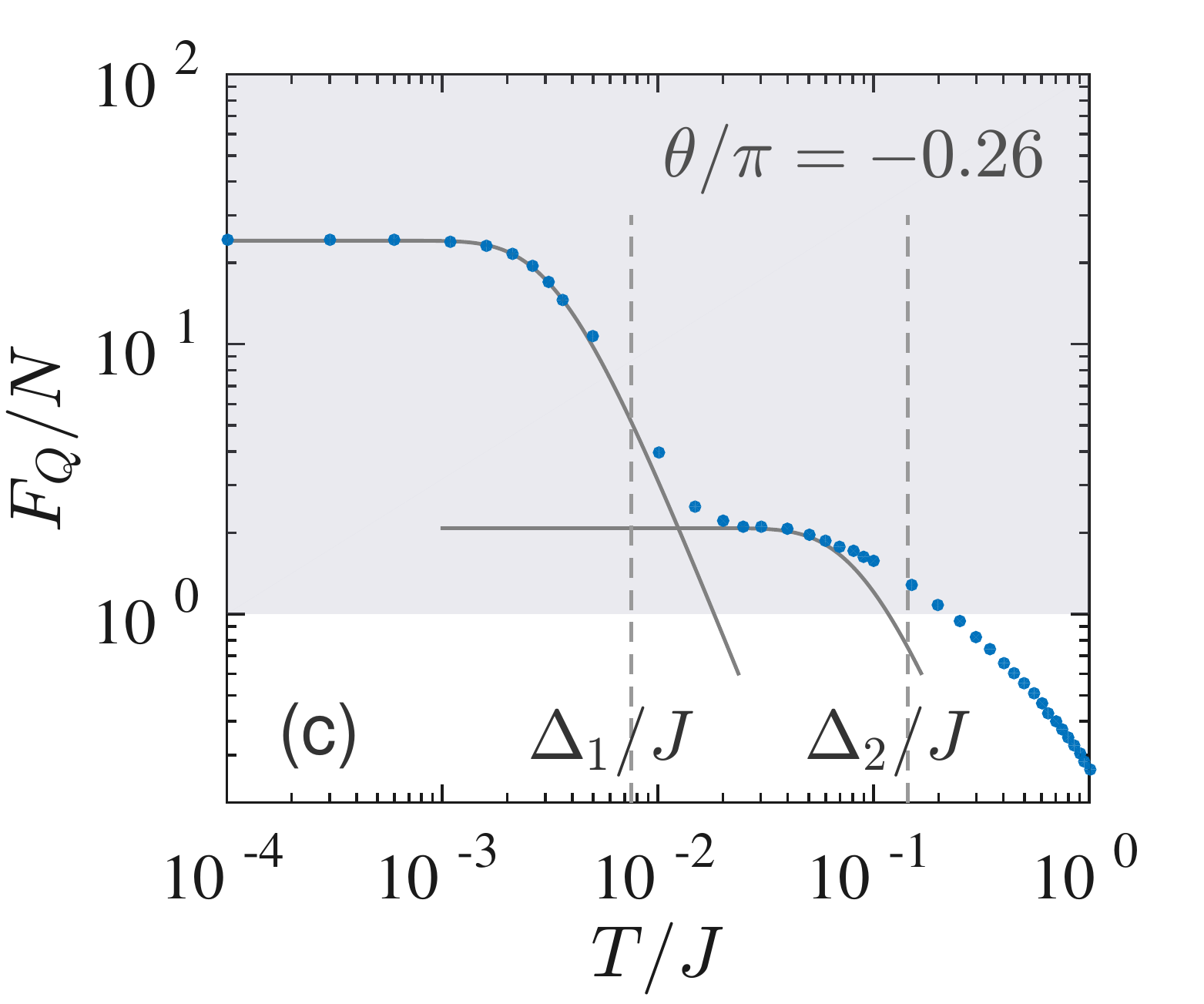} 
\caption{\textbf{(a)} Fisher density $F_Q[\hat{\rho}_T]/N$ for a chain of $N=50$ spins, 
calculated using the optimal operator $\hat{J}_z$, as a function of $\theta$ and $T$. 
In the coloured region we find $F_Q>N$, which corresponds to multipartite entanglement.
\textbf{(b,\,c)} Cuts of the density plot in panel (a) for fixed parameter $\theta>\thetac$ and $\theta<\thetac$, respectively,
showing the dependence on $T$. We compare numerical calculations (dots) to the two-mode low-temperature decay 
$\tanh^2(\DeltaE/2T)$ and $\tanh^2(\DeltaEE/2T)$ as in Eq.~(\ref{QFIfactorization}) (solid lines); 
the vertical dashed lines indicate $T=\DeltaE$ and $T=\DeltaEE$.} 
\label{fig:IsingThermic1}
\end{figure}

\begin{figure}[t!]
\centering
\includegraphics[width=0.55\textwidth]{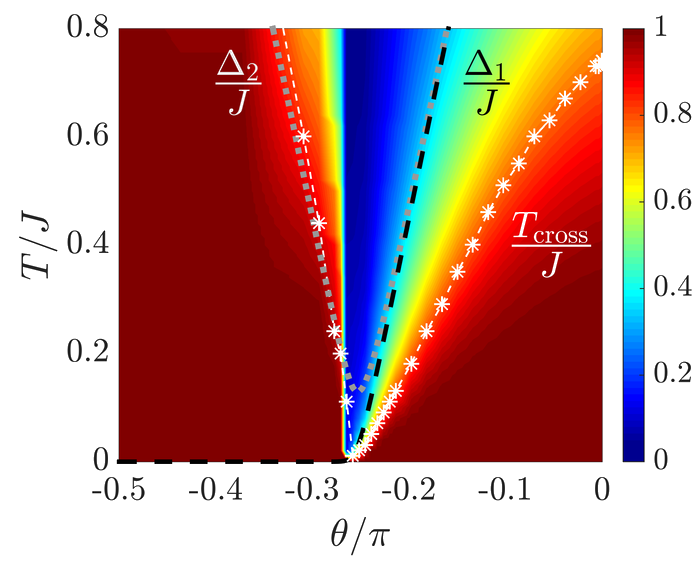}
\vspace{-9pt}
\caption{Normalized Fisher information $F_Q[\hat{\rho}_T]/F_Q[\hat{\rho}_0]$ as a function of $\theta$ and $T$ for $N=50$. 
The white dots flag a maximum of $\partial{F_Q}/\partial{T}$ 
and the resulting fan-shaped curve is taken as a finite-temperature crossover.
The first and second energy separations $\DeltaE$ (dashed line) and $\DeltaEE$ (dotted line) are also shown.} 
\label{fig:IsingThermic2}
\end{figure}

\paragraph{Results}
The optimal operator at $T>0$ is numerically found to be $\hat{J}_z$, that is the order parameter at $T=0$.
Thus $\,F_Q[\hat{\rho}_T] \equiv \max_\varrho F_Q\big[\hat{\rho}_T,\hat{J}_\varrho\big] = F_Q\big[\hat{\rho}_T,\hat{J}_z\big]\,$ 
and $\,\pazocal{S}_{\varrho\varrho}(\omega) = \pazocal{S}_{zz}(\omega)$. 
The main results are shown in Figs.~\ref{fig:IsingThermic1} and \ref{fig:IsingThermic2}.\footnote{\ We expect all the results to be symmetric for the antiferromagnetic coupling $\theta>0$ using the operator $\hat{J}_z^{\rm (st)}$.\\[-9pt]}
%%%%%%%%
\\[-6pt]

%%%%%%%%%%%%%%%%%%%%%%%%%%%%%%%%%%%%%%%%%%%%%%%%%%%%%%%%%%%%%%%%%%
%%%%%%%%%%%%%%%%%%%%%%%%%%%%%%%%%%%%%%%%%%%%%%%%%%%%%%%%%%%%%%%
%%%%%%%%%%%%%%%%%%%%%%%%%%%%%%%%%%%%%%%%%%%%%%%%%%%%%%%%%%%%%%%
%%%%%%%%%%%%%%%%%%%%%%%%%%%%%%%%%%%%%%%%%%%%%%%%%%%%%%%%%%%%%%%

\noindent \tripieno{MySky} \ At finite temperature, the competition between thermal population of excited states 
and genuinely quantum correlations of the ground state results in the phase diagram in Fig.~\ref{fig:IsingThermic1}\Panel{a}, 
where the maximal Fisher density $f_Q[\hat{\rho}_T]=F_Q[\hat{\rho}_T]/N$ is plotted as a function 
of control parameter $\theta$ and temperature $T$. 
%%%%%%%
\\[9pt]
%%%%%%%
\trivuoto{MyBlue} \ The coloured area $f_Q[\hat{\rho}_T]>1$ corresponds to entangled thermal states: 
quantum correlations survives against thermal fluctuations in a broad region 
around the ferromagnetic critical point $\thetac\equiv\thetaFM$.
Albeit at $T=0$ and $\theta=\thetac$ the power-law scaling $f_Q[\ket{\psi_{0}}] \sim N^{3/4}$ 
implies a divergent multipartiteness in the thermodynamic limit, 
at $T>0$ multipartiteness does not scale with $N$: the Fisher density is intensive $f_Q[\hat{\rho}_T]\sim\pazocal{O}(1)$ for large $N$
(we will discuss in the following the meaning of ``large $N$'').
Nevertheless, thermal states for $\theta\approx\thetac$ contains the maximum fraction of entangled particles also at $T>0$.
%%%%%%%
\\[9pt]
%%%%%%%
\trivuoto{MyBlue} \ The Fisher density monotonically decreases from the zero-temperature value $f_Q[\ket{\psi_{0}}]$ as $T$ increases,
painting a characteristic shape for the region of the witnessed multipartite entanglement in the $\theta$--$T$ plane:
it fans out from the zero-temperature noninteracting corner $(\theta=0,T=0)$,
reaches a maximum extension $\Tmax$ at $\theta_{\rm max}\approx0.4\,J$,
then suddenly shrinks for $\theta\lesssim\thetac$. 
Yet, for $\theta^*<\theta<\thetac$, with $\theta^*\approx2/3$, thermal entanglement survives.
For lower field $\theta<\theta^*$, multipartite entanglement is no more witnessed for any $T>0$.
%%%%%%%
\\[9pt]
%%%%%%%
\trivuoto{MyBlue} \ The multipartite entanglement detected by the QFI survives up to $\Tmax\approx0.6\,J$ 
for a suitable choice of the parameter $\theta$ and up to $T\approx0.3\,J$ at $\theta=\thetac$.
Interestingly, the latter threshold is consistent with the temperature up to which the effects of quantum criticality persist 
in the Ising chain, as reported both theoretically (about $0.35\,J$)~\cite{KoppNATPHYS2005} 
and experimentally (about $0.28\,J$)~\cite{KinrossPRX2014}. 
This finding is remarkable because it suggests to link two different concepts: 
the thermal robustness of multipartite entanglement and the boundary of the universal part of the quantum critical region in the phase diagram.
%%%%%%%%
\\[-6pt]

\noindent The \emph{thermal phase diagram} of the Ising model on the $\theta$--$T$ plane first appeared in Ref.~\cite{HaukeNATPHYS2016}. 
Our plot~\ref{fig:IsingThermic1}\Panel{a} is in perfect agreement with this earlier one.
%%%%%%%%
\\[-6pt]

\noindent \tripieno{MySky} \ The typical behaviour of $F_Q[\hat{\rho}_T]$ as a function of temperature is shown in Figs.~\ref{fig:IsingThermic1}\Panel{b,\,c}. It is intimately connected to the structure of the low-energy spectrum
and its dependence on the chain size $N$.
%%%%%%%
\\[9pt]
%%%%%%%
\trivuoto{MyBlue} \ At $\theta>\thetac$, the first energy separation is finite $\DeltaE>0$: 
the QFI is approximately constant and equal to $F_Q[\ket{\psi_{0}}]$ for low temperatures $T\ll\DeltaE$ 
and smoothly decays for higher temperatures $T\gtrsim\DeltaE$ according to the qualitative law 
$F_Q[\hat{\rho}_T] = F_Q[\hat{\rho}_0] \tanh^2(\DeltaE/2T)$, 
where $\hat{\rho}_0 = \ket{\psi_0}\bra{\psi_0}$ is the ground state of the system.
%%%%%%%
\\[9pt]
%%%%%%%
\trivuoto{MyBlue} \ At $\theta<\thetac$, the first energy gap $\DeltaE$ becomes smaller and smaller when increasing $N$:
a finite-size analysis extended to $N=10\div10^3$ reveals that the gap closes exponentially $\DeltaE \propto \neper^{-N/n_\DeltaE}$, 
with $n_\DeltaE=a|1-\cot\theta|^b$ ($a\approx1.5$, $b\approx-0.85$).
The decay of $F_Q[\hat{\rho}_T]$ from the zero-temperature value occurs around a finite 
(exponentially small in $N$) temperature $T\lesssim\DeltaE$, according to
$F_Q[\hat{\rho}_T] = F_Q[\ket{\psi_0}] \tanh^2(\DeltaE/2T)$ as predicted by Eq.~(\ref{QFIfactorization}). 
Multipartite entanglement witnessed by the QFI in ferromagnetic phases results to be thermally very fragile:
the extensive multipartiteness hosted by the ground state is lost already at extremely small temperatures for a sufficiently large chain.
In the limit $N\to\infty$, the energy gap $\DeltaE$ vanishes, the ground state becomes doubly degenerate ($\mu=2$) 
and so does the first excited state ($\nu=2$): the first finite energy separation in the many-spin spectrum is given by 
the difference $\DeltaEE=E_3-E_2$ between the energies of these two degenerate states.
% Accounting for the double degeneracy of the ground ($\mu=2$) and first excited ($\nu=2$) states, 
A slower decay $F_Q[\hat{\rho}_T] = F_Q[\hat{\rho}_0] \tanh^2(\DeltaEE/2T)$ 
takes place for $\DeltaE \ll T \lesssim \DeltaEE$, where 
$\hat{\rho}_0 = \frac{1}{2}\big(\ket{\psi_0}\bra{\psi_0} + \ket{\psi_1}\bra{\psi_1}\big)$. 
Note that in the thermodynamic limit there is no continuity between 
the zero-temperature value $F_Q[\ket{\psi_0}] \sim N^2$ 
and the finite-temperature plateau $F_Q[\hat{\rho}_0] \sim N$ at $T\to0^+$.
%%%%%%%
\\[9pt]
%%%%%%%
\trivuoto{MyBlue} \ At criticality $\theta=\thetac$, the first energy gap closes as a power law $\DeltaE \simeq J\,N^{-1}$.
For low temperatures $T\ll\DeltaE$, the system is effectively confined in the ground state
and the zero-temperature scaling $F_Q \sim N^{7/4}$ is still proven to hold~\cite{HaukeNATPHYS2016}.
For intermediate temperatures $\DeltaE \ll T \ll J$ the QFI follows the power law $F_Q \sim T^{-3/4}$, 
that can been rigorously determined from a scaling ansatz~\cite{HaukeNATPHYS2016,Gabbrielli2018a}.
We notice that this scaling behaviour is a characteristic feature of the whole quantum critical region 
$(\theta>\thetac,T\gg\DeltaE)\,\vee\,(\theta<\thetac,T\gg\DeltaEE)$, not only of the critical point.
%%%%%%%%
\\[-6pt]

%%%%%%%%%%%%%%%%%%%%%%%%%%%%%%%%%%%%%%%%%%%%%%%%%%%%%%%%%%%%%%%%%
%%%%%%%%%%%%%%%%%%%%%%%%%%%%%%%%%%%%%%%%%%%%%%%%%%%%%%%%%%%%%%%
%%%%%%%%%%%%%%%%%%%%%%%%%%%%%%%%%%%%%%%%%%%%%%%%%%%%%%%%%%%%%%%
%%%%%%%%%%%%%%%%%%%%%%%%%%%%%%%%%%%%%%%%%%%%%%%%%%%%%%%%%%%%%%%

\noindent Essentially, we confirm former predictions~\cite{NakataPRA2009} about the existence of two regimes of parameter~$\theta$: 
one where multipartiteness at zero temperature is low (intensive) but thermally robust 
and another one where multipartiteness at zero temperature is high (extensive) but thermally fragile. 
%%%%%%%%
\\[-6pt]

\noindent \tripieno{MySky} \ Figure~\ref{fig:IsingThermic2} is a density plot of $F_Q[\hat{\rho}_T]/F_Q[\hat{\rho}_0]$ in the $\theta$--$T$ plane.
We normalize over the QFI at ``low temperature''\footnote{\ We denote with the label ``$\,T=0^+\,$'' the range of temperatures 
$\DeltaEE \gg T \gg \DeltaE \xrightarrow[]{N\gg1} 0$, sufficiently high to generate an incoherent mixture of the quasi-degenerate states 
of energies $E_1,\,E_2$ but low enough not to populate the state at energy $E_3$.\\[-9pt]} 
\be
F_Q[\hat{\rho}_0] = 
\left\{
\begin{array}{ll} 
F_Q[\ket{\psi_0}] \equiv F_Q(T=0) & {\rm for} \ \theta>\thetac \\ 
F_Q\big[\frac{1}{2}(\ket{\psi_0}\bra{\psi_0} + \ket{\psi_1}\bra{\psi_1})\big]
\equiv F_Q(T=0^+) & {\rm for} \ \theta<\thetac
\end{array} 
\right.
\ee 
because here we are not interested in capturing the fast decay from the zero-temperature entanglement in the ordered phase, 
but rather in highlighting the slower decay governed by the second energy separation $\DeltaEE$.
The diagram displays the V-shaped structure radiating from the critical point as in the paradigmatic Fig.~\ref{fig:thermalDecay}.
The entanglement response to thermal fluctuations $\partial{F_Q}/\partial{T}$ displays a maximum\footnote{\ The maximum of $\partial F_Q/\partial T$ as a function of $T$ does not turn into a sharp infinite peak for $N\to\infty$, so it is profoundly different from the maximum of the entanglement response to the control parameter $\partial{F_Q}/\partial{\theta}$ examined in paragraph~\ref{subsec:IsingMEinGS}, whose divergent character signals the QPT. 
%The crossover temperature $\Tcross$ can be equivalently identified by the inflection point $\partial^2F_Q/\partial{T}^2=0$.
\\[-9pt]} at the crossover temperature $\Tcross(\theta)$.
This is presented as a white dotted line, signalling the loss of the low-temperature entanglement.
In other words, multipartite entanglement witnessed in $\hat{\rho}_0$ survives for $T\lesssim\Tcross$, while
for $T\gg\Tcross$ the system crosses over into a region of smooth decay of entanglement.
%%%%%%%
\\[9pt]
%%%%%%%
\trivuoto{MyBlue} \ For $\theta>\thetac$, the crossover temperature $\Tcross$ follows the energy gap $\DeltaE$. 
In particular, we numerically find a constant ratio $\DeltaE/\Tcross\approx2.54$, 
remarkably close to the one we will obtain in chapter~\ref{ch:LMG} for a different spin model. 
%%%%%%%
\\[9pt]
%%%%%%%
\trivuoto{MyBlue} \ For $\theta<\thetac$, the crossover temperature $\Tcross$ mimics the energy gap $\DeltaE$. 
Numerical data are not neat in this region, owing to a truncation in the calculation of the integral~(\ref{IsingDynamicFactor}) 
that has been necessary due to computational resources.
%%%%%%%%
\\[-6pt]

\noindent Thus, $\DeltaE$ and $\DeltaEE$ act as energy scales ruling the fabric texture of the phase diagram 
not only in vicinity of the ferromagnetic critical point.

\section{Chain with arbitrary-range interaction} \label{sec:IsingArbitrary}
In the last decades, long-range interactions in spin models have received considerable attention~\cite{CardyJPA1981, DuttaPRB2001, LahayeRPP2009, KoffelPRL2012, VodolaNJP2016} inasmuch they generate exotic new phenomena, 
such as interaction-induced frustration~\cite{IslamSCIENCE2013}, 
violation of the area law for bipartite entanglement~\cite{KoffelPRL2012, VodolaNJP2016}
and nonlocal propagation of quantum correlations~\cite{RichermeNATURE2014}.
Moreover, the possibility to engineer Ising-like models with tens spins and long-range interactions
on both one-dimensional and two-dimensional lattices materialized very recently~\cite{BrittonNATURE2012, SchneiderRPP2012, IslamSCIENCE2013, JurcevicNAT2014}; in particular, with reference to the Hamiltonian introduced in Eq.~(\ref{IsingHam}), 
the interval of decay power $0\lesssim\alpha\lesssim3$ is currently within reach. 
All the aforesaid motivations encouraged us to extend the investigation on the Ising chain 
keeping into account spin-spin interactions having arbitrary decay range with the distance.
We limited our study to the ground state; 
% along the lines of Refs.~\cite{KoffelPRL2012,VodolaNJP2016}, ... but providing ...;
we incidentally mention here that some attempts to describe the finite-temperature scenario also appeared in literature~\cite{DuttaPRB2001}.
As far as we know, no analogous work on multipartite entanglement in the ground state of the long-range Ising model 
for both ferromagnetic and antiferromagnetic coupling exists in literature.

\paragraph{Method}
The Ising Hamiltonian~(\ref{IsingHam}) with arbitrary interaction range is not treatable analytically. 
Therefore, we rely only on numerical results. 
For short chains $N\leq20$ we performed an exact brute-force diagonalization to find the energy levels and the ground state,
from which an easy calculation of the energy gap, order parameter, fidelity susceptibility and quantum Fisher information was possible. 
For $N>20$, we utilized an algorithm based on the density-matrix renormalization group (DMRG)~\cite{WhitePRL1992,WhitePRB1993} 
and the Lanczos scheme~\cite{WuSIAM2000}.
The latter procedure\footnote{\ In our case, the DMRG method aims to approximate the actual (unknown) $2^N$-dimensional pure ground state 
via a $m \times m$ reduced density matrix through the iteration of $n$ steps, 
during which the approximated density-matrix candidate is updated. 
The overall accuracy is estimated by the \emph{truncation error}, 
which is nothing but the sum of the $2^N-m$ eigenvalues of the discarded states; or, in other words,
the residue of the trace of the density matrix after the final truncation.
Some brief technical details: for all simulations we used a number of basis states in the interval $m\approx100\div150$
and a total number of sweeps $n=5$, to get an accuracy in the determination of the ground-state energy smaller than 
the prescribed tolerance of $10^{-9}$. The truncation error was always less than $10^{-7}$.\\[-9pt]} 
-- an iterative, variational technique optimized for the convergence of the ground state -- 
provided us with all the spin-spin correlations, by dint of which we could evaluate the QFI via the relation~(\ref{IsingfQCorr}) 
up to about $N\approx150$.

\subsection{Entanglement phase diagram}
\begin{figure}[p!]
\centering
\framebox{$\alpha=10$} \\
\includegraphics[width=0.72\textwidth]{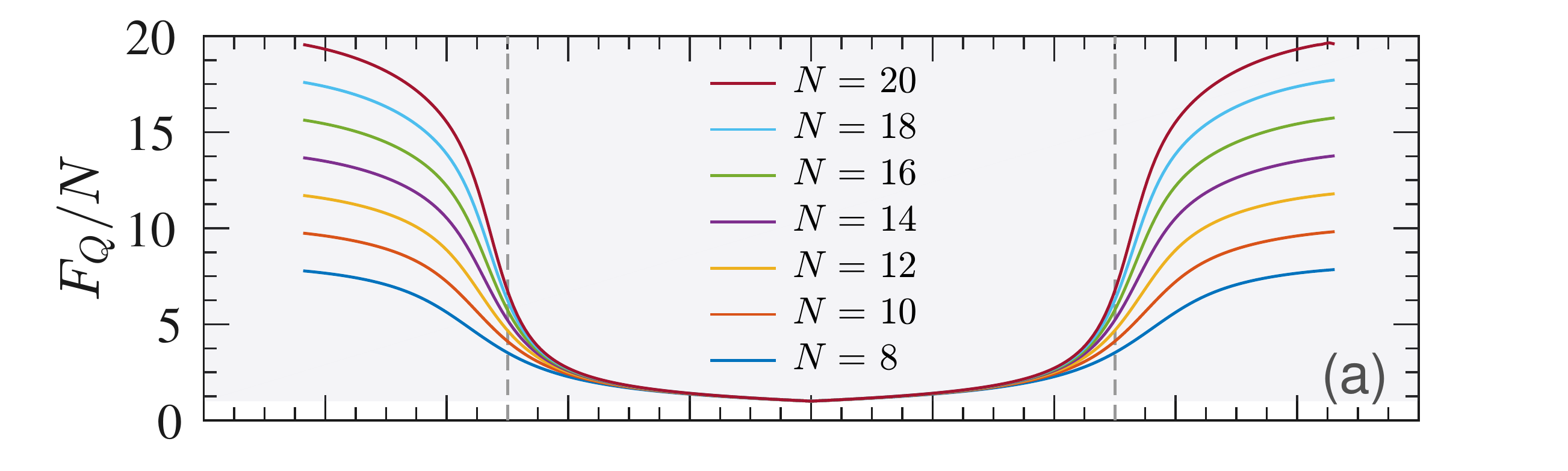} \hspace{-20pt}
\includegraphics[width=0.29\textwidth]{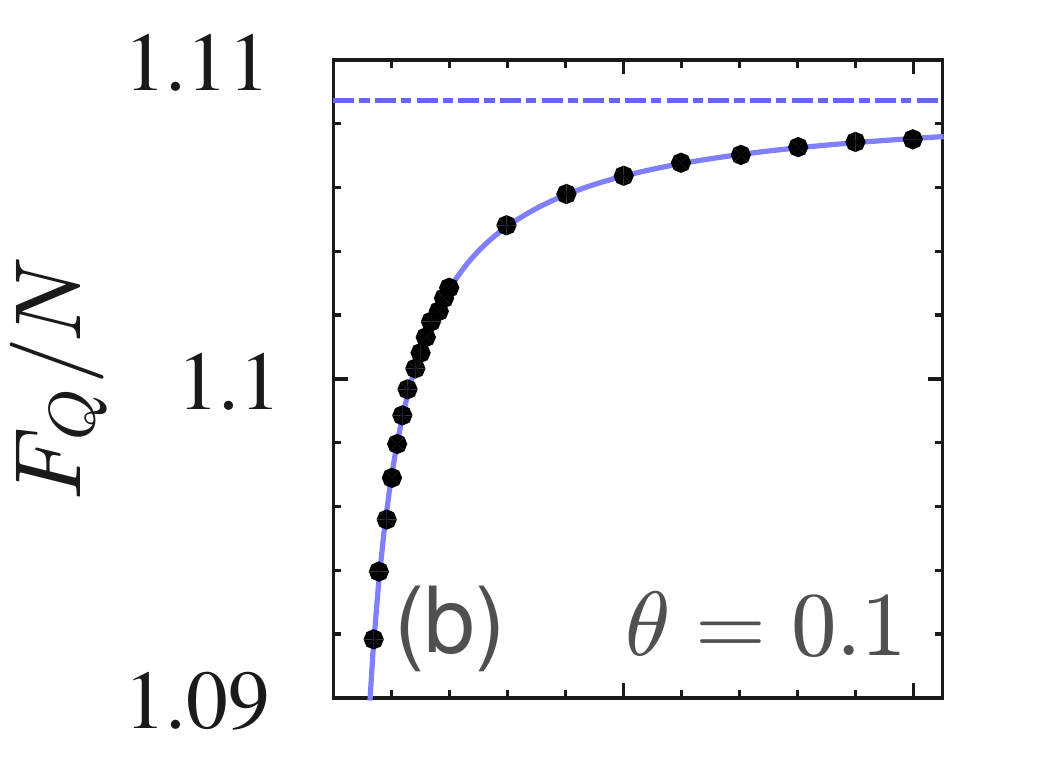} \\[-6pt]
\includegraphics[width=0.72\textwidth]{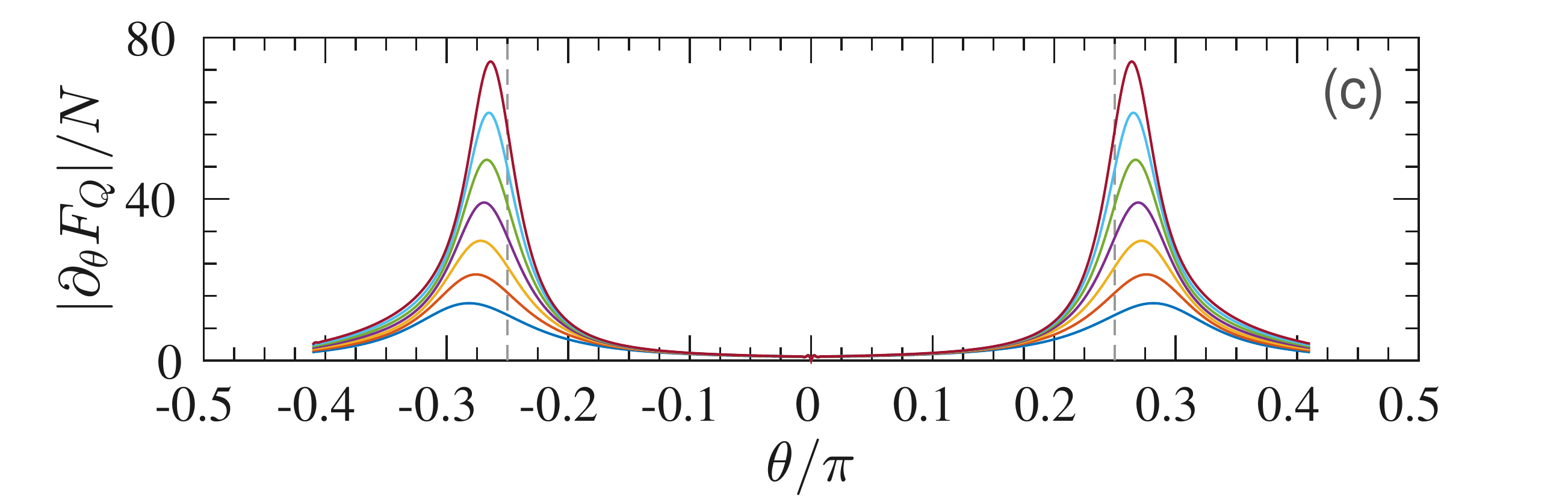} \hspace{-20pt}
\includegraphics[width=0.29\textwidth]{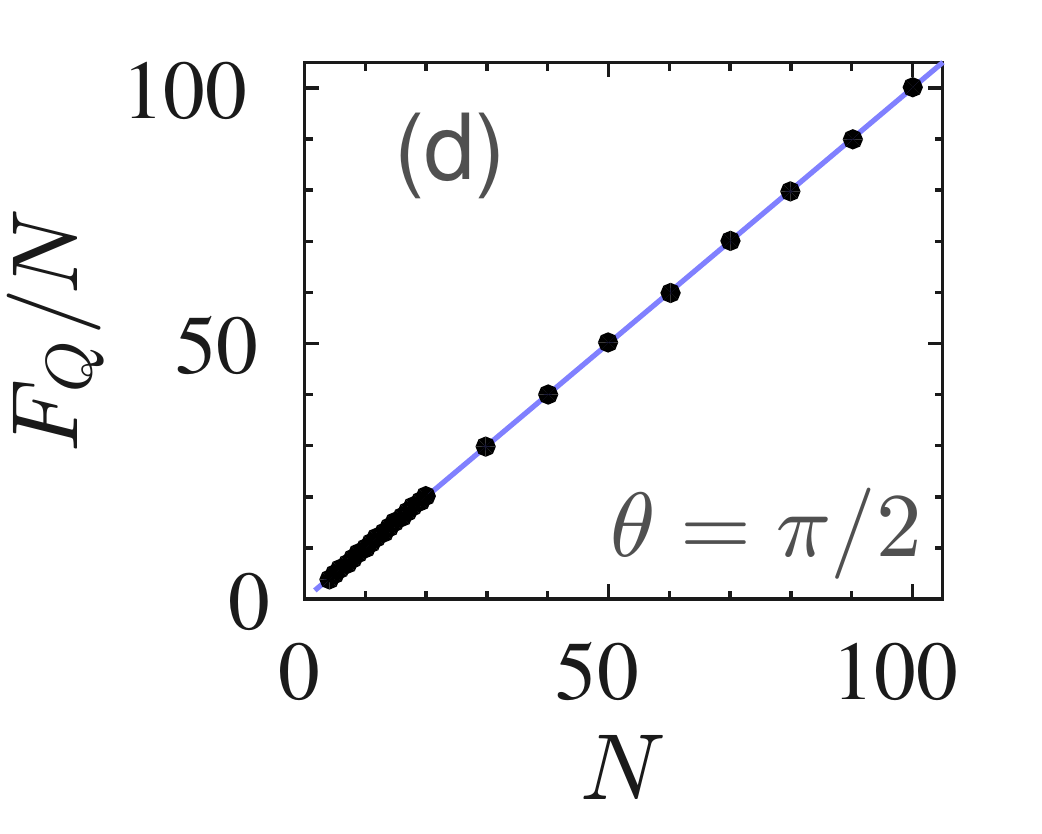} \\
\framebox{$\alpha=1$} \\
\includegraphics[width=0.72\textwidth]{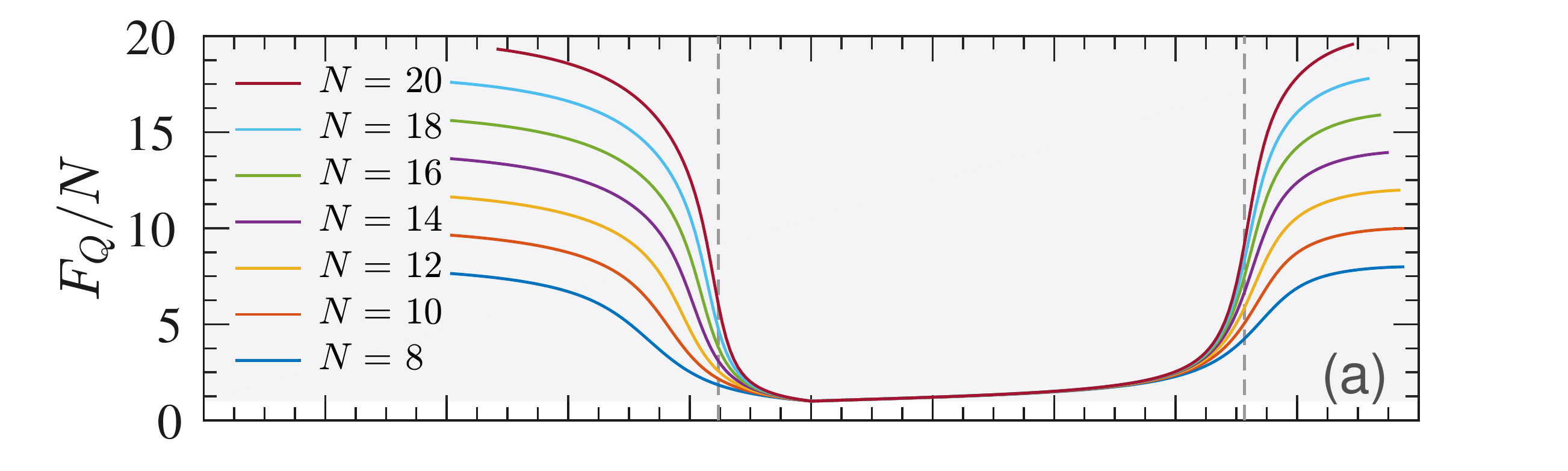} \hspace{-20pt}
\includegraphics[width=0.29\textwidth]{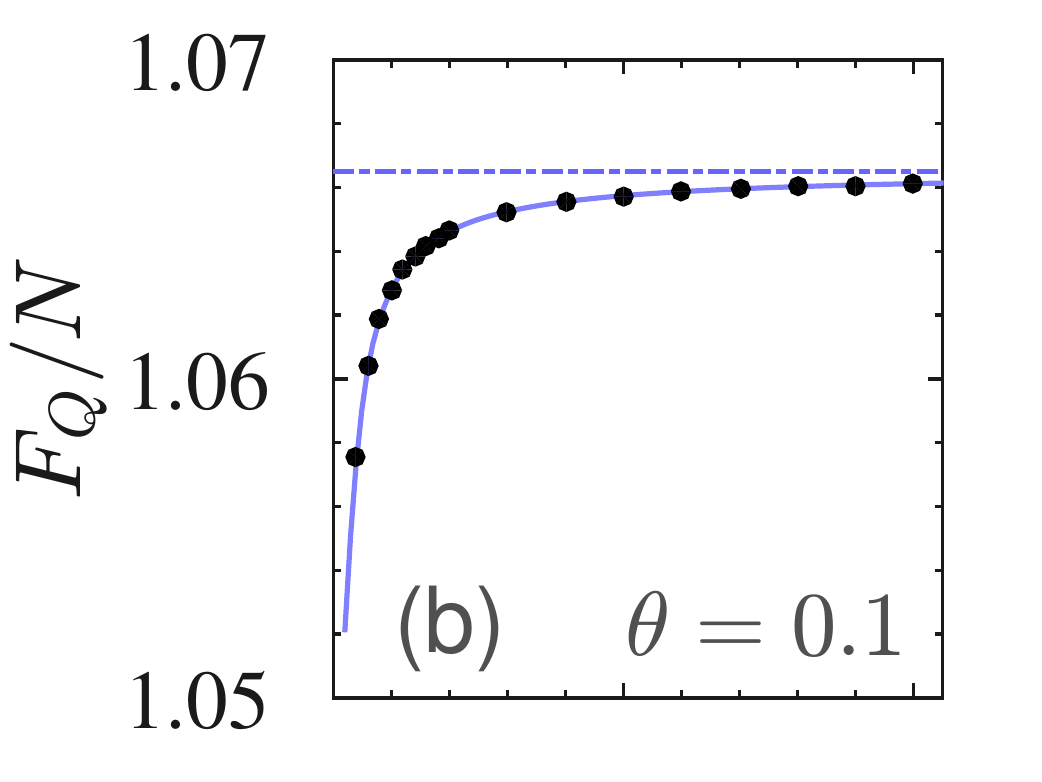} \\[-6pt]
\includegraphics[width=0.72\textwidth]{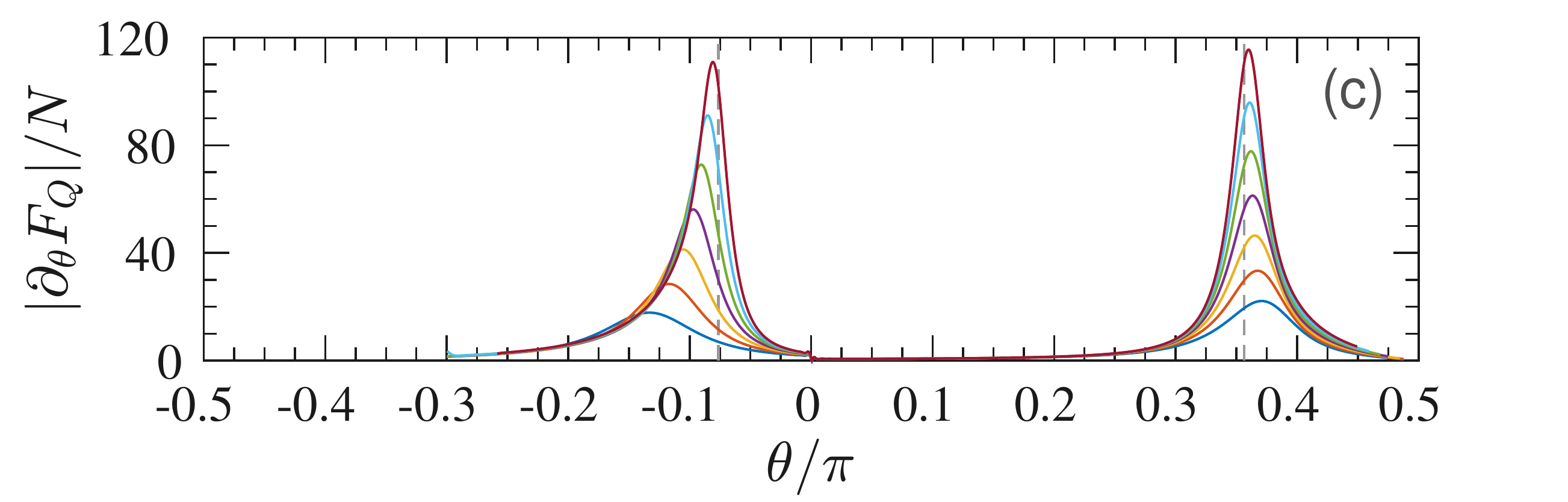} \hspace{-20pt}
\includegraphics[width=0.29\textwidth]{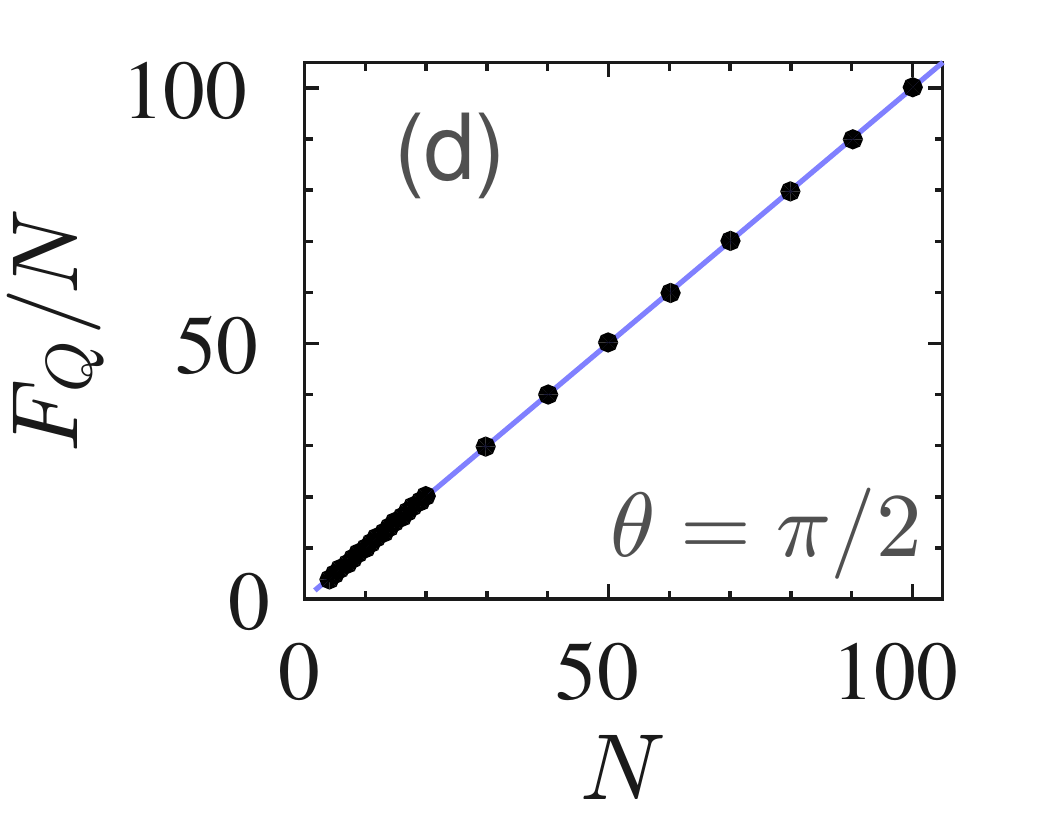} \\
\framebox{$\alpha=0.1$} \\
\includegraphics[width=0.72\textwidth]{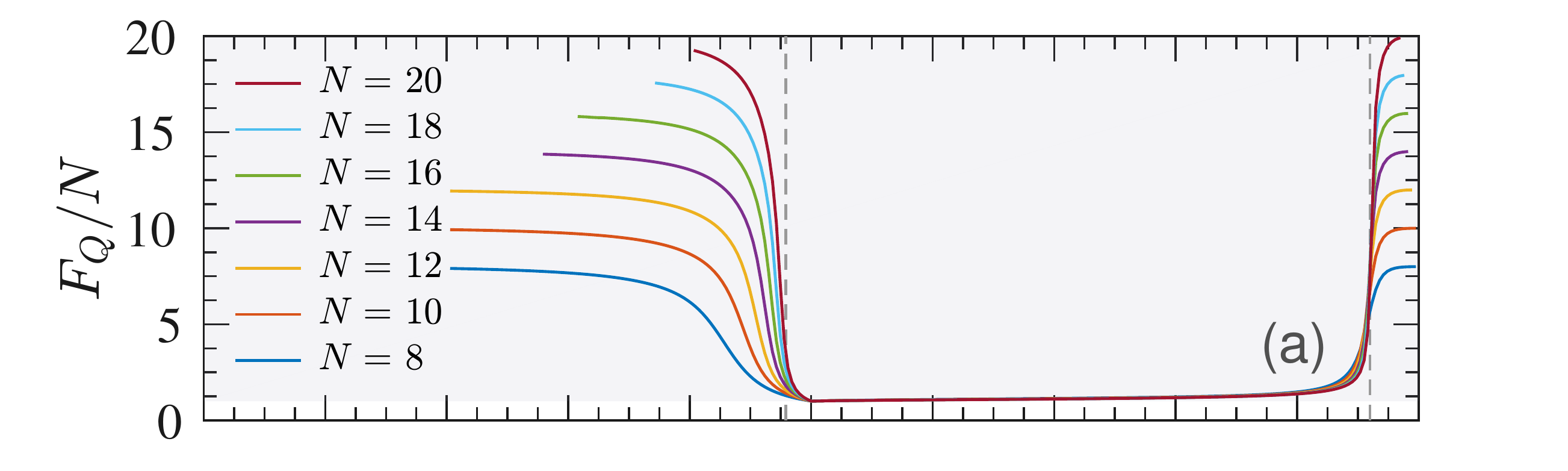} \hspace{-20pt}
\includegraphics[width=0.29\textwidth]{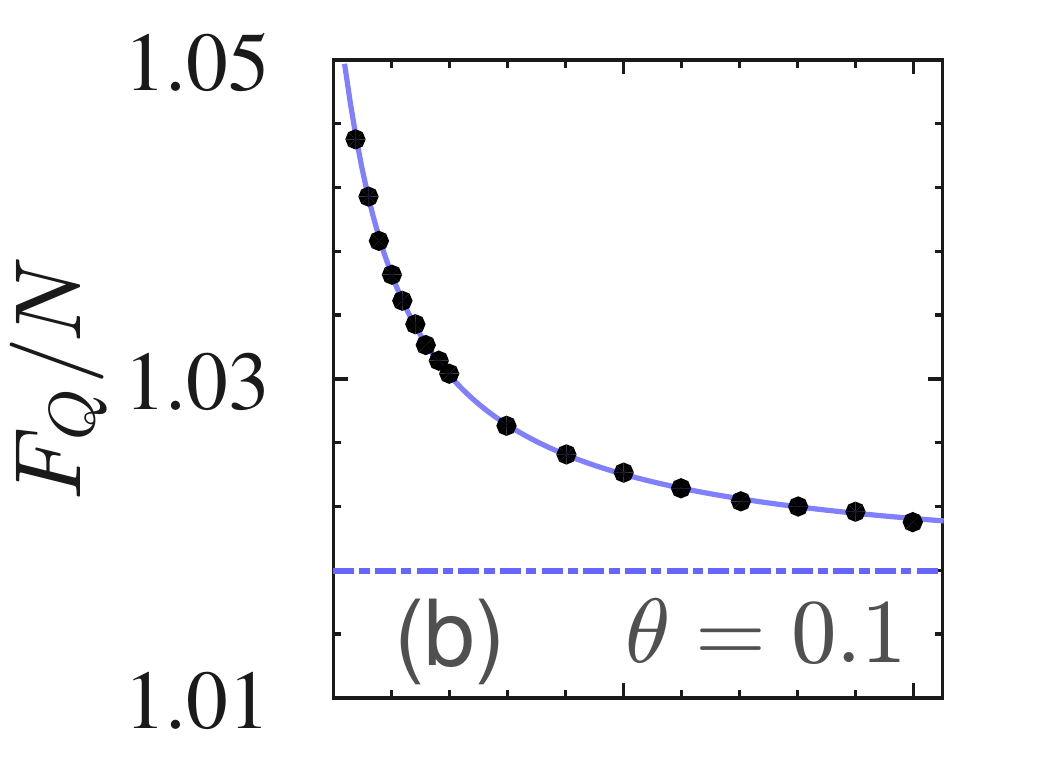} \\[-6pt]
\includegraphics[width=0.72\textwidth]{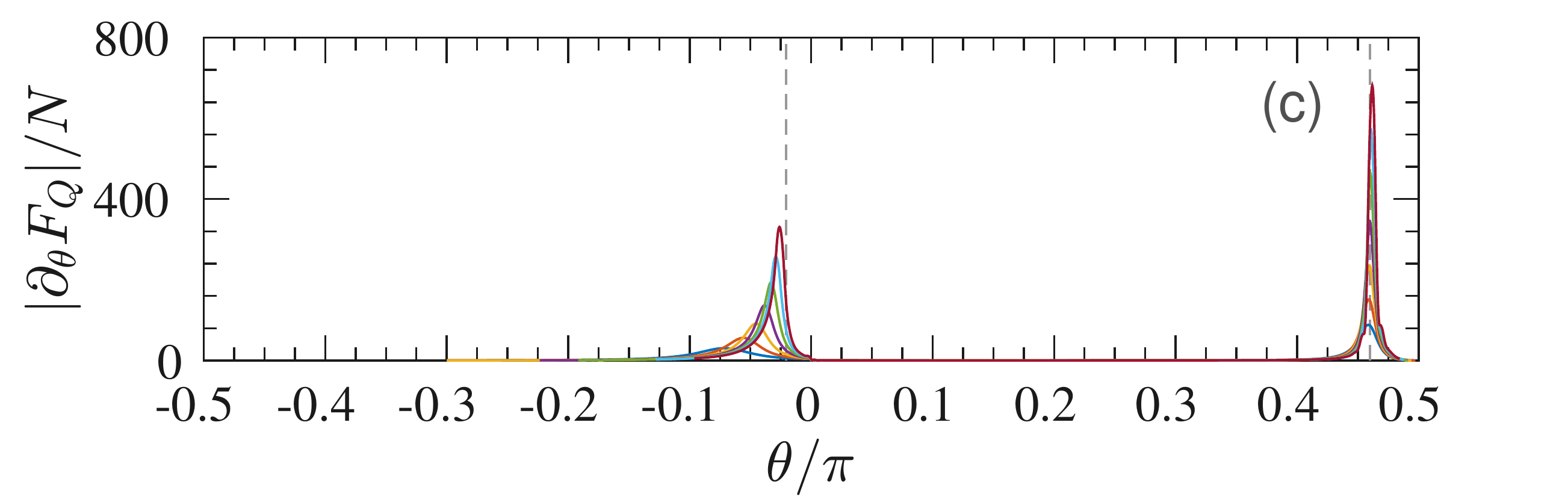} \hspace{-20pt}
\includegraphics[width=0.29\textwidth]{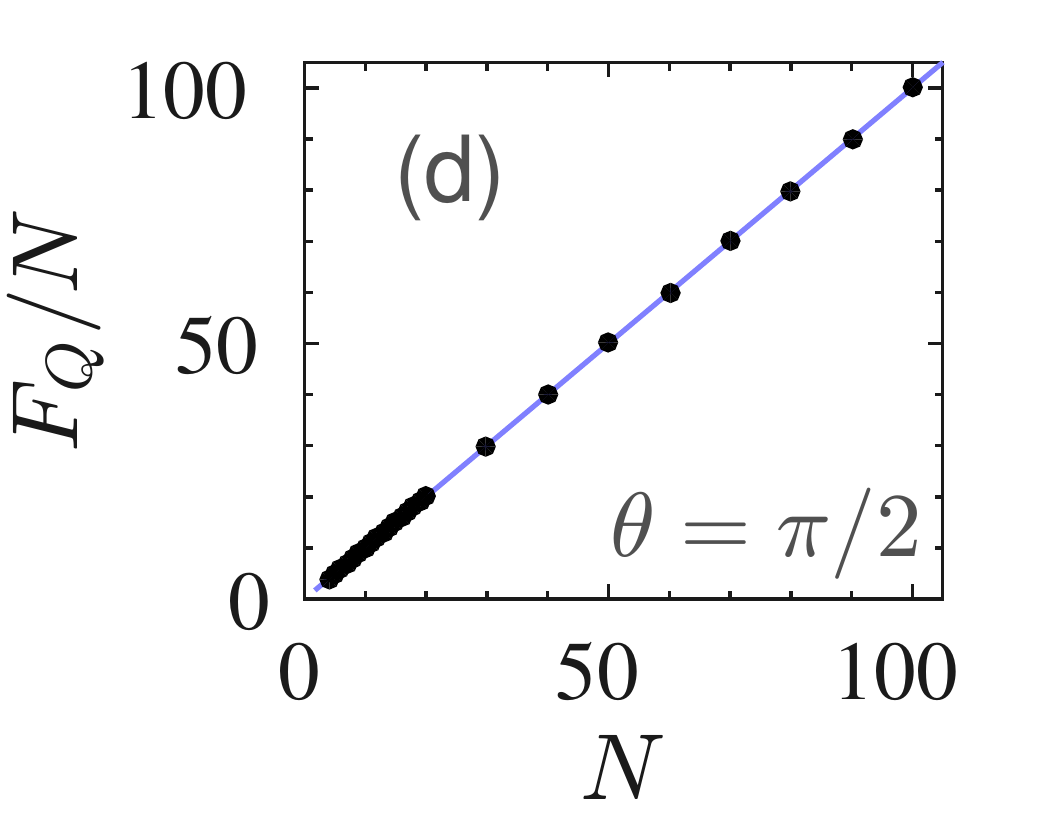} \\[-6pt]
\caption{Behaviour of the optimal Fisher density $f_Q[\ket{\psi_0}]$ for $\alpha=10,\,1,\,0.1$: 
\textbf{(a)} $f_Q$ as a function of $\theta$ for several values of $N$ 
(different colours: the cutoff is imposed by numerical instability);
\textbf{(c)} first derivative of $f_Q$ with respect to $\theta$ (colours as in panels a); 
%(b,\,d) $f_Q[\ket{\psi_0}]$ at $\theta=0.1$ and $\frac{\pi}{2}$, respectively, for increasing $N$ 
%(solid lines are fit curves, dashed lines are bounds).
\textbf{(b)} $f_Q$ at $\theta=0.1$ for increasing $N$ (solid lines are rational fit curve bounded by the dashed lines);
\textbf{(d)} $f_Q$ at $\theta=\frac{\pi}{2}$ for increasing $N$ (solid lines are guides to the eye $f_Q=N$).
}
\label{fig:IsingAlphaOpt}
\end{figure}

\begin{figure}[t!]
\centering
\includegraphics[width=0.49\textwidth]{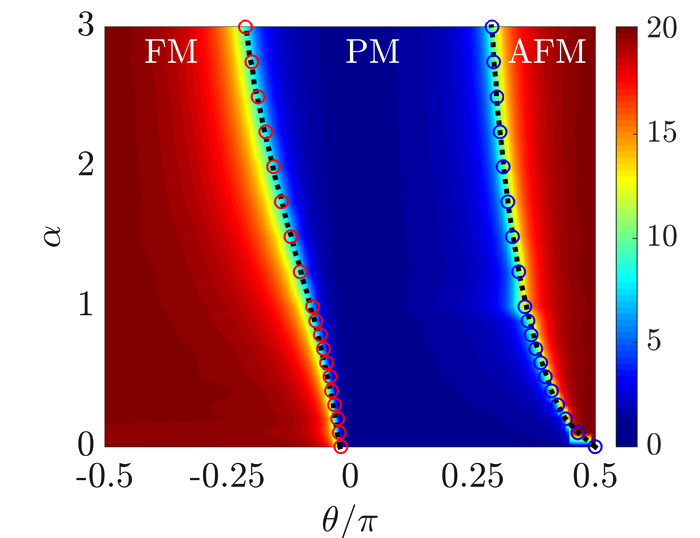}
\includegraphics[width=0.49\textwidth]{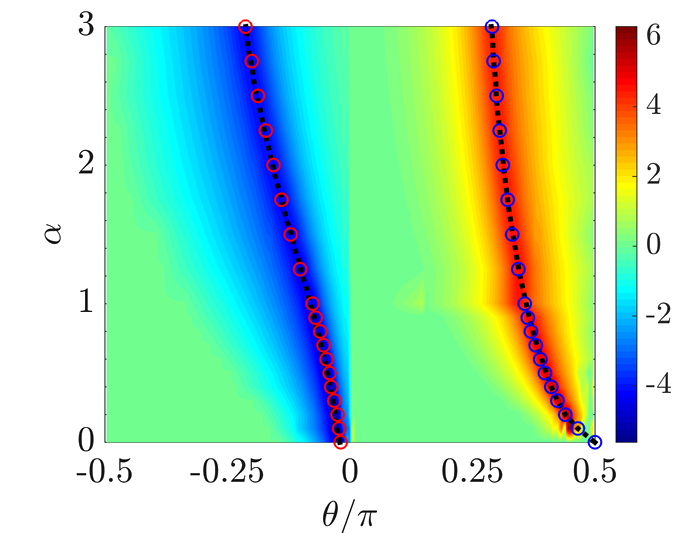}
\caption{Phase diagram of the transverse Ising chain with $N=20$ spins on the plane $\theta$--$\alpha$, 
based on the content of multipartite entanglement witnessed by the QFI. 
\textbf{(Left)} Density plot of the Fisher density $f_Q[\ket{\psi_{0}}]$: 
the FM and AFM orders are recognized by the extensivity $f_Q \sim N$, whereas 
in the PM phase multipartiteness does not grow with the size $f_Q \sim \pazocal{O}(1)$.
Plots in panels~\ref{fig:IsingAlphaOpt}\Panel{a} are cuts at fixed $\alpha$'s. 
\textbf{(Right)} Density plot of the first derivative $\partial_\theta f_Q[\ket{\psi_{0}}]$: 
the peaks denote proximity to QPTs, as certified by the comparison to the critical line 
obtained from the maxima of fidelity susceptibility (dashed lines); 
a sample of the points where the energy gap vanishes is also superimposed (circles).
The peaks become sharper and sharper towards the thermodynamic limit.
} 
\label{fig:IsingPhaseDiagram20}
\end{figure}

Along the lines of the foregoing section, we quantify the multipartite entanglement by computing the QFI 
and comparing it to the well-known bound for $\kappa$-partite entanglement (paragraph~\ref{QFIbounds}). 
The choice of the optimal observables mirrors the choice of the order parameters made in paragraph~\ref{subsec:IsingMEinGS}: 
$\hat{J}_z$ for FM coupling ($\theta<0$) and $\hat{J}_z^{\rm (st)}$ for AFM coupling ($\theta>0$). 
Several checks at various values of parameters $\theta$ and $\alpha$ 
were performed at low spin number $N\lesssim20$ for confirming this stance.
Figures~\ref{fig:IsingAlphaOpt} and \ref{fig:IsingPhaseDiagram20} summarizes the main findings
and we discuss them in details.

%%%%%%%%%%%%%%%%%%%%%%%%%%%%%%%%%%%%%%%%%%%%%%%%%%%%%%%%
%%%%%%%%%%%%%%%%%%%%%%%%%%%%%%%%%%%%%%%%%%%%%%%%%%%%%%%%
%%%%%%%%%%%%%%%%%%%%%%%%%%%%%%%%%%%%%%%%%%%%%%%%%%%%%%%%
%%%%%%%%%%%%%%%%%%%%%%%%%%%%%%%%%%%%%%%%%%%%%%%%%%%%%%%%

%%%%%%%%%%%%%%%%%%%%%%%%%%%%%%%%%%%%%%%%%%%%%%%%%%%%%%%%%%%%%%%
%%%%%%%%%%%%%%%%%%%%%%%%%%%%%%%%%%%%%%%%%%%%%%%%%%%%%%%%%%%%%%%
%%%%%%%%%%%%%%%%%%%%%%%%%%%%%%%%%%%%%%%%%%%%%%%%%%%%%%%%%%%%%%%
%%%%%%%%%%%%%%%%%%%%%%%%%%%%%%%%%%%%%%%%%%%%%%%%%%%%%%%%%%%%%%%

\paragraph{Distinguishing different phases}
In panels~\ref{fig:IsingAlphaOpt}\Panel{a} we plot the behaviour of the optimal QFI as a function of the control parameter $\theta$ 
for increasing chain length $N$ and some representative \emph{finite} values of the decay range:
$\alpha=10$ (short range), $\alpha=1$ (unshielded Coulomb potential) and $\alpha=0.1$ (long range).
This quantity is able to detect the ordered phases in terms of superextensive scaling and to discriminate them 
from the disordered phase, characterized by a pure extensive behaviour.
%%%%%%%%
\\[-6pt]

\noindent \tripieno{MySky} \ Ferromagnetic and antiferromagnetic orders persist for all values $\alpha>0$. 
%%%%%%%
\\[9pt]
%%%%%%%
\trivuoto{MyBlue} \ The \emph{disordered} PM phase at $\thetaFM<\theta<\thetaAFM$ 
is dominated by the local part of Hamiltonian~(\ref{IsingHam}), 
provided by the transverse magnetic field. At $\theta\simeq0$, the ground state is approximately given by the $x$-polarized coherent state
weakly perturbed by the presence of local spin-flip excitations.
In contrast to the very-short range $\alpha\to\infty$ discussed in section~\ref{sec:IsingNN}, 
where correlation functions decay with distance according to a simple exponential law, 
for small $\alpha\lesssim2$ the correlator $C_{zz}^{(i,j)} = \bra{\psi_{0}}\hat{\sigma}_z^{(i)}\hat{\sigma}_z^{(j)}\ket{\psi_{0}}$ 
shows a \emph{hybrid} decay, that is exponential at short distances and algebraic at long distances~\cite{VodolaNJP2016}:
\be \label{HybridCzz}
C_{zz}^{(i,j)} \sim \left\{
\begin{array}{ll} 
\neper^{-|i-j|/\Xi_\alpha} & {\rm for} \ |i-j| \ll R_\alpha \\ 
|i-j|^{-\Upsilon_\alpha} & {\rm for} \ |i-j| \gg R_\alpha
\end{array}
\right. \, ,
\ee
where the decay length $\Xi_\alpha$, decay power $\Upsilon_\alpha$ and crossover distance $R_\alpha$ depend on $\alpha$. 
When $\alpha\lesssim1$, the decay appears to become purely algebraic since $R_\alpha\approx1$.
As $\Upsilon_\alpha>1$ for all values of $\alpha$, Eq.~(\ref{IsingfQCorr}) still guarantees an intensive Fisher density $f_Q$ for large $N$,
as we observe indeed: see panels~\ref{fig:IsingAlphaOpt}\Panel{b} for an example.
%%%%%%%
\\[9pt]
%%%%%%%
\trivuoto{MyBlue} \ The two \emph{ordered} phases are ruled by the nonlocal exchange term: 
the order is displayed in the form of a GHZ state (N\'eel state) along the $z$ direction at $\theta<\thetaFM$ ($\theta>\thetaAFM$), 
perturbed by pairs of domain walls that interpolate between the two degenerate ground states with spontaneous (staggered) magnetization along the $z$ and $-z$ directions.
It has been shown numerically~\cite{VodolaNJP2016} that the connected correlation function 
$\langle\hat{\sigma}_z^{(i)}\hat{\sigma}_z^{(j)}\rangle - \langle\hat{\sigma}_z^{(i)}\rangle\langle\hat{\sigma}_z^{(j)}\rangle$
conserves a power-law tail, ensuring a finite correlator $C_{zz}^{(i,j)}$.
Hence, the Fisher density is a superextensive quantity $f_Q \sim N$ in virtue of relation~(\ref{IsingfQCorr}). 
At the borders $\theta=\pm\frac{\pi}{2}$, the ground state is a maximally entangled state of $N$ spins 
and $f_Q=N$: %grows strictly with the system size: 
examples are provided in panels~\ref{fig:IsingAlphaOpt}\Panel{d}.
%%%%%%%%
\\[-6pt]

\noindent \tripieno{MySky} \ For interactions with infinite range $\alpha=0$, the antiferromagnetic order disappears. 
%%%%%%%
\\[9pt]
%%%%%%%
\trivuoto{MyBlue} \ As long as all the spins interact with the others via a constant coupling regardless of the mutual distance,
the actual one-dimensional geometry of the lattice is no more perceived: 
the doubly-degenerate N\'eel state at $\theta=\frac{\pi}{2}$ dissolves into a (exponentially degenerate) ground state 
made of all possible configurations of $N/2$ spins up and $N/2$ spins down. 
For $0<\theta<\frac{\pi}{2}$, the AFM order is destroyed and the system only experiences the PM phase.
%%%%%%%
\\[9pt]
%%%%%%%
\trivuoto{MyBlue} \ For FM all-to-all coupling $\theta<0$, the ground state is uniquely the twofold degenerate GHZ state: 
no PM phase exists and the system is necessarily ordered.
%%%%%%%%
\\[-6pt]

\noindent \tripieno{MySky} \ Multipartite entanglement $f_Q[\ket{\psi_{0}}]>1$ is witnessed for all values of $\alpha$ and any $\theta\neq0$, 
both in the ordered and disordered phases.

\paragraph{Signatures of the QPTs}
As seen above and in section~\ref{sec:IsingModel}, for any $\alpha>0$ the ground state undergoes two second-order QPTs.
Figures~\ref{fig:IsingAlphaOpt}\Panel{c} and \ref{fig:IsingPhaseDiagram20} illustrate how the first derivative of the optimal QFI 
with respect to the control parameter $\theta$ can be regarded as a response function sensitive to criticality.
%%%%%%%%
\\[-6pt]

\noindent \tripieno{MySky} \ We computed the derivative of the QFI for fixed $N$ and several values of the parameter $\alpha$: 
it exhibits strongly localized peaks at $\theta_{\rm max} = {\rm arg\,max_\theta}\,\partial_\theta F_Q(\theta)$
and $\theta_{\rm min} = {\rm arg\,min_\theta}\,\partial_\theta F_Q(\theta)$; these peaks turn into divergences in the thermodynamic limit.
We fit $\theta_{\rm max}$ and $\theta_{\rm min}$ as functions of $N$ in the interval $N=4\div20$
and extrapolate the asymptotic values for larger $N$.
The experience gained in section~\ref{sec:IsingNN} prompts us to identify 
the approximated asymptotic positions of these peaks as the actual critical points: 
$\thetaFM\approx\lim_{N\to\infty}\theta_{\rm min}\,$ and $\,\thetaAFM\approx\lim_{N\to\infty}\theta_{\rm max}$. 
A cross comparison to the singular behaviour of the fidelity susceptibility and the energy gap confirms this guess, 
within a tolerance imposed by the numerical extrapolation.
%%%%%%%
\\[9pt]
%%%%%%%
\trivuoto{MyBlue} \ We evaluated the fidelity susceptibility $\chi_\theta$ for small chains and repeated the same procedure 
to find the asymptotic location of its maxima: the relative deviation $\epsilon$ from the aforementioned results is
$\epsilon<4\,\%$ on the AFM side of the phase diagram and $\epsilon<10\,\%$ on the FM side.
%%%%%%%
\\[9pt]
%%%%%%%
\trivuoto{MyBlue} \ We also evaluated the energy separation $\DeltaE$ between the ground state and the first excited state
for medium-sized chains and looked for the maximum of the second derivative $\big|\partial^2_\theta\,\DeltaE(\theta)\big|$ 
(indicating the point where the energy gap closes and the gapped disordered phase makes way for the quasi-gapless ordered phases):
the relative deviation from the locations of critical points extracted from the QFI and the closing gap is
$\epsilon<4\,\%$ for AFM coupling and $\epsilon<3\,\%$ for the FM one.
%%%%%%%%
\\[-6pt]

\noindent This approach allowed us to approximately find the dependence of the critical points $\thetac^{\,\pm}$ on the parameter $\alpha$. 
%%%%%%%%
\\[-6pt]

\noindent \tripieno{MySky} \ The symmetry under change of interaction sign $\theta\mapsto-\theta$, that characterizes the nearest-neighbour chain 
$\alpha\to\infty$ (see paragraph~\ref{subsec:IsingMEinGS} and Fig.~\ref{fig:IsingQFIT0}), is lost for finite $\alpha$.
In particular, the transition values from the PM phase to the FM or AFM order turn out to be larger than for nearest-neighbour interaction: 
$\thetaFM(\alpha)>-\frac{\pi}{4}$ and $\thetaAFM(\alpha)>+\frac{\pi}{4}$.
%%%%%%%
\\[9pt]
%%%%%%%
\trivuoto{MyBlue} \ For AFM coupling ($\theta>0$), the physical explanation should be sought 
in the interplay between long-range ordering and frustration. 
The smaller the decay power $\alpha$ is, the slower the two-spin interaction decays and the larger its strength remains on longer distances.
Since each spin tends to force antiparallel allignment onto not only neighbouring spins but also farther spins, 
the system becomes more and more frustrated due to the long-range exchange interaction: 
frustration entails a preference for the system to endure in the lower-energy quasi--$x$-polarized state even for lower magnetic fields. Consequently, the transition point shifts towards larger values of $\theta$ for smaller and smaller $\alpha$.
In particular, at $\alpha\to0$, the system becomes completely frustrated: 
the AFM order has a vanishing extension and it coincides with the critical point $\thetaAFM(\alpha\to1)=\frac{\pi}{2}$.
%%%%%%%
\\[9pt]
%%%%%%%
\trivuoto{MyBlue} \ For FM coupling ($\theta<0$), the strengthening of the two-spin interaction on longer distances for smaller $\alpha$ 
can only help in establishing the FM order: no kind of frustration is present. 
Thus $\thetaFM(\alpha)$ increases for decreasing $\alpha$. 
%%%%%%%%
\\[-6pt]

\noindent The collection of all the data obtained for $N=20$ 
results in the \emph{entanglement phase diagram} on the plane $\theta$--$\alpha$ of Fig.~\ref{fig:IsingPhaseDiagram20}.
It is limited in the interval $0\leq\alpha\leq3$, from infinite-range interactions to dipolar interactions. 
The coloured background represents the values of the QFI (panel a) and its first derivative with respect to $\theta$ (panel b): 
the maxima or minima of the QFI signal the close presence of QPTs.
We superimpose the lines of critical points as detected by the fidelity susceptibility and the energy gap: 
all these complementary methods give consistent results up to finite-size effects.

This diagram, obtained from considerations on the entanglement content and distinguishability of the ground state, 
should be compared to the phase diagram dictated by the order parameter and presented in Fig.~\ref{fig:IsingOrderParameter}.
There is an excellent qualitative agreement between the two diagrams; discrepancies are only due to finite-size effects: 
% the former is realized for $N=20$ spins, the latter is asymptotic in the thermodynamic limit.

To our best knowledge, none of the standard references to the arbitrary-range Ising model~\cite{DuttaPRB2001,KoffelPRL2012,VodolaNJP2016}
provides a complete characterization of the ferromagnetic region, 
even if some work in the mean-field approximation and by one-loop renormalization group has been carried out~\cite{DuttaPRB2001}.
We believe that our preliminary discussion, condensed in Figs.~\ref{fig:IsingOrderParameter} and \ref{fig:IsingPhaseDiagram20},
can be a good spark for further research~\cite{Gabbrielli2018b}.

\paragraph{Entanglement entropy}
In the seminal work of Ref.~\cite{KoffelPRL2012}, the critical points has been identified probing the maxima of the half-chain 
von Neumann entropy $\pazocal{S}_{N/2}=-\tr\big(\hat{\rho}_{N/2}\log\hat{\rho}_{N/2}\big)$, 
measuring the bipartite entanglement between two segments of $\frac{N}{2}$ contiguous spins.
Moreover, the von Neumann entropy was found to scale logarithmically with the chain length $N$ within the PM phase for $\alpha\lesssim1$;
in other words, the long-range character of interactions induces a violation of the area law even in the gapped phase.

This observation motivates the introduction of an effective ``central charge'' $c$ by way of the relation 
$\pazocal{S}_{N/2}\sim\frac{c}{6}\log{N}$, in analogy to critical gapless systems~\cite{CalabreseJSM2004}, 
for which the constant $c$ determines the universality class of the QPT. 
The central charge $c$ has also been used\footnote{\ Actually Ref.~\cite{VodolaNJP2016} exploits a different bipartition of the system: there, the von Neumann entropy $\pazocal{S}_{\ell}=-\tr\big(\hat{\rho}_{\ell}\log\hat{\rho}_{\ell}\big)$ measures the entanglement between two subsets of $\ell$ and $N-\ell$ contiguous spins; the violation of the area law is monitored by enlarging the size $\ell$ of one subsystem at the expense of the other, keeping $N$ fixed.\\[-9pt]} as a tool for sensing the phase diagram of the transverse Ising chain~\cite{VodolaNJP2016}: it is finite only in the PM long-range ($\alpha\lesssim1$) phase. 
Moreover the values of $\alpha$ and $\theta$ for which $c$ does not depend on $N$, 
leading to a pure logarithmic correction to the area law, correspond to the critical points. 

The phase diagram dictated by the multipartite entanglement in Fig.~\ref{fig:IsingPhaseDiagram20} 
agrees with the phase diagrams extracted from the bipartite entanglement and presented in literature~\cite{KoffelPRL2012,VodolaNJP2016}.
We underline that these two methods -- remarkably precise for a wide interval of values of $\alpha\gtrsim0.3$ -- 
% break down / fail to reproduce / are inadequate    % $\alpha\,{\small\gtrsim}\,0.3$
could not give exhaustive results for $\alpha\lesssim0.3$.
On the contrary, the singularities in the fidelity susceptibility and QFI allow for an adequate determination of the critical points 
even at very small values of $\alpha$.

\subsection{Detection of the interaction range} \label{subsec:IsingDetection}
Corrections to the area law for the entanglement entropy in the paramagnetic phase of the Ising chain~\cite{KoffelPRL2012}, 
as long as the hybrid decays of correlation functions~\cite{VodolaNJP2016}, 
can be helpful diagnostics for the long-range character of the interaction.
We wish to outline here how the QFI associated to the ground state can also be regarded as a convenient detector for the change
of nature of the paramagnetic phase occurring at $\alpha=1$.

\begin{figure}[p!]
\centering
\framebox{$\alpha=10$} \\
\includegraphics[width=0.38\textwidth]{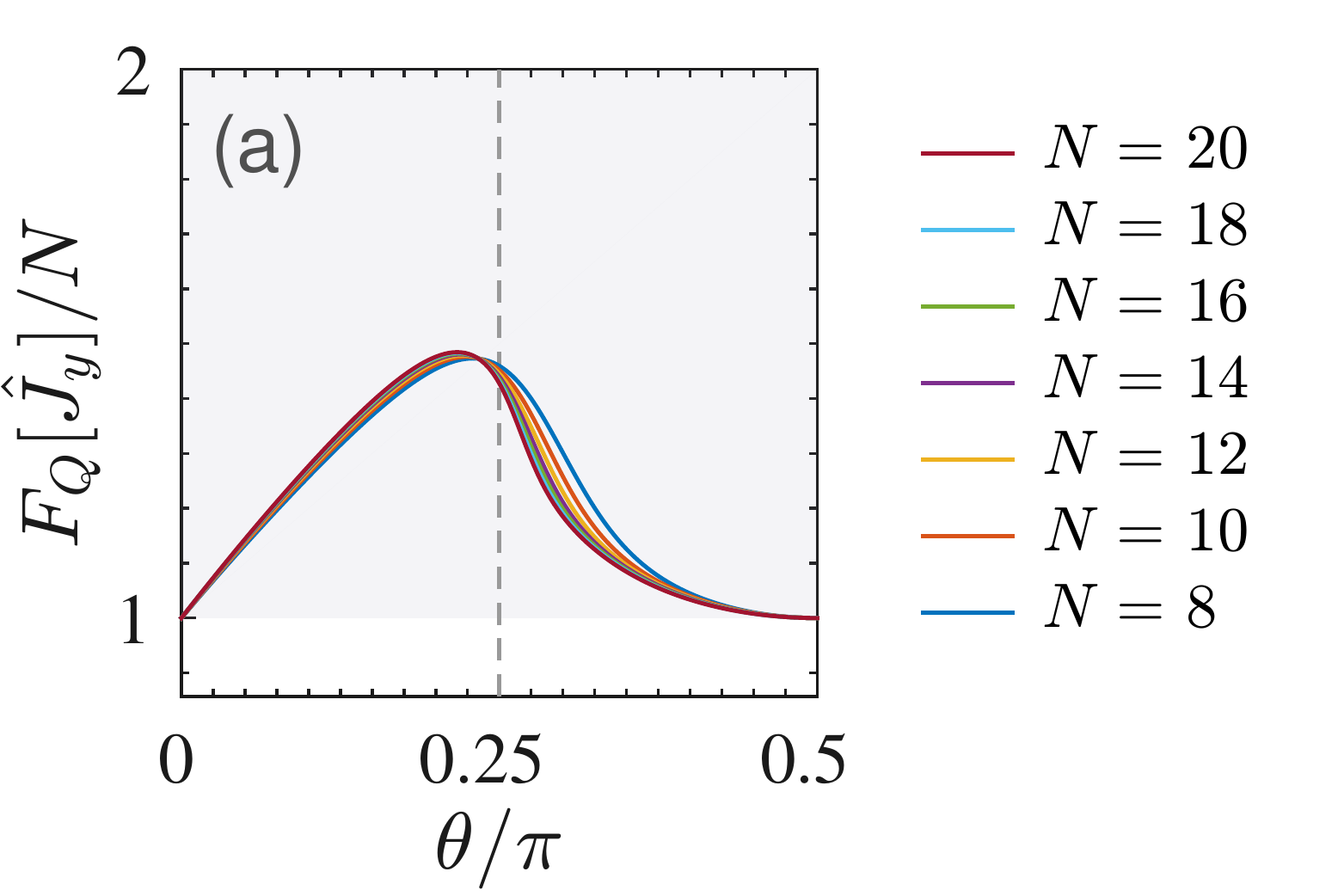} \hspace{-9pt}
\includegraphics[width=0.305\textwidth]{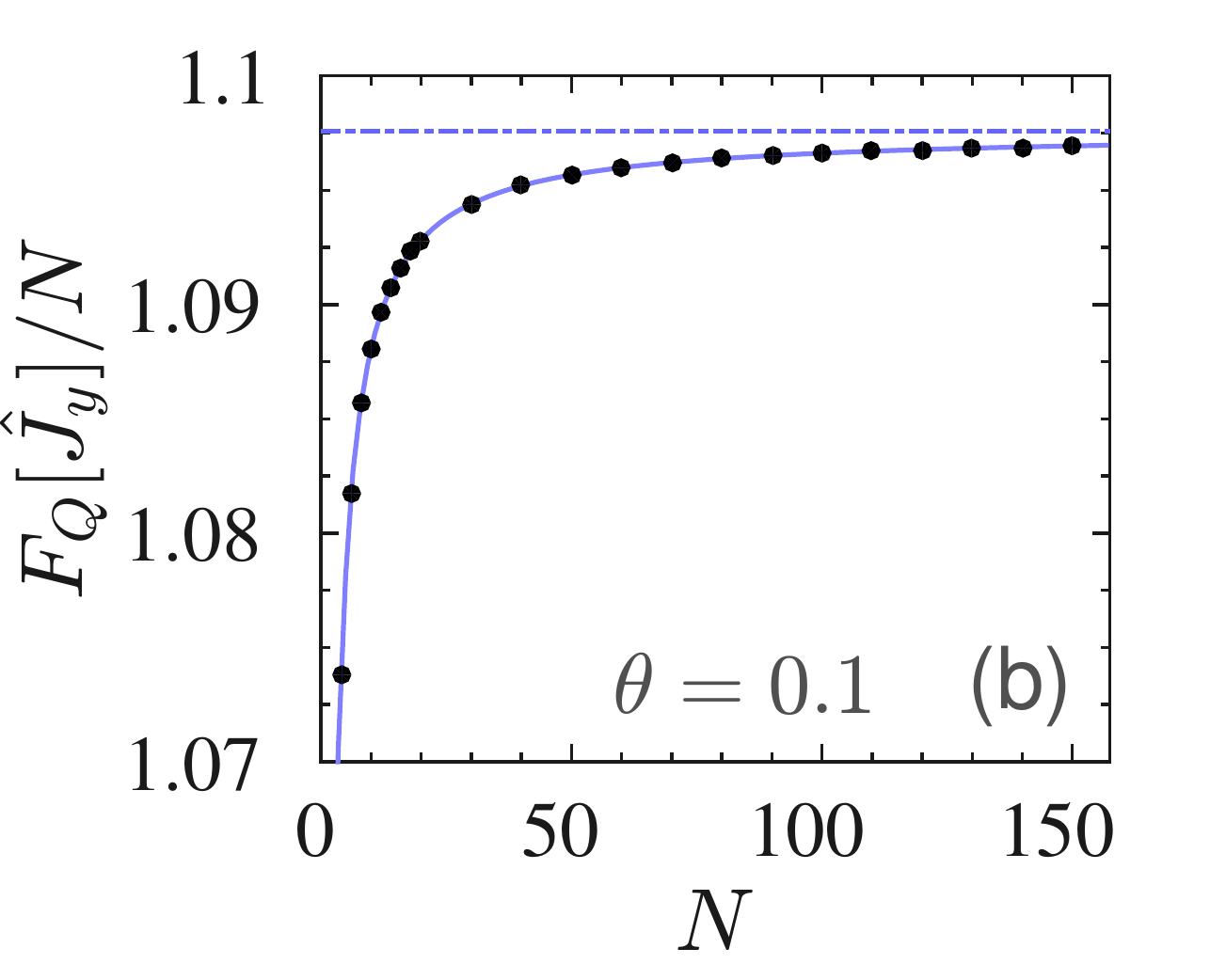} \hspace{-12pt}
\includegraphics[width=0.305\textwidth]{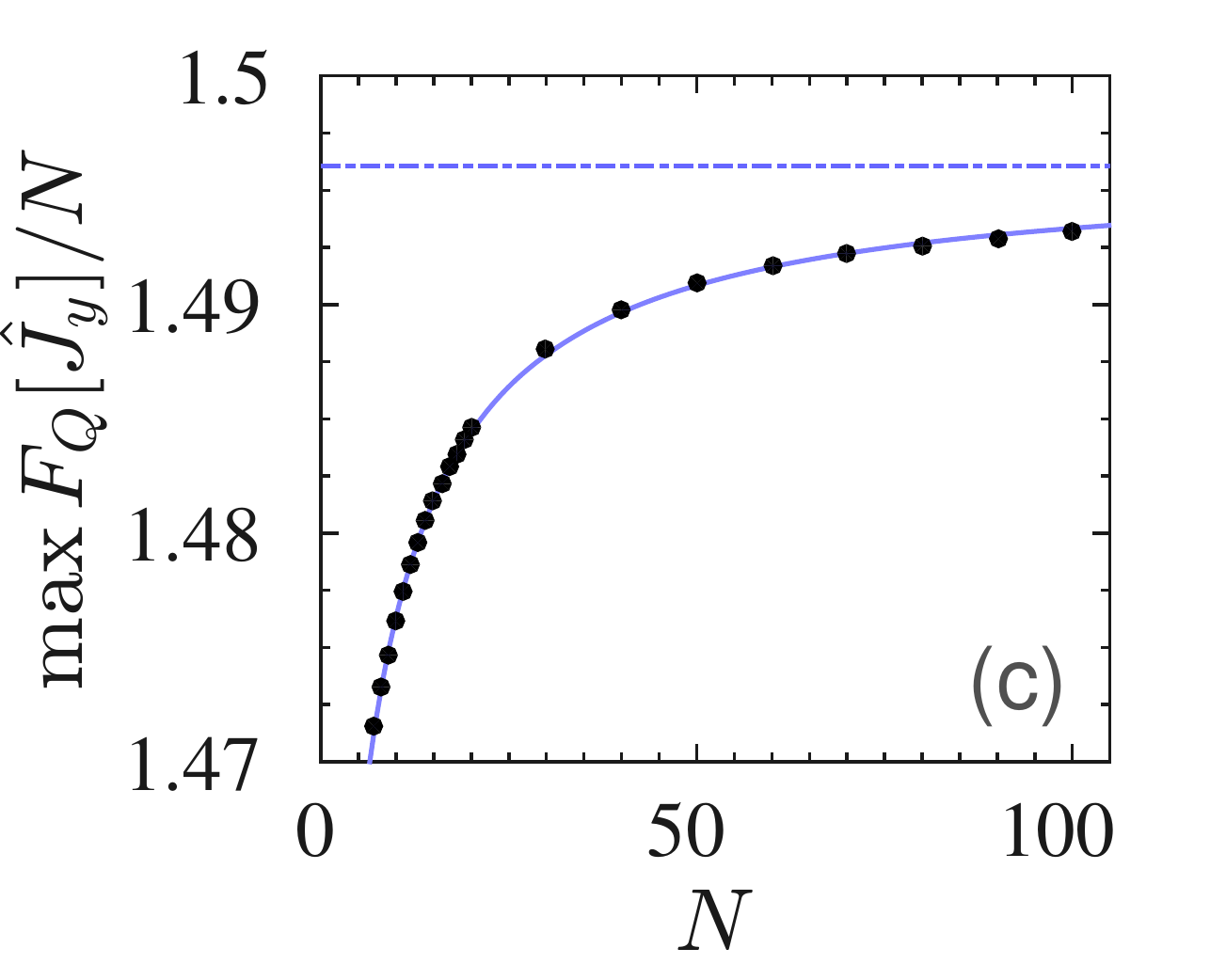} \\[3pt]
\framebox{$\alpha=1$} \\
\includegraphics[width=0.38\textwidth]{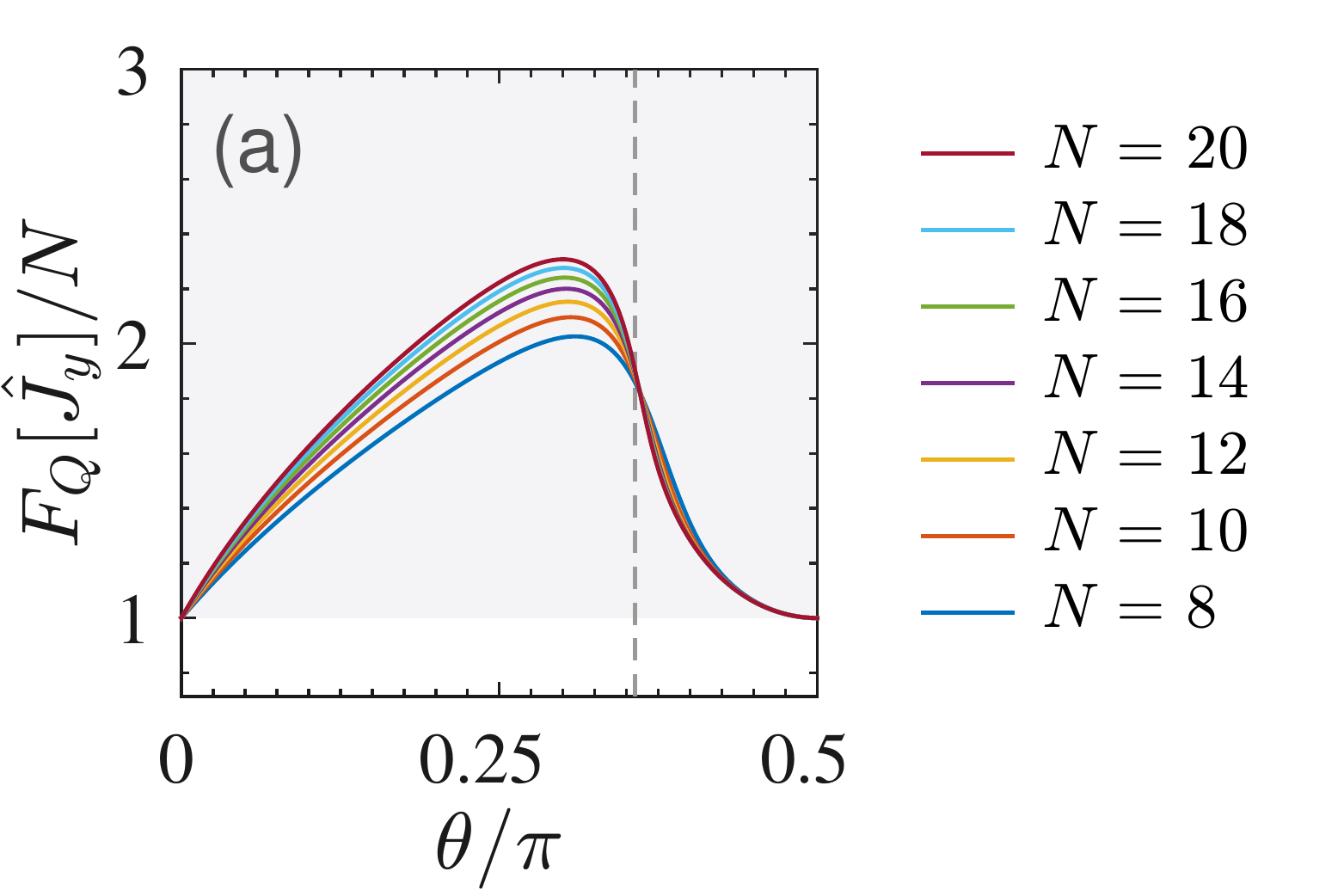} \hspace{-6pt}
\includegraphics[width=0.305\textwidth]{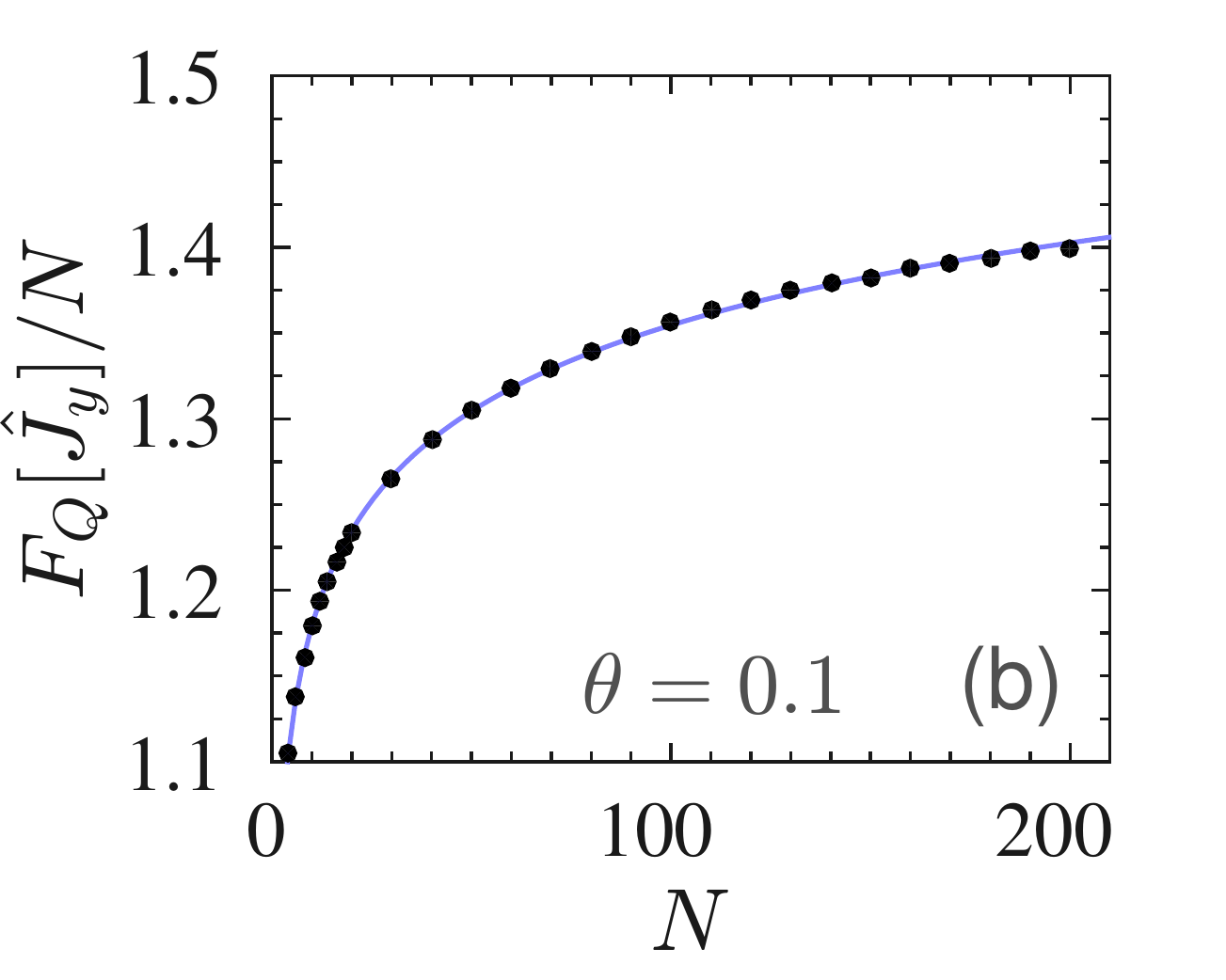} \hspace{-12pt}
\includegraphics[width=0.305\textwidth]{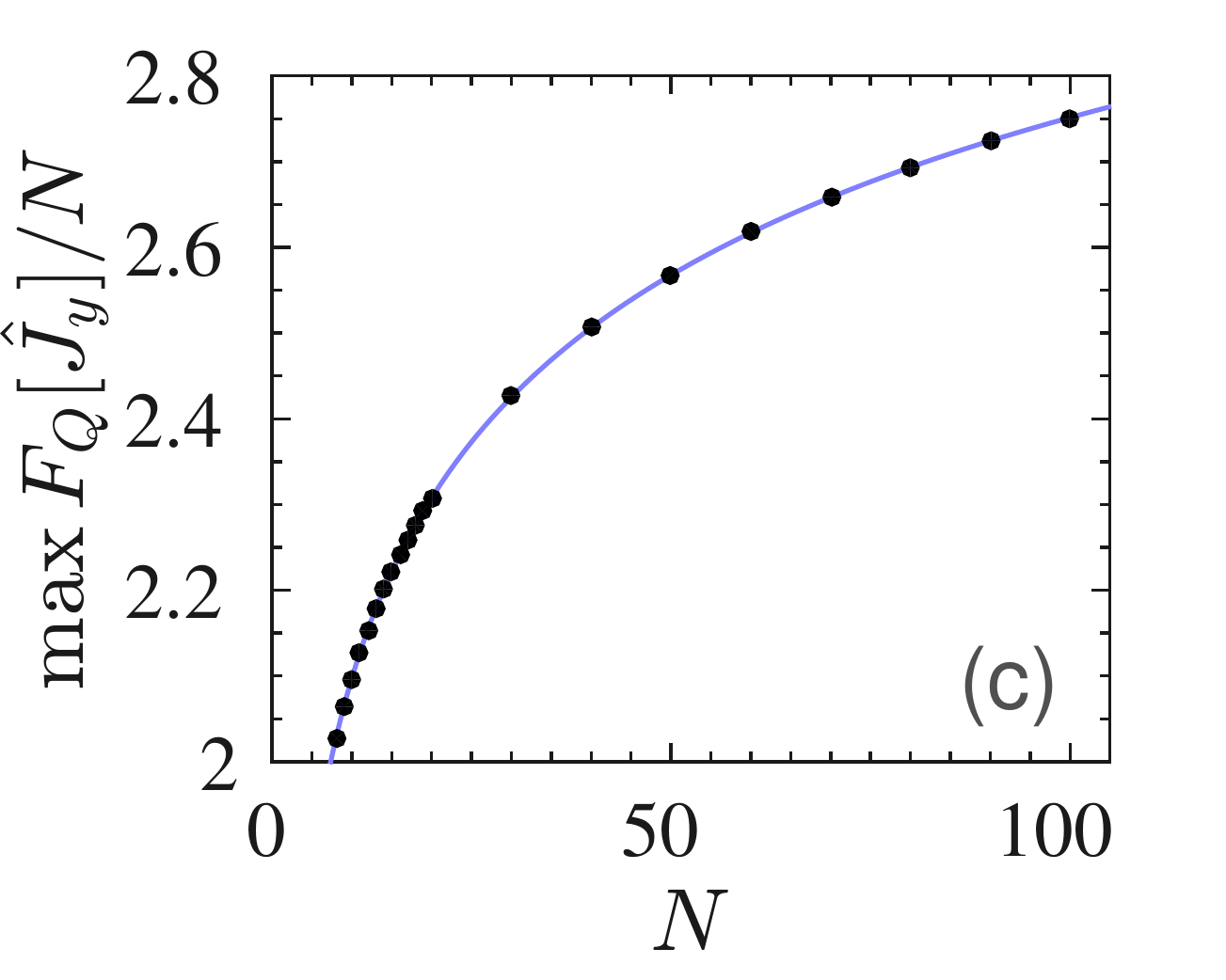} \\[3pt]
\framebox{$\alpha=0.1$} \\
\includegraphics[width=0.38\textwidth]{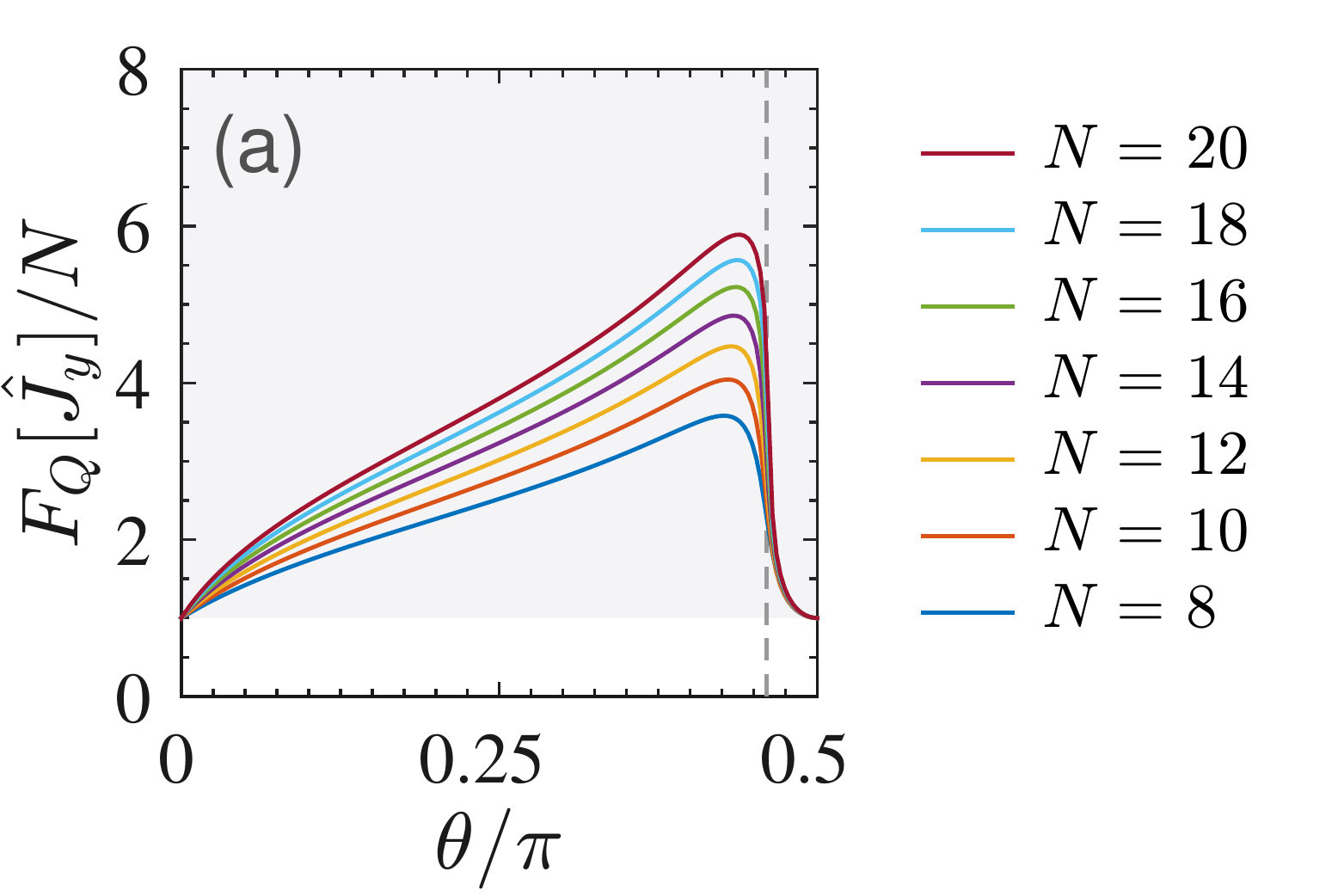} \hspace{-3pt}
\includegraphics[width=0.305\textwidth]{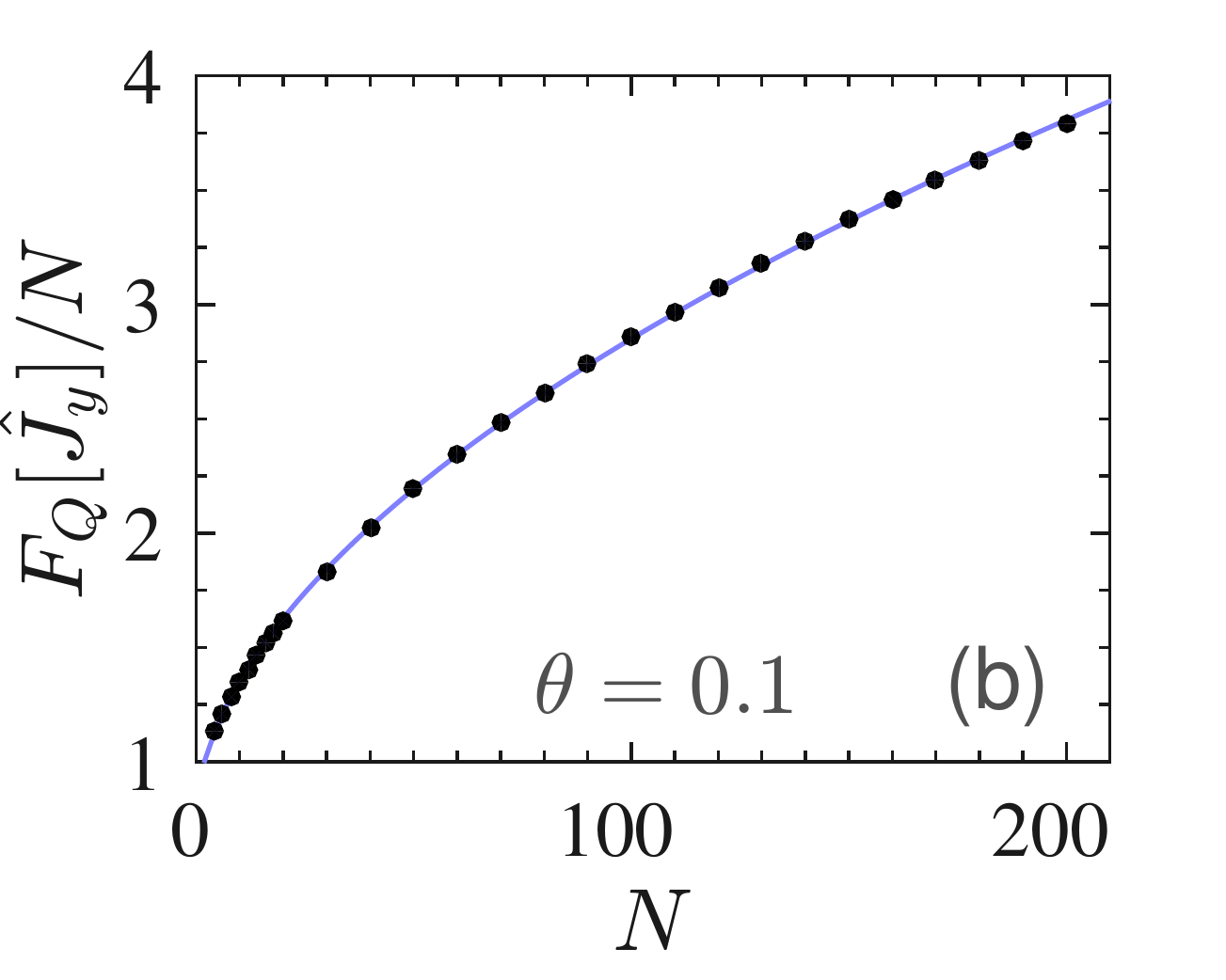} \hspace{-12pt}
\includegraphics[width=0.305\textwidth]{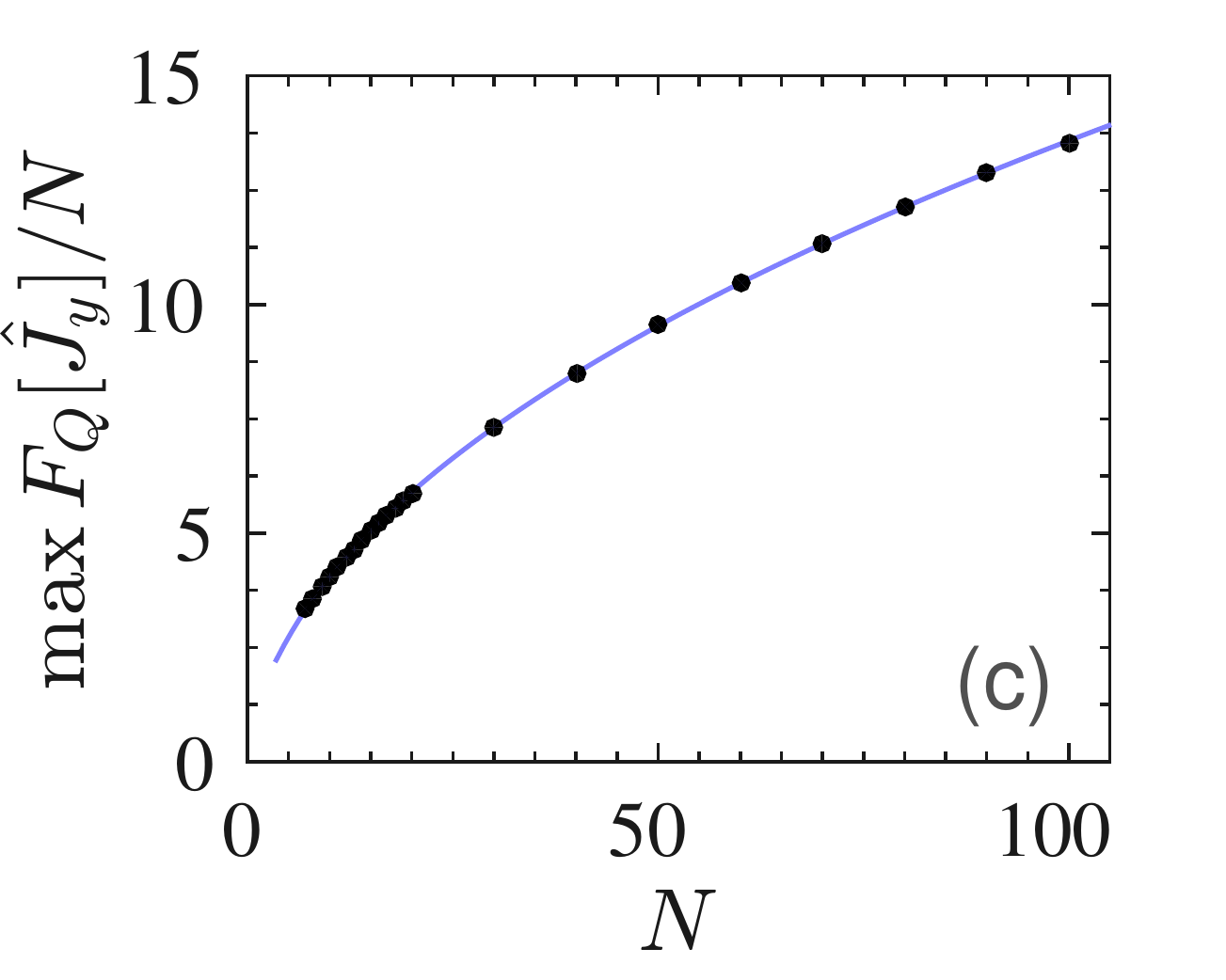} \\
\caption{Behaviour of the Fisher density $f_Q[\ket{\psi_0},\hat{J}_y]$ evaluated using a transverse magnetization 
for the ground state of Hamiltonian~(\ref{IsingHamNN}) without the addition of symmetry-breaking terms for ${\alpha=10,}\,1,\,0.1$.
\textbf{(a)} Fisher density as a function of the control parameter $\theta>0$. 
Different colours refer to different chain lengths~$N$. 
Shaded area signals the region where entanglement is witnessed $f_Q[\ket{\psi_0}]>1$.
Dashed gray lines indicate the location of inflection points extrapolated for $N\gg1$, coinciding with the critical points. 
\textbf{(b)} Finite-size scaling of the Fisher density in the paramagnetic phase $\theta=0.1$. 
Dots are numerical data, solid lines are fit curves: a rational bounded function for $\alpha=10$ (the dashed line is the bound),
a logarithm for $\alpha=1$ and a power law for $\alpha=0.1$, 
as reported in Tab.~\ref{tab:IsingAlphaDetection}.
\textbf{(c)} Finite-size scaling of the maximum of the Fisher density in the interval $0<\theta<\frac{\pi}{2}$.
Dots are numerical data, solid lines are fit curves, of the same form as in panel (b).}
\label{fig:IsingAlphaDetection}
\end{figure}

We already detailed about the QFI calculated via the pertinent order parameter,
its utility for recognizing the ordered phases and its extensivity in the PM phase on both FM and AFM sides of the phase diagram.
Suppose now to employ the three local observables $\hat{J}_\varrho$'s 
to calculate the QFI for the Ising chain with AFM interaction ($0<\theta<\frac{\pi}{2}$), 
whose proper order parameter is instead the staggered magnetization $\hat{J}_z^{\rm (st)}$.
On these terms, the transverse magnetization $\hat{J}_y$ is found to maximize the witness:
$F_Q\big[\ket{\psi_{0}},\hat{J}_y\big] = \max_\varrho F_Q\big[\ket{\psi_{0}},\hat{J}_\varrho\big]$. 
Figures~\ref{fig:IsingAlphaDetection} show some examples for the representative cases of short range ($\alpha=10$),
Coulombic decay ($\alpha=1$) and long range ($\alpha=0.1$).
%%%%%%%%
\\[-6pt]

\noindent \tripieno{MySky} \ Multipartite entanglement is still witnessed $f_Q>1$ 
for every interaction strength $\theta\neq0$, even if not optimally in the AFM ordered phase: 
$f_Q\big[\ket{\psi_{0}},\hat{J}_y\big]<f_Q\big[\ket{\psi_{0}},\hat{J}^{\rm (st)}_z\big]$ at $\thetaAFM<\theta<\frac{\pi}{2}$. 
On the other hand, in the PM phase the transverse magnetization witnesses a larger depth of entanglement than the order parameter: 
$f_Q\big[\ket{\psi_{0}},\hat{J}_y\big]>f_Q\big[\ket{\psi_{0}},\hat{J}^{\rm (st)}_z\big]$ at $0<\theta<\thetaAFM$.
%%%%%%%%
\\[-6pt]

\noindent \tripieno{MySky} \ The QFI is not a monotonically increasing function of $\theta$: 
there exist a maximum $\max_\theta f_Q$ at $\theta_{\rm max}<\thetaAFM$. 
The concavity of $f_Q\big[\ket{\psi_{0}},\hat{J}_y\big]$ is opposite to the one exhibited by the counterpart 
$f_Q\big[\ket{\psi_{0}},\hat{J}^{\rm (st)}_z\big]$ (see Fig.~\ref{fig:IsingAlphaOpt}): 
it is downward for $\theta<\thetaAFM$ and upward for $\theta>\thetaAFM$.
Once more, the inflection point with vertical tangent reveals the position of the critical point 
for any $\alpha$ in the thermodynamic limit. Panels~\ref{fig:IsingAlphaDetection}\Panel{a} clarify this behaviour.
%%%%%%%%
\\[-6pt]

\noindent \tripieno{MySky} \ The most surprising result is the unexpected relation between the range of interactions 
and the scaling of witnessed multipartite entanglement in the disordered phase: 
while in the AFM phase the Fisher density is invariably upper bounded by a constant independent of $N$, 
in the PM phase the long range induces a nontrivial dependence on the chain size, 
as depicted in panels~\ref{fig:IsingAlphaDetection}\Panel{b,\,c}.
%%%%%%%
\\[9pt]
%%%%%%%
\trivuoto{MyBlue} \ For $\alpha>1$ and $\theta<\thetaAFM$, when increasing $N$ we distinctly find the saturation to a constant value 
$f_Q\sim\pazocal{O}(1)\,$ as well as $\,\max_\theta f_Q \sim \pazocal{O}(1)$.
%%%%%%%
\\[9pt]
%%%%%%%
\trivuoto{MyBlue} \ For $\alpha=1$ and $\theta<\thetaAFM$, we encounter a logarithmic scaling $f_Q\sim\log{N}\,$ 
and $\,\max_\theta f_Q \sim \log{N}$.
%%%%%%%
\\[9pt]
%%%%%%%
\trivuoto{MyBlue} \ For $\alpha<1$ and $\theta<\thetaAFM$, we infer a power law $f_Q \sim N^b$, 
with $b(\theta,\alpha)\simeq g(\theta)(1-\alpha)$; an exact determination of $g(\theta)$ was not possible from the available data: 
the only constraint we can pose, consistently with the perturbative treatment carried out hereinafter, is $g(\theta\to0)=1$. 
Also $\max_\theta f_Q$ is compatible with a similar algebraic scaling, even though with a slightly different exponent $b$.
%%%%%%%%
\\[-6pt]

%%%%%%%%%%%%%%%%%%%%%%%%%%%%%%%%%%%%%%%%%%%%%%%%%%%%%%%%%
%%%%%%%%%%%%%%%%%%%%%%%%%%%%%%%%%%%%%%%%%%%%%%%%%%%%%%%%
%%%%%%%%%%%%%%%%%%%%%%%%%%%%%%%%%%%%%%%%%%%%%%%%%%%%%%%%
%%%%%%%%%%%%%%%%%%%%%%%%%%%%%%%%%%%%%%%%%%%%%%%%%%%%%%%%%

\noindent We underline that the correct laws are difficult to unveil unambiguously, because the system sizes are too small to observe 
a well-defined asymptotic scaling. Although we performed an analysis up to $N\approx150$, our data are not conclusive: 
they only suggest a logarithmic growth for $\alpha=1$ and a superlogarithmic growth for lower $\alpha$; 
it is not clear if the latter concurs with a pure power law or a logarithmic correction is required.
Moreover, it remains an open question whether the QFI eventually saturates for larger system sizes.
All these issues, along with the accurate determination of the powers, would deserve further investigation.

%%%%%
\newpage
%%%%%

Details aside, \emph{the emergence of a growth of multipartiteness for larger and larger chains is a remarkable signature of long range}. 
The overall conjectured scenario is summarized in Tab.~\ref{tab:IsingAlphaDetection}.
Notice that the above considerations are \emph{not} transferable in the PM phase of the transverse Ising chain with FM interaction ($\thetaFM<\theta<0$) 
by simply swapping the role of observables $\hat{J}_y$ $\leftrightarrow$ $\hat{J}^{\rm (st)}_y$ 
and $\hat{J}_z$ $\leftrightarrow$ $\hat{J}^{\rm (st)}_z$~\cite{Gabbrielli2018b}.

\paragraph{Perturbative approach}
We are in a position to provide an intuitive justification to the assortment of scalings 
advanced above and listed in the first column of Tab.~\ref{tab:IsingAlphaDetection}.
Restricting to low external field, we apply perturbation theory for a small positive coupling parameter $\theta\approx0^{+}$ 
and look for the asymptotic behaviour of entanglement depth in the thermodynamic limit.
For a chain of weakly interacting spins, the Hamiltonian~(\ref{IsingHam}) takes the form
\be
\frac{\hat{H}_{\alpha}}{J} \ \simeq \ 
% J\sin\theta\,\displaystyle \sum_{i=1}^N \sum_{j>i}^N \frac{\,\hat{\sigma}_z^{(i)} \hat{\sigma}_z^{(j)}}{\,|i-j|^\alpha} 
% \, + \, J\cos\theta\,\displaystyle\sum_{i=1}^N \hat{\sigma}_x^{(i)} \simeq 
\displaystyle\sum_{i=1}^N \hat{\sigma}_x^{(i)} \, + \, 
\theta\,\displaystyle \sum_{i=1}^N \sum_{j>i}^N \frac{\,\hat{\sigma}_z^{(i)} \hat{\sigma}_z^{(j)}}{\,|i-j|^\alpha} \ \equiv \ 
\hat{H}^{(0)} \, + \, \theta \, \hat{H}_\alpha^{(1)} \, .
\ee
The eigenstates of the unperturbed chain are 
$\ket{\psi_n}^{(0)}=\binom{N}{n}^{-1/2}\sum\spinleft^{\otimes N-n}\spinright^{\otimes n}$,
% $\ket{\psi_n(\theta)}^{(0)}\equiv\ket{\psi_n(0)}=\binom{N}{n}^{-1/2}\sum\spinleft^{\otimes N-n}\spinright^{\otimes n}$
where the sum extends over all $\binom{N}{n}$ permutations of $N$ spins among which $n$ are aligned along the field and 
$N-n$ are aligned along the opposite direction; in particular, the ground state is 
the product state $\ket{\psi_0}^{(0)}=|\!\leftarrow\leftarrow\cdots\leftarrow\rangle$,
while the $n$th excited state contains $n$ excitations in the form of local spin flips.
The energy levels of the unperturbed chain, in units of~$J$, are $E_n^{(0)}=-(N-2n)$.

%%%%%%%%%%%%%%%%%%%%%%%%%%%%%%%%%%%%%%%%%%%%%%%%%%%%%%%%%%
%%%%%%%%%%%%%%%%%%%%%%%%%%%%%%%%%%%%%%%%%%%%%%%%%%%%%%%%%
\begin{table}[!b] 
\begin{center}
\begin{tabular}{ccccccccccccc}
\toprule 
 &&& \hspace{-45pt} \textsf{PM} $ \ 0<\theta<\thetaAFM$ \hspace{-45pt} &&&& \hspace{-45pt} \textsf{criticality} $ \ \theta=\thetaAFM$ \hspace{-45pt} &&&& \hspace{-45pt} \textsf{AFM} $\ \thetaAFM<\theta\leq\frac{\pi}{2}$ \hspace{-45pt} & \\[3pt]
\cmidrule{3-13}
\parbox[b][15pt][b]{\widthof{\textsf{range}}}{\textsf{range}} && $\quad \hat{J}_y \quad$ \hspace{-0pt} && \hspace{-10pt} $\quad \hat{J}^{\rm (st)}_z \quad$ && $\quad \hat{J}_y \quad$ \hspace{-0pt} && \hspace{-10pt} $\quad \hat{J}^{\rm (st)}_z \quad$ && $\quad \hat{J}_y \quad$ \hspace{-0pt} && \hspace{-10pt} $\quad \hat{J}^{\rm (st)}_z \quad$ \\[3pt]
\hline\hline % \midrule
$\alpha>1$ && \parbox[c][30pt][c]{\widthof{$\pazocal{O}(1)$}}{$\pazocal{O}(1)$} && \hspace{-10pt} \parbox[c][30pt][c]{\widthof{$\pazocal{O}(1)$}}{$\pazocal{O}(1)$} && \parbox[c][30pt][c]{\widthof{$\pazocal{O}(1)$}}{$\pazocal{O}(1)$} && \hspace{-10pt} \parbox[c][30pt][c]{\widthof{$\sim N^a$}}{$\sim N^a$} && \parbox[c][30pt][c]{\widthof{$\pazocal{O}(1)$}}{$\pazocal{O}(1)$} && \hspace{-10pt} \parbox[c][30pt][c]{\widthof{$\sim N$}}{$\sim N$} \\[3pt] \midrule
$\alpha=1$ && \parbox[c][30pt][c]{\widthof{$\sim\log{N}$}}{$\sim\log{N}$} && \hspace{-10pt} \parbox[c][30pt][c]{\widthof{$\pazocal{O}(1)$}}{$\pazocal{O}(1)$} && \parbox[c][30pt][c]{\widthof{\textsl{?}}}{\textsl{?}} && \hspace{-10pt} \parbox[c][30pt][c]{\widthof{$\sim N^\upsilon$}}{$\sim N^a$} && \parbox[c][30pt][c]{\widthof{$\pazocal{O}(1)$}}{$\pazocal{O}(1)$} && \hspace{-10pt} \parbox[c][30pt][c]{\widthof{$\sim N$}}{$\sim N$} \\[3pt] \midrule
$0\leq\alpha<1$ && \parbox[c][30pt][c]{\widthof{$\sim N^b$}}{$\sim N^b$} && \hspace{-10pt} \parbox[c][30pt][c]{\widthof{$\pazocal{O}(1)$}}{$\pazocal{O}(1)$} && \parbox[c][30pt][c]{\widthof{\textsl{?}}}{\textsl{?}} && \hspace{-10pt} \parbox[c][30pt][c]{\widthof{$\sim N^a$}}{$\sim N^a$} && \parbox[c][30pt][c]{\widthof{$\pazocal{O}(1)$}}{$\pazocal{O}(1)$} && \hspace{-10pt} \parbox[c][30pt][c]{\widthof{$\sim N$}}{$\sim N$} \\[3pt]
\bottomrule
\end{tabular} \\[9pt]
\caption{Scenarios for the leading scaling of the Fisher density $f_Q[\ket{\psi_{0}},\hat{O}]$ for large Ising chains 
with antiferromagnetic coupling ($\theta>0$), calculated using either the standard SU(2) operators $\hat{O}=\hat{J}_\varrho$ 
or the staggered operators $\hat{O}=\hat{J}_\varrho^{\rm (st)}$ and optimized over the three spatial directions $\varrho\in\{x,y,z\}$. 
Whereas the order parameter $\hat{J}_z^{\rm (st)}$ allows for a neat discrimination of the ordered phase from the disordered one,
the observable $\hat{J}_y$ permits to claim if a certain paramagnetic ground state 
pertains to a short-range model or a long-range one. The power $b(\theta,\alpha)\propto1-\alpha$ shows a merely weak dependence on $\theta$. 
The power $a(\alpha)\leq1$ satisfies $a\to\frac{3}{4}$ for $\alpha\to\infty$ and $a\to1$ for $\alpha\to0$; 
for an in-depth numerical scrutiny of the behaviour of $a$ and how it can be related to the breakdown of conformal invariance, see Ref.~\cite{Gabbrielli2018b}. 
The question marks indicate the impossibility to extract a scaling from the available data.} \label{tab:IsingAlphaDetection}
\end{center}
\end{table}
%%%%%%%%%%%%%%%%%%%%%%%%%%%%%%%%%%%%%%%%%%%%%%%%%%%%%%%%%
%%%%%%%%%%%%%%%%%%%%%%%%%%%%%%%%%%%%%%%%%%%%%%%%%%%%%%%%%

%%%%%
%\newpage
%%%%%

We find that the (not normalized) first-order perturbed ground state in the paramagnetic phase reads
\be \label{IsingPerturbationState}
\ket{\psi_{0}(\theta>0)}^{(1)} \simeq \ket{\psi_0}^{(0)} - \theta\,\mathcal{G}_N(\alpha)\,\ket{\psi_2}^{(0)} \, , 
\qquad {\rm with} \quad \mathcal{G}_N(\alpha) = \frac{N\,{\rm H}_{N,\,\alpha}-{\rm H}_{N,\,\alpha-1}}{\sqrt{8N(N-1)}} \,,
\ee
where ${\rm H}_{N,\,\alpha}$ is the $N$th generalized harmonic number of order $\alpha$.
In the limit $N\to\infty$, the calculation of $\mathcal{G}_N(\alpha)$ involves handling hyperharmonic series, 
whose convergence is only attained for $\alpha>1$; nevertheless, for $\alpha\leq1$ we can find out the asymptotic behaviour for large $N$: 
\be \label{IsingPerturbationFunction}
\mathcal{G}_N(\alpha) \stackrel{\small N\,\gg\,1}{^{\textcolor{white}{o}}\approx^{\textcolor{white}{o}}} \frac{1}{\sqrt{8}}\cdot
\left\{
\begin{array}{ll} 
\zeta(\alpha) & {\rm for} \ \alpha>1 \\ 
\log N & {\rm for} \ \alpha=1 \\ 
% \frac{N^{1-\alpha}}{(1-\alpha)(2-\alpha)}
\frac{1}{(1-\alpha)(2-\alpha)}\,N^{1-\alpha}   & {\rm for} \ 0\leq\alpha<1
\end{array}
\right.
\ee
The symbol $\zeta$ indicates the Riemann zeta function: $\zeta(\infty)=1$ and $\zeta(1^+)\rightarrow+\infty$. 
The viability condition of perturbation theory $\,\theta\,\mathcal{G}_N(\alpha)\ll1\,$ sets an upper limit 
for the validity of Eq.~(\ref{IsingPerturbationFunction}): 
for fixed $\theta\ll1$, the finite-size scalings are only guaranteed when $N\ll\exp(\theta^{-1})$ for $\alpha=1$ 
and $N\ll\theta^{-1/(1-\alpha)}$ for $\alpha<1$. Notwithstanding, the above-mentioned results are a useful analytical evidence
for the numerical findings that inspired Tab.~\ref{tab:IsingAlphaDetection}.
% $N\ll\sqrt[\alpha-1]{\theta}$ for $\alpha<1$ 
In fact, as a consequence of Eqs.~(\ref{IsingPerturbationState}) and (\ref{IsingPerturbationFunction}), 
the QFI can be accessed and, at first order in $\theta$, is given by
\be
f_Q\big[\ket{\psi_{0}(\theta>0)}, \hat{J}_y\big] = 1 \,+\, \sqrt{8}\,\theta\,\mathcal{G}_N(\alpha) \,+\, \pazocal{O}(\theta^2)\,.
\ee
Accordingly, note that panel~\ref{fig:IsingAlphaDetection}\Panel{b} shows that, for $\alpha=10$ and $\theta=0.1$, 
the Fisher density saturates $f_Q=1+0.1\,\zeta(10)\approx1.1$ when $N\gg1$.

\section*{Summary}
Even more than fifty years after its first appearance, the Ising chain in a transverse field is still a widely investigated model. 
Chiefly, it exhibits continuous quantum phase transitions that can be captured by the Ginzburg-Landau formalism.
Nevertheless, for arbitrary-range interactions, an in-depth analysis of the amount of multipartite entanglement 
hosted by the ground and thermal states is still far to be exhaustive.

We have shown that the quantum Fisher information computed through the order parameter is a tool
to detect multipartite entanglement among the spins. 
In particular, the ground state as well as thermal states of the chain contains a potential resource 
for enhanced quantum metrology around the critical points.
Moreover, multipartite entanglement can adequately mimic the phase diagram of the model.
Ordered phases, characterized by a nonvanishing order parameter, are revealed by a superextensive scaling of the quantum Fisher information,
implying the size of the largest cluster of entangled spins becomes larger and larger when enlarging the chain length.
Disordered phases, on the contrary, do not show a diverging multipartiteness.
Furthermore, the occurrence of phase transitions is uncovered by the divergence of the derivative of the quantum Fisher information 
with respect to the control parameter. 
A comprehensive characterization of the phase diagram in term of multipartite entanglement, 
both on the ferromagnetic and antiferromagnetic side, was missing: our work tried to fill this lack.

Finally, we have proven that a suitable choice of observables lead to clearly distinguish the range of the interactions:
more in general, we believe that for an any spin system -- where the nature of the inner cooperative interaction among spins is unknown --
the quantum Fisher information can discriminate between short range and long range in terms of extensive or superextensive behaviour 
in the limit of a large system.
We can thus conclude that the quantum Fisher information, aside from witnessing quantum correlations useful for metrological applications, 
can be regarded as a reliable locator of critical points, for both short-range and long-range interactions, 
and a faithful detector for the range of microscopic interactions.

\chapter[Lipkin-Meshkov-Glick model]{A semiclassical example: \\ the Lipkin-Meshkov-Glick model} \label{ch:LMG}

{\sl % We investigate the behaviour of multipartite entanglement as quantified by the quantum Fisher information, 
We witness multipartite entanglement, both at zero and finite temperature, 
in the vicinity of critical points of the fully-symmetric anisotropic Lipkin-Meshkov-Glick model,
an exemplary infinite-range spin model hosting a second-order and first-order quantum phase transition. 
We demonstrate that multipartite entanglement detects the critical points displaying a discontinuous behaviour in the thermodynamic limit.
In the classical limit of a very large collective spin, exact analytical calculations are also possible.}

\section*{Introduction}
% Though the curtain went up on classical phase transitions in the historical theatre of condensed matter, 
% Though the spotlight was originally put on classical phase transitions in the historical theatre of condensed matter,
Though condensed-matter physics is the historical theatre where the spotlight was originally put on classical phase transitions,
also nuclear physics has been an important stage for their appearance and understanding, 
since nuclear matter is known to show sudden changes in quadrupole-deformed shapes or momenta of inertia 
driven by the temperature or rotational frequency~\cite{CejnarPPNP2009}.
Starting from the Sixties, quantum manifestations of phase transitions has been observed in isolated nuclei as well: 
for instance, microscopic signatures of an abrupt change of the nuclear ground state have been predicted and observed 
when crossing critical lines on the chart of nuclides \cite{ThoulessNP1961, DieperinkPRL1980}.
Such transitions are induced by a variation of the effective interaction strength among nucleons 
and can be studied theoretically within certain phenomenological collective models of nuclear structure. 

One of these test models is the so-called Lipkin-Meshkov-Glick model, 
firstly introduced in its simpler form in the seminal works of Ref.~\cite{LMG1965} 
and later generalized to comprise an enriched variety of phenomena~\cite{FengPRC1981}.
%This is a spin model in all respects: the spin representation permits to effectively describe the nucleus in a two-mode approximation, 
%where each nucleon can occupy only the lower or the upper single-particle energy levels.
Only afterwards it was adopted by the community working on condensed matter 
as an archetype of infinitely-coordinated integrable system~\cite{BotetPRB1983} and it is ubiquitous in several fields.
In all respects, this is a spin model involving nonlinear interactions: 
the spin representation permits to effectively describe a large variety of systems 
(from nucleons to trapped ions to bosons in a lattice) where particles can occupy only two single-particle energy levels.

We devote the present chapter to characterize the critical points hosted by the Lipkin-Meshkov-Glick model 
in the light of multipartite entanglement. 
We actually deal with a restriction of the model into a permutationally symmetric sector of the Hilbert space: 
this common choice is motivated by the experimental possibility to implement such a spin system using ultracold degenerate bosons.
The persisting appeal of this full-symmetric specie of the Lipkin-Meshkov-Glick model is ascribable to at least two reasons: 
on one side, the mapping from the ensemble of interacting two-level bosons to a continuous single-particle picture,
providing useful insights as long as the number of spins is large~\cite{Shchesnovich2008,Buonsante2012};
on the other side, the important role it plays in the arena of quantum-enhanced interferometry~\cite{FattoriPRL2008,TrenkwalderNATPHYS2016,MuesselPRA2015}.
Owing to its huge metrological significance, we also adopt the Wineland spin-squeezing parameter 
to enrich the treatment based on the quantum Fisher information 
for quantifying the maximum amount of metrologically useful entanglement in the system.

After introducing the model and its applications (section~\ref{sec:LMGModel}),
we show that the witnessed multipartite entanglement adequately detects many-body correlations arising at null temperature
in the vicinity of quantum phase transitions, both of the second order (section~\ref{sec:symmetricLMG}) and first order (section~\ref{sec:asymmetricLMG}).
We extend the investigation to finite temperature, in order to disclose the effect of thermal fluctuations on entanglement
and gain information about the robustness of atom interferometric schemes against thermal noise~\cite{FattoriPRL2008}.
In the ordered symmetric phase, we show that the ground state takes the form of a Schr\"odinger-cat--like state
and we find that the large amount of multipartite entanglement therein is extremely delicate to temperature 
and external symmetry-breaking perturbations. 
This result agrees with the experimental difficulty in preparing such highly-entangled states 
and protecting them for long times~\cite{TrenkwalderNATPHYS2016}.
In the disordered phase, we stress that the large squeezing of the ground state allows for quantum-enhanced parameter estimation,
and we point out the existence of a crossover from shot-noise scaling to the ultimate Heisenberg limit.
Furthermore, entanglement is outstandingly robust against temperature, as caught by a basic harmonic model: 
this robustness explains the experimental observations of Ref.~\cite{EsteveNATURE2008}, 
where ground-state entanglement was detected at finite temperature in a Bose-Einstein condensate with repulsive interatomic interactions.

Literature already covers a large variety of perspectives on the subject. 
As regards the spectral properties of the model around the critical points, 
we complement the existing research by supplying a detailed analysis of finite-size behaviours.
Concerning entanglement, we go beyond the well-established study of bipartite entanglement and
-- even if the usage of the quantum Fisher information and spin-squeezing parameter is not new -- 
we extensively investigate multipartite entanglement at finite temperature, both numerically and analytically, 
providing a novel point of view on the thermal phase diagram of the model.
Moreover, we are not aware of any previous work focused on entanglement in the first-order phase transition.
The original results of this chapter are contained in the paper~\cite{Gabbrielli2018a}.

\section{The model} \label{sec:LMGModel}
The Lipkin-Meshkov-Glick (LMG) model describes an ensemble of $N$ spins $\sfrac{1}{2}$ 
mutually interacting via a pairwise infinite-range coupling 
and subject to homogeneous transverse and longitudinal magnetic fields:
\be \label{HamLMG1}
\hat{H}_{\rm LMG} = \chi\,\hat{J}_z^2 - \Omega\,\hat{J}_x + \zeta\,\hat{J}_z \, .
\ee
The Hamiltonian\footnote{\ Several Authors~\cite{DusuelPRB2005,barthel2006,RibeiroPRL2007,orus2008,Caneva2009} 
studied a more general version where the interaction term is written as $\hat{J}_z^2+\gamma\hat{J}_y^2$, 
with $\gamma$ the XY anisotropy coefficient. The model we consider is the fully-anisotropic case for $\gamma=0$.
It is worth to note that for the isotropic model ($\gamma=1$) 
a complete exact analytical solution is easily available~\cite{DusuelPRB2005}.\\[-9pt]} 
is formulated entirely in terms of collective pseudospin operators 
$\hat{J}_\varrho = \frac{1}{2} \sum_{i=1}^N \hat \sigma_\varrho^{(i)}$ ($\varrho=x,y,z$) 
obeying the SU(2) algebra $\,\big[\hat{J}_\varrho,\hat{J}_\varsigma\big]=\ii\,\varepsilon_{\varrho\varsigma\tau}\hat{J}_\tau\,$ 
and $\,\big[\hat{J}_\varrho,\hat{\mathbf{J}}^2\big]=0 \ \forall\,\varrho$, with $\hat{\mathbf{J}}^2=\hat{J}_x^2+\hat{J}_y^2+\hat{J}_z^2$.
%The spin representation permits to effectively describe the nucleus in a two-mode approximation, 
%where each nucleon can occupy only the lower or the upper single-particle energy levels.
Here, $\hat{J}_z^2$ describes the interaction of each spin with every other spin along the quantization direction $z$
and $\chi$ is the strength of the uniform all-to-all spin interaction; 
$\hat{J}_x$ and $\hat{J}_z$ accounts for the local coupling to the external transverse and longitudinal fields 
of magnitude $\Omega$ and $\zeta$ respectively.
For simplicity, in the following we assume $N$ even and $\Omega>0$, with no loss of generality.

Due to its fully-connected nature, the LMG model describes spin systems having an \emph{infinite coordination number} 
(in the thermodynamic limit), regardless of their actual spatial dimensionality.
In these systems, each spin equally interacts with all the others: 
no matter if they are arranged on a chain, a planar lattice or a three-dimensional structure. 
Effectively, the whole system can be thought as localized in one point, 
because the linear lattice spacing \textsl{a} can be made arbitrarily small.
Therefore, the usual concept of ``correlation length'' -- whose divergence $\xi\gg\textsl{a}$ signals a critical point 
and whose scaling around criticality plays an essential role in determining the universality class of the critical system -- 
loses its meaning. 
% any length is much larger than the system size.

The basis of the complete Hilbert space contains $2^N$ microscopical spin configurations, most of which are highly degenerate.
Owing to their relevance for applications, we restrict our attention to states that are invariant under permutation of couples of spin, 
as done in many standard references~\cite{vidal2004, DusuelPRB2005, barthel2006, RibeiroPRL2007, orus2008, Kwok2008, MaPRA2009, Caneva2009, Durkin2016, HaukeNATPHYS2016}. 
% In other words, we perform a change of basis and restrict the Hilbert space to the fully-symmetric subspace.
Note that in the common eigenstates of $\hat{\mathbf{J}}^2$ and $\hat{J}_z$, the Hamiltonian~(\ref{HamLMG1}) 
decomposes into disconnected blocks, each one corresponding to a different length $J$ of the pseudospin vector;
our restriction naturally leads to select only the \emph{largest} of these blocks, having maximum value of the spin $J=\frac{N}{2}$
and hosting $N+1$ permutational symmetric states, labelled by the eigenvalues $m=-J,-J+1,\dots\,J-1,J$ of $\hat{J}_z$.
The benefits following from the restriction of the Hilbert space to the \emph{fully-symmetric subspace} 
are enormous: the dimension of the basis (and thus the complexity of the problem) is reduced 
from exponentially large to linearly growing with the population $N$; 
moreover, it is possible to introduce a generalized Bloch sphere of radius $\frac{N}{2}$ 
on which the LMG eigenstates can be visualized in terms of quasi-probability distribution.
Obviously, we must have good reasons for such a strong assumption: as we outline below, 
it is justified by the bosonic nature of the particles involved in real systems commonly described by the LMG model.
% The bosonic nature of the particles (???) allow to restrict the Hilbert space to the fully-symmetric subspace,
% that corresponds to the sector having largest value of the spin $J=\frac{N}{2}$.
Henceforth, we will often interchange the spin picture with the more realistic view of bosons in two modes.

\subsection{Applications}
Although it was originally introduced to account for the transition between spherical and deformed (oblate or prolate) shapes in nuclei,
as a spin model the LMG model effectively outlines a large variety of two-mode physical systems in different fields, 
encompassing the internal Josephson effect in $^3$He-A~\cite{WebbPRL1974},
specific ferroelectrics~\cite{Chakrabarti1996}, small magnetic molecules~\cite{Garanin1998},
degenerate Bose gases in two hyperfine levels coupled by a microwave field~\cite{ZiboldPRL2010} 
or in a double-well potential in the tunnelling regime~\cite{Milburn1997}. 
Very recently, it has attracted large attention in connection to quantum annealing \cite{Caneva2009,Bapst2012,Durkin2016}.

%The LMG Hamiltonian is also well-known in the solid-state community as a special case of the familiar Bose-Hubbard model.
%In fact, by virtue of the Schwinger representation $\hat{J}_x = \frac{1}{2}\big(\hat{a}^\dag\hat{b}+\hat{b}^\dag\hat{a}\big)$, 
%$\hat{J}_y = \frac{1}{2\ii}\big(\hat{a}^\dag\hat{b}-\hat{b}^\dag\hat{a}\big)$, 
%$\hat{J}_z = \frac{1}{2}\big(\hat{a}^\dag\hat{a}-\hat{b}^\dag\hat{b}\big)$,
%dropping a positive prefactor and constant terms containing the total number $\hat{N}=\hat{a}^\dag\hat{a}+\hat{b}^\dag\hat{b}$,
%Eq.~(\ref{HamLMG1}) can be cast into the form of a ``dimer''~\cite{Franzosi2000,Buonsante2012}
%\be
%\hat{H}_{\rm BJJ} = \chi\,\big(\hat{a}^\dag\hat{a}^\dag\hat{a}\hat{a}+\hat{b}^\dag\hat{b}^\dag\hat{b}\,\hat{b}\big) 
%- \Omega\,\big(\hat{a}^\dag\hat{b}+\hat{b}^\dag\hat{a}\big) + \zeta\,\big(\hat{a}^\dag\hat{a}-\hat{b}^\dag\hat{b}\big) \, ,
%\ee
%describing $N$ interacting bosons hopping on a two-site lattice, where $\Omega$ is the kinetic energy, 
%$\chi$ is the on-site interaction strength and $\zeta$ allows for a local energy offset between the two sites.

\begin{figure}[t!]
\centering
\includegraphics[width=1\textwidth]{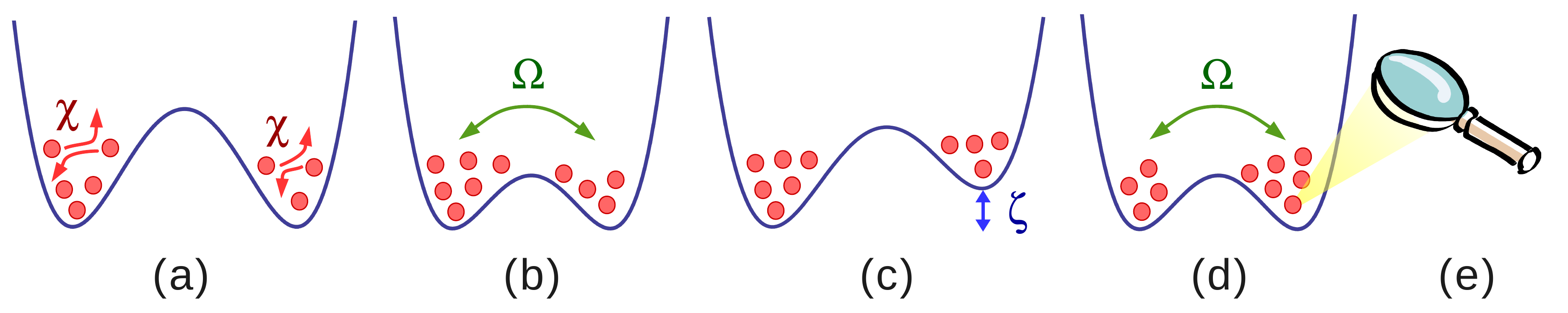} 
\caption{A BEC trapped in a potential with double-well profile is a practical realization of the bosonic Josephson junction 
and can sustain a local Mach-Zehnder--like interferometric process. 
\textbf{(a)} State preparation: the potential is symmetric ($\zeta=0$) 
and the barrier is sufficiently high to forbid tunnelling ($\Omega=0$); 
entanglement in the input state can be created via interactions $\chi$
either by dynamical evolution or adiabatic deformation starting from an initial separable state of uncorrelated particles, 
that is a coherent condensate delocalized over the two wells.
\textbf{(b)} First beam splitter: interactions are switched off ($\chi=0$), 
the barrier is lowered and the splitting is realized by waiting a time $t_{\rm BS}$ 
under tunnel effect $\Omega$ such that $\Omega\,t_{\rm BS}\approx\frac{\pi}{2}$.
\textbf{(c)} Phase imprint: the barrier is raised to suppress tunnelling ($\Omega=0$), 
a energy shift $\zeta$ between the two local minima of the potential is introduced 
and a free evolution of time $t_{\rm PS}$ generates the phase shift $\phi\simeq\zeta\,t_{\rm PS}$.
\textbf{(d)} Second beam splitter: the asymmetry is removed ($\zeta=0$), 
the barrier is lowered in order for the tunnelling $\Omega$ to act again for the same time $t_{\rm BS}$.
\textbf{(e)} Measurement: finally, particle number is detected by imaging the atomic clouds; 
from the measure of the number imbalance a phase estimation of $\phi$ can be carried through.}
\label{fig:DWInterferometer}
\end{figure}

\paragraph{Bosonic Josephson junction} 
Generally speaking, the restriction to the fully-symmetric sector permits to map the LMG model into the 
``bosonic Josephson junction'' (BJJ),\footnote{\ The name descends from the strong analogy with the superconducting Josephson junction, where Cooper pairs tunnel between two superconductors through a layer of insulating material.\\[-9pt]} 
describing $N$ interacting bosons constrained to occupy two coupled modes only~\cite{Cirac1998,Gordon1999}.
In fact, by virtue of the Schwinger representation $\hat{J}_x = \frac{1}{2}\big(\hat{a}^\dag\hat{b}+\hat{b}^\dag\hat{a}\big)$, 
$\hat{J}_y = \frac{1}{2\ii}\big(\hat{a}^\dag\hat{b}-\hat{b}^\dag\hat{a}\big)$, 
$\hat{J}_z = \frac{1}{2}\big(\hat{a}^\dag\hat{a}-\hat{b}^\dag\hat{b}\big)$~\cite{Biederharn1981},
dropping a positive prefactor and constant terms containing the total number $\hat{N}=\hat{a}^\dag\hat{a}+\hat{b}^\dag\hat{b}$,
Eq.~(\ref{HamLMG1}) can be cast into the form 
\be \label{HamBJJ}
\hat{H}_{\rm BJJ} = \chi\,\big(\hat{a}^\dag\hat{a}^\dag\hat{a}\hat{a}+\hat{b}^\dag\hat{b}^\dag\hat{b}\,\hat{b}\big) 
- \Omega\,\big(\hat{a}^\dag\hat{b}+\hat{b}^\dag\hat{a}\big) + \zeta\,\big(\hat{a}^\dag\hat{a}-\hat{b}^\dag\hat{b}\big) \, ,
\ee
where the operators $\hat{a}^\dag$ and $\hat{b}^\dag$ create one boson in modes \textsf{a} and \textsf{b}. 
If the two modes are \emph{external} spatially-separated motional states of a trapping potential, 
the Hamiltonian~(\ref{HamBJJ}) is equivalent to the familiar Bose-Hubbard model for a two-site lattice 
-- also named ``dimer'' -- where $\Omega$ is the kinetic energy, $\chi$ is the on-site interaction strength 
and $\zeta$ allows for a local energy offset between the two sites~\cite{Franzosi2000}.
Such a system has been implemented using a dilute Bose-Einstein condensate (BEC) 
trapped in a (magnetic or optical) double-well potential~\cite{berrada2013,TrenkwalderNATPHYS2016}: 
the nonlinear interaction is generated by the contact collisions and $\chi$ is proportional to the s-wave scattering length,
the coupling $\Omega$ is provided by the tunnel effect between the two wells, 
while the energy imbalance $\zeta$ is related to the asymmetry of the double well (see also Fig.~\ref{fig:DWInterferometer}).
An \emph{internal} variant of the BJJ has been also realized using BECs, 
exploiting two internal degrees of freedom, such as hyperfine atomic states, 
coupled by a driving laser field $\Omega$ that coherently transfers atom population from one level to the other,
with a detuning $\zeta$ resulting from the mismatch between 
the driving frequency and the actual atomic energy separation~\cite{ZiboldPRL2010,MuesselPRA2015}.
The rich scenery of exotic dynamical phenomena occurring in all these systems has been successfully captured 
by means of the LMG model~\cite{RaghavanPRA1999,Franzosi2000,ZiboldPRL2010}.

%%%%%%%%%%%%
%% FIGURA %%
%%%%%%%%%%%%

\paragraph{Double-well interferometer} 
Elastic two-body collisions are naturally present in ultracold atoms and represents a well-established method to generate 
metrologically useful entanglement. 
The experimental capability in tuning this nonlinear interactions, 
along with the high control of the parameters of usual optical and magnetic confinement, 
makes a BEC trapped in a double-well potential an optimal platform to implement 
a two-mode atom interferometer (Fig.~\ref{fig:DWInterferometer}) 
formally equivalent to the Mach-Zehnder interferometer (Fig.~\ref{figMZR}).

Since this system is a particular BJJ with external degree of freedom, 
both the preparation of the input state and the dynamical evolution towards the output state 
can be described by means of the anisotropic fully-symmetric LMG model. 
In this context, the interaction parameter $\chi$ can be switched on and off at will; 
its absolute value can be controlled by tuning the scattering length, 
an operation which is usually performed by precisely varying the applied magnetic field near to a Feshbach resonance;
its sign is even adjustable by changing the nature of the interaction 
($\chi>0$ for repulsive interaction, $\chi<0$ for attractive interaction).
The tunnelling parameter $\Omega$ is tunable by raising or lowering the barrier height separating the two wells, 
while the asymmetry parameter $\zeta$ can be varied by tilting the energy potential.

The ground state of the LMG model with parity symmetry ($\zeta=0$) contains a certain amount of quantum correlations, 
that we will discuss in sections~\ref{sec:symmetricLMG}. 
Therefore, it candidates to be a good input state for performing quantum interferometry 
and it can be experimentally addressed\footnote{\ We focus here on the state preparation through
the adiabatic deformation of the ground state: starting from a noninteracting BEC ($\chi=0$), 
the target input state is accessed by adiabatically tuning the interatomic interaction strength $\chi$ or the tunnelling parameter $\Omega$. 
Instead, for a review on the dynamical generation of entanglement by dint of the LMG model, see Ref.~\cite{PezzeRMP}.\\[-9pt]}~\cite{TrenkwalderNATPHYS2016}. 
The ``beam splitting'' is operated in the noninteracting gas ($\chi=0$) 
by the tunnel effect $\Omega$ between the two symmetric wells ($\zeta=0$). 
The ``phase imprint'' is operated in the noninteracting gas ($\chi=0$) 
by the energy shift $\zeta$ between the two local minima when the tunnel effect is suppressed by a high barrier ($\Omega=0$). 
The tilt $\zeta$ in the potential, producing the phase difference between the wave functions of the two clouds, 
is generated by some external physical perturbation -- such as gravity or other inertial forces. 
In principle, thanks to the initial multipartite entanglement, suitable statistical inference 
permits finally to estimate the phase difference with sensitivity 
outperforming any other classical interferometer~\cite{FattoriPRL2008,PezzeRMP,ChwedenczukNJP2012}.
The sequence of all these steps should be performed without providing excitations to the system~\cite{TrenkwalderNATPHYS2016}.
The ideal interferometric evolution is sketched in Fig.~\ref{fig:DWInterferometer}.

\paragraph{Important remarks}
Since $J$ is a constant of motion, the total number of spins $N=2J$ is conserved.
As a model for nuclei, Hamiltonian~(\ref{HamLMG1}) has been often used with a \emph{finite} number of constituents. 
Instead, for an \emph{infinite} number of degree of freedom it has been exploited to describe dense gases, 
such as in the spin van der Waals model~\cite{Niemeijer1970}. 
However, an increase in the number of interacting particles makes the total energy grow faster than linearly with $N$, 
due to the two-spin part that increases quadratically $\big\langle\hat{J}_z^2\big\rangle \sim N^2$.
In the limit $N\gg1$ an important issue arises: the energy is not an extensive quantity.\footnote{\ The same issue arises in the Ising model for long-range interaction $\alpha\leq1$ discussed in section~\ref{sec:IsingArbitrary}: 
the energy of the ground state is found to be extensive in the paramagnetic phase and superextensive in the ordered phases.
This is the very reason the perturbative approach of paragraph~\ref{subsec:IsingDetection} 
was not allowed in the thermodynamic limit. 
Certainly, also the Ising model is amenable to an appropriate readjusting of the interaction strength $\sin\theta \mapsto N^{-1}\sin\theta$
that permits to switch to a modified extensive Hamiltonian for $\alpha\leq1$: 
this rescaling would not modify the distinct scaling behaviours of multipartite entanglement in the ordered and disordered phases, 
but it would yield a different location of the critical points.
Moreover, the capability of the quantum Fisher information in detecting the range of the interaction is expected to be lost.\\[-9pt]}
In order to guarantee the convergence of the free energy per spin in the thermodynamic limit, 
we need to damp the interaction constant through a prefactor $N^{-1}$. 
Thence, it is customary to define the LMG Hamiltonian as
\be \label{HamLMG2}
\hat{H}_{\rm LMG} = \frac{\Lambda}{N}\,\hat{J}_z^2 - \,\hat{J}_x + \delta\,\hat{J}_z \, ,
\qquad {\rm with} \quad \Lambda=\frac{N\chi}{\Omega} \ \ {\rm and} \ \ \delta=\frac{\zeta}{\Omega} \, .
\ee
where we are measuring energies in units of $\Omega>0$.
The sign of $\Lambda$ determines the character of the interaction among the spins: 
it is ferromagnetic if $\Lambda<0$, antiferromagnetic if $\Lambda>0$; the case $\Lambda=0$ corresponds to noninteracting spins.
Equation~\ref{HamLMG2} is the form of the LMG Hamiltonian we will consider in the following when discussing the multipartite entanglement. 

We underline here that taking the \emph{thermodynamic limit} of Eq.~(\ref{HamLMG2}) means to increase the number of spins $N\to\infty$
and, concurrently, weaken the spin-spin strength $\chi\to0$ so that the interaction constant $\Lambda$ is kept fixed.
The meaning of thermodynamic limit is thus different from the one settled in ordinary short-range lattice models
(like the nearest-neighbour Ising model discussed in section~\ref{sec:IsingNN}), 
where the condition $N\to\infty$ is sufficient and no according decrease of the coupling among the particles is required.

\subsection{Semiclassical approach} \label{subsec:semiclassicalLMG}
The LMG model is an example of exactly solvable many-particle system: 
in principle, the diagonalization of the Hamiltonian can be accomplished by solving the Bethe-ansatz equation~\cite{pan1999}.
Nevertheless, this is a very costly operation; numerical means must intervene at some point~\cite{morita2006}.
An exact quantum phase model was developed~\cite{AnglinPRA2001}, which consists in projecting the LMG Hamiltonian 
over the Bargmann's states in the phase space: it provides an exact determination of the spectrum in a quite simple fashion, 
but the price to pay is hidden in the overcompleteness of Bargmann's basis, that 
can bring serious complications in computation and notable counterfeits as a result.

The infinite-range nature of the interactions justifies a mean-field approximation of the LMG model.
In general, there exist several ways to build a semiclassical treatment of this problem: 
essentially, they all can be regarded somehow as expansions in powers of $N^{-1}$ in the limit of large~$N$.

\paragraph{Mean-field approach}
For example, the ground-state properties can be easily accessed using 
a zero-order variational method~\cite{BotetPRL1982,DusuelPRB2005,CastanosPRB2006}:
it comprises the projection of the LMG Hamiltonian (\ref{HamLMG2}) on a classical coherent spin state $|\vartheta,\varphi\rangle$ and 
the determination of the wave function by minimizing the energy with respect to the free parameters,
which are the fractional population difference $z=\cos\vartheta$ and the relative phase $\varphi$ between the two modes
\textsf{a} and \textsf{b}. 
In the thermodynamic limit, this variational approach is straightforwardly equivalent to treating the spin operator 
$\hat{\mathbf{J}}$ as a classical vector 
$\langle\mathbf{J}\rangle\mapsto\frac{N}{2}\big(\sin\vartheta\cos\varphi,\sin\vartheta\sin\varphi,\cos\vartheta\big)$~\cite{BotetPRB1983}.

A zero-order description of the dynamics of the BJJ on a classical phase space is gained
by replacing the pseudospin operators with real numbers 
$\hat{J}_x\mapsto\frac{N}{2}\sqrt{1-z^2}\cos\varphi$ and $\hat{J}_z\mapsto\frac{N}{2}z$~\cite{smerzi1997},
and then studying the trajectories\footnote{\ The number imbalance $z$ and the relative phase $\varphi$ play the role 
of canonical conjugate variables: the equation of motion are simply the classical Hamilton equations 
$\dot{z}=\partial_\varphi H(z,\varphi)$ and $\dot{\varphi}=\partial_z H(z,\varphi)$.\\[-9pt]} induced by the classical Hamiltonian
\be \label{HamRaghavan}
H(z,\varphi) = \frac{\Lambda}{2}z^2-\sqrt{1-z^2}\cos\varphi + \delta z \, .
\ee
Equivalently, the same results can be obtained from a time-dependent variational principle 
based on Glauber's coherent states~\cite{BuonsantePRA2005} 
or from the Gross-Pitaevskii equation in a two-mode approximation of the wave function~\cite{RaghavanPRA1999}. 
Here the global nonlinear interactions of each particle 
with the others in the same mode are taken into account as a mean-field perturbation.

The simplest method\footnote{\ Alternative attempts exist: they are based on the Holstein-Primakoff representation of the spin operators 
or on the technique of continuous unitary transformations~\cite{DusuelPRB2005}.\\[-9pt]} 
to go one step beyond the mean-field analysis relies 
on a naive quantization of the classical Hamiltonian~\cite{Leggett1991}: 
if one expands Eq.~(\ref{HamRaghavan}) to second order around the minimum point $(z=0,\varphi=0)$, 
replaces the two conjugate number and phase variables with operators 
$z\mapsto\hat{z},\,\varphi\mapsto\hat{\varphi}=-\frac{2}{N}\,\ii\,\partial_z$ 
and imposes a standard commutation rule $[\hat{z},\hat{\varphi}]=\frac{2}{N}\ii$, 
an equivalent quantum harmonic-oscillator Hamiltonian is derived~\cite{TrenkwalderNATPHYS2016}: 
\be \label{HamHarmonicOscillator}
H_z = -\frac{2}{N^2}\,\frac{\partial^2}{\partial z^2} \,+\, \frac{1+\Lambda}{2}z^2 \,+\, \delta z \, .
\ee
Equation~(\ref{HamHarmonicOscillator}) has the merit to predict a Gaussian wave function for the ground state. 
Unfortunately, a rigorous phase operator is missing~\cite{LynchPR1995} 
and the difficulties arising from the definition of two conjugated variables in this case are renowned~\cite{Susskind1964}. 
Moreover, the above procedure cannot lead to any physical result when $\Lambda\leq-1$.

A much more rigorous way to compute first-order corrections in the $N^{-1}$ expansion
has been elaborated in several slightly different variants~\cite{Shchesnovich2008,Buonsante2012}.
Eigenstates of the full Hamiltonian (\ref{HamLMG2}) can be expanded on the Fock basis of the site occupation numbers
$\ket{\psi_n}=\sum_{k=1}^N c^{(k)}_n\ket{k,N-k}$; in the limit of large total population $N\gg1$, 
the coefficients of the expansion can be regarded as smooth functions $c^{(k)}_n\simeq\psi_n(z)$ of a continuous variable 
$\frac{k}{N} \mapsto z=\frac{1}{N}(N_\mathsf{a}-N_\mathsf{b})\in[-1,+1]$ 
expressing the fractional imbalance between spin up (mode \textsf{a}) and spin down (mode \textsf{b}); 
a series expansion of the stationary Schr\"odinger equation in the coordinate representation 
to the lowest order in $N^{-1}$ finally leads to\footnote{\ The result in Eq.~(\ref{ShcrodingerlikeLMG}) 
-- that represents an improvement over the naive Eq.~(\ref{HamHarmonicOscillator}) -- 
can be also obtained by a rigorous application of the Fock-space Wentzel-Kramers-Brillouin method 
to the discrete Schrödinger equation~\cite{Shchesnovich2008}.\\[-9pt]}
%\be 
%H(z) = -\frac{2}{N^2}\,\frac{\partial}{\partial z}\sqrt{1-z^2}\frac{\partial}{\partial z} \,+\, V(z)\,,
%\qquad {\rm with} \quad V(z) = \frac{\Lambda}{2}z^2 \,-\, \sqrt{1-z^2} \,+\, \delta z \, .
%\ee
\be \label{ShcrodingerlikeLMG}
\frac{N}{2}\bigg[-\frac{2}{N^2}\,\frac{\partial}{\partial z}\sqrt{1-z^2}\frac{\partial}{\partial z} \,+\, V(z)\bigg]\,\psi_n(z)=E_n\psi_n(z),
\quad {\rm with} \ V(z) = \frac{\Lambda}{2}z^2 \,-\, \sqrt{1-z^2} \,+\, \delta z \, .
\ee
Equation~(\ref{ShcrodingerlikeLMG}) establishes a clear mapping from the ensemble of interacting two-level bosons 
to a \emph{continuous single-particle picture}: the many-body spectrum can be simply obtained 
by solving a stationary Schr\"odinger-like equation for one particle having position-dependent mass 
and evolving in an effective continuous one-dimensional potential.
The  factor $N^{-1}$ in the kinetic term can be regarded as an effective Planck constant: 
hence, the thermodynamic limit $N\rightarrow\infty$ is formally identical to the semiclassical limit $\hbar\to0$.
This is the reason such a system is often introduced as a semiclassical model, in spite of its full quantum nature. 
In this semiclassical picture, the energy $V(z)$ plays the role of a Ginzburg-Landau potential, 
useful to describe both the second-order and first-order transitions occurring (see sections~\ref{sec:symmetricLMG} and \ref{sec:asymmetricLMG}).

\section{The balanced model} \label{sec:symmetricLMG}
\begin{figure}[p!]
\centering
\includegraphics[width=0.34\textwidth]{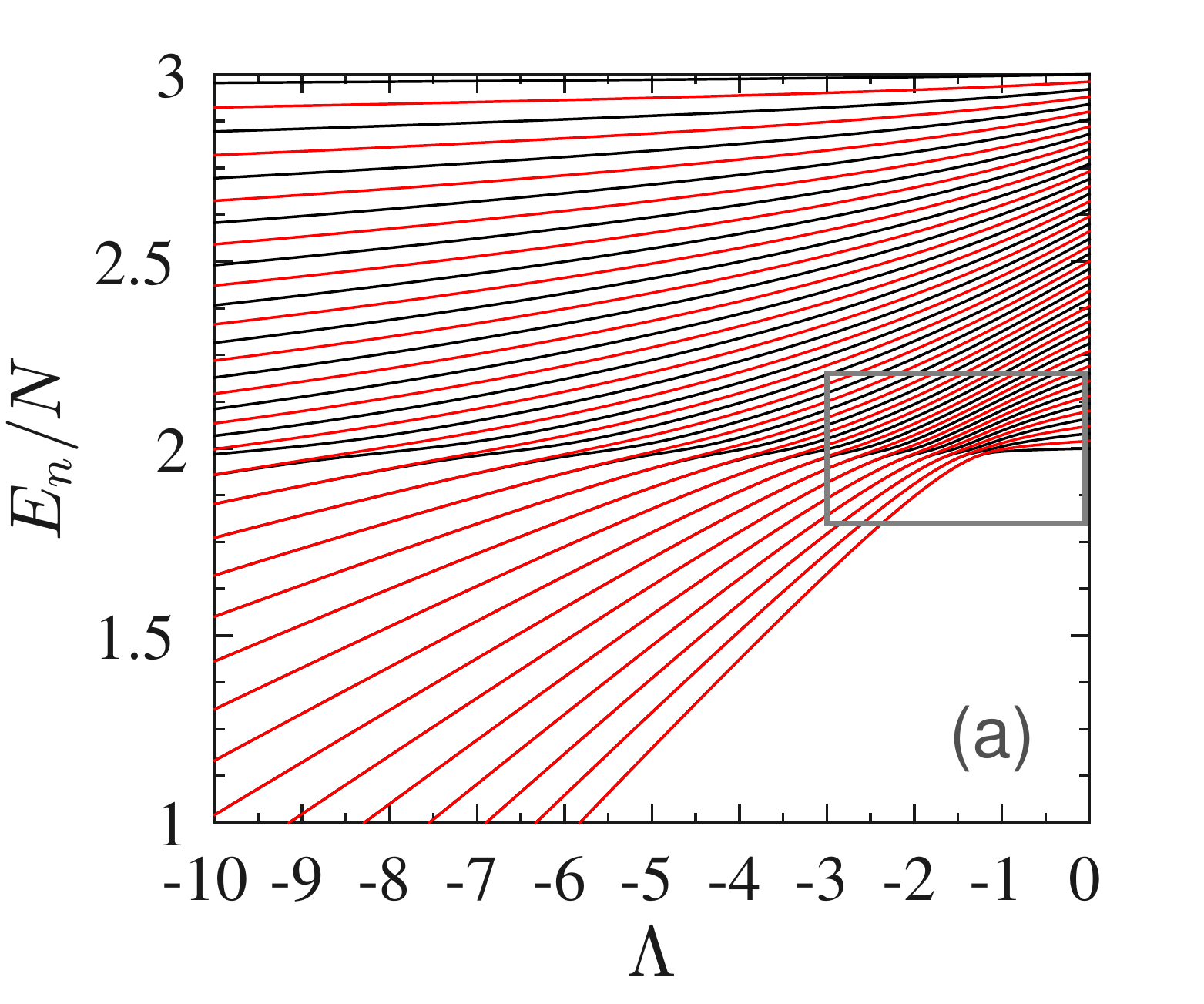} \hspace{-15pt}
\includegraphics[width=0.34\textwidth]{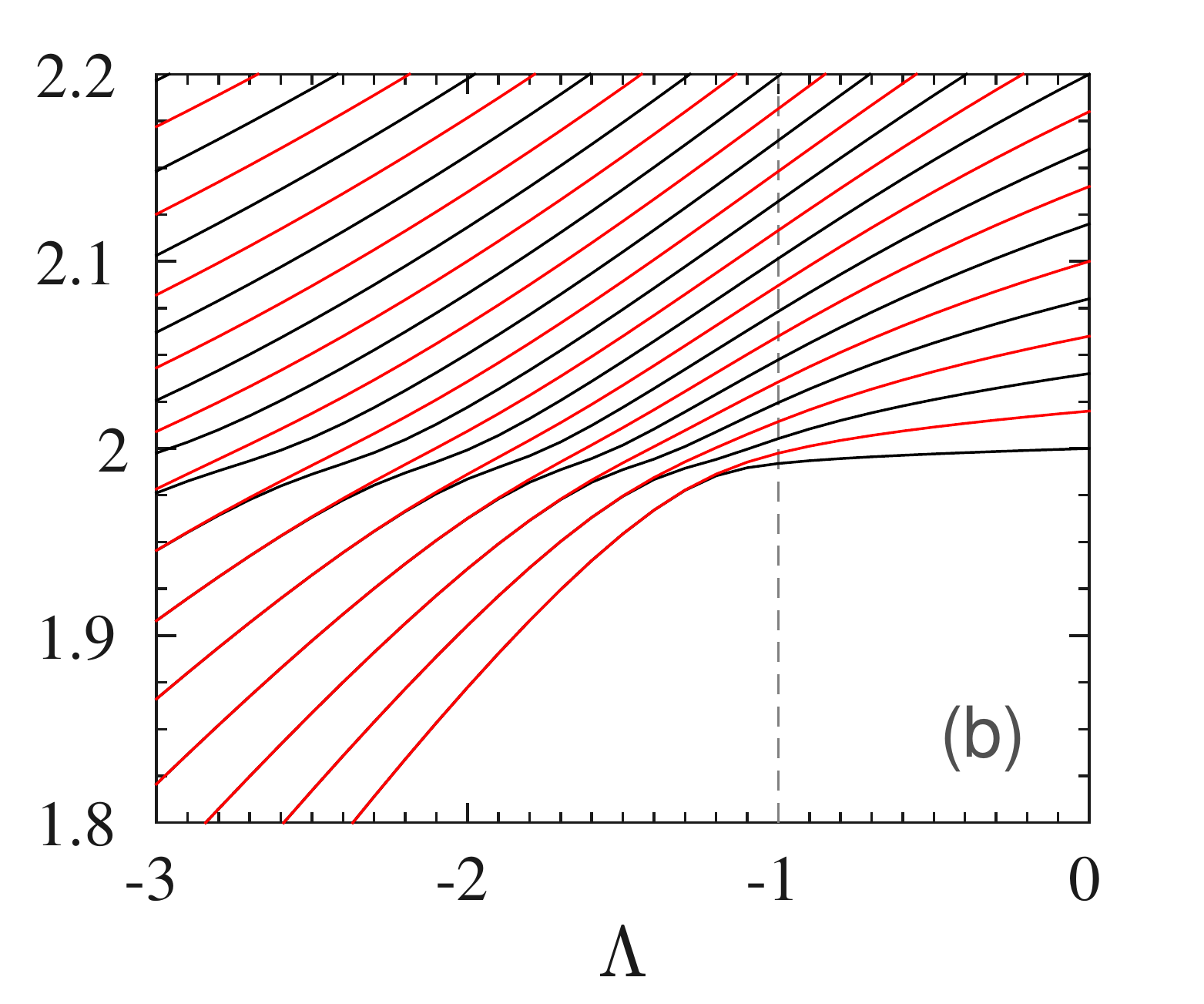} \hspace{-15pt}
\includegraphics[width=0.34\textwidth]{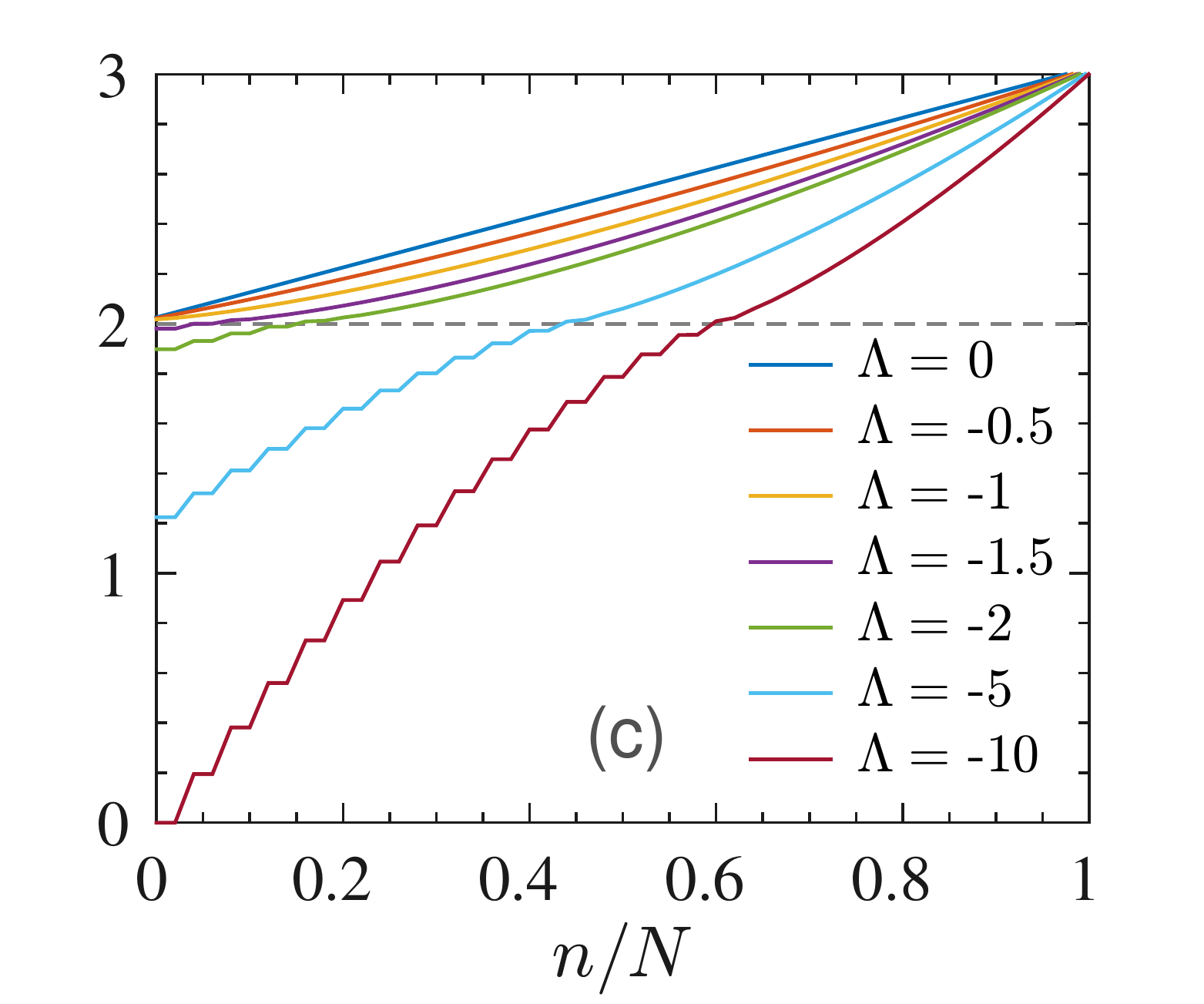} \\
\includegraphics[width=0.34\textwidth]{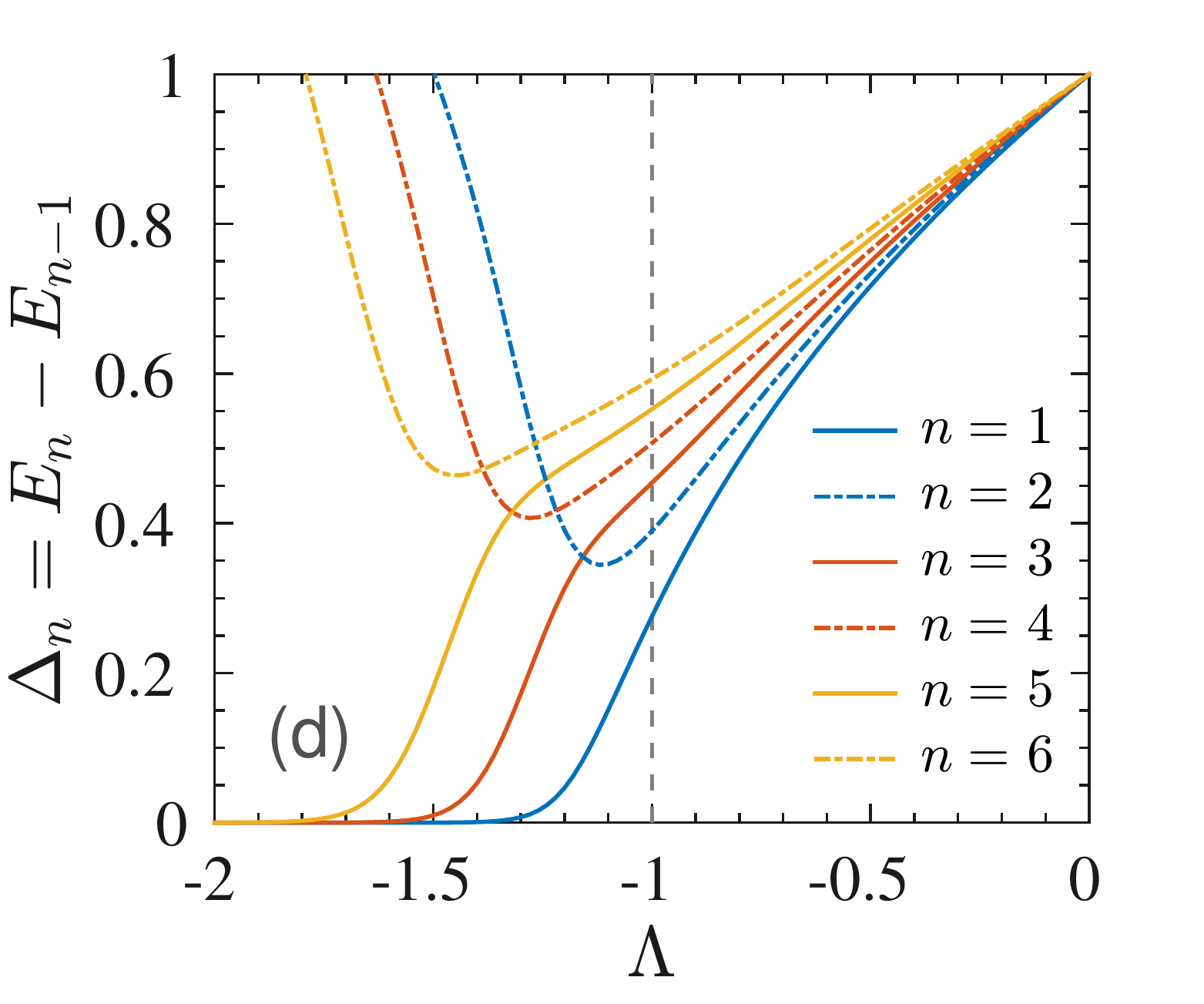} \hspace{-15pt}
\includegraphics[width=0.34\textwidth]{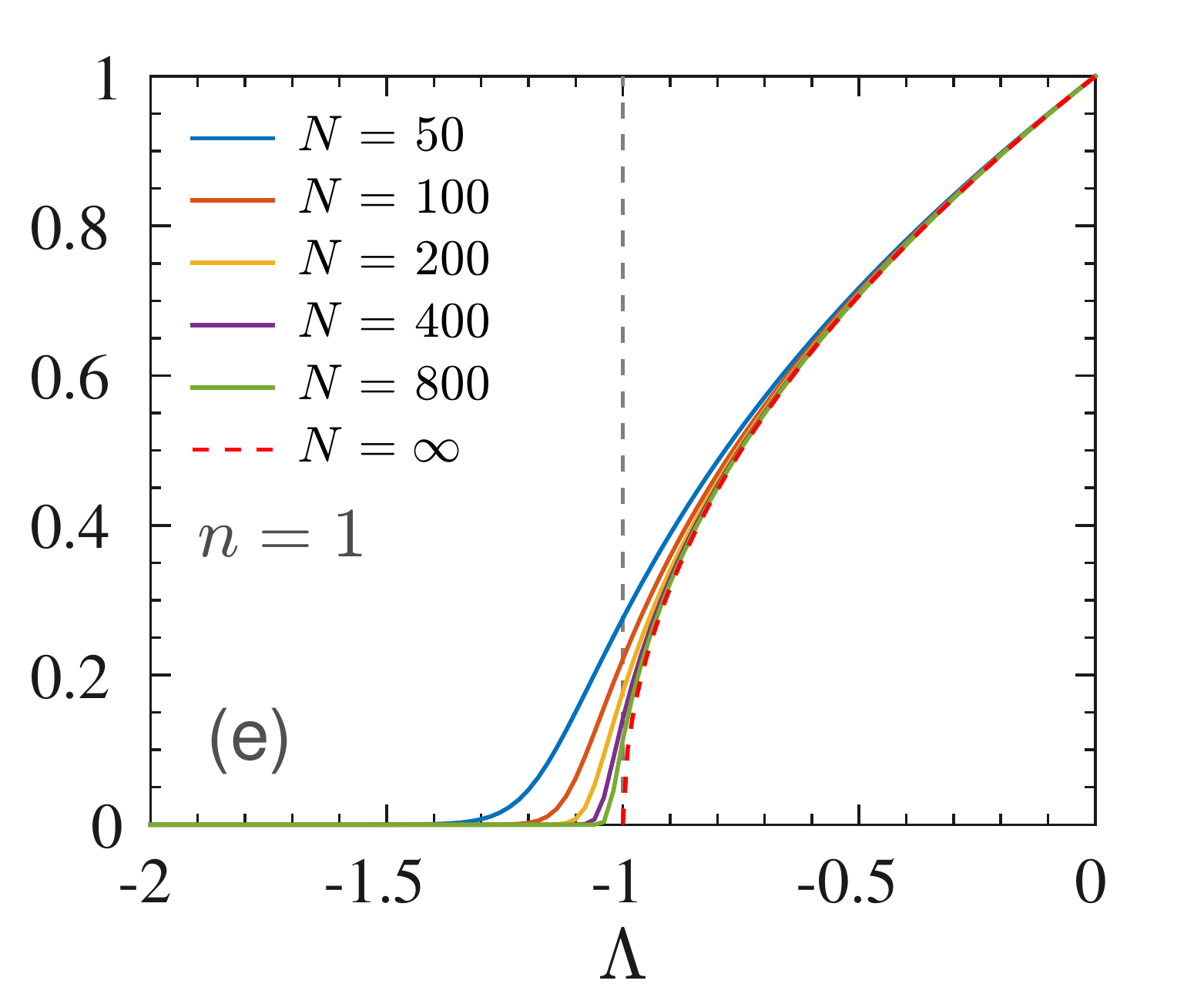} \hspace{-15pt}
\includegraphics[width=0.34\textwidth]{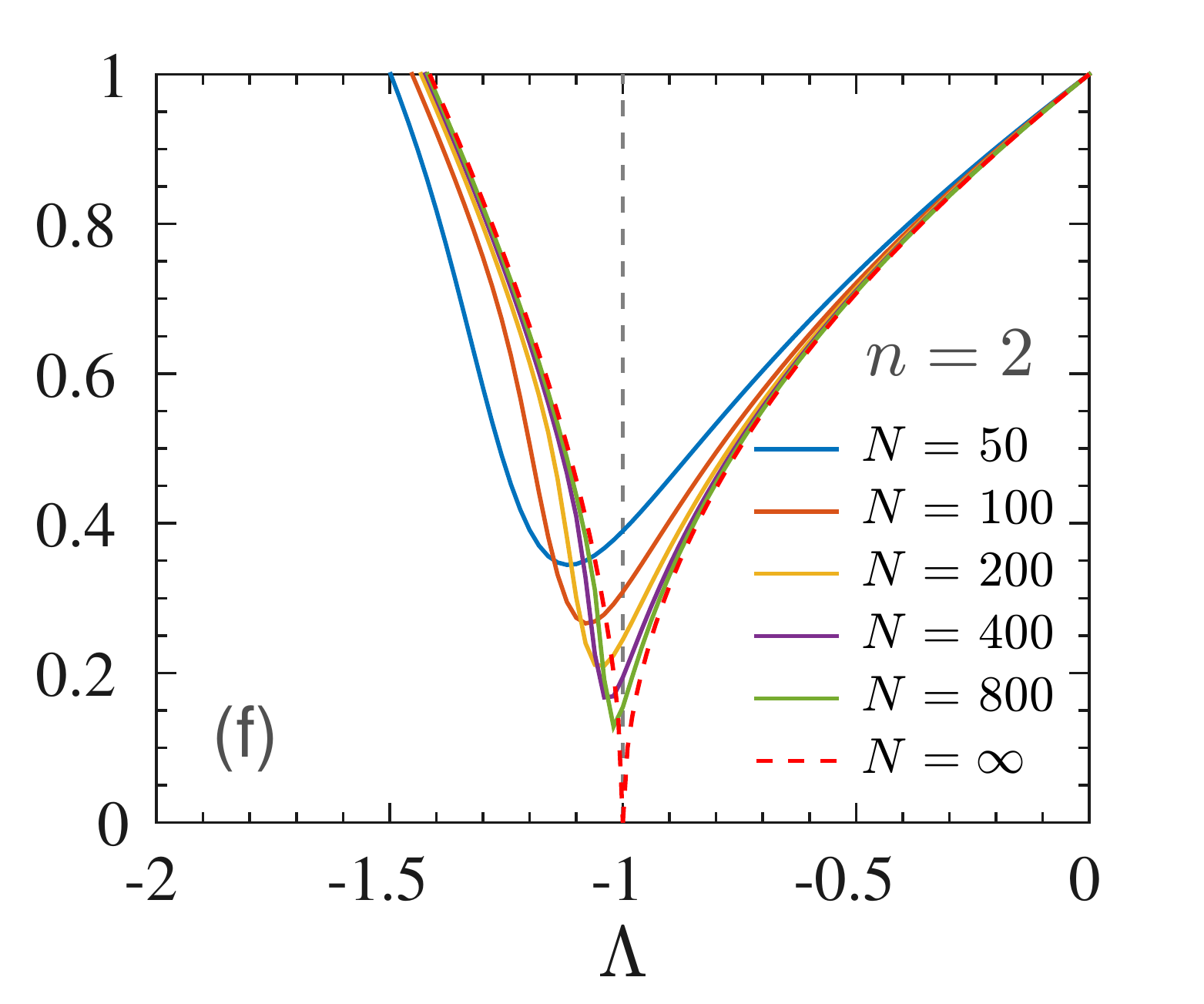} \\
\includegraphics[width=0.34\textwidth]{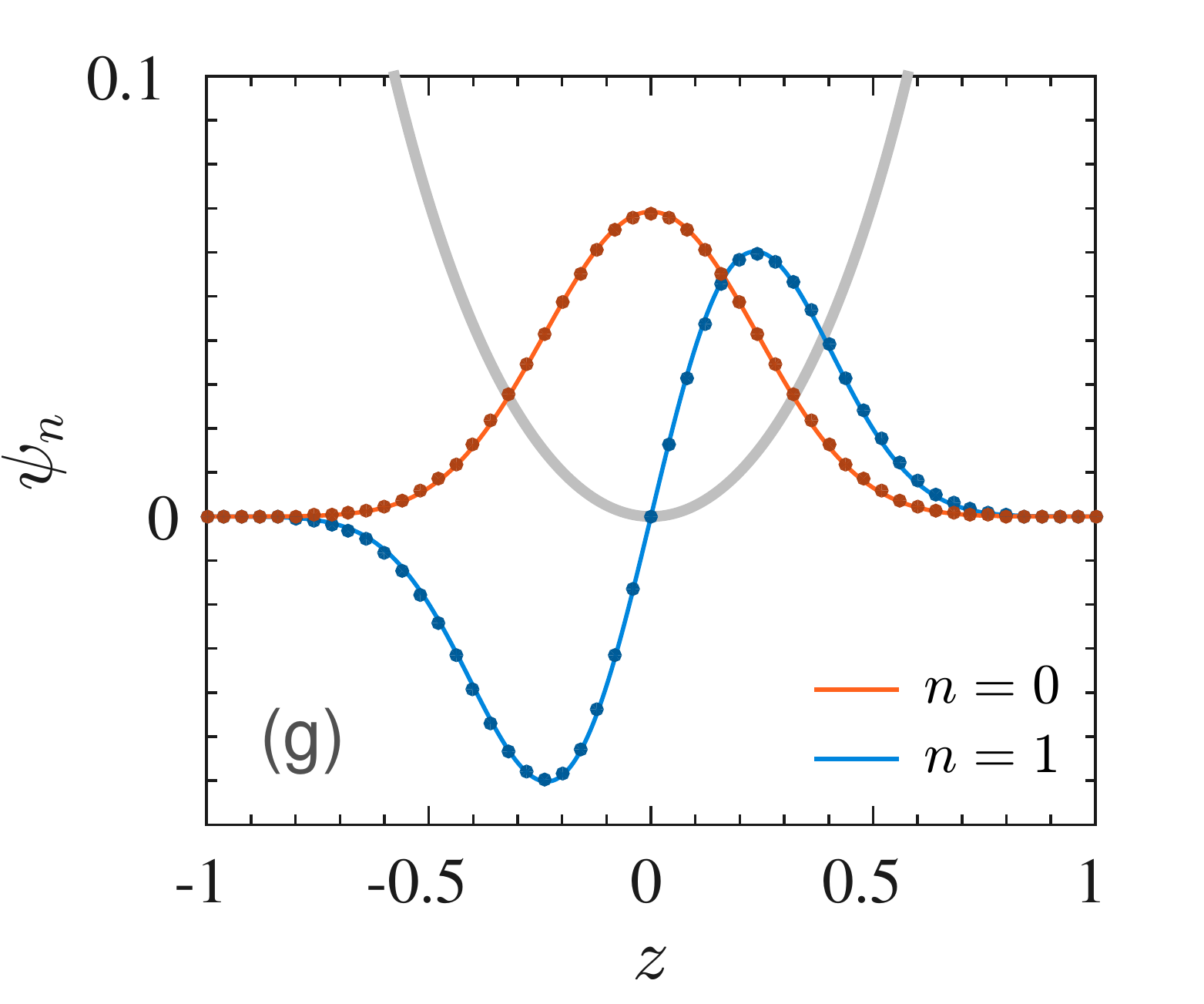} \hspace{-15pt}
\includegraphics[width=0.34\textwidth]{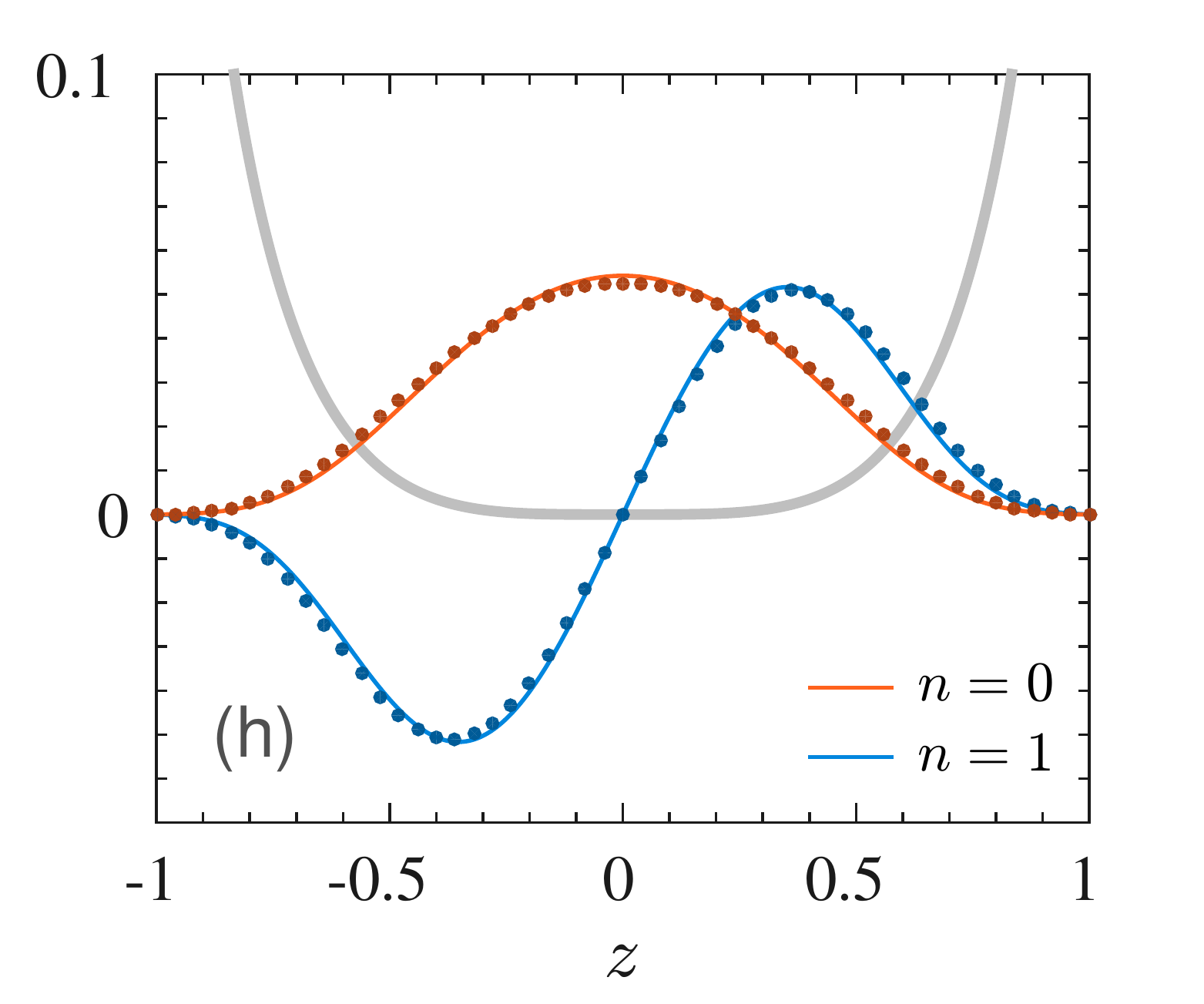} \hspace{-15pt}
\includegraphics[width=0.34\textwidth]{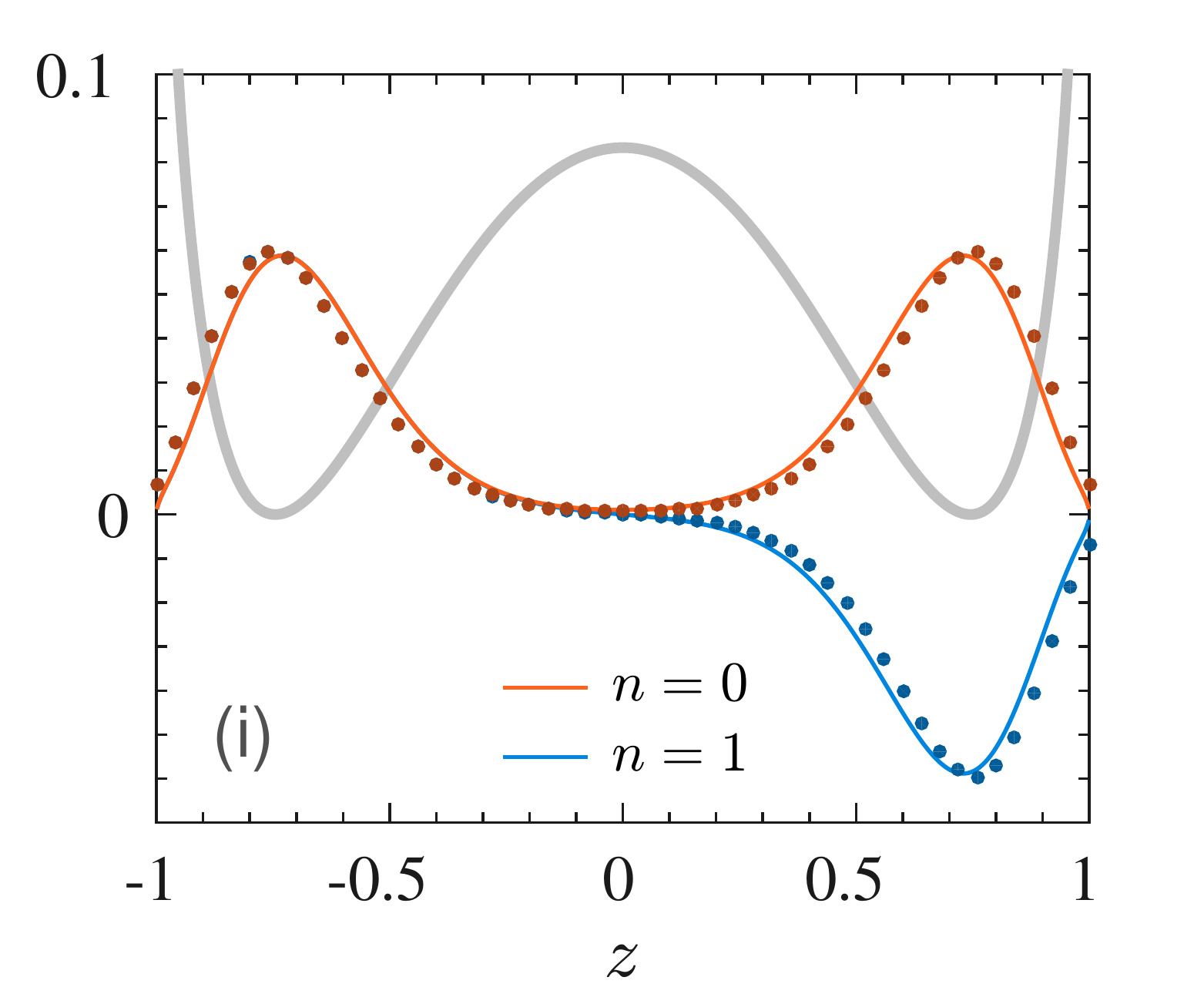} 
\caption{Characteristic features of the many-body balanced LMG spectrum. 
\textbf{(a)} Dependence of the energy levels on the control parameter $\Lambda$ for $N=50$. 
In blue (red) the eigenvalues associated to eigenfunctions with even (odd) parity $n=0,2,4...$ ($n=1,3,5...$).
The rectangle marks the region magnified in the next panel.
\textbf{(b)} Zoom around the critical value of the control parameter (flagged by the dashed vertical line); 
the bunching of the levels into doublets is made clear.
\textbf{(c)} Dependence of the energy levels on the quantum number $n\in[0,N]$ labelling the eigenstates,
for several values of $\Lambda$ and $N=50$. For $\Lambda<-1$ the doublet formation is evident. 
The horizontal dashed line indicates the energy $E_s$ associated to the maximum of the effective potential $V(z)$ for $\Lambda<-1$, 
namely the height of the barrier.
\textbf{(d)} Energy separations in the LMG spectrum as a function of $\Lambda$ for $N=50$. 
For $-1\lesssim\Lambda<0$ the system has the equispaced spectrum typical of a harmonic oscillator; 
at $\Lambda\approx-1$ the first energy gap $\Delta_1=E_1-E_0$ vanishes (and remains closed for any $\Lambda\lesssim-1$) 
while the second energy gap $\Delta_2=E_2-E_1$ reaches a minimum (then opens again). 
The finite-size scenario may lead to some misunderstanding, in case promptly banished by the next two panels.
\textbf{(e)} Energy gap over the ground state $\Delta_1$ for increasing population $N$. 
The red dashed line is the analytical result $\sqrt{1+\Lambda}$ found in the semiclassical approximation.
\textbf{(f)} Second energy separation $\Delta_2$ for increasing population $N$. 
The red dashed line for $\Lambda<-1$ is the analytical result $\sqrt{\Lambda^2-1}$.
\textbf{(g,\,h,\,i)} Lowest-energy eigenfunctions (ground state and first excited state) in the semiclassical approximation (solid lines)
compared to the results from exact diagonalization (dots) for $N=50$ 
and, respectively, $\Lambda=-0.5$ (supercritical), $\Lambda=-1$ (critical) and $\Lambda=-1.5$ (subcritical); 
the agreement between analytical and numerical methods improves for larger values of $N$.
The thick gray lines reproduce the shape of the effective potential $V(z)$.}
\label{fig:spectrumLMG}
\end{figure}

In case of null longitudinal field $\delta=0$, the LMG Hamiltonian
\be \label{HamLMG3}
\hat{H}_{\rm LMG} \ = \ \frac{\Lambda}{N}\,\hat{J}_z^2 \, - \, \hat{J}_x 
\ee
acquires a $Z_2$ symmetry 
under time reversal $\hat{J}_z\mapsto-\hat{J}_z$. 
Limited to the permutational symmetric sector with $J=\frac{N}{2}$, it describes two-mode bosonic systems that are invariant 
$[\hat{\Pi},\,\hat{H}_{\rm LMG}]=0$ after exchange of modes $\hat{\Pi}:\mathsf{a}\leftrightarrow\mathsf{b}$. 
Relevant examples are a BEC trapped in the motional ground and first-excited states 
of a spatially-symmetric double well~\cite{TrenkwalderNATPHYS2016}
or a spinor BEC confined in two hyperfine levels of the electronic ground state and 
invested by a zero-detuned resonant light~\cite{MuesselPRA2015}.
Such a model hosts a ground-state transition and we are interested to unveil its spectral (Fig.~\ref{fig:spectrumLMG})
and entanglement properties (Fig.~\ref{fig:QFIT0LMG}).

\paragraph{Methods}
In general, an exact diagonalization of Hamiltonian~(\ref{HamLMG3}) in the subspace % with $J=\frac{N}{2}$ 
spanned by the basis vectors $|J,m\rangle$ (with $m=-J,-J+1,\dots\,J-1,J$) 
is practicable up to very high population number $N$ and provides all the $N+1$ energy levels and eigenfunctions: 
hence, momenta and expectation values of the pseudospin operators can be efficiently accessed 
and entanglement at both zero and finite temperature unveiled. 
Due to the $Z_2$ invariance, each eigenstates has a definite parity under $m\mapsto-m$, 
allowing for a reorganization the basis in terms of vectors that are either symmetric or antisymmetric 
$|J,m\rangle_{\pm}=\big(|J,m\rangle\pm|J,-m\rangle\big)/\sqrt{2}$.
By virtue of this representation~\cite{FranzosiPRA2001}, 
we implemented an optimized numerical routine to exactly diagonalize $\hat{H}_{\rm LMG}$ for $N$ up to $10^4$.

For very large $N$, it is reasonable to adopt the semiclassical approach presented in paragraph~\ref{subsec:semiclassicalLMG}: 
this means to numerically solve the Schr\"odinger-like equation (\ref{ShcrodingerlikeLMG}) with no other approximations. 
The procedure gives reliable results only for the lowest part of the spectrum,\footnote{\ Energy levels $E_n$ and wave functions $\psi_n(z)$
provided by Eq.~(\ref{ShcrodingerlikeLMG}) have physical meaning as long as the eigenfunctions do not experience the effect of the borders
of the finite domain, where the potential $V(z)$ is substituted by infinite repulsive walls $V(z=\pm1)=\infty$. 
In other words, only the lowest-lying levels $n \ll n_{\rm max}$ are not affected by the finiteness of the domain of $z\in[-1,+1]$.
A simple analytical calculation based on the knowledge of the energy gap in the thermodynamic limit Eqs.~(\ref{LMGenergygap1}) 
and (\ref{LMGenergygap2}) permits to state the condition $n_{\rm max}=\frac{N}{2}\min\{1,\mathcal{L}(\Lambda)\}$, 
with $\mathcal{L}(\Lambda)\simeq(\Lambda+2)\big/2\sqrt{1+\Lambda}$ for $\Lambda>-1$ and 
$\mathcal{L}(\Lambda)\simeq1\big/|\Lambda|\sqrt{\Lambda^2-1}$ for $\Lambda<-1$.\\[-9pt]}
but it permits to extend our analysis up to $N\approx10^5$.
Thus, it is an excellent tool for better addressing finite-size corrections and asymptotic behaviours at low and zero temperature.
A comparison between the exact diagonalization and the semiclassical technique is offered in Figs.~\ref{fig:spectrumLMG}\Panel{g,\,h,\,i}: 
the agreement is good even for a small number of spins.

The semiclassical method expresses all its might in the thermodynamic limit, where it provides exact analytical results.
Due to the term $N^{-2}$ in the kinetic term of Eq.~(\ref{ShcrodingerlikeLMG}), 
for large populations $N$ the ground state and low-energy excited states  
results to be sharply localized around the minima of the potential energy (see Fig~\ref{fig:LMGdeformationQPT}).
Thus, Eq.~(\ref{ShcrodingerlikeLMG}) can be further simplified by expanding the Hamiltonian
and retaining only the quadratic terms in $z$, that are the first nonvanishing terms 
(except for $\Lambda=-1$, where the leading term is quartic in $z$). 
For $\Lambda>-1$, the effective potential $V(z)$ takes the form of a single harmonic well 
centred at $z_{\rm min}=0$ with angular frequency $\propto\sqrt{1+\Lambda}$ 
(we are measuring energy in units of frequency, $\hbar=1$):
\be \label{HarmonicSuperCriticalLMG}
\frac{N}{2}\bigg[-\frac{2}{N^2}\,\frac{\ud^2}{\ud z^2} \,+\, \frac{1+\Lambda}{2}\,z^2\bigg]\,\psi_n(z)=E_n\,\psi_n(z).
\ee 
For $\Lambda<-1$, it becomes the superposition of two independent harmonic wells, separated by an infinite barrier and
centred at $z_{\rm min}=\pm\sqrt{1-1/\Lambda^2}$ with angular frequency $\propto\sqrt{\Lambda^2-1}$: 
\be \label{HarmonicSubCriticalLMG}
\frac{N}{2}\bigg[-\frac{2}{N^2|\Lambda|}\,\frac{\ud^2}{\ud z^2} \,+\, \frac{|\Lambda|(\Lambda^2-1)}{2}\,\big(z-z_{\rm min}\big)^2\bigg]\,\psi_n(z)=E_n\,\psi_n(z).
\ee 
This \emph{harmonic} approximation allows for explicit calculations of the spectrum, the eigenfunctions and the entanglement quantifiers,
and becomes exact when $N\to\infty$. So, it provides precious physical insights, both at zero and finite temperature, 
for a vast region of the parameter $\Lambda$. 
We will often compare the numerical data for finite $N$ to the analytical expressions valid for infinite $N$.
Note that the harmonic approximation does not hold for $\Lambda=-1$, 
where all the eigenfunctions experience the quartic deformation in the expansion of the effective potential $V(z)$.

%%%%%%%%%%%%
%% FIGURA %%
%%%%%%%%%%%%

\paragraph{Energy spectrum}
The spectrum of the balanced LMG model is depicted in Figs.~\ref{fig:spectrumLMG}\Panel{a,\,b,\,c},
and an illustration of the first energy separations $\Delta_n=E_{n}-E_{n-1}$ (for $n\geq1$) 
is reported in Figs.~\ref{fig:spectrumLMG}\Panel{d,\,e,\,f}.
When tuning the interaction parameter $\Lambda$ from zero to large negative values, two distinctive behaviours leap out: 
levels with different parity cluster into doublets and levels with equal parity repel each other.
%%%%%%%%
\\[6pt]
%%%%%%%%
\trivuoto{MyBlue} \ For $\Lambda=0$, the model reduces to $\hat{H}_{\rm LMG}=-\hat{J}_x$, 
schematizing a linear coupling of the spins to the transverse field: the eigenstates are Dicke states~\cite{DickePR1954}
and the levels are equispaced by the quantity $\Delta_n\equiv\Delta_1=1$ (in units of $\Omega$).
%%%%%%%%
\\[6pt]
%%%%%%%%
\trivuoto{MyBlue} \ For $\Lambda>-1$, the system is still gapped above the ground state. 
%Even though the spacing is no more uniform, the energy separations are $\Delta_n\simeq\Delta_1$. In fact, 
For sufficiently large $N$, the spacing is still uniform: $\Delta_n=\sqrt{1+\Lambda}$. 
In fact, for $N\to\infty$ the semiclassical description dictates for the energy levels 
\be \label{LMGenergygap1}
E_n = \sqrt{1+\Lambda}\,\Big(n+\tfrac{1}{2}\Big),
\ee
while the wave functions are simply the eigenstates of the quantum harmonic oscillator
\be \label{HarmonicEigenfunctions}
\hspace{-3pt} \psi_n(z) \equiv \Psi_n(z) = \sqrt{\textrm{\footnotesize$\frac{N}{2^{n+1}n!}$}}\left(2\pi\sigma^2\right)^{-\frac{1}{4}} \,
\exp\left[-\left(\textrm{\footnotesize$\frac{N}{4\sigma}z$}\right)^2\right] \, % \neper^{-\left(\frac{N}{4\sigma}z\right)^2} 
H_n\Big(\textrm{\footnotesize$\frac{N}{2\sqrt{2}\sigma}z$}\Big),
\ee
where $H_n$ is the $n$th Hermite polynomial and the square width of the ground-state Gaussian wave function is given by 
\be
\sigma^2\textrm{\small$(\Lambda>-1)$} = \frac{N}{4}\frac{1}{\sqrt{1+\Lambda}}.
\ee
In particular, the ground state is antisqueezed along the $z$ direction. 
For finite $N$, this qualitative structure is preserved only in the low-lying part of the spectrum: 
for low $n$ the spacing is almost uniform $\Delta_n\simeq\Delta_1$, whereas for higher quantum numbers $n$ 
the energy levels scale as $E_n \sim n^{4/3}$ 
due to the hardening of the potential induced by the quartic term in $V(z)$~\cite{GilmorePRA1986}.
The crossover from $E_n \sim n$ to $E_n \sim n^{4/3}$ depends on the quantitative details of the potential.
%%%%%%%%
\\[6pt]
%%%%%%%%
\trivuoto{MyBlue} \ Around $\Lambda=-1$, the system becomes gapless in the thermodynamic limit: 
the lowest two levels clump together and the first energy gap $\Delta_1$ closes and 
the first nonvanishing energy separation is $\Delta_2$, that here reaches a minimum value.
At finite size, the degeneracy between the ground state and the first excited state is lifted: 
we numerically find that $\Delta_1$ vanishes as a power law $N^{-1/3}$, 
in agreement with the original finding of Ref.~\cite{BotetPRB1983} and the scaling ansatz of Ref.~\cite{BuonsantePRA2005};
the values of $\Lambda$ at which $\Delta_1$ closes and $\Delta_2$ reaches its minimum approach $\Lambda=-1$ algebraically fast in $N$, 
specifically $\Lambdac - \arg\max_\Lambda |\partial_\Lambda\Delta_1| \sim N^{-2/3}$ 
and
$\Lambdac - \arg\min_\Lambda \Delta_2 \sim N^{-2/3}$ 
from a study limited to the range $N=200\div1000$.
%%%%%%%%
\\[6pt]
%%%%%%%%
\trivuoto{MyBlue} \ For $\Lambda<-1$, the formation of doublets involves pairs of levels at higher and higher energy, 
each pair containing a symmetric eigenstate (even $n$) and an antisymmetric one (odd $n$).
In the harmonic-oscillator picture, the energy levels are given by 
\be \label{LMGenergygap2}
E_n = \sqrt{\Lambda^2-1}\,\Big(\Big\lfloor\textrm{\footnotesize$\frac{n}{2}$}\Big\rfloor\!+\tfrac{1}{2}\Big) 
\ee
and the spectrum is only composed by twofold degenerate levels, since $\Delta_{n}\to0$ for odd $n$, 
while the repulsion between doublets is finite and increases with $|\Lambda|$: $\Delta_{n}=\sqrt{\Lambda^2-1}$ for even $n$. 
The wave functions are expressed as a superposition -- with definite parity -- of two nonoverlapping harmonic solutions $\Psi$, 
each one localized in one well,
\be
\psi_n(z) = \frac{\Psi_{\lfloor\frac{n}{2}\rfloor}{\textrm{\small$(z-z_{\Lambda}^{-})$}} + (-1)^{\big\lfloor\frac{n+1}{2}\big\rfloor} 
\Psi_{\lfloor\frac{n}{2}\rfloor}{\textrm{\small$(z-z_{\Lambda}^{+})$}}}{\sqrt{2}},
\ee
where $z_{\Lambda}^\pm=\pm\sqrt{1-1/\Lambda^2}$ and the function $\Psi$ is defined in Eq.~(\ref{HarmonicEigenfunctions}) with
\be
\sigma^2\textrm{\small$(\Lambda<-1)$} = \frac{N}{4}\frac{1}{|\Lambda|\sqrt{\Lambda^2-1}}.
\ee
For finite but large $N$, each level is \emph{almost} twofold degenerate: 
the gaps $\Delta_n$ between the doublets (odd $n$) is exponentially small for increasing $N$. 
In particular, $\Delta_1\sim\exp(-N/n_{\Delta_1})$ was predicted by Ref.~\cite{newman1977} 
and observed in Refs.~\cite{BotetPRL1982,BotetPRB1983}; 
here we supply the dependence $n_{\Delta_1}\propto|1+\Lambda|^{-4/3}$.
A simple analysis of the ground state of Hamiltonian (\ref{HamLMG3}) on the Dicke basis for $|\Lambda|\gg1$ 
provides a level anticrossing $\Delta_n=\frac{N-n+1}{N}|\Lambda|$ (even $n$).

\subsection{Continuous quantum phase transition} \label{subsec:LMGcontinuousQPT}
The avoided level crossing around $\Lambda\approx-1$ strongly suggests the presence of 
an abrupt change in the structure of the ground state of the many-body system.
Indeed, in the thermodynamic limit, the \emph{symmetric} LMG model with ferromagnetic coupling ($\Lambda<0$) 
exhibits a second-order QPT when tuning the strength of the spin-spin coupling $\Lambda$ across the critical point $\Lambdac=-1$. 
% from zero to large negative values 
The critical properties of the system Eq.~(\ref{HamLMG3}) are determined by the competing effects 
of the internal interactions and the external transverse field;
intuitively  the QPT occurs when the energy scales associated to the two contributions are equal.
The critical point $\Lambdac$ separates a \emph{disordered} phase ($-1<\Lambda<0$), 
where the spins prefer to align along the transverse field, 
from an \emph{ordered} phase ($\Lambda<-1$), where macroscopic clusters of spins point along a favoured direction, 
selected by a symmetry-breaking mechanism\footnote{\ The $Z_2$ symmetry of the Hamiltonian and its eigenfunctions 
is broken by letting the external longitudinal field $\delta$ to be finite but small ($\delta\sim\Delta_1$) 
and then taking the limit $\delta\rightarrow0^{\pm}$. For more details, see chapter~\ref{ch:Ising}.\\[-9pt]} 
not dissimilar to the one of the Ising model discussed in chapter~\ref{ch:Ising}. 

In the permutational symmetric subspace spanned by the Dicke states $|J,m\rangle$, 
the QPT is well understood in terms of a macroscopic change of the ground-state structure:
as $\Lambda$ varies from zero to large negative values crossing $\Lambdac$ 
the ground-state wave function gradually changes from a coherent spin state ($\Lambda=0$),
or more in general a unimodal Gaussian phase-squeezed distribution ($-1<\Lambda<0$),
to a bimodal non-Gaussian distribution ($\Lambda<-1$), superposition of two Gaussian wave functions,
approaching the ideal NOON profile~\cite{LeeJMO2002} in the limit $\Lambda\rightarrow-\infty$.
This transition has been experimentally observed in a spinor degenerate gas of ultracold $^{87}$Rb atoms~\cite{ZiboldPRL2010} and 
in a dilute degenerate gas of attractive ultracold $^{39}$K atoms~\cite{TrenkwalderNATPHYS2016}. 
The latter experiment, in particular, faces the issue of the stability of the condensate affecting strongly-attractive bosons: 
low atom density prevents the collapse of the BEC.

%%%%%%%%%%%%
%% FIGURA %%
%%%%%%%%%%%%
\begin{figure}[t!]
\centering
\hfill
\includegraphics[width=0.28\textwidth]{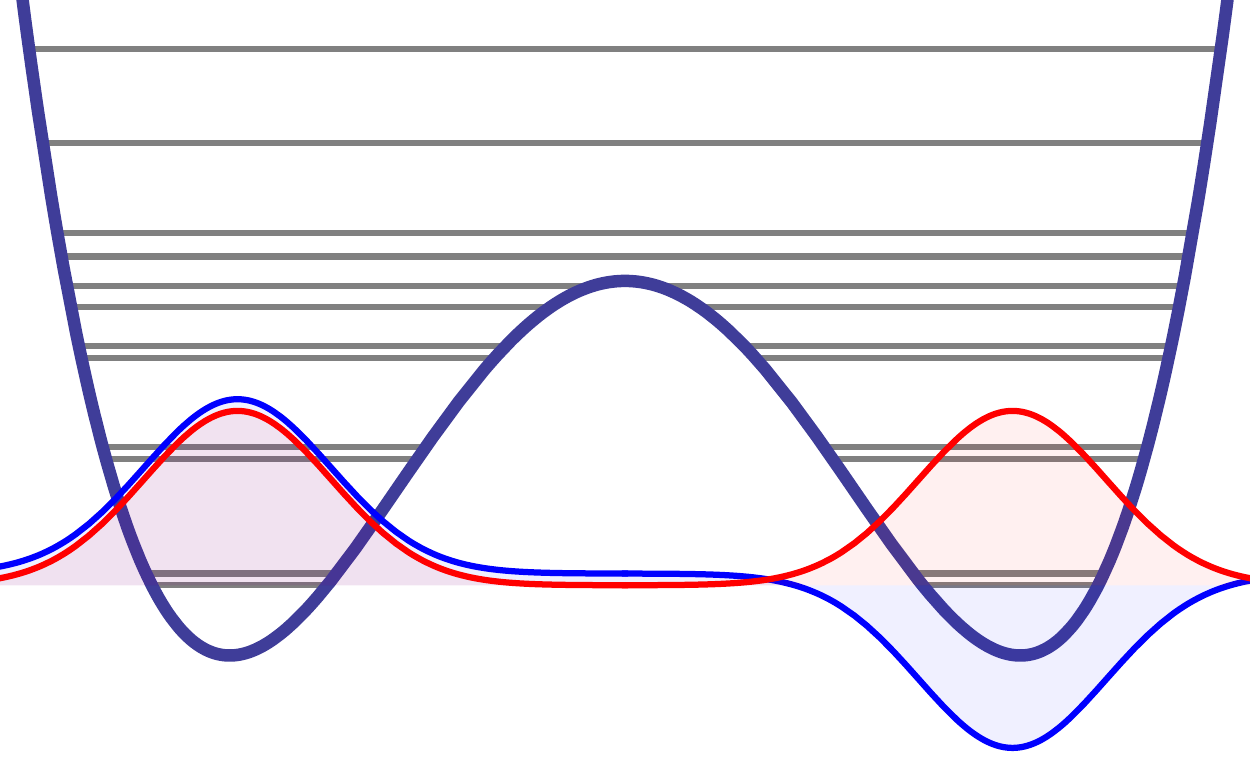} \hfill 
\includegraphics[width=0.28\textwidth]{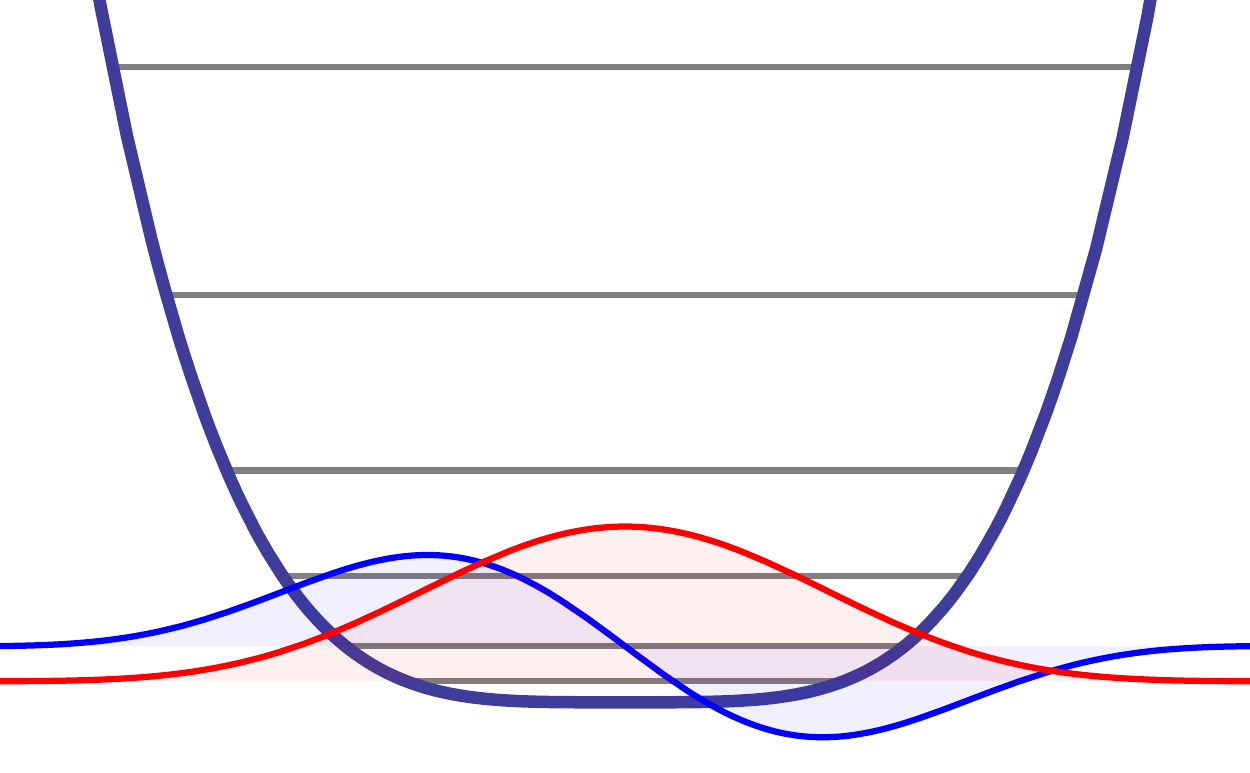} \hfill
\includegraphics[width=0.28\textwidth]{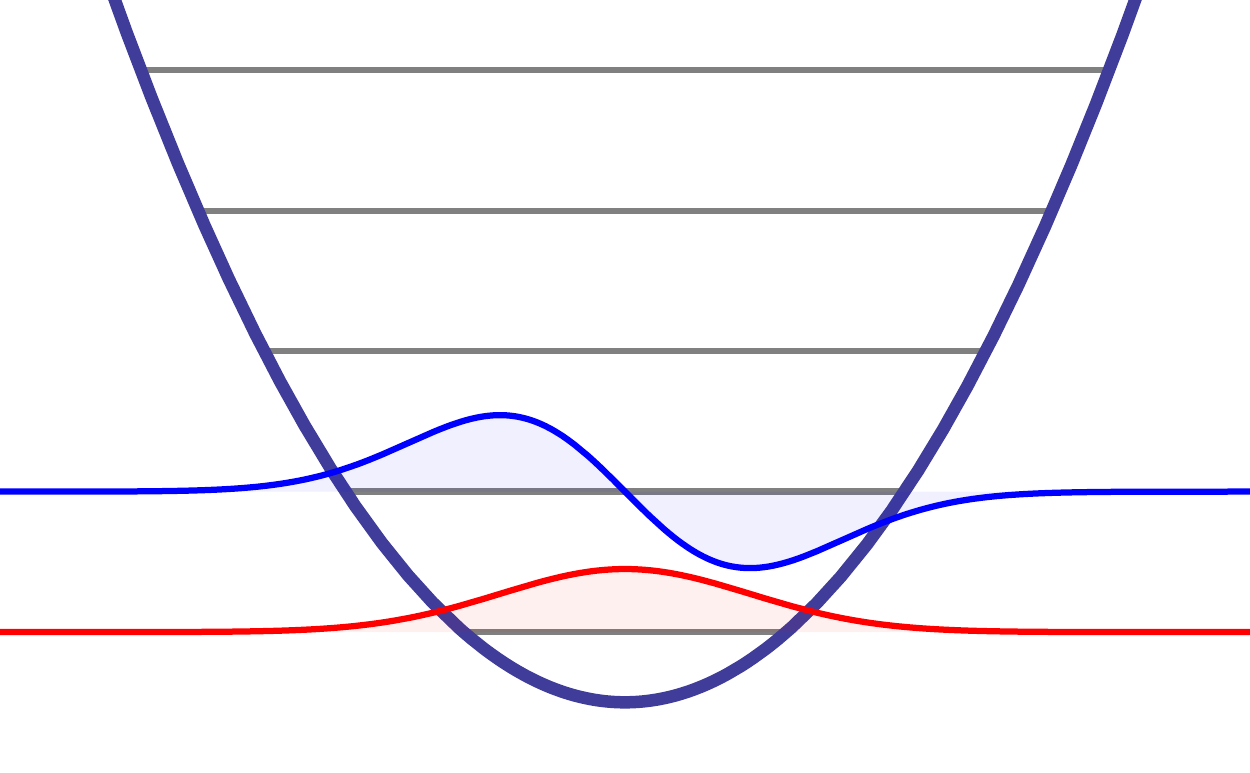} \hfill \\
\textcolor{white}{.} \hfill \quad \ \ \ \fcolorbox{MyGreen}{MyLightGreen}{\textcolor{MyDarkGreen}{$\Lambda<-1$}} \hfill \textcolor{white}{.} \hfill \quad \quad \ \ \ \fcolorbox{MyGreen}{MyLightGreen}{\textcolor{MyDarkGreen}{$\Lambda=-1$}} \hfill \textcolor{white}{.} \hfill \quad \quad \quad \fcolorbox{MyGreen}{MyLightGreen}{\textcolor{MyDarkGreen}{$\Lambda>-1$}} \hfill \textcolor{white}{.}
\caption{Description of the second-order quantum phase transition in the balanced LMG model 
in terms of deformation of the mean-field Ginzburg-Landau potential,
as discussed in the main text. Nestled in the potential profile, the qualitative distribution of the energy levels (gray) as well as
the amplitude of the single-particle symmetric ground state (red) and antisymmetric first-excited state (blue).}
\label{fig:LMGdeformationQPT}
\end{figure}

\paragraph{Semiclassical picture}
For $N\gg1$, the QPT can be visualized by way of the deformation of the effective potential $V(z)$
from a single harmonic well with a global minimum at $z=0$ ($-1<\Lambda\leq0$) 
to a double harmonic well with two minima at $z=z_\Lambda^\pm$ ($\Lambda<-1$), 
passing through a quartic profile ($\Lambda=-1$). Figure~\ref{fig:LMGdeformationQPT} depicts the transition.
The critical region is characterized by mean-field critical exponents~\cite{BotetPRL1982,BotetPRB1983,DusuelPRL2004,DusuelPRB2005,BuonsantePRA2005,HaukeNATPHYS2016}.
% \think{in particular, the dynamical exponent and scaling dimensions are computed analytically}~\cite{BotetPRB1983}.
Table~\ref{tab:exponents} compares some of these exponents to the ones for the Ising chain of chapter~\ref{ch:Ising}.

\begin{table}[!b] 
\begin{center}
\begin{tabular}{cllcc}
\toprule 
\textbf{Exponent} & \textbf{Quantity} & \textbf{Definition} & \textbf{1D Ising} & \textbf{LMG} \\[3pt]
\hline\hline \\
$\beta$ & order parameter $\varphi=\langle\hat{\sigma}_z\rangle$ & $\varphi\textrm{\,\small$(\delta=0)$} \propto (\lambda-\lambda_{\rm c})^\beta $ & 1/8 & 1/2 \\[6pt]
$\gamma$ & susceptibility $\chi=\ud\varphi/\ud\delta$ & $\chi\textrm{\,\small$(\delta=0)$}\propto|\lambda-\lambda_{\rm c}|^{-\gamma}$ & 7/4 & 1 \\[6pt]
$\nu$ & deviation from critical point $\lambda_{\rm c}$ & $|\lambda_{\rm c}^{\textrm{\tiny$(N)$}}-\lambda_{\rm c}|\propto N^{-1/(\nu d)}$ & 1 & 3/2 \\[6pt]
$z$ & energy gap $\DeltaE=E_2-E_1$ & $\DeltaE\textrm{\small$(\lambda=\lambda_{\rm c},\,\delta=0)$}\propto N^{-z/d}$ & 1 & 1/3 \\[12pt]
\bottomrule
\end{tabular}
\caption{Short list of critical exponents associated with the second-order quantum phase transition occurring in typical $d$-dimensional spin systems
modelled by $\hat{H}=-\lambda\sum_{i,j}J_{ij}\hat{s}_z^{\textrm{\tiny$(i)$}}\hat{s}_z^{\textrm{\tiny$(j)$}} - \sum_i\hat{s}_x^{\textrm{\tiny$(i)$}} - \delta \sum_i\hat{s}_z^{(i)}$, 
where $\hat{s}_\varrho^{\textrm{\tiny$(i)$}}$ is the $i$th spin operator ($i=1,\dots N$) along the $\varrho$ direction, 
$\lambda>0$ is the control parameter driving the transition,
$\delta$ is the longitudinal source field. 
The definitions of $\beta$ and $\gamma$ are given in the thermodynamic limit; instead, the definitions of $\nu$ and $z$ are here based on finite-size behaviours.
The actual critical point is denoted as $\lambda_{\rm c}$, 
whereas $\lambda_{\rm c}^{\textrm{\tiny$(N)$}}$ indicates the position of the emergent signatures of the transition. 
When $d=1$ and $J_{ij}=1$, the model reduces to the ferromagnetic nearest-neighbour Ising chain in a transverse field. 
When $d=1$, $J_{ij}=N^{-1}$ and $\sum_{\varrho,i,j}\hat{s}_\varrho^{\textrm{\tiny$(i)$}}\hat{s}_\varrho^{\textrm{\tiny$(j)$}}=\frac{N}{2}(\frac{N}{2}+1)\identity$, 
the model becomes the fully-symmetric Lipkin-Meshkov-Glick model.
In both cases, the critical point is $\lambda_{\rm c}=1$. 
The two models evidently belongs to distinct universality classes: 
the range of the interaction is different.} 
\label{tab:exponents}
\end{center} 
\end{table}

It is known that, in the limit of infinite population, the second-order QPT at $\Lambda=-1$ corresponds to a pitchfork \emph{bifurcation} 
in the energy landscape of the system. At the semiclassical level, for high fields there is only one stable fixed point, 
namely a unique configuration minimizing the semiclassical energy; for fields below a critical threshold, 
the stable fixed point bifurcates into two symmetric stable points and the original stable point becomes unstable. 
In the full quantum counterpart, the threshold corresponds to the critical point 
and the two degenerate minima of the energy correspond to the superposition of the two degenerate eigenstates.
% that the spontaneous symmetry-breaking mechanism breaks into a non-symmetric configurations
Classical bifurcations occur whenever a smooth small change of some driving parameter in a classical dynamical system 
leads to a sudden qualitative change in its behaviour. They frequently appear in relation to chaos \cite{Gutzwiller1991} 
and their manifestations are encountered also in quantum systems within the limit they can be described by a classical theory.
In principle, increasing the number of spins $N$ permits to observe the crossover between 
the signatures of the ground-state transition in the genuinely quantum LMG model 
to the emergence of the dynamical bifurcation in the semiclassical limit.
A discussion of the relation between the QPT and the bifurcation in the LMG model is provided in Refs.~\cite{BuonsantePRA2005,gessner2016}.
A detailed theoretical and experimental analysis of the trajectories in the classical phase space is presented, respectively, 
in Refs.~\cite{CastanosPRB2006,ZiboldPRL2010}.

%%%%%%%%%%%%
%% FIGURA %%
%%%%%%%%%%%%
\begin{figure}[t!]
\centering
\includegraphics[width=0.496\textwidth]{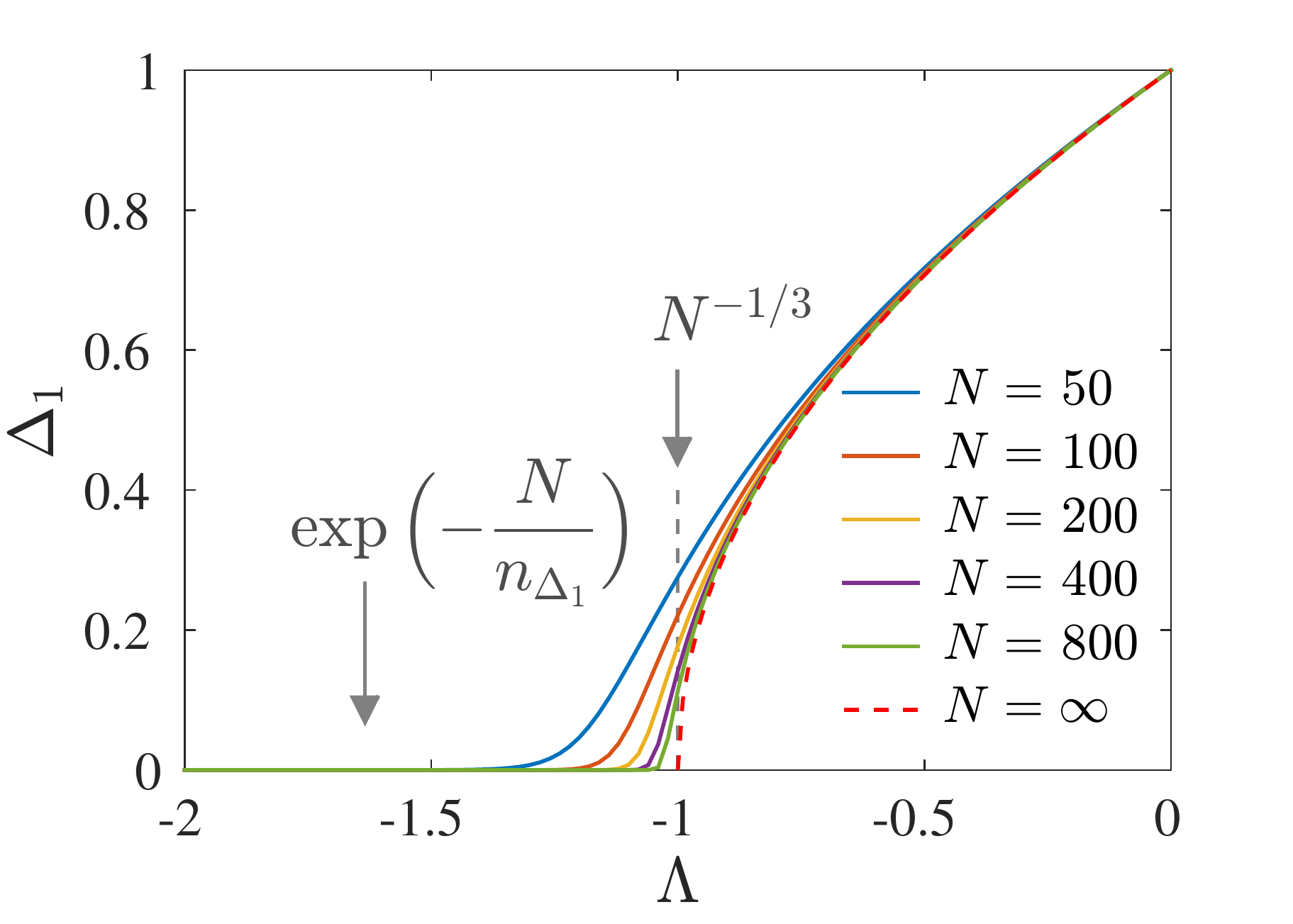} 
\includegraphics[width=0.496\textwidth]{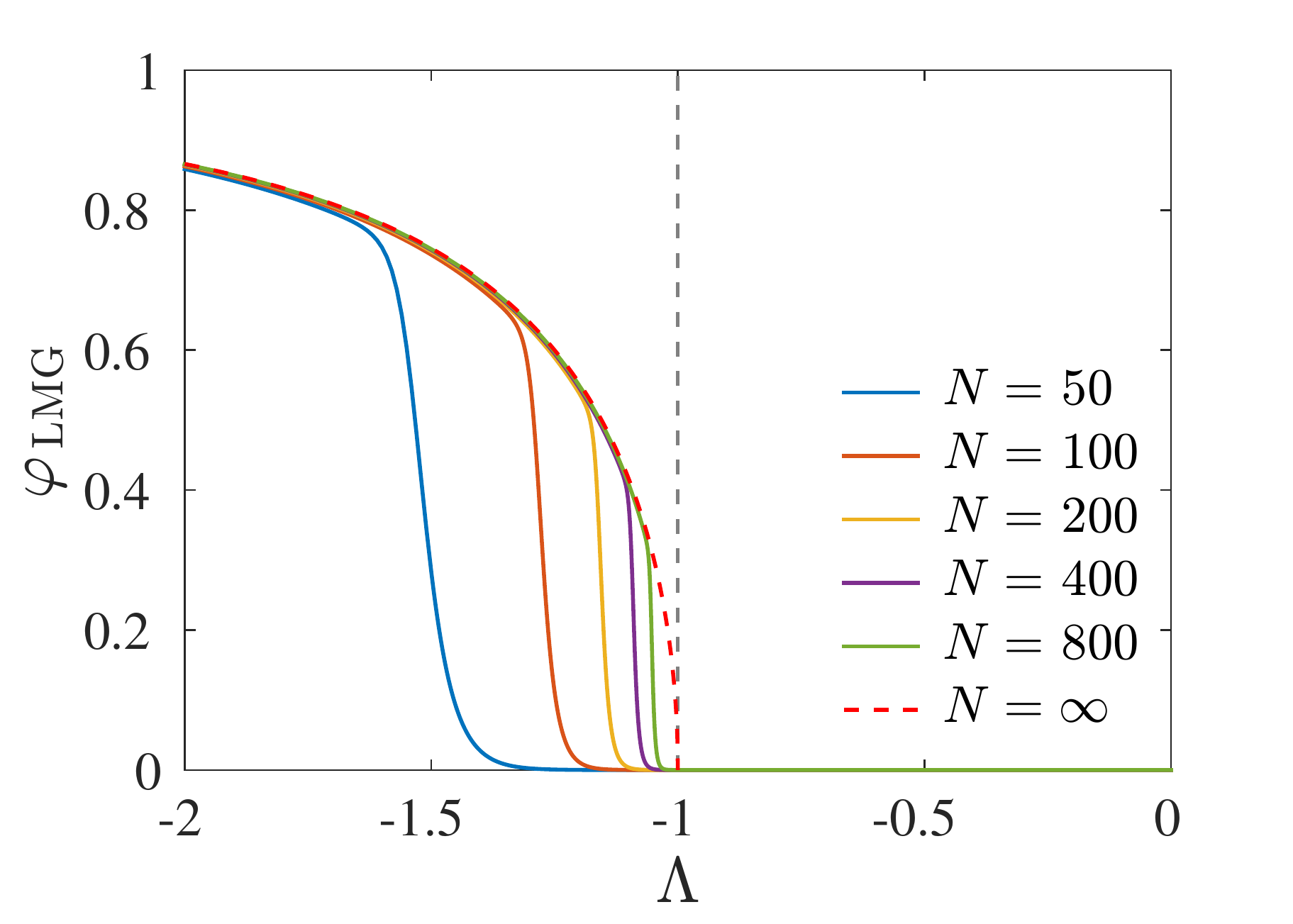} \\
\includegraphics[width=0.34\textwidth]{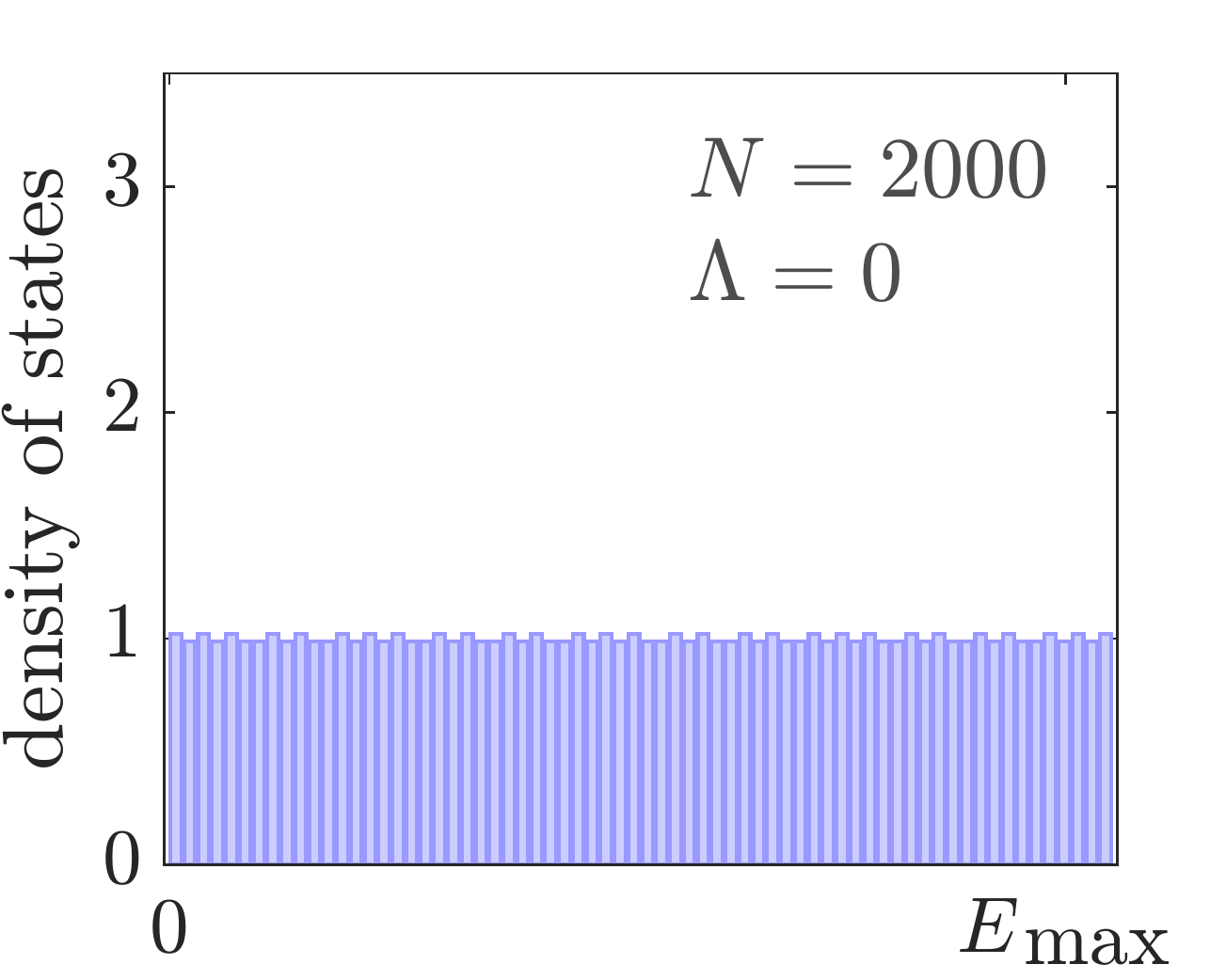} \hspace{-12pt}
\includegraphics[width=0.34\textwidth]{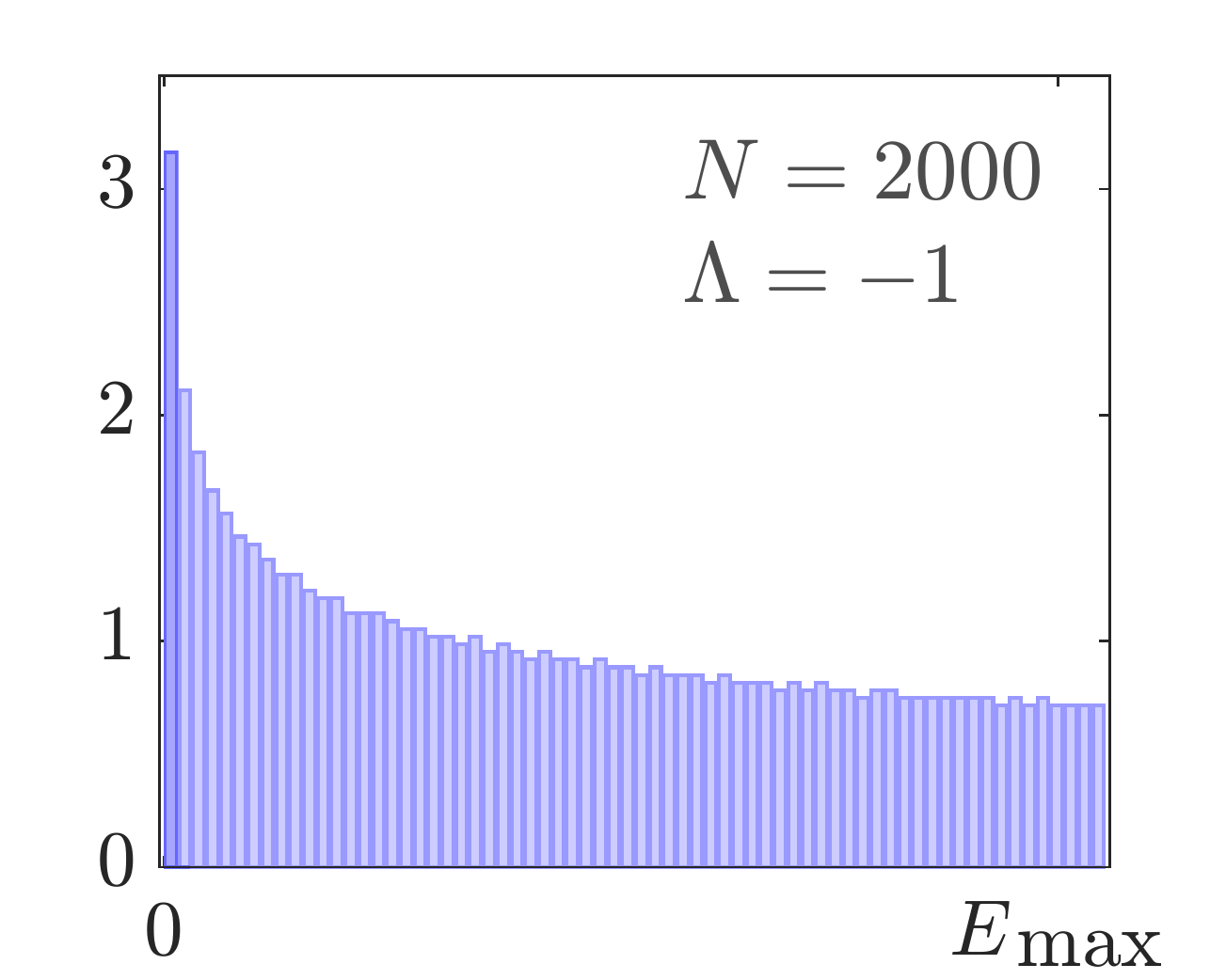} \hspace{-12pt}
\includegraphics[width=0.34\textwidth]{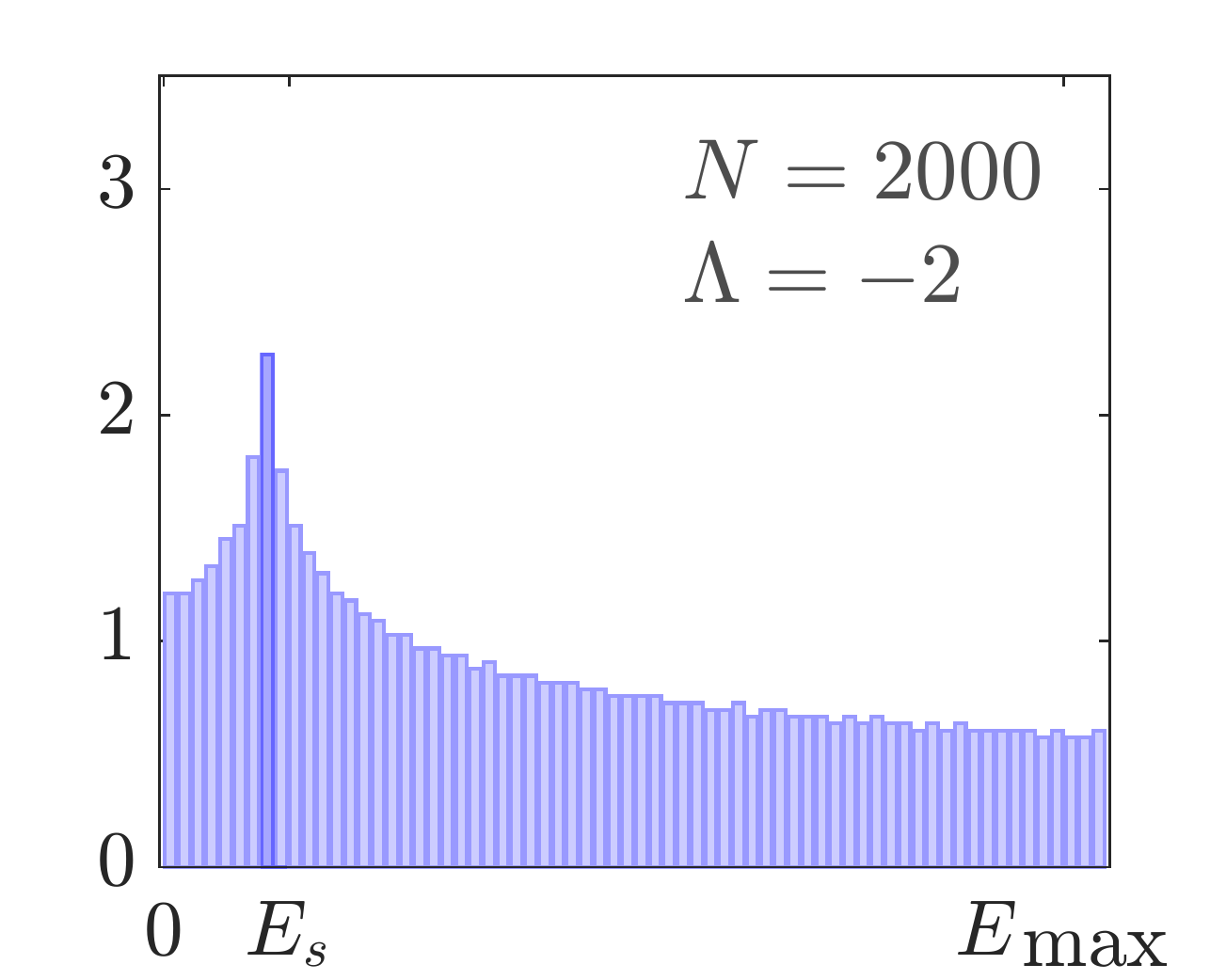} \\
\caption{Signatures of the second-order quantum phase transition in the balanced LMG model. 
\textbf{(Top left)} Energy gap between the ground state and first excited state as a function of the control parameter $\Lambda$
for several values of $N$ (numerical data, solid lines) and in the limit $N\to\infty$ 
(analytical curve $\Delta_1=\sqrt{1+\Lambda}$, dashed line).
The scaling for increasing $N$ at criticality ($\Lambda=-1$) and in the ordered phase ($\Lambda<-1$) are shown.
\textbf{(Top right)} Order parameter $\varphi_{\rm LMG}=2\langle\hat{J}_z\rangle/N$ as a function of $\Lambda$.
Here the symmetry is broken by a small longitudinal field $\delta=-10^{-6}$ in units of $\Omega$.
The dashed line is the analytical curve $\varphi_{\rm LMG}=\sqrt{1-1/\Lambda^2}$.
\textbf{(Bottom)} Density of states for $N=2000$ spins in the supercritical ($\Lambda>-1$), critical ($\Lambda=-1$)
and subcritical ($\Lambda<-1$) regime. For $\Lambda\leq-1$ the peak centred at the saddle energy $E_s$, 
i.e.~the energy level corresponding to the height of the barrier, is highlighted. 
The peak becomes a divergence in the limit $N\to\infty$.}
\label{fig:signaturesLMG}
\end{figure}

\paragraph{Signatures of the QPT}
The order parameter that recognizes the QPT is the total longitudinal magnetization: $\varphi_{\rm LMG}=2\langle\hat{J}_z\rangle/N$,
i.e. the population imbalance between the ``up'' spins and ``down'' spins. 
It continuously increases from zero (in the symmetric phase and at criticality) to a finite value (in the broken-symmetry phase): 
in the semiclassical limit, we find
\be
\varphi_{\rm LMG} = \langle z \rangle = \left\{
\begin{array}{ll} 
0 & {\rm for} \ \Lambda>-1 \\ 
\pm\sqrt{1-\dfrac{1}{\Lambda^2}} & {\rm for} \ \Lambda<-1
\end{array}
\right. \, , 
\ee
that implies a critical exponent $\beta=\frac{1}{2}$. 
Size corrections on the order parameter are reported in Fig.~\ref{fig:signaturesLMG}.
In particular, we highlight that $\Lambdac - \arg\max_\Lambda|\partial_\Lambda \varphi_{\rm LMG}| \sim N^{-2/3}$.

Moreover, as previously hinted, at $\Lambda=-1$ the first energy gap $\Delta_1$ vanishes 
and the density of states $\rho=\delta\pazocal{N}/\delta{E}$ diverges at low energies because of the bunching of the levels.
The behaviour of $\rho$ for the LMG model was previously studied within the catastrophe formalism~\cite{GilmorePRA1986}
and by means of a rigorous perturbative expansion in powers of $N^{-1}$~\cite{LeyvrazPRL2005,RibeiroPRL2007}.
These theoretical studies and our numerical simulations reveal that the density of states becomes discontinuous in the ordered phase.
In the thermodynamic limit, $\rho$ is uniform for $\Lambda>-1$ and it is predicted to diverge at $E_s$ for $\Lambda\leq-1$, 
being $E_s$ the saddle energy associated to the unstable point of the semiclassical potential $V(z)$, 
namely the height of the barrier separating the two degenerate minima of the energy.
At finite population, these features are still present even if the discontinuity is mitigated into a smooth behaviour: 
for $\Lambda>-1$, the density of states displays a slow monotonic decrease as $\sim E^{-1/4}$ for large energies, 
induced by the nonuniform spacing in the highest-part of the spectrum; 
for $\Lambda\leq-1$ the divergence is rounded off to a sharp peak. 
Figure~\ref{fig:signaturesLMG} collects all these spectral signatures of the QPT.

\begin{figure}[t!]
\centering
\includegraphics[height=5cm]{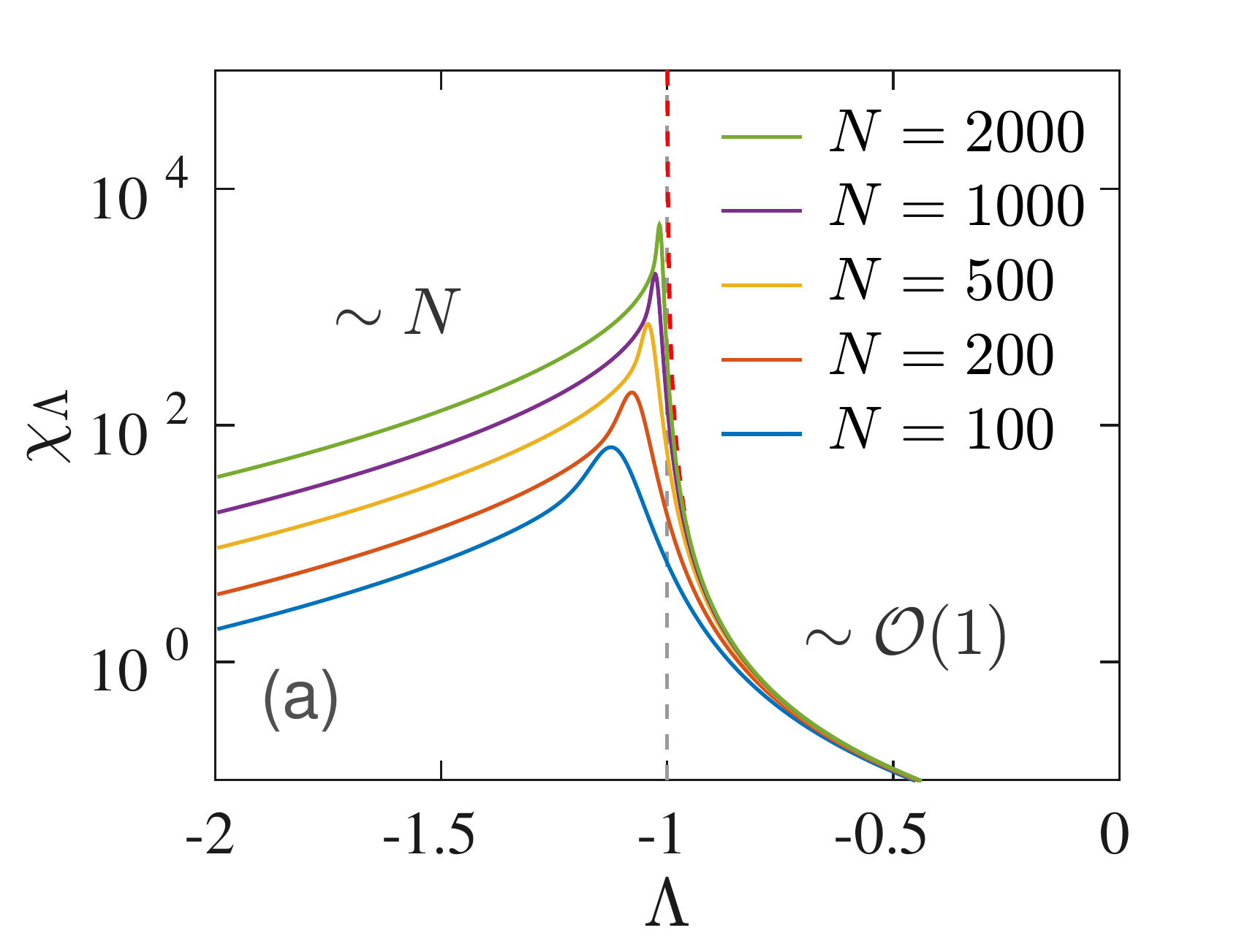} \hspace{-15pt}
\includegraphics[height=5cm]{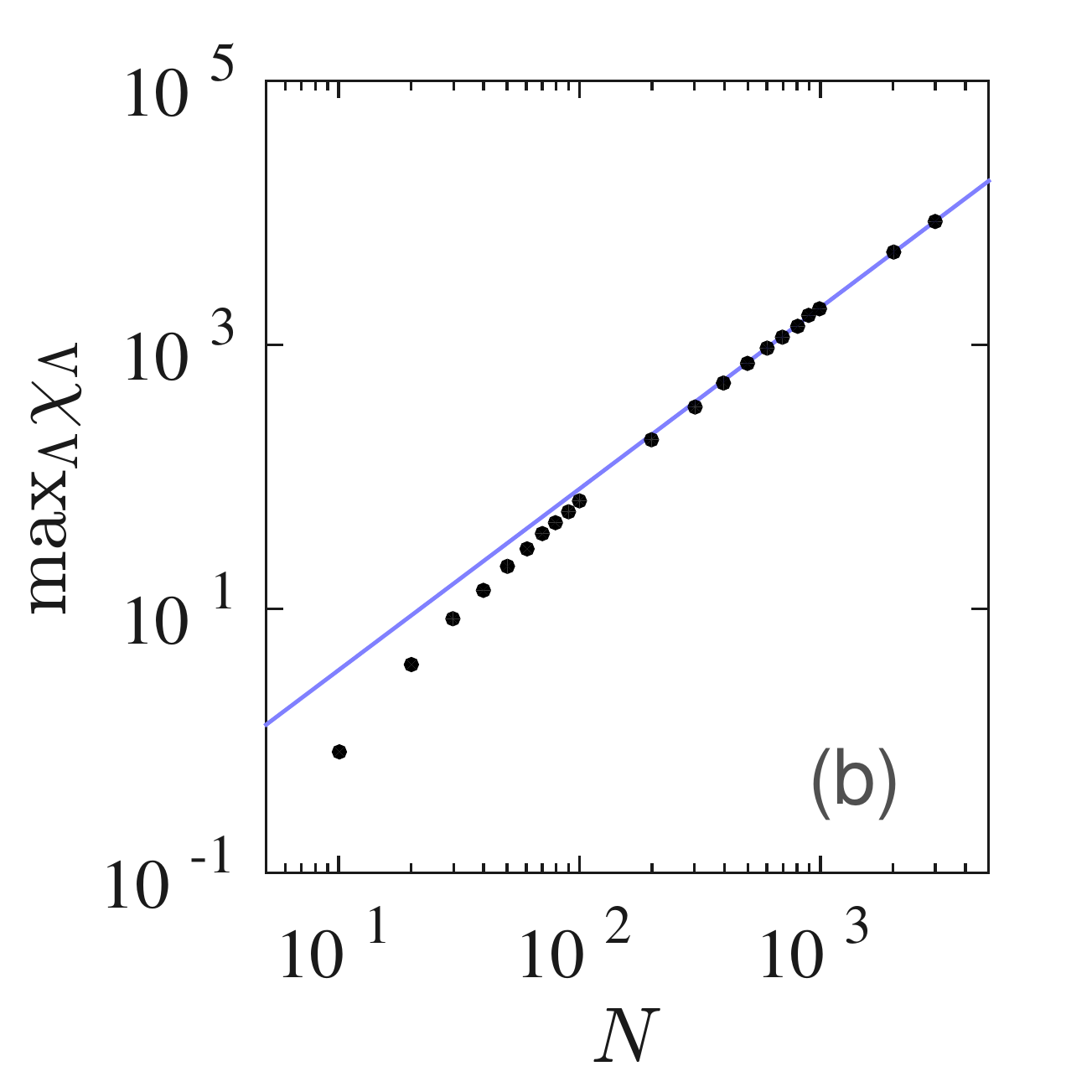} \hspace{-15pt}
\includegraphics[height=5cm]{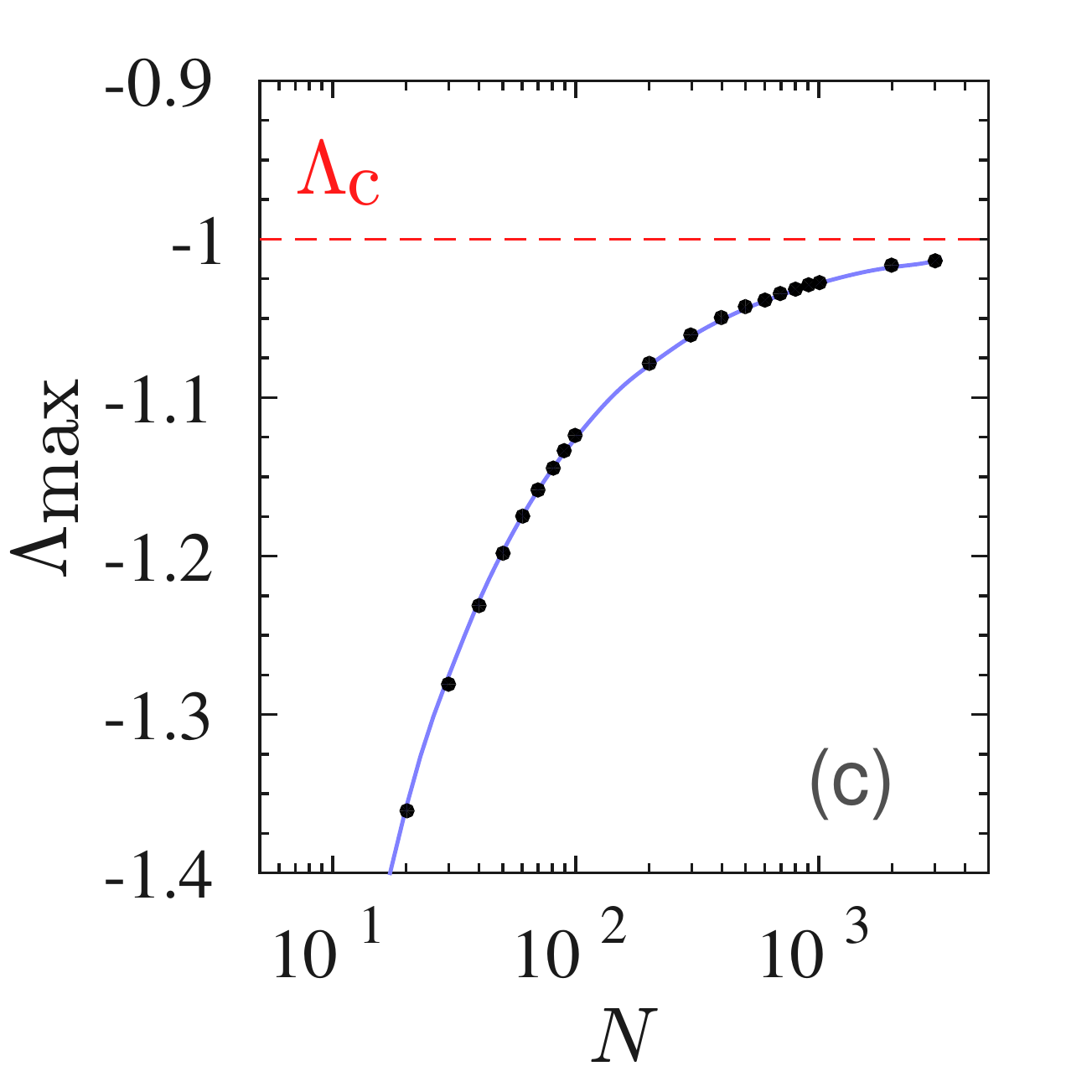}
\caption{Fidelity susceptibility $\chi_\Lambda$ for the balanced LMG model.
\textbf{(a)} Dependence of $\chi_\Lambda$ on the parameter $\Lambda$ driving the second-order phase transition 
for increasing population $N$. 
Numerical results (solid lines) agree with the analytical curve Eq.~(\ref{LMGfidelity}) 
(red dashed line, shown in the disordered region $\Lambda>-1$ only). 
The extensive (intensive) behaviour in the ordered (disordered) phase is emphasized.
\textbf{(b)} Superextensive scaling of the peak of $\chi_\Lambda$ (dots). The line is a fit to the data: $\chi_\Lambda \sim N^{4/3}$.
\textbf{(c)} Finite-size corrections to the position of the peak $\Lambda_{\rm max}=\arg\max_\Lambda\chi_\Lambda$ (dots)
with respect to the asymptotic critical value $\Lambdac=-1$ (red dashed line).
The line is a fit to the data: $\Lambdac-\Lambda_{\rm max}\sim N^{-2/3}$.}
\label{fig:FidelityLMG}
\end{figure}

\paragraph{Fidelity susceptibility}
Clearly, we expect that also the fidelity susceptibility -- quantifying the response of the ground-state structure
to small variations of the driving parameter $\Lambda$ -- provides a visible mark of the QPT. 
The critical properties of the LMG model have been extensively studied in terms of the fidelity susceptibility~\cite{Kwok2008,Gu2010,Buonsante2012}.
An analytical calculation at first order in $N^{-1}$ yields
\be \label{LMGfidelity}
\chi_\Lambda \!\! \stackrel{\ \small N\,\gg\,1}{^{\textcolor{white}{o}}\approx^{\textcolor{white}{o}}} \! 
\left\{
\begin{array}{ll} 
\dfrac{1}{32}\dfrac{1}{(1+\Lambda)^2} & {\rm for} \ -1<\Lambda<0 \\[12pt]
\dfrac{N}{4} \dfrac{1}{|\Lambda|^3\sqrt{\Lambda^2-1}}  & {\rm for} \ \Lambda<-1
\end{array}
\right.
\ee
indicating that the susceptibility diverges around the critical point in the thermodynamic limit: see Fig.~\ref{fig:FidelityLMG}\Panel{a}. 
The behaviour at criticality cannot be captured by any semiclassical method, hence numerical investigation is needed:
we find $\chi_\Lambda\textrm{\small$(\Lambda=\Lambdac)$} \sim N^{4/3}$ 
as well as $\max_\Lambda\chi_\Lambda \sim N^{4/3}$ and $\Lambdac - \arg\max_\Lambda\chi_\Lambda \sim N^{-2/3}$.
Panels~\ref{fig:FidelityLMG}\Panel{b,\,c} make these scalings clear.

Moreover, $\chi_\Lambda$ is found to be intensive $\chi_\Lambda\sim\pazocal{O}(1)$ in the gapped phase
and extensive $\chi_\Lambda \sim N$ in the gapless phase. 
This asymmetric behaviour is prominently different from the one flaunted by the long-range Ising model of chapters~\ref{ch:Ising}
and the one we will encounter in the long-range Kitaev model of chapter~\ref{ch:Kitaev}, 
where the susceptibility grows extensively on both sides of the critical point.
However, the common link among all the models is that the superextensive peak at criticality emerges from an (at most) extensive background: 
this is interpreted as the actual smoking gun of the QPT.
% An explicit symmetry-breaking term shall remove the degeneracy of the ground state and lead to a sharp peak 

%%%%%%%%%%%%
%% FIGURA %%
%%%%%%%%%%%%

\paragraph{Bipartite entanglement}
Bipartite entanglement -- especially at null temperature -- have already been studied in the LMG model 
using a plethora of entanglement measures or indicators, like one-tangle~\cite{vidal2004}, 
concurrence~\cite{vidal2004, DusuelPRL2004, DusuelPRB2005}, entanglement entropy \cite{barthel2006}, 
global geometric entanglement \cite{orus2008}, single-copy entanglement \cite{orus2008}, 
logarithmic negativity \cite{wichterich2010} and mutual information~\cite{WilmsJSM2012}.
All these works have emphasized a maximal entanglement close to the quantum critical point. 

Both the one-tangle, that measures the entanglement of a selected spin with all the others, 
and the concurrence, that measures pairwise quantum correlations between two selected spins, 
exhibit an apparent discontinuity in their derivative with respect to the control parameter~\cite{vidal2004,DusuelPRB2005}: 
the derivative can be seen as a proper response function for the detection of the critical point.
Nevertheless, these two measures are only partially reliable for quantifying entanglement in the ordered phase~\cite{vidal2004}.
All the other aforementioned quantities diverge logarithmically in the spin population $N$ at the critical point,
but they still capture only the entanglement between two selected spins or two partitions of the system
and they are blind to the richness of many-body quantum correlations.

\subsection{Multipartite entanglement in the ground state}
Multipartite entanglement was studied by using the quantum Fisher information 
both in the LMG model at zero temperature~\cite{MaPRA2009} and the dissipative LMG model~\cite{huang2016}. 
So far, the analysis has been limited to the ferromagnetic regime ($\Lambda<0$), the one hosting the second-order QPT.

\begin{figure}[t!]
\centering
\includegraphics[height=4.5cm]{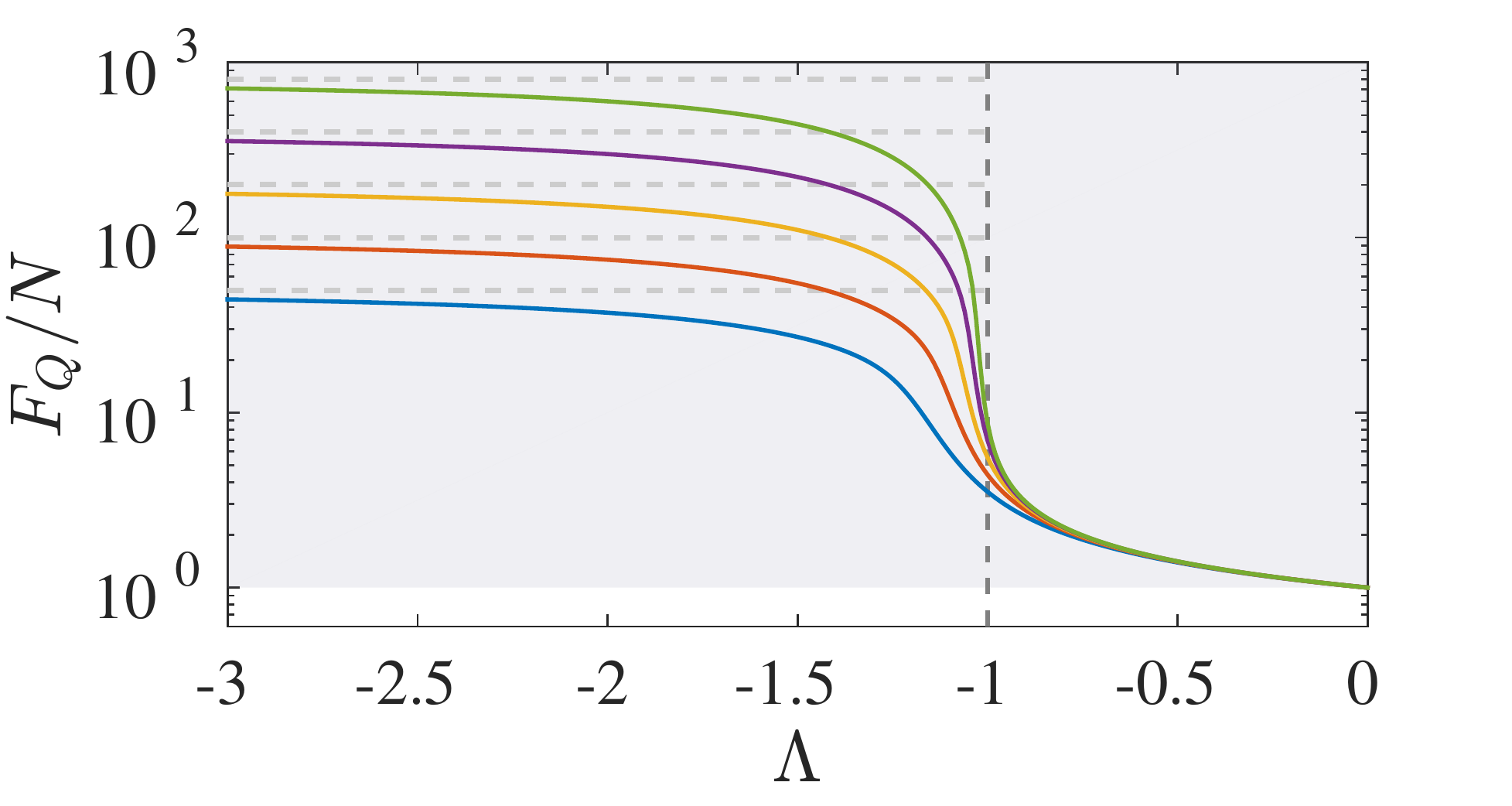} \hspace{-15pt}
\includegraphics[height=4.5cm]{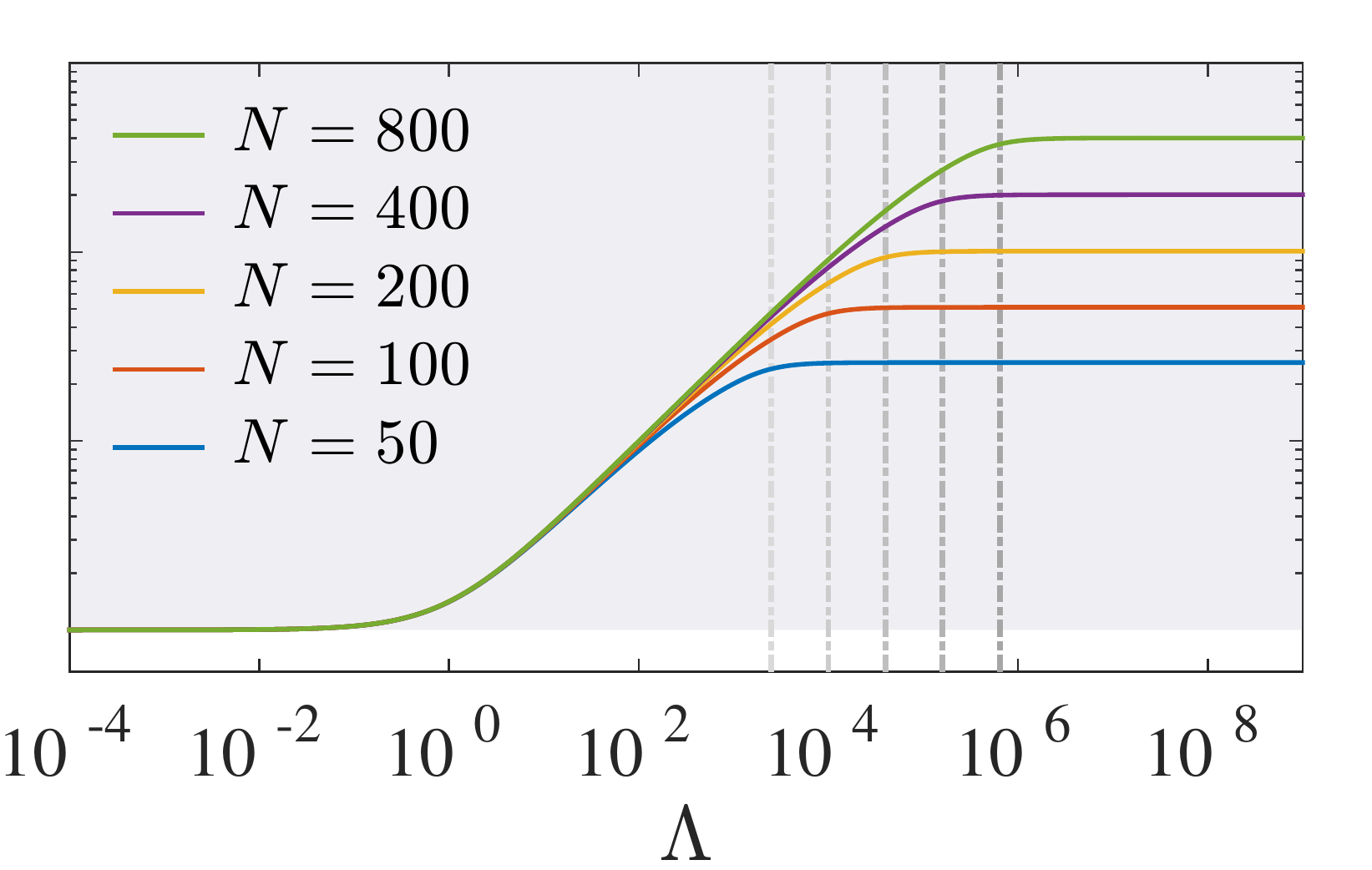} \\
\includegraphics[width=0.15\textwidth]{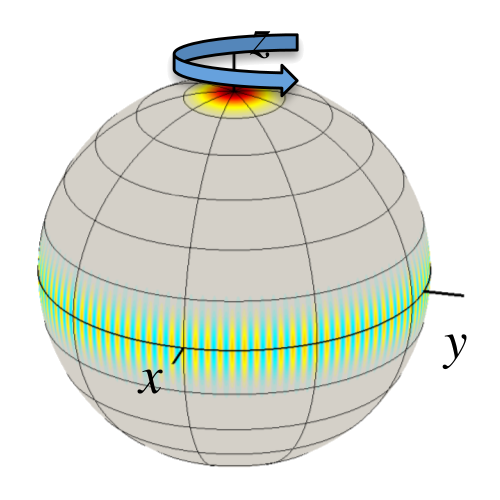}
\includegraphics[width=0.15\textwidth]{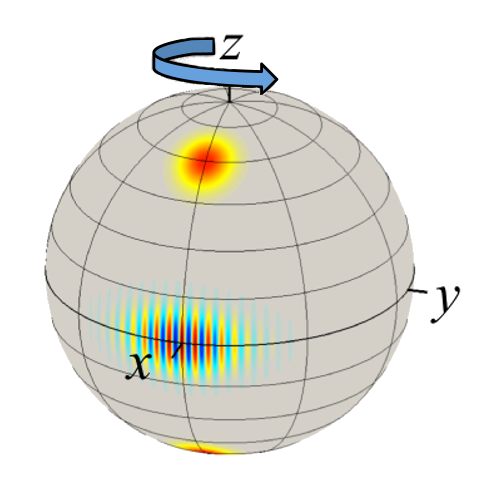}
\includegraphics[width=0.15\textwidth]{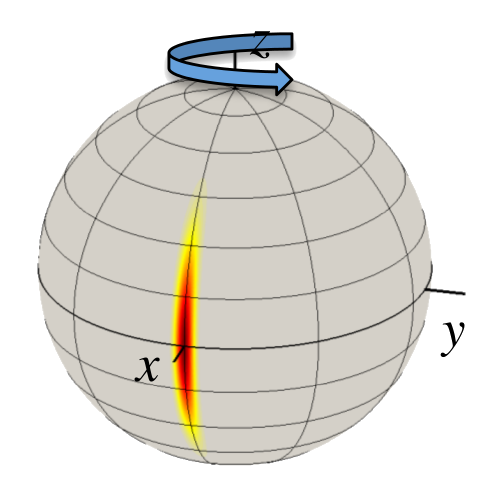}
\includegraphics[width=0.15\textwidth]{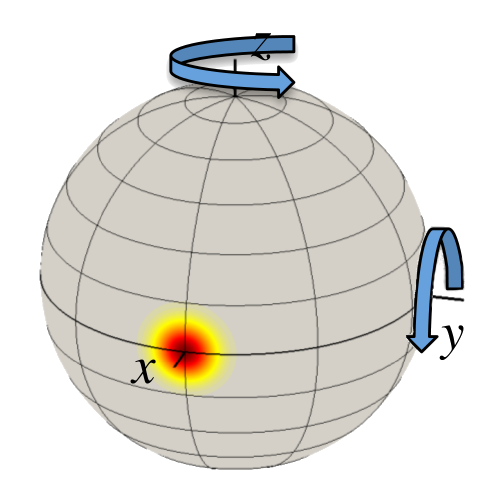}
\includegraphics[width=0.15\textwidth]{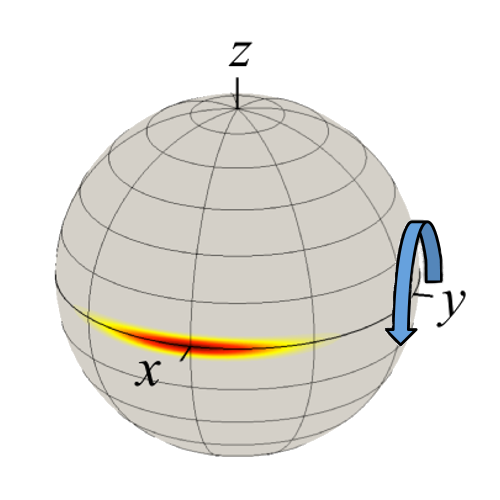}
\includegraphics[width=0.15\textwidth]{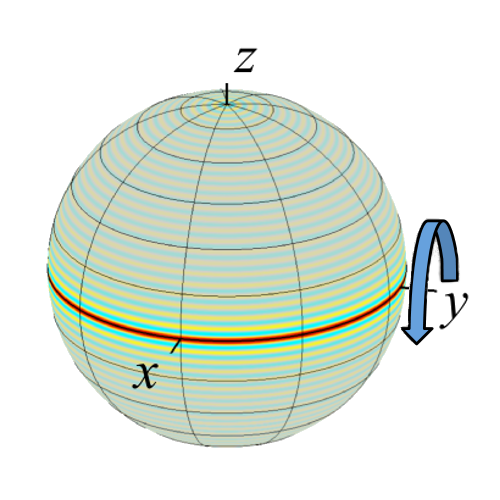} \\
\begin{small}
\textcolor{white}{.} \hspace{13pt} \fcolorbox{MyGreen}{MyLightGreen}{\textcolor{MyDarkGreen}{$\Lambda\to-\infty$}} \hfill 
\textcolor{white}{.} \hspace{-6pt} \fcolorbox{MyGreen}{MyLightGreen}{\textcolor{MyDarkGreen}{$\Lambda=-2$}} \hfill 
\textcolor{white}{.}  \hspace{-3pt} \fcolorbox{MyGreen}{MyLightGreen}{\textcolor{MyDarkGreen}{$\Lambda=-0.5$}} \hfill
\textcolor{white}{.}  \hspace{-3pt} \fcolorbox{MyGreen}{MyLightGreen}{\textcolor{MyDarkGreen}{$\Lambda=0$}} \hfill
\textcolor{white}{.}  \hspace{3pt} \fcolorbox{MyGreen}{MyLightGreen}{\textcolor{MyDarkGreen}{$\Lambda=N$}} \hfill
\textcolor{white}{.}  \hspace{-0pt} \fcolorbox{MyGreen}{MyLightGreen}{\textcolor{MyDarkGreen}{$\Lambda\gg N$}} \hfill 
\textcolor{white}{.}
\end{small}
\caption{\textbf{(Top)} Optimal Fisher density for the ground state of the balanced LMG model 
as a function of the interaction parameter with different population numbers. 
On the ferromagnetic side (left panel), the vertical dashed line indicates the critical point $\Lambda=-1$
and the horizontal dashed lines mark the asymptotic values $f_Q=N$.
On the antiferromagnetic side (right panel), the vertical dashed lines at $\Lambda=N^2$ signal the conventional location of the
crossover between the low-interaction and high-interaction regimes.
In both panels, the shaded area highlights the parameter region where multipartite entanglement is witnessed.
\textbf{(Bottom)} Wigner quasi-probability distribution (normalized to its maximum value) of the ground state 
plotted on the generalized Bloch sphere for some representative values of $\Lambda$.
The squeezed nature of the ground state for $-1<\Lambda<0$ and $0<\Lambda\lesssim N$ is crystal clear, 
along with the non-Gaussianity of the ground state 
for strong enough ferromagnetic $\Lambda<-1$ and antiferromagnetic $\Lambda\gg N$ interaction.
The blue arrow indicates the direction of the collective rotation that takes advantage of the structure of the ground state 
and makes it maximally distinguishable even for small angles; 
to wit, the optimal interferometric transformation used to evaluate the Fisher information shown in the top panels.
Image adapted from Ref.~\cite{PezzeRMP}.}
\label{fig:QFIT0LMG}
\end{figure}

\begin{figure}[t!]
\centering
\includegraphics[height=4.8cm]{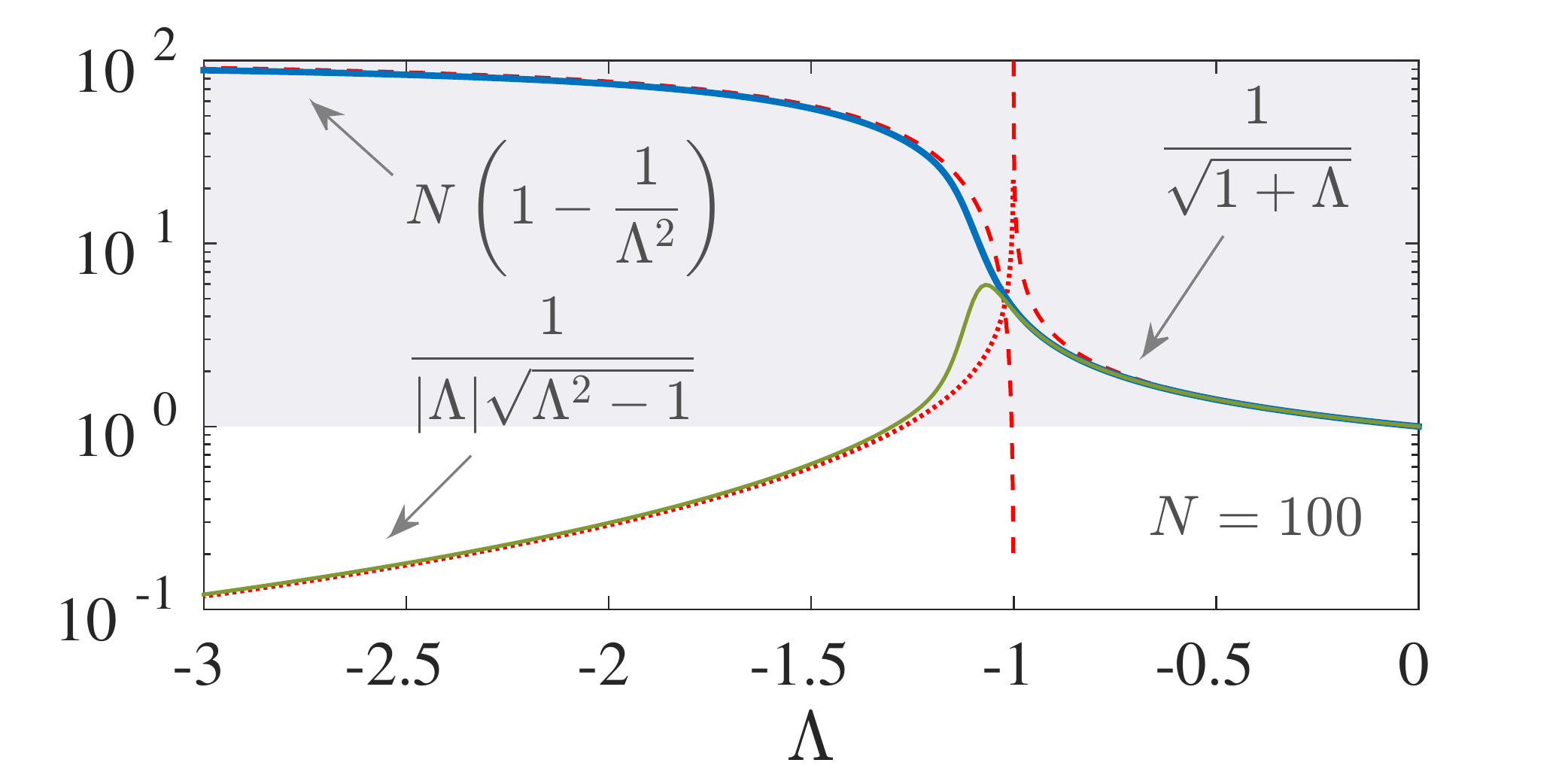} \hspace{-24pt}
\includegraphics[height=4.8cm]{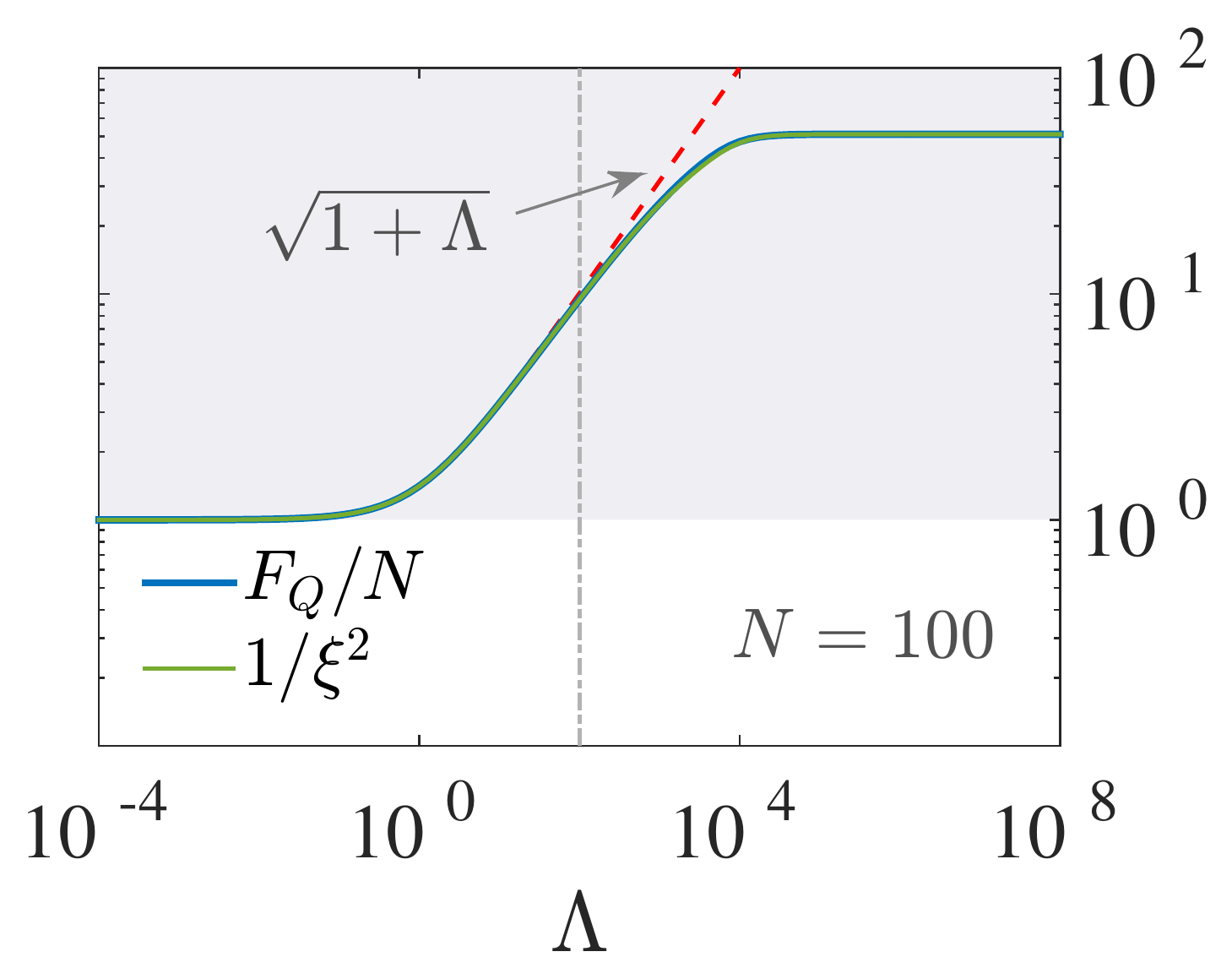} \\
\caption{Optimal Fisher density (solid blue line) and Wineland spin squeezing (solid green line) 
of the ground state of the balanced LMG model as a function of the interaction parameter $\Lambda$ for $N=100$ spins.
The analytical curves obtained in the semiclassical harmonic fremework -- exact in the thermodynamic limit -- 
for the Fisher density (spin squeezing) are superimposed as red dashed (dotted) lines.}
\label{fig:QFIeWSST0LMG}
\end{figure}

In the following we reproduce the main results by numerical means. 
In addition, we provide a complete analytical characterization of the quantum Fisher information 
according to the simple semiclassical method described in paragraph~\ref{subsec:semiclassicalLMG} 
and compare the results to the behaviour of the spin squeezing, another indicator of multipartite entanglement with relevance 
for metrological applications.
We extend the investigation to the antiferromagnetic region, where we identify a smooth crossover in the ground-state structure 
that is actually captured by the quantum Fisher information. 
All the results are summarized in Figs.~\ref{fig:QFIT0LMG} and \ref{fig:QFIeWSST0LMG}.

\paragraph{Quantum Fisher information}
Physical intuition, encouraged by the form of the Hamiltonian (\ref{HamLMG3}) 
and experimental evidences~\cite{LuckeSCIENCE2011,StrobelSCIENCE2014,LuckePRL2014}, 
suggests to employ the collective pseudospin operators along the three spatial directions $\hat{J}_\varrho$ ($\varrho=x,y,z$)
as local observables to probe the distinguishability of the ground state under unitary transformation and witness 
multipartite entanglement via the QFI. 
We maximize the QFI to obtain the optimal value $F_Q[\ket{\psi_{0}}]=\max_\varrho F_Q\big[\ket{\psi_{0}},\hat{J}_{\varrho}\big]$.
Essentially, we are applying again the SU(2) optimization procedure that has a large importance for the field of linear quantum interferometry~\cite{pezze2014,PezzeRMP}, since collective rotations generated by the pseudospin operators in the fully-symmetric subspace can be easily implemented in atomic ensambles of ultracold degenerate bosons~\cite{EsteveNATURE2008,ZiboldPRL2010,LuckeSCIENCE2011,berrada2013,StrobelSCIENCE2014,LuckePRL2014}.

Thanks to the $Z_2$ symmetry of the model, the ground state $\ket{\psi_{0}}$ is such that
$\langle\hat{J}_y\rangle = \allowbreak \langle\hat{J}_z\rangle = \allowbreak
\langle\hat{J}_x\hat{J}_y\rangle = \allowbreak \langle\hat{J}_x\hat{J}_z\rangle = \allowbreak 
\langle\{\hat{J}_y,\hat{J}_z\}\rangle = \allowbreak  0$. 
Therefore, we just need to calculate the variances of the observables:
$F_Q[\ket{\psi_{0}}]=4\max\{(\Delta\hat{J}_x)^2,\,\langle\hat{J}_y^2\rangle,\,\langle\hat{J}_z^2\rangle\}$.
Numerical simulations certify that the optimal directions are $\textbf{z}$ for the ferromagnetic coupling 
and $\textbf{y}$ for the antiferromagnetic coupling,
as pictorially notified by the Wigner distribution of the ground state on the Bloch sphere (see cartoons in Fig~\ref{fig:QFIT0LMG}).
This statement is completely confirmed by the semiclassical formulation of the Fisher matrix, that becomes exact in the thermodynamic limit. 
In the supercritical region we have
\be \label{FisherMatrixLMG1}
\mathbb{F}_Q\textrm{\small$(\Lambda>-1)$} = \begin{pmatrix}
\textrm{\footnotesize$\pazocal{O}(1)$} & 0 & 0 \\
0 & N\textrm{\footnotesize$\sqrt{1+\Lambda}$} & 0 \\
0 & 0 & N\frac{1}{\sqrt{1+\Lambda}}
\end{pmatrix},
\ee
where $\pazocal{O}(1)$ denotes a quantity depending on $\Lambda$ but not scaling with $N$, and $F_{yy}>F_{zz}\,\Leftrightarrow\,\Lambda>0$;
whereas in the subcritical region
\be \label{FisherMatrixLMG2}
\mathbb{F}_Q\textrm{\small$(\Lambda<-1)$} = \begin{pmatrix}
N\sqrt{1-\frac{1}{\Lambda^2}} & 0 & 0 \\
0 & N\sqrt{1-\frac{1}{\Lambda^2}} & 0 \\
0 & 0 & N^2\left(1-\frac{1}{\Lambda^2}\right)
\end{pmatrix},
\ee
where $F_{zz}>F_{xx}=F_{yy}$.
These considerations permit to unambiguously define the zero-temperature optimal Fisher information per particle as 
$f_Q[\ket{\psi_{0}}] \equiv F_{zz}/N = 4\langle\hat{J}_z^{2}\rangle/N$ for $\Lambda\leq0$ and
$f_Q[\ket{\psi_{0}}] \equiv F_{yy}/N = 4\langle\hat{J}_y^{2}\rangle/N$ for $\Lambda\geq0$.

Numerical results are shown in Fig.~\ref{fig:QFIT0LMG}, where we plot the optimal QFI as a function of the control parameter $\Lambda$ 
for several spin numbers $N$. The analytical curves for large $N$ are superimposed to numerical data in Fig.~\ref{fig:QFIeWSST0LMG}.

\paragraph{Spin squeezing}
Due to the metrological relevance of the LMG model, we introduce now the Wineland spin-squeezing (WSS) parameter\footnote{\ We also analyzed the Kitagawa-Ueda spin-squeezing parameter $\xi_{\rm S}^2$, 
that differs from the Wineland's one for the normalization with respect to the radius of the Bloch sphere 
instead of the length of the mean spin direction~\cite{ma2011}. 
Since there exist no direct criterion for metrologically-useful multipartite entanglement involving $\xi_s^2$, 
we avoid to account for the analysis in this venue.\\[-9pt]}
$\WSS = N(\Delta\hat{J}_{\mathbf{n}_\perp})^2\big/\langle\hat{J}_{x}\rangle^2$~\cite{Wineland1994,MaPRA2009,ma2011}, 
where $\langle\hat{J}_x\rangle$ is the sole nonzero component of the pseudospin $\langle\hat{\mathbf{J}}\rangle$ and 
$\mathbf{n}_\perp$ is a versor on the orthogonal plane $y$-$z$.
The inequality $\WSS<1$ defines spin-squeezed states and also guarantees at least bipartite entanglement among the spins.
In this context, it is customary to label as \emph{phase squeezed} any state having $N(\Delta\hat{J}_y)^2<\langle\hat{J}_x\rangle^2$ 
and \emph{number squeezed} states having $N(\Delta\hat{J}_z)^2<\langle\hat{J}_x\rangle^2$. 
This jargon is borrowed from the semiclassical picture, where the phase between the two modes ``spin up'' and ``spin down''
is visualized as the azimuthal angle of the mean spin direction on the generalized Bloch sphere 
and the number difference is the polar coordinate of the mean spin direction.

An optimization over all possible directions --~necessary for a comparison to the optimal QFI~-- 
provides $\mathbf{n}_\perp=\mathbf{y}$ for $\Lambda\leq0$ and $\mathbf{n}_\perp=\mathbf{z}$ for $\Lambda\geq0$.
% and $\mathbf{n}_\parallel=\mathbf{x}$ for any $\Lambda$.
In the limit of large population, we find 
\be
\WSS = \left\{
\begin{array}{ll} 
|\Lambda|\sqrt{\Lambda^2-1} & {\rm for} \ \Lambda<-1 \\ 
\sqrt{1+\Lambda} & {\rm for} \ -1<\Lambda<0 \\ 
\dfrac{1}{\sqrt{1+\Lambda}} & {\rm for} \ \Lambda>-1
\end{array}
\right. \, .
\ee
Figure~\ref{fig:QFIeWSST0LMG} shows a comparison between the QFI and the WSS parameter for finite $N$,
and clarifies the limit of validity of the analytical results in the limit of large $\Lambda>0$.

\paragraph{Results}
The ground state of the LMG model contains useful entanglement for any interaction strength $\Lambda\neq0$.
The QFI detects the criticality thanks to a singular behaviour in the thermodynamic limit
and recognizes the ordered phase in terms of a superextensive growth.
The optimal operator for witnessing entanglement at the critical point and the ordered phase coincides with the order parameter.
%%%%%%%%
\\[-6pt]

\noindent \tripieno{MySky} \ For a noninteracting system $\Lambda=0$, 
the ground state is a coherent spin state polarized along the transverse field 
$\ket{\psi_{0}} \equiv|\vartheta=\frac{\pi}{2},\varphi=0\rangle = 2^{-N/2}\,(\spinup_z+\spindown_z)^{\otimes N}$,
easily written on the eigenbasis $|m,J\rangle$ of $\hat{\mathbf{J}}^2$ and $\hat{J}_z$ as
\be 
\ket{\psi_{0}} = \frac{1}{\sqrt{2}^N}\sum_{m=-N/2}^{N/2}\begin{pmatrix} N \\ N/2+m \end{pmatrix}^{\frac{1}{2}} |m,J\rangle ,
\ee 
with a binomial occupation distribution for the two modes.
Being the coherent states the most classical quantum states, it does not surprise that we find $F_Q=N$,
i.e. a saturation of the shot-noise limit, for any operator $\hat{J}_\varrho$. 
No quantum correlations are present in the ground state (nor in the excited states).
Exactly as for the Ising model, the discontinuity in the first derivative of the QFI at $\Lambda=0$ 
is only due to the sudden change in the optimal operator from $\hat{J}_z$ ($\Lambda<0$) to $\hat{J}_y$ ($\Lambda>0$).
%%%%%%%%
\\[-6pt]

%%%%%%%%%%%%
%% FIGURA %%
%%%%%%%%%%%%

\begin{figure}[t!]
\centering
\includegraphics[width=0.40\textwidth]{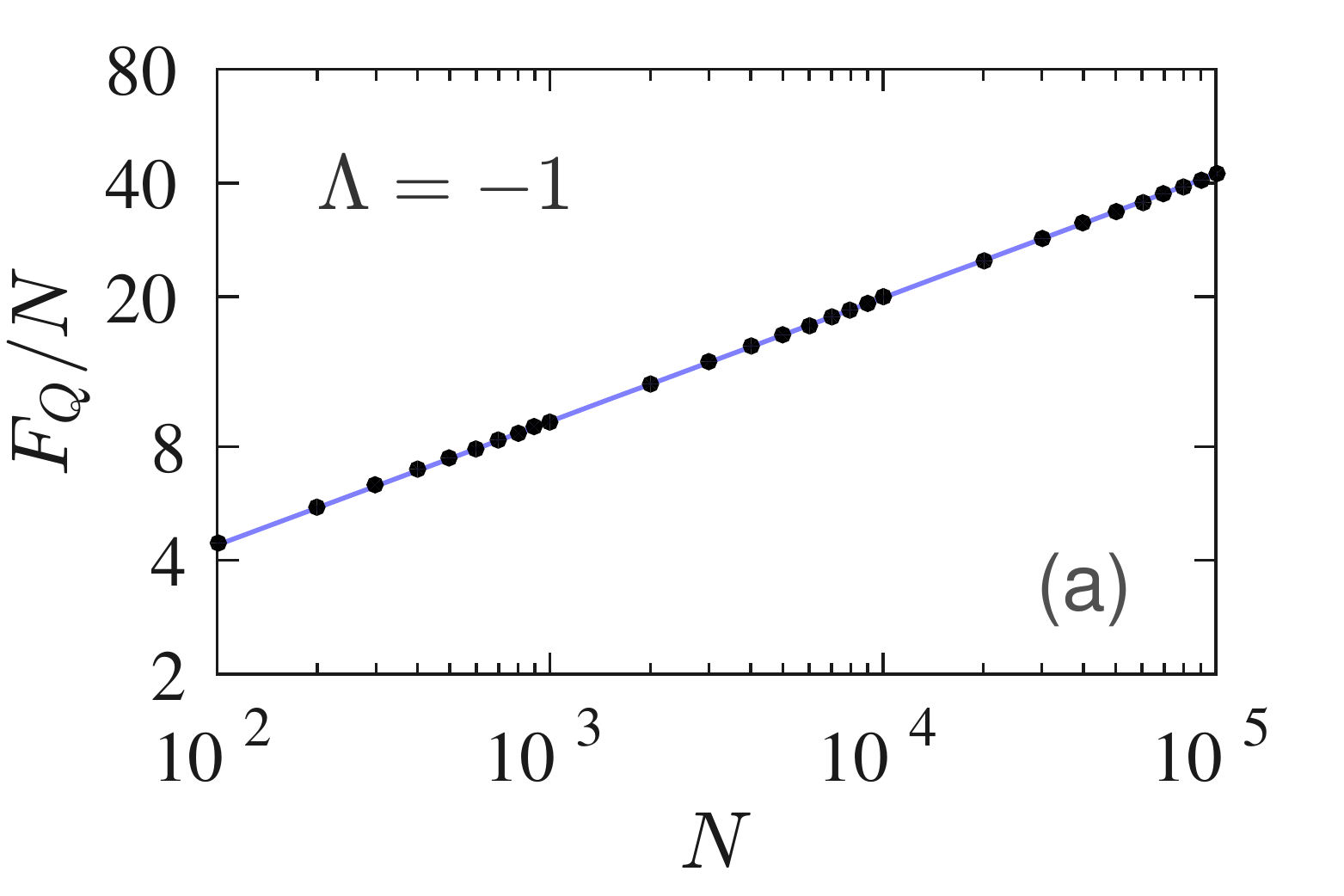} 
\includegraphics[width=0.40\textwidth]{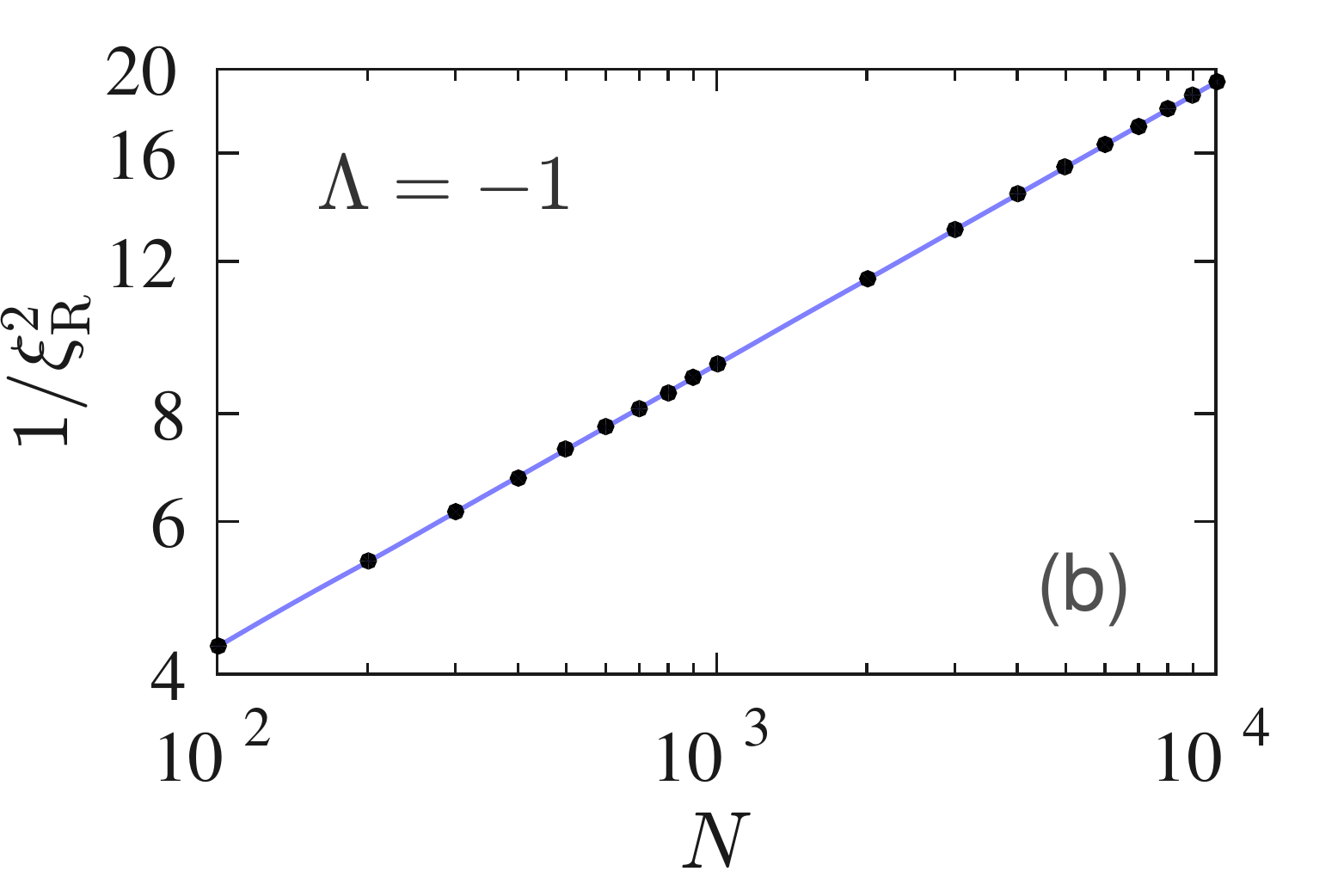} \\ 
\hspace{-9pt}
\includegraphics[width=0.43\textwidth]{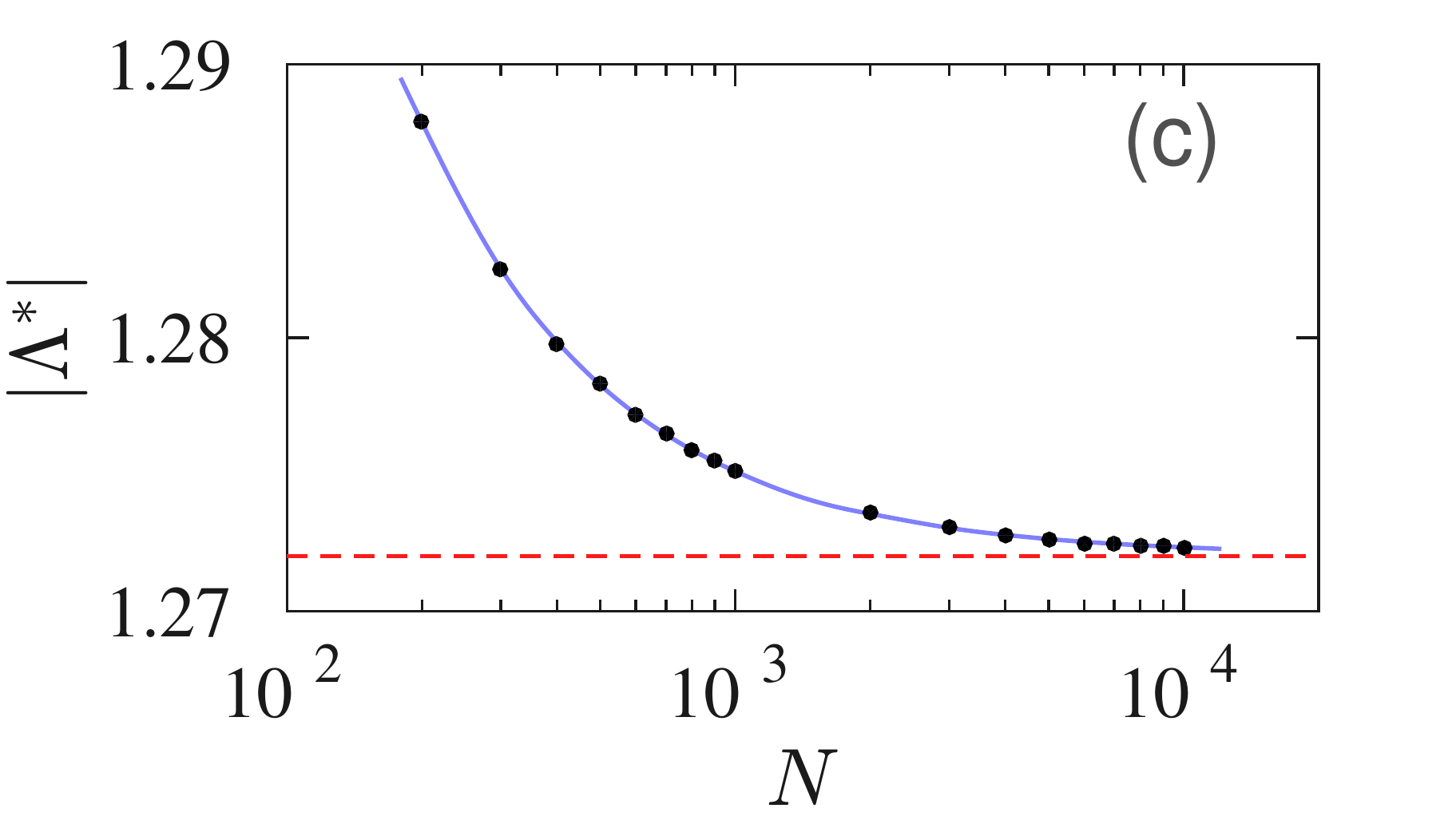} \hspace{-15pt}
\includegraphics[width=0.43\textwidth]{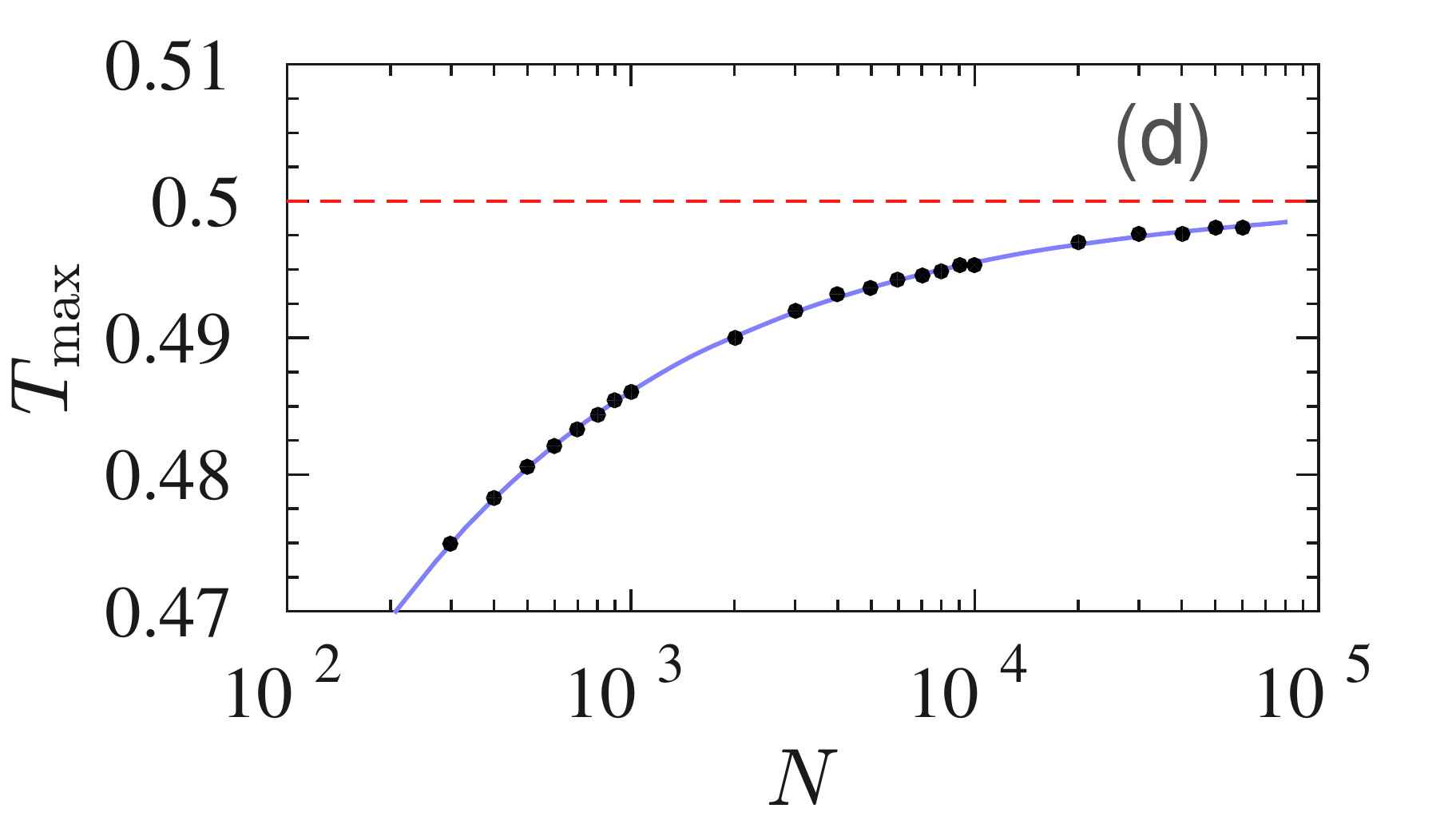}
\caption{Relevant finite-size scalings in the balanced LMG model: dots are numerical data, blue solid lines are power-law fit curves,
red dashed lines indicate analytical asymptotic values in the thermodynamic limit. 
\textbf{(a)} Critical behaviour of the Fisher density $f_Q(\Lambdac) \sim N^{1/3}$ for the ground state.
\textbf{(b)} Critical behaviour of the spin squeezing $\WSS(\Lambdac) \sim N^{-1/3}$ for the ground state.
\textbf{(c)} Correction to the semiclassical value $|\LambdaStar| - \sqrt{\varphi} = \pazocal{O}(1/N)$ 
for the threshold parameter below which the ground state ceases to be phase squeezed ($\varphi$ is the golden number).
\textbf{(d)} Correction to the semiclassical value $\big|T_{\rm max} - 0.5 \big| = \pazocal{O}(1/\sqrt{N})$
for the maximum extension of multipartite entanglement over temperature (in unit of $\Omega$).}
\label{fig:ScalingLMG}
\end{figure}

\noindent \tripieno{MySky} \ For ferromagnetic coupling $\Lambda<0$, 
the monotonic behaviour of the Fisher density of the LMG model is very similar to the one exhibited by the infinite-range Ising model:
Fig.~\ref{fig:QFIT0LMG} can be compared to Fig.~\ref{fig:IsingAlphaOpt}\Panel{a} of section~\ref{sec:IsingArbitrary}. 
Nevertheless, we underline that the two models are fundamentally incompatible for a major difference: 
the restriction of the Hilbert space operated in the former model.
Moreover, the prefactor $N^{-1}$ in the LMG Hamiltonian~(\ref{HamLMG3}),
besides ensuring a finite energy density, permits to fix the critical point at a nonvanishing value of $\Lambda$:
dropping the prefactor $N^{-1}$ in Hamiltonian~(\ref{HamLMG3}) would result into a new location of the critical point at $\Lambda=0$.
%%%%%%%
\\[9pt]
%%%%%%%
\noindent \trivuoto{MyBlue} \ For weak interaction $-1<\Lambda<0$, 
$\ket{\psi_{0}}$ is a Gaussian state characterized by a squeezed phase distribution
$\Delta\hat{J}_y<\sqrt{N}/2$ at the expense of enhanced number-difference fluctuation $\Delta\hat{J}_z>\sqrt{N}/2$. 
For large $N$ we find 
\be
f_Q\big[\textrm{\small$-1<\Lambda<0$},\hat{J}_z\big] = \frac{1}{\sqrt{1+\Lambda}} \equiv \frac{1}{\WSS} \, .
\ee
Thus, the ground state is useful for sensing rotations around the $z$ axis (Fig.~\ref{fig:QFIeWSST0LMG}).
Witnessed multipartite entanglement is intensive in the disordered phase.
%%%%%%%
\\[9pt]
%%%%%%%
\noindent \trivuoto{MyBlue} \ We already discussed the reasons why the harmonic-oscillator approximation 
breaks down at the critical point $\Lambda=-1$. 
Here we just mention that it predicts a divergent number-difference fluctuation as 
$(\Delta\hat{J}_z)^2=\frac{N}{4}(1+\Lambda)^{-1/2}\xrightarrow{\Lambda\to-1}\infty$. 
At criticality, numerical calculation provides the finite-size scaling (Fig.~\ref{fig:ScalingLMG})
\be
f_Q\big[\textrm{\small$\Lambda=-1$},\hat{J}_z\big] \sim \frac{1}{\WSS} \sim N^{1/3} \, ,
\ee
that entails a diverging multipartiteness.\footnote{\ The divergence $F_Q \sim N^{4/3}$ for the LMG model is softer than the one 
$F_Q \sim N^{7/4}$ paraded by the short-range transverse Ising model of chapter~\ref{ch:Ising}. 
This disparity has been read as a sign of the mean-field nature of the fully-connected model~\cite{HaukeNATPHYS2016}.\\[-9pt]}
Remarkably, the critical divergence of multipartite entanglement substantially differs from the logarithmic behaviour 
exhibited by the bipartite entanglement: the former faces a much faster growth for increasing spin population.
Moreover, as sketched in Fig.~\ref{fig:QFIT0LMG}, at $\Lambda=\Lambdac$ the QFI displays a singular behaviour in the thermodynamic limit: 
its derivative \emph{diverges} according to $\partial_\Lambda F_Q(\Lambdac) \sim N$. 
The size correction to the position of the emergent divergence is numerically found to be 
$\Lambdac - \arg\max_\Lambda|\partial_\Lambda F_Q| \sim N^{-2/3}$.
%%%%%%%
\\[9pt]
%%%%%%%
\noindent \trivuoto{MyBlue} \ In the ordered phase $\Lambda<-1$, 
the ground state is actually characterized by a macroscopic superposition\footnote{\ The interactions favours localization of particles in one mode, in the sense that any infinitesimal field $\delta\to0^\pm$ along the longitudinal axis breaks the $Z_2$ symmetry (i.e. the parity of the potential $V$) and causes the collapse of the ground state into a coherent state. The quantum superposition is extremely delicate against external perturbations: a quantitative discussion will be supplied in section~\ref{sec:asymmetricLMG}.\\[-9pt]}
of quantum states~\cite{Cirac1998,ho2004}. The Gaussianity of the ground state is lost,
the nonzero pseudospin component vanishes as $\langle\hat{J}_x\rangle=\frac{N}{2}|\Lambda|^{-1}\xrightarrow{\Lambda\to-\infty}0$ 
and the spin squeezing definitively ceases to witness entanglement for $\Lambda\leq\LambdaStar$.
We find that the mean-field value of $\LambdaStar$ is given by the square root of the golden ratio 
$\LambdaStar = -\sqrt{\frac{1+\sqrt{5}}{2}} \approx -1.272$ 
and a numerical scaling points out finite-size corrections of the order of $N^{-1}$
(Fig.~\ref{fig:ScalingLMG}). Even if the ground state is not spin squeezed, the QFI still detect quantum resources 
for enhanced phase estimation:
\be
\frac{1}{\WSS} = \frac{1}{|\Lambda|\sqrt{\Lambda^2-1}} \ < \ f_Q\big[\textrm{\small$\Lambda<-1$},\hat{J}_z\big] = N\left(1-\frac{1}{\Lambda^2}\right) \, .
\ee
This is a neat example of the capacity of the QFI to witness multipartite entanglement 
for non-Gaussian quantum states~\cite{StrobelSCIENCE2014}, for which the entire class of spin-squeezing parameters 
recognizes no entanglement~\cite{pezze2014}.
Finally, we cannot omit the extensive character of multipartite entanglement in the ordered phase.
%%%%%%%
\\[9pt]
%%%%%%%
\noindent \trivuoto{MyBlue} \ For strong nonlinearities $|\Lambda|\gg1$, the ground state is well approximated by a NOON state 
% $\ket{\psi_{0}} = (\spinup_z^{\otimes N} + \spindown_z^{\otimes N})/\sqrt{2}$~\cite{bollinger1996},
$\ket{\psi_{0}} = \frac{1}{\sqrt{2}}\big(|\!+\!J,J\rangle+|\!-\!J,J\rangle\big) \equiv 
\frac{1}{\sqrt{2}}\big(|N\rangle_\mathsf{a}|0\rangle_\mathsf{b}+|0\rangle_\mathsf{a}|N\rangle_\mathsf{b}\big)$~\cite{bollinger1996},
that is a symmetric superposition of states localized in the two wells of the effective potential $V(z)$, 
also known as a specimen of the family of Schr\"odinger's cat states.
The QFI reaches its ultimate bound $f_Q\big[\textrm{\small$\Lambda\to-\infty$},\hat{J}_z\big]=N$, namely the Heisenberg limit: 
the ground state is maximally useful for probing rotations around the $z$ axis.
%%%%%%%%
\\[-6pt]

%%%%%%%%%%%%%
%% FIGURA %%
%%%%%%%%%%%%

\noindent \tripieno{MySky} \ Antiferromagnetic coupling $\Lambda>0$ makes number imbalance between the two modes energetically unfavourable: 
the number-difference fluctuation is reduced $\Delta\hat{J}_z<\sqrt{N}/2$ 
at expense of increased phase fluctuation $\Delta\hat{J}_y>\sqrt{N}/2$.
On this side of the phase diagram no critical point is encountered. Nevertheless, we want to underline a clear crossover from 
an intensive to extensive behaviour of multipartite entanglement, signalling a structural adiabatic (not sudden) change 
in the ground state, known in literature as the ``Josephson crossover'' between weak and strong interaction~\cite{PezzeRMP}
located at $\Lambda \approx N^2$.
%%%%%%%
\\[9pt]
%%%%%%%
\noindent \trivuoto{MyBlue} \ For $0<\Lambda\lesssim N$, the system is effectively thought as weakly interacting,
because its physics is governed by the transverse field, that is responsible for a well-defined relative phase between the two modes:
$(\Delta\hat{J}_y)^2=\frac{N}{4}\sqrt{1+\Lambda}$. Since $\langle\hat{J}_x\rangle \approx \frac{N}{2}$,
the ground state is number squeezed $N(\Delta\hat{J}_z)^2<\langle\hat{J}_x\rangle^2$ (Fig.~\ref{fig:QFIeWSST0LMG}).
For large $N$, the semiclassical potential can still support a Gaussian lowest-energy eigenfunction of finite width: 
the harmonic description gives
\be
f_Q\big[\textrm{\small$0<\Lambda\lesssim N$},\hat{J}_y\big] = \sqrt{1+\Lambda} \equiv \frac{1}{\WSS} \, .
\ee
The dependence on $\Lambda$ is entirely explained by the squeezed nature of the ground state.
Apart from the details, the most important aspect is that multipartiteness is constant in the thermodynamic limit: $f_Q\sim\pazocal{O}(1)$.
%%%%%%%
\\[9pt]
%%%%%%%
\noindent \trivuoto{MyBlue} \ For increasing $\Lambda$, number-difference variance $(\Delta\hat{J}_z)^2=\frac{N}{4}(1+\Lambda)^{-1/2}$ 
is further reduced, while the indeterminacy on the phase increases. Somewhere between $\Lambda\approx N$ and $\Lambda\approx N^2$ 
the approximation leading to the harmonic-oscillator picture in Eq.~(\ref{HarmonicSuperCriticalLMG}) ceases to be valid:
the phase fluctuation $\Delta\hat{J}_y\approx N$ becomes extremely large, of the order of the radius of the Bloch sphere, 
so that the very definition of a relative phase between the two modes is ill-posed.
In this regime, the QFI does not follow the semiclassical prediction anymore (Fig.~\ref{fig:QFIeWSST0LMG}):
as a function of $\Lambda$, it responds with a change of concavity around $\Lambda \approx N$
and a peculiar geometrically intermediate scaling between the shot-noise limit and the Heisenberg limit: 
$f_Q\big[\textrm{\small$N\ll\Lambda\ll N^2$},\hat{J}_y\big]\sim \sqrt{N}$.
%%%%%%%
\\[9pt]
%%%%%%%
\noindent \trivuoto{MyBlue} \ For very strong interaction $\Lambda \gg N^2$, 
the tunnelling imposed by the external field becomes negligible: 
the number-difference fluctuation $\Delta\hat{J}_z$ vanishes and the ground state is well approximated by the twin Fock state
$\ket{\psi_{0}} = |m\!=\!0,J\rangle \equiv \big|\frac{N}{2}\big\rangle_\mathsf{a}\big|\frac{N}{2}\big\rangle_\mathsf{b}$
(a Dicke state with the same number of spins up and spins down~\cite{HollandPRL1993}).
%Hence, at $\Lambda\approx N^2$ we encounter a crossover regime from ... to ....
The crossover regime around $\Lambda\approx N^2$ is marked by a flattening of the QFI on a constant asymptotic value
\be
f_Q\big[\textrm{\small$\Lambda\gg N^2$},\hat{J}_y\big] = \frac{N+2}{2} \equiv \frac{1}{\WSS} \, ,
\ee
scaling superextensively in the guise of the Heisenberg bound.
Note that, in the limit $\Lambda\gg N^2$, number-difference variance $(\Delta\hat{J}_z)^2$ 
and pseudospin length $\langle\hat{J}_x\rangle$ vanish, but their ratio is finite: 
so, even if a high spin squeezing survives, it proves to be highly sensitive to any noise.

%%%%%%%%%%%%
%% FIGURA %%
%%%%%%%%%%%%

\subsection{Multipartite entanglement at finite temperature} 
There exist only few works about entanglement in the LMG model at finite temperature,
and they mainly focused on concurrence~\cite{matera2008} and mutual information~\cite{WilmsJSM2012}, 
which only quantify correlations between two parties. 
Concerning multipartite entanglement, spin squeezing has been studied close to the quantum critical point~\cite{FrerotArxiv2017}, 
while the use of the QFI for thermal states was initiated by Ref.~\cite{HaukeNATPHYS2016}, where 
a universal scaling for the QFI at criticality was found.
Surprisingly, comprehensive results involving the QFI are still missing in literature.
The present part of the chapter aims to fill this unexpected lack: 
the main motivation is gaining an insight about the robustness of quantum correlations to thermal noise in the vicinity (and not only) 
of the quantum critical point, in order to better understand the feasibility of 
a two-mode linear interferometer based on a BEC~\cite{FattoriPRL2008,cronin2009} interacting with the thermal fraction of the gas.

\begin{figure}[p!]
\centering
\includegraphics[width=0.51\textwidth]{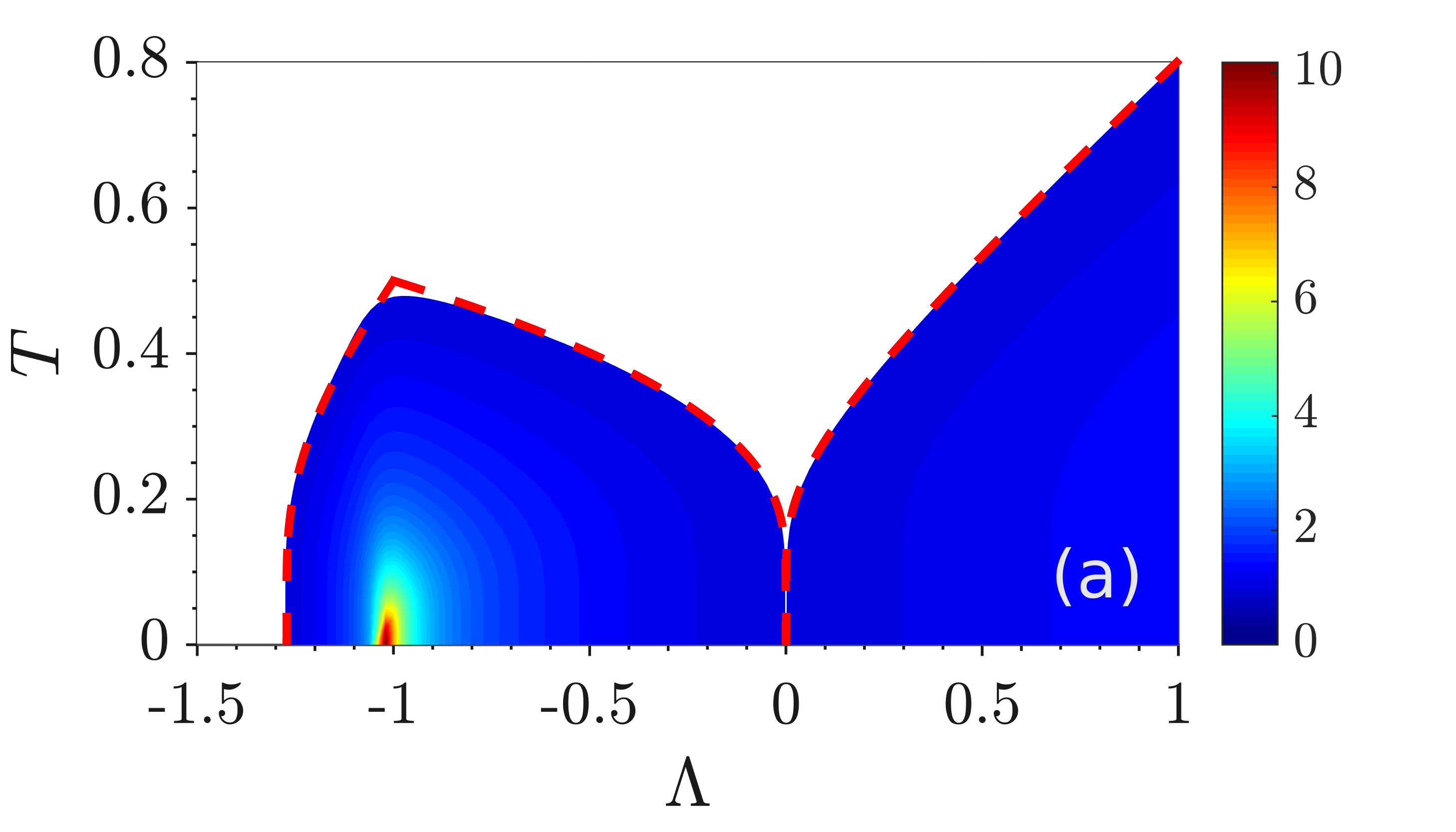} \hspace{-18pt}
\includegraphics[width=0.51\textwidth]{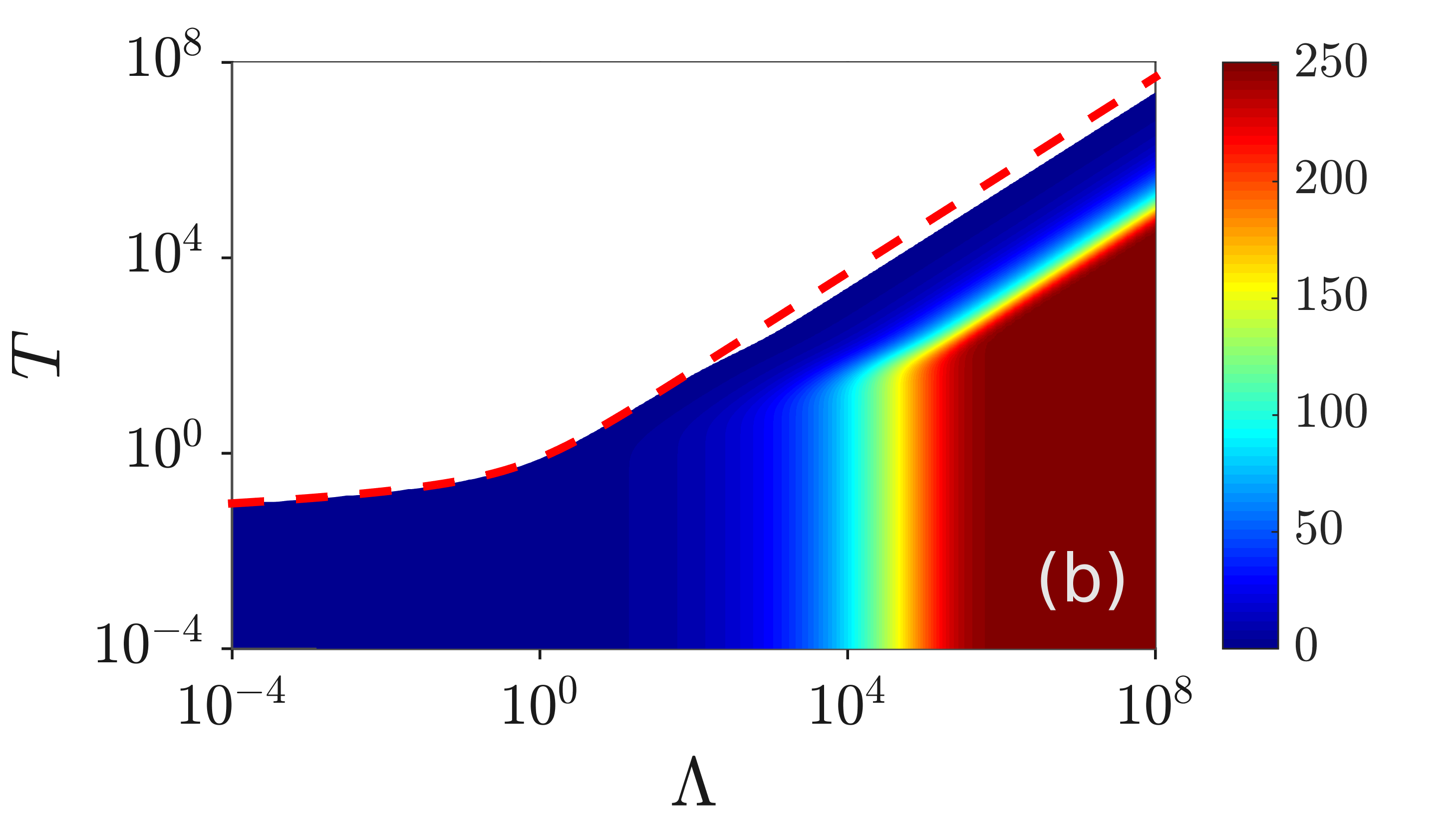} \\
\includegraphics[width=0.51\textwidth]{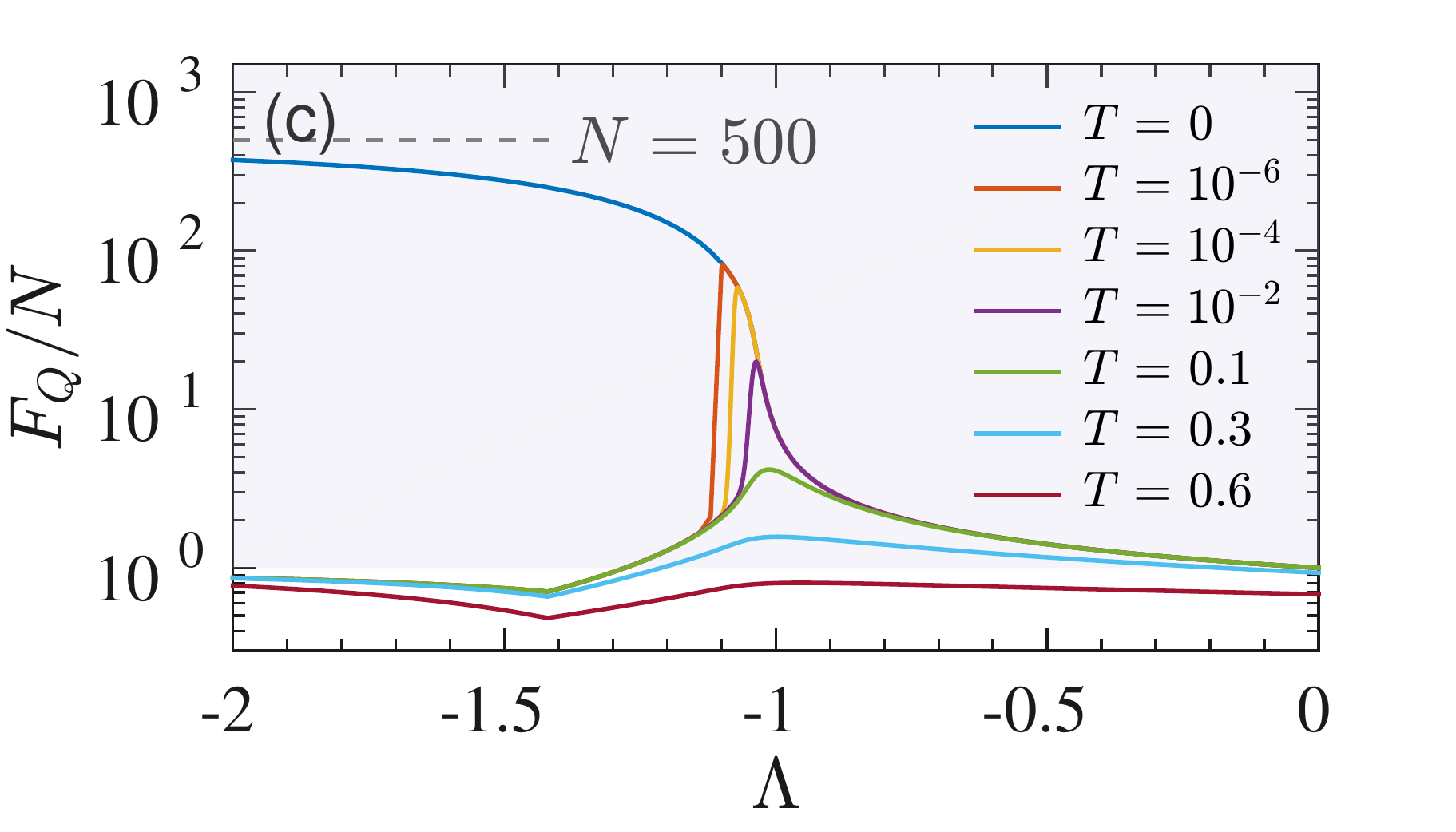} \hspace{-18pt}
\includegraphics[width=0.51\textwidth]{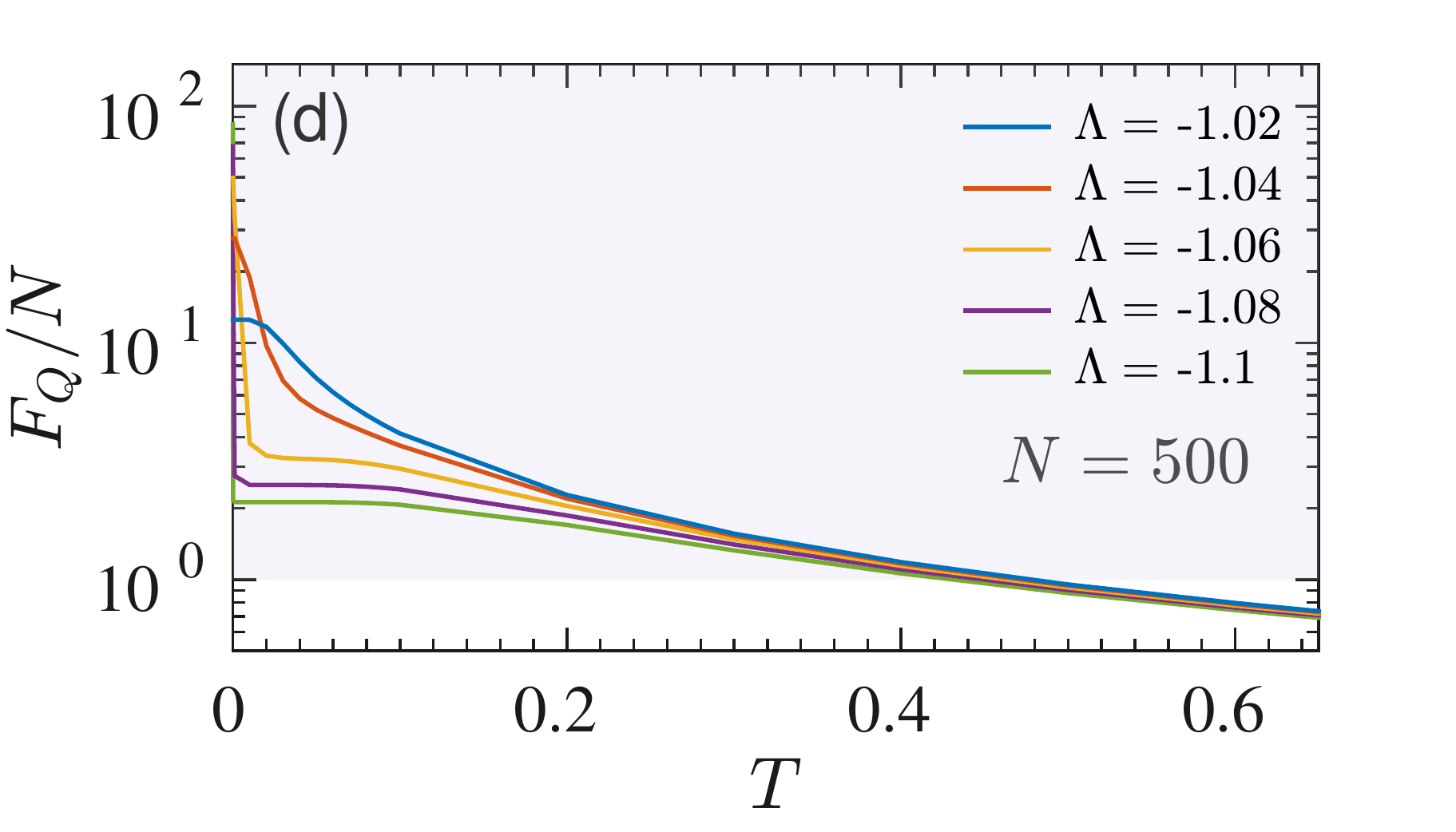} \\
\includegraphics[width=0.498\textwidth]{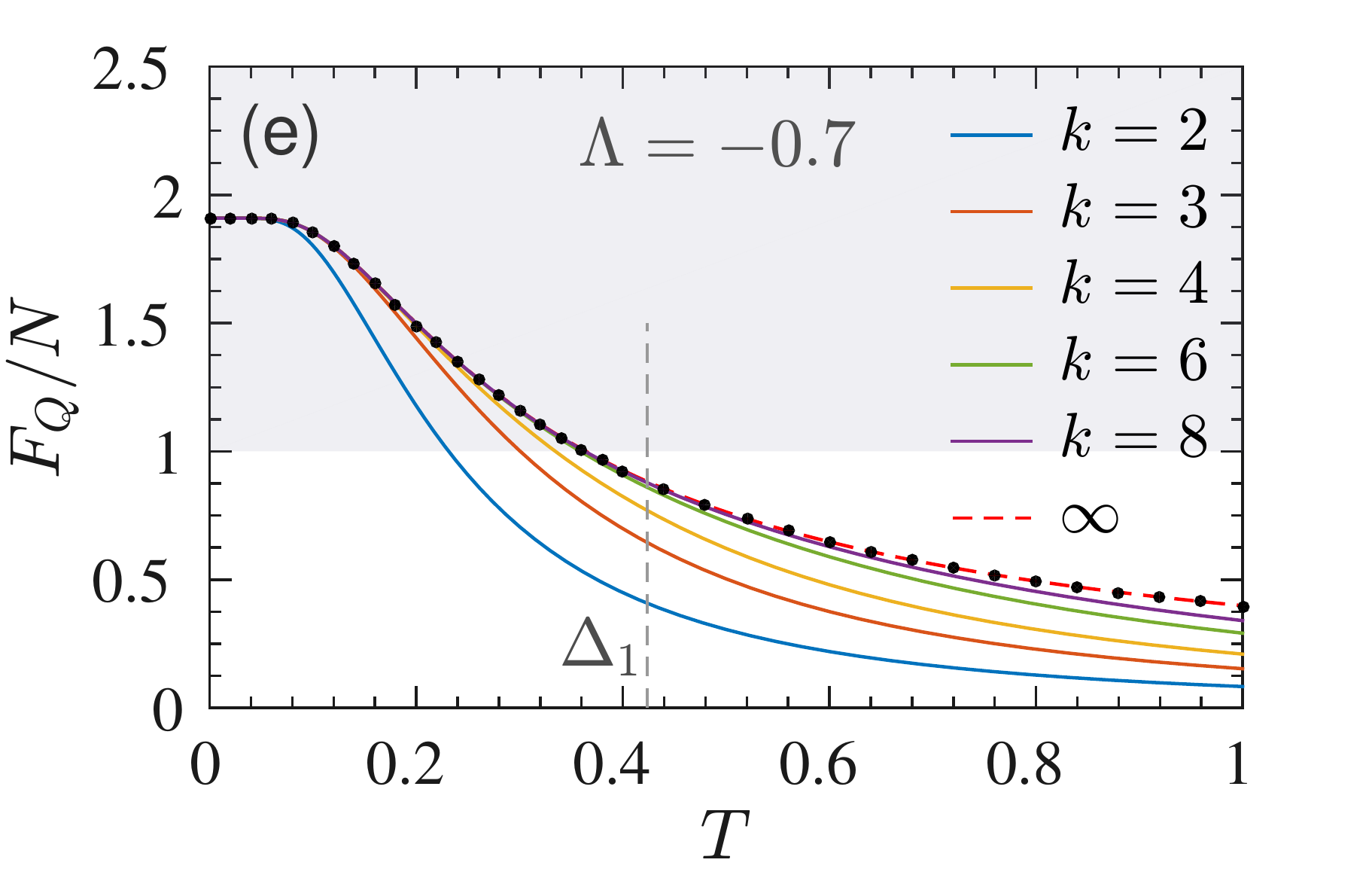}  \hspace{-12pt}
\includegraphics[width=0.498\textwidth]{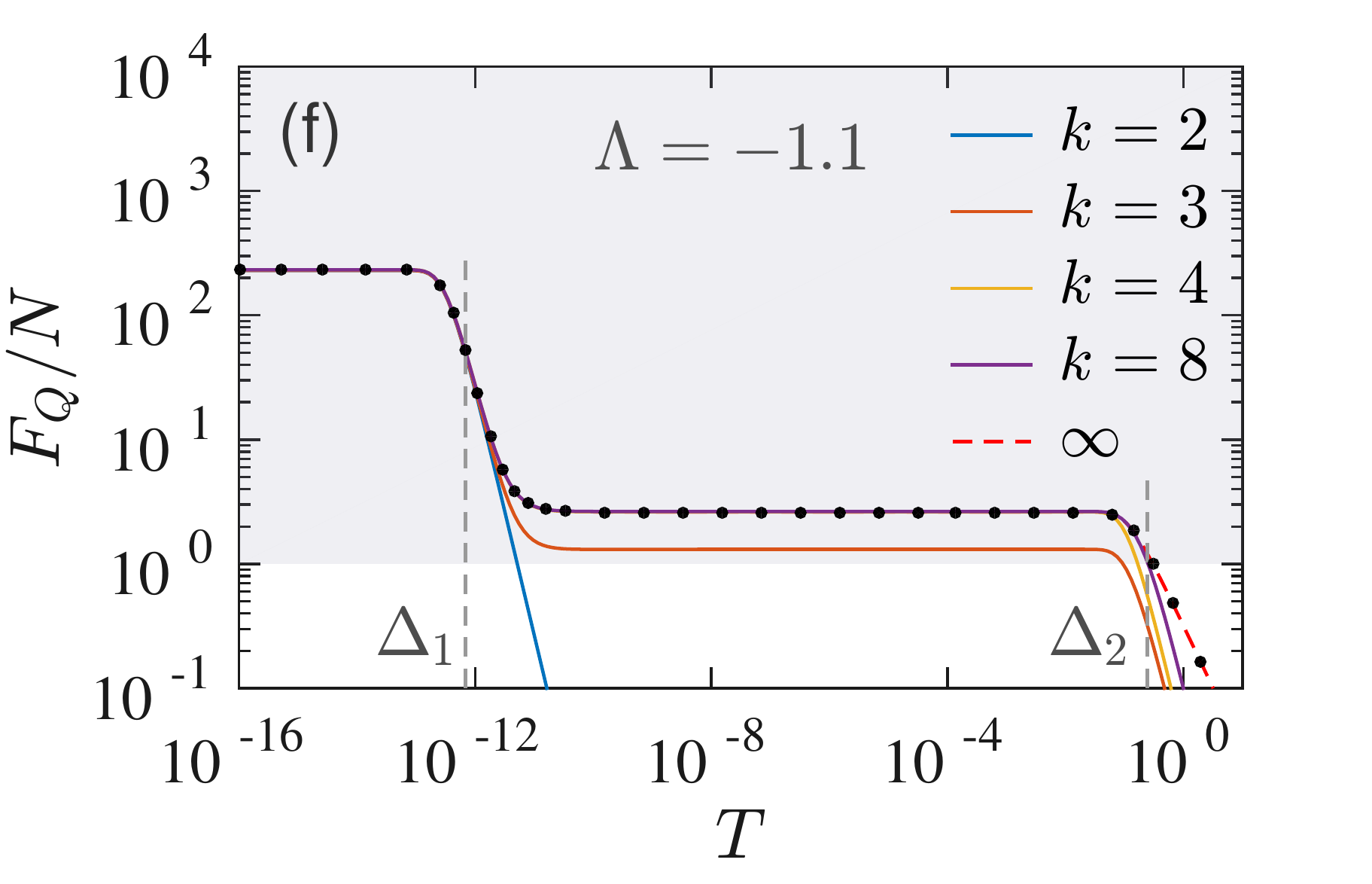}
\caption{Quantum Fisher information for thermal states of the balanced LMG model. 
\textbf{(a)} Fisher density $F_Q[\hat{\rho}_T]/N$ as a function of driving parameter $\Lambda$ and temperature $T$ for $N=500$.
Multipartite entanglement is witnessed in the coloured region, where $F_Q>N$. 
The red dashed line is the analytical boundary $F_Q=N$ of thermal entanglement in the thermodynamic limit, 
given by Eq.~(\ref{LMGboundaryAntiferr}) for $\Lambda>0$, Eq.~(\ref{LMGboundarySuper}) for $-1<\Lambda<0$
and Eq.~(\ref{LMGboundarySub}) for $\Lambda<-1$, with $\kappa=1$.
\textbf{(b)} Antiferromagnetic side of the phase diagram in panel (a) on a dilated logarithmic chart.
The thermal region of quantum correlations expands for increasing $\Lambda$.
Moreover, an outstanding region of extensive multipartiteness at finite temperature for $\Lambda\gg N^2$ stands out. 
\textbf{(c,\,d)} Cuts of the density plot in panel (a) for several fixed $T$ and $\Lambda<-1$ respectively.
\textbf{(e,\,f)} Detailed decay of $F_Q[\hat{\rho}_T]/N$ with $N=2000$ as a function of $T$, 
respectively for ferromagnetic supercritical ($0>\Lambda>\Lambdac$) and subcritical ($\Lambda<\Lambdac$) parameter.
Dots are numerical data. 
Solid lines are analytical curves Eqs.~(\ref{QFIkmodesSuper}) and (\ref{QFIkmodesSub}) for different values of the cutoff $k$. 
The vertical dashed lines indicate $T=\DeltaE$ and $T=\DeltaEE$.
The full semiclassical description, provided by Eqs.~(\ref{LMGparaQFIfull}) and (\ref{LMGferroQFIfull}), appears as a dashed line. 
In panels (c,\,d,\,e,\,f) the shaded area marks multipartite entanglement.}
\label{fig:LMGthermic1}
\end{figure}

\begin{figure}[t!]
\centering
\includegraphics[width=0.49\textwidth]{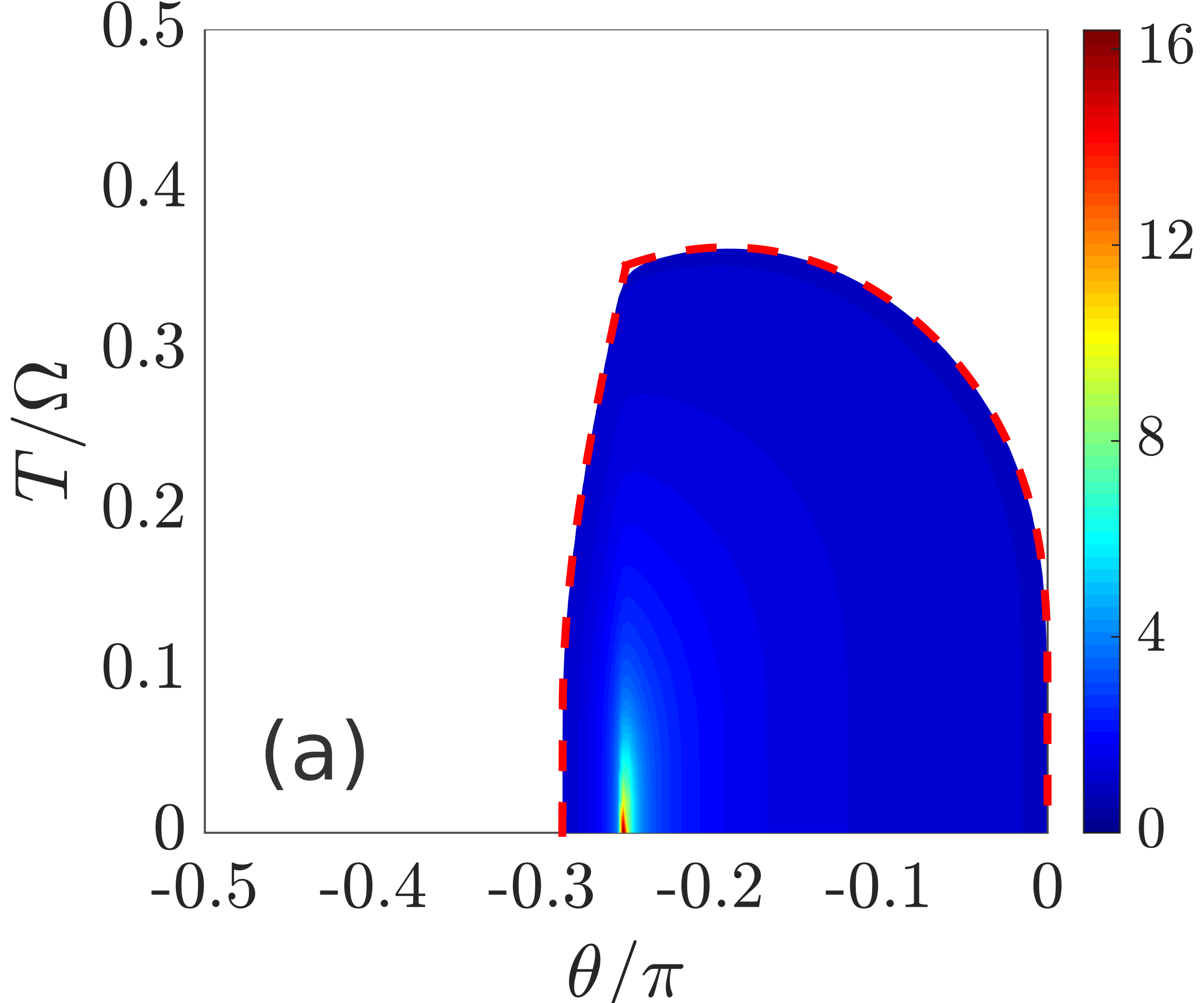}  \hspace{-0pt}
\includegraphics[width=0.49\textwidth]{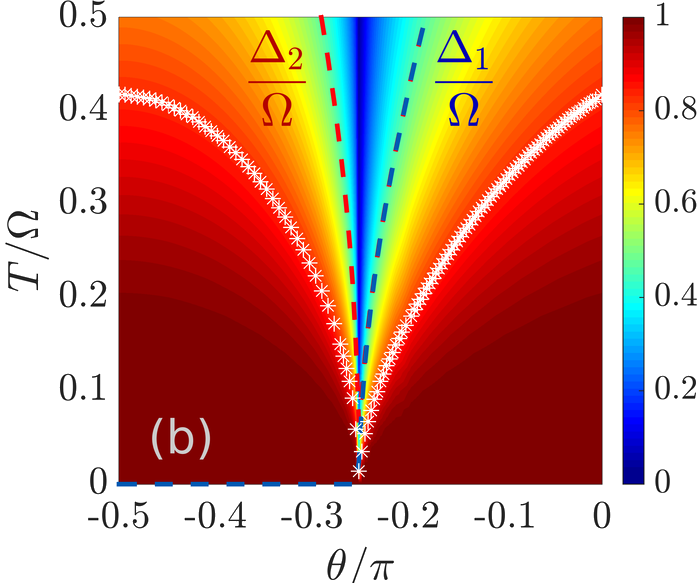}
\caption{Thermal phase diagram of the balanced LMG in the goniometric formulation~(\ref{HamLMGtheta}) for $N=2000$. 
\textbf{(a)} Fisher density $F_Q[\hat{\rho}_T]/N$ on the plane $\theta$--$T$.
The analytical result (dashed line) perfectly agrees with numerical calculations even at finite $N$.
\textbf{(b)} Quantum criticality expressed in terms of the normalized Fisher density $F_Q[\hat{\rho}_T]/F_Q[\hat{\rho}_0]$.
The low-temperature behaviour survives below a crossover temperature $T \lesssim \Tcross$ (white dots) 
following the first nonvanishing energy gap (dashed lines). 
Here the crossover is obtained as the set of inflection points $\partial^2 F_Q/\partial T^2=0$.}
\label{fig:LMGthermic2}
\end{figure}

We consider the condensed fraction as a closed lossless system in equilibrium with a thermal reservoir: 
at temperature $T$ (in units of tunnelling energy $\Omega$), 
the occupation probability for the $n$th energy level is the canonical Boltzmann's weight 
$p_n=\pazocal{Z}^{-1}\neper^{-E_n/T}$, $E_n$ being the energy of the level 
and $\pazocal{Z}=\sum_n\neper^{-E_n/T}$ being the partition function, where the sum is extended over all the states.
Thus, for a thermal state $\hat{\rho}_T=\sum_n p_n \ket{\psi_n}\bra{\psi_n}$, 
the general definition of the QFI for mixed states in Eq.~(\ref{SU2optimization}) reduces to diagonalizing the Fisher matrix
\be \label{QFIthermalLMG}
\mathbb{F}_Q[\hat{\rho}_T] = \frac{2}{\pazocal{Z}} \left[ \sum_{n,m} \frac{\big(\neper^{-E_n/T} - \neper^{-E_m/T}\big)^2}{\neper^{-E_n/T} + \neper^{-E_m/T}} \,\bra{\psi_n} \hat{J}_\varrho \ket{\psi_m} \bra{\psi_m} \hat{J}_\varsigma \ket{\psi_m} \right]_{\varrho,\,\varsigma\,=\,x,\,y,\,z}
\ee
and looking for the maximum eigenvalue, that can be done both numerically and through analytical calculations in the limit $N\gg1$.
We also propose to study multipartite entanglement using the extended definition of the WSS parameter
$\WSS = \min_{\mathbf{n}_\perp} N(\Delta\hat{J}_{\mathbf{n}_\perp})^2\big/\langle\hat{J}_{\mathbf{n}_\parallel}\rangle^2$,
where $\langle\,\cdot\,\rangle=\tr(\,\cdot\,\,\hat{\rho}_T)$ denotes ensemble average,
$\mathbf{n}_\parallel$ is the direction of the mean-spin vector $\langle\hat{\mathbf{J}}\rangle$
and $\mathbf{n}_\perp$ is an arbitrary orthogonal direction. 
Figures~\ref{fig:LMGthermic1} and \ref{fig:LMGthermic2} gather all the results relative to QFI and WSS parameter at finite temperature.

\paragraph{Thermal multipartiteness and squeezing}
The competition between thermal population of excited states and genuinely quantum correlations of the ground state 
results in panels~\ref{fig:LMGthermic1}\Panel{a,\,b}, where we plot the optimal Fisher density $F_Q[\hat{\rho}_T]/N$
as a function of the interaction parameter $\Lambda$ and temperature $T$.
The coloured area in the density plot corresponds to entangled thermal states ($F_Q>N$).
Loosely speaking, all the states corresponding to values of $\Lambda$ and $T$ in the coloured region 
contain quantum correlations useful for interferometry:
they allow for sub--shot-noise sensitivity in phase estimation when chosen as input states of an ideal linear interferometer.
The peak around the quantum critical point $\Lambda=-1$ in panel~\ref{fig:LMGthermic1}\Panel{a} 
indicates a large entanglement depth, inherited by the divergence of multipartiteness at $T=0$. 
The thermal phase diagram is qualitatively similar to the one of the Ising model in Fig.~\ref{fig:IsingThermic1}.
%%%%%%%%
\\[-6pt]

\noindent \tripieno{MySky} \ Multipartiteness is an intensive quantity at finite temperature: it does not grow with the population $N$. 
This feature prefigures an abrupt discontinuity in the regime of strong ferromagnetic coupling,
whereby multipartiteness diverges at null temperature.
%%%%%%%%
\\[-6pt]

\noindent \tripieno{MySky} \ Interestingly, at finite temperature the QFI detects the same amount of entanglement 
detected by the spin-squeezing parameter: $1/\WSS=F_Q/N$ for any $\Lambda$ in the limit $N\gg1$.
Therefore, multipartite entanglement only results from the squeezing of the thermal state. 
%%%%%%%
\\[9pt]
%%%%%%%
\noindent \trivuoto{MyBlue} \ When increasing the temperature, 
the witnessed multipartite entanglement decays from its maximum value at $T=0$: 
thermal noise is responsible for a loss of coherence entailing a spread of spin fluctuations in any direction.
%%%%%%%
\\[9pt]
%%%%%%%
\noindent \trivuoto{MyBlue} \ The boundary of the region of witnessed entangled thermal states on the $\Lambda$--$T$ plane 
can be fully understood in the large-$N$ limit: it indicates the set of parameters for which the minimum spin fluctuation
becomes comparable to that one of the coherent spin state: $\min_\varrho(\Delta\hat{J}_\varrho)^2 \approx \frac{N}{4}$.
The analytical curve for the boundary is superimposed to the density plot in panels~\ref{fig:LMGthermic1}\Panel{a,\,b} 
as a thick dashed line.
%%%%%%%%
\\[-6pt]

%%%%%%%%%%%%
%% FIGURA %%
%%%%%%%%%%%%

\noindent \tripieno{MySky} \ For antiferromagnetic interaction $\Lambda>0$ in panel~\ref{fig:LMGthermic1}\Panel{b}, the direction of maximal squeezing is $\mathbf{n}_\perp=\mathbf{z}$ and the operator giving the optimal QFI is $\hat{J}_y$ at all values of $T$. 
Multipartite entanglement detected by the QFI smoothly degrades for increasing temperature.
%%%%%%%
\\[9pt]
%%%%%%%
\noindent \trivuoto{MyBlue} \ In the limit $1\ll \Lambda\lesssim N$ we find
\be \label{LMGQFISuper}
F_Q[\hat{\rho}_T] = F_Q[\ket{\psi_0}] \,\tanh\Big(\frac{\Delta_1}{2T}\Big) \equiv \frac{1}{\WSS} \, , \quad {\rm with} \ \DeltaE=\sqrt{1+\Lambda} \, .
\ee
Consequently, $(\kappa\!+\!1)$-partite entanglement survives against thermal fluctuations for temperatures
\be \label{LMGboundaryAntiferr}
T < T_\kappa\textrm{\small$(\Lambda>0)$} = \frac{\sqrt{1+\Lambda}}{2\,\textrm{arctanh} \left(\pazocal{D}_\kappa/\sqrt{1+\Lambda}\right)} 
\, ,
\ee
with $\pazocal{D}_\kappa=\left\lfloor\frac{N}{\kappa}\right\rfloor\frac{\kappa^2}{N}+\big(\sqrt{N} - \left\lfloor\frac{N}{\kappa}\right\rfloor\frac{\kappa}{\sqrt{N}}\big)^2$ a rational number, see Eq.~(\ref{boundKpartite}).
The red dashed curve in panels~\ref{fig:LMGthermic1}\Panel{b}, tracing the boundary of thermal entanglement, 
is nothing else but $T_\kappa(\Lambda)$ with $\kappa=1$.
Equation~(\ref{LMGboundaryAntiferr}) is an outstanding results itself: multipartite entanglement in number-squeezed states 
is extremely robust against thermal decoherence. In particular, if the interaction is strong enough $\Lambda\gg1$, 
quantum correlations are found up to remarkably high temperature $T \simeq \frac{1}{2}\Lambda$, in excellent agreement with numerical data.
%%%%%%%
\\[9pt]
%%%%%%%
\noindent \trivuoto{MyBlue} \ This stability is not surprising: it follows from the uniform spacing of the harmonic potential $\Delta_n=\Delta_1\simeq\sqrt{\Lambda}$ and from the number fluctuation on the excited states 
$\bra{\psi_n}\hat{J}_z^2\ket{\psi_n} \simeq \frac{2n+1}{4}\frac{N}{\sqrt{\Lambda}}$.
In~fact, the boundary $T_{\kappa=1}(\Lambda)$ of the region of quantum correlations can be easily explained 
as the minimum temperature required to populate excited states $n$ having fluctuation comparable with the shot noise:
$T \approx n\,\Delta_1$ such that $\bra{\psi_n}\hat{J}_z^2\ket{\psi_n} \approx \frac{N}{4}$.
%%%%%%%
\\[9pt]
%%%%%%%
\noindent \trivuoto{MyBlue} \ For $\Lambda\gg N$, the approximation leading to the boundary in Eq.~(\ref{LMGboundaryAntiferr}) breaks down. 
Nevertheless, Eq.~(\ref{LMGboundaryAntiferr}) still qualitatively mimics the actual thermal loss of entanglement: 
see panel~\ref{fig:LMGthermic1}\Panel{b}.
%%%%%%%%
\\[-6pt]

\noindent \tripieno{MySky} \ For $-1<\Lambda\leq0$, it is instructive to calculate the Fisher matrix~(\ref{QFIthermalLMG}) 
taking into account only the first $k\geq2$ states in the harmonic framework. We obtain a plain factorization 
between zero-temperature entanglement and thermal decay:
\be \label{QFIkmodesSuper}
F_Q[\hat{\rho}_T] = F_Q[\ket{\psi_0}] \, \tanh\Big(\frac{\DeltaE}{2T}\Big) \cdot \bigg( 1 - k\,\frac{\neper^{\,\DeltaE/T}-1}{\neper^{\,k\DeltaE/T}-1} \bigg) \, .
\ee
%%%%%%%%
\\[-3pt]
%%%%%%%%
\noindent \trivuoto{MyBlue} \ For two states only, we recover the two-mode approximation leading to the fundamental Eq.~(\ref{QFIfactorization}): 
\be 
F_Q[\hat{\rho}_T] = F_Q[\ket{\psi_0}]\,\tanh^2\Big(\frac{\DeltaE}{2T}\Big) \qquad {\rm with} \ k=2 \, . 
\ee
%%%%%%%%
\\[-3pt]
%%%%%%%%
\noindent \trivuoto{MyBlue} \ Considering all the states sustained by the harmonic potential, we obtain 
\be \label{LMGparaQFIfull}
F_Q[\hat{\rho}_T] = F_Q[\ket{\psi_0}] \tanh \Big(\frac{\DeltaE}{2T}\Big) \quad \quad {\rm with} \ k\to\infty \, ,
\ee
that is the actual decay behaviour shown by the LMG model, as depicted in panel~\ref{fig:LMGthermic1}\Panel{e}.
%%%%%%%
\\[9pt]
%%%%%%%
\trivuoto{MyBlue} \ Let's examine the asymptotic behaviour for high temperature: if $T\gg k\DeltaE$ a series expansion provides
\be 
F_Q[\hat{\rho}_T] \simeq F_Q[\ket{\psi_0}] \bigg[(k-1)\Big(\frac{\DeltaE}{2T}\Big)^2+\pazocal{O}\Big(\frac{1}{T^3}\Big)\bigg]
\ee
The inverse quadratic decay with respect to $T$ is valid as long as $k\,F_Q[\hat{\rho}_T] \ll F_Q[\ket{\psi_0}]$.
This means that, in the limit $k\to\infty$, the quadratic behaviour is lost at all temperatures: 
in fact, in this limit the decrease is purely inverse $F_Q[\hat{\rho}_T]\propto\DeltaE/T$.
% which decreases \think{\emph{linearly}} for increasing $T$.
%%%%%%%
\\[9pt]
%%%%%%%
\trivuoto{MyBlue} \ Conversely, at low enough temperature $T\ll\DeltaE$, the QFI can be expanded in powers of the small quantity $\neper^{-\DeltaE/T}$:
\be 
F_Q[\hat{\rho}_T] \simeq F_Q[\ket{\psi_0}] \,\cdot\, \left\{\begin{array}{l} 1-4\,e^{-\DeltaE/T}+\pazocal{O}\big[(e^{-\DeltaE/T})^2\big] \quad {\rm for} \ k=2 \\ 1-2\,e^{-\DeltaE/T}+\pazocal{O}\big[(e^{-\DeltaE/T})^2\big] \quad {\rm for} \ k>2 \end{array} \right.
\ee
Thus, low temperatures feature an exponentially fast saturation of the zero-temperature value $F_Q[\ket{\psi_0}]$, 
but the behaviour is different for $k=2$ and $k>2$, in which cases the expansion is the same as $\tanh^2(\Delta/2T)$ or $\tanh(\Delta/2T)$, respectively. In particular, for $k\neq2$, the result is independent of $k$.
Only in the limit $T\to0$ the two-mode approximation and the multimode approximation agree.
%%%%%%%
\\[9pt]
%%%%%%%
\trivuoto{MyBlue} \ Multipartite entanglement with depth larger than $\kappa$ is found at temperatures $T<T_\kappa$, where 
\be \label{LMGboundarySuper}
T_\kappa\textrm{\small$(-1<\Lambda<0)$} = \frac{\sqrt{1+\Lambda}}{2\,\textrm{arctanh}\left(\pazocal{D}_\kappa\sqrt{1+\Lambda}\right)} \, .
\ee 
In particular, $T_{\kappa=1}$ is interpreted as the minimum temperature 
at which the thermal state exhibits spin fluctuation above the shot-noise limit $\tr[\hat{\rho}_T\hat{J}_y^2] \gtrsim N$.
%%%%%%%
\\[9pt]
%%%%%%%
\trivuoto{MyBlue} \ The region where genuinely quantum correlations are not destroyed by thermal fluctuations 
has a maximum extension over temperature $\Tmax = \max_T T_{\kappa=1}\textrm{\small$(\Lambda<0)$} = \frac{1}{2}$ 
around $\Lambda\approx\Lambdac$, as shown in panel~\ref{fig:ScalingLMG}\Panel{d}. 
The first size correction reads $\frac{1}{2}-\Tmax\sim\sqrt{N}$.
%%%%%%%%
\\[-6pt]

\noindent \tripieno{MySky} \ For $\Lambda=-1$, the QFI can be calculated in accordance with a scaling ansatz~\cite{HaukeNATPHYS2016}
and it was found to satisfy the universal law $F_Q=N^{4/3}\phi(T N^{1/3})$, with $\phi$ a suitable real function. 
An accurate data collapse confirms this universal behaviour and a fit on our numerical data supplies 
the actual profile $\phi(x)=\tanh\frac{1}{2x}$.
%%%%%%%
\\[9pt]
%%%%%%%
\noindent \trivuoto{MyBlue} \ In particular, $F_Q \sim N^{4/3}$ independent of $T$ at low enough temperature $T\ll N^{-1/3}$. 
We remind that $\DeltaE\sim N^{-1/3}$, so the latter condition simply means $T\ll\DeltaE$.
%%%%%%%
\\[6pt]
%%%%%%%
\noindent \trivuoto{MyBlue} \ On the other hand, $F_Q \simeq \frac{1}{2}\,T^{-1}$ as a function of temperature at $T\gtrsim N^{-1/3}$.
We remark that this is a true asymptotic behaviour at any large $T$,
differently from the case of the short-range Ising in chapter~\ref{ch:Ising} 
-- where the power-law decay $F_Q \sim T^{-3/4}$ holds only within an upper-bounded interval $T\ll J$~\cite{Gabbrielli2018a}.
%%%%%%%%
\\[-6pt]

\noindent \tripieno{MySky} \ The behaviour of witnessed multipartite entanglement dramatically changes for $\Lambda<-1$, 
because of the vanishing gap: see panel~\ref{fig:LMGthermic1}\Panel{c,\,d}.
It is convenient to distinguish the case of finite $N$, where the energy gap $\DeltaE$ between the ground state and the first excited state is small but finite, and the thermodynamic limit $N\to\infty$, where $\DeltaE=0$.
%%%%%%%
\\[9pt]
%%%%%%%
\trivuoto{MyBlue} \ At finite $N$, the large amount of entanglement hosted by the ground state is lost 
even at an exponentially small temperature: for $T\lesssim\Delta_1\sim\exp(-N/n_{\Delta_1})$, the QFI undergoes a rapid decay
\be \label{ferroQFI2modes} 
F_Q[\hat{\rho}_T] = F_Q[\ket{\psi_0}]\,\tanh^2 \Big(\frac{\DeltaE}{2T}\Big)\,.
\ee
Multipartite entanglement detected by the QFI in the gapless ordered phase is extremely fragile against thermal fluctuations.
%%%%%%%
\\[9pt]
%%%%%%%
\trivuoto{MyBlue} \ For a temperature larger than the gap $\DeltaE \ll T \lesssim \DeltaEE$ and sufficient to populate the second doublet, 
the system behaves as an effective two-mode model with a gap $\DeltaEE$: according to Eq.~(\ref{QFIfactorization}) we have 
\be \label{ferroQFI4modes}
F_Q[\hat{\rho}_T] = F_Q[\hat{\rho}_0]\,\tanh^2 \Big(\frac{\DeltaEE}{2T}\Big) \, , \quad {\rm with} \ \DeltaEE=\sqrt{\Lambda^2-1} \, ,
\ee
where $\mu=2$ and $\nu=2$, % we account for the double degeneracy of the ground state ($\mu=2$) and the first excited state ($\nu=2$),
$\hat{\rho}_0 = \frac{1}{2}\big(\ket{\psi_0}\bra{\psi_0} + \ket{\psi_1}\bra{\psi_1}\big)$ is the 
thermal incoherent mixture of the two quasi-degenerate states $\ket{\psi_0},\,\ket{\psi_1}$ of the first doublet, and 
\be \label{ferroQFIplateau}
F_Q[\hat{\rho}_0] = 2 \sum_{n=0,\,1} \Big[\,\bra{\psi_n}\hat{J}_z^2\ket{\psi_n} - \sum_{m=0,\,1}\big|\bra{\psi_n}\hat{J_z}\ket{\psi_m}\big|^2\,\Big]
% \sum_{m=0,\,1} \big| \bra{\psi_n} \hat{J}_z \ket{\psi_m} \big|
= \frac{N}{|\Lambda|\sqrt{\Lambda^2-1}} \, .
\ee
It should be noticed that Eq.~(\ref{ferroQFIplateau}) is a factor $N$ smaller than the zero-temperature value $F_Q[\ket{\psi_0}]$.
In the thermodynamic limit $N\to\infty$, we thus have a discontinuous jump of the QFI at finite temperature.
%%%%%%%
\\[9pt]
%%%%%%%
\trivuoto{MyBlue} \ For $T\gg\Delta_1$ we can calculate the QFI using a $k$-mode approximation (taking the first $k$ states 
in each harmonic well the potential $V(z)$ is made up of):
\be \label{QFIkmodesSub} 
F_Q[\hat{\rho}_T] = F_Q[\hat{\rho}_0]\,\tanh \Big(\frac{\DeltaEE}{2T}\Big) \cdot
\bigg( 1 - \frac{k}{2} \frac{\neper^{\,\Delta_2/T}-1}{\neper^{\,k\Delta_2/2T}-1} \bigg),
\ee
that reduces to Eq.~(\ref{ferroQFI4modes}) when taking into account only the first two doublets $k=4$, and becomes
\be \label{LMGferroQFIfull}
F_Q[\hat{\rho}_T] = F_Q[\hat{\rho}_0] \tanh \Big( \frac{\Delta_2}{2T} \Big)
\ee
when taking the limit $k \to \infty$.
The QFI in the different regimes is illustrated in panel~\ref{fig:LMGthermic1}\Panel{f}, 
where we underline a very good agreement between the analytical predictions and the numerical results. 
Also in this case, for large enough temperature $T \gg \Delta_2$, we recover $F_Q[\hat{\rho}_T] \propto 1/T$.
%%%%%%%
\\[9pt]
%%%%%%%
\trivuoto{MyBlue} \ For $\LambdaStar<\Lambda<-1$, the region where thermal entanglement is witnessed shrinks. 
Yet, thermal $(\kappa\!+\!1)$-partite entanglement survives at $T<T_\kappa$, with 
\be \label{LMGboundarySub}
T_\kappa\textrm{\small$(\LambdaStar<\Lambda<-1)$} = \frac{\sqrt{\Lambda^2-1}}{2\,\textrm{arctanh}\Big({\pazocal{D}_k|\Lambda|\sqrt{\Lambda^2-1}}\Big)} \, .
\ee
For $\Lambda<\LambdaStar=-\sqrt{\varphi}\approx-1.272$, not only spin squeezing cannot characterize the ground state, 
but even no entanglement is detected at any finite temperature in the large-$N$ limit, 
as depicted in panels~\ref{fig:ScalingLMG}\Panel{a,\,c}.
A sort of \emph{sudden evaporation} of the capability of the QFI to detect multipartite entanglement 
takes place passing from $T=0$ to arbitrary small $T>0$. 
A similar intriguing phenomena occurs for the Ising model.
%%%%%%%
\\[9pt]
%%%%%%%
\trivuoto{MyBlue} \ We notice that the nonanalyticity of the QFI in $\Lambda=-\sqrt{2}\approx-1.414$ 
(the derivative $\partial_\Lambda F_Q[\hat{\rho}_T]$ experience a finite discontinuity at all values $T\neq0$) 
is not a signal of criticality at all, but just a consequence of the sudden change of the optimal interferometric direction  
from the $z$ axis ($-\sqrt{2}<\Lambda<0$) to a direction on the $x$-$y$ plane ($\Lambda<-\sqrt{2}$).

%%%%%%%%%%%%
%% FIGURA %%
%%%%%%%%%%%%

\paragraph{Universal thermal phase diagram}
Finally, we report on the quantum critical region at finite temperature in the light of persistence of multipartite entanglement.
In order to level out the thermal phase diagrams and allow for a direct comparison 
with the one for the Ising model of chapter~\ref{ch:Ising} (and for the Kitaev chain of chapter~\ref{ch:Kitaev}), 
we adopt -- only temporarily -- an angular parametrization of the LMG Hamiltonian: 
\be \label{HamLMGtheta}
\hat{H}_{\rm LMG} = \frac{\Omega}{N}\sin\theta\,\hat{J}_z^2 \, + \, \Omega\cos\theta\,\hat{J}_x \, .
\ee
We believe that the phase diagram consequent from the latter formulation also helps in avoiding possible misunderstandings,
because it has the advantage to map the unbounded set of values for the driving parameter $\Lambda\in\mathbb{R}$ 
into the bounded interval $\theta\in\big[-\frac{\pi}{2},+\frac{\pi}{2}\big]$.
The critical point is located at $\thetac=-\frac{\pi}{4}$ 
and we limit the next discussion to ferromagnetic coupling $-\frac{\pi}{2}<\theta<0$.

Figure~\ref{fig:LMGthermic2}\Panel{a} reports the multipartiteness diagram. 
This plot can be straightforwardly obtained from the one in panel~\ref{fig:LMGthermic1}\Panel{a} 
performing the transformation $\Lambda\mapsto\tan\theta$ and $T\mapsto\sec\theta\;T/\Omega$.
Note that multipartite entanglement detected by the QFI survives up to $\Tmax=\max_\theta{T_{\kappa=1}}\approx0.4\,\Omega$. 
This threshold is lower than the one for the Ising model (see section~\ref{subsec:IsingthermalME}), 
due to the more classical character of this all-to-all model.

Figure~\ref{fig:LMGthermic2}\Panel{b} shows the phase diagram $F_Q[\hat{\rho}_T]\big/F_Q[\hat{\rho}_0]$ in the $\theta$--$T$ plane,
where $\hat{\rho}_0=\ket{\psi_0}\bra{\psi_0}$ for $-\frac{\pi}{4}<\theta<0$ and 
$\hat{\rho}_0 = \frac{1}{2}\big(\ket{\psi_0}\bra{\psi_0} + \ket{\psi_1}\bra{\psi_1}\big)$ for $-\frac{\pi}{2}<\theta<-\frac{\pi}{4}$.
Witnessed multipartite entanglement in the ground state survives at $T \lesssim \Tcross(\theta)$: 
the crossover temperature follows the first nonvanishing energy gap, $\DeltaE$ for for $-\frac{\pi}{4}<\theta<0$
or $\DeltaEE$ for $-\frac{\pi}{2}<\theta<\frac{\pi}{4}$, apart from a constant multiplication factor $\DeltaE/\Tcross\approx2.40$. 
The diagram has the characteristic V-shape illustrated in the fundamental Fig.~\ref{fig:thermalDecay}.

\section{The unbalanced model} \label{sec:asymmetricLMG}
In case of longitudinal field $\delta$ acting on the spins is not zero, an imbalance between the two modes ``spin up'' and ``spin down''
in introduced and the general LMG Hamiltonian, that we propose again here for convenience
\be
\hat{H}_{\rm LMG} = \frac{\Lambda}{N}\,\hat{J}_z^2 - \,\hat{J}_x + \delta\,\hat{J}_z \, ,
\ee
is no more symmetric under spin flip. In the fully-symmetric subspace, it can been successfully implemented using either 
a polarized BEC confined in an asymmetric optical double-well potential~\cite{TrenkwalderNATPHYS2016}
or a spinor BEC invested by laser light far-detuned from resonance~\cite{MuesselPRA2015}.

%%%%%%%%%%%%
%% FIGURA %%
%%%%%%%%%%%%
\begin{figure}[t!]
\centering
\hfill
\includegraphics[width=0.28\textwidth]{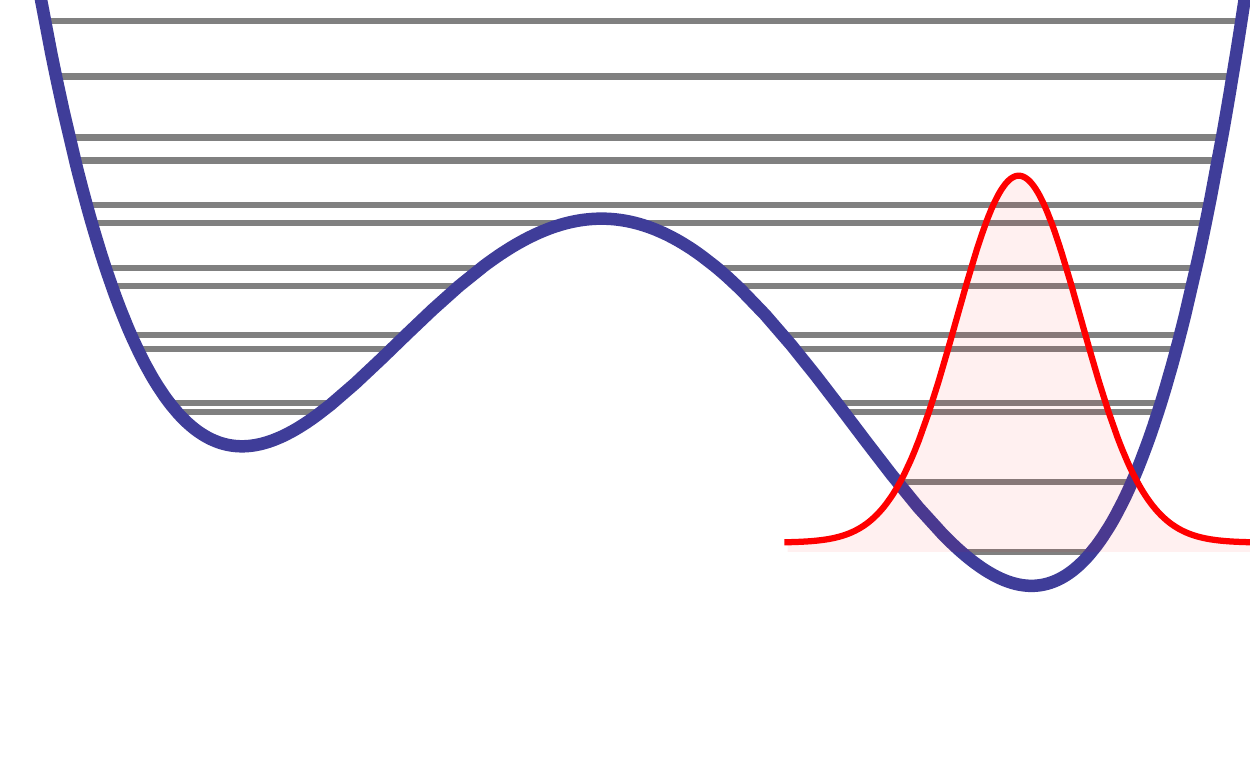} \hfill 
\includegraphics[width=0.28\textwidth]{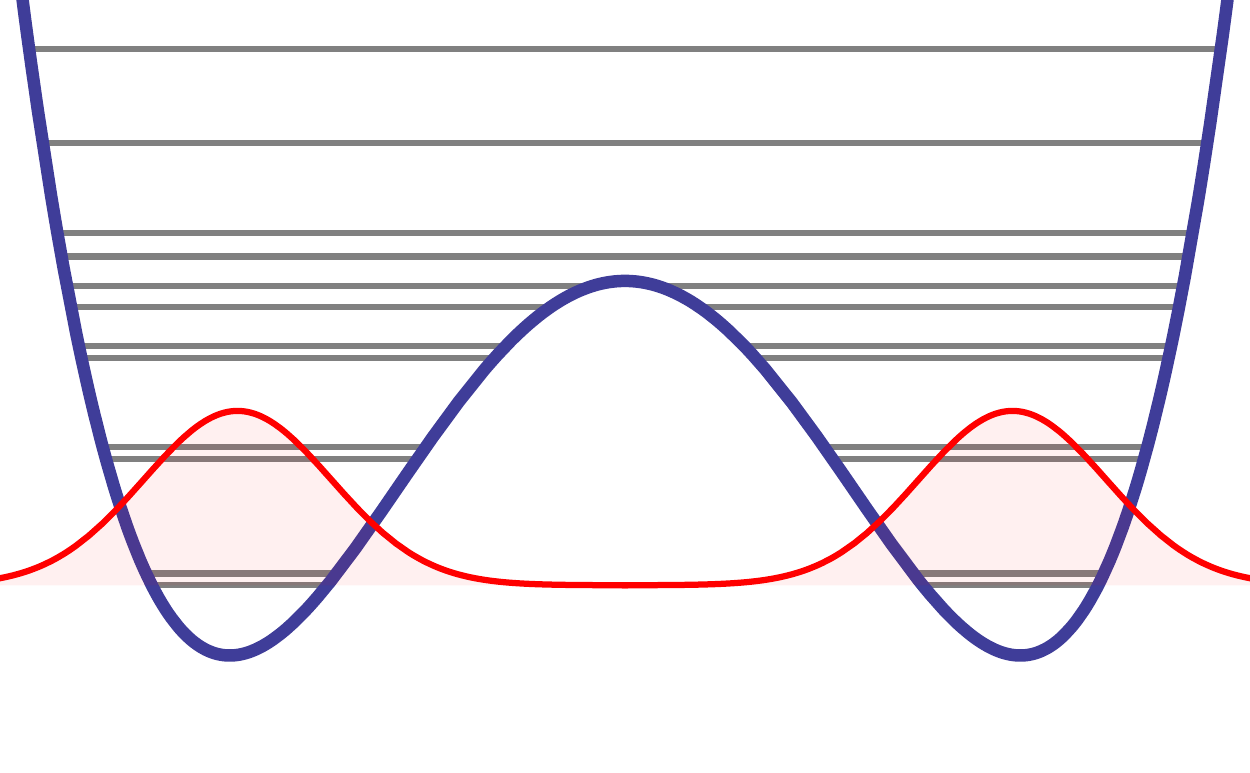} \hfill
\includegraphics[width=0.28\textwidth]{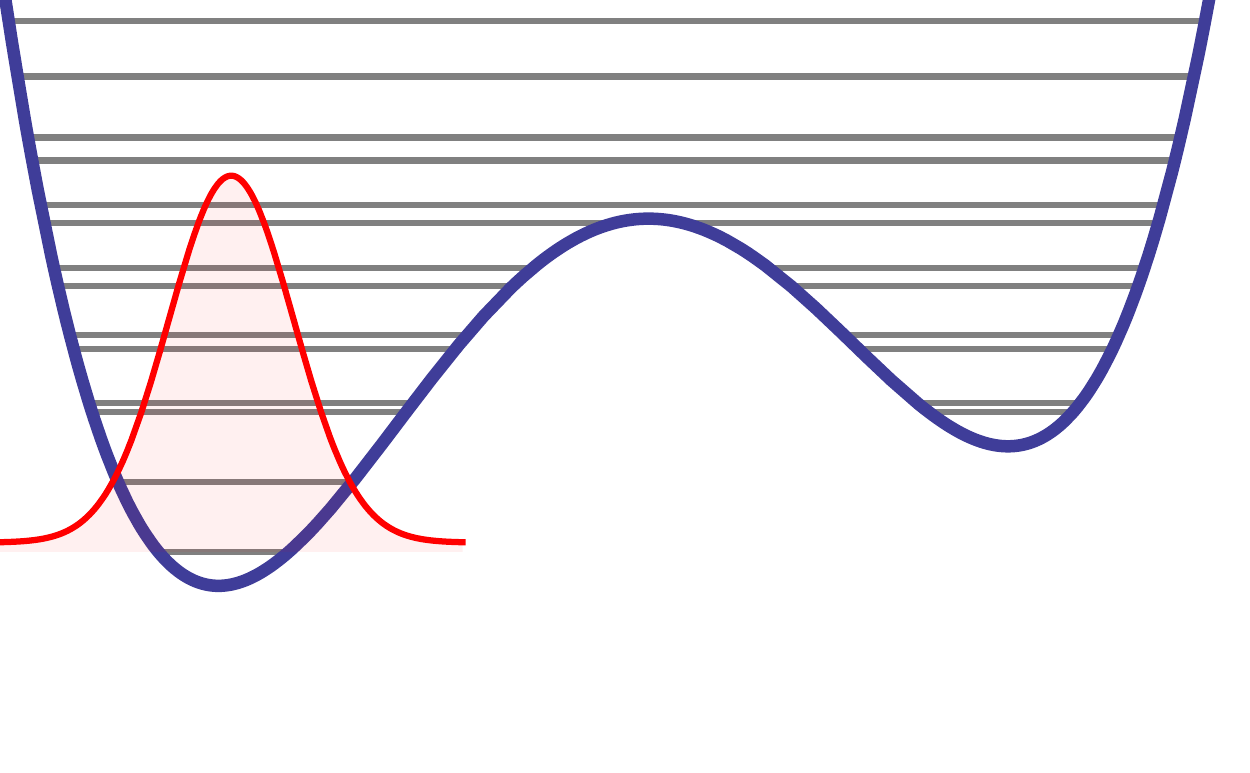} \hfill \\
\textcolor{white}{.} \hfill \quad \ \ \ \fcolorbox{MyBlue}{MyLightBlue}{\textcolor{Navy}{$\delta<0$}} \hfill \textcolor{white}{.} \hfill \quad \quad \ \ \ \fcolorbox{MyBlue}{MyLightBlue}{\textcolor{Navy}{$\delta=0$}} \hfill \textcolor{white}{.} \hfill \quad \quad \quad \fcolorbox{MyBlue}{MyLightBlue}{\textcolor{Navy}{$\delta>0$}} \hfill \textcolor{white}{.}
\caption{Description of the first-order quantum phase transition in the unbalanced LMG model 
in terms of the effective semiclassical potential (blue). 
The qualitative distribution of the energy levels (gray) as well as the amplitude of the single-particle ground state (red) are shown.}
\label{fig:LMGdeformationQPTdelta}
\end{figure}

\subsection{Discontinuous quantum phase transition} 
The additional control of the strength $\delta$ of the longitudinal field 
permits to drive a first-order QPT within the ferromagnetic phase $\Lambda<-1$ (Fig.~\ref{fig:LMGdeformationQPTdelta}): 
the order parameter $\varphi_{\rm LMG}=\frac{2}{N}\langle\hat{J}_z\rangle$ 
shows a finite jump from positive to negative values at the critical point $\deltac=0$.
In the thermodynamic limit, $\varphi_{\rm LMG}\equiv\langle z \rangle$ can be regarded as the center of mass of the 
global ground-state wave function $\psi_0(z)$ according to the Ginzburg-Landau potential in Eq.~(\ref{ShcrodingerlikeLMG}).
The parameter $\delta$ acts as a controllable symmetry-breaking term that breaks the symmetry under exchange of modes
$\hat{\Pi}:\mathsf{a}\leftrightarrow\mathsf{b}$: in fact $[\hat{H}_{\rm LMG},\hat{\Pi}]\neq0 \ \Leftrightarrow \ \delta\neq0$.

\paragraph{Signatures}
Once again, the origin of QPT has an important manifestation in the spectral structure. 
In the limit $N\to\infty$, for any $\delta\neq0$ the spectrum is gapped even for $\Lambda\leq-1$,
but the first energy gap $\DeltaE$ vanishes for $\delta$ approaching the critical value $\deltac=0$: 
$\DeltaE\textrm{\small$(\delta\neq0)$}\sim\pazocal{O}(1)$ and $\DeltaE\textrm{\small$(\delta=0)$}=0$. 
However, at finite $N$ we find\footnote{\ At the special point $\Lambda=-1$, where the behaviour of the order parameter is continuous, 
we find $\DeltaE(\delta)\sim{N}^{-1/3}$ for $\delta\in(-\Delta\delta_\Lambda,+\Delta\delta_\Lambda)$
and the width of the interval vanishes as $\Delta\delta_\Lambda\sim{N}^{-1}$.\\[-9pt]} 
$\DeltaE(\delta)\sim\exp(-N/n_{\DeltaE})$ within an interval 
$\delta\in(-\Delta\delta_\Lambda,+\Delta\delta_\Lambda)$ whose width itself vanishes as $\Delta\delta_\Lambda\sim\exp(-N/n_\delta)$,
with a characteristic decay scale $n_\delta(\Lambda)\propto|1+\Lambda|^{-1}$. 
Here $n_{\DeltaE}$ is the characteristic decay scale for the energy gap of the balanced LMG model at $\Lambda<1$, 
as defined in section~\ref{sec:symmetricLMG}.

Moreover, in the thermodynamic limit the density of states features a divergent character 
in correspondence of the energy barrier $E_s$ of the semiclassical potential $V(z)$
and a discontinuous jump at the energy imbalance $E_m \simeq 2\delta\sqrt{\Lambda^2-1}/|\Lambda|$ 
between the two tilted wells of $V(z)$~\cite{GilmorePRA1986,RibeiroPRL2007}.

\begin{figure}[t!]
\textcolor{white}{.} \hfill
\includegraphics[width=0.48\textwidth]{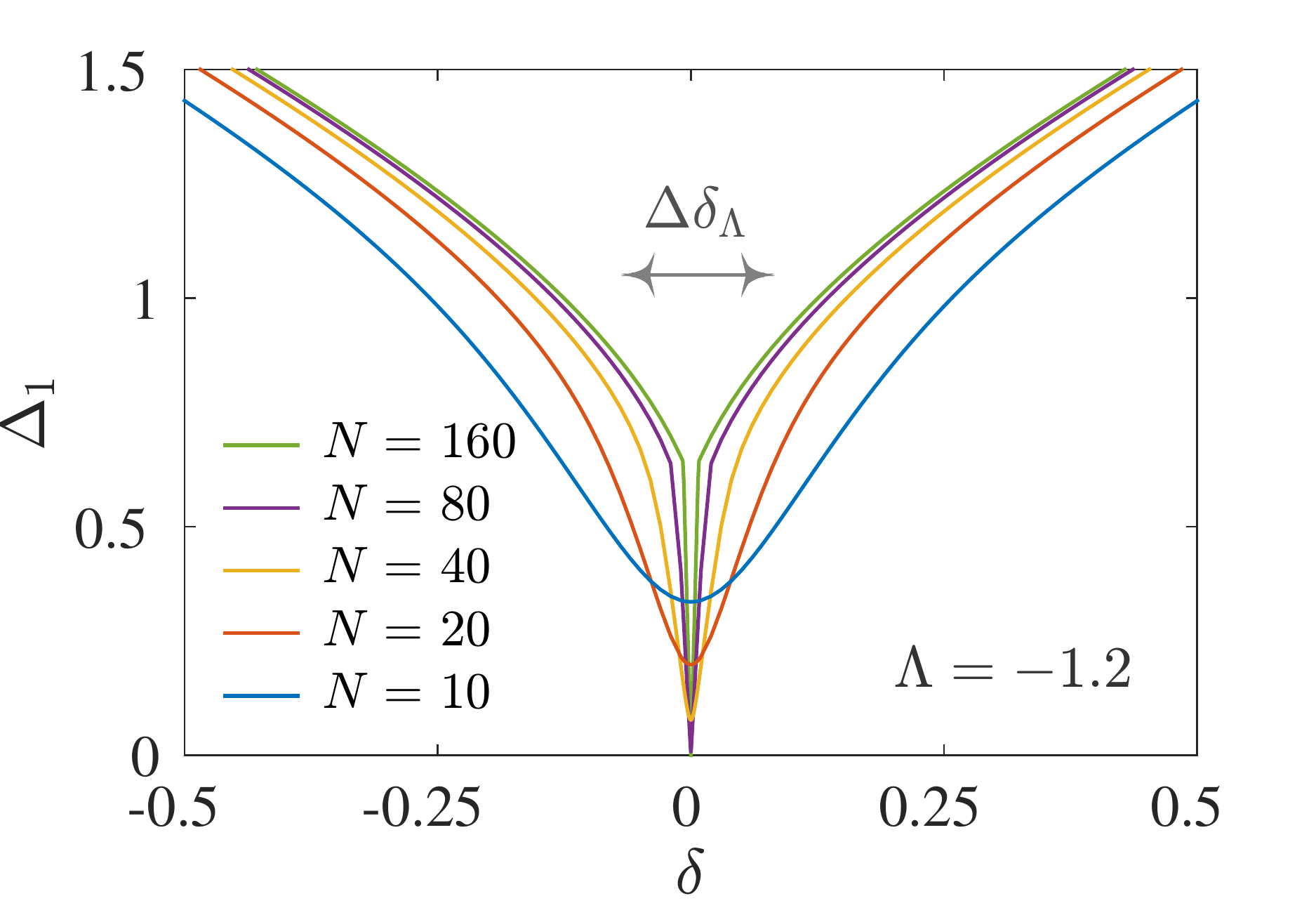} \hspace{-15pt}
\includegraphics[width=0.48\textwidth]{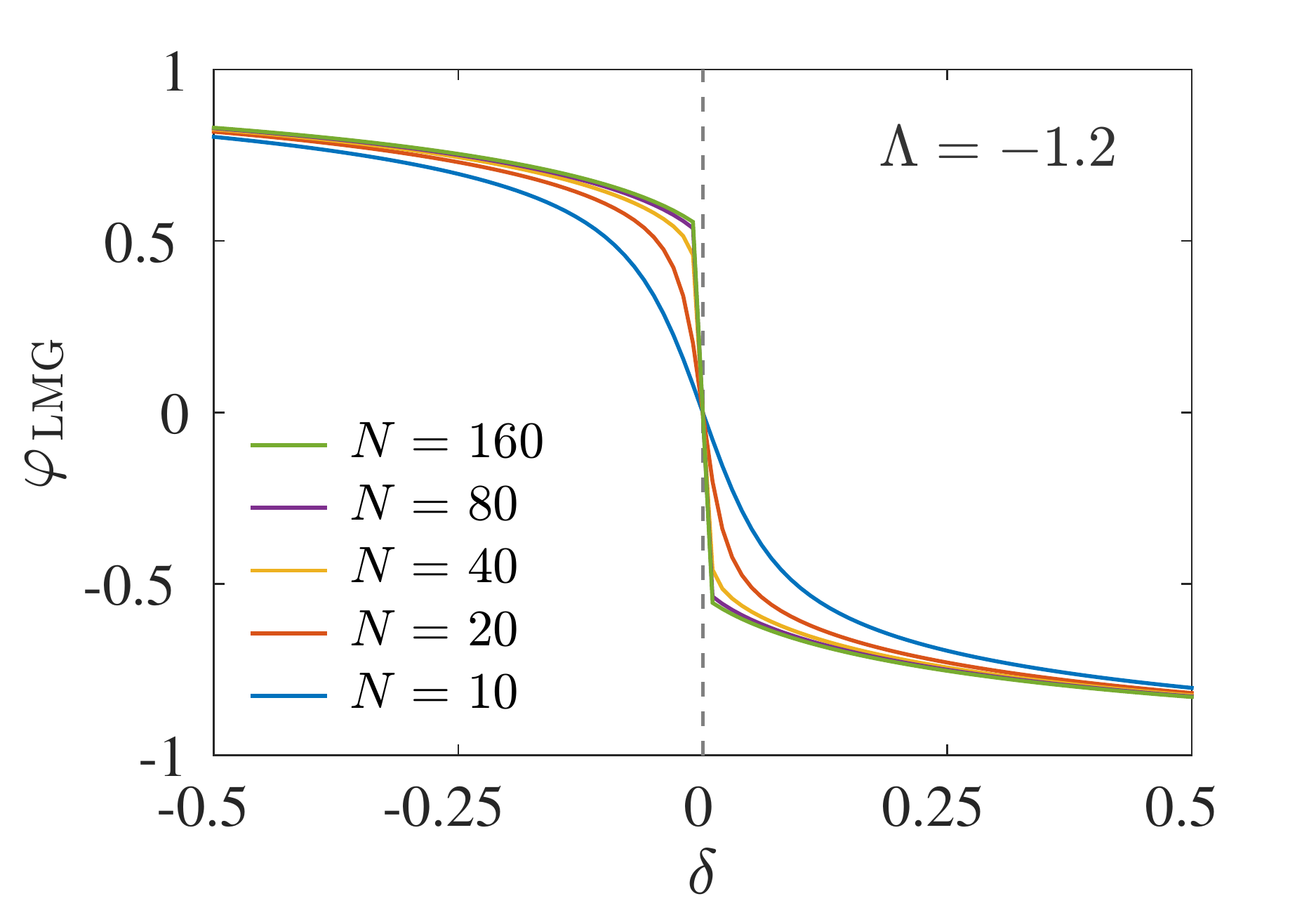} \hfill \textcolor{white}{.} \\
\textcolor{white}{.} \hspace{1pt}
\includegraphics[width=0.445\textwidth]{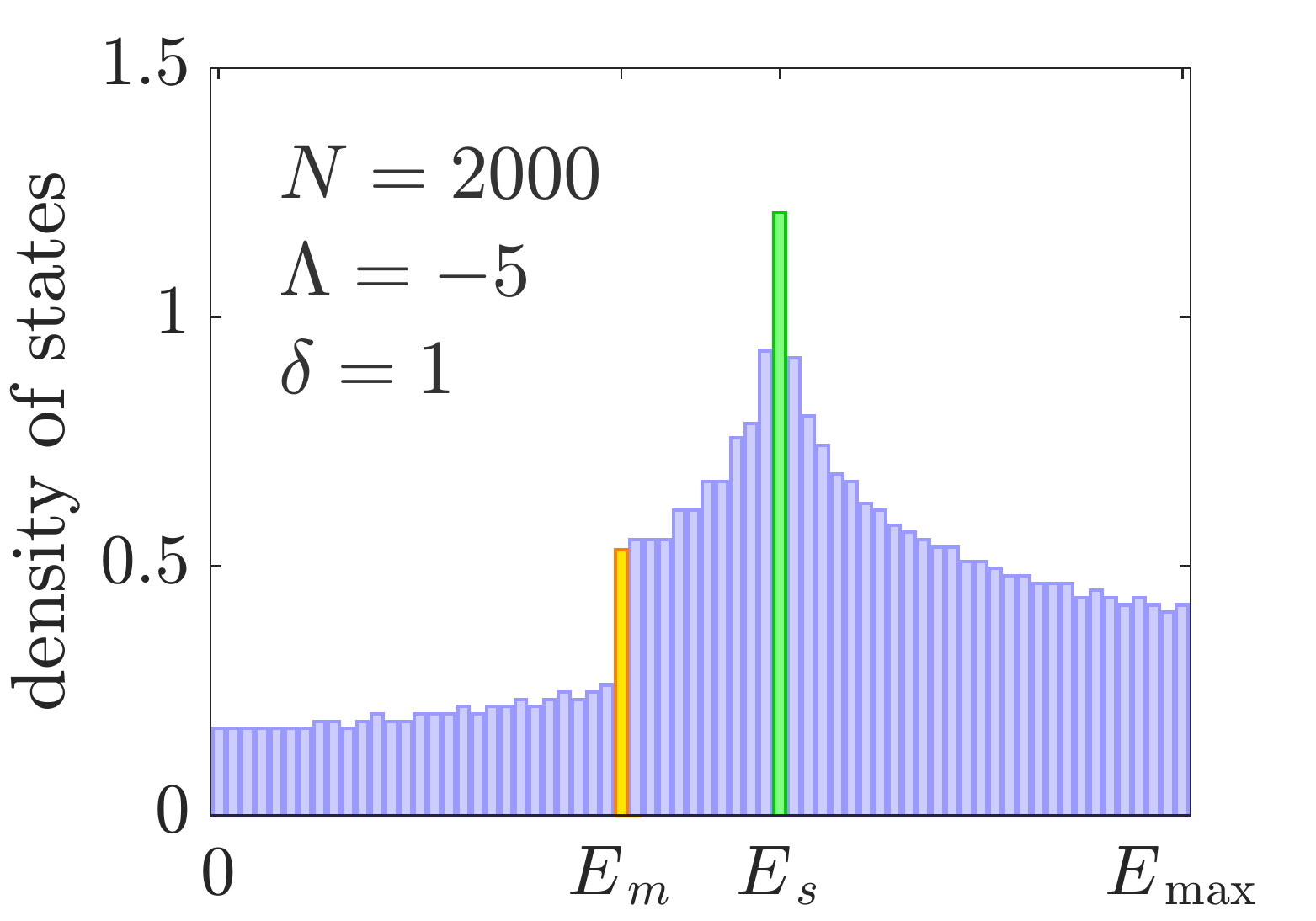} \hspace{42pt}
\includegraphics[width=0.35\textwidth]{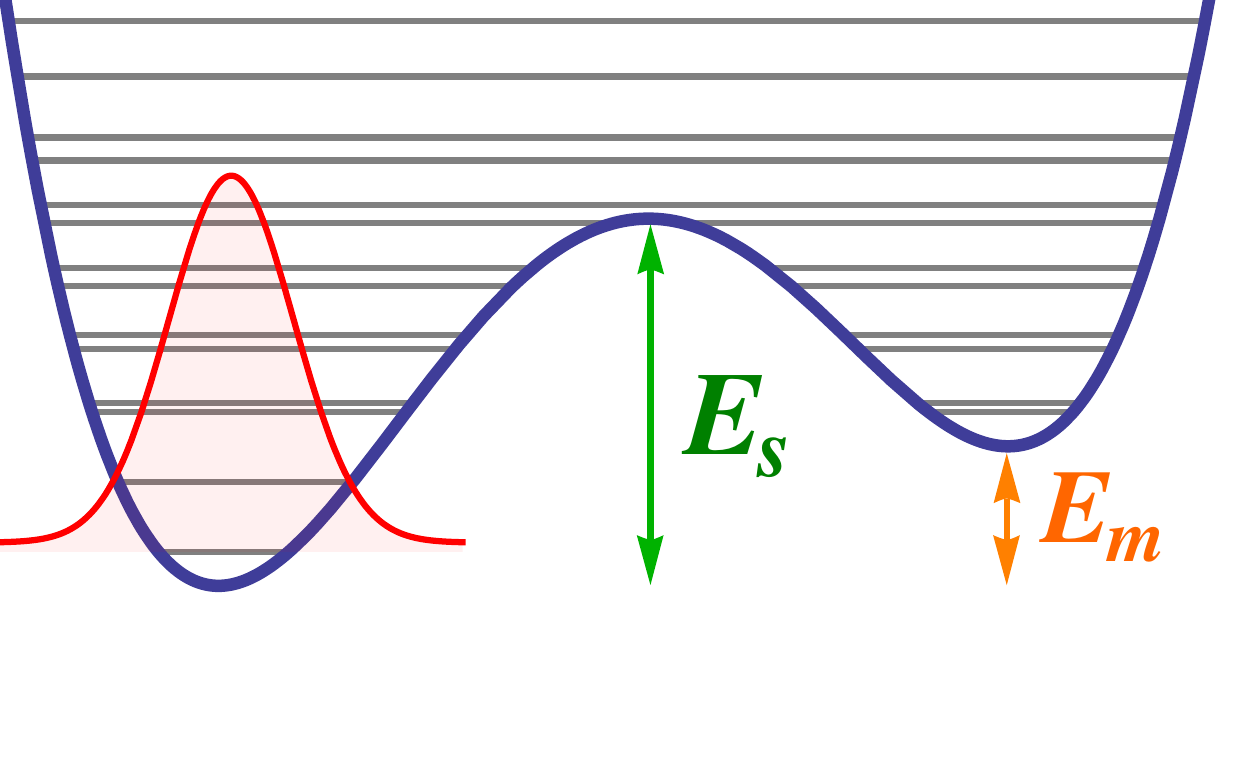} \hfill \textcolor{white}{.}
\caption{Signatures of the first-order quantum phase transition in the unbalanced LMG model in the subcritical regime $\Lambda<\Lambdac$. 
\textbf{(Top left)} Energy gap between the ground state and first excited state as a function of the control parameter $\delta$
for several values of $N$. At $\delta=0$, the gap closes $\DeltaE\to0$ exponentially fast for increasing $N$ 
and remains very small within an interval of width $\Delta\delta_\Lambda$ around $\delta=0$.
Even the width $\Delta\delta_\Lambda$ vanishes exponentially fast for increasing $N$.
\textbf{(Top right)} Order parameter $\varphi_{\rm LMG}=2\langle\hat{J}_z\rangle/N$ as a function of $\delta$ for several values of $N$; 
the vertical dashed line flags the critical point $\deltac=0$.
\textbf{(Bottom left)} Density of states for $N=2000$ spins and $\delta>0$; 
the finite jump at the metastable energy $E_m$ and the divergence at the saddle energy $E_S$ are highlighted.
\textbf{(Bottom right)} Pictorial description of the metastable energy $E_m$ as the lower level of the metastable band 
and the saddle energy $E_s$ as the level corresponding to the barrier height of the effective potential $V(z)$.}
\label{fig:signaturesLMGdelta}
\end{figure}

The collection of signatures of the first-order QPT is displayed in Fig.~\ref{fig:signaturesLMGdelta}.
Incidentally, we note that the fidelity approach is not viable here: 
the differentiation providing the fidelity susceptibility is not well defined at $\delta=\deltac$, neither at finite spin number.

%%%%%%%%%%%%
%% FIGURA %%
%%%%%%%%%%%%

\paragraph{Metastability}
Experimentally, this transition has been recently observed in an ultracold gas with tunable attractive interactions, 
trapped in an optical double-well potential with tunable energy imbalance between the two wells~\cite{TrenkwalderNATPHYS2016}
and it is associated to metastability and hysteresis. 
With reference to Fig.~(\ref{fig:DWInterferometer}) for a simple visualization of the physical parameters,
the strong attractive interaction $\Lambda<-1$ between atoms forces the BEC to remain localized in the well
where it was prepared, even if its energy minimum is raised above the other well.
When $|\delta|$ exceeds a threshold value $\delta_{\rm max}$, 
an instability steers the gas down to the absolute minimum of the trapping potential.

A physical insight on metastability and the hysteretic behaviour it produces 
is provided by the mean-field approach and directly follows from the one-dimensional nature of the effective potential.
The degeneracy between the two stable points of $V(z)$ is lifted by adding an energy shift $\delta$: 
the new absolute and local minima, whose positions $z_{\rm min}$ are given by the condition
\be
\sqrt{1-z_{\rm min}^2} = - \frac{z_{\rm min}}{\delta+\Lambda\,z_{\rm min}} > 0\, ,
\ee
are the stable and metastable points of the semiclassical model. 
% *** PARTE SPERIMENTALE ***
%At zero temperature, the confinement of the system persist as long as 
%the interwell tunnelling rate of the single atoms in the physical double-well trap
%is larger than the quantum tunnelling rate of the semiclassical wave function through the barrier of the effective potential 
%(depending on the energy splitting between the two lowest levels of the metastable band).
The metastable state can be found expanding the Schr\"odinger-like equation (\ref{ShcrodingerlikeLMG}) around $z_{\rm min}$: 
\be
\frac{N}{2}\bigg[\!-\frac{2}{N^2}\bigg(\!\!-\frac{z_{\rm min}}{\delta+\Lambda\,z_{\rm min}}\bigg)\frac{\ud^2}{\ud z^2} \,+\, \frac{1}{2}\bigg(\Lambda+\frac{1}{(1-z_{\rm min}^2)^{3/2}}\bigg)\big(z-z_{\rm min}\big)^2\bigg]\psi_0(z)=E_0\,\psi_0(z).
\ee 
Inasmuch as the frequency of the harmonic well ought be positive, we can find a metastable point only if 
$|\delta|<\delta_{\rm max}=(\Lambda^{2/3}-1)^{3/2}$.

\subsection{Multipartite entanglement}
As far as we know, no preceding work on entanglement in the balanced LMG model was conducted.
Here we discover that multipartite entanglement is strikingly sensitive not only to second-order QPTs -- as unveiled so far -- 
but also to first-order QPTs, at least in so far as this simple spin model suggests.

%%%%%%%%%%%%%
%% FIGURA %%
%%%%%%%%%%%%
\begin{figure}[t!]
\centering
\includegraphics[width=0.95\textwidth]{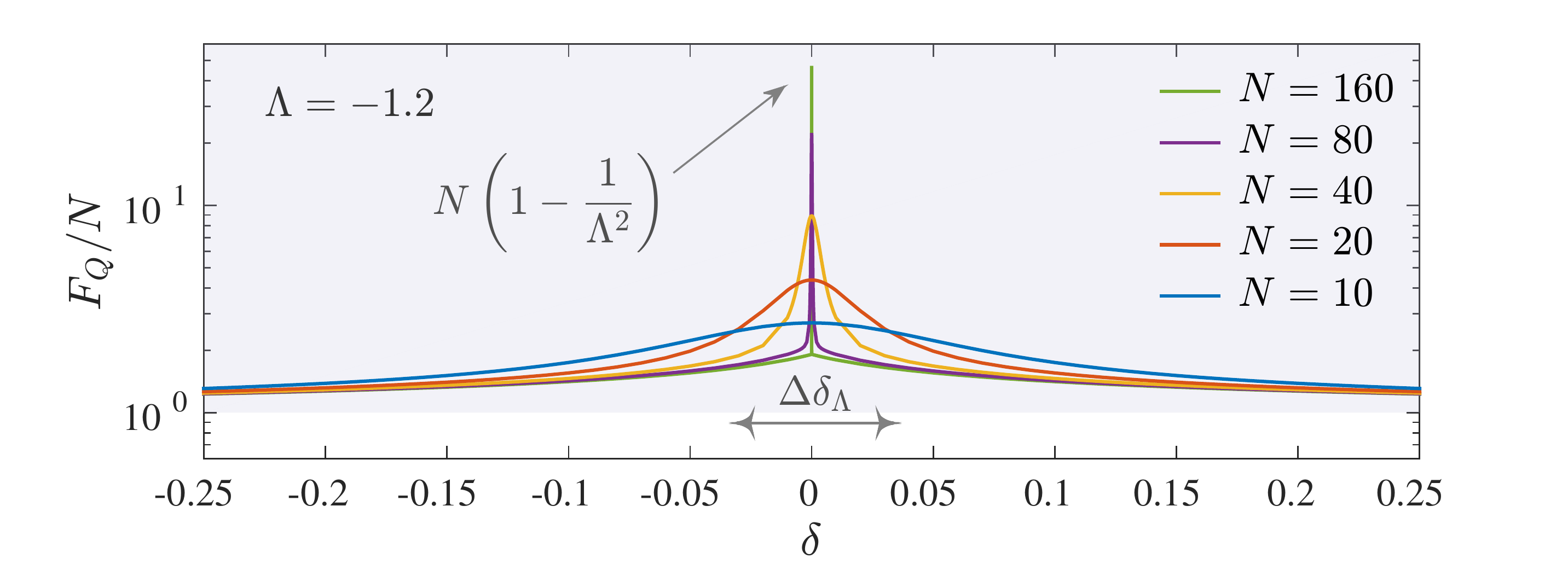}
\caption{Fisher density $F_Q[\ket{\psi_{0}}]/N$ as a function of the longitudinal field $\delta$ for finite size $N$ 
and fixed subcritical interaction parameter $\LambdaStar<\Lambda<\Lambdac$. 
The symbol $\Delta\delta_\Lambda$ denotes the full width at half maximum of the sharp peak around $\delta=0$.
The shaded area indicates multipartite entanglement.} 
\label{fig:LMGT0delta}
\end{figure}

\paragraph{Ground state}
Figure~\ref{fig:LMGT0delta} shows the typical behaviour of multipartite entanglement in the ground state as witnessed by the QFI,
according to an optimal collective rotation whose direction depends nontrivially on $\delta$: the results are obtained via
a numerical exact diagonalization and optimization procedure.
The QFI and the WSS parameter, as well as any spectral property, are even functions of $\delta$: 
it is an obvious consequence of the invariance of the Hamiltonian under the combined transformations of 
field flip $\delta\mapsto-\delta$ and spin flip $\hat{\Pi}:\mathsf{a}\leftrightarrow\mathsf{b}$.

At finite size $N$, the QFI is maximum at the critical point $\delta=0$ and rapidly decay for increasing field strength $\delta$:
the region of large amount of entanglement around $\delta\approx0$ quickly shrinks for increasing~$N$.
Intriguingly, the decay width decreases exponentially fast with increasing~$N$.  
An accurate finite-size analysis reveals that the half-height width of $F_Q[\ket{\psi_{0}}]$ as a function of~$\delta$
is proportional to the width $\Delta\delta_\Lambda\sim\exp(-N/n_\delta)$ of the region where the gap $\DeltaE$ is vanishing.
The interpretation is sharp and simple: at zero temperature, the system conserves the zero-field properties 
(in particular, the superextensive amount of entanglement) inasmuch as the imbalance introduced by the longitudinal field
does not perturb the spectrum spacing, that is for  $|\delta|\ll\DeltaE$. 
For larger values of the longitudinal field, the effective double-well potential is tilted so much that 
the first almost-degenerate doublet is splitted and the system perceive the existence of a new local minimum of the energy:
the ground state deforms from a highly-entangled non-Gaussian quantum superposition of macroscopic wave functions 
to a unimodal squeezed distribution localized in one well: Fig.~\ref{fig:LMGdeformationQPTdelta}. 
Entanglement endures even for high values of $\delta$ and it is due to the squeezing of the state.

In the thermodynamic limit, multipartiteness detected by the QFI is extensive only at the critical point: 
$F_Q[\ket{\psi_{0}}]/N=N\big(1-1/\Lambda^2\big)$ at $\delta=0$, while $F_Q[\ket{\psi_{0}}]/N\sim\pazocal{O}(1)$ elsewhere.
As a consequence, we predict a singular point of $F_Q[\ket{\psi_{0}}]$ at $\deltac$ for any $\Lambda<-1$,
the divergent character of the first derivative $\partial F_Q/\partial \delta$ being a signature of the phase transition itself.
This observation reveals a large sensitivity of metrologically useful entanglement to small fluctuations of $\delta$ around $\delta=0$:
in experimental implementations, it translates to an exceptional delicateness of a double-well atom interferometer 
against the asymmetry of the optical potential during the adiabatic preparation of the entangled input state 
(stage \textsf{a} in Fig.~\ref{fig:DWInterferometer}, to clarify). 

On the contrary, the spin squeezing $1/\WSS$ at $\delta=0$ still exhibits a sharp peak,
even if saturating a constant value for large $N$: it undergoes no divergence.
For large enough $|\delta|$, the two witnesses coincide: $F_Q\approx1/\WSS$.

%%%%%%%%%%%%%
%% FIGURA %%
%%%%%%%%%%%%
\begin{figure}[t!]
\centering
\includegraphics[width=0.45\textwidth]{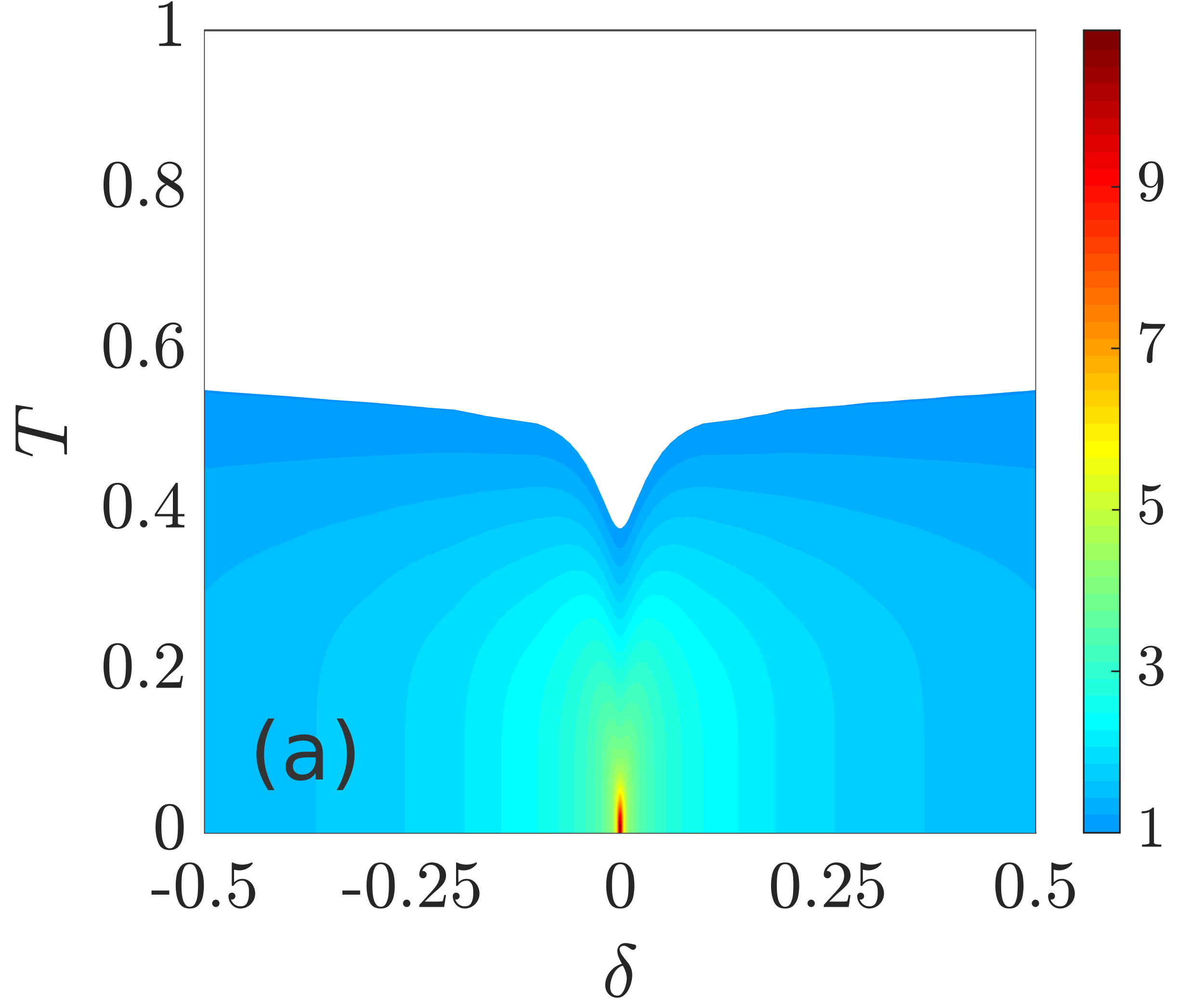}
\includegraphics[width=0.45\textwidth]{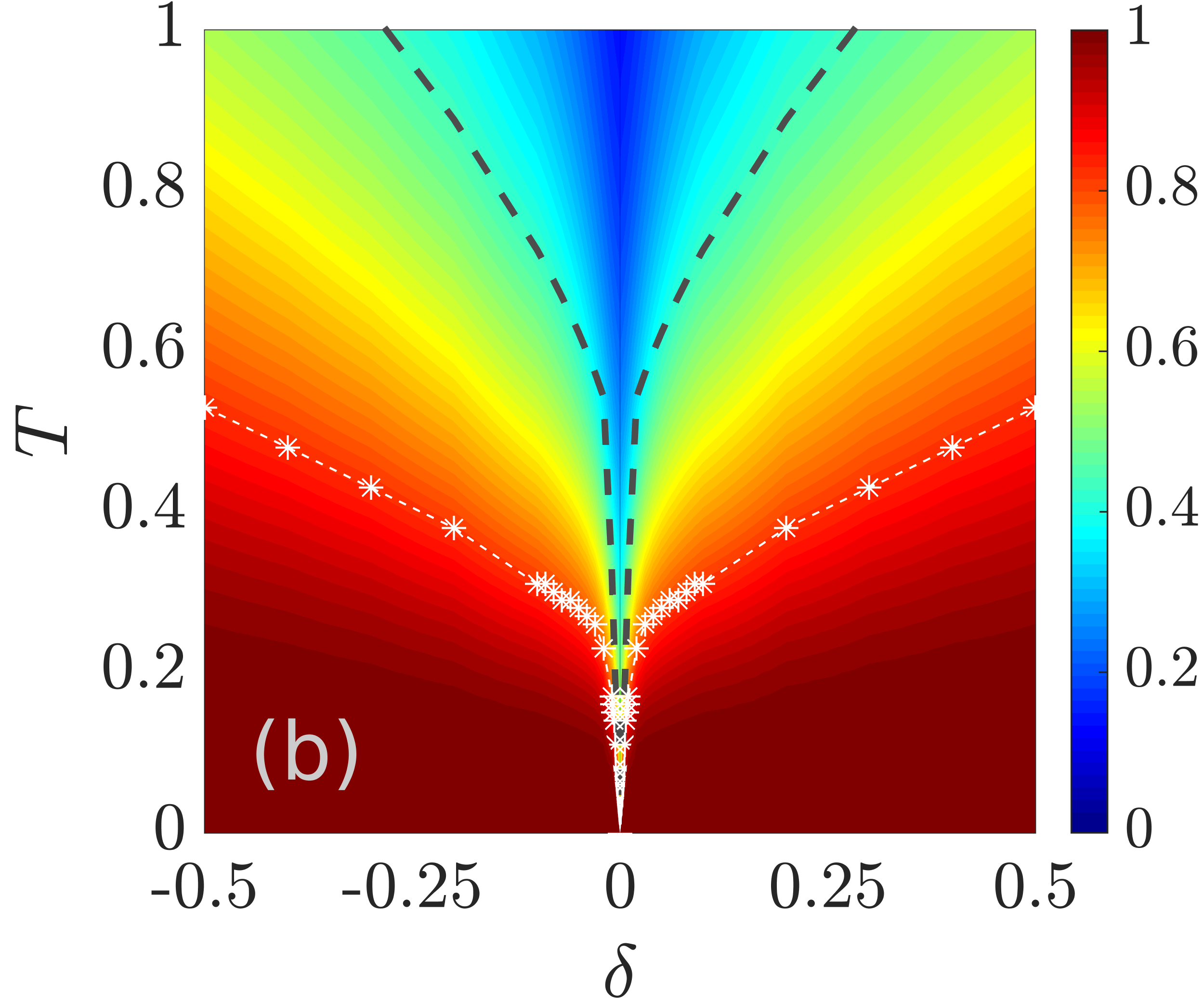} 
\includegraphics[width=0.45\textwidth]{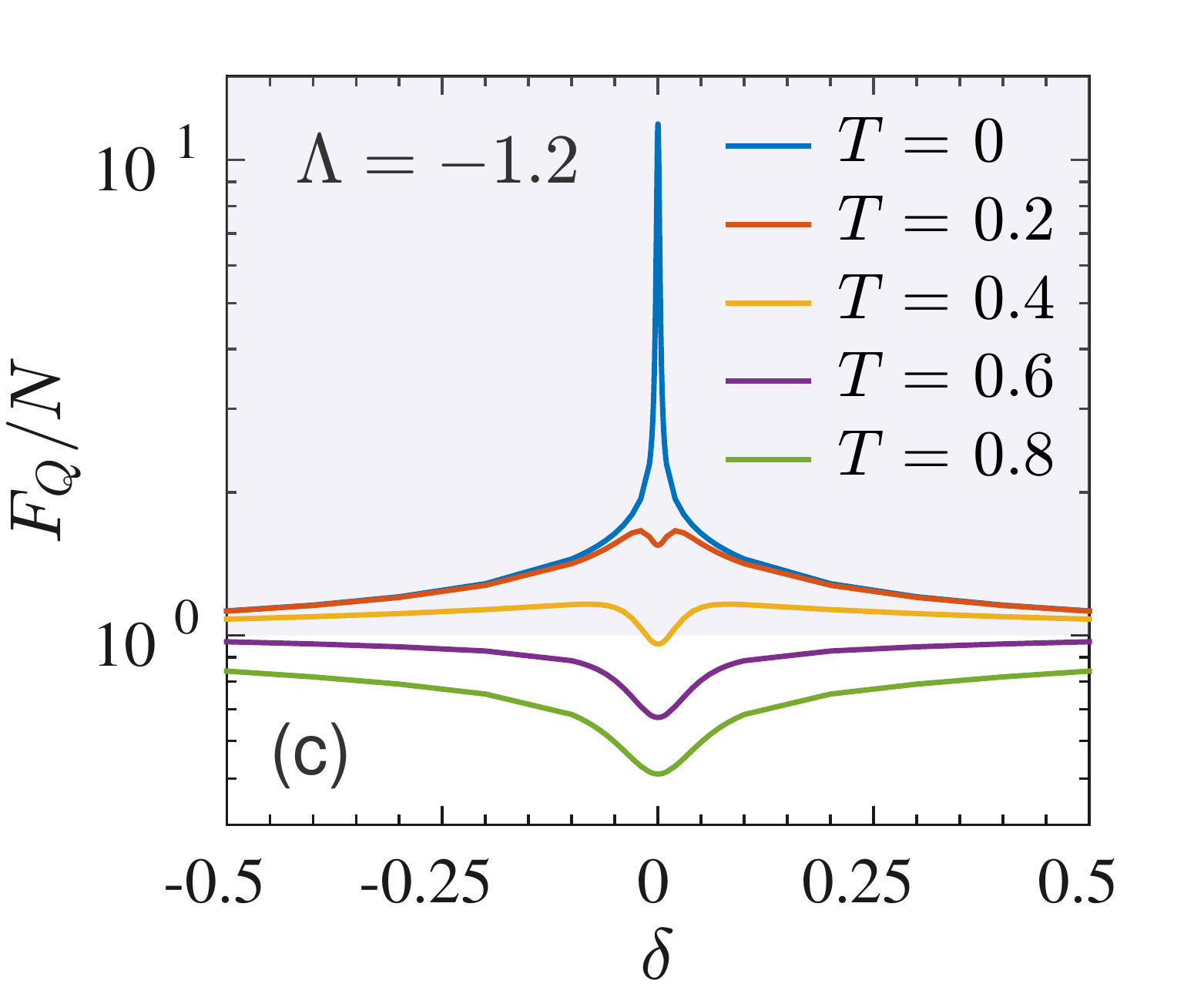}
\includegraphics[width=0.45\textwidth]{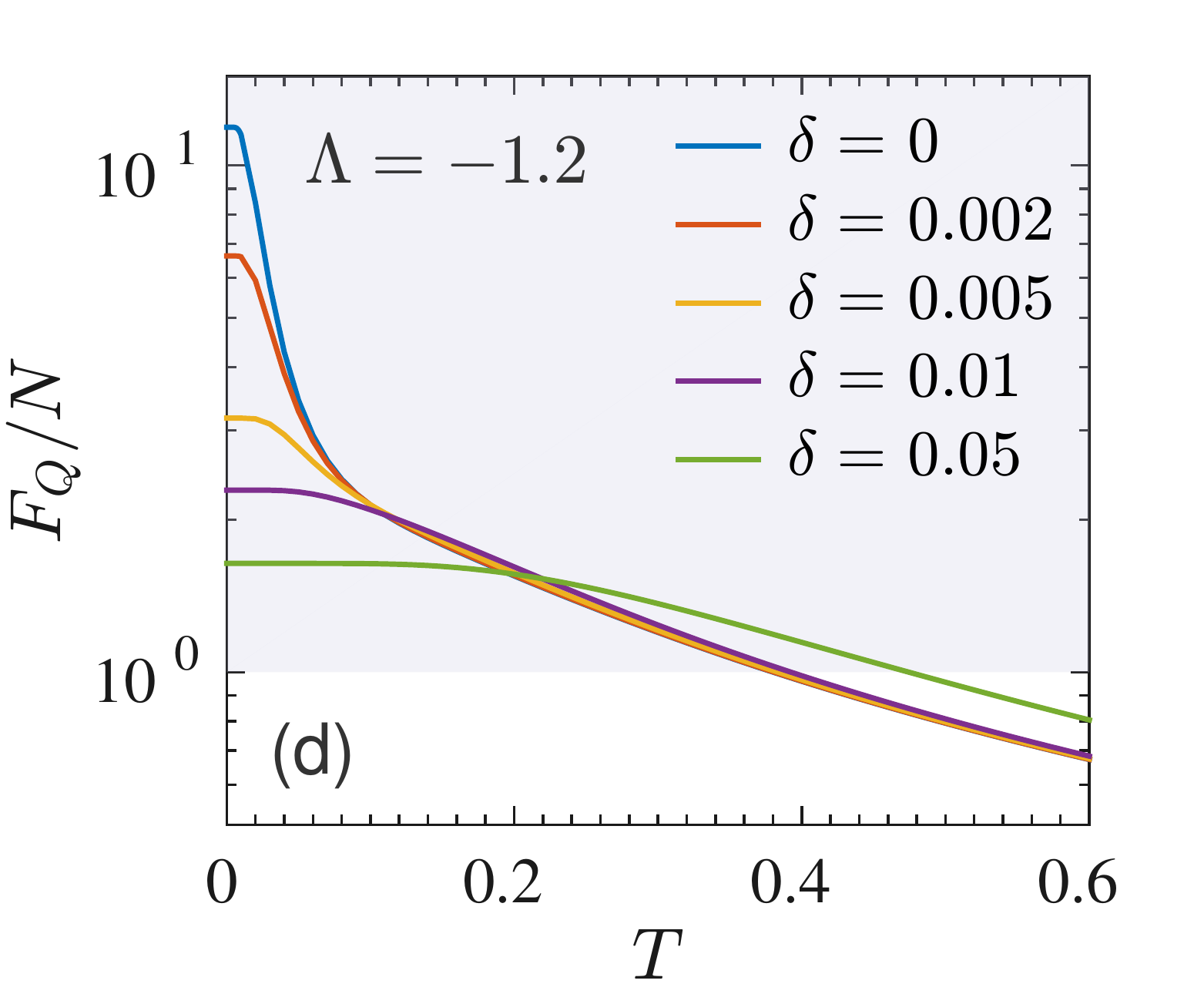}
\caption{Thermal multipartite entanglement witnessed by the QFI in the unbalanced LMG model for $N=50$ spins at $\Lambda=-1.2$. 
\textbf{(a)} Multipartiteness $F_Q[\hat{\rho}_T]/N$ on the plane $\delta$--$T$:
as usual, inside the white region no entanglement is witnessed by the QFI.
\textbf{(b)} Phase diagram of $F_Q[\hat{\rho}_T]/F_Q[\ket{\psi_{0}}]$ highlighting the quantum critical region around $\deltac$.
The crossover temperature $T=\Tcross$ is marked by the white dots; 
the first energy gap $T=\DeltaE$ is traced by the dashed line.
\textbf{(c,\,d)} Cuts of panel (a) at fixed temperature $T$ and asymmetry $\delta$, respectively.} 
\label{fig:thermicLMGdelta}
\end{figure}

\paragraph{Finite temperature}
Figure~\ref{fig:thermicLMGdelta} shows the effect of finite temperature on witnessed multipartite entanglement.
The density plot in panel~\ref{fig:thermicLMGdelta}\Panel{a} illustrates the thermal multipartiteness $F_Q[\hat{\rho}_T]/N$
as a function of $\delta$ in the ferromagnetic phase $\Lambda<-1$; see also supplementary panels~\ref{fig:thermicLMGdelta}\Panel{c,\,d}.
Astonishingly, the robustness of entanglement at finite temperature is slightly enhanced by the asymmetry term $\delta$.
This peculiar phenomenon is due to the combined action of two causes: 
on the one hand, the asymmetry term $\delta$ generates a localization of the low-lying states around the absolute minimum
of one well of the (asymmetric) double-well profile; 
on the other hand, large interactions $\Lambda<-1$ simultaneously generate squeezing in the ground state 
and provide negligible probability of populating the first excited state, since the energy gap scales as $\DeltaE\sim|\Lambda|$.
For large enough values of $|\delta|$, numerics prompt us that $F_Q\approx1/\WSS$ at any $T$ and~$\Lambda$. 

For sufficiently weak asymmetry and low temperature\footnote{\ We quickly clarify these two important, but rather technical, hypotheses
that permit to attempt the simple analytical approach described in the text. 
The longitudinal field must satisfy the constraint $|\delta|\lesssim\sqrt{\Lambda^2-1}$, 
so that the quadratic double-well shape of the potential $V(z)$ in Eq.~(\ref{ShcrodingerlikeLMG}) is not excessively perturbed 
by the linear term and its minima can be taken equal to the ones of the unperturbed ($\delta=0$) double well: 
$z_{\rm min} \approx z_\Lambda^\pm = \pm\sqrt{1-1/\Lambda^2}$. 
Moreover, temperature should be low enough to populate only the levels in the stable well (centred around the absolute minimum),
leaving the probability to populate levels in the metastable well (centred around the local minimum) negligible: 
a quantitative requirement can be written as $T \lesssim N|\delta|\sqrt{1-1/\Lambda^2}$. 
In the limit $N\to\infty$, the latter conditions holds for all temperatures. 
With the numbers in Fig.~\ref{fig:thermicLMGdelta}, for our considerations to be valid we need $|\delta|<0.66$ and $T<18$ 
(both in units of tunnelling rate $\Omega$), conditions abundantly fulfilled in the figure.\\[-9pt]}
an analytical calculation of the WSS parameter is possible 
in terms of unimodal harmonic well centred around $z_{\rm min}=\pm\sqrt{1-1/\Lambda^2}$: 
\be
\frac{1}{\WSS} = \frac{|\Lambda|}{\sqrt{\Lambda^2-1}}\,\tanh\Big(\frac{\DeltaEE}{2T}\Big) \, .
\ee
Then, the approximated location of the temperature delimiting the frontier of witnessed quantum correlations is 
\be
T_{\kappa=1} = \frac{\sqrt{\Lambda^2-1}}{2\,\textrm{arctanh}\left(\frac{\sqrt{\Lambda^2-1}}{|\Lambda|}\right)} \, .
\ee
At this very simple level, the modelling is quite rough and provides no dependence on $\delta$.
Nonetheless, it allows for some quantitative assessment: for instance, with reference to Fig.~\ref{fig:thermicLMGdelta} 
it predicts entanglement constantly for $T\lesssim0.53$, in noticeable agreement with numerical data. 
More importantly, the evaporation point introduced in section~\ref{sec:symmetricLMG} disappears: at fixed finite $T$, 
increasing $|\delta|$ allows to access entangled thermal states even for $\Lambda<\LambdaStar$.

Panels~\ref{fig:thermicLMGdelta}\Panel{c,\,d} further explore multipartiteness at finite temperature.
A dip appears in $F_Q[\hat{\rho}_T]$ at $\delta=0$: at finite temperature, witnessed entanglement is first lost around the critical point, 
while it survives for larger $|\delta|$. 
For sufficiently high temperature $T>T_{\kappa=1}(\Lambda)$, witnessed entanglement is lost for any $\delta$. 
At high temperature, $F_Q[\hat{\rho}_T]$ is a monotonic increasing function of $|\delta|$.
As a function of temperature, $F_Q[\hat{\rho}_T]$ always decays -- slower and slower for increasing $|\delta|$.
We remark that for $|\delta|\gg|\Lambda|$, the model reduces to $\hat{H}_{\rm LMG} \approx \delta \hat{J}_z$,
its eigenstates are Dicke states and the spectrum is approximately equispaced $\Delta_n\approx|\delta|$: 
at $T\ll|\delta|$ just the ground state (which is a coherent spin state) is populated, resulting in $F_Q/N\approx1$.

Finally, another amazing (but not surprising at all) result is reported in panel~\ref{fig:thermicLMGdelta}\Panel{b}.
Also a first-order QPT, characterized by avoided crossing at $\delta=\deltac$ and level repulsion at $\delta\neq\deltac$, 
is entirely able to induce a V-shaped region on the phase diagram control parameter versus temperature:
ground-state entanglement is accessible for finite temperature, up to a crossover where thermal decay occurs.
The crossover temperature is found to follow the energy gap faithfully: $\DeltaE/\Tcross\approx2.40$.

%\paragraph{Multipartitenessnt spectrum // spctrum of multipartite entanglement}
%figure (anche 3d)
%direzione
%punti particolari

\section*{Summary}
We have studied the quantum many-body correlations in the Lipkin-Meshkov-Glick model, at zero and finite temperature, 
for the metrologically relevant case in which the spins $\sfrac{1}{2}$ are practically implemented by two-level indistinguishable bosons.
We mainly focused on the quantum Fisher information and showed that its first derivative with respect to the control parameter
can be regarded as an ``entanglement susceptibility'', diverging at the critical point of both first-order and second-order transition.
Again, a nonanalyticity of the quantum Fisher information signals the onset of a quantum phase transition.

We believe our work sheds some light on the robustness of a two-mode linear atom interferometer and 
supplies a novel comprehension of the recent experimental results on parity-symmetry breaking observed 
in ultracold atoms with tunable \emph{attractive} interactions trapped in an optical double-well potential~\cite{TrenkwalderNATPHYS2016}.
Decoherence poses a significant threat to robust ground-state preparation: 
the tailoring of metrologically-useful entangled states has revealed a big challenge for future quantum technologies. 
Here we have shown that the useful entanglement in a canonical thermal state of the LMG model
is quite robust against temperature in the vicinity of critical points, 
though its appealing Heisenberg scaling is very fragile in the ordered phase.
In general, whenever the probability to populate the first-excited state is nonzero 
or the degeneracy of the ground state is lifted by an external perturbation, 
the intensive amount of multipartiteness useful for applications is lost. 
Yet, we highlighted that entanglement for \emph{repulsively} interacting atoms, in particular its scaling \`a la Heisenberg,
is not affected dramatically by temperature and noise.
% possesses an astonishing robust against temperature, as caught by a basic harmonic model: 

\chapter[Kitaev model]{A topological model: the Kitaev chain} \label{ch:Kitaev} 

{\sl Using the quantum Fisher information, we witness multipartite entanglement in the Kitaev chain, 
a benchmark model for one-dimensional topological superconductor with arbitrary range of the superconducting pairing.
Only nontrivial topological phases, both for short-range and long-range pairing, are 
characterized by a power-law diverging finite-size scaling of multipartite entanglement. 
Moreover, for arbitrary pairing range, topological quantum phase transitions are identified by a sudden variation
of multipartite entanglement, even in the absence of a closing energy gap. 
The effect of temperature on entanglement is also considered and a surprising resistance against thermal noise is reported.}

\section*{Introduction}
Topological insulators are electronic materials hosting conducting edge states in a insulating bulk~\cite{KanePRL2005,HasanRMP2010}. 
In topological superconductors, instead, the bulk itself is superconducting. 
These novel phases of matter exhibit nontrivial symmetry-protected topological order~\cite{GuPRB2009,PollmannPRB2012}, 
characterized by a topological invariant acting as a nonlocal order parameter that is insensitive 
to smooth changes in the parameters of the material, unless a quantum critical point is crossed.
Thus, the transitions from topologically different phases cannot be described 
by the conventional Ginzburg-Landau theory of symmetry-breaking quantum phase transitions~\cite{ThoulessPRL1982,WenAP1995}.

Such marvellous systems have been extensively studied in terms of topological properties 
and a classification has been provided~\cite{KitaevAIP209}.
Besides their fundamental interest, topological insulators and superconductors 
may find application in photonics and quantum computation~\cite{HasanRMP2010},
not to mention that they can provide a new venue for realizing anyons, 
exotic particles whose statistics is neither fermionic nor bosonic~\cite{ReadPRB2000}.
Long-range entanglement has also been probed in topologically ordered phases, especially by means of entanglement spectrum~\cite{FidkowskiPRL2010} and entanglement entropy~\cite{JiangNATPHYS2012}.

In this chapter we focus on a simple paradigmatic example of one-dimensional topological superconductor: the Kitaev chain~\cite{Kitaev2001}.
% The Kitaev closed chain is a toy model describing a one-dimensional topological insulator with boundary conditions.
We start reviewing the basic bulk properties of this exotic toy model; they all are well known in literature~\cite{AliceaRPP2012}, 
even if we present some interesting analytical technical details,
not found elsewhere to our best knowledge, especially on the finite-size scaling of the vanishing energy gap.
In addition, we shortly comment on the experimental platforms able to realize the model. 
Then we discuss the anatomy of the phase diagram and 
examine the assortment of topological phase transitions that the model hosts, 
making use of analytical results regarding the diverging fidelity susceptibility: 
the treatment is only partially presented in Ref.~\cite{ZanardiPRL2007} and is here expanded.
Finally, we turn to study the multipartite entanglement 
and present a surprising connection between the phase diagram and the finite-size scaling of entanglement.

While a large body of literature has focused upon bipartite entanglement in the ground state of the Kitaev chain
as measured by the entanglement entropy~\cite{VodolaPRL2014,VodolaNJP2016,LeporiArxiv2016}, 
much less works has been devoted to addressing the problem of multipartite entanglement in this model.
As far as we know, the one presented here is the first comprehensive exposition on the subject, 
with the exception of the recent seminal Ref.~\cite{HaukeNATPHYS2016} -- where no entanglement was detected
since local operators were considered for calculating the quantum Fisher information. 
We calculate the quantum Fisher information using nonlocal operators --
suitably chosen treasuring the experience gained in chapter~\ref{ch:Ising} --
that are able to detect diverging multipartite entanglement in topological phases: 
Furthermore, we do not limit our discussion to the idealized case of zero temperature 
and carry out an attempt to better understand the interplay between entanglement in topological phases 
and finite temperature.
These original results are collected in papers~\cite{Gabbrielli2018a} and \cite{Pezze2017}.

\section{The model} \label{sec:KitaevModel}
%The Kitaev closed chain is a toy model describing a one-dimensional topological \think{insulator} with boundary conditions.
%We start reviewing the basic bulk properties of this exotic model 
%and shortly discuss about the experimental platform able to realize it. 
%We then discuss the anatomy of the phase diagram and 
%examine the assortment of topological phase transitions that this model hosts, 
%making use of analytical results on the fidelity susceptibility.
%Finally, we present a surprising connection between the phase diagram and the finite-size scaling of multipartite entanglement.
The closed Kitaev chain is a tight-binding model for spinless fermions on a cyclic lattice\footnote{\ A discussion of the continuum counterpart of this lattice model can be found in Ref.~\cite{vonOppen2017}.\\[-9pt]} 
with both tunnelling and superconducting pairing. 
This model was originally studied for nearest-neighbour pairing~\cite{Kitaev2001} and recently extended 
to variable-range pairing \cite{VodolaPRL2014, VodolaNJP2016}.
Here we introduce the minimal Hamiltonian describing tight-bound spinless fermions that hop on a one-dimensional $\sites$-site chain
and interact via a p-wave pairing with arbitrary range:
\be \label{HamKitaev}
\hat{H}_\alpha = - \frac{J}{2} \sum_{j=1}^{\sites} \left(\hat{a}^\dagger_j \hat{a}_{j+1} + \mathrm{H.c.}\right)  - \mu \sum_{j=1}^\sites 
\bigg( \hat{a}^\dagger_j\hat{a}_j - \frac{\identity_j}{2} \bigg) 
+\frac{\DD}{2} \sum_{j=1}^\sites \,\sum_{\ell=1}^{\sites-j} d_\ell^{-\alpha} \big( \hat{a}_j \hat{a}_{j+\ell} + \mathrm{H.c.} \big).
\ee
The operator $\hat{a}^\dagger_j$ creates a spinless fermion at the $j$-th site of the chain 
and satisfies the anticommutation relation $\{\hat{a}_i,\hat{a}_j^\dagger\}=\delta_{ij}$.
The real positive parameter $J$ is the hopping amplitude, quantifying the nearest-neighbour tunnelling rate. 
The superconducting gap\footnote{\ In literature, the superconducting gap is usually named $\Delta$~\cite{AliceaRPP2012,FranzNATNANO2013,VodolaPRL2014}. For consistency of notation, we adopt the symbol $\DD$, preserving the symbol $\Delta$ to indicate the energy separations in the many-body spectrum as introduced in chapter~\ref{ch:Intro}.\\[-9pt]}  
$\DD$ is the strength of the p-wave pairing, proportional to the pairing energy,
namely the energy required to create a Cooper pair; 
$\DD$ is assumed real for definiteness: the corresponding superconducting phase is set to zero.
The chemical potential $\mu$, or Fermi level, 
is essentially the energy necessary to add one particle to the superconductor from a reservoir: 
this parameter can be experimentally tuned by applying a voltage difference in a superconducting metal or 
tuning the doping in a superconducting semiconductor;
if $\mu>0$, the associated term in the Hamiltonian favours the filling of the chain sites.
The offset $-1/2$ in the middle term just generates an overall shift of the energy:
its use will be clarified in a moment.
The number $d_\ell>0$ is a site-to-site distance, while $\alpha\geq0$ specifies the range of the pairing: 
$\alpha\to\infty$ corresponds to nearest-neighbour pairing (Cooper pairs only form from fermions in nearest-neighbour sites), 
while $\alpha\to0$ corresponds to the infinite-range pairing (Cooper pairs can form from arbitrarily distant fermions).

In writing Eq.~(\ref{HamKitaev}) we have assumed a \emph{closed} chain (also known as ``Kitaev ring''):
in order to avoid cancellations between terms of the form $\hat{a}_{i}\hat{a}_{j}$ and $\hat{a}_{j}\hat{a}_{i+\sites}$, 
we adopted antiperiodic boundary conditions $\hat{a}_{j+\sites}=-\hat{a}_j$, 
that correctly preserve the translational invariance of the Hamiltonian.
% even though they erase the edge effects and only leave bulk properties. 
Obviously, wrapping the chain into a loop erases the edge effects and only leaves bulk properties.
Considering a lattice with unit spacing $\textsl{a}=1$ and even number of sites $\sites$ for avoiding complications in the notation, 
we have $d_\ell=\ell$ for $1\leq\ell\leq\frac{\sites}{2}$ and $d_\ell=\sites-\ell$ for $\frac{\sites}{2}\leq\ell<\sites$.

% What does ``spinless'' and ``p-wave pairing'' mean?
The \emph{spinless} nature of the fermions described by Eq.~(\ref{HamKitaev}) is evident: 
operators $\hat{a}_j$ and $\hat{a}^\dagger_j$ do not carry any spin index. 
This important assumption deserves a comment. 
In metallic superconductors, electrons pair into Cooper pairs and 
Pauli's exclusion principle only allows for \emph{antisymmetric} wavefunction for each Cooper pair. 
In a standard s-wave superconductor, this requirement is satisfied 
because the two electrons are in a symmetric orbital s-state, while their spin part is an antisymmetric singlet configuration. 
For spinless fermions, there is no spin part and the antisymmetry of the Cooper-pair wavefunction concerns the orbital part only. 
Thus, the exclusion principle prevents Cooper pairing with even parity: 
the major attribute of spinless supercondutors -- namely systems with only one active fermionic mode rather than two -- 
is the odd parity symmetry of the orbital wavefunction.
The pairing can no longer be s-wave, and the lowest-order option is the p-wave pairing.

\paragraph{Mapping to the XY model}
We note that for nearest-neighbours pairing ($\alpha=\infty$) the Kitaev Hamiltonian maps, via the inverse Jordan-Wigner transformation 
$\hat{s}_+^{(j)} = \hat{a}_j^\dag \exp\big(\ii\pi \sum_{l=1}^{j-1} \hat{a}_l^\dag \hat{a}_l\big)$, 
to the short-range XY model with transverse field~\cite{Lieb1961}
\be \label{HamXY}
\hat{H}_{\infty} = - \sum_{j=1}^{\sites} \left[\big(J+\DD\big)\hat{s}_x^{(j)}\,\hat{s}_x^{(j+1)} + 
\big(J-\DD\big)\hat{s}_y^{(j)}\,\hat{s}_y^{(j+1)}\right]
 - \mu \sum_{j=1}^\sites \hat{s}_z^{(j)},
\ee
%describing a collection of spin-1/2 on a cyclic one-dimensional lattice ($\hat{s}_k^{(L+1)}\equiv\hat{s}_k^{(1)}$, $k\in\{x,y,z\}$), 
describing a one-dimensional cyclic chain of spins~$\sfrac{1}{2}$ 
($\hat{s}_\varrho^{(L+1)}\equiv\hat{s}_\varrho^{(1)}$, $\varrho\in\{x,y,z\}$), 
that interact through a nearest-neighbour coupling (of strength $2J$ and anisotropy coefficient $\DD/J$), 
subject to a uniform transverse magnetic field of magnitude $\mu$. 
The offset $-1/2$ in Hamiltonian~(\ref{HamKitaev}), inessential for the physics of the system, 
is tailored to lead to Eq.~(\ref{HamXY}).
When $\DD=0$, Eq.~(\ref{HamXY}) reduces to the fully-isotropic XX model; 
when $\DD=J$, it turns into the standard Ising model. 
The equivalence between the spin and fermionic models suggests that the tunelling and the pairing are both 
responsible for the entanglement in the Kitaev model.

This mapping does not hold for pairing involving more than nearest neighbours ($\alpha<\infty$).
Moreover, here boundary conditions do matter: the mapping rigorously holds only for the open chain. 
For a closed chain, additional terms arise but they count less and less when increasing the size: 
explicitly, they give a contribution of order $\pazocal{O}(\sites^{-1})$ to macroscopic physical quantities
and can be neglected in the thermodynamic limit.

\paragraph{Insert: Majorana modes in the open chain}
% The original Kitaev chain was an open one, with free ends, in this context better known as Kitaev wire.
Even if we will deal solely with the cyclic chain, we remark that
Kitaev's original proposed model was an \emph{open} chain -- in this context better known as ``Kitaev wire'' -- 
still described by Eq.~(\ref{HamKitaev}) but with free ends.
The main motivation for introducing the open chain was somehow related to the long quest for Majorana particles, 
namely fermions that are their own antiparticles~\cite{Majorana1937}, 
formally characterized by the equality between the particle creation and annihilation operators: $\hat{\gamma}=\hat{\gamma}^\dag$.
There have been several attempts to find such particles in Nature, 
but no experiment in particle physics has yet provided definite proof of their existence. 
The enormous interest for Kitaev model lies in the discovery that the open wire hosts \emph{Majorana bound edge modes},
that are nondispersive zero-energy states localized at the edges of the wire.
Having no dispersion as a function of the momentum, these quasiparticle excitations present no dynamics of their own: 
in this sense, the exotic condensed-matter Majorana quasiparticles are different from the high-energy particle counterpart, 
but still satisfy the condition $\hat{\gamma}=\hat{\gamma}^\dag$.

In the Kitaev wire, two Majorana modes appear, spatially separated by the entire length of the lattice: 
they are associated to a highly-nonlocal doubly-degenerate many-body ground state 
and in principle could be probed independently. 
A sensational feature is found when exchanging two Majorana modes: 
the many-body ground state of the system neither remains unchanged (as for bosons), nor it changes sign (as for fermions), 
nor it acquires a generic phase factor (as for abelian anyons), but rather it experiences a transformation 
belonging to a group obeying no commutation law.
Thus, the quantum statistics of Majorana modes is \emph{non-abelian}~\cite{ReadPRB2000}.

\paragraph{Experimental implementations} 
The discovery that spinless p-wave superconductors host Majorana excitations 
has encouraged the realization of such systems, even though large conceptual and technical challenges are involved in this search.
On the one hand, electrons carry spin-$\sfrac{1}{2}$: how can they act as platforms for spinless superconductors?
The spin degree of freedom must be frozen, so that the system appears as effectively spinless: 
magnetic field is usually employed to polarize the electronic spin in the chain. 
% A spin-polarized electron systems are a close relative to spinless fermions.
On the other hand, the majority of superconductors are s-wave ones and only few p-wave superconductors are known: 
p-wave pairing in the wire must be induced by proximity to a conventional s-wave superconductor via spin-orbit coupling. 

%Spin-orbit coupling. 
%It might be simplest to locate the spin-orbit coupling in the superconductor
%rather than the one-dimensional system. Then, orbital angular momentum is no longer
%a good quantum number in the superconductor and there can be a small p-wave ad-
%mixture to the s-wave pairing. Unlike the s-wave correlations, the p-wave correlations
%can transfer to the spin-polarized system. As a result, the one-dimensional system
%effectively develops p-wave superconducting correlations by proximity.

Thus, the basic architecture required to engineer the Kitaev chain consists of a wire 
(made of either a one-dimensional semiconductor with strong spin-orbit coupling, as InSb and InAs, 
or a three-dimensional topological insulator, as Bi$_2$Se$_3$),
subject to a low uniform magnetic field 
and in which superconductivity is inherited from a parent bulk s-wave superconductor 
via spin-orbit coupling (that can be activated either on the parent or directly in the wire)~\cite{FranzNATNANO2013}.
Recent advances in nanofabrication and control of the spin-orbit coupling in solids 
allowed to experimentally realize suitable physical platforms for implementing the Kitaev wire 
and observing signatures of the existence of Majorana fermions~\cite{MourikSCI2012,RokhinsonNATPHYS2012,DasNATPHYS2012,DengNANO2012,FinkPRL2013}.
The properties of the cyclic chain coincides with the bulk properties of the open chain for a large lattice
and can be studied using the same setups for the open wire.

\subsection{Quasiparticle picture} \label{subsec:quasiparticle}
The exact diagonalization of the quadratic Kitaev Hamiltonian, 
leading to the spectrum of elementary excitations and to the analytical expression of the ground state of the system, is discussed.

\paragraph{Diagonalization}
Following Refs.~\cite{Lieb1961}, we show how the Hamiltonian (\ref{HamKitaev}) 
can be exactly diagonalized by means of a Bogoliubov transformation. 
% through/via the consecutive application of the Fourier and Bogoliubov transformations. The Fourier-transformed operators 
The Fourier transform on the local fermionic operators and its inverse read
\be \label{KitaevFourier}
\hat{a}_k^\dag = \frac{1}{\sqrt{\sites}} \sum_{j=1}^{\sites} \neper^{-{\ii} k j} \hat{a}_{j}^\dag \qquad \textrm{and} \qquad
\hat{a}_j^\dag = \frac{1}{\sqrt{\sites}} \sum_{k} \neper^{{\ii} k j} \hat{a}_k^\dag
\ee
where the quasimomenta are\footnote{\ Since $k$ depends on the integer $n$, the right notation would be $k_n$, 
but we omit the label $n$ wherever for simplicity's sake. Moreover, if the lattice spacing from site to site was $\textsl{a}\neq1$, the quasimomentum would look like $k = \frac{2\pi}{\textsl{a}\sites}\big(n + \frac{1}{2}\big)$.\\[-9pt]}
$k = \frac{2\pi}{\sites} \big( n + \frac{1}{2}\big)$, with quantum number $n=0,\,1,\,2,\,...,\,\sites-1$.
The discrete variable $k$ varies from $k_{\rm min}=\frac{\pi}{\sites}$ to $k_{\rm max}=2\pi-\frac{\pi}{\sites}$ by step $\Delta{k}=\frac{2\pi}{\sites}$.
The thermodynamic limit on the real lattice $\sites\to\infty$ corresponds, as usual, to the continuum limit in the momentum space 
% (inverse lattice)
$\Delta{k}\to0$ and $k\in[0,2\pi)$.
It is easy to verify that the transformed operators $\hat{a}_k,\,\hat{a}_k^\dag$ are still fermionic operators: 
$\big\{\hat{a}_k,\hat{a}_{k'}^\dag\big\}=\delta_{kk'}$ and 
$\big\{\hat{a}_k,\hat{a}_{k'}\big\}=\big\{\hat{a}^\dag_k,\hat{a}^\dag_{k'}\big\}=0$.
%\be
%k = \frac{2\pi}{\sites} \bigg( n + \frac{1}{2}\bigg), \quad n=0,1,2,...,L-1.
%\ee
%The single bricks composing the Hamiltonian (\ref{HamKitaev}) transform into
%\beq
%- \frac{J}{2} \sum_{j=1}^{\sites} \left(\hat{a}^\dagger_j \hat{a}_{j+1} + \mathrm{h.c.}\right)  - \mu \sum_{j=1}^\sites \left(\hat{n}_j - \frac{1}{2}\right) 
%= - \sum_{k} \bigg( \frac{J}{2} \cos k + \frac{\mu}{2} \bigg) \big( \hat{a}^\dagger_k \hat{a}_k +  \hat{a}^\dagger_{-k} \hat{a}_{-k} \big) + \frac{\mu \sites}{2},
%\eeq
%and 
%\be 
%\sum_{j=1}^\sites \,\sum_{\ell=1}^{\sites-j} \frac{\hat{a}_j \hat{a}_{j+\ell} }{d_\ell^{\alpha}}
% = \frac{1}{2} \sum_{j=1}^\sites \,\sum_{\ell=1}^{\sites-1} \frac{\hat{a}_j \hat{a}_{j+\ell} }{d_\ell^{\alpha}} 
% = \frac{1}{4} \sum_{j=1}^\sites \,\sum_{\ell=1}^{\sites-1} \frac{\hat{a}_j \hat{a}_{j+\ell} -  \hat{a}_{j+\ell}\hat{a}_j }{d_\ell^{\alpha}}
% = \frac{{\rm i}}{2} \sum_k f_{\alpha}(k) \hat{a}_k \hat{a}_{-k}
%\ee

%%%%%%%%%%%%%%%%%%%%%%%%%%%%%%%%%%%%%%%%%%%%%%%%%%%%%%%%%
%%%%%%%%%%%%%%%%%%%%%%%%%%%%%%%%%%%%%%%%%%%%%%%%%%%%%%%%%
\begin{figure}[t!]
\centering
\hfill 
\includegraphics[width=0.45\textwidth]{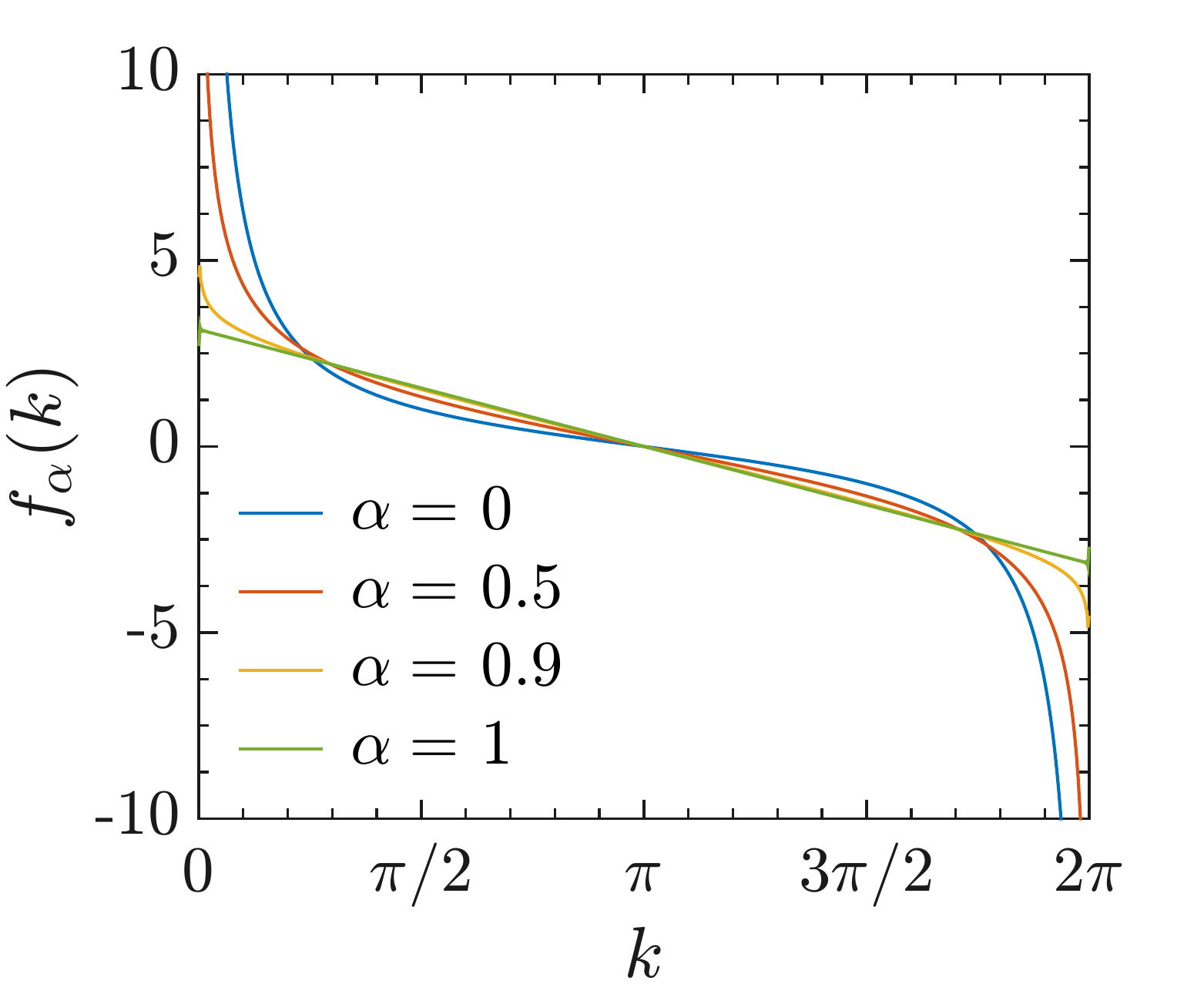} \hfill
\includegraphics[width=0.45\textwidth]{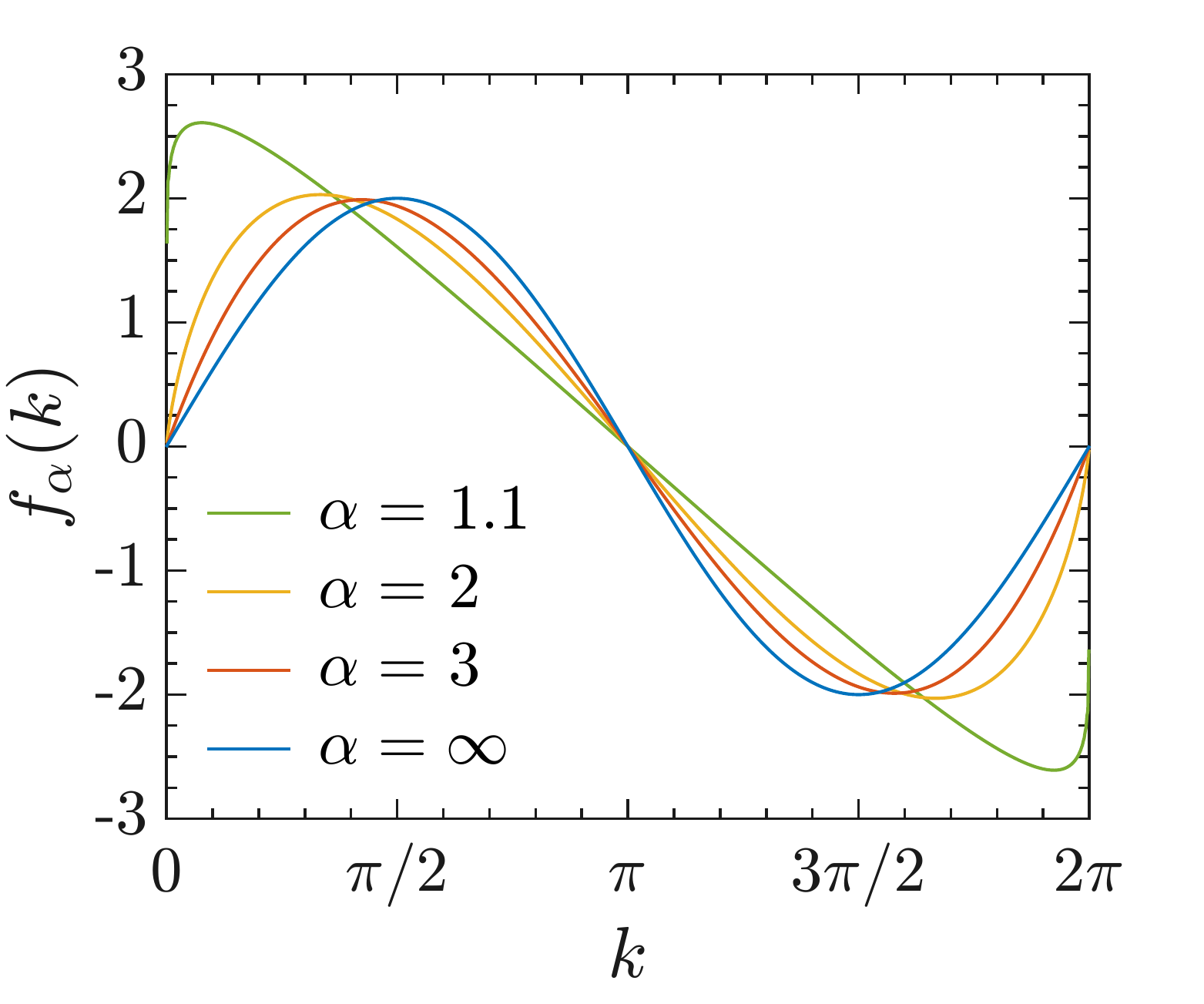} \hfill
\caption{Behaviour of $f_\alpha$ as a function of quasimomentum $k$ for long-range pairing ($\alpha\leq1$, \textbf{left}) 
and  short-range pairing ($\alpha>1$, \textbf{right}). 
% For $\alpha=1$, $f_k$ is still diverging at the boundaries of the Brillouin zone. 
The curves for $\alpha=0$ and $\alpha=\infty$ are analytical: $f_0(k)=\cot\frac{k}{2}$ and $f_\infty(k)=2\sin{k}$. 
The other curves are obtained numerically for a lattice with $\sites = 1000$.}
\label{fig:functionfk}
\end{figure}
%%%%%%%%%%%%%%%%%%%%%%%%%%%%%%%%%%%%%%%%%%%%%%%%%%%%%%%%%
%%%%%%%%%%%%%%%%%%%%%%%%%%%%%%%%%%%%%%%%%%%%%%%%%%%%%%%%%

After introducing the two-component Nambu spinor $\hat{\mathcal{C}}_k=(\hat{a}_k \,\ \hat{a}_{-k}^\dag)^T$, 
the total Hamiltonian (\ref{HamKitaev}) can be written in the standard Bogoliubov--de\,Gennes form
(where we drop a constant term)
%\be \label{Ham}
%\hat{H} = \sum_k \hat{\mathcal{C}}_k^\dag \, \mathcal{H}_k \, \hat{\mathcal{C}}_k
%= \sum_k \big( \hat{a}_k^\dag \,\ \hat{a}_{-k} \big)
%\mathcal{H}_k 
%\begin{pmatrix}
%\hat{a}_k \\
%\hat{a}_{-k}^\dag
%\end{pmatrix}, 
%\ee
%where
%\be
%\mathcal{H}_k = \begin{pmatrix}
%-(\tfrac{J}{2}\cos k + \tfrac{\mu}{2}) & {\rm i} \tfrac{\DD}{4} f_\alpha(k) \\
%-{\rm i} \tfrac{\DD}{4} f_\alpha(k) & (\tfrac{J}{2}\cos k + \tfrac{\mu}{2})\\
%\end{pmatrix}
%=- \frac{\DD}{4} f_\alpha(k) \hat{\sigma}_y -\bigg(\frac{J}{2}\cos k + \frac{\mu}{2}\bigg) \hat{\sigma}_z,
%\ee
\be
\hat{H}_\alpha = \frac{1}{2} \sum_k \hat{\mathcal{C}}_k^\dag \, \mathcal{H}_k \, \hat{\mathcal{C}}_k
= \frac{1}{2} \sum_k \big( \hat{a}_k^\dag \,\ \hat{a}_{-k} \big)
\mathcal{H}_k 
\begin{pmatrix}
\hat{a}_k \\
\hat{a}_{-k}^\dag
\end{pmatrix}, 
\ee
with
\be \label{HamBdeG}
\mathcal{H}_k = \begin{pmatrix}
-(J\cos k + \mu) & {\rm i} \tfrac{\DD}{2} f_\alpha(k) \\
-{\rm i} \tfrac{\DD}{2} f_\alpha(k) & J\cos k + \mu\\
\end{pmatrix}
= -\frac{\DD}{2} f_\alpha(k) \hat{\sigma}_y -\big(J\cos k + \mu\big) \hat{\sigma}_z \, .
\ee
Here $\hat{\sigma}_y$ and $\hat{\sigma}_z$ are Pauli matrices, while
\be \label{functionfk}
f_\alpha(k) = \sum_{\ell=1}^{\sites-1} \frac{\sin k \ell}{d_\ell^{\alpha}}
\ee
is a real odd\footnote{\ Since $f_\alpha(-k)=-f_\alpha(k)$, 
the off-diagonal terms in the Bogoliubov-deGennes Hamiltonian $\mathcal{H}_k$ are odd in $k$:
this is a direct consequence of the p-wave nature of the pairing.\\[-9pt]}
function that plays an essential role in determining the topological phases 
and the entanglement content of the ground state of the system. 
In fact, its behaviour at the boundaries of the Brillouin zone permits to discriminate between the two cases of 
short-range ($\alpha>1$) and long-range ($\alpha\leq1$) pairing: 
when increasing the lattice size, $f_\alpha(k_{\rm min})$ vanishes for $\alpha>1$, 
it is constant for $\alpha=1$ 
and diverges for $\alpha<1$.\footnote{\ Specifically, numerical investigation in the range $\sites=10\div10^4$ provides
the power laws $f_\alpha(k_{\rm min})\sim\sites^{1-\alpha}$ for $\alpha<1$, 
$f_\alpha(k_{\rm min})\approx2.742=\pazocal{O}(1)$ for $\alpha=1$, 
$f_\alpha(k_{\rm min})\sim\sites^{1-\alpha}$ for $1<\alpha\lesssim2$ and
$f_\alpha(k_{\rm min})\sim\sites^{-1}$ for $\alpha\gtrsim2$. 
Same finite-size scalings are found at $k_{\rm max}$.
A further -- more delicate -- inspection with analytical expansion in powers of $\/\sites$ confirms these scaling 
and reveals that, at leading orders, 
$f_\alpha(k_{\rm min})=2\pi^{\alpha-1}{\rm Re}(\ii^\alpha)\Gamma(1-\alpha)\sites^{1-\alpha}$ for $0\leq\alpha<2$, 
$f_\alpha(k_{\rm min})=2\pi[1+\log(\sites/\pi)]\sites^{-1}$ for $\alpha=2$,
and $f_\alpha(k_{\rm min})=2\pi\zeta(\alpha-1)\sites^{-1}$ for $\alpha>2$.\\[-9pt]}
In particular, closed expressions can be found in the limit cases $\alpha\to\infty$ and $\alpha=0$: 
namely, $f_\infty(k)=2\sin{k}$ and $f_0(k) = \cot\frac{k}{2}$.
In the limit $\sites\to0$, Eq.~(\ref{functionfk}) becomes 
$f_\alpha(k)=-\ii\big[\textrm{Li}_\alpha(\neper^{\ii k})-\textrm{Li}_\alpha(\neper^{-\ii k})\big]$,
where the special function $\textrm{Li}_\alpha(z)=\sum_{n=0}^\infty n^{-\alpha} z^n$ is the polylogarithm of order $\alpha$.
This function vanishes at $k=0,\,\pi,\,2\pi$ when $\alpha>1$, and only at $k=\pi$ when $\alpha\leq1$; 
it displays singularities at $k=0,\,2\pi$ when $\alpha<1$.
Figure~\ref{fig:functionfk} shows the behaviour of $f_\alpha(k)$ for several values of $\alpha$.

Finally, we apply the following Bogoliubov transformation 
\be \label{KitaevBogoliubov}
\begin{pmatrix}
\hat{a}_k \\
\hat{a}_{-k}^\dag
\end{pmatrix} = 
\mathcal{U}_k
\begin{pmatrix}
\hat{\eta}_k \\
\hat{\eta}_{-k}^\dag
\end{pmatrix} 
\qquad {\rm with} \quad
\mathcal{U}_k = 
\begin{pmatrix}
\cos\frac{\Theta_k}{2} & \ii\,\sin\frac{\Theta_k}{2} \\
\ii\,\sin\frac{\Theta_k}{2} & \cos\frac{\Theta_k}{2} \\
\end{pmatrix}, \quad \mathcal{U}_k^\dag\mathcal{U}_k=\mathds{1} ,
\ee
towards new fermionic quasiparticle operators $\hat{\eta}_k^\dag,\,\hat{\eta}_k$. 
By suitably defining 
\be \label{defThetak}
\epsilon_k\,\sin\Theta_k \equiv - \frac{\DD}{2}f_\alpha(k)
\quad {\rm and} \quad
\epsilon_k\,\cos\Theta_k \equiv - (J \cos k + \mu),
\ee
the matrix in Eq.~(\ref{HamBdeG}) transforms into $\mathcal{U}_k^\dag\,\mathcal{H}_k\,\mathcal{U}_k=\hat{\sigma}_z$ 
and the Hamiltonian~(\ref{HamKitaev}) takes the diagonal form
\be
\hat{H}_\alpha = \sum_k \epsilon_k \bigg(\hat{\eta}_k^\dag \hat{\eta}_k - \frac{\identity}{2}\bigg),
\ee
where the spectrum of excitations reads
\be \label{BogoliubovSpectrum}
\epsilon_{k} = \sqrt{ \big(J \cos k + \mu \big)^2 + \Big(\tfrac{\DD}{2}\,f_\alpha(k)\Big)^2} \, .
\ee

\paragraph{Closure of the gap}
The energy separation between the ground state and the first excited state 
in the many-body spectrum of the full Hamiltonian Eq.~(\ref{HamKitaev}) 
coincides with the smallest energy gap in the quasiparticle spectrum Eq.~(\ref{BogoliubovSpectrum}), $\DeltaE=\min_k\epsilon_k$,
and corresponds to the energy necessary to create only one excitation with quasimomentum $k={\rm arg\,min_k}\,\epsilon_k$.
A vanishing quasiparticle spectrum means a gap closure in the many-body spectrum and signals a second-order QPT. 

\begin{figure}[t!]
\centering
\includegraphics[width=0.205\textwidth]{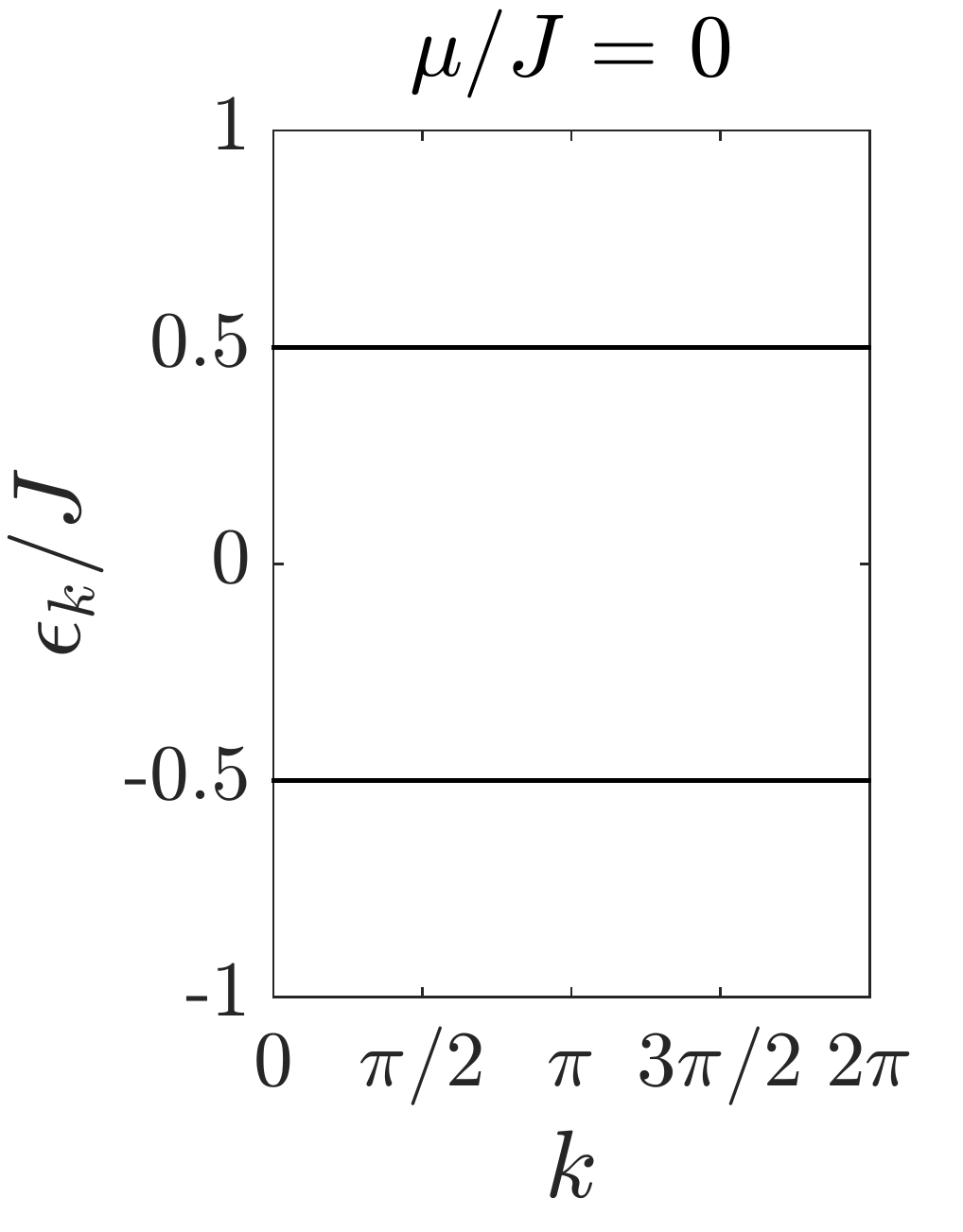} \hspace{-12pt}
\includegraphics[width=0.205\textwidth]{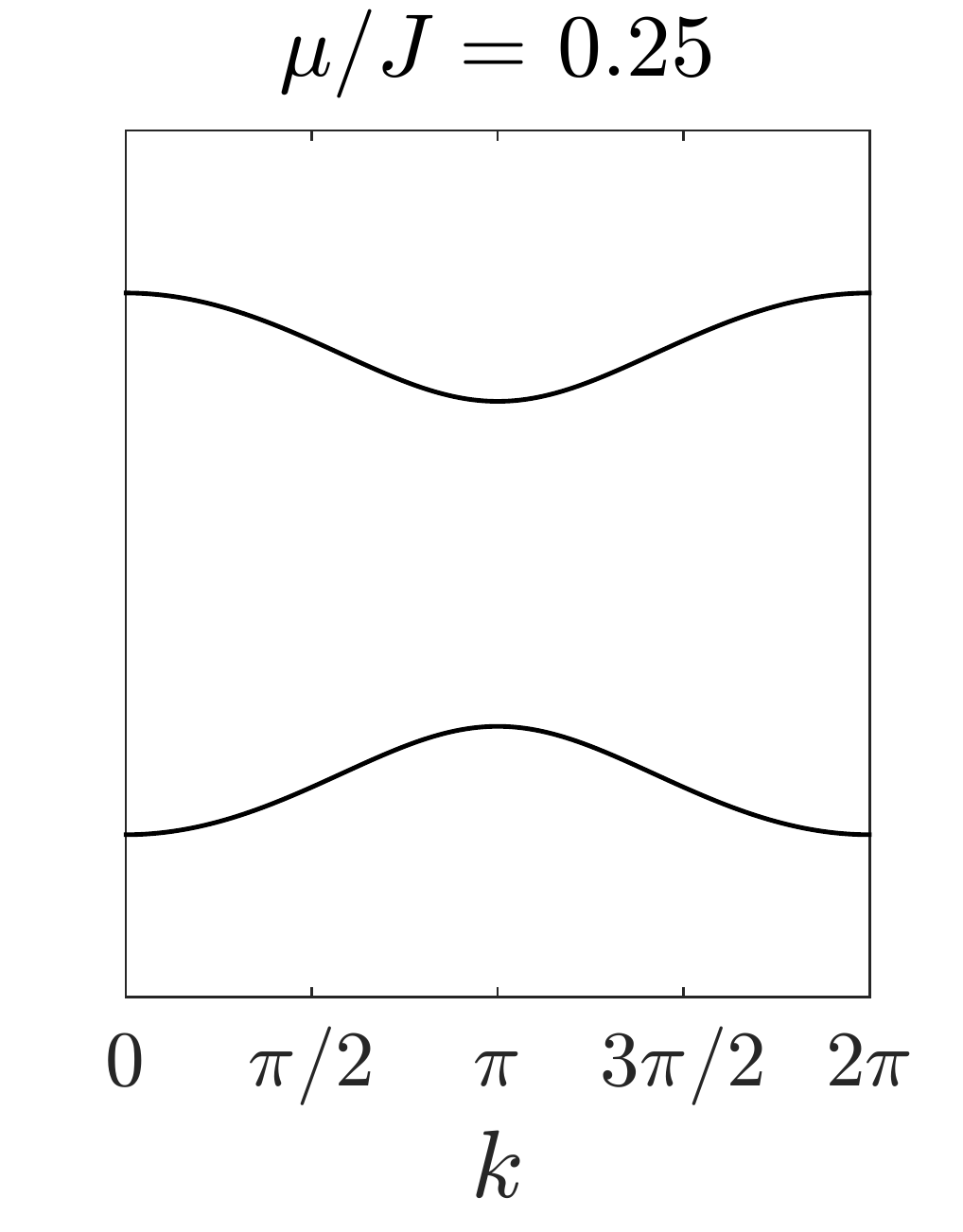} \hspace{-12pt}
\includegraphics[width=0.205\textwidth]{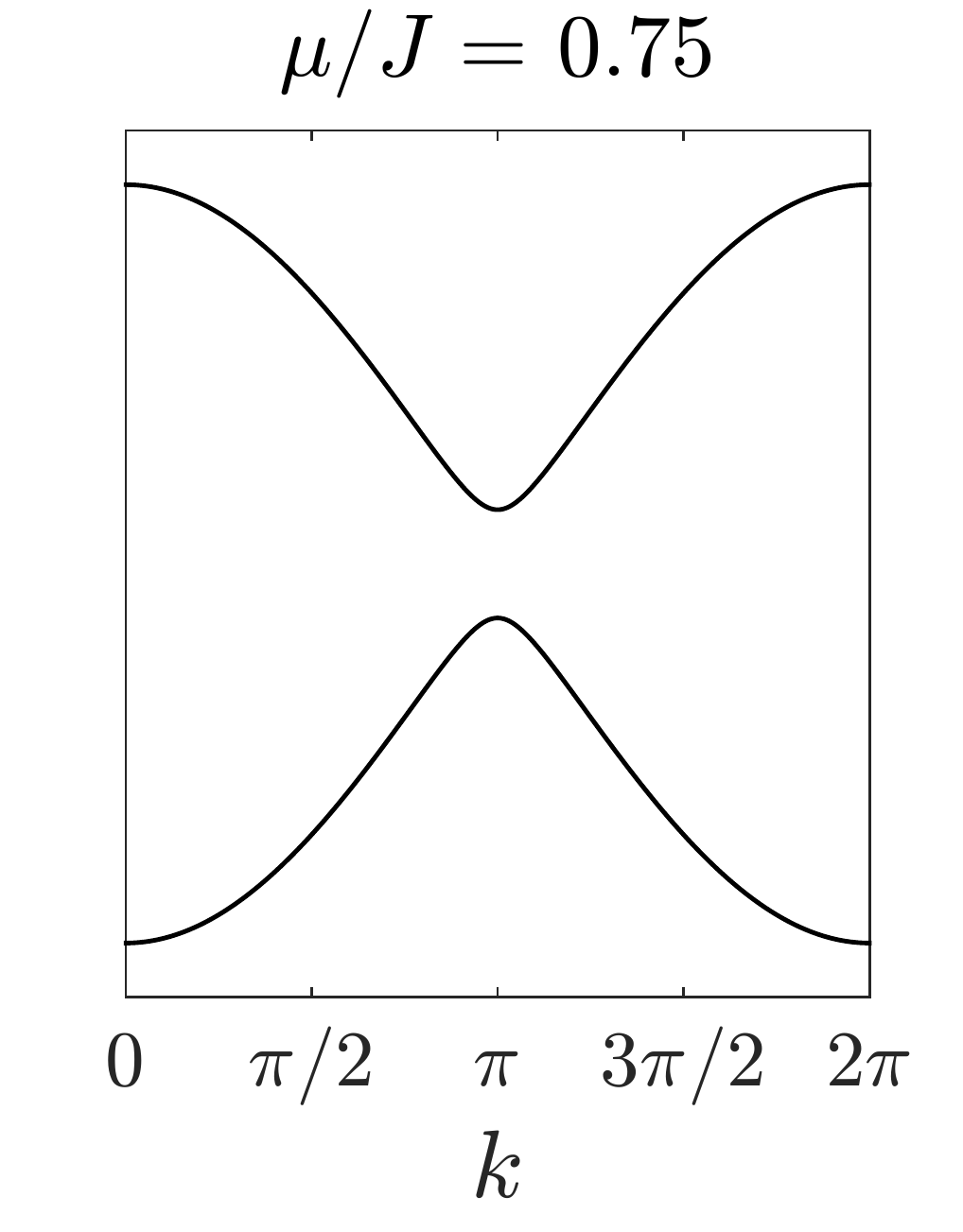} \hspace{-12pt}
\includegraphics[width=0.205\textwidth]{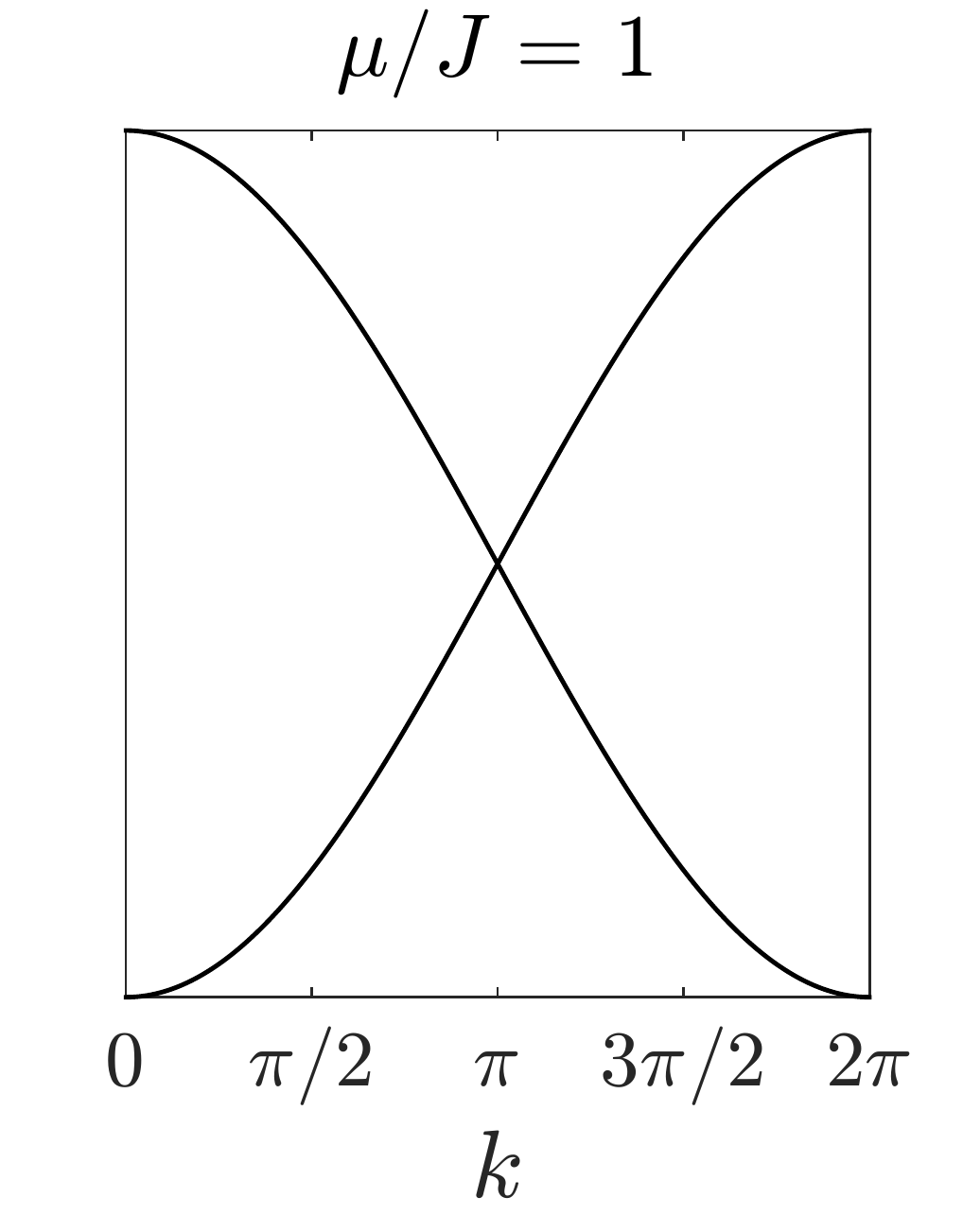} \hspace{-12pt}
\includegraphics[width=0.205\textwidth]{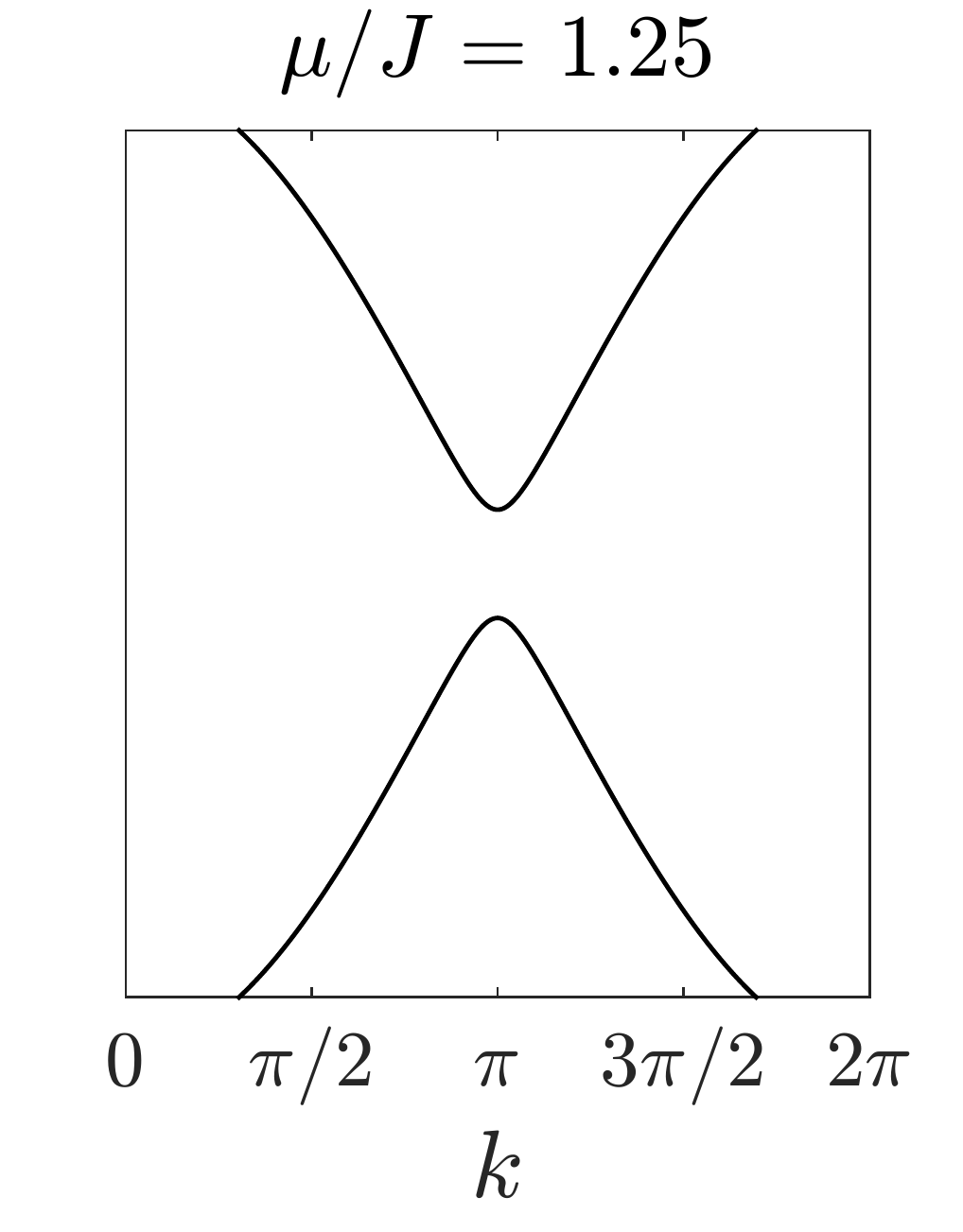} \\[12pt]
\includegraphics[width=0.205\textwidth]{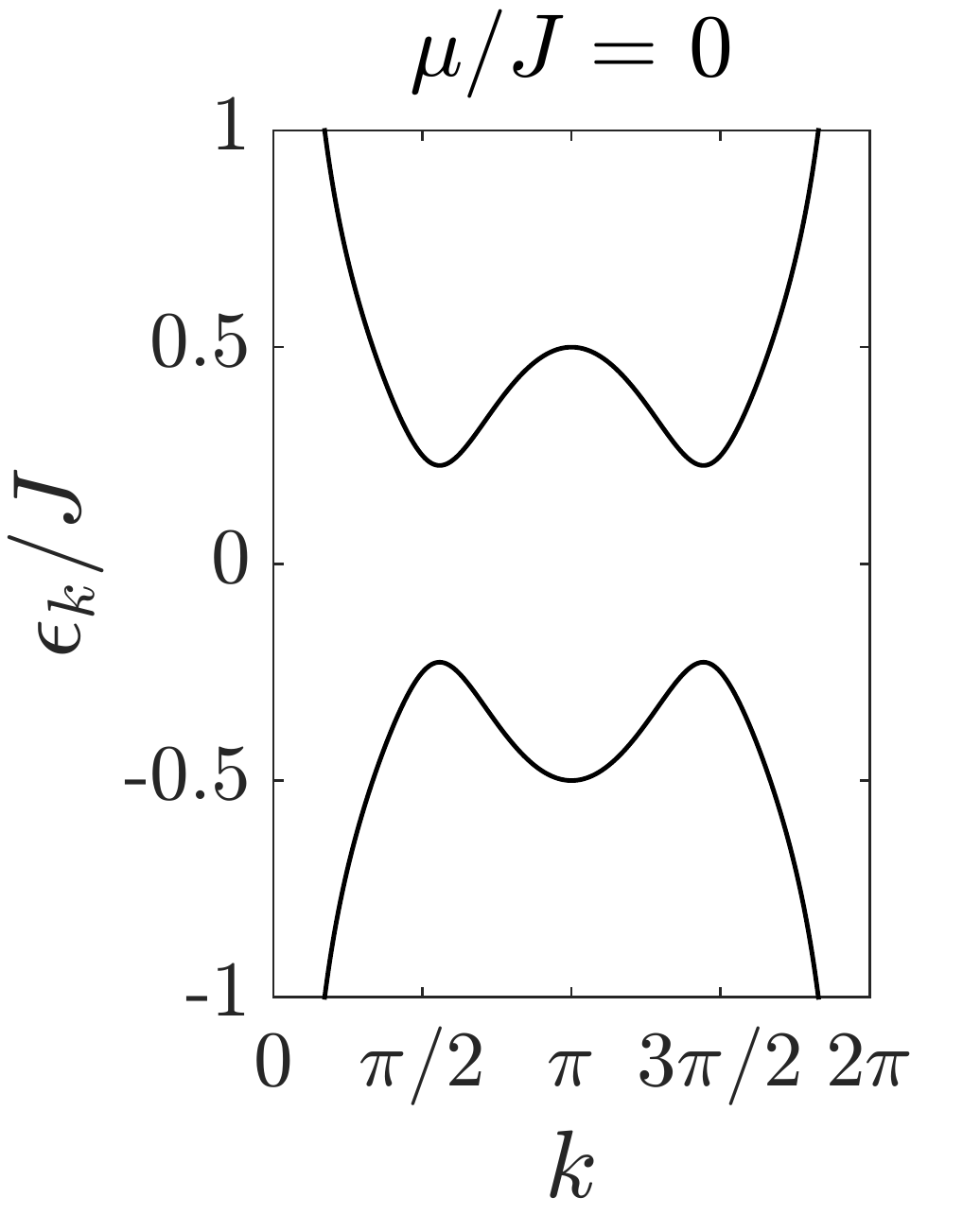} \hspace{-12pt}
\includegraphics[width=0.205\textwidth]{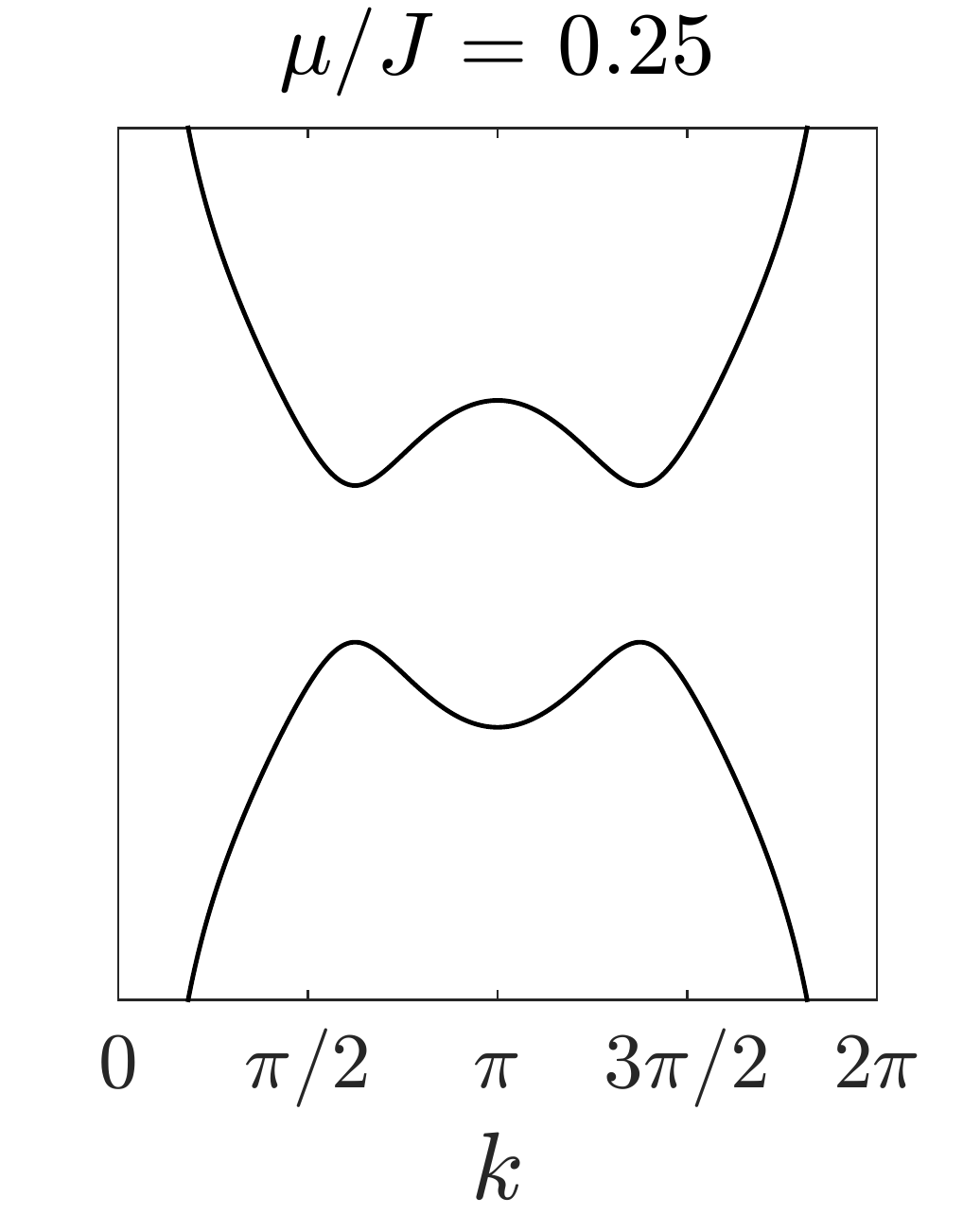} \hspace{-12pt}
\includegraphics[width=0.205\textwidth]{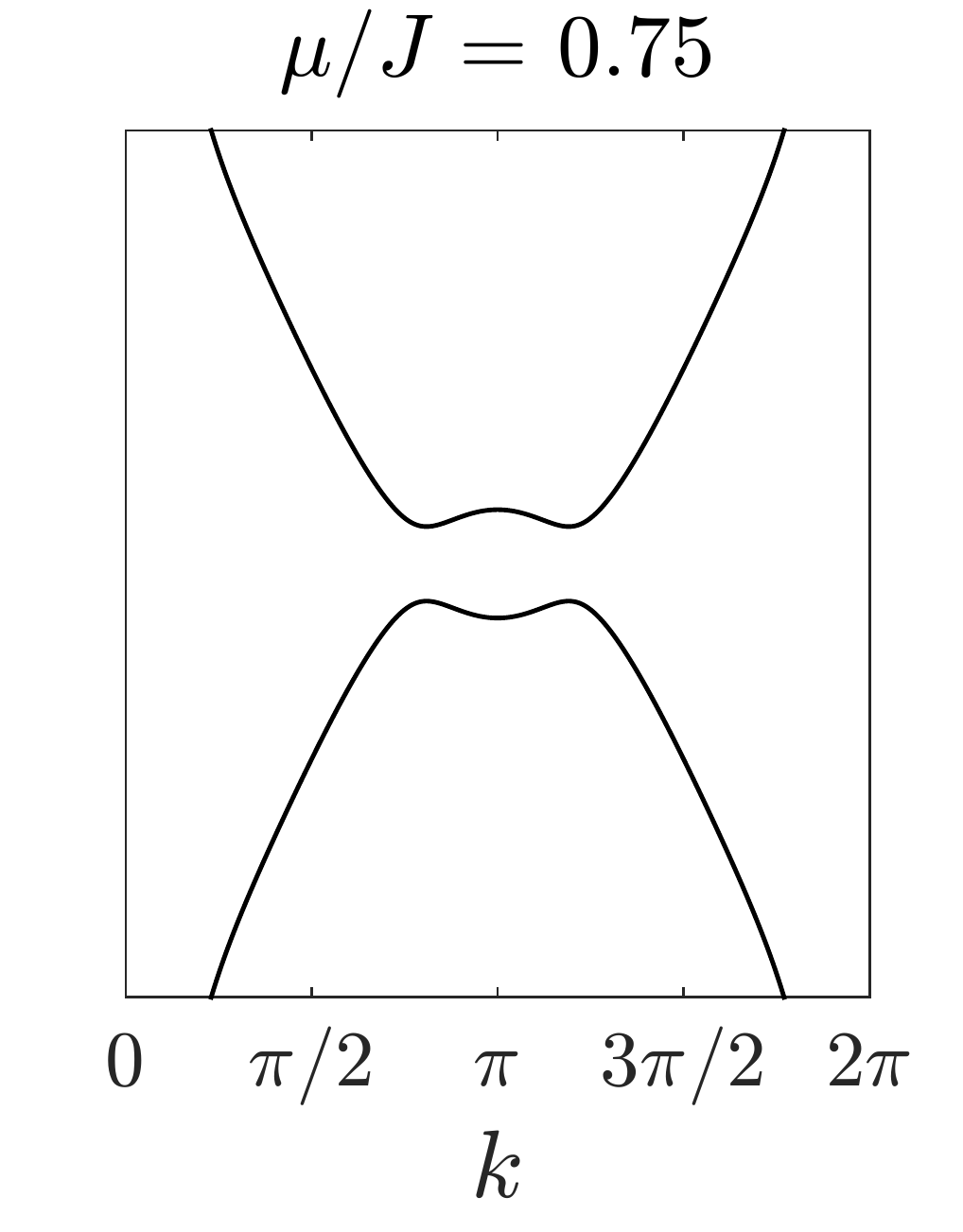} \hspace{-12pt}
\includegraphics[width=0.205\textwidth]{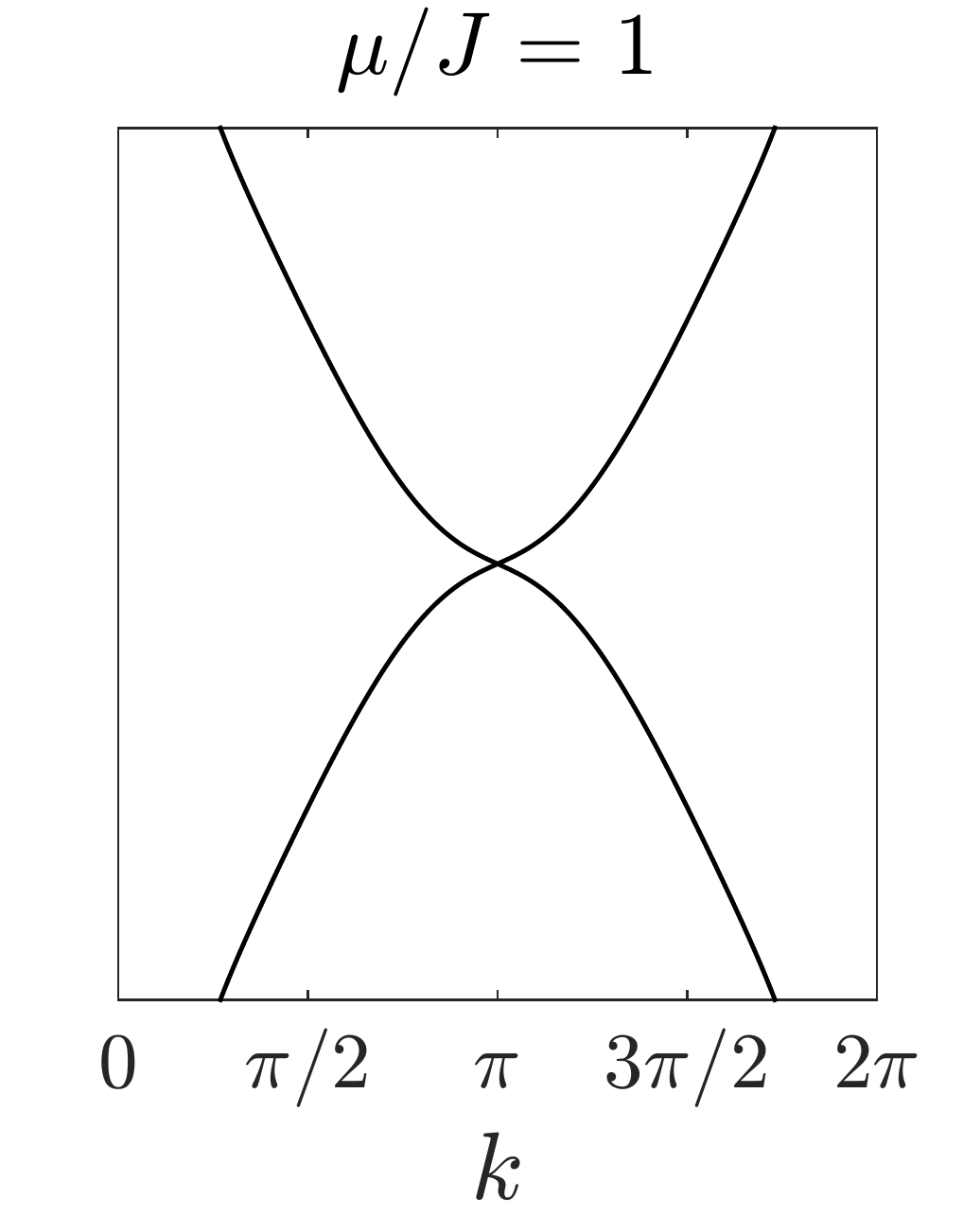} \hspace{-12pt}
\includegraphics[width=0.205\textwidth]{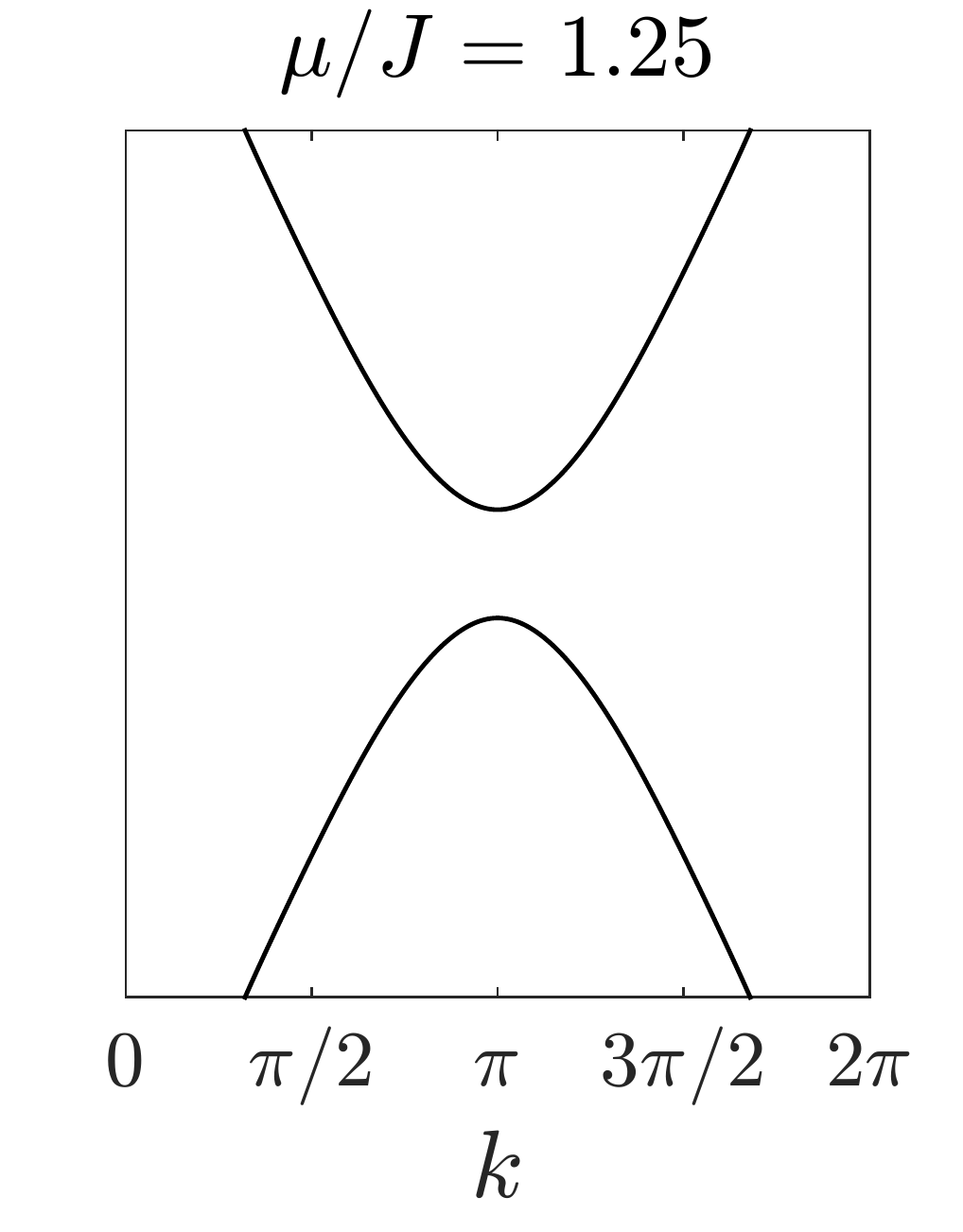} 
\caption{The excitation dispersion relation $\epsilon_k$ in the one-dimensional Brillouin zone as in Eq.~(\ref{BogoliubovSpectrum}), 
displayed such that the center of the gap lies at zero energy in all the panels.
We show the subcritical ($0<\mu<J$), critical ($\mu=J$) and supercritical ($\mu>J$) regimes, for $\DD=J$,
in the limit cases $\alpha=\infty$ \textbf{(top)} and $\alpha=0$ \textbf{(bottom)}.
In both cases, the closure of the gap is evident at $k=\pi$ when $\mu$ is finely tuned through $\mu=J$.
}
\label{fig:quasiSpectrum}
\end{figure}

Let's first focus on the thermodynamic limit $\sites\to\infty$. 
For null pairing $\DD=0$, the quasiparticle spectrum vanishes at $k={\rm arccos}(-\mu/J)$ 
and the system is gapless for any $|\mu| \leq J$.
For $\DD\neq0$, the gap can close only if $f_\alpha(k)=0$ for some quasimomentum: 
when $\alpha>1$, this happens at $k=0$ (in which case $\epsilon_k=0$ for $\mu/J=-1$) 
and $k=\pi$ (in which case $\epsilon_k=0$ for $\mu/J=1$);
when $\alpha \leq 1$, this happens only at $k=\pi$ (and $\epsilon_k=0$ for $\mu/J=1$).
Thus, the system is gapless for $\mu/J = 1$ for all the values of $\alpha$, 
as well as for $\mu/J = -1$ when $\alpha>1$ only.
Figure~\ref{fig:KitaevTopology} highlights the set of parameters for which the closure of the gap occurs.
Figure~\ref{fig:quasiSpectrum} shows the vanishing of $\epsilon_k$ in a band-like fashion.

%%%%%%%%%%%%%%%%%%%%%%%%%%%%%%%%%%%%%%%%%%%%%%%%%%%%%%%%%
%%%%%%%%%%%%%%%%%%%%%%%%%%%%%%%%%%%%%%%%%%%%%%%%%%%%%%%%%
%%%%%%%%%%%%%%%%%%%%%%%%%%%%%%%%%%%%%%%%%%%%%%%%%%%%%%%%%
%%%%%%%%%%%%%%%%%%%%%%%%%%%%%%%%%%%%%%%%%%%%%%%%%%%%%%%%%

A finite-size study allows to disclose the sensitivity of the gap to increasing $\sites$. 
For instance, considering $\mu=J=\DD$ (case of interest for the subsequent discussion), 
the quasiparticle spectrum reaches a \emph{finite} minimum at $k=\pi+\frac{\pi}{\sites}\to\pi$ 
(given by $n=\frac{\sites}{2}$) for any $\alpha$.
A series expansion in $\sites^{-1}$ provides, at leading order, the power-law scaling
$\epsilon_k/J = A(\alpha)\sites^{-1}$, with 
$A(\alpha) \allowbreak = \allowbreak \pi \allowbreak \big|(2^{2-\alpha}-1) \cdot \allowbreak \zeta(\alpha-1)\big|$.
In particular, $A(\infty)=\pi$, $A(1)=\frac{\pi}{2}$ and $A(0)=\frac{\pi}{4}$.

%quasiparticle energy
%quasiparticle excitation spectrum 

% Numerically, (for $\sites=10\div1000$) we find $\DeltaE/J=A\sites^{-b}$, 
% with prefactor $A(\alpha)\approx3.14-2.35\exp(-\alpha/2.25)$ and exponent $b=1.00$ independently of $\alpha$ (for $\alpha=0\div100$).

% We can thus conclude that the power-law scaling $\DeltaE\propto\sites^{-1}$ holds for any $\alpha$

\paragraph{Ground state} 
The normalized ground state of the Kitaev chain Eq.~(\ref{HamKitaev}) can be written in the explicit form \cite{VodolaPRL2014}
\be \label{KitaevGS}
\ket{\psi_{0}} = \prod_{0\,<\,k\,<\,\pi} \bigg(\cos\frac{\Theta_k}{2} - \ii\,\sin\frac{\Theta_k}{2} \, \hat{a}_k^\dag\hat{a}_{-k}^{\dag} \bigg) \ket{0},
\ee 
%\be
%\ket{\psi_{0}} = \prod_{0\,<\,k\,<\,\pi} \Big(\cos\tfrac{\Theta_k}{2} - \ii\,\sin\tfrac{\Theta_k}{2} \, \hat{a}_k^\dag\hat{a}_{-k}^{\dag} \Big) \ket{0},
%\ee 
where the notation $0<k<\pi$ is a shorthand for $n=0,\,1,\,...,\,\frac{\sites}{2}-1$ and means spanning over half of the Brillouin zone;
the angle $\Theta(k)$ is defined in Eq.~(\ref{defThetak}) in terms of the physical parameters; 
and $\ket{0}$ is the vacuum of $\hat{a}_k$, so a state with no fermions in the chain.

Using the definitions Eqs.~(\ref{KitaevFourier}) and (\ref{KitaevBogoliubov}), 
considering that the ground state in the quasiparticle picture is the Bogoliubov vacuum for the excitations $\hat{\eta}_k$'s
and exploiting the Wick's theorem, 
we can analytically access the following expectations values and correlations functions on the ground state:
\begin{align} \label{KitaevCorrelations}
\begin{split}
& \!\! \big\langle\hat{a}_i\big\rangle = \big\langle\hat{a}_i^\dag\big\rangle = 0 \\ 
& \!\! \big\langle\hat{a}_i\hat{a}_j\big\rangle = -\big\langle\hat{a}_i^\dag\hat{a}_j^\dag\big\rangle^* 
= \frac{1}{2\ii\sites} \sum_k \neper^{-\ii k(i-j)}\sin\Theta_k \\ 
& \!\! \big\langle\hat{a}_i^\dag\hat{a}_j\big\rangle = \delta_{ij} - \big\langle\hat{a}_i\hat{a}_j^\dag\big\rangle^*
= \frac{1}{\sites} \sum_k \neper^{\ii k(i-j)}\sin^2\tfrac{\Theta_k}{2} \\ 
& \!\! \big\langle\hat{a}_i^\dag\hat{a}_i\hat{a}_j^\dag\hat{a}_j\big\rangle = 
\frac{1}{\sites^2} \bigg[ \Big(\sum_k\sin^2\tfrac{\Theta_k}{2}\Big)^2 + 
\sum_{k\,k'}\neper^{\ii(k-k')(i-j)}\sin^2\tfrac{\Theta_k}{2}\cos^2\tfrac{\Theta_{k'}}{2} - 
% \Big(\sum_k\neper^{\ii k(i-j)}\sin^2\tfrac{\Theta_k}{2}\Big) \Big(\sum_k\neper^{-\ii k(i-j)}\cos^2\tfrac{\Theta_k}{2}\Big) -
\Big|\sum_k\neper^{\ii k(i-j)}\tfrac{\sin\Theta_k}{2}\Big|^2 \bigg]
\end{split}
\end{align}
from which a variety of bulk properties at zero temperature can be calculated. 

For instance, the average total number of particles is 
\be
\langle\hat{N}\rangle = \sum_{i=1}^\sites \langle\hat{a}_i^\dag\hat{a}_i\rangle = \sum_k \sin^2\tfrac{\Theta_k}{2} =
\frac{1}{2} \sum_k \left( 1 + \frac{J \cos k + \mu}{\sqrt{(J \cos k + \mu)^2 + \big(\tfrac{\DD}{2}\,f_\alpha(k)\big)^2}} \right).
\ee
The number of particles on the $\sites$-site chain is not fixed, but strongly depends on the chemical potential:
as expected from Eq.~(\ref{HamKitaev}), for $\mu\to-\infty$ the chain is completely empty $\langle\hat{N}\rangle=0$, 
while for $\mu\to+\infty$ the chain is fully occupied $\langle\hat{N}\rangle=\sites$.
Clearly, being the fermions spinless, the average occupation number per site is $0\leq\langle\hat{a}_i^\dag\hat{a}_i\rangle\leq1$.
For a nearest-neighbour system with $\DD=J$, the meaning of null chemical potential (Fermi level) is clear: 
$\mu=0$ is such that $\langle\hat{N}\rangle=\sites/2$.

\subsection{Topological phases} \label{subsec:topology}
\begin{figure}[t!]
\centering
\includegraphics[width=1\textwidth]{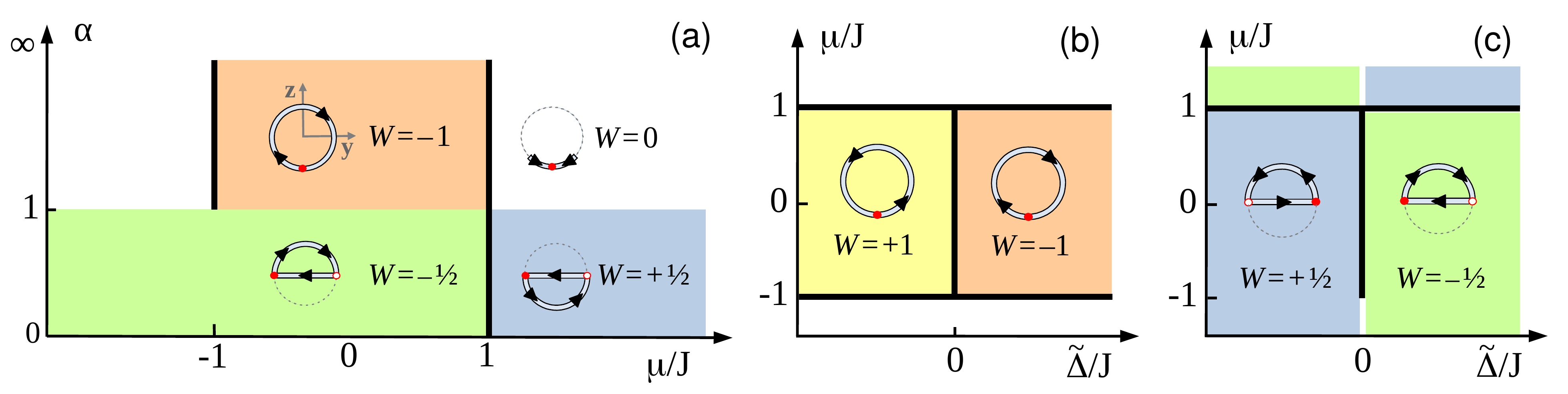}
\caption{Sketches of the phase diagram for the Kitaev chain in the thermodynamic limit: 
\textbf{(a)}~$\mu/J$--$\alpha$ plane for $\DD>0$; 
\textbf{(b)}~$\DD/J$--$\mu/J$ plane for nearest-neighbor pairing ($\alpha=+\infty$);
\textbf{(c)}~$\DD/J$--$\mu/J$ plane for infinite-range pairing ($\alpha=0$).
The thick lines mark a vanishing gap in the quasiparticle spectrum.
Different phases are highlighted by different colours and winding number $W$.
In each phase, solid lines shows the trajectory of the Anderson vector $\vect{h}(k)$ in the unit circle on the $y$-$z$ plane (dotted line) 
as $k$ varies from $0$ (red filled circle) to $2\pi$ (red dot).}
\label{fig:KitaevTopology}
\end{figure}

The schematic zero-temperature phase diagrams in Fig.~\ref{fig:KitaevTopology} display the topological phases and their boundaries 
for the Kitaev chain as a function of the Hamiltonian parameters $\mu$ and $\DD$ (measured in units of $J$) and the pairing range $\alpha$.
%Figure~\ref{fig:KitaevTopology} displays the topological phases and their boundaries on the phase diagrams of the Kitaev chain.
Regions with different colours refer to phases with different constant values of a suitable topological invariant. 
There are several ways in which one can express the topological invariant~\cite{AliceaRPP2012}, 
that is the property able to distinguish between phases with different topology, 
therefore acting as an ``order parameter'' for the topological QPTs.
We will rely on the simple concept of ``winding number'' as defined in the following~\cite{NiuPRB2012}.

%%%%%%%%%%%%%%%%%%%%%%%%%%%%%%%%%%%%%%%%%%%%%%%%%%%%%%%%%
%%%%%%%%%%%%%%%%%%%%%%%%%%%%%%%%%%%%%%%%%%%%%%%%%%%%%%%%%
%%%%%%%%%%%%%%%%%%%%%%%%%%%%%%%%%%%%%%%%%%%%%%%%%%%%%%%%%
%%%%%%%%%%%%%%%%%%%%%%%%%%%%%%%%%%%%%%%%%%%%%%%%%%%%%%%%%

\paragraph{Winding number}
It is convenient to express the $2\times2$ Hamiltonian matrix in Eq.~(\ref{HamBdeG}) 
in terms of the Pauli vector and the normalized Anderson pseudospin vector~\cite{AndersonPR1958}: 
$\mathcal{H}_k = \epsilon_k \, \vect{h}(k)\cdot\hat{\mbox{\boldmath$\sigma$}}$, with 
\be 
h_x(k)=0, \quad h_y(k)=-\frac{\frac{\DD}{2} f_\alpha(k)}{\epsilon_k}\equiv\sin\Theta_k, \quad h_z(k)=-\frac{J\cos k + \mu}{\epsilon_k}\equiv\cos\Theta_k
\ee
satisfying $h_{x,y}(-k)=-h_{x,y}(k)$ and $h_z(-k)=h_z(k)$.
Supposing that the excitation spectrum is gapped $\epsilon_k\neq0$ throughout the whole Brillouin zone, 
then $\vect{h}(k)$ is always well-defined and provides a mapping from the reciprocal space to the unit circle on the $x$-$y$ plane.
As $k$ sweeps from $0$ to $2\pi$, the vector $\vect{h}(k)$ winds on the circle: 
the topological invariant we seek is the \emph{winding number}
\be
W \, = \, \int_{\rm BZ} \frac{\ud\Theta_k}{2\pi} \ = \ \frac{1}{2\pi} \int_{0}^{2\pi} \frac{\ud \Theta_k}{\ud k} \ud k 
\ee
which counts the number of times the vector winds around the center of the circle.
If $\vect{h}(k)$ performs full (half) turns, $W$ is integer (semi-integer);
the sign of $W$ distinguishes different rotation directions.
Supposing, instead, that the excitation spectrum is gapless $\epsilon_k=0$ somewhere in the Brillouin zone,
then $\vect{h}(k)$ is ill-defined and a discontinuous change in $W$ (i.e., a topological QPT) is expected~\cite{ReadPRB2000}.

Coloured regions in Fig.~\ref{fig:KitaevTopology} mark topological phases with $W\neq0$; 
white regions correspond to topologically trivial phases, having $W=0$; 
the winding of $\vect{h}(k)$ is also indicated for each phase. 
The value of $W$ changes when the system crosses the boundary between two phases -- we say that a topological QPT takes place. 
Chiefly, such a change occurs when crossing the thick lines, that mark the set of parameters for which the system is gapless ($\DeltaE=0$).
However, Fig.~\ref{fig:KitaevTopology} demonstrate that gaplessness is only a \emph{sufficient} condition: 
there exist topological QPTs that take place with no gap closure in the energy spectrum ($\DeltaE>0$), 
as will be discussed afterward in this paragraph. 
The main conclusion here is that all the points on the black lines in Fig.~\ref{fig:KitaevTopology} are topological critical points; 
but they are not the only ones.

Panel~\ref{fig:KitaevTopology}\Panel{a} shows the phase diagram in the $\mu/J$--$\alpha$ plane for fixed $\DD>0$.\footnote{\ In the case $\DD<0$ the phase diagram is obtained just changing $W\mapsto-W$.\\[-9pt]}
Panels~\ref{fig:KitaevTopology}\Panel{b,\,c} show the phase diagram in the $\DD/J$--$\mu/J$ plane 
for $\alpha=\infty$ and $\alpha=0$, respectively.
For short-range pairing ($\alpha>1$), $W$  assumes only integer values: $W=0$ for $|\mu|/J>1$ and $W=\pm1$ for $|\mu|/J<1$.
For long-range pairing ($\alpha\leq 1$), semi-integer values $W=\pm1/2$ appear for $\mu/J\gtrless1$. 
These fractional numbers are due to the singularity of $f_\alpha(k)$ for $\alpha<1$ that occurs at $k=0,\,2\pi$ in the thermodynamic limit:
the corresponding phases are characterized by a violation of the area law for the Von Neumann entropy~\cite{VodolaPRL2014,AresPRA2015}
and we indicate them as ``long-range'' phases~\cite{ViyuelaPRB2016,LeporiArxiv2016}.

\paragraph{Nearest-neighbour scenario}
In the case $\alpha=\infty$ depicted in panel \ref{fig:KitaevTopology}\Panel{b}, 
the system has a trivial phase for $|\mu|>J$ and a topological phase for $|\mu|<J$.
These two phases are both gapped, but a physical insight reveals them to be quite different. 
In the former regime, the ground state is adiabatically connected to 
the trivial vacuum where no fermions are present (upon taking $\mu\to-\infty$)
or the fully occupied band ($\mu\to+\infty$).
In contrast, in the latter regime the system cannot connect smoothly to the vacuum (of particles or holes)
because there is always a closing gap when deforming the Hamiltonian from the topological phase towards the
vacuum limit.

Moreover, loosely interpreting $\psi_{\rm C}(k)=-\ii\tan\Theta_k$ as the eigenfunction 
of the Cooper pair formed by fermions with momenta $k$ and $-k$ like suggested by Eq.~(\ref{KitaevGS}),
a distinctive difference between the two phases emerges in the real space~\cite{ReadPRB2000,AliceaRPP2012}. 
For $|\mu|>J$, it was found $\psi_{\rm C}(r)\sim\neper^{-r/\zeta}$ 
so Cooper pairs form from two fermions tightly bound in position over a length scale $\zeta$. 
On the contrary, for $|\mu|<J$ the spatial wavefunction has a long tail for $r\to\infty$: 
the size of the Cooper pairs can be much larger than the lattice spacing.

Finally, the phase diagram is found to be symmetric with respect to $\mu=0$ for any fixed~$\DD$,
as we expected from the particle-hole symmetry~\cite{AliceaRPP2012} of Hamiltonian (\ref{HamKitaev}). 
This symmetry is conserved for all finite $\alpha>1$ in the thermodynamic limit, but it is lost in the ``long-range'' regime $\alpha<1$:
see panel \ref{fig:KitaevTopology}\Panel{a}.

\paragraph{Transitions without gap closure}
We comment here on topological QPTs in the Kitaev chain that are less studied in literature:
the ones that happen when the excitation spectrum does not vanish ($\epsilon_k\neq0 \ \forall \, k\in{\rm BZ}$), 
implying no closing gap in the many-body spectrum ($\DeltaE>0$).
A first example is found for fixed $\alpha<1$ and $|\mu|/J>1$: 
we have a transition between $W=+1/2$ and $W=-1/2$ driven by the control parameter $\DD$
when it crosses the critical point $\DD=0$, at which we find trivial topology $W=0$. 
The singular line of critical points is highlighted by the white lines $\big(\DD=0,|\mu|>J\big)$ in panel~\ref{fig:KitaevTopology}\Panel{c}. 

Another curious -- even if not completely unexpected~\cite{VodolaPRL2014,LeporiArxiv2016} -- family of transitions
is present along the line $\alpha=1$: between two phases with $W=-1/2$ and $W=0$ for $\mu/J<-1$, 
between $W=-1/2$ and $W=1$ for $-1<\mu/J<1$, and between $W=1/2$ and $W=0$ for $\mu/J>1$.
They are associated  with various discontinuities, 
for instance in the mutual information and in the decay exponents for the two-point correlations 
$\langle \hat{a}_{i}^{\dagger} \hat{a}_j \rangle$, $\langle \hat{a}_{i} \hat{a}_j \rangle$ \cite{VodolaPRL2014,LeporiArxiv2016}.
Notably, the change of behaviour at $\alpha=1$ is not sensed~\cite{VodolaPRL2014} by the half-chain von Neumann entropy 
$S_{\sites/2}=-\tr\big(\rho_{\sites/2}\log\rho_{\sites/2}\big)$.
% $S\big(\frac{\sites}{2}\big)=-\tr\big(\rho_{\frac{\sites}{2}}\log\rho_{\frac{\sites}{2}}\big)$.
In the following we will show that, similarly to the QPTs at $\mu/J = \pm 1$ and $\DD=0$, 
the transition at $\alpha=1$ is suggested by the divergence of the fidelity susceptibility, 
as well as by the divergence of some derivative of the QFI with respect to $\alpha$.

%%%%%%%%%%%%%%%%%%%%%%%%%%%%%%%%%%%%%%%%%%%%%%%%%%%%%%%%%
%%%%%%%%%%%%%%%%%%%%%%%%%%%%%%%%%%%%%%%%%%%%%%%%%%%%%%%%%
\begin{figure}[p!]
\centering
\hfill \includegraphics[width=0.45\textwidth]{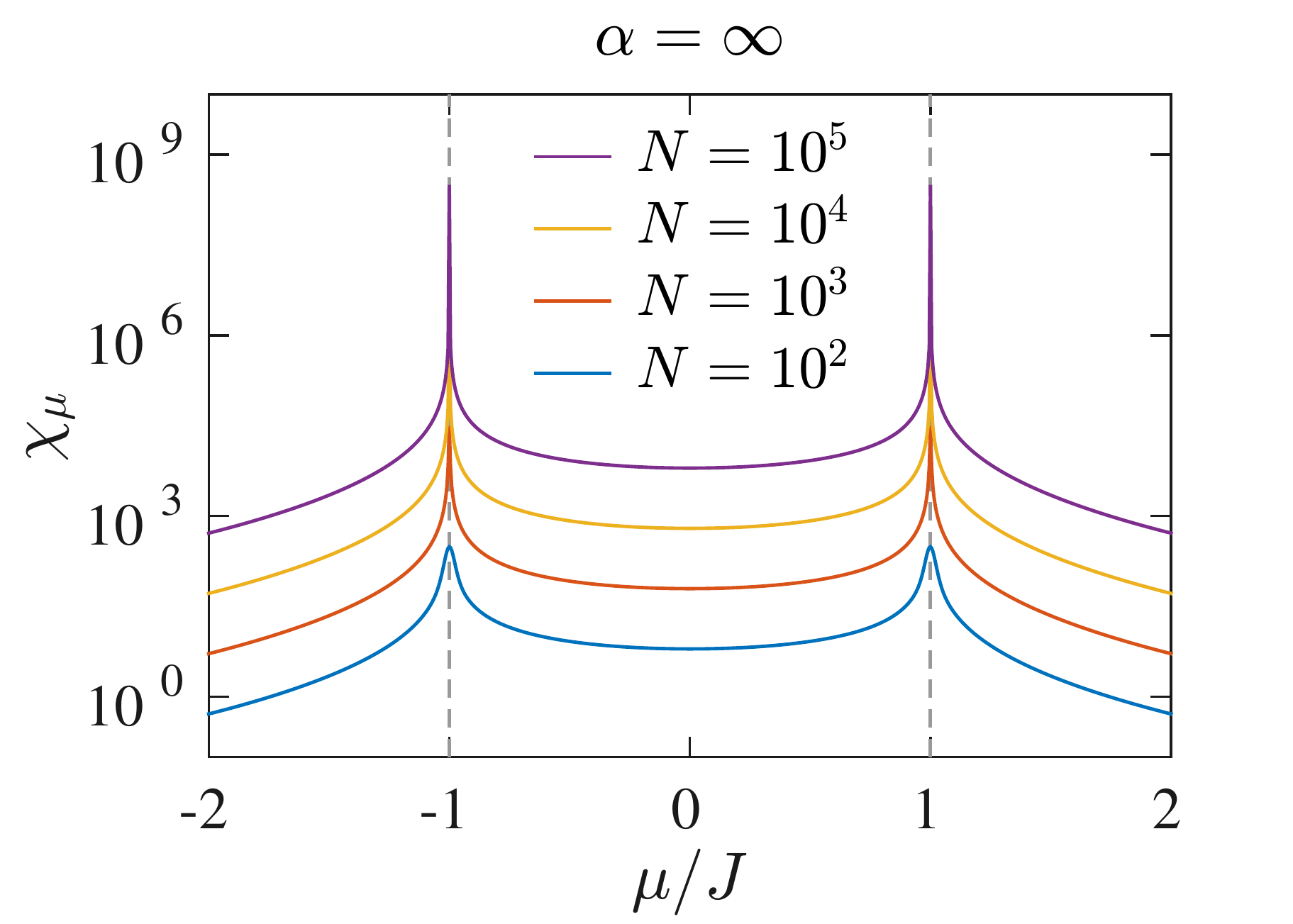} \hfill
\includegraphics[width=0.45\textwidth]{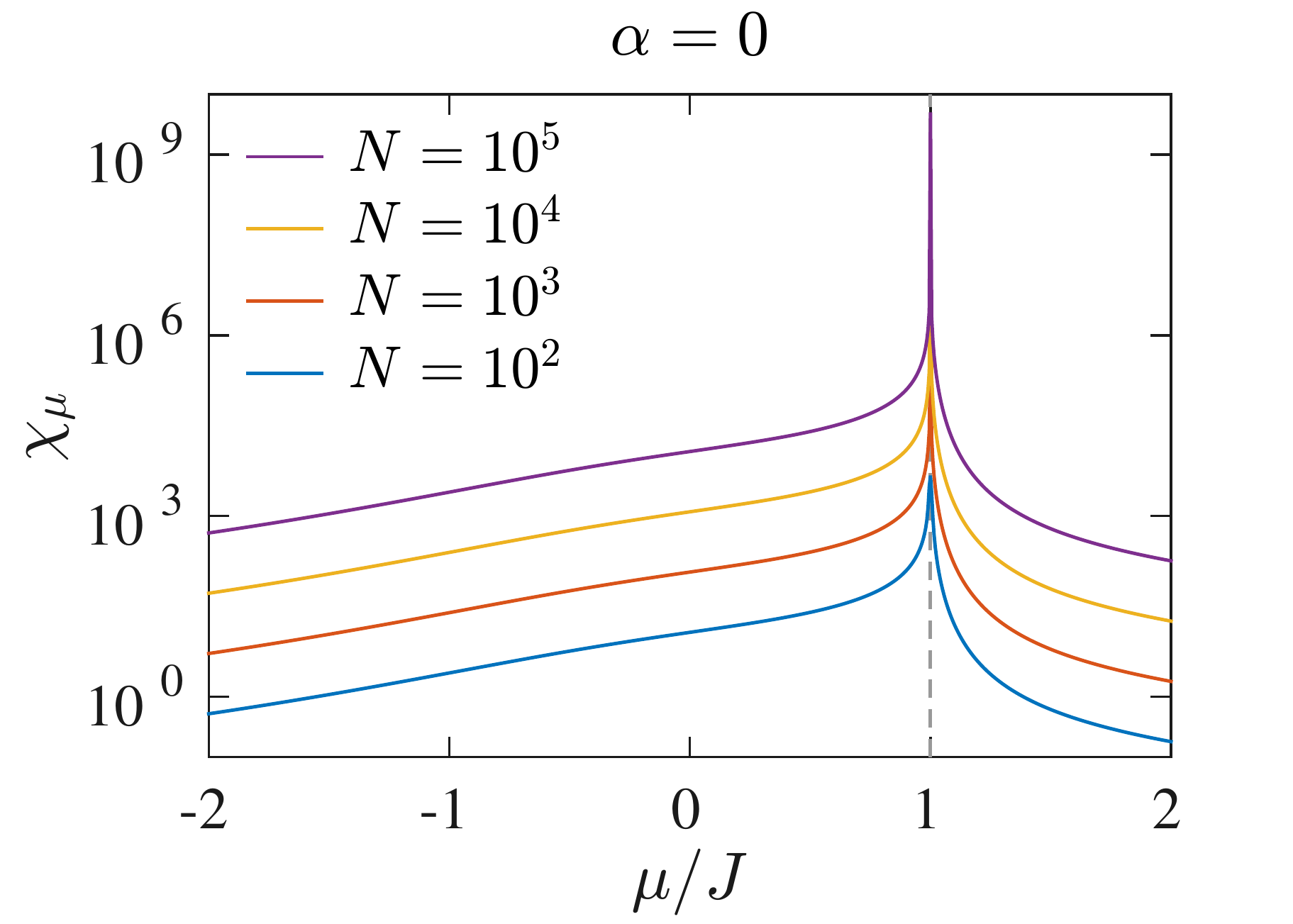} \hfill
\caption{Fidelity susceptibility Eq.~(\ref{chiMu}) as a function of $\mu$ for $\alpha=100$ \textbf{(left)} and $\alpha=0$ \textbf{(right)}. 
Here $\DD=J$. The dashed gray lines signal the critical points.
The maximum of $\chi_\mu$ occurs in correspondence of the critical points, signalled by dashed vertical lines. 
In both cases, we find a superextensive scaling $\max_\mu\chi_\mu \sim \sites^2$.}
\label{fig:FidelitySusceptibilityMu}
\end{figure}
%%%%%%%%%%%%%%%%%%%%%%%%%%%%%%%%%%%%%%%%%%%%%%%%%%%%%%%%%
%%%%%%%%%%%%%%%%%%%%%%%%%%%%%%%%%%%%%%%%%%%%%%%%%%%%%%%%%

%%%%%%%%%%%%%%%%%%%%%%%%%%%%%%%%%%%%%%%%%%%%%%%%%%%%%%%%%%
%%%%%%%%%%%%%%%%%%%%%%%%%%%%%%%%%%%%%%%%%%%%%%%%%%%%%%%%%
\begin{figure}[p!]
\centering
\hfill \includegraphics[width=0.45\textwidth]{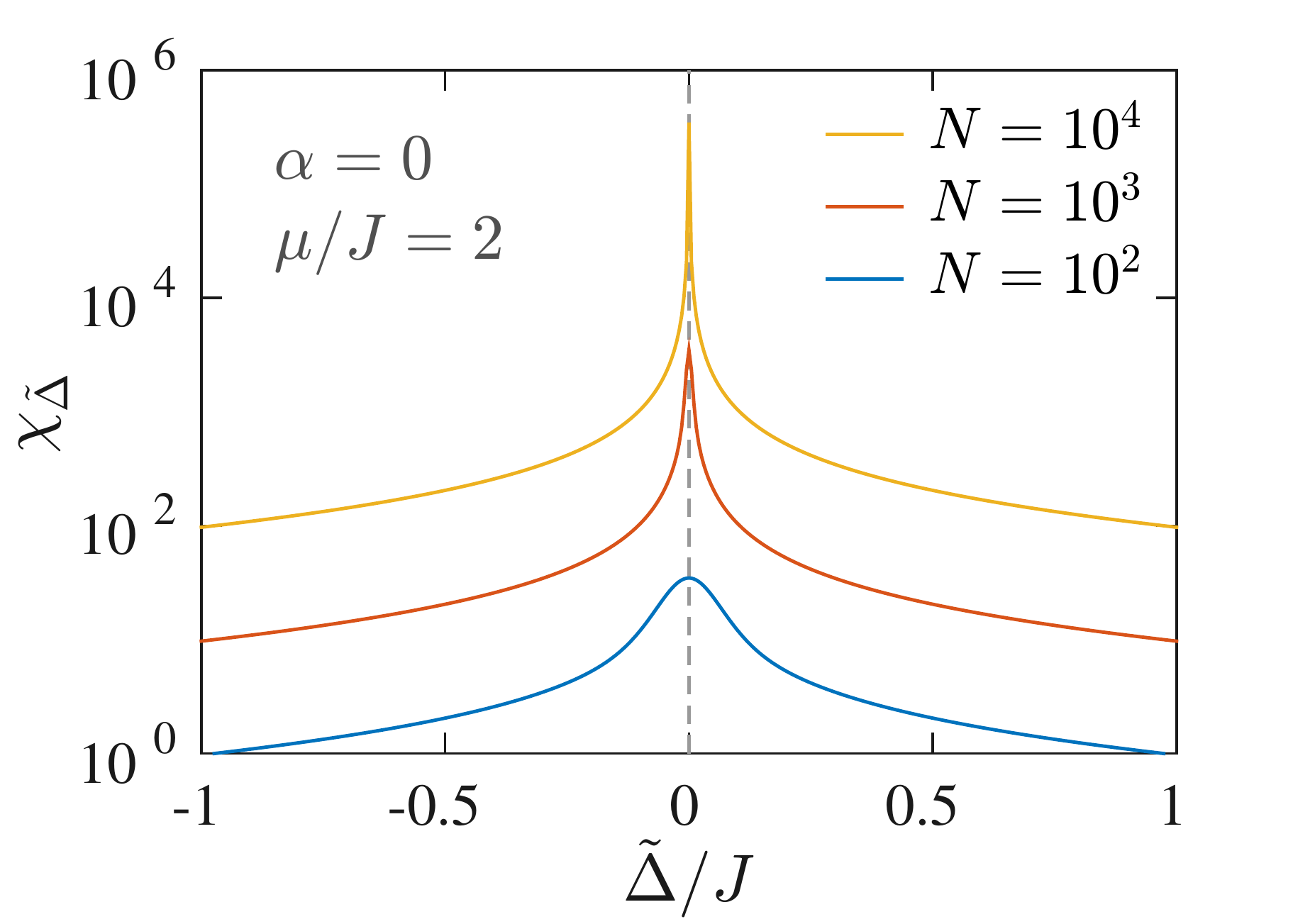} \hfill
\includegraphics[width=0.45\textwidth]{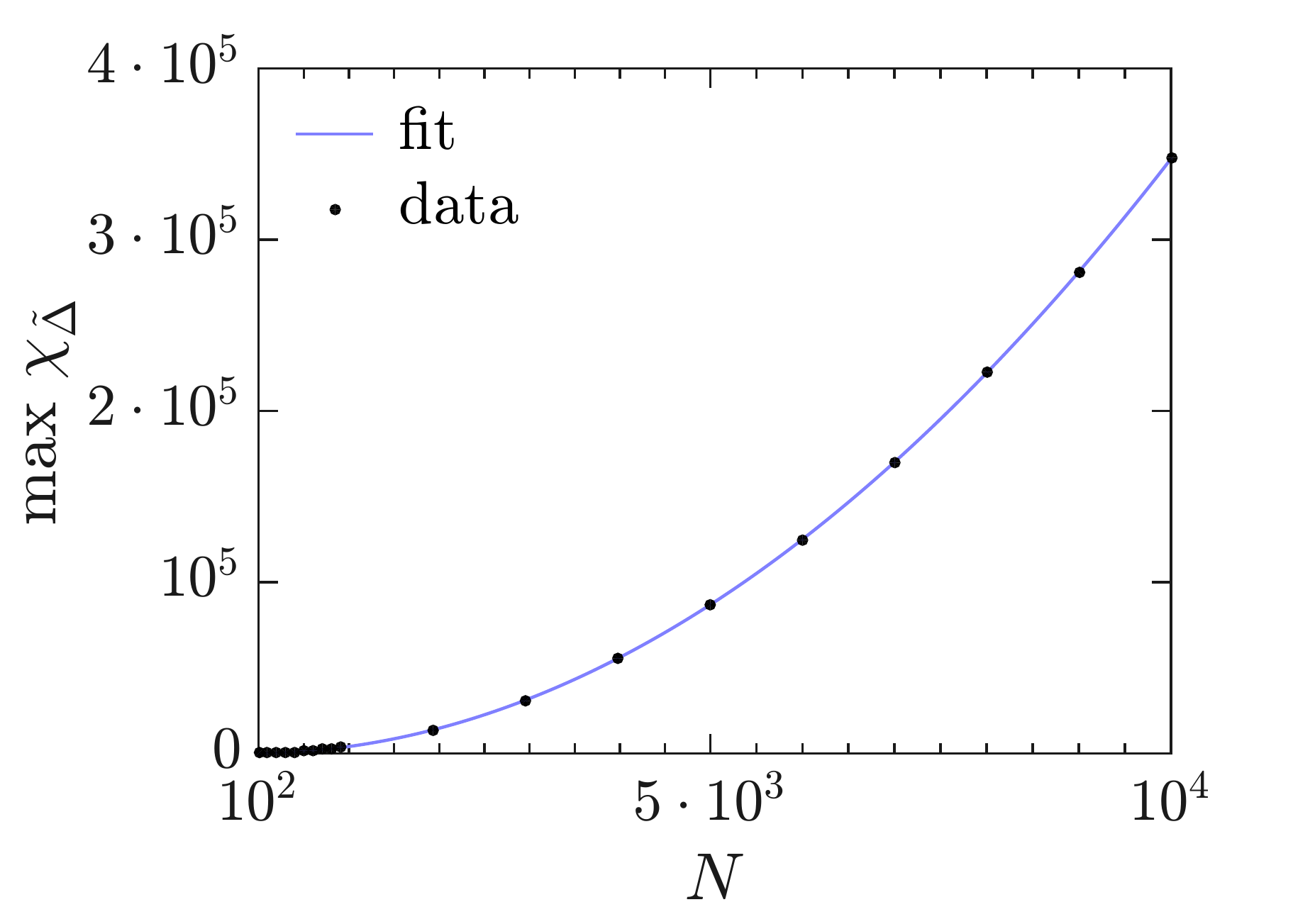} \hfill
\caption{\textbf{(Left)} Fidelity susceptibility Eq.~(\ref{chiDelta}) as a function of $\DD$ for $\alpha=0$ and $\mu/J>1$.
Different solid lines refer to different values of $\sites$ in the limit $\sites\to\infty$. 
The dashed gray line signals the critical point $\DD=0$, not associated to any closing gap.
\textbf{(Right)} Scaling of the maximum of $\chi_{\DD}$ for increasing $\sites$. 
The solid line is a fit curve $\chi_{\DD}(0) \propto \sites^2$.}
\label{fig:FidelitySusceptibilityDelta}
\end{figure}
%%%%%%%%%%%%%%%%%%%%%%%%%%%%%%%%%%%%%%%%%%%%%%%%%%%%%%%%%
%%%%%%%%%%%%%%%%%%%%%%%%%%%%%%%%%%%%%%%%%%%%%%%%%%%%%%%%%

%%%%%%%%%%%%%%%%%%%%%%%%%%%%%%%%%%%%%%%%%%%%%%%%%%%%%%%%%
%%%%%%%%%%%%%%%%%%%%%%%%%%%%%%%%%%%%%%%%%%%%%%%%%%%%%%%%%
\begin{figure}[p!]
\centering
\hfill \includegraphics[width=0.45\textwidth]{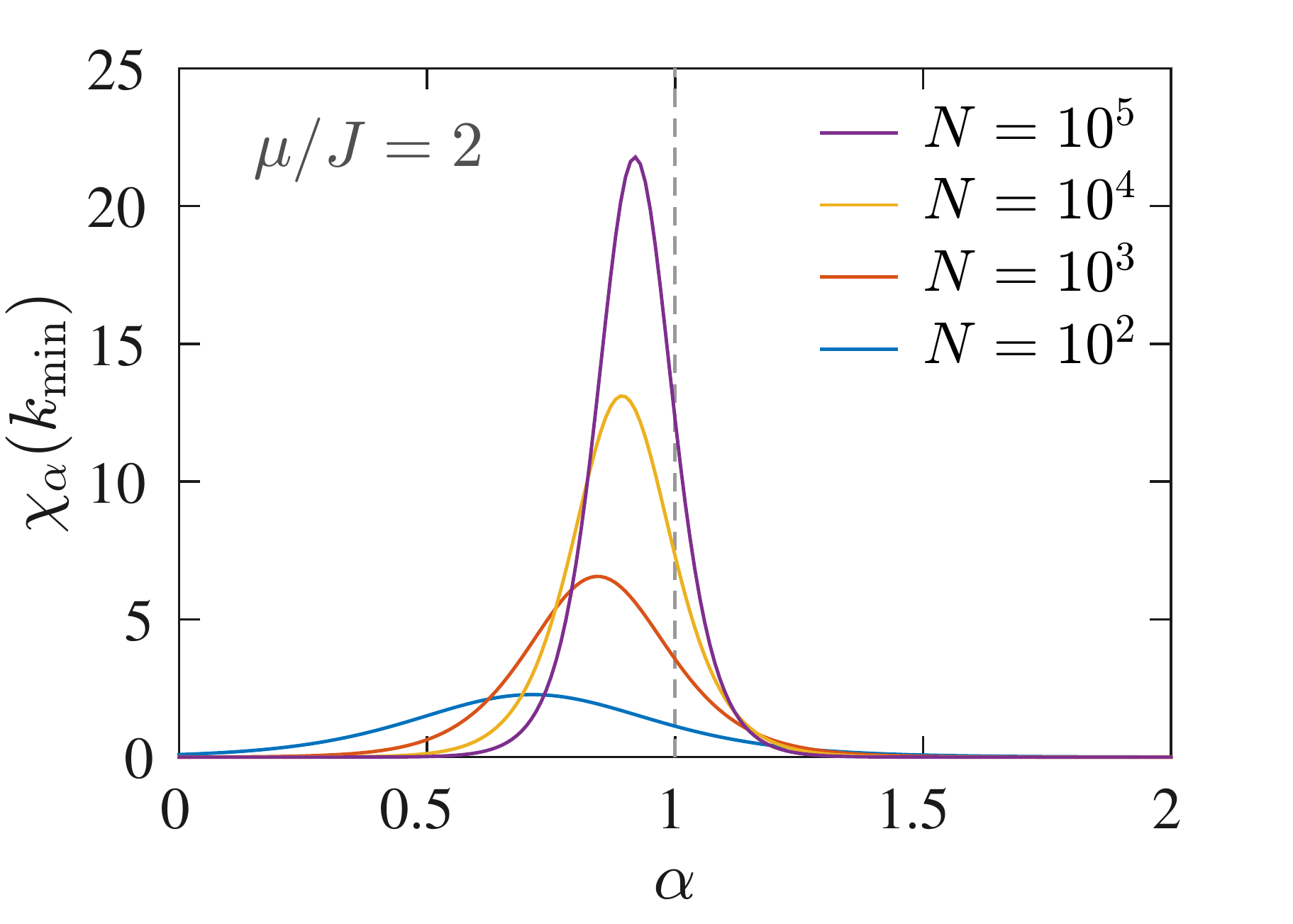} \hfill
\includegraphics[width=0.45\textwidth]{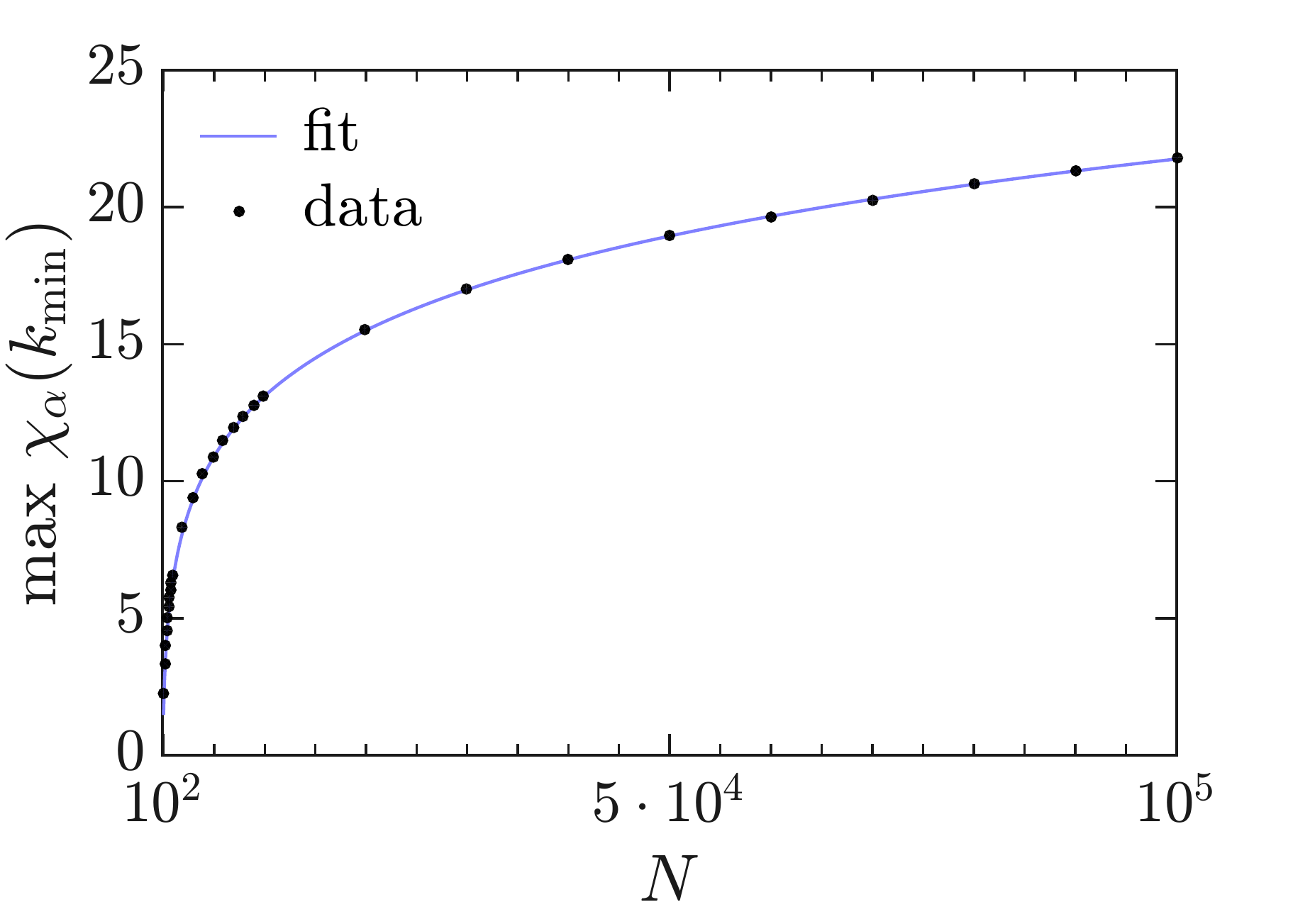} \hfill
\caption{\textbf{(Left)} Plot of $\chi_\alpha(k_{\rm min})$ in Eq.~(\ref{chiAlphaK}) around $\alpha=1$ (dashed vertical line). 
Different solid lines refer to different values of $\sites$ in the limit $\sites\to\infty$.
Here $\DD=J$ and $\mu/J=2$ (similar results can be found for other values of $\mu$).
\textbf{(Right)} Scaling of the maximum of $\chi_\alpha(k_{\rm min})$ for increasing $\sites$. 
The solid line is a fit curve $\max_\alpha\chi_\alpha(k_{\rm min}) \propto (\log{\sites})^2 + \mathrm{const}$.}
\label{fig:FidelitySusceptibilityAlpha}
\end{figure}
%%%%%%%%%%%%%%%%%%%%%%%%%%%%%%%%%%%%%%%%%%%%%%%%%%%%%%%%%
%%%%%%%%%%%%%%%%%%%%%%%%%%%%%%%%%%%%%%%%%%%%%%%%%%%%%%%%%

\section{Fidelity susceptibility}
The ground-state fidelity is given by %~\cite{Pezze2017}
\be
\pazocal{F} =  \big| \bra{ \psi_{0} } \widetilde{\psi}_{0} \rangle \big| = 
\prod_{0\,<\,k\,<\,\pi} \bigg| \cos\bigg(\frac{\Theta_k - \widetilde{\Theta}_k}{2} \bigg) \bigg|,
\ee
where $\ket{\psi_{0}}=\ket{\psi_{0}(\mbox{\boldmath$\eta$})}$ 
and $\ket{\tilde{\psi}_{0}}=\ket{\psi_{0}(\tilde{\mbox{\boldmath$\eta$}})}$ 
are to the ground states for different values of the parameters $\mbox{\boldmath$\eta$}=(J,\mu,\DD,\alpha)$.
Expanding in Taylor series for $\widetilde{\Theta}_k\approx\Theta_k$, 
up to the second order in $\delta\Theta_k = \widetilde{\Theta}_k-\Theta_k\approx0$, we find
\be
\pazocal{F} \approx 1 - \frac{1}{8} \sum_{0\,<\,k\,<\,\pi} (\delta\Theta_k)^2 \, .
\ee
The fidelity susceptibility with respect to the variation of a generic parameter $\eta$ is defined as 
\be
\chi_\eta = - \frac{\ud^2 \pazocal{F}}{\ud \eta^2} = \sum_{0\,<\,k\,<\,\pi} \bigg(\frac{1}{2}\frac{\ud \Theta_k}{\ud \eta}\bigg)^2,
\ee
where 
\be
\Theta_k = \arccos \frac{-(J\cos k + \mu)}{\sqrt{(J \cos k + \mu)^2 + \big(\tfrac{\DD}{2}\,f_\alpha(k)\big)^2}} \, .
\ee
The explicit expression of the fidelity susceptibility considering the variation of different parameters follows:
\\[-9pt]
%%%%%%%%

\noindent \tripieno{MySky} tuning $\mu$:
\be \label{chiMu}
\chi_\mu = \sum_{0\,<\,k\,<\,\pi}
\frac{1}{\Big[(J \cos k + \mu)^2 + \big(\tfrac{\DD}{2}\,f_\alpha(k)\big)^2\Big]^2} \bigg(\frac{\DD}{4}\,f_\alpha(k)\bigg)^2
\ee 
\\[-9pt]
%%%%%%%%

\noindent \tripieno{MySky} tuning $\DD$:
\be \label{chiDelta}
\chi_{\DD} = \sum_{0\,<\,k\,<\,\pi}
\frac{(J \cos k + \mu)^2}{\Big[(J \cos k + \mu)^2 + \big(\tfrac{\DD}{2}\,f_\alpha(k)\big)^2\Big]^2} \bigg(\frac{f_\alpha(k)}{4}\bigg)^2
\ee
\\[-9pt]
%%%%%%%%

\noindent \tripieno{MySky} tuning $\alpha$:
\be \label{chiAlpha}
\chi_\alpha = \sum_{0\,<\,k\,<\,\pi}
\frac{(J \cos k + \mu)^2}{\Big[(J \cos k + \mu)^2 + \big(\tfrac{\DD}{2}\,f_\alpha(k)\big)^2\Big]^2}
\bigg(\frac{\DD}{4} \frac{\ud f_\alpha(k)}{\ud\alpha} \bigg)^2
\ee
\\[-9pt]
%%%%%%%%

\begin{figure}[t!]
\centering
\includegraphics[width=0.9\textwidth]{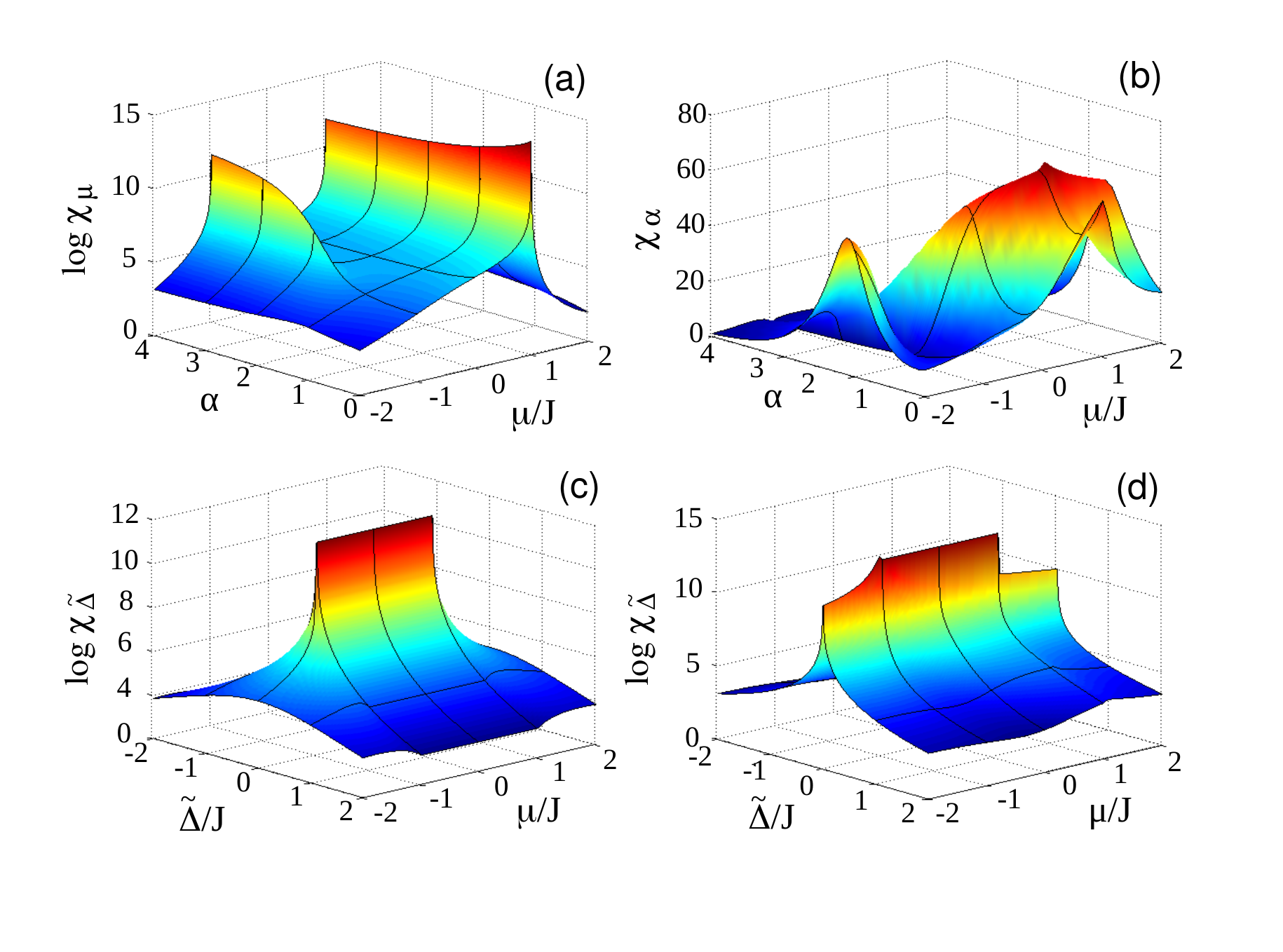}
\vspace{-30pt}
\caption{Fidelity susceptibilities $\chi_{\mu}$ with $\DD=J$ \textbf{(a)}, $\chi_{\alpha}$ with $\DD=J$ \textbf{(b)}, 
$\chi_{\DD}$ with $\alpha=100$ \textbf{(c)}, and $\chi_{\DD}$ with $\alpha=0$ \textbf{(d)}.
%All plots have been obtained for $\sites=1000$ sites. 
%In panels (a,\,b) $\DD=J$; in panels (c) $\alpha=1000$, while in (d) $\alpha=0$.
The singularity at $\alpha=1$ for $\chi_\alpha$ develops very slowly in the system size.}
\label{fig:FidelitySusceptibility3D}
\end{figure}

\noindent In particular, Eq.~(\ref{chiMu}) agrees with the result previously given in Ref.~\cite{ZanardiPRL2007}
and it is shown in Fig.~\ref{fig:FidelitySusceptibilityMu} for the limit cases $\alpha=\infty$ and $\alpha=0$. 
Complete plots of the fidelity susceptibilities are shown in Fig.~\ref{fig:FidelitySusceptibility3D}.

%%%%%%%%%%%%%%%%%%%%%%%%%%%%%%%%%%%%%%%%%%%%%%%%%%%%%%%%%
%%%%%%%%%%%%%%%%%%%%%%%%%%%%%%%%%%%%%%%%%%%%%%%%%%%%%%%%%
%%%%%%%%%%%%%%%%%%%%%%%%%%%%%%%%%%%%%%%%%%%%%%%%%%%%%%%%%
%%%%%%%%%%%%%%%%%%%%%%%%%%%%%%%%%%%%%%%%%%%%%%%%%%%%%%%%%

\paragraph{Signatures of the QPTs}
Fidelity susceptibility succeeds in detecting the topological QPTs of the Kitaev chain: 
every QPT driven by the control parameter $\eta$ 
is signalled by a \emph{divergence} of $\chi_\eta$ in the thermodynamic limit $\sites\to\infty$. 
Figure~\ref{fig:FidelitySusceptibilityMu} reports the divergence of $\chi_\mu$ at the critical points 
$\mu/J=+1$ (when $\alpha\geq0$) and $\mu/J=-1$ (when $\alpha>1$), where a closing gap shows up.
Figure~\ref{fig:FidelitySusceptibilityDelta} reports the divergence of $\chi_{\DD}$ at the critical point
$\DD=0$ (when $\alpha<1$); in particular, the plot refers to the case $\mu>J$, where no closing gap arises.

The divergence of $\chi_\alpha$ at $\alpha=1$ is thornier to seize numerically.
Since the sum in Eq.~(\ref{chiAlpha}) contains solely positive terms, it is sufficient to find that one term in the sum diverges.
We calculate 
\be \label{chiAlphaK}
\chi_\alpha(k_{\rm min}) = \frac{(J \cos k_{\rm min} + \mu)^2}
{\Big[(J \cos k_{\rm min} + \mu)^2 + \big(\tfrac{\DD}{2}\,f_\alpha(k_{\rm min})\big)^2\Big]^2} 
\bigg(\frac{\DD}{4} \frac{\ud f_\alpha(k_{\rm min})}{\ud \alpha} \bigg)^2
\ee 
in the limit $k_{\rm min} = \pi/\sites\to0$.
The numerical result is shown in Fig.~\ref{fig:FidelitySusceptibilityAlpha}. 
Very slowly in $\sites$, $\chi_\alpha(k_{\rm min})$ develops a peak at $\alpha=1$ growing as $(\log\sites)^2$.
The divergence of $\chi_\alpha (k)$ justifies the emergence of new ``long-range'' phases below $\alpha=1$, 
even if no vanishing gap in the Bogoliubov spectrum $\epsilon_k$ occurs at this threshold. 

More complete plots of the fidelity susceptibility as a function of the parameters are shown in Fig.~\ref{fig:FidelitySusceptibility3D}.
They should be compared with the phase diagram of Fig.~\ref{fig:KitaevTopology}.

\section{Multipartite entanglement in the ground state}
We want to quantify the entanglement in the topological phases of the Kitaev model 
in terms of the lower bound on the entanglement depth provided by the quantum Fisher information. 
Moreover, guided by the experience gained in the previous chapters, 
we wish that quantum Fisher information can flag the boundaries between different phases.

Initially, we rely on local observables, along the lines of what we arranged for the models previously examined,
and we learn that the quantum critical points are points of nonanalyticity for the QFI. 
Unfortunately, the local probing is not suitable for revealing the high (namely, superextensive) entanglement content 
both at the critical points and in the nontrivial phases.
For this task, we need to employ suitable nonlocal observables: 
with their aid, the QFI can discriminate between trivial and nontrivial phases
and still exhibits singularities at the critical points.

It is important to clarify that here multipartite entanglement pertains to the lattice sites, 
and not the actual fermions (whose number is not conserved): 
the sites of the chain are the physical distinct subsystems, arising from the partition of the Hilbert space, 
among which the tunnelling and pairing generate quantum correlations. 
% that we are interested to witness by means of the QFI.
This is a form of \emph{mode entanglement} rather than particle entanglement,
emergent in the second-quantization picture.
% rather than in the single-particle picture

\subsection{Local probing operators}
In continuity with the procedure we followed in Chapters~\ref{ch:Ising} and \ref{ch:LMG} for the other quadratic models,
we attempt here to detect multipartite entanglement by evaluating the quantum Fisher information with local probing operators.
In order to do so, we identify each lattice site with an effective spin\footnote{\ Notice that this object is not the spin obtained from an inverse Jordan-Wigner transformation: it is just a fictitious two-level system built up from the elementary fermionic operators.\\[-9pt]}
associating ``spin up'' to the occupied site and ``spin down'' to the empty site, 
thus constructing ladder operators for the $j$th spin: $\hat{\sigma}_+^{(j)}=2\hat{a}^\dag_j$ and $\hat{\sigma}_-^{(j)}=2\hat{a}_j$.
Hence, $\big[\hat{\sigma}_+^{(j)},\hat{\sigma}_-^{(j)}\big]=4\hat{\sigma}_z^{(j)}$ 
with $\hat{\sigma}_z^{(j)}=2\hat{a}_j^\dag\hat{a}_j-\identity_j$.
% $\hat{\sigma}_x^{(j)}=\hat{a}_j^\dag+\hat{a}_j$, 
% $\hat{\sigma}_y^{(j)}=-\ii(\hat{a}_j^\dag-\hat{a}_j)$ and
% $\hat{\sigma}_z^{(j)}=2\hat{a}_j^\dag\hat{a}_j-1$.
% Again, we introduce effective collective spin operators
We then introduce collective operators
\be \label{KitaevLocalProbes}
\hat{G}_x = \sum_{j=1}^\sites \frac{\hat{a}_j^\dag+\hat{a}_j}{2}, \quad 
\hat{G}_y = \sum_{j=1}^\sites \frac{\hat{a}_j^\dag-\hat{a}_j}{2\ii}, \quad 
\hat{G}_z = \sum_{j=1}^\sites \frac{\,\hat{\sigma}_z^{(j)}}{2} \, . 
%\hat{G}_z = \sum_{j=1}^\sites \frac{2\hat{a}_j^\dag\hat{a}_j-1}{2} \, .
\ee
They are \emph{local} operators in the chain sites, acting independently on each fermionic mode. 
Evaluating the QFI by means of such observables means to look for the degree of distinguishability of the ground state 
after a modification of the occupation number in each site according to Eqs.~(\ref{KitaevLocalProbes}).
%after a Ramsey rotation has been applied to all of the $\sites$ sites with the same amplitude; 
%that is, after carrying out $\sites$ equal beam splitters that modify the average occupation number in each site.
%These operators satisfy the usual angular-momentum algebra 
Notice that the above operators are \emph{not} collective spin operators, 
since they do not satisfy the usual angular-momentum algebra: 
$\big[\hat{G}_\varrho,\hat{G}_\varsigma\big]\neq\ii\,\varepsilon_{\varrho\varsigma\tau}\,\hat{G}_\tau$.
% $\big[\hat{\sigma}_\alpha^{(j)},\hat{\sigma}_\beta^{(j)}\big]=2\,\ii\,\varepsilon_{\alpha\beta\gamma}\hat{\sigma}_\gamma^{(j)}$. 
% $\big[\hat{\sigma}_\varrho^{(j)},\hat{\sigma}_\varsigma^{(j)}\big]=2\,\ii\,\varepsilon_{\varrho\varsigma\tau}\hat{\sigma}_\tau^{(j)}$.
%$\big[\hat{\sigma}_\lambda^{(j)},\hat{\sigma}_\mu^{(j)}\big]=2\,\ii\,\varepsilon_{\lambda\mu\nu}\hat{\sigma}_\nu^{(j)}$.

%%%%%%%%%%%%%%%%%%%%%%%%%%%%%%%%%%%%%%%%%%%%%%%%%%%%%%%%%%
%%%%%%%%%%%%%%%%%%%%%%%%%%%%%%%%%%%%%%%%%%%%%%%%%%%%%%%%%
\begin{figure}[t!]
\centering
\includegraphics[height=0.34\textwidth]{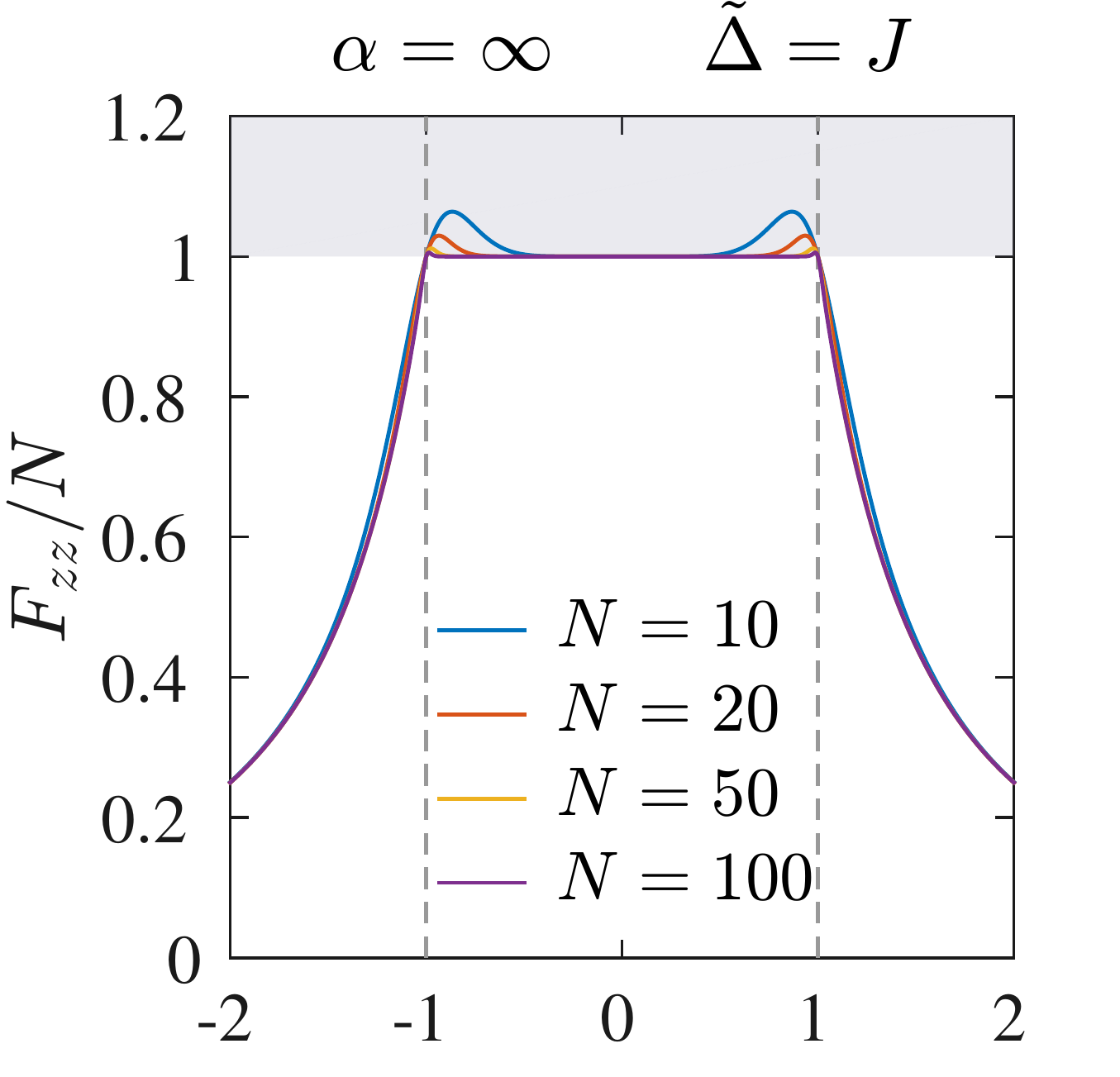} \hspace{-12pt} 
\includegraphics[height=0.34\textwidth]{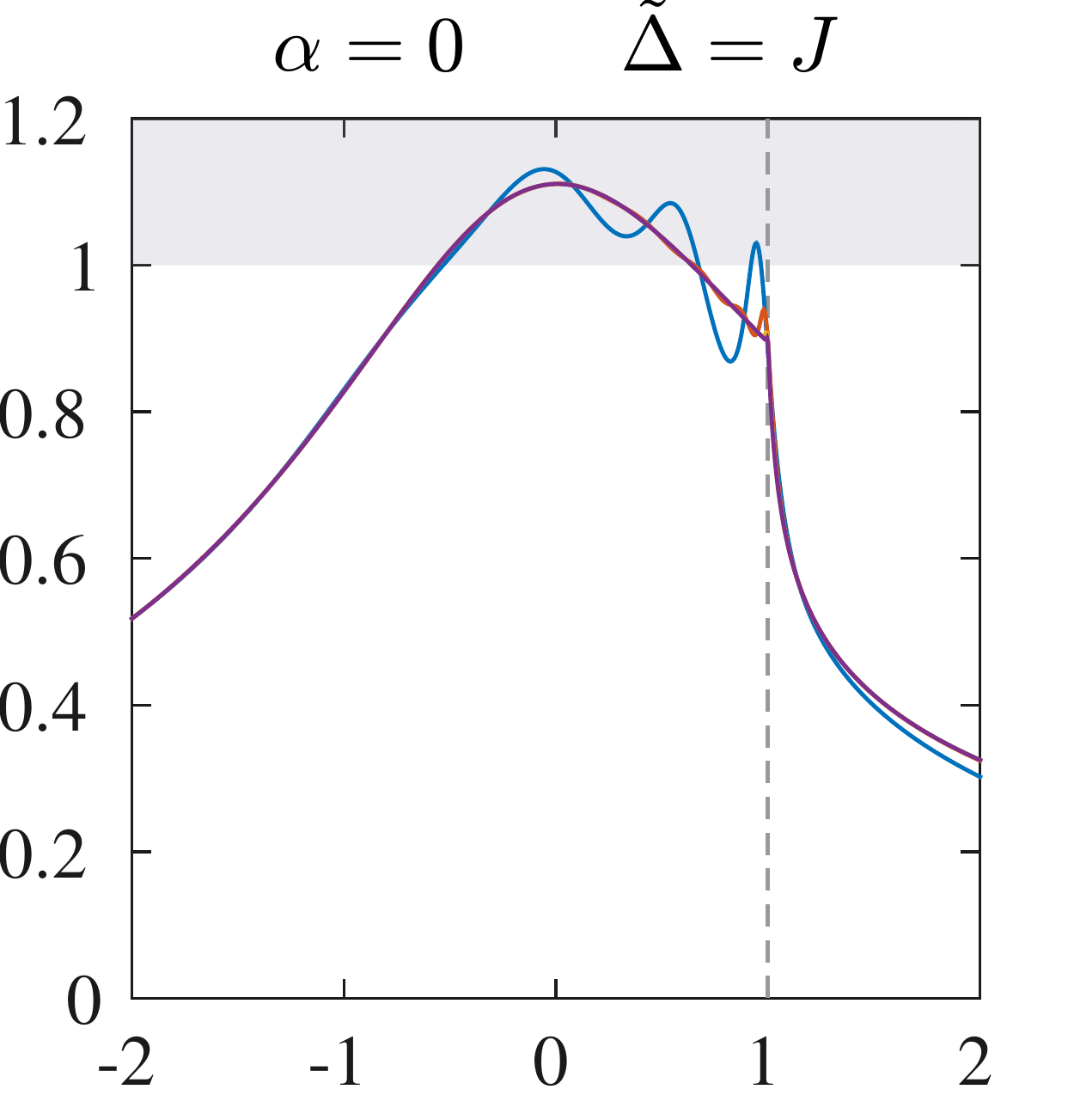} \hspace{-12pt}
\includegraphics[height=0.34\textwidth]{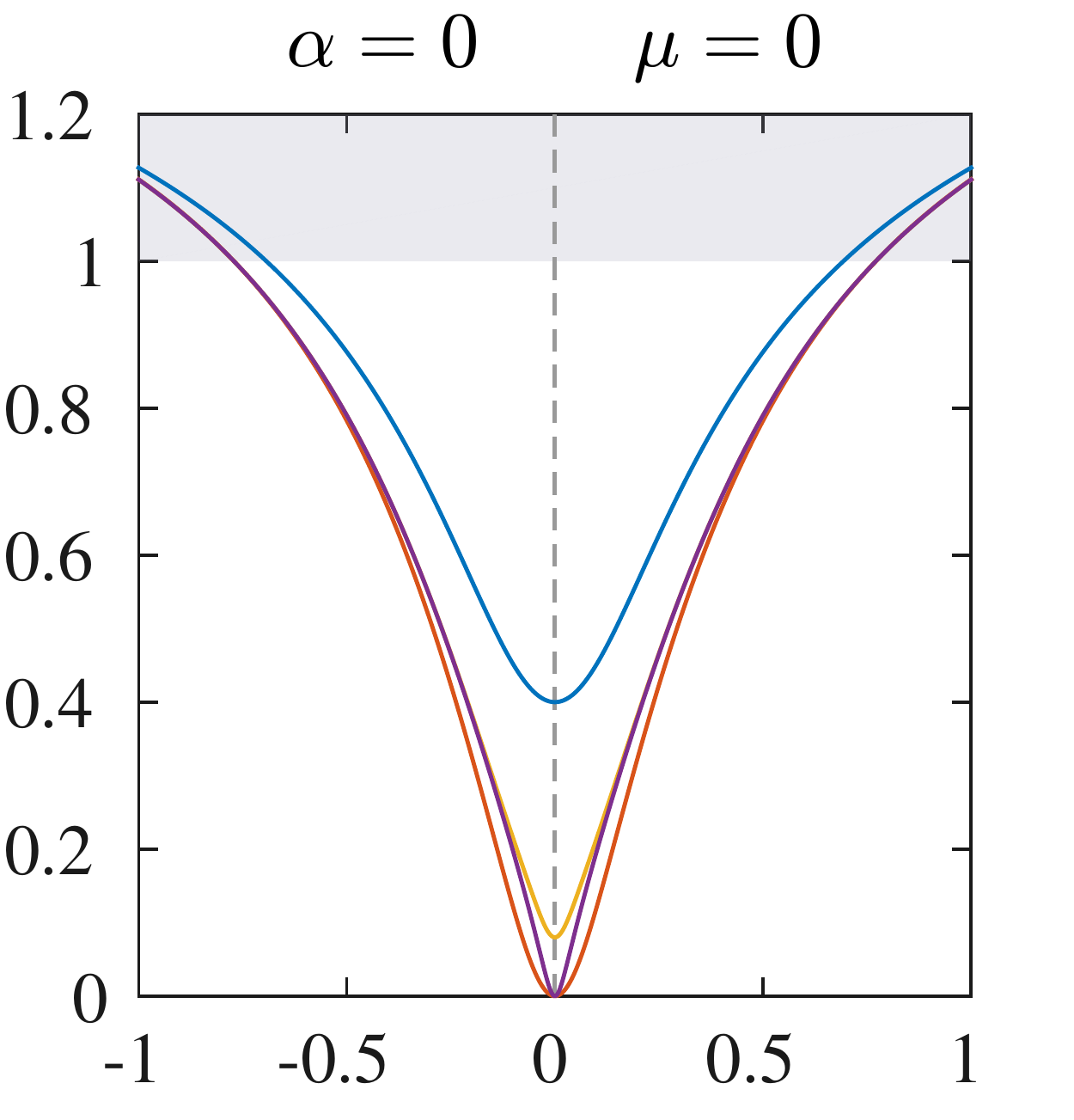} \\
\includegraphics[height=0.34\textwidth]{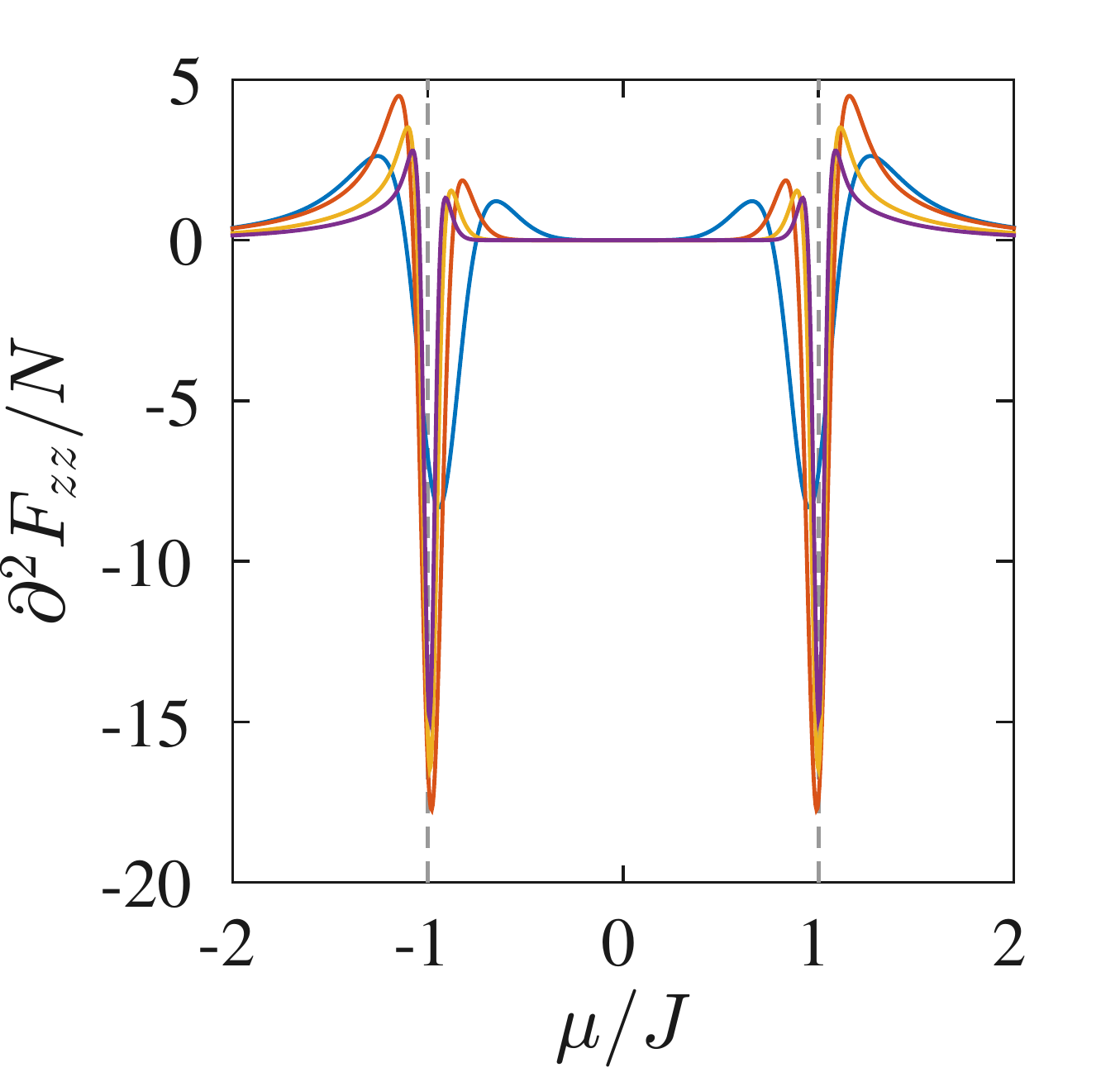} \hspace{-14pt}
\includegraphics[height=0.34\textwidth]{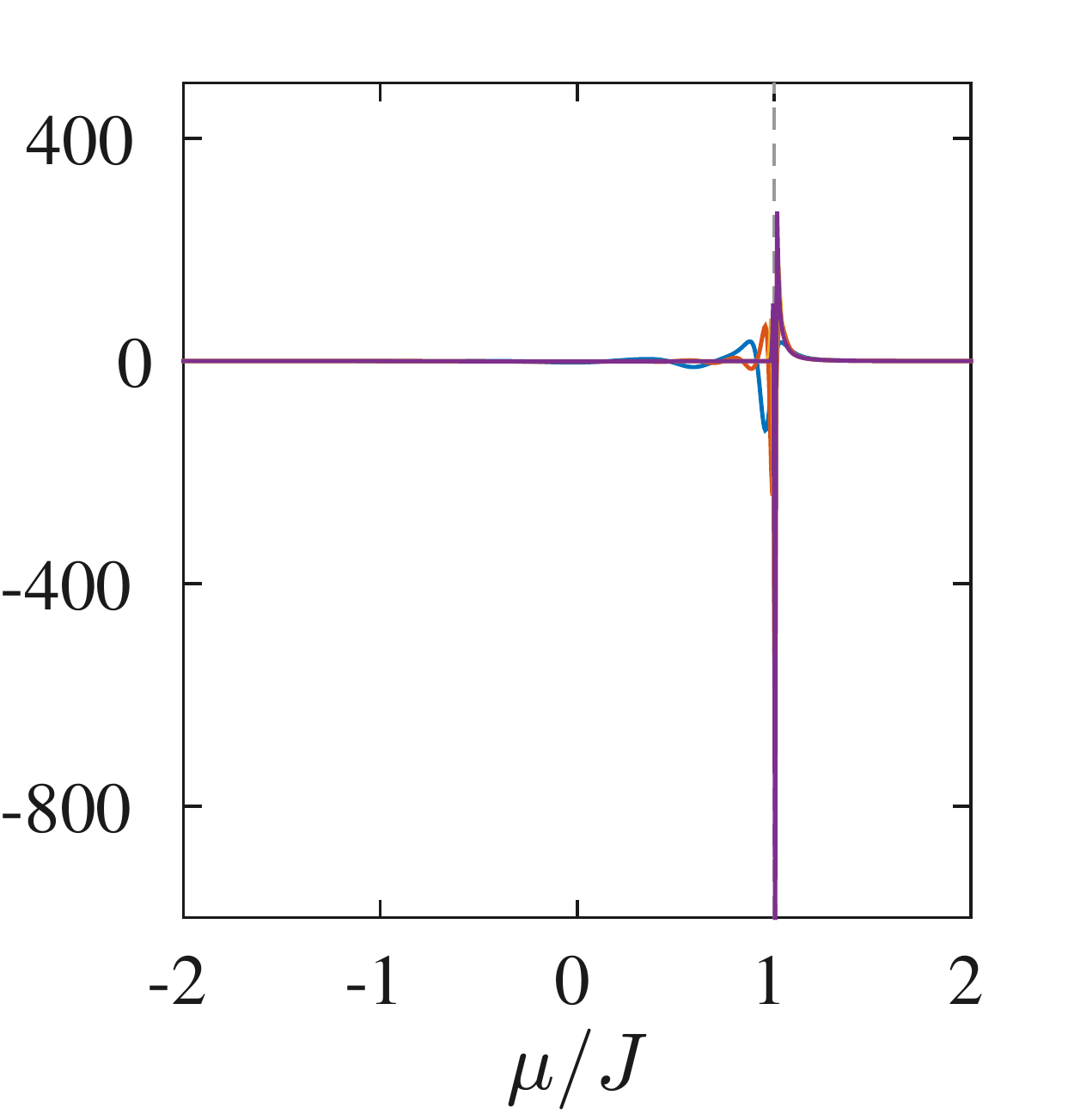} \hspace{-12pt}
\includegraphics[height=0.34\textwidth]{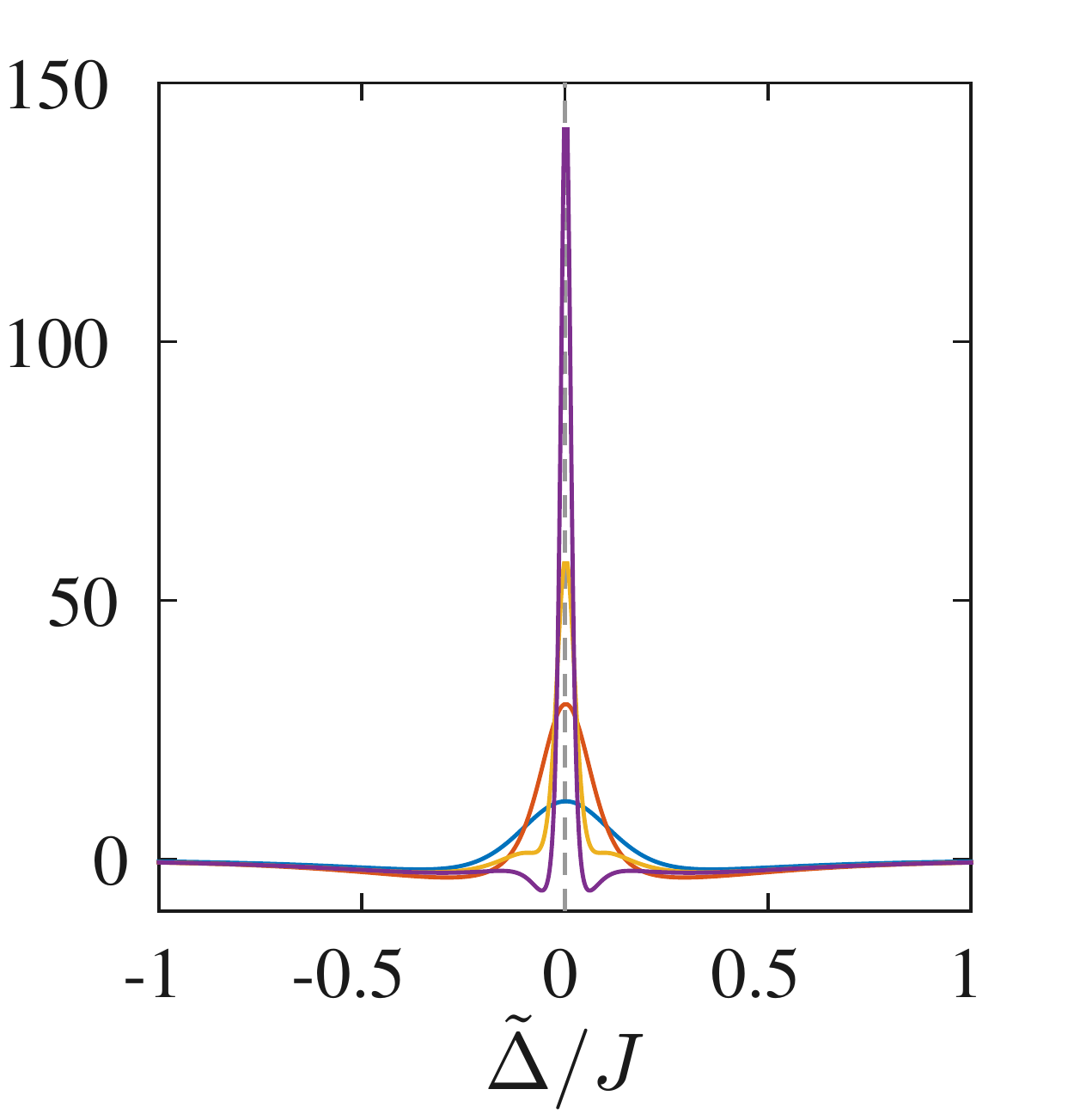} \\
\caption{\textbf{(Top)} Quantum Fisher information $F_Q\big[\ket{\psi_{0}},\hat{G}_z\big]\equiv F_{zz}$ 
as a function of $\mu$ for $\alpha=100$ and $\DD>0$ (left); as a function of $\mu$ for $\alpha=0$ and $\DD>0$; 
as a function of $\DD$ for $\alpha=0$ and $\mu=0$ (right). 
\textbf{(Bottom)} Second derivative of the upper plots with respect to the proper parameter.
The coloured solid lines refer to four system size $\sites=10,\,20,\,50,\,100$.
The dashed gray lines indicate the critical points. 
In all these examples, the critical points are associated to closing gap.
Shaded area is where entanglement is found.}
\label{fig:KitaevLocalFisher}
\end{figure}
%%%%%%%%%%%%%%%%%%%%%%%%%%%%%%%%%%%%%%%%%%%%%%%%%%%%%%%%%
%%%%%%%%%%%%%%%%%%%%%%%%%%%%%%%%%%%%%%%%%%%%%%%%%%%%%%%%%

\paragraph{Quantum Fisher information} 
Using the results in Eqs.~(\ref{KitaevCorrelations}), we find out that 
${\big\langle\hat{G}_\varrho\big\rangle=0}$ ${\forall\varrho\in\{x,y,z\}}$. 
Moreover, the variances $\big\langle\hat{G}_\varrho^2\big\rangle$ 
and covariances $\big\langle\hat{G}_\varrho\hat{G}_\varsigma\big\rangle$ of the collective local operators
can also be calculated analytically with ease.
The quantum Fisher information matrix Eq.~(\ref{SU2optimization}) for the local probing operators 
(\ref{KitaevLocalProbes}) turns out to be 
\be \label{KitaevQFIlocal}
\mathbb{F}_Q = \begin{pmatrix}
\sites & 0 & 0 \\
0 & \sites & 0 \\
0 & 0 & F_{zz}
\end{pmatrix}, \qquad \mathrm{with} \quad 
F_{zz} = 2 \sum_k \frac{\big(\tfrac{\DD}{2}\,f_\alpha(k)\big)^2}{(J \cos k + \mu)^2 + \big(\tfrac{\DD}{2}\,f_\alpha(k)\big)^2} \, .
\ee 
The Fisher information $F_{xx}=F_Q\big[\ket{\psi_{0}},\hat{G}_x\big]=\sites$ was previously obtained in Ref.~\cite{HaukeNATPHYS2016}
and the constant behaviour as a function of the chemical potential was reported.
An optimization over the three spatial directions leads to 
$\bf{n}_{\rm opt}=\bf{z}$ when $F_{zz}/\sites>1$ and $\bf{n}_{\rm opt}\in(\bf{x},\bf{y})$ otherwise. 
Here we focus only on $F_{zz}=F_Q\big[\ket{\psi_{0}},\hat{G}_z\big]$, 
since it is the sole component that can overcome the shot-noise limit and witness entanglement, 
and the only one to display some dependency on the parameters.

\paragraph{Signatures of the QPTs}
Equation~(\ref{KitaevQFIlocal}) looks qualitatively different from $\chi_\mu$ in Eq.~(\ref{chiMu})
owing to the presence of $\epsilon_k^2$ instead of $\epsilon_k^4$ at the denominator. 
This prevents $F_{zz}$ to diverge even if $\epsilon_k\to0$, 
the singularity in $k=\pi$ being removed by the same-order infinitesimal $f_\alpha\to0$ at the numerator.
Instead, Eq.~(\ref{KitaevQFIlocal}) predicts a \emph{divergence of the second derivative} of the QFI when the gap closes. 
So, the QFI with a local probe detects the critical points associated to a closing energy gap: 
examples are reported in Fig.~\ref{fig:KitaevLocalFisher}.
In the QPTs driven by $\mu$, the divergence of $\partial_\mu^2F_{zz}$ is associated to an angle in $F_{zz}$. 
In the QPTs driven by $\DD$, the divergence is associated to an increasing concavity of $F_{zz}$ in the thermodynamic limit, 
even when the gap is open.
An apparent signal of divergence is also found in $\partial_\alpha^2F_{zz}$ around $\alpha=1$ for the guessed QPTs driven by $\alpha$, 
even though we are not in a position to infer the actual finite-size scalings, 
since values of $\sites$ higher than the ones we reached would be necessary.

Nevertheless, the local probing seems not to be the optimal procedure to unveil the entanglement content in the system:
since $\big(\tfrac{\DD}{2}\,f_\alpha(k)\big)^2 \leq (J \cos k + \mu)^2 + \big(\tfrac{\DD}{2}\,f_\alpha(k)\big)^2$, 
we have $F_{zz}/\sites\leq2$ and $F_{zz}/\sites\to2$ only in the limit $\DD\to\pm\infty$.
Thus, no superextensivity of the QFI is expected from local probing.

\subsection{Nonlocal probing operators}
Which operators can be suitable for disclosing the amount of entanglement hosted by the ground state of the Kitaev chain
in its topological phases? 
An answer can be prompted by the exact Jordan-Wigner mapping between the nearest-neighbour Kitaev model and the nearest-neighbour XY model,
discussed in Sec.~\ref{sec:KitaevModel}.
The XY model reduces to the Ising model for $\DD=J$: 
in virtue of the experience gained in chapter~\ref{ch:Ising}, 
we argue that a good choice for the probing operator --~at least for $\DD=J$ and $\alpha=\infty$~-- is the local collective spin operator 
$\hat{J}_x=\frac{1}{2}\sum_{j=1}^\sites\hat{\sigma}_x^{(j)}$ 
transformed via the Jordan-Wigner transformation into a collective fermionic operator.
Different values of the anistropy $\DD$ leads to taking into account also the Jordan-Wigner transformation of 
$\hat{J}_y=\frac{1}{2}\sum_{j=1}^\sites\hat{\sigma}_y^{(j)}$ 
and maximizing over the two orthogonal components for each set of Hamiltonian parameters. 
As expected, we find this choice to be optimal for $\alpha=\infty$. 
We decide to extend it to all the values of $\alpha$ and collect all the results in the following.

\paragraph{Quantum Fisher information} 
We witness multipartite entanglement in the ground state of the Kitaev model (\ref{HamKitaev}) using the QFI
\be 
\label{KitaevQFInonlocal}
F_Q\big[\ket{\psi_{0}}, \hat{O}_{\varrho}\big] = \bra{\psi_{0}} \hat{O}_{\varrho}^2 \ket{\psi_{0}} - \bra{\psi_{0}} \hat{O}_{\varrho} \ket{\psi_{0}}^2,
\ee
calculated with respect to the operator $\hat{O}_{\varrho} = \sum_{j=1}^\sites \hat{o}_{\varrho}^{(j)}$ ($\varrho\in\{x,y\}$), where 
\be \label{KitaevNonlocalProbes}
\hat{o}_{\varrho}^{(j)}= 
(-{\rm i})^{\delta_{\varrho y}}  \big( \hat{a}_j^\dag e^{i \pi \sum_{l=1}^{j-1} \hat{n}_l} + 
(-1)^{\delta_{\varrho y}} 
e^{-i \pi \sum_{l=1}^{j-1} \hat{n}_l} \hat{a}_j \big) \ \stackrel{^{\pazocal{ \ JW \ }}}{\longleftrightarrow} \ \hat{\sigma}^{(j)}_\varrho
\ee
and $\delta_{\varrho y} = 1$ for $\varrho=y$, $\delta_{\varrho y} = 0$ otherwise.
We also calculate the QFI with respect to the staggered operator $\hat{O}_{\varrho}^{({\rm st})} = \sum_{j=1}^\sites (-1)^j \hat{o}_{\varrho}^{(j)}$, namely $F_Q\big[\ket{\psi_{0}}, \hat{O}_{\varrho}^{(\rm st)}\big]$, $\varrho\in\{x,y\}$.
We generically indicate as $F_Q[\ket{\psi_{0}}]$ the maximum of Eq.~(\ref{KitaevQFInonlocal}) with
respect to the four operators above: 
$F_Q[\ket{\psi_{0}}] = \max_\varrho\big\{ F_Q\big[\ket{\psi_{0}}, \hat{O}_{\varrho}\big], \, 
F_Q\big[\ket{\psi_{0}}, \hat{O}_{\varrho}^{(\rm st)}\big] \big\}$.
Note that all these operators are highly \emph{nonlocal} in the fermionic modes. 
An experimental realizations of similar observables seems hardly reachable. 

We deliberately neglect the Jordan-Wigner transformation on the third orthogonal direction
$\hat{J}_z = \frac{1}{2}\sum_{j=1}^\sites\hat{\sigma}_z^{(j)} \stackrel{^{\pazocal{ \ JW \ }}}{\longrightarrow} 
\sum_{j=1}^\sites\big(\hat{a}_j^\dag\hat{a}_j-\frac{1}{2}\big)$ 
because it is one of the local probes in Eq.~(\ref{KitaevLocalProbes}). 
It would yield the component $F_{zz}$ in the Fisher matrix, that was already examined in Eq.~(\ref{KitaevQFIlocal}),
and found to be at most extensive $F_{zz}\leq2\sites$.

We will use interchangeably the QFI or the Fisher density 
$f_Q\big[\ket{\psi_{0}}, \hat{O}_{\varrho}\big] \equiv F_Q\big[\ket{\psi_{0}}, \hat{O}_{\varrho}\big]/\sites$, 
where we properly divide by the number of sites (and not by the mean number of fermions $\langle\hat{N}\rangle$)
because this is the number of independent parties whose correlation we are concerned about.
Noticing that $\bra{\psi_{0}} \hat{O}_{\varrho} \ket{\psi_{0}}=\bra{\psi_{0}} \hat{O}_{\varrho}^{({\rm st})} \ket{\psi_{0}}= 0$, we can rewrite the Fisher density  for the closed chain as
\be \label{QFIcorr}
f_Q\big[\ket{\psi_{0}}, \hat{O}_{\varrho}\big] = 1 + \frac{2}{\sites} \, \sum_{\ell=1}^{\sites-1} \ell\,C_{\varrho}(\ell) = 1 + \sum_{l=1}^{\sites-1}  C_{\varrho}(\ell) \, ,
\ee
where $C_{\varrho}(\ell)=\langle \psi_{0} \vert \hat{o}_\varrho^{(1)} \hat{o}_\varrho^{(1+\ell)} \vert \psi_{0} \rangle$, 
and analogous expressions hold true for $\hat{O}_{\varrho}^{(\rm st)}$, with 
$C_{\varrho}^{(\rm st)}(\ell)=\mean{\psi_{0} \vert  (-1)^\ell\hat{o}_\varrho^{(1)} \hat{o}_\varrho^{(1+\ell)} \vert \psi_{0}}$. 
Equation (\ref{QFIcorr}) directly links the QFI to the connected \emph{correlation function} for the operator $\hat{O}_{\varrho}$.

\paragraph{Numerical recipe}
The Jordan-Wigner string operators $\exp\big(\pm\ii\,\pi \sum_{l=1}^{j-1} \hat{a}_l^\dag \hat{a}_l\big)$ 
prevent a simple analytical treatment of Eq.~(\ref{KitaevQFInonlocal}).
Thus, we have numerically calculated Eq.~(\ref{QFIcorr}) for different values of the parameters of Kiteav chain
following a standard exact procedure~\cite{Lieb1961}.
In fact, the quadratic Hamiltonian (\ref{HamKitaev}) can be cast in the general form 
\be 
\hat{H}_\alpha = \sum_{i,\,j=1}^\sites \hat{a}^\dag_i A_{ij} \hat{a}_j \ + \, 
\frac{1}{2} \sum_{i,\,j=1}^\sites \big( \hat{a}^\dag_i B_{ij} \hat{a}^\dag_j + {\rm H.c.} \big)
\ee
with the definitions 
\begin{align} \label{KitaevAandB}
\begin{split}
& A_{ij} = -\mu\,\delta_{ij} - \frac{J}{2} \big(\delta_{i,\,j+1}+\delta_{i,\,j-1}-\delta_{|i-j|,\,\sites-1} \big) \\ 
& B_{ij} = {\rm sign}(i-j)\,\frac{\DD}{2} \Big[ |i-j|^{-\alpha}\,\big(1-\delta_{|i-j|,\,\sites-1}\big) + \delta_{|i-j|,\,\sites-1} \Big]
\end{split}
\end{align}
for the closed chain.
%$ \, A_{ij} = -\mu\,\delta_{ij} - \frac{J}{2} \big(\delta_{i,\,j+1}+\delta_{i,\,j-1}-\delta_{|i-j|,\,\sites-1} \big) \, $ 
%as well as 
%$ \, B_{ij} = {\rm sign}(i-j)\,\frac{\DD}{2} \big[ |i-j|^{-\alpha}\,(1-\delta_{|i-j|,\,\sites-1}) + \delta_{|i-j|,\,\sites-1} \big] $
%$A_{ij} = -\mu\,\delta_{ij} - \frac{J}{2} \big(\delta_{i,\,j+1}+\delta_{i,\,j-1}-\delta_{|i-j|,\,\sites-1} \big)$ 
%and \\ $B_{ij} = {\rm sign}(i-j)\,\frac{\DD}{2} \big[ |i-j|^{-\alpha}\,{\rm H}_{\rm n} + (1-{\rm H}_{\rm n}) \big]$
%for the closed chain, where ${\rm H}_{\rm n}$ is the Heaviside's step function of the discrete variable ${\rm n}=\sites-2-|i-j|$:
%${\rm H}_{\rm n}=1$ if $|i-j|\neq N-1$, ${\rm H}_{\rm n}=0$ else.
$\mathbb{A}$ and $\mathbb{B}$ are $\sites\times\sites$ real matrices: 
$\mathbb{A}$ is symmetric ($\mathbb{A}^T=\mathbb{A}$), $\mathbb{B}$ is antisymmetric ($\mathbb{B}^T=-\mathbb{B}$).
The singular value decomposition $\mathbb{A}+\mathbb{B}=\mathbf{\Phi}^T \mathbb{V} \mathbf{\Psi}$ gives
the diagonal matrix of eigenvalues $\mathbb{V}$ and the matrices of left and right normalized eigenstates $\mathbf{\Phi}$, $\mathbf{\Psi}$.
The QFI can be immediately calculated from the matrix $\mathbb{G}=-\mathbf{\Psi}^T\mathbf{\Phi}$, since
\be \label{KitaevCorrelators}
C_x(\ell) = 
\det \begin{pmatrix}
G_{1,\,2} & \dots & G_{1,\,\ell+1} \\
\vdots & \ddots & \\
G_{\ell,\,2} & & G_{\ell,\,\ell+1} \\
\end{pmatrix} \, ,
\quad C_y(\ell) = 
\det \begin{pmatrix}
G_{2,\,1} & \dots & G_{2,\,\ell} \\
\vdots & \ddots & \\
G_{\ell+1,\,1} & & G_{\ell+1,\,\ell} \\
\end{pmatrix} 
\ee
and $C_z(\ell) = G_{1,\,1}G_{\ell+1,\,\ell+1} - G_{1,\,\ell+1}G_{\ell+1,\,1}$.

\paragraph{Entanglement among sites}
The relation between the QFI in Eq.~(\ref{KitaevQFInonlocal}) and multipartite entanglement is obtained by exploiting the 
convexity properties of the QFI \cite{PezzePRL2009, HyllusPRA2012, TothPRA2012},
that holds even when the QFI is calculated with respect to a nonlocal operator~\cite{PezzePNAS2016}.
We can then use well known bounds of $\kappa$-particle entanglement 
derived for the QFI of $\sites$ spin-$\sfrac{1}{2}$ particles as bounds for $\kappa$-partite entanglement 
between the $\sites$ fermionic modes: see discussion in paragraph~\ref{QFIbounds}.
Specifically, the condition 
\be \label{KitaevQFIineq}
f_Q\big[\ket{\psi_{0}}, \hat{O}_{\varrho}\big] > \kappa, 
\ee 
signals $(\kappa\!+\!1)$-partite entanglement ($1\leq \kappa\leq \sites-1$) between sites of the fermionic chain.
In particular, separable pure states $\ket{\psi_{\rm sep}} = \bigotimes_{j=1}^\sites \ket{n_j}$, 
where $n_j$ is the occupation number of the $j$-th site 
(that can assume only values $0$ or $1$, due to the spinless nature of the fermions), 
satisfy $f_Q\big[\ket{\psi_{\rm sep}} , \hat{O}_{\varrho} \big] \leq 1$.
Moreover, states $\vert \psi \rangle$
with $\sites-1 < f_Q\big[\vert \psi \rangle, \hat{O}_{\varrho}\big] \leq \sites$ are genuinely $\sites$-partite entangled, 
$f_Q\big[\vert \psi \rangle, \hat{O}_{\varrho}\big] = \sites$ being the ultimate (Heisenberg) bound.

%%%%%%%%%%%%%%%%%%%%%%%%%%%%%%%%%%%%%%%%%%%%%%%%%%%%%%%%%
%%%%%%%%%%%%%%%%%%%%%%%%%%%%%%%%%%%%%%%%%%%%%%%%%%%%%%%%%
\begin{figure}[t!]
\centering
\includegraphics[width=1.02\textwidth]{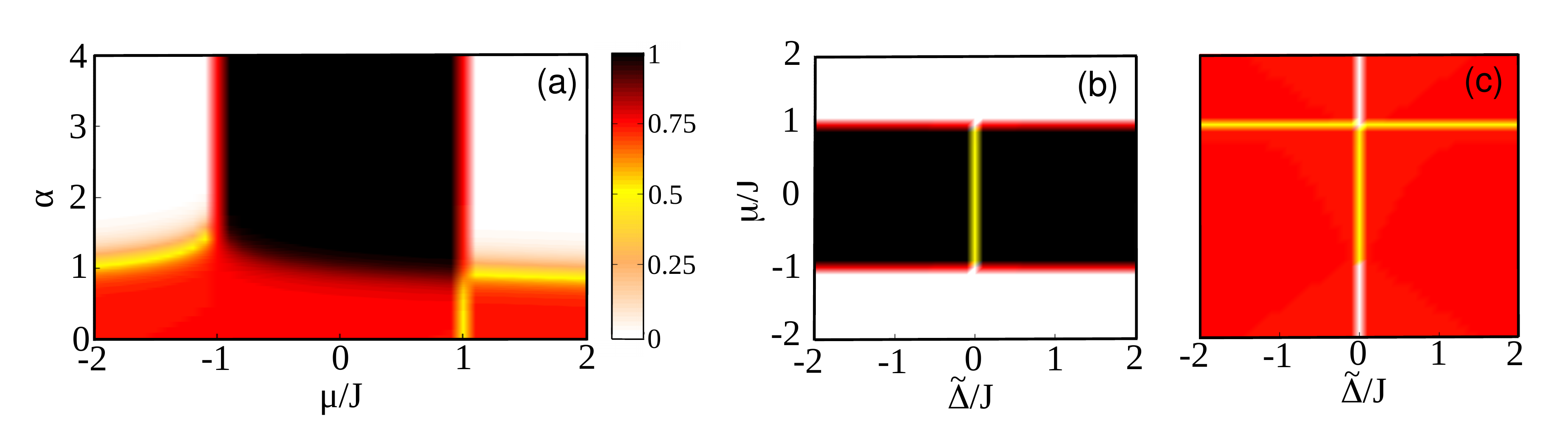}
\caption{Phase diagram of the Kitaev chain obtained numerically from the scaling of 
the Fisher density as a function of the system size $\sites$, $f_Q = 1 + c\sites^b$, Eq.~(\ref{KitaevQFIscaling}).
The color scale shows the scaling coefficient $b$ in the $\mu/J$--$\alpha$ plane for $\DD = J$ \textbf{(a)} and 
in the $\DD/J$--$\mu/J$ plane for short-range pairing $\alpha=100$ \textbf{(b)} and infinite-range $\alpha=0$ pairing \textbf{(c)}.
The polynomial fits are based on data up to $\sites\approx10^3$.}
\label{fig:PhaseDiagram}
\end{figure}
%%%%%%%%%%%%%%%%%%%%%%%%%%%%%%%%%%%%%%%%%%%%%%%%%%%%%%%%%
%%%%%%%%%%%%%%%%%%%%%%%%%%%%%%%%%%%%%%%%%%%%%%%%%%%%%%%%%

\paragraph{Finite-size scaling behaviour}
We find the QFI follows the asymptotic power-law scaling 
\be \label{KitaevQFIscaling} 
f_Q\big[\ket{\psi_{0}}, \hat{O}_{\varrho}\big] = 1 + c \, \sites^b \, ,
\ee
where the coefficients $b$ and $c$ depend on $\mu/J$, $\DD/J$, $\alpha$ and $\varrho$, but are independent of $\sites$. 
The scaling coefficient $b$ is directly related to the behavior of the correlation function.
For instance, an exponentially decaying correlation, $C_{\varrho}(\ell)=\neper^{-d_\ell/\xi}$ on a ring, 
with $\xi>0$ independent of $\sites$, 
gives $b=0$ and $c=2(\neper^{1/\xi}-1)^{-1}$ in the thermodynamic limit.
In this case, the QFI is \emph{extensive}: $\kappa$-partite entanglement is witnessed for 
$\xi^{-1} <\log[(\kappa+1)/(\kappa-1)]$ and the entanglement depth remains constant when increasing the system size.  
On the contrary, when $b>0$, the QFI is \emph{superextensive}: 
the larger $\sites$, the larger the witnessed $\kappa$-partite entanglement.

In particular, a power-law decay $C(\ell) \sim 1/\ell^{1-b}$ ($0<b<1$) can be related to a rescaling of the correlation functions with the system size, $C_{\varrho}(\ell)=\sites^{b-1} c_{\varrho}(\ell/\sites)$. 
This rescaling gives rise to a data collapse of curves 
$\sites^{1-b} C_{\varrho}(\ell)$ as a function of $\ell/\sites$ and for different values of $\sites$, 
the collapse curve being $c_{\varrho}(\ell/\sites)$. 
We checked the data collapse for some values of the physical parameters. 
% see Supplementary Materials of Ref.~\cite{Pezze2017} for two examples.
Computing Eq.~(\ref{QFIcorr}) gives
\be
f_Q\big[\vert \psi_{0}\rangle, \hat{O}_{\varrho}\big] = 1 + \sites^{b-1} \sum_{l=1}^{\sites-1} c_{\varrho}(\ell/\sites) \, 
\xrightarrow{\sites\to\infty} \, 1 + \sites^b \int_{0}^{1} dx \, c_{\varrho}(x) \, .
\ee
and we can identify $c = \int_{0}^1 d x \, c_{\varrho}(x) = {\rm const}$.

\begin{figure}[t!]
\centering
\includegraphics[width=1.02\textwidth]{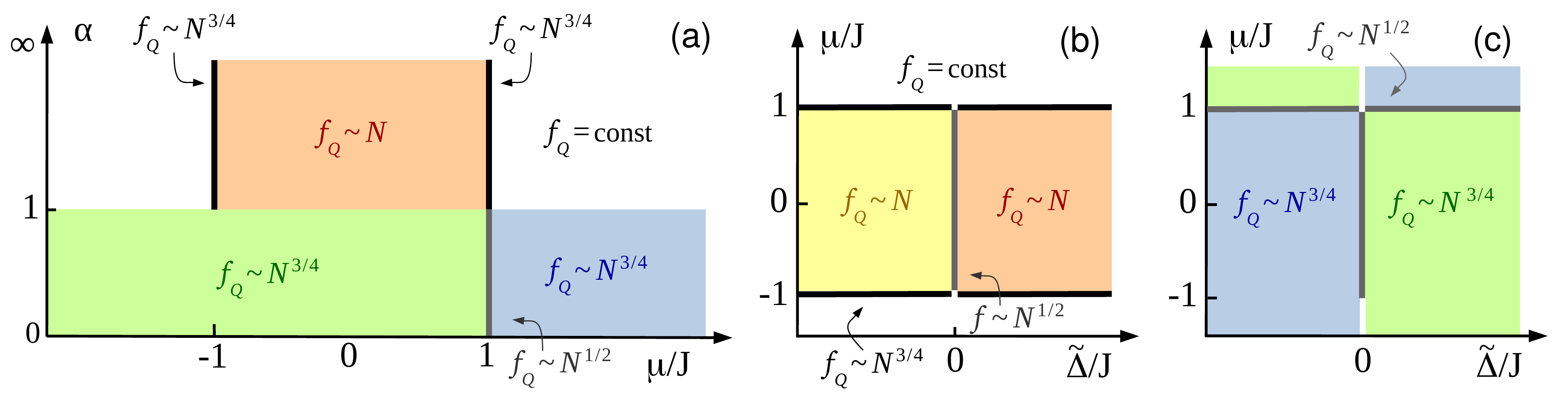}
\caption{Sketch of the phase diagrams for the Kitaev chain in the thermodynamic limit, showing the superposition of the
topological properties (as in Fig.~\ref{fig:KitaevTopology}) 
and the scaling of the Fisher density with system length $\sites$ (as in Fig.~\ref{fig:PhaseDiagram}).}
\label{fig:EntanglementPhaseDiagram}
\end{figure}

\begin{figure}[p!]
\centering
\includegraphics[height=0.34\textwidth]{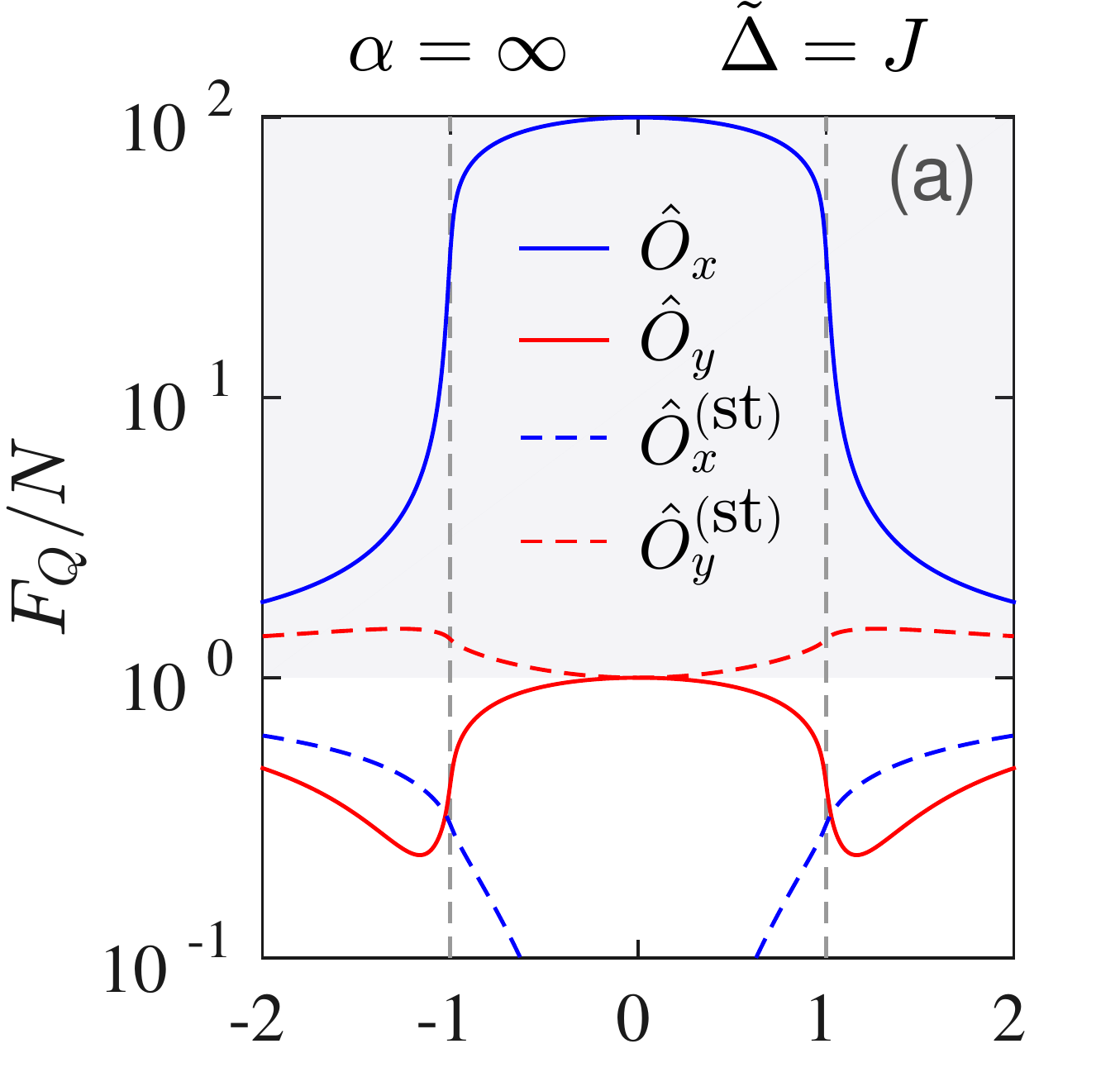} \hspace{-12pt}
\includegraphics[height=0.34\textwidth]{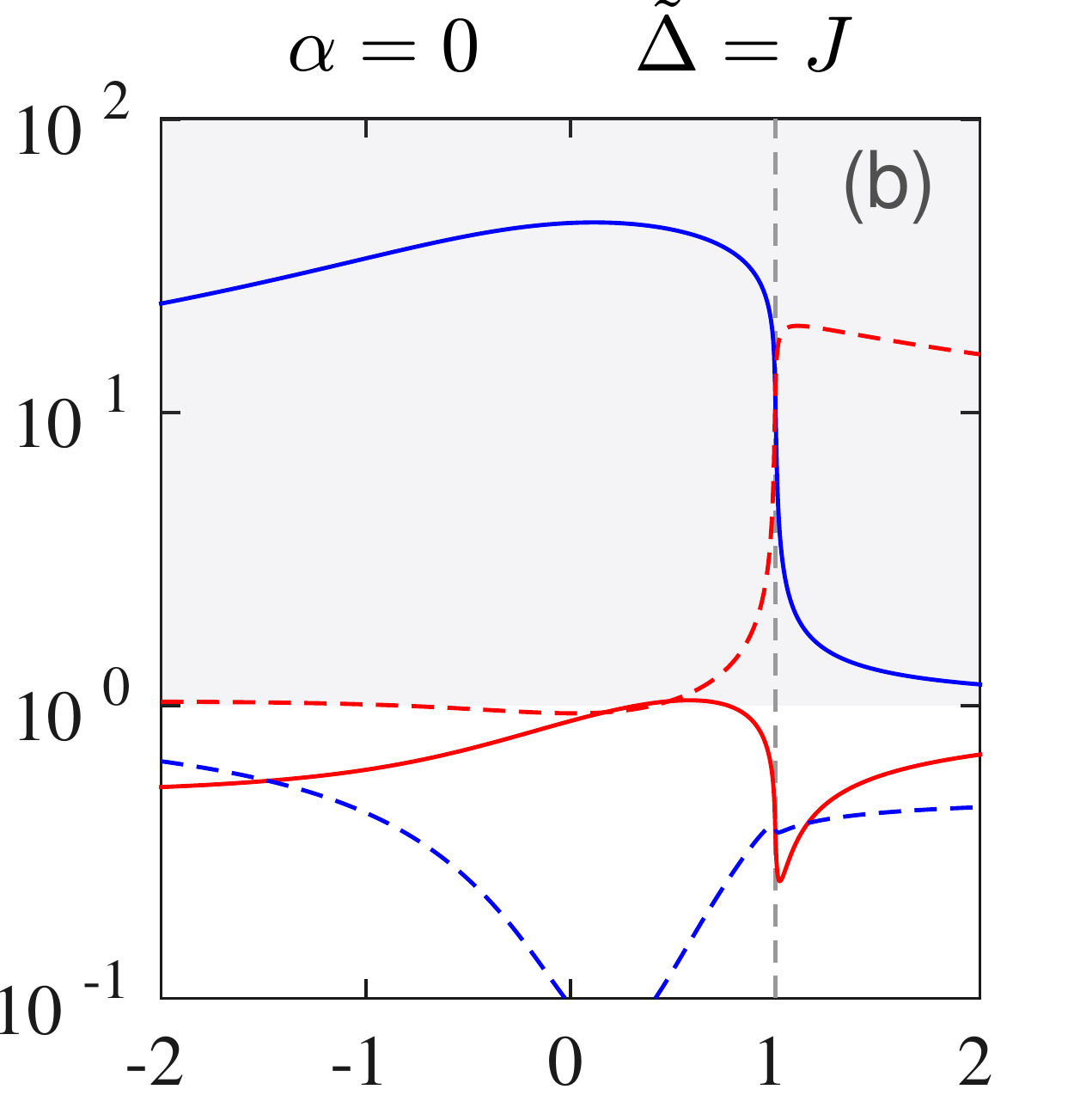} \hspace{-12pt} 
\includegraphics[height=0.34\textwidth]{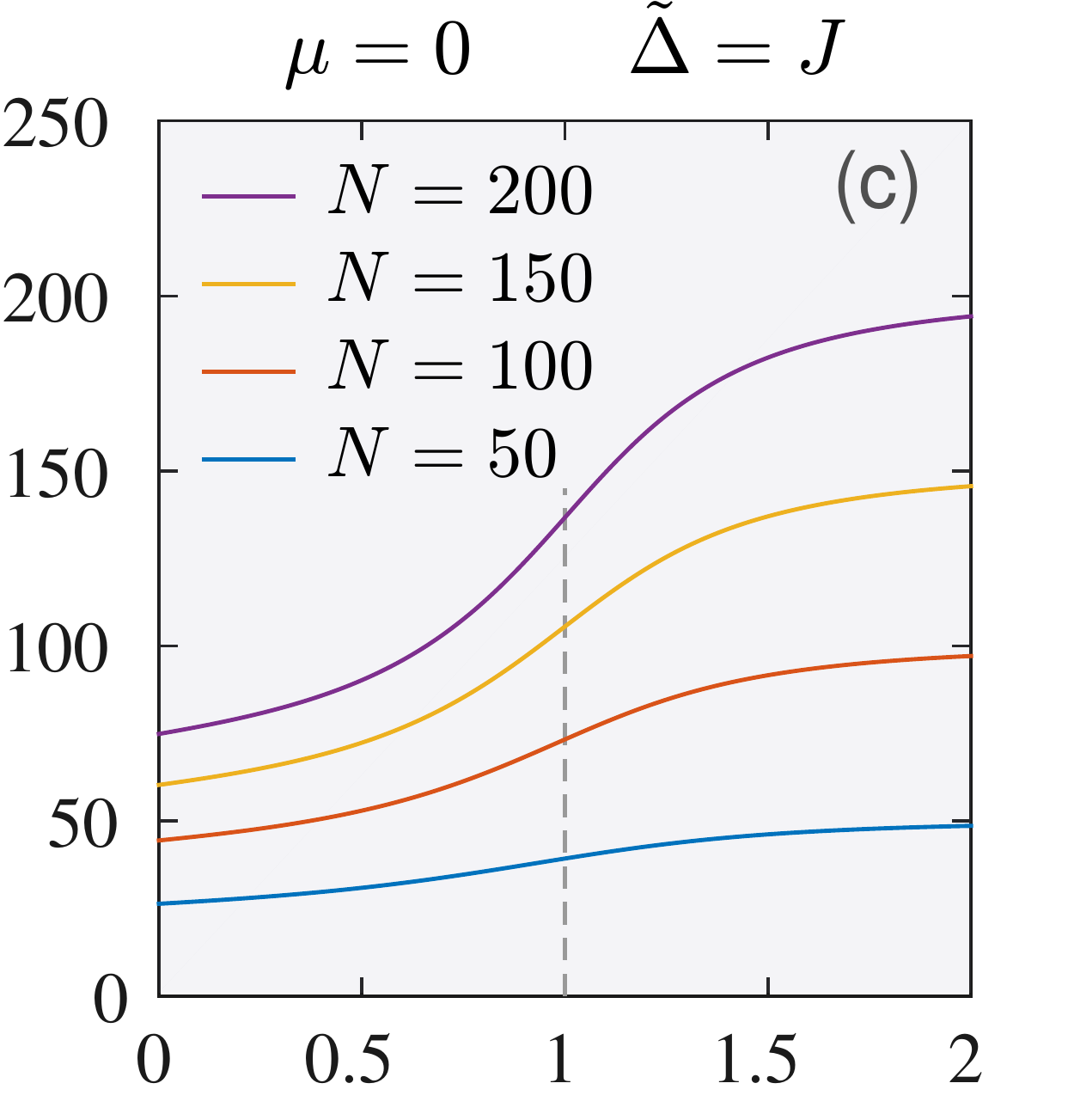} \\
\includegraphics[height=0.34\textwidth]{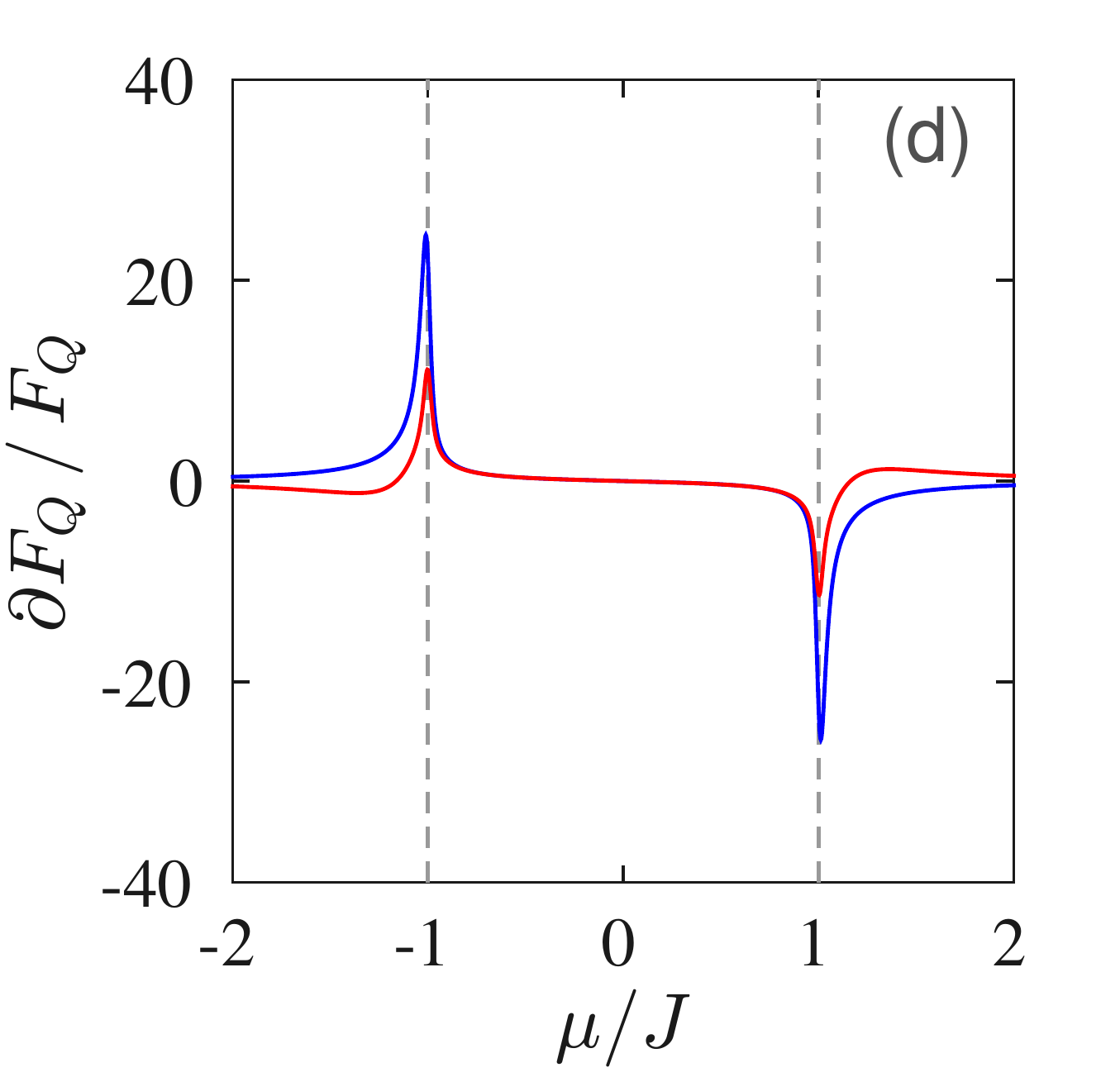} \hspace{-10pt}
\includegraphics[height=0.34\textwidth]{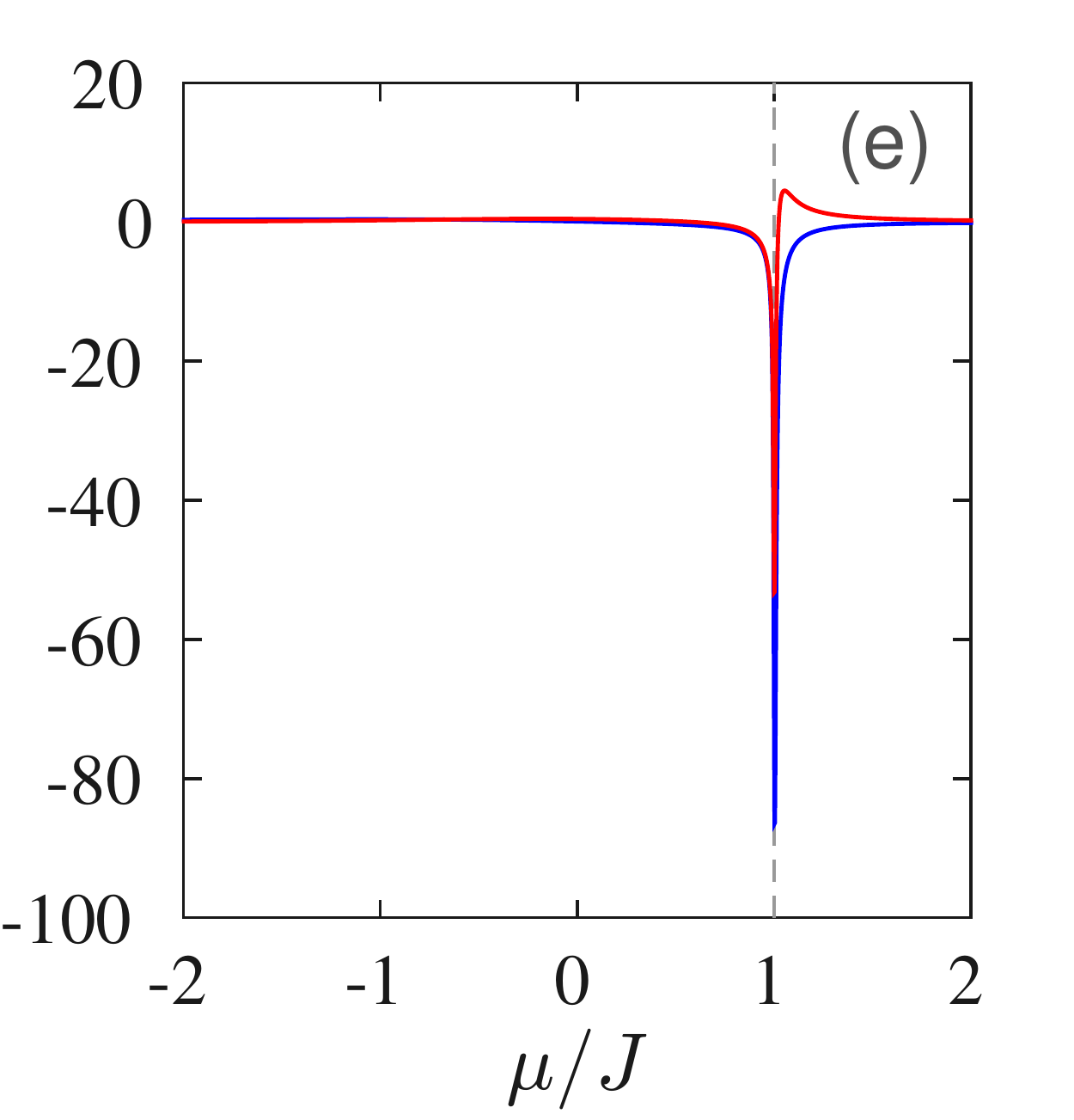} \hspace{-10pt}
\includegraphics[height=0.34\textwidth]{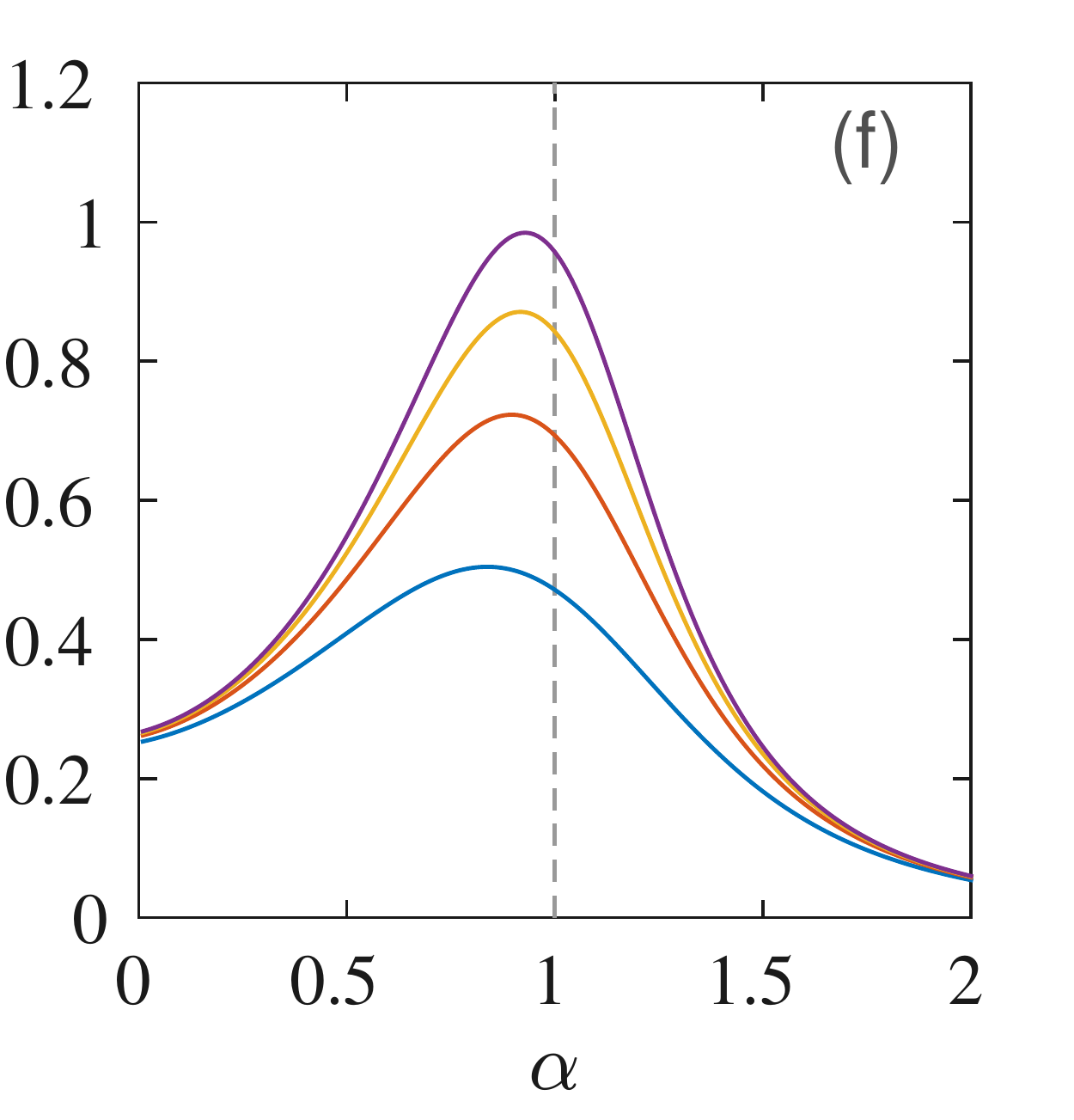} \\
\caption{Two-dimensional plots of the QFI and its first derivative as a function of $\mu$, 
calculated via the four nonlocal operators introduced in the main text for $\sites=100$ \textbf{(left and center)};
and as a function of $\alpha$, calculated via the optimal operator $\hat{O}_x$ for several values of $\sites$ \textbf{(right)}. 
Shaded area indicates multipartite entanglement.}
\label{fig:KitaevNonlocalFisher}
\end{figure}

\paragraph{Entanglement phase diagram}
We have fitted $f_Q[\vert \psi_{0} \rangle]$ as a function of the size $\sites$ according to $f_Q[\vert \psi_{0} \rangle] = 1 + c\sites^b$, 
for the operator $\hat{O}_\varrho$ maximizing the QFI.\footnote{\ We point out once again that the optimal operator for the QFI 
is, in general, different in the different regions of the phase diagram in Fig.~\ref{fig:KitaevTopology}.
See for instance panels~\ref{fig:KitaevNonlocalFisher}\Panel{a,\,b}.\\[-9pt]}
Here we use the scaling coefficient $b$ to characterize the different phases of the Kitaev model. 
Figure~\ref{fig:PhaseDiagram} shows numerical results obtained up to $\sites\approx10^3$, 
while the scaling behaviours are summarized schematically in Fig.~\ref{fig:EntanglementPhaseDiagram}.
These plots should be compared to the phase diagram of Fig.~\ref{fig:KitaevTopology}.
 
We see an extraordinary \emph{correspondence}\footnote{\ The mild discrepancy between the theoretic picture in Fig.~\ref{fig:KitaevTopology} and the numerics in Fig.~\ref{fig:PhaseDiagram} is due to the finite range of exploration $\sites \lesssim 10^3$.\\[-9pt]} 
between the phases of the Kitaev chain (determined uniquely by the values of the invariant $W$, which is a topological property) 
and the finite-size scaling of multipartite entanglement witnessed in the chain itself 
(given by the exponent $b$ of the power law obeyed by the Fisher information). 
It seems that the QFI manages to discriminate topological phases from trivial phases in terms of superextensivity;
furthermore different scaling implies different topology. 
This is a remarkable and intriguing result at the verge of quantum information and topological many-body systems.

%%%%%%%%%%%%%%%%%%%%%%%%%%%%%%%%%%%%%%%%%%%%%%%%%%%%%%%%%
%%%%%%%%%%%%%%%%%%%%%%%%%%%%%%%%%%%%%%%%%%%%%%%%%%%%%%%%% 
%%%%%%%%%%%%%%%%%%%%%%%%%%%%%%%%%%%%%%%%%%%%%%%%%%%%%%%%%
%%%%%%%%%%%%%%%%%%%%%%%%%%%%%%%%%%%%%%%%%%%%%%%%%%%%%%%%%

In particular, in panel~\ref{fig:PhaseDiagram}\Panel{a} we plot the phase diagram 
of $b$ in the $\mu/J$--$\alpha$ plane. 
In panels~\ref{fig:PhaseDiagram}\Panel{b,\,c} we plot $b$ in the $\DD/J$--$\mu/J$ plane for nearest-neighbour pairing ($\alpha=\infty$) 
and infinite-range pairing ($\alpha=0$) respectively.
Notice that the regime $\DD<0$ is mapped to the one with $\DD>0$ by the phase redefinition $\hat{a}_j \to \pm i \, \hat{a}_j$. 
This operation also interchanges the optimal operators
$ \, f_Q\big[\vert \psi_{0} \rangle, \hat{O}_x\big] \leftrightarrow f_Q\big[\vert \psi_{0} \rangle, \hat{O}_y\big] \, $ and 
$ \, f_Q\big[\vert \psi_{0} \rangle, \hat{O}_x^{(\rm st)}\big] \leftrightarrow f_Q\big[\vert \psi_{0} \rangle, \hat{O}_y^{(\rm st)}\big]$.
A detailed discussion of Figs.~\ref{fig:PhaseDiagram} and \ref{fig:EntanglementPhaseDiagram} follows.
%%%%%%%%
\\[-6pt]

\noindent \tripieno{MySky} \ For $\alpha=\infty$ and $\DD\neq 0$, we have $b=1$ for $|\mu|/J <1$ and $b=0$ elsewhere.
Therefore, a superextensive QFI is observed only in the topologically nontrivial phase. 
The optimal choice of operators is $\hat{O}_x$ for $\DD\geq0$ and $\hat{O}_y$ for $\DD\leq0$.
The correlation function of the $\hat{O}_{x,\,y}$ operator is exponentially decaying for $|\mu|/J >1$ and 
constant for $|\mu|/J <1$: genuine $\sites$-partite entanglement is witnessed at $\mu=0$ where $C_x(\ell)=1$ $\forall\,\ell$.
We can justify these findings having a glance at the ground state. 
For $\mu\to+\infty$, it is a separable state of all occupied sites $\ket{\psi_0}=\ket{1}^{\otimes\sites}$, 
while for $\mu\to-\infty$ it is a separable state of all empty sites $\ket{\psi_0}=\ket{0}^{\otimes\sites}$,
where $\{\ket{n}_i\}$ is the occupation basis and $n\in\{0,1\}$ is the occupation number at the $i$th site.
In contrast, at $\mu=0$ the ground state is the coherent superposition of a fully-occupied chain and a empty chain
$\ket{\psi_0}=\big(\ket{1}^{\otimes\sites}+\ket{0}^{\otimes\sites}\big)/\sqrt{2}$, saturating $F_Q[\ket{\psi_0}]=\sites^2$.
See also Fig.~\ref{fig:KitaevNonlocalFisher}\Panel{a}.
For $\DD=0$, instead, the Kitaev chain maps to the XX model and it is known~\cite{BarouchPRA1971} 
that the ground state is a product state for $|\mu|/J>1$: accordingly, we find $c=0$, whereas $b=1/2$ for $|\mu|/J <1$.
%%%%%%%%
\\[-6pt]

\noindent \tripieno{MySky} \ For short-range pairing ($\alpha > 1$) and $\DD\neq 0$, we find 
again a superextensive scaling of the QFI in correspondence of phases with nonzero winding numbers. 
On the critical lines $|\mu| / J = 1$, we have $b=3/4$, that is 
associated to the algebraic asymptotic decay of the correlation functions 
$C_x(\ell \to \sites) \sim \sites^{1/4}$, implied at $\alpha>2$ by conformal invariance~\cite{Mussardo,LeporiANN2016}.
%%%%%%%%
\\[-6pt]

\noindent \tripieno{MySky} \ For $\alpha \approx 1$ the numerical calculations are affected by finite-size effects. 
Yet, when $\alpha=1$, we can argue that $b=1/2$ for $|\mu|/J>1$ and $b=3/4$ for $|\mu|/J<1$.
%%%%%%%%
\\[-6pt]

\noindent \tripieno{MySky} \ For long-range pairing ($\alpha<1$), $b=3/4$ for $\mu/J\neq 1$ and $b=1/2$ for $\mu/J= 1$.
These scalings refer for every $\alpha$ to the QFI relative to an optimal choice of operators, 
that is $\hat{O}_x$ for $\mu/J\leq 1$ and $\hat{O}_y^{(\rm st)}$ for $\mu/J\geq1$.
%%%%%%%%
\\[-6pt]

\noindent \tripieno{MySky} \ For infinite-range pairing ($\alpha=0$) we find $b=3/4$ everywhere, 
except at the phase boundary: $b=1/2$ for $\mu/J=1$ and $\DD\neq0$, as well as for $\DD=0$ and $|\mu|/J <1$, 
while $b=0$ for $\DD=0$ and $|\mu|/J\geq1$.
We can distinguish four regions in the phase diagram of Fig.~\ref{fig:PhaseDiagram}\Panel{c}, 
singled out by the operators optimizing the QFI.
For $\mu/J < 1$, the optimal operators are $\hat{O}_x$ for $\DD>0$ and $\hat{O}_y$ for $\DD<0$; 
for $\mu/J > 1$, the optimal operators are $\hat{O}_y^{(\rm st)}$ for $\DD>0$ and $\hat{O}_x^{(\rm st)}$ for $\DD<0$.

\paragraph{Signatures of the QPTs}
It is clear from Fig.~\ref{fig:PhaseDiagram} and Fig.~\ref{fig:EntanglementPhaseDiagram} that 
the multipartite entanglement witnessed by the QFI varies suddenly at QPTs. 
In particular, we find an abrupt change of the scaling exponent $b$ whenever we cross the boundaries between different topological phases,
even when the energy gap in the quasiparticle spectrum does not close.
Next, we show that the QPT between phases characterized by different winding numbers 
are signaled by a divergence of the derivative of the QFI
with respect to the control parameter.   
Taking the derivative of Eq.~(\ref{KitaevQFIscaling}) 
with respect to a parameter $\eta$ of the model ($\eta = \mu$, $\DD$, $\alpha$) gives
\be \label{dFQ}
\frac{1}{f_Q}\frac{\partial f_Q}{\partial \eta} = \frac{c \, \sites^b}{1+c \, \sites^b} \, \cdot \, \bigg( \frac{1}{c} \frac{\partial c}{\partial \eta} + \frac{\partial b}{\partial \eta} \, \log \sites \bigg).
\ee
In the interesting case $c\neq 0$ and $b\geq 0$, Eq.~(\ref{dFQ}) diverges in the thermodynamic limit $\sites\to\infty$
either because of a divergence of $\partial_\eta c / c$ or, even when $\partial_\eta c$ is smooth, 
because of $\partial_\eta b \neq 0$.

\begin{figure}[p!]
\centering
\includegraphics[width=0.75\textwidth]{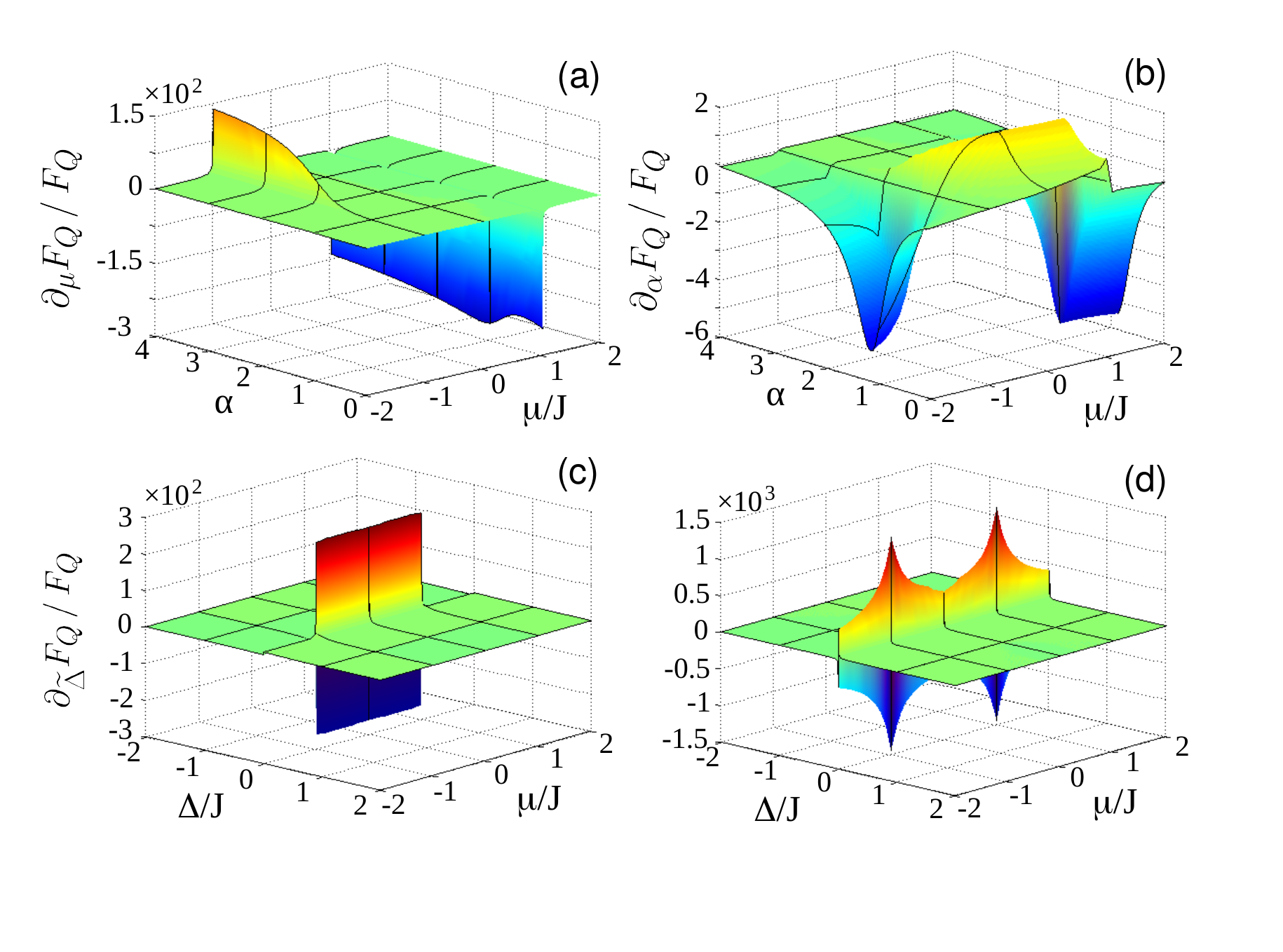}
\vspace{-30pt}
\caption{Weighted derivative of the Fisher density, $f_Q^{-1} \, \partial f_Q/\partial\eta$, 
with respect to $\eta=\mu/J$ for ${\DD=J}$ \textbf{(a)}, with respect to $\eta=\alpha$ for $\DD=J$ \textbf{(b)},
with respect to $\eta=\DD/J$ for $\alpha=100$ \textbf{(c)} 
and with respect to $\eta=\DD/J$ for $\alpha=0$ \textbf{(d)}.
All plots have been obtained for $\sites=1000$ sites.}
\label{fig:KitaevQFI3D}
\end{figure}

This is illustrated in Fig.~\ref{fig:KitaevQFI3D}, where we plot the weighted derivative of the QFI
with respect to $\mu$, $\alpha$ and $\DD$.
The plots, obtained with $\sites = 1000$, shows that $\partial_\mu f_Q/f_Q$ and $\partial_{\DD} f_Q/f_Q$ 
vary sharply at the critical points.
From the analysis of the scaling coefficients $b$ and the prefactor $c$, it is evident that the 
$\partial_\mu F_Q$ diverges at $\mu/J=1$ when $\alpha\geq 0$ and at $\mu/J=-1$ when $\alpha>1$, in correspondence of the QPTs.
It is less evident, from the numerical data, that $\partial_\alpha F_Q$  diverges at $\alpha=1$ and any value of $\mu$: 
$\partial_\alpha f_Q/f_Q$ varies smoothly as a function of $\alpha$.
Yet, a finite-size scaling analysis shows that, in the limit $\sites\to\infty$ (up to $\sites\approx1000$ in our numerics), 
$\partial_\alpha f_Q/f_Q$ tends to peak around $\alpha=1$, even if very slowly with increasing $\sites$: 
panels~\ref{fig:KitaevNonlocalFisher}\Panel{c,\,f} account for this.
Therefore, a fast change of the QFI is able to detect the QPT at $\alpha=1$
-- associated to a change of the winding number -- even if it occurs without closing of the gap in the quasiparticle spectrum.

%%%%%%%%%%%%%%%%%%%%%%%%%%%%%%%%%%%%%%%%%%%%%%%%%%%%%%%%%
%%%%%%%%%%%%%%%%%%%%%%%%%%%%%%%%%%%%%%%%%%%%%%%%%%%%%%%%%
%%%%%%%%%%%%%%%%%%%%%%%%%%%%%%%%%%%%%%%%%%%%%%%%%%%%%%%%%
%%%%%%%%%%%%%%%%%%%%%%%%%%%%%%%%%%%%%%%%%%%%%%%%%%%%%%%%%

%%%%%%%%%%%%%%%%%%%%%%%%%%%%%%%%%%%%%%%%%%%%%%%%%%%%%%%%%
%%%%%%%%%%%%%%%%%%%%%%%%%%%%%%%%%%%%%%%%%%%%%%%%%%%%%%%%%
%%%%%%%%%%%%%%%%%%%%%%%%%%%%%%%%%%%%%%%%%%%%%%%%%%%%%%%%%
%%%%%%%%%%%%%%%%%%%%%%%%%%%%%%%%%%%%%%%%%%%%%%%%%%%%%%%%%

\section{Multipartite entanglement at finite temperature} 
A thermodynamic description of topological insulators and superconductors have recently been provided 
for simple models in various spatial dimensions~\cite{QuellePRB2016,KempkesSR2016}.
For nontrivial electronic matter, thermodynamic signatures of both the topological edge structure 
and phase transitions of the edge not manifested in the bulk have been found. 
Furthermore, it seems possible to categorize these signatures according to universality classes: 
all models exhibit the same universal behaviour in the order of the topological phase transition, depending on the spatial dimension. Moreover, a topological phase diagram at finite temperature have been obtained. 
Other works have focused on the stability of topological properties against thermal noise~\cite{Viyuela2DM2014,Viyuela2DM2015}. 
They studied the resistance of the topological invariant\footnote{\ At finite temperature the topological invariant cannot be the winding number any more, even for one-dimensional systems, because the state is described by a density matrix: the Uhlmann phase is usually chosen for characterizing topological insulators and superconductors at both zero and finite temperatures. Notably, for one-dimensional two-band models such as the Kitaev chain, at the limit of zero temperature the Uhlmann phase recovers the usual notion of topological order as given by the winding number.\\[-9pt]} at finite temperature 
and revealed that a critical temperature exists above which the topological behaviour abruptly ceases to exist.

Much less attention has been devoted to the effect of temperature on the entanglement in topological phases and near to topological QPTs. 
We examine here the finite-temperature entanglement phase diagram of the Kitaev model 
and find interesting similarities with the ones discussed for ordinary QPTs in chapters~\ref{ch:Ising} and \ref{ch:LMG}.

\paragraph{Method} 
Hamiltonian~(\ref{HamKitaev}) introduced at the beginning of the present chapter represents a very general family 
of Kitaev chains and its parameters have a well-defined physical meaning. 
Its only drawback is that a description of the limit case of dominant chemical potential $|\mu| \gg J$ and $|\mu|\gg|\DD|$ 
cannot be mathematically reached if not in the limit $\mu\to\pm\infty$. 
Exactly as we have done for the LMG model in chapter~\ref{ch:LMG}, in treating the effect of temperature 
we renounce the general notation, specialize $\DD=J$ and compress all the unbound range of the chemical potential $\mu\in(-\infty,+\infty)$ into a compact interval $\theta\in\big[-\frac{\pi}{2},+\frac{\pi}{2}\big]$:
\be \label{HamKitaevThermic}
\frac{\hat{H}_\alpha}{J} = - \displaystyle \cos\theta \sum_{j=1}^\sites \big( \hat{a}^\dagger_j \hat{a}_{j+1} + {\rm H.c.} \big) 
- 2\sin\theta \sum_{j=1}^\sites \bigg(\hat{a}^\dagger_i\hat{a}_j - \frac{1}{2}\bigg) + \displaystyle \cos\theta \sum_{j=1}^\sites \,\sum_{\ell=1}^{\sites-j} d_\ell^{-\alpha} \big( \hat{a}_j \hat{a}_{j+\ell} + {\rm H.c.} \big) 
\ee
where $2J\cos\theta$ is the hopping amplitude equal to the strength of the p-wave superconducting pairing
and $2J\sin\theta$ is the chemical potential (we set $J>0$).
The critical points of this model are located at $\thetac=\pm\frac{\pi}{2}$ for short-range pairing ($\alpha>1$)
or $\thetac=+\frac{\pi}{2}$ for long-range pairing ($\alpha\leq1$): 
in other words, when the weights of the chemical potential and the hopping amplitude are equal.\footnote{\ We incidentally note that the energy spectrum for the elementary excitations of Eq.~(\ref{HamKitaevThermic}) is given by $\epsilon_{k} = J \sqrt{(2\cos\theta\cos k + 2\sin\theta)^2 + (f_\alpha(k)\cos\theta)^2}$.\\[-9pt]}

We assume that the chain is in contact with a thermal bath at temperature $T$. 
% and that the thermalization process preserves the particle number. 
In this case, the canonical equilibrium thermal state is given by
$\hat{\rho}_T = \pazocal{Z}^{-1}\,\neper^{-\hat{H}_\alpha/T}$, where $\pazocal{Z}$ is the partition function ($k_{\rm B}=1$).
The technique used here for evaluating the QFI is the same as for the transverse Ising chain discussed in chapter~\ref{ch:Ising}
and exploits the time-dependent correlation functions as given in Ref.~\cite{DerzhkoPRB1997}.
Contrary to the optimization procedure performed for the above case of zero temperature, 
we calculate the QFI by fixing\footnote{\ Changing the operator according to the parameter $\theta$ and temperature $T$ would have appeared costly from the numerical point of view as well as cumbersome from an empirical point of view.\\[-9pt]}
the nonlocal observable $\hat{O}_x$ as defined in Eq.~(\ref{KitaevNonlocalProbes}) 
for every $\theta$ and $T$: we are looking at $F_Q\big[\hat{\rho}_T,\hat{O}_x\big]$ and not the maximal $F_Q[\hat{\rho}_T]$.
The operator $\hat{O}_x$ is not always optimal in the explored regimes, 
but it is still the optimal one in the topological phase containing the null chemical potential $\theta=0$ at $T=0$.

%%%%%%%%%%%%%%%%%%%%%%%%%%%%%%%%%%%%%%%%%%%%%%%%%%%%%%%%%%%%%%%%
%%%%%%%%%%%%%%%%%%%%%%%%%%%%%%%%%%%%%%%%%%%%%%%%%%%%%%%%%%%%%%%
\begin{figure}[p!]
\begin{center}
\hfill \quad \framebox{$\alpha=\infty$} \hfill \textcolor{white}{.} \hfill \framebox{$\alpha=1$} \hfill \textcolor{white}{.} \hfill \framebox{$\alpha=0$} \hfill \textcolor{white}{.}
\end{center}
\centering
~\\[-23pt]
\includegraphics[width=0.32\textwidth]{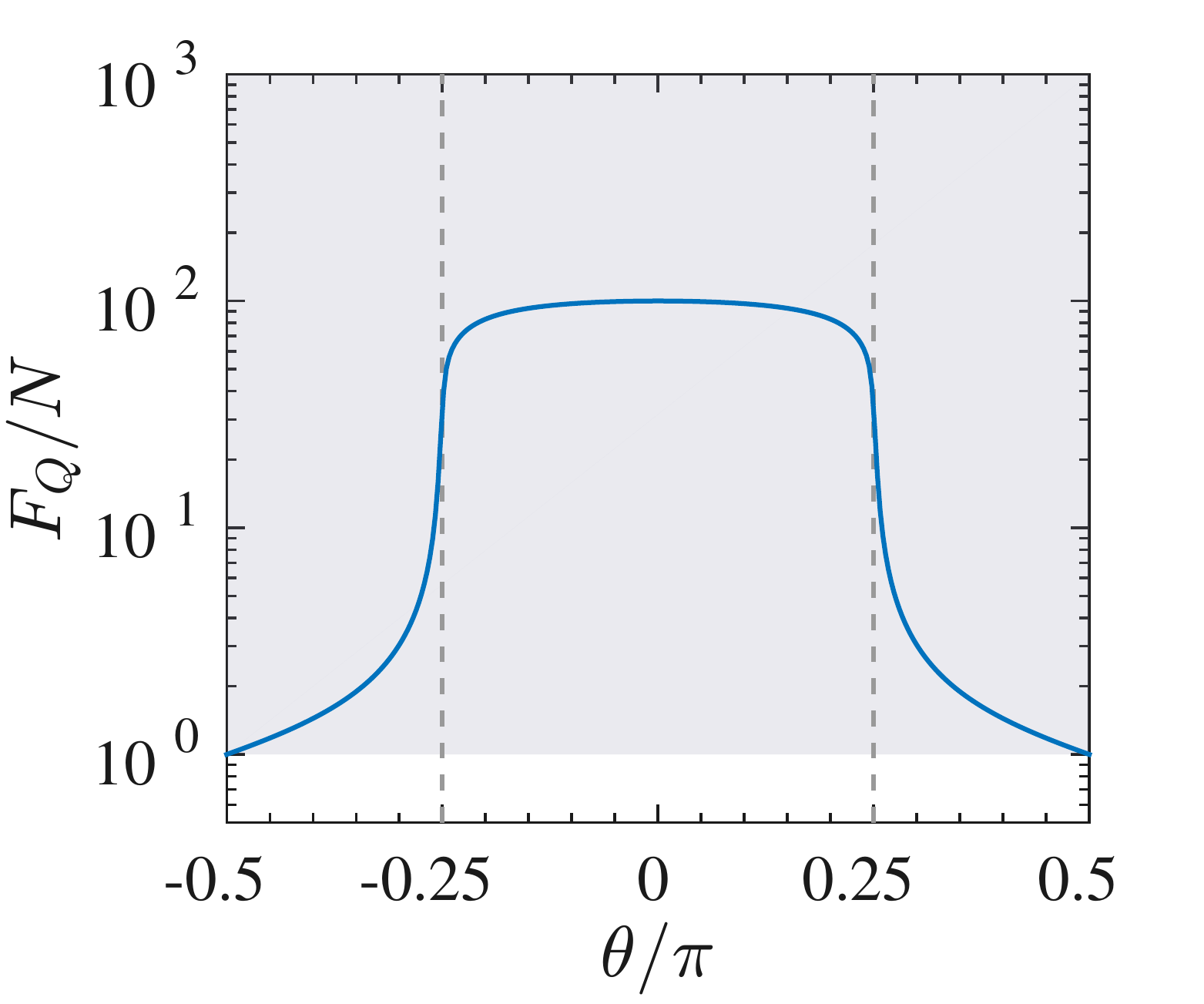} 
\includegraphics[width=0.32\textwidth]{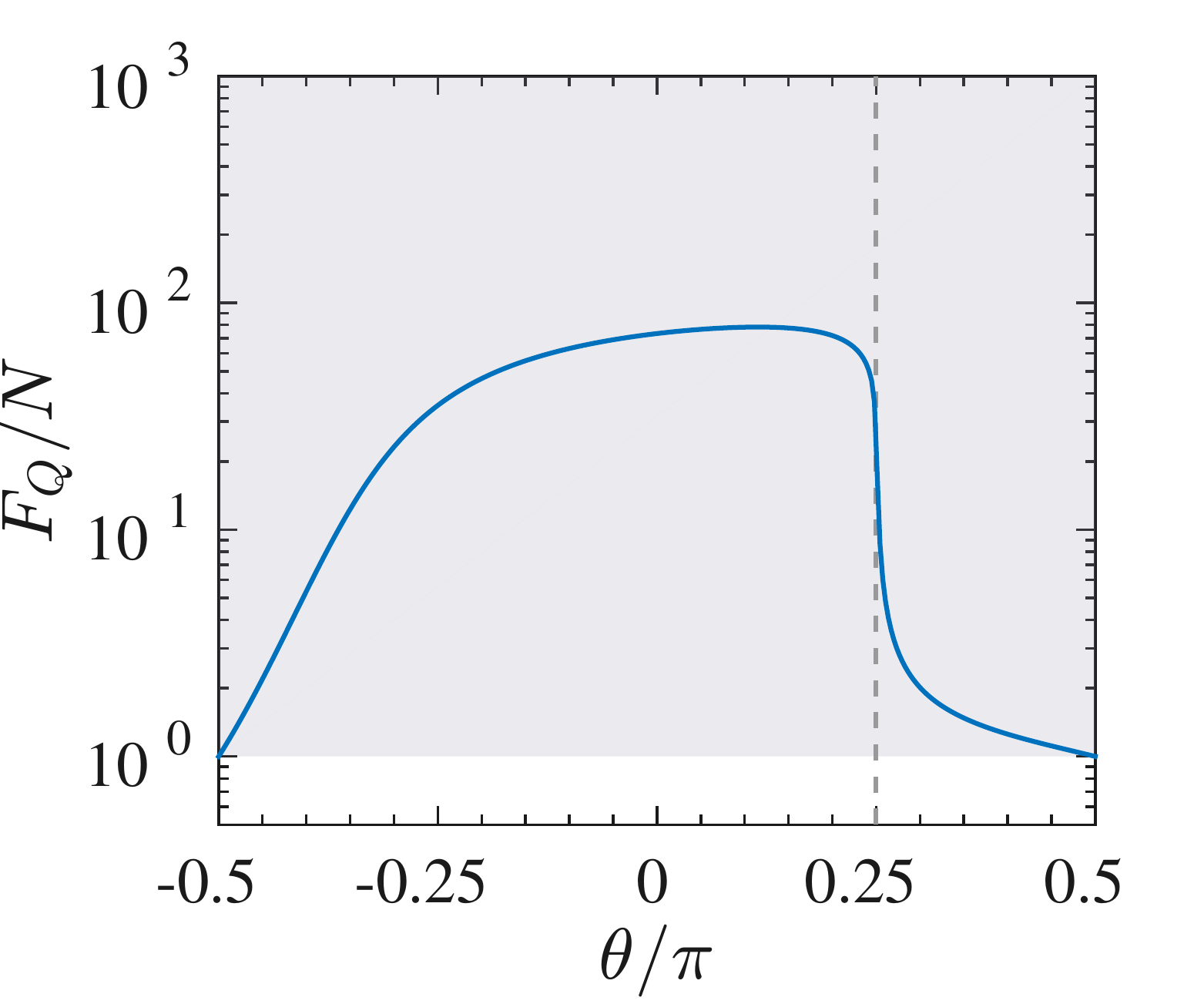} 
\includegraphics[width=0.32\textwidth]{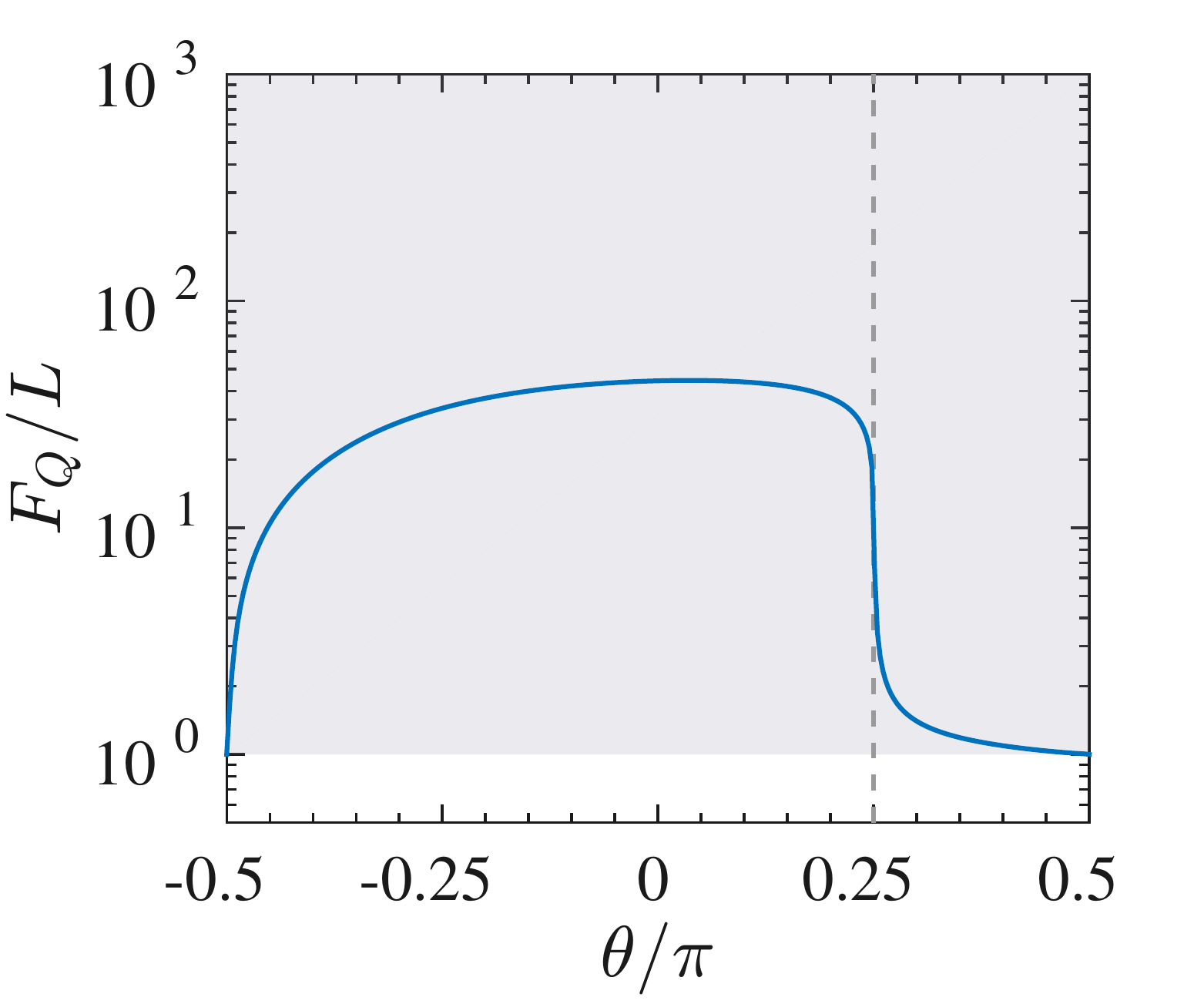} \\[-3pt]
\includegraphics[width=0.33\textwidth]{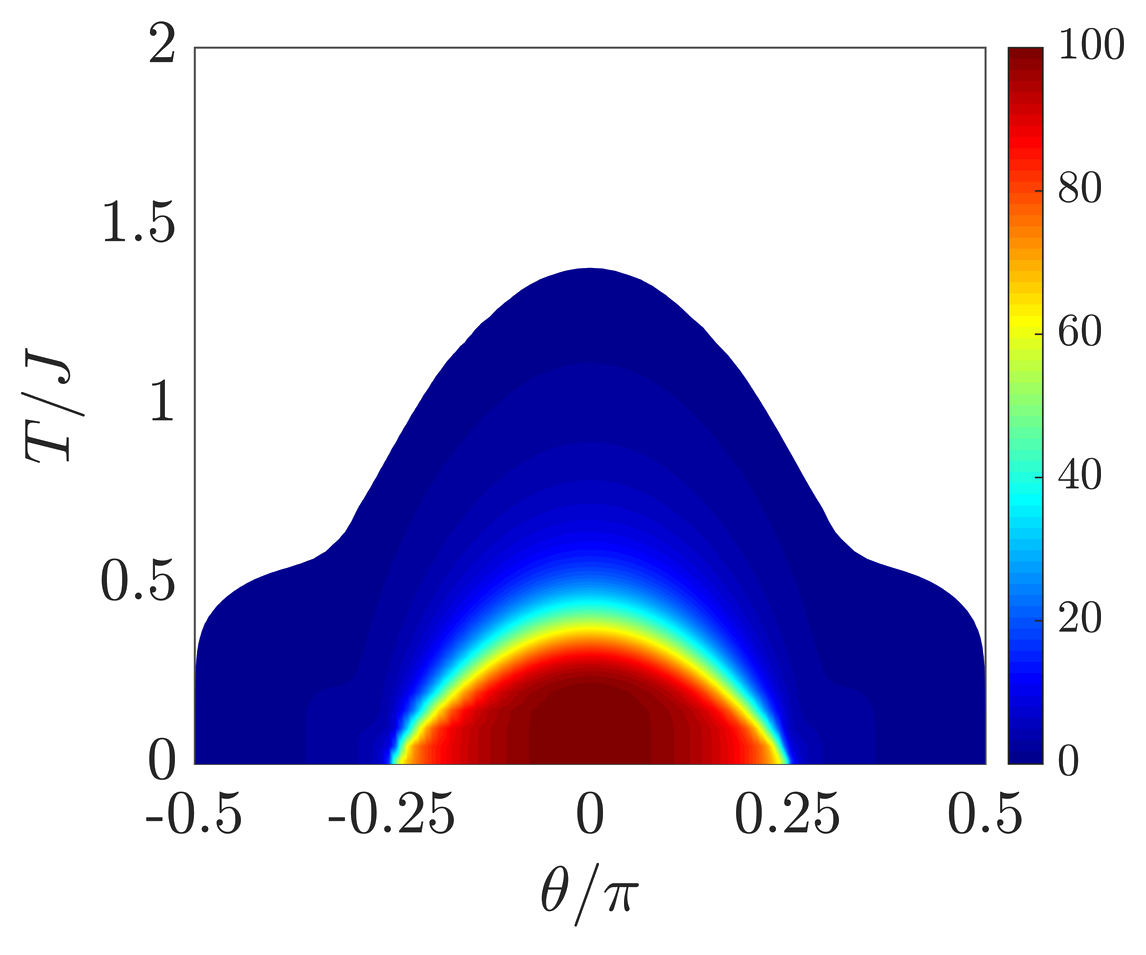} \hspace{-6pt}
\includegraphics[width=0.33\textwidth]{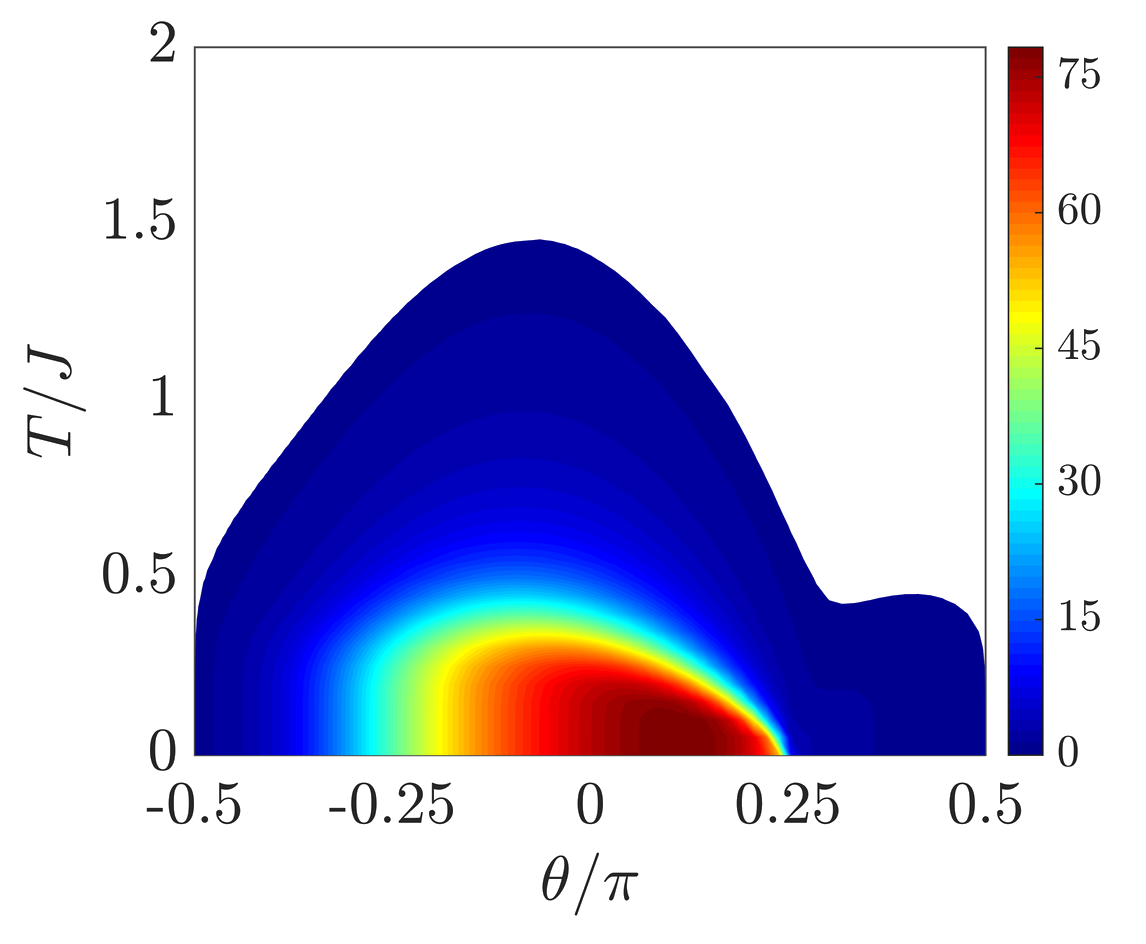} \hspace{-6pt}
\includegraphics[width=0.33\textwidth]{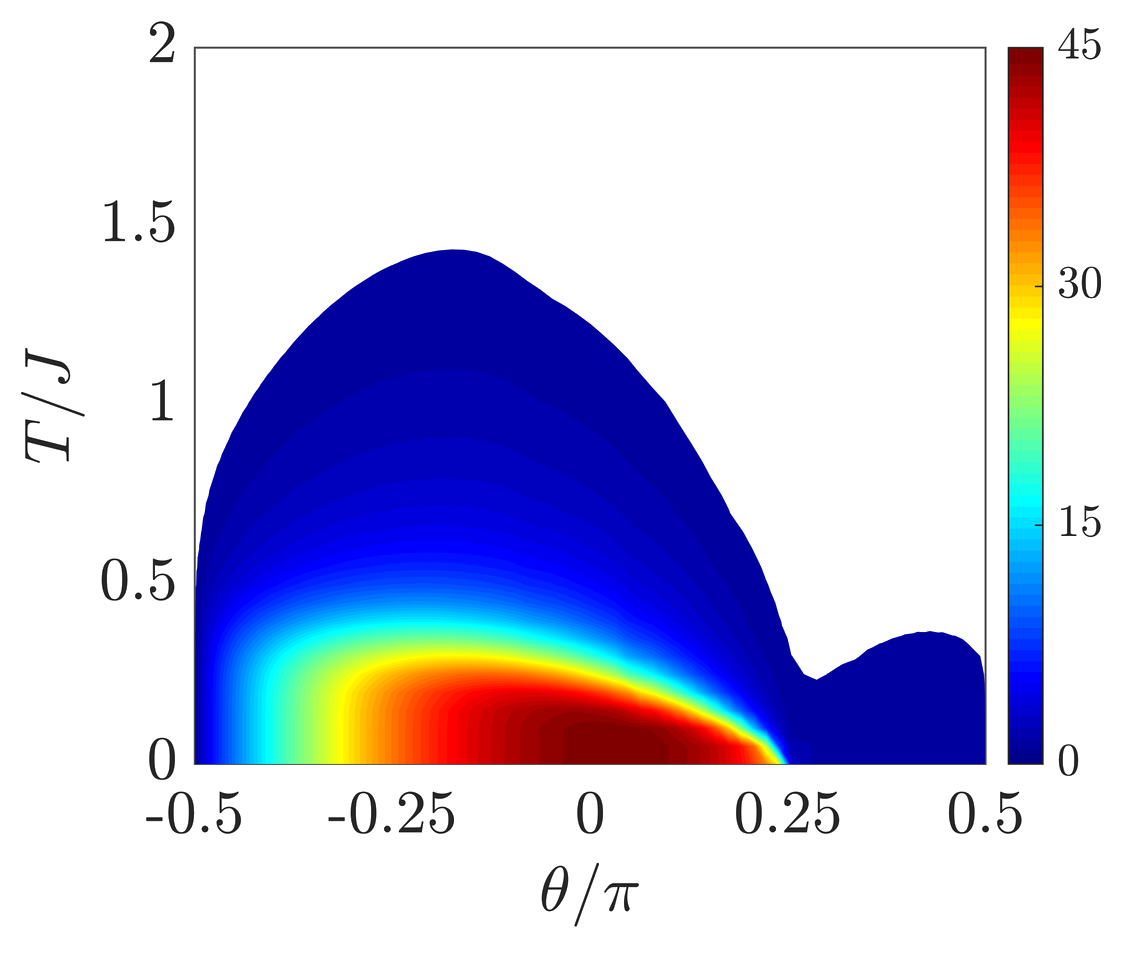} \\[-6pt]
\includegraphics[width=0.33\textwidth]{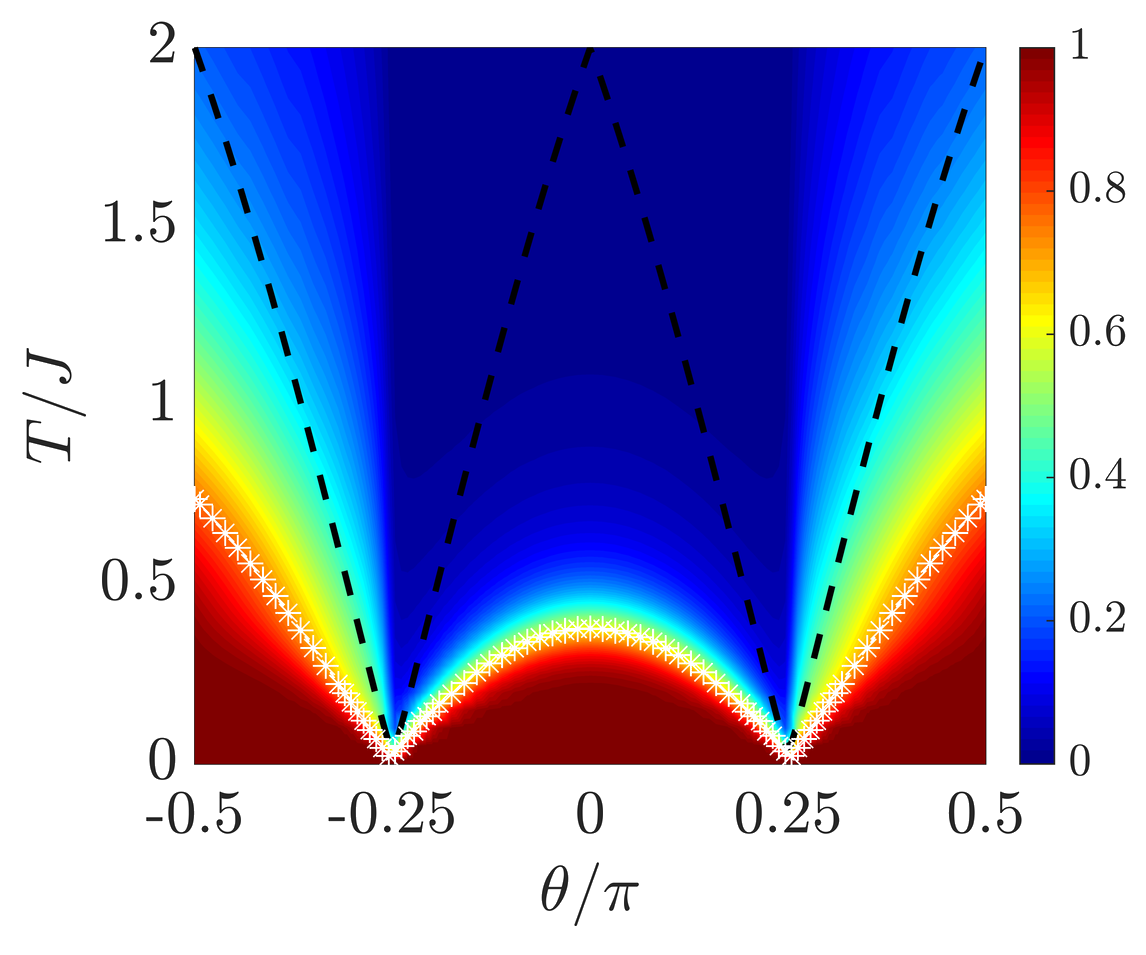} \hspace{-6pt}
\includegraphics[width=0.33\textwidth]{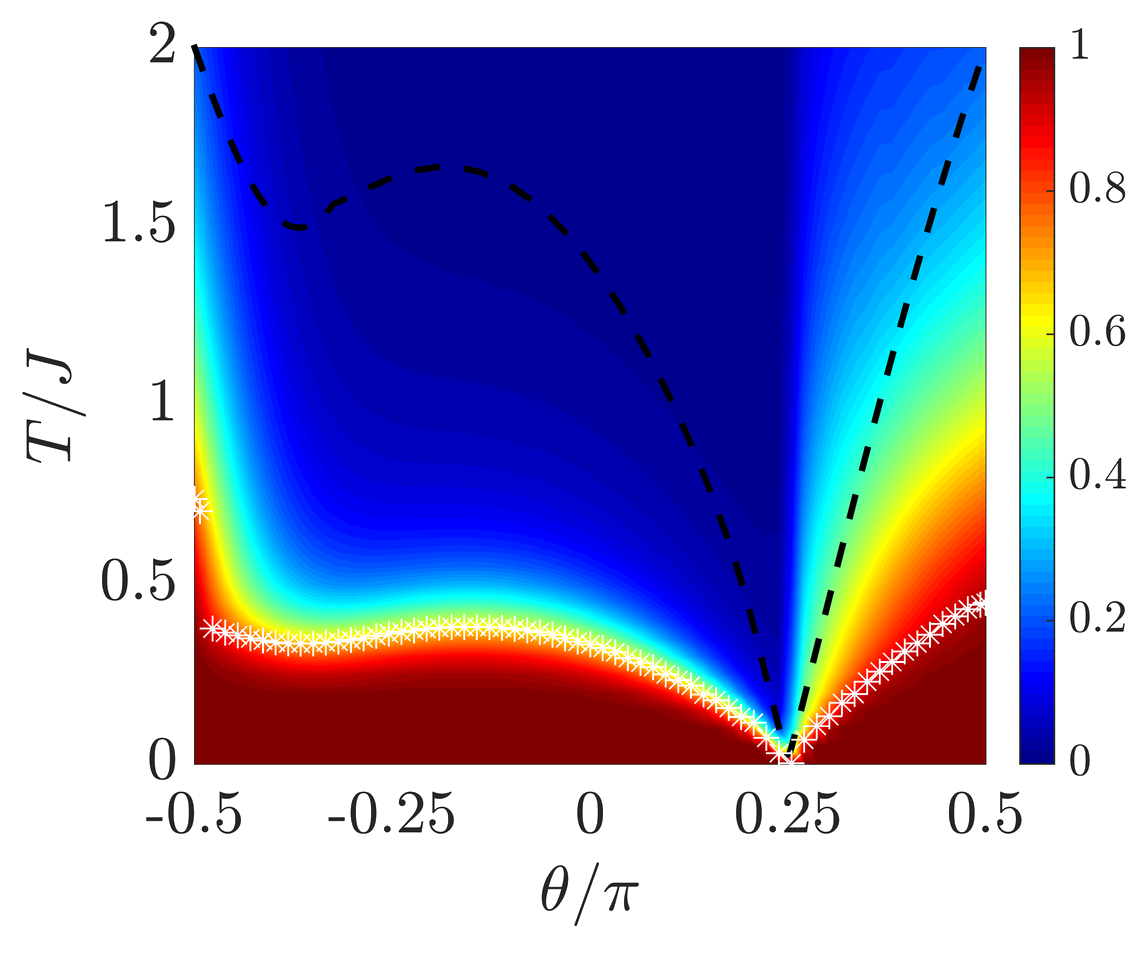} \hspace{-6pt}
\includegraphics[width=0.33\textwidth]{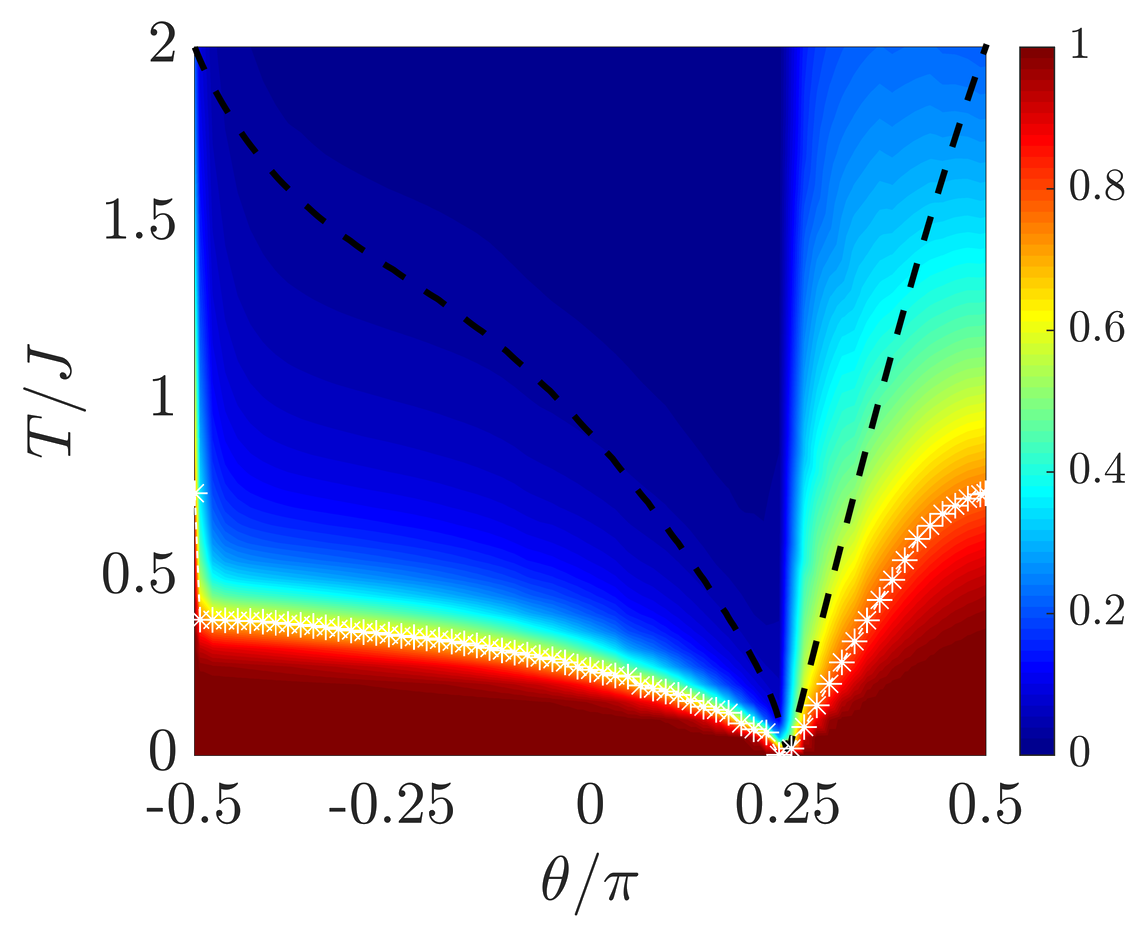} % \\[-15pt]
\caption{Phase diagrams of the Kitaev chain at finite temperature as a function of $\theta$ and $T$, 
for $\alpha=100$ \textbf{(left)}, $\alpha=1$ \textbf{(center)} 
and $\alpha=0$ \textbf{(right)}.
The upper row contains $F_Q[\ket{\psi_0},\hat{O}_x]/\sites$; the shaded area signals multipartite entanglement.
The central row shows density plots of $F_Q[\hat{\rho}_T,\hat{O}_x]/\sites$; the coloured area indicates multipartite entanglement. 
The lower row presents the normalized function $F_Q[\hat{\rho}_T,\hat{O}_x]/F_Q[\ket{\psi_0},\hat{O}_x]$; 
the black lines represent the first energy separation $T=\DeltaE$ in the many-body spectrum, 
the white dots flag the crossover temperature $\Tcross$.} 
\label{fig:Thermic1}
\end{figure}
%%%%%%%%%%%%%%%%%%%%%%%%%%%%%%%%%%%%%%%%%%%%%%%%%%%%%%%%%%%%%%%
%%%%%%%%%%%%%%%%%%%%%%%%%%%%%%%%%%%%%%%%%%%%%%%%%%%%%%%%%%%%%%%
%~\\[-10pt] % spacing
%%%%%%%%%%%%%%%%%%%%%%%%%%%%%%%%%%%%%%%%%%%%%%%%%%%%%%%%%%%%%%%
%%%%%%%%%%%%%%%%%%%%%%%%%%%%%%%%%%%%%%%%%%%%%%%%%%%%%%%%%%%%%%%
\begin{figure}[p!]
\centering
\textcolor{white}{.} \hfill
\includegraphics[height=0.3\textwidth]{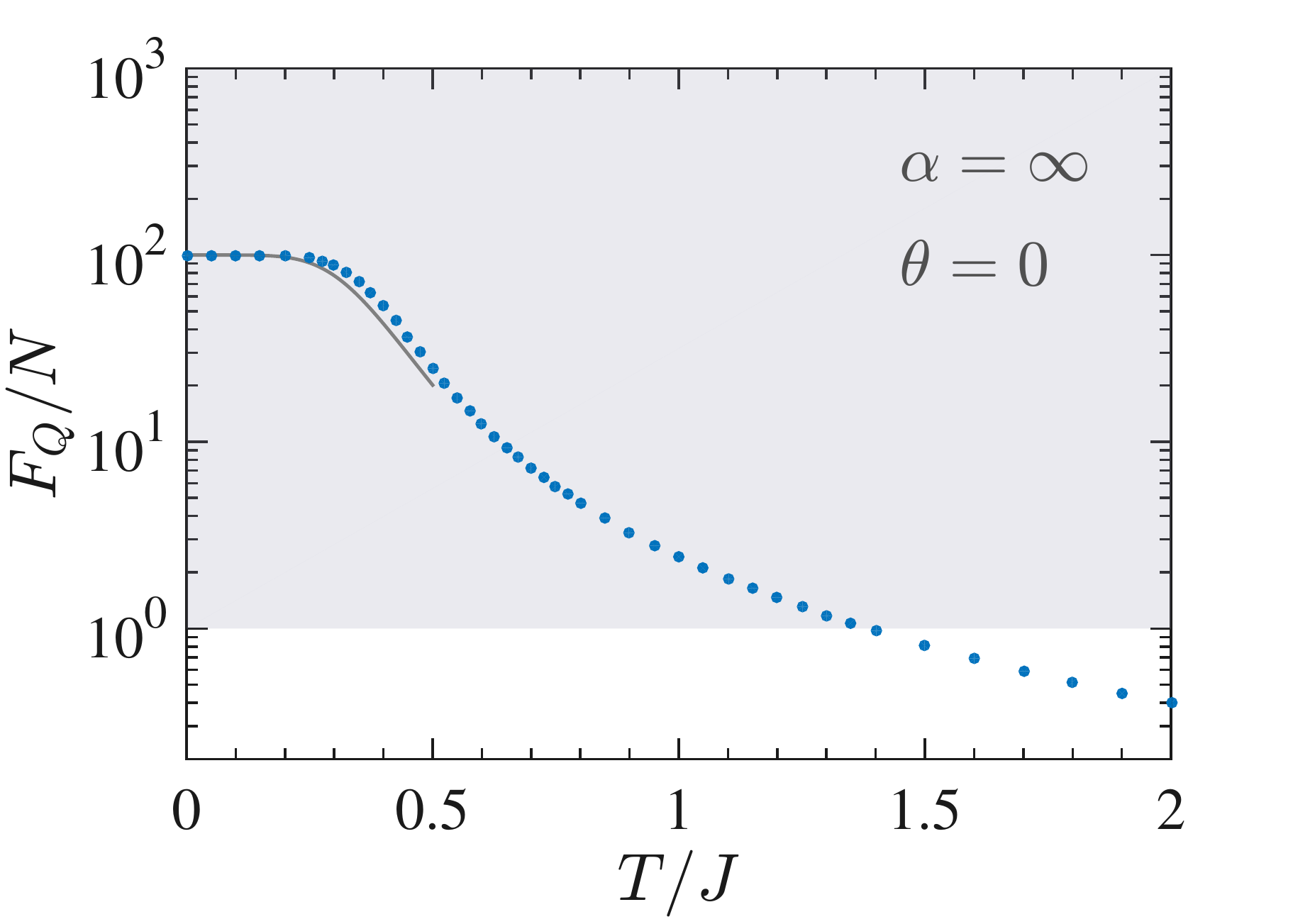} \hfill
\includegraphics[height=0.3\textwidth]{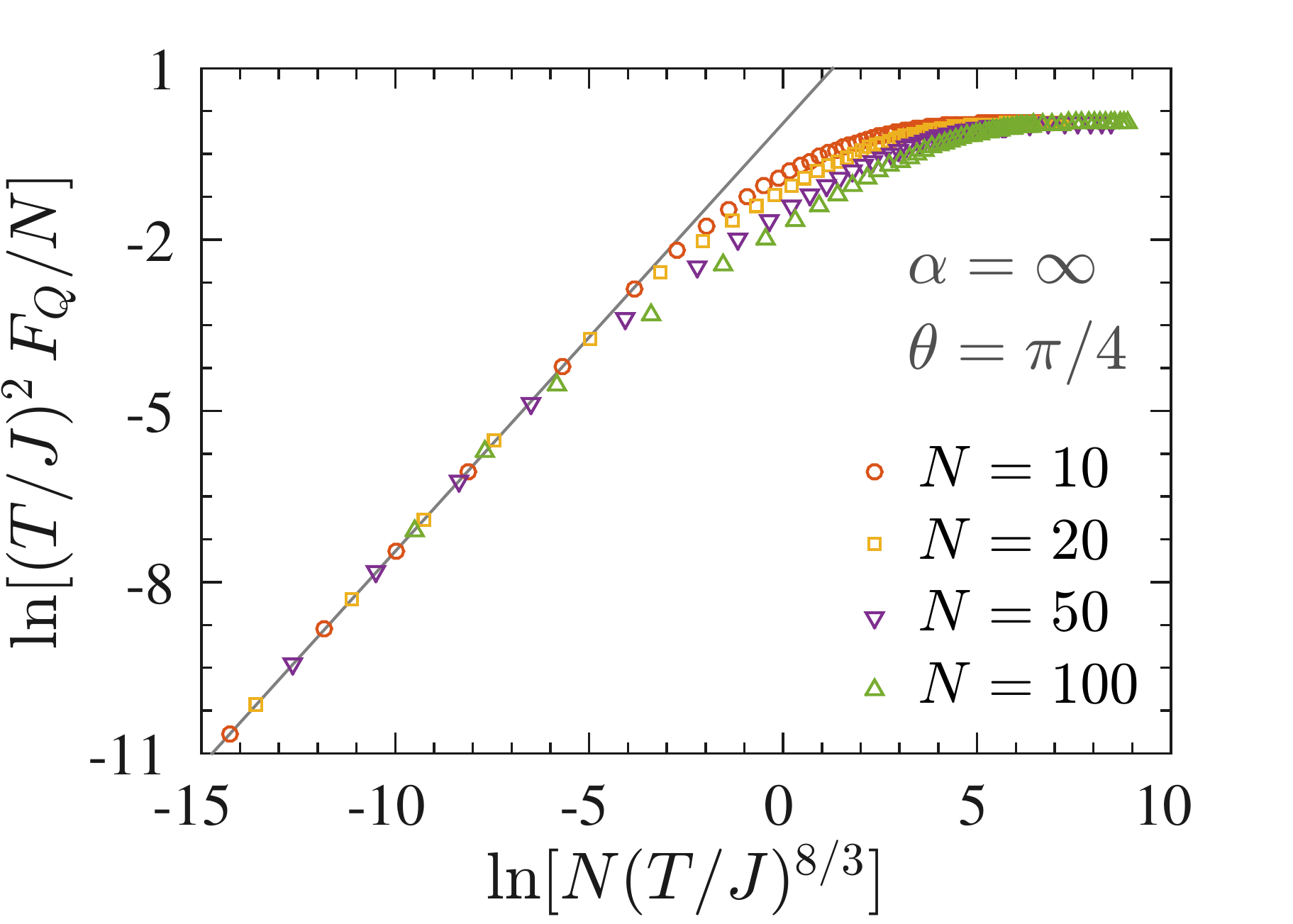} \hfill \textcolor{white}{.}
%~\\[-12pt] % spacing
\caption{\textbf{(Left)} Multipartiteness $F_Q[\hat{\rho}_T]/\sites$ as a function of $T$ for $\alpha=100$ at $\theta=0$; 
dots are numerical data and the solid line is $\tanh^2\frac{J}{T} \, (1+\neper^{-2J/T})/(1+100\,\neper^{-2J/T})$, 
namely Eq.~(\ref{QFIfactorization}) with $\DeltaE=2J$, $\mu=1$ and $\nu=\sites$.
\textbf{(Right)} Finite-temperature and finite-size scaling at criticality $\theta=\pi/4$ for $\alpha=100$ and different values of $\sites$; 
the gray line is the power law $\sites^{3/4}$.}
\label{fig:Thermic2}
\end{figure}
%%%%%%%%%%%%%%%%%%%%%%%%%%%%%%%%%%%%%%%%%%%%%%%%%%%%%%%%%%%%%%%
%%%%%%%%%%%%%%%%%%%%%%%%%%%%%%%%%%%%%%%%%%%%%%%%%%%%%%%%%%%%%%%

\paragraph{Entanglement phase diagram} 
The main results for thermal entanglement in the Kitaev chain Eq.~(\ref{HamKitaevThermic}) 
are illustrated in Fig.~\ref{fig:Thermic1} for three specific ranges $\alpha=\infty$, $\alpha=1$ and $\alpha=0$.
The upper panels report, for completeness, 
the large amount of multipartiteness $F_Q\big[\ket{\psi_0},\hat{O}_x\big]/\sites$
witnessed by the QFI at $T=0$ in the nontrivial phases $|\theta|<\frac{\pi}{4}$ 
(where $F_Q\big[\hat{\rho}_T,\hat{O}_x\big]/\sites\sim\sites$) for $\alpha=\infty$,
and $0<\theta<\frac{\pi}{4}$ (where $F_Q\big[\hat{\rho}_T,\hat{O}_x\big]/\sites\sim\sites^{3/4}$) for $\alpha=1$ and $\alpha=0$.
These plots should be compared with the graphs for the ground state already presented in Fig.~\ref{fig:KitaevNonlocalFisher}.
In the trivial phase $\frac{\pi}{4}<|\theta|<\frac{\pi}{2}$ for $\alpha=\infty$, instead, 
we have $F_Q\big[\hat{\rho}_T,\hat{O}_x\big]/\sites\sim\pazocal{O}(1)$, 
as well as in the nontrivial phase $\frac{\pi}{4}<\theta<\frac{\pi}{2}$ for $\alpha=1$ and $\alpha=0$ 
because no optimization on the operator has been accomplished.
\\[-6pt]
%%%%%%%%

\noindent \tripieno{MySky} \ Multipartiteness $F_Q\big[\hat{\rho}_T,\hat{O}_x\big]/\sites$ at finite temperature 
is shown in the central panels of Fig.~\ref{fig:Thermic1}:
the continuity $F_Q\big[\ket{\psi_0},\hat{O}_x\big] \equiv \lim_{\,T\to0} F_Q\big[\hat{\rho}_{T},\hat{O}_x\big]$ 
prove the entanglement to be preserved at low temperature. At most, it is witnessed up to high temperature $\Tmax\approx1.5\,J$. 
Moreover, the superextensivity inherited from $T=0$ survives up to $\Tcross$, 
as suggested by the lower panels of Fig.~\ref{fig:Thermic1}, 
where we show density plots of $F_Q\big[\hat{\rho}_T,\hat{O}_x\big]/F_Q\big[\ket{\psi_0},\hat{O}_x\big]$.
This robustness against thermal noise is due to the nondegenerate nature of the ground state for any $\theta\neq\thetac$
and represents a remarkable difference with respect to the symmetry-breaking QPTs presented in chapters~\ref{ch:Ising} and \ref{ch:LMG}, 
where the superextensive entanglement at $T=0$ is extremely fragile against thermal noise,
due to the closure of the first energy separation $\DeltaE$.
The present result opens the possibility to exploit the great amount of zero-temperature entanglement also at finite $T$ for metrological tasks.
\\[-6pt]
%%%%%%%%

\noindent Remarkably, the current diagrams in Fig.~\ref{fig:Thermic1} have extraordinary similarities with the ones presented in the chapter~\ref{ch:Ising} and \ref{ch:LMG}, even if the nature of the zero-temperature QPT is different here. 
\emph{The general features of quantum criticality discussed in Sec.~\ref{sec:MEinQPT} seems to stand also for topological QPTs.}
The variable ratio $2.3\lesssim\DeltaE(\theta)/\Tcross(\theta)\lesssim5.3$ is due to the complex structure 
of the low-lying many-body energy spectrum, as suggested by the fundamental Eq.~(\ref{QFIfactorization}).
For $\alpha=\infty$ at $\theta=0$, the $\sites$-fold degeneration of the first excited state results in 
the peculiar thermal decay shown in Fig.~\ref{fig:Thermic2} and a maximum ratio $\DeltaE/\Tcross\approx5.3$.
The two-level prediction Eq.~(\ref{QFIfactorization}), with $\DeltaE=2J$ and $\mu=1,\nu=\sites$, agrees with data for $T \ll J$.
\\[-6pt]
%%%%%%%%

\noindent \tripieno{MySky} \  An insight on multipartite entanglement at criticality for $\alpha=\infty$ 
is given in Fig.~\ref{fig:Thermic2}, 
where we present the finite-temperature behaviour of the QFI for different system sizes $\sites$ at $\theta=\thetac$.
The finite-size scaling $F_Q\big[\hat{\rho}_T,\hat{O}_x\big]/\sites \sim \sites^{3/4}$ provides a good data collapse 
within the low-temperature regime $T \ll J/\sites$,
while a non universal behaviour holds at higher temperatures.
Since at criticality the gap vanishes as $\min_k\epsilon_k\sim\sites^{-1}$ for all values of $\alpha$ 
(the redefinition of the Hamiltonian in terms of $\theta$ changes the scaling reported in paragraph~\ref{subsec:quasiparticle} 
only for a factor $\sqrt{2}$ in the prefactor), it is intuitive that the system effectively lies in the ground state if $T\ll\DeltaE=\min_k\epsilon_k\sim\sites^{-1}$.

\section*{Summary}
We have shown how the quantum Fisher information can be used to detect multipartite entanglement 
among the fermionic sites in the ground state of the Kitaev chain.
Topologically nontrivial phases, characterized by nonvanishing winding numbers, are recognized by 
a superextensive scaling of the quantum Fisher information,
meaning a divergence of multipartite entanglement when increasing the chain size.
Since topological phases cannot be portrayed according to the standard Ginzburg-Landau equipment 
and no local order parameter serves as a guide,  
the choice of the observables plays a crucial and delicate role: 
while the quantum Fisher information relative to local operators is unable to properly capture entanglement in this model, 
we justified the use of a class of nonlocal operators showing long-range correlations.
%%%
An extension of this procedure to finite temperature reveals that witnessed entanglement is preserved up to high temperatures, 
defined by the tight-binding lattice. Especially, divergent multipartiteness can survive at low temperature.

Furthermore, we have demonstrated that the occurring topological phase transitions are sharply marked by 
the divergence of the derivative of the quantum Fisher information with respect to different control parameters, 
even when the transition is not associated to a closing gap in the excitation spectrum. 
Then, the quantum Fisher information is a sensitive detector for any change in topology, 
at least in the simple one-dimensional model investigated.

All these results provide a clear evidence of multipartite entanglement in topological free-fermion models. 
Its analysis and extraction poses both theoretical and experimental challenges for future prospects. 
Contriving some clever way to exploit the large amount of entanglement at low temperature 
can be extremely useful for enhanced metrological applications; 
nevertheless, this is a demanding task, since the implementation of nonlocal operators on the dichotomous variable 
presence/absence of a fermion on each site is yet far to be actualized.
Also, we believe that looking for possible rigorous relations between the topological invariant
and multipartite entanglement -- here only disclosed at a phenomenological level -- can be greatly intriguing 
from a fundamental theoretical point of view.

\thispagestyle{empty}

\chapter{Outlook and conclusions}
In recent years, the considerable amount of knowledge accumulated in the study of quantum systems 
with the tools and from the perspective of quantum information science led to a fundamental as well as technological revolution: 
entanglement -- the quintessence of quantum world -- has turned from a conceptually troublesome source of paradoxes
into a useful resource, that allows to outperform classical tasks in information processing or even accomplish tasks that cannot be 
carried out through classical means. This change of perspective justifies the growing attention it has been attracting.

Several clues suggest that entanglement can also be intimately related to quantum phase transitions.
In this work, we have selected some representative models hosting quantum phase transitions
and analyzed their ground states and thermal states under the light of multipartite entanglement 
witnessed by the quantum Fisher information. 
Since detailed conclusions on each model have been presented in the ending sections 
of chapters~\ref{ch:Ising}, \ref{ch:LMG} and \ref{ch:Kitaev},
we summarize here the most important results only. 

We found that the quantum Fisher information at zero temperature is a reliable detector of criticality: 
singular behaviour in its rate of variation when tuning the parameter driving the transition 
sharply marks the boundary between different phases, 
not only in symmetry-breaking first-order and second-order transitions, but also in topological transitions.
This remarkable feature suggests an enhanced susceptibility of multipartite entanglement 
to any minute change of the driving parameter across the quantum critical points, 
as long as the observable employed as a probe recognizes the emerging order.
The order arises either from a spontaneous symmetry breaking or from a symmetry-protected topological invariance:
in the former case, we showed that the order parameter plays the role of the optimal operator for probing multipartite entanglement,
whereas in the latter case a wise model-dependent ansatz must be formulated according to the nature of the phase transition itself.
Consequently, our work addressed an open problem in the literature, namely the detection of multipartite entanglement 
at topological transitions, and provided an extension of the standard protocol to nonlocal probe operators
for characterizing topological quantum phases.

For systems where the identification of the local order parameter is a difficult task, because the broken symmetry is unknown,
or even impossible, as in topological systems, no systematic method for identifying the optimal probe operator has been established yet, 
and its choice still is a theoretical open issue.
This is probably the major drawback in using the quantum Fisher information as an entanglement witness, 
especially if compared to the metric approach based on the fidelity susceptibility -- 
which does not directly quantify entanglement, though.

Nevertheless, once the suitable probe operator has been singled out -- based on some \emph{a priori} knowledge of the system -- 
the quantum Fisher information not only signals the onset of a dominant order, 
but also discerns the whole ordered phase by means of a superextensive scaling: 
the size of the largest cluster of entangled particles becomes larger and larger when the system size increases. 
In models exhibiting a symmetry breaking, this finding is naturally related to the high degree of statistical distinguishability 
of the ground state under collective local unitary transformations.
In topological models, the result is probably more astonishing because it implies that 
multipartite entanglement is able to distinguish between trivial phases and topological phases.

Thermal fluctuations and noise pose a significant threat to the preparation, protection and usage of highly-entangled states
and constitute a big challenge for future quantum technologies. 
We showed that thermal states of paradigmatic spin models contains intensive multipartiteness around the critical points: 
entanglement is outstandingly robust in the disordered phase, 
where it is protected by the energy gap and survives up to high temperatures,
while it is extremely fragile to any perturbation within the ordered phase.
On the contrary, the extensive degree of multipartiteness in the nontrivial phases of topological free-fermion models 
is a potential resource exploitable even at finite temperature: 
it may eventually pave the way towards enhanced topological quantum metrology, 
providing an estimation of Hamiltonian coupling constants that is accurate beyond any classical limit 
and robust against local noise sources.

The study of the thermal region surrounding quantum critical points 
also revealed the existence of a universal decay law of multipartite entanglement detected by the quantum Fisher information 
at sufficiently low temperature.
This observation suggests the exciting conjecture that, for a generic model, the quantum critical region on the phase diagram
-- where criticality strongly affects the behaviour of the system --
might be related to (or emerge from) the underlying entanglement content. 
Such a guess makes no claim to be definitive at all and a conclusive understanding is currently out of our reach.
\\[0pt]

{\sl \noindent To sum up, a direct connection between multipartite entanglement and quantum phase transitions 
-- two somewhat abstract though experimentally accessible concepts -- has been established and explored. 
We gathered evidences that the phase diagram of simple one-dimensional critical models, 
both at zero and finite temperature, can be portrayed by dint of the scaling behaviour of the 
multipartite entanglement witnessed by the quantum Fisher information: 
in fact, it successfully recognizes different symmetries and different topologies.
Moreover, we found that the range of interactions among the microscopic constituents does not qualitatively modify the overall scenario.
%Actually, we demonstrated that the quantum Fisher information can discriminate between short range and long range  
%prior a suitable choice of the probe operator
Yet, at the very moment we have no clue about the fundamental link between entanglement and topological invariant (if it exists at all),
neither we know how to generalize our approach to an arbitrary system whose inner order is captured by nonlocal observables.
Fairly, this is just the first step of the research in this direction. 
We hope our results will help in opening new routes for the investigation and characterization of quantum critical matter 
and shall be of some interest to the community working on strongly-correlated systems and quantum information.}

\backmatter
\phantomsection

\chapter{Acknowledgements}
%\addcontentsline{toc}{chapter}{Acknowledgements}

\noindent Foremost, I wish to thank all the people I worked with throughout this three-year experience.
%First of all, 
%coloro che mi hanno seguito durante ...  e aiutato astendere la tesi.
\\[6pt]
I thank Augusto Smerzi for supervising the entire project and encouraging me 
in the most fervid periods as much as in the idle ones: 
his zestful approach to new problems and his eager way of looking for in-depth large-scale insights
stimulated me to refine my physical intuition.
\\[6pt]
%His way to approach new problems with zest 
%and the eager way of both asking questions and looking for intuitive in-depth large-scale/ answers insights
%has stimulated me to 
%and strongly based on physical intuition to 
%targeting / covering far-reaching solutions problems
I thank Alessandro Cuccoli 
% for the readiness to act as a second supervisor, 
for helping me to understand relevant conceptual points 
and for the exceptional care he devoted to % into 
proofreading the thesis.
% when writing down my ... during the last part of the of the present work
\\[6pt]
Many thanks to Luca Pezz\`e for his precious daily support: 
%his explanation
%he explained me several aspect ;
he provided invaluable aid % in solving practical problems
in writing numerical codes, checking out analytical calculations and clarifying several aspects of the present work. 
%several details / aspects of my work that now I 
%throw (some) light on (something)
\\[6pt]
I also acknowledge worthy discussions with Luca Lepori, who collaborated with the group during the third year
and introduced me to topological transitions.
\\[6pt]
Finally, I am in debt of gratitude to Fabio Ortolani,
who wrote the DMRG code fruitfully employed in the investigation of the Ising chain. 
%the use of / employing / the fruitful employment of his DMRG code. 
\\[12pt]

%He transmitted me 
%His zestful euristic way to approach unknown new problems / with enthusiasm  eagerness zest passion vigour vim keenness vitality
%and the way of asking questions and lookign for analyzing the general aspect from bird's eye 
%stimulated me to 
%His approach to physics
%and strongly based on physical intuition to 
%has taught me 
%inspired me and was a good education for me as a physicist??!
%
%I thank Alessandro Cuccoli for helping me to understand several relevant conceptual aspects 
%during the writing /
%when writing down my ... during the last part of the of the present work
%and for his scrupulous / careful and the exceptional care he devoted into proofreading the thesis.
%
%Many thanks to Luca Pezz\`e for his precious helpfulness, patience and the material invaluable help 
%in writing codes and checking out the analytical calculations.
%He provided invaluable assistance/support in problems
%and for sproning me , day by day / day after day / every day.
%
%I owe my thanks also to 
%I also acknowledge the worthy helpfulness shown by Luca Lepori, with whom I collaborated during the last year.
%Finally, I am in debt of gratitude to Fabio Ortolani for the use of / employing / the fruitful employment of his DMRG code. 
\noindent Furthermore, my gratefulness is due to all the fellows % companions colleagues 
that shared some time with me at the institute on the hill of Arcetri, where most of the research activity was carried out.
\\[6pt]
I am doubly thankful to Manuel, for his friendly company as well as the interesting discussions we had about Physics, % and much more.
especially when I approached the new topic of phase transitions for the first time.
%helping me with many questions.
\\[6pt]
I owe my thanks to Roberto and Pierfrancesco, who contributed to create an enjoyable environment % (at the institute)
and kindly advised me in delicate moments.
\\[6pt]
A special mention goes to Giacomo, with whom I shared the office during the first year: 
his curiosity always spurred me to learn more;  
in addition, his admirable care for graphics played a precious part in the overall % aesthetic 
appearance of the present thesis.
\\[6pt]
I also feel grateful to the people that, for one reason or another, visited the institute over time: 
especially, I recall with pleasure the joyful conversations with Fabian, Polina, Beatrice, Tommaso and Artur.
% during the lunch time and not only.
\\[6pt]
Thanks to all the companions at the Astrophysical Observatory in Arcetri for welcoming %/accepting/receiving/reception
me in their amusing clique.
% / their company.
\\[24pt]
%It has been a good time to chat about Physics %and everything else /
%and much more with you all.

%his love for ...
%his curiosity always spurred/stimulated/sparked/led to me to learn more;  his love for ...
%and his constant desire/longing/yearning for 
%we shared 

%   Special/Huge are reserved for
%
%precious hand
%
%for interesting discussion 
%constant support and help 
%
%from whom I learnt many new different things
%
%everyone
%
%special
%
%I also ... 
%
%I also greatly thank the other students who shared in this experience
%In particular, I mention for the joyful
%
%I thank you all for the many amusement / entertainment / good time that have kept me sane through the PhD and in the place...
%
%I express my gratefulness 
%

%I remind 
%
% 
%I greet all the inhabitants of the istitute.
%
%Thank you for the your joyful company and support.
%
%Ricordo con piacere le distrazioni e le lazy discussions con Sophie e Tomasz durante le pause pranzo al Qstar.
%
%Tommaso
%Beatrice
%Michele
%(Jagoba)
%(Safoura) 
%(Sophie)
%
%A person 
%
%All of you contributed to 
%
%Lastly,  I wish to thank my family 

\noindent Being a PhD student can be troublesome at times: thanks to everyone that understood it and helped me; 
and thanks to everyone that did not understand it but wanted to talk about it anyway.
\\[6pt]
Thanks to the colleagues at the Physics Departments for the happy moments shared together: %shared after seminars and during poster sessions:
it has been a good time to chat about Physics %and everything else /
and much more with you. 
\\[6pt]
A person without whom this thesis there would probably not have been is Michele, 
%who indubitably/certainly contributed  to probe first 
who accompanied me into the field of many-body physics and quantum information for the first time. 
%some years ago.
\\[6pt]
Most of the entertainment as a PhD student certainly came from visiting new places: % and discovering new things: 
in particular, I acknowledge delightful travelling experiences with Giulia and Lorenzo.
\\[6pt]
During schools and conferences in Italy and abroad I met fantastic people, 
whose enthusiasm and critical sense motivated me to appreciate % more my work and its value.
my work and its value even more.
\\[6pt]
I thank the whole crew of the organizing committee of the PhD Day in Florence: 
the energy we put in realizing the event turned out to be a great source of fun that contributed to keep me sane. 
\\[6pt]
Lastly, I would like to reserve a special word of gratitude to all the friends and companions for the generous help over these years:
chatting with you % has been more helpful than 
has been worth as much as thousands of lessons.
\\[12pt]

\begin{figure}[!h]
\begin{center}
\includegraphics[width=0.2\textwidth]{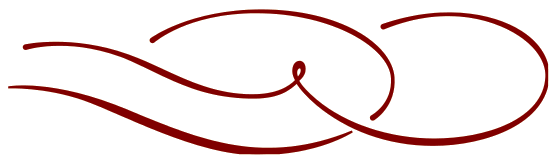}
\end{center}
\end{figure}

%\begin{figure}[!b]
%\begin{center}
%\includegraphics[width=0.1\textwidth]{./Ringraziamenti/ornamento6.png}
%\end{center}
%\end{figure}
%
%\begin{figure}[!b]
%\begin{center}
%\includegraphics[width=0.1\textwidth]{./Ringraziamenti/ornamento5.png}
%\end{center}
%\end{figure}
%
%
%\begin{figure}[!b]
%\begin{center}
%\includegraphics[width=0.3\textwidth]{./Ringraziamenti/ornamento4.png}
%\end{center}
%\end{figure}
%
%\begin{figure}[!b]
%\begin{center}
%\includegraphics[width=0.3\textwidth]{./Ringraziamenti/ornamento2.png}
%\end{center}
%\end{figure}
%
%\begin{figure}[!b]
%\begin{center}
%\includegraphics[width=0.3\textwidth]{./Ringraziamenti/ornamento1.png}
%\end{center}
%\end{figure}
%
%\hrulefill\hspace{0.2cm} 
%{\Huge $\star$}
%\hspace{0.2cm} \hrulefill

\end{document}